\documentclass[preprint2]{aastex62}
\usepackage{natbib}
\usepackage{graphicx}
\usepackage{color}
\usepackage{xspace}
\usepackage{xcolor}
\usepackage{import}

\newcommand{\referee}{} 
\newcommand{\nc}[2]{\newcommand{#1}{\ensuremath{#2}\xspace}}

\nc{\Kepler}{ \textit{Kepler} }
\nc{\Ktwo}{ \textit{K2} }
\nc{\Kp}{ \textit{Kp} }
\nc{\SpecMatch}{\mathrm{SpecMatch}} 
\nc{\dAICc}{\Delta\mathrm{AICc}}

\nc{\rsun}{R_{\odot}}
\nc{\msun}{M_{\odot}}
\nc{\rearth}{R_{\oplus}}
\nc{\mearth}{M_{\oplus}}
\nc{\fearth}{F_{\oplus}}
\nc{\arstar}{a/R_\star}
\nc{\rprstar}{R_P/R_\star}
\nc{\Rp}{R_P}
\nc{\Mp}{M_P}
\nc{\msini}{M_P \sin i}
\nc{\Rstar}{R_\star} 
\nc{\Mstar}{M_\star}

\newcommand{\shk}{\ensuremath{S_\mathrm{HK}}}
\newcommand{\rphk}{\ensuremath{R'_{\mbox{\scriptsize HK}}}}
\newcommand{\lrphk}{\ensuremath{\log{\rphk}}}

\nc{\teff}{T_{\rm eff}}
\nc{\logg}{\log{g}}
\nc{\feh}{[\mbox{Fe}/\mbox{H}]}
\nc{\vsini}{V \sin i}

\nc{\kms}{\text{km\,s}^{-1}}
\nc{\ms}{\text{m\,s}^{-1}}
\nc{\msy}{\text{m\,s}^{-1}\text{yr}^{-1}}
\nc{\gmc}{\text{g\,cm}^{-3}}

\newcommand{\jjhiPNAMEone}{K2-222 b\xspace}
\newcommand{\fcabPNAMEone}{K2-236 b\xspace}
\newcommand{\eidfPNAMEone}{EPIC 229004835 b\xspace}
\newcommand{\hehhPNAMEone}{K2-277 b\xspace}
\newcommand{\iahiPNAMEone}{K2-261 b\xspace}
\newcommand{\aiggPNAMEone}{K2-100 b\xspace}
\newcommand{\jgjjPNAMEone}{K2-31 b\xspace}
\newcommand{\hhedPNAMEone}{K2-39 b\xspace}
\newcommand{\befbPNAMEtwo}{K2-229 c\xspace}
\newcommand{\befbPNAMEone}{K2-229 b\xspace}

\newcommand{\eaccPNAMEone}{K2-111 b\xspace}
\newcommand{\dcijPNAMEone}{K2-99 b\xspace}
\newcommand{\bejgPNAMEone}{K2-265 b\xspace}
\newcommand{\bcgdPNAMEtwo}{K2-38 c\xspace}
\newcommand{\bcgdPNAMEone}{K2-38 b\xspace}
\newcommand{\fffdPNAMEone}{K2-73 b\xspace}
\newcommand{\fffdPNAMEtwo}{K2-73 c\xspace}
\newcommand{\dcbjPNAMEone}{K2-66 b\xspace}
\newcommand{\ddeiPNAMEone}{K2-36 b\xspace}
\newcommand{\ddeiPNAMEtwo}{K2-36 c\xspace}
\newcommand{\fdijPNAMEone}{K2-105 b\xspace}
\newcommand{\gafePNAMEone}{K2-214 b\xspace}
\newcommand{\bhiiPNAMEone}{K2-220 b\xspace}
\newcommand{\bbggPNAMEone}{K2-110 b\xspace}
\newcommand{\dbfaPNAMEone}{WASP-47 e\xspace}
\newcommand{\dbfaPNAMEtwo}{WASP-47 b\xspace}
\newcommand{\dbfaPNAMEthree}{WASP-47 d\xspace}
\newcommand{\dbfaPNAMEfour}{WASP-47 c\xspace}
\newcommand{\ccdhPNAMEone}{K2-79 b\xspace}
\newcommand{\eicdPNAMEone}{K2-106 b\xspace}
\newcommand{\eicdPNAMEtwo}{K2-106 c\xspace}
\newcommand{\gcidPNAMEone}{EPIC 213546283 b\xspace}
\newcommand{\baeiPNAMEone}{EPIC 245991048 b\xspace}
\newcommand{\bjiePNAMEone}{K2-32 b\xspace}
\newcommand{\bjiePNAMEtwo}{K2-32 c\xspace}
\newcommand{\bjiePNAMEthree}{K2-32 d\xspace}
\newcommand{\ggacPNAMEtwo}{K2-62 c\xspace}
\newcommand{\ggacPNAMEone}{K2-62 b\xspace}
\newcommand{\egijPNAMEtwo}{K2-189 c\xspace}
\newcommand{\egijPNAMEone}{K2-189 b\xspace}
\newcommand{\hadfPNAMEone}{K2-10 b\xspace}
\newcommand{\hidfPNAMEone}{EPIC 201357835 b\xspace}
\newcommand{\bebbPNAMEone}{K2-216 b\xspace}
\newcommand{\ecdiPNAMEone}{K2-280 b\xspace}
\newcommand{\gedgPNAMEone}{K2-37 b\xspace}
\newcommand{\gedgPNAMEtwo}{K2-37 c\xspace}
\newcommand{\gedgPNAMEthree}{K2-37 d\xspace}
\newcommand{\jgbhPNAMEone}{K2-180 b\xspace}
\newcommand{\ghijPNAMEone}{K2-27 b\xspace}
\newcommand{\fdecPNAMEone}{K2-181 b\xspace}
\newcommand{\deffPNAMEone}{EPIC 245943455 b\xspace}
\newcommand{\eiadPNAMEone}{K2-61 b\xspace}
\newcommand{\ifgjPNAMEone}{K2-121 b\xspace}
\newcommand{\cffcPNAMEone}{K2-18 b\xspace}
\newcommand{\jjhiPERone}{\ensuremath{15.3863\pm0.0014}\xspace}
\newcommand{\fcabPERone}{\ensuremath{19.4910\pm0.0007}\xspace}
\newcommand{\eidfPERone}{\ensuremath{16.1388\pm0.0016}\xspace}
\newcommand{\hehhPERone}{\ensuremath{6.32677\pm0.00018}\xspace}
\newcommand{\iahiPERone}{\ensuremath{11.63395\pm0.00021}\xspace}
\newcommand{\aiggPERone}{\ensuremath{1.673890\pm0.000017}\xspace}
\newcommand{\jgjjPERone}{\ensuremath{1.25784906\pm0.00000098}\xspace}
\newcommand{\hhedPERone}{\ensuremath{4.60547\pm0.00047}\xspace}
\newcommand{\befbPERtwo}{\ensuremath{8.3262\pm0.0015}\xspace}
\newcommand{\befbPERone}{\ensuremath{0.584272\pm0.000017}\xspace}

\newcommand{\eaccPERone}{\ensuremath{5.35232\pm0.00035}\xspace}
\newcommand{\dcijPERone}{\ensuremath{18.24901\pm0.00062}\xspace}
\newcommand{\bejgPERone}{\ensuremath{2.36906\pm0.00008}\xspace}
\newcommand{\bcgdPERtwo}{\ensuremath{10.56104\pm0.00059}\xspace}
\newcommand{\bcgdPERone}{\ensuremath{4.01668\pm0.00032}\xspace}
\newcommand{\fffdPERone}{\ensuremath{7.49556\pm0.00029}\xspace}
\newcommand{\fffdPERtwo}{\ensuremath{1000\pm100}\xspace}
\newcommand{\dcbjPERone}{\ensuremath{5.06939\pm0.00053}\xspace}
\newcommand{\ddeiPERone}{\ensuremath{1.422586\pm0.000029}\xspace}
\newcommand{\ddeiPERtwo}{\ensuremath{5.340786\pm0.000073}\xspace}
\newcommand{\fdijPERone}{\ensuremath{8.26726\pm0.00018}\xspace}
\newcommand{\gafePERone}{\ensuremath{8.59656\pm0.00051}\xspace}
\newcommand{\bhiiPERone}{\ensuremath{13.68186\pm0.00095}\xspace}
\newcommand{\bbggPERone}{\ensuremath{13.86368\pm0.00019}\xspace}
\newcommand{\dbfaPERone}{\ensuremath{0.789570\pm0.000034}\xspace}
\newcommand{\dbfaPERtwo}{\ensuremath{4.159152\pm0.000013}\xspace}
\newcommand{\dbfaPERthree}{\ensuremath{9.03101\pm0.00037}\xspace}
\newcommand{\dbfaPERfour}{\ensuremath{592.5^{+2.6}_{-2.5}}\xspace}
\newcommand{\ccdhPERone}{\ensuremath{10.99497\pm0.00045}\xspace}
\newcommand{\eicdPERtwo}{\ensuremath{13.33918\pm0.00099}\xspace}
\newcommand{\eicdPERone}{\ensuremath{0.571282\pm0.000015}\xspace}
\newcommand{\bggePERone}{\ensuremath{10.13693\pm0.00046}\xspace}
\newcommand{\gcidPERone}{\ensuremath{9.77058\pm0.00035}\xspace}
\newcommand{\baeiPERone}{\ensuremath{8.58289\pm0.00041}\xspace}
\newcommand{\bjiePERone}{\ensuremath{8.99196\pm0.00007}\xspace}
\newcommand{\bjiePERtwo}{\ensuremath{20.66093\pm0.00078}\xspace}
\newcommand{\bjiePERthree}{\ensuremath{31.7169\pm0.0011}\xspace}
\newcommand{\ggacPERtwo}{\ensuremath{16.19697\pm0.00065}\xspace}
\newcommand{\ggacPERone}{\ensuremath{6.67199\pm0.00018}\xspace}
\newcommand{\egijPERtwo}{\ensuremath{6.67932\pm0.00021}\xspace}
\newcommand{\egijPERone}{\ensuremath{2.58812\pm0.00011}\xspace}
\newcommand{\hadfPERone}{\ensuremath{19.30553\pm0.00047}\xspace}
\newcommand{\hidfPERone}{\ensuremath{11.89376\pm0.00083}\xspace}
\newcommand{\bebbPERone}{\ensuremath{2.174789\pm0.000056}\xspace}
\newcommand{\ecdiPERone}{\ensuremath{19.89500\pm0.00035}\xspace}
\newcommand{\gedgPERone}{\ensuremath{4.44340\pm0.00032}\xspace}
\newcommand{\gedgPERtwo}{\ensuremath{6.42966\pm0.00018}\xspace}
\newcommand{\gedgPERthree}{\ensuremath{14.09229\pm0.00055}\xspace}
\newcommand{\jgbhPERone}{\ensuremath{8.8656\pm0.0004}\xspace}
\newcommand{\ghijPERone}{\ensuremath{6.771347\pm0.000059}\xspace}
\newcommand{\fdecPERone}{\ensuremath{6.89410\pm0.00043}\xspace}
\newcommand{\deffPERone}{\ensuremath{6.33932\pm0.00013}\xspace}
\newcommand{\eiadPERone}{\ensuremath{2.57336\pm0.00016}\xspace}
\newcommand{\ifgjPERone}{\ensuremath{5.185738\pm0.000014}\xspace}
\newcommand{\cffcPERone}{\ensuremath{32.94112\pm0.00081}\xspace}
\newcommand{\jjhiTCone}{\ensuremath{2457399.0652\pm0.0042}\xspace}
\newcommand{\fcabTCone}{\ensuremath{2457158.82659\pm0.00092}\xspace}
\newcommand{\eidfTCone}{\ensuremath{2457613.7661\pm0.0019}\xspace}
\newcommand{\hehhTCone}{\ensuremath{2457221.2301\pm0.0012}\xspace}
\newcommand{\iahiTCone}{\ensuremath{2457906.84055\pm0.00054}\xspace}
\newcommand{\aiggTCone}{\ensuremath{2457140.71966\pm0.00042}\xspace}
\newcommand{\jgjjTCone}{\ensuremath{2456893.598952\pm0.000036}\xspace}
\newcommand{\hhedTCone}{\ensuremath{2456985.4268\pm0.0038}\xspace}
\newcommand{\befbTCtwo}{\ensuremath{2457611.3227\pm0.0033}\xspace}
\newcommand{\befbTCone}{\ensuremath{2457605.08611\pm0.00078}\xspace}

\newcommand{\eaccTCone}{\ensuremath{2457067.9628\pm0.0024}\xspace}
\newcommand{\dcijTCone}{\ensuremath{2457233.8255\pm0.0012}\xspace}
\newcommand{\bejgTCone}{\ensuremath{2456981.6450\pm0.0014}\xspace}
\newcommand{\bcgdTCtwo}{\ensuremath{2456900.4751\pm0.0021}\xspace}
\newcommand{\bcgdTCone}{\ensuremath{2456896.8707\pm0.0035}\xspace}
\newcommand{\fffdTCone}{\ensuremath{2456987.6726\pm0.0012}\xspace}
\newcommand{\fffdTCtwo}{\ensuremath{2456905^{+62}_{-99}}\xspace}
\newcommand{\dcbjTCone}{\ensuremath{2456984.0075\pm0.0037}\xspace}
\newcommand{\ddeiTCone}{\ensuremath{2456827.96295\pm0.00068}\xspace}
\newcommand{\ddeiTCtwo}{\ensuremath{2456812.84096\pm0.00059}\xspace}
\newcommand{\fdijTCone}{\ensuremath{2457147.98867\pm0.00079}\xspace}
\newcommand{\gafeTCone}{\ensuremath{2457396.6009\pm0.0023}\xspace}
\newcommand{\bhiiTCone}{\ensuremath{2457401.2729\pm0.0025}\xspace}
\newcommand{\bbggTCone}{\ensuremath{2457233.73861\pm0.00048}\xspace}
\newcommand{\dbfaTCone}{\ensuremath{2456981.3436\pm0.0014}\xspace}
\newcommand{\dbfaTCtwo}{\ensuremath{2456982.9772\pm0.0001}\xspace}
\newcommand{\dbfaTCthree}{\ensuremath{2456988.3079\pm0.0011}\xspace}
\newcommand{\dbfaTCfour}{\ensuremath{2455993^{+6}_{-7}}\xspace}
\newcommand{\ccdhTCone}{\ensuremath{2457070.2428\pm0.0015}\xspace}
\newcommand{\eicdTCtwo}{\ensuremath{2457405.7330\pm0.0019}\xspace}
\newcommand{\eicdTCone}{\ensuremath{2457393.4405\pm0.0012}\xspace}
\newcommand{\bggeTCone}{\ensuremath{2457145.9796\pm0.0016}\xspace}
\newcommand{\gcidTCone}{\ensuremath{2457312.1249\pm0.0012}\xspace}
\newcommand{\baeiTCone}{\ensuremath{2457742.782\pm0.002}\xspace}
\newcommand{\bjieTCone}{\ensuremath{2456909.91884\pm0.00023}\xspace}
\newcommand{\bjieTCtwo}{\ensuremath{2456961.4065\pm0.0012}\xspace}
\newcommand{\bjieTCthree}{\ensuremath{2456903.7860\pm0.0014}\xspace}
\newcommand{\ggacTCtwo}{\ensuremath{2456991.5453\pm0.0013}\xspace}
\newcommand{\ggacTCone}{\ensuremath{2456982.685\pm0.001}\xspace}
\newcommand{\egijTCtwo}{\ensuremath{2457223.4182\pm0.0013}\xspace}
\newcommand{\egijTCone}{\ensuremath{2457222.1466\pm0.0019}\xspace}
\newcommand{\hadfTCone}{\ensuremath{2456819.57944\pm0.00089}\xspace}
\newcommand{\hidfTCone}{\ensuremath{2457611.3378\pm0.0017}\xspace}
\newcommand{\bebbTCone}{\ensuremath{2457394.0417\pm0.0011}\xspace}
\newcommand{\ecdiTCone}{\ensuremath{2457327.47611\pm0.00046}\xspace}
\newcommand{\gedgTCone}{\ensuremath{2456893.6753\pm0.0033}\xspace}
\newcommand{\gedgTCtwo}{\ensuremath{2456898.8548\pm0.0012}\xspace}
\newcommand{\gedgTCthree}{\ensuremath{2456907.2331\pm0.0014}\xspace}
\newcommand{\jgbhTCone}{\ensuremath{2457143.3945\pm0.0023}\xspace}
\newcommand{\ghijTCone}{\ensuremath{2456812.84470\pm0.00036}\xspace}
\newcommand{\fdecTCone}{\ensuremath{2457143.7948\pm0.0025}\xspace}
\newcommand{\deffTCone}{\ensuremath{2457741.79456\pm0.00088}\xspace}
\newcommand{\eiadTCone}{\ensuremath{2456983.2140\pm0.0022}\xspace}
\newcommand{\ifgjTCone}{\ensuremath{2457143.5607\pm0.0001}\xspace}
\newcommand{\cffcTCone}{\ensuremath{2456836.17187\pm0.00057}\xspace}
\newcommand{\jjhiRPone}{\ensuremath{2.41^{+0.17}_{-0.08}}\xspace}
\newcommand{\fcabRPone}{\ensuremath{5.50^{+0.19}_{-0.08}}\xspace}
\newcommand{\eidfRPone}{\ensuremath{2.18^{+0.13}_{-0.07}}\xspace}
\newcommand{\hehhRPone}{\ensuremath{2.23^{+0.14}_{-0.06}}\xspace}
\newcommand{\iahiRPone}{\ensuremath{10.32^{+0.17}_{-0.08}}\xspace}
\newcommand{\aiggRPone}{\ensuremath{3.57^{+0.10}_{-0.04}}\xspace}
\newcommand{\jgjjRPone}{\ensuremath{46\pm8}\xspace}
\newcommand{\hhedRPone}{\ensuremath{6.40^{+0.41}_{-0.17}}\xspace}
\newcommand{\befbRPtwo}{\ensuremath{2.04^{+0.21}_{-0.07}}\xspace}
\newcommand{\befbRPone}{\ensuremath{1.260^{+0.082}_{-0.034}}\xspace}

\newcommand{\eaccRPone}{\ensuremath{1.184^{+0.056}_{-0.034}}\xspace}
\newcommand{\dcijRPone}{\ensuremath{12.37^{+0.18}_{-0.10}}\xspace}
\newcommand{\bejgRPone}{\ensuremath{1.676^{+0.084}_{-0.039}}\xspace}
\newcommand{\bcgdRPtwo}{\ensuremath{2.18^{+0.15}_{-0.06}}\xspace}
\newcommand{\bcgdRPone}{\ensuremath{1.55^{+0.11}_{-0.05}}\xspace}
\newcommand{\fffdRPone}{\ensuremath{2.58^{+0.13}_{-0.06}}\xspace}
\newcommand{\fffdRPtwo}{---\xspace}
\newcommand{\dcbjRPone}{\ensuremath{2.75^{+0.16}_{-0.10}}\xspace}
\newcommand{\ddeiRPone}{\ensuremath{1.291^{+0.081}_{-0.038}}\xspace}
\newcommand{\ddeiRPtwo}{\ensuremath{2.41^{+0.25}_{-0.07}}\xspace}
\newcommand{\fdijRPone}{\ensuremath{3.40^{+0.12}_{-0.05}}\xspace}
\newcommand{\gafeRPone}{\ensuremath{2.51^{+0.14}_{-0.07}}\xspace}
\newcommand{\bhiiRPone}{\ensuremath{2.37^{+0.14}_{-0.07}}\xspace}
\newcommand{\bbggRPone}{\ensuremath{2.558^{+0.085}_{-0.037}}\xspace}
\newcommand{\dbfaRPone}{\ensuremath{1.79^{+0.18}_{-0.06}}\xspace}
\newcommand{\dbfaRPtwo}{\ensuremath{12.251^{+0.043}_{-0.033}}\xspace}
\newcommand{\dbfaRPthree}{\ensuremath{3.58^{+0.32}_{-0.10}}\xspace}
\newcommand{\dbfaRPfour}{---\xspace}
\newcommand{\ccdhRPone}{\ensuremath{3.99^{+0.11}_{-0.07}}\xspace}
\newcommand{\eicdRPtwo}{\ensuremath{3.01^{+0.15}_{-0.09}}\xspace}
\newcommand{\eicdRPone}{\ensuremath{1.87^{+0.11}_{-0.05}}\xspace}
\newcommand{\bggeRPone}{\ensuremath{4.93^{+0.13}_{-0.07}}\xspace}
\newcommand{\gcidRPone}{\ensuremath{3.4^{+0.2}_{-0.1}}\xspace}
\newcommand{\baeiRPone}{\ensuremath{2.33^{+0.12}_{-0.06}}\xspace}
\newcommand{\bjieRPone}{\ensuremath{5.038^{+0.096}_{-0.045}}\xspace}
\newcommand{\bjieRPtwo}{\ensuremath{3.01^{+0.13}_{-0.06}}\xspace}
\newcommand{\bjieRPthree}{\ensuremath{3.33^{+0.22}_{-0.07}}\xspace}
\newcommand{\ggacRPtwo}{\ensuremath{1.99^{+0.19}_{-0.07}}\xspace}
\newcommand{\ggacRPone}{\ensuremath{2.02^{+0.17}_{-0.06}}\xspace}
\newcommand{\egijRPtwo}{\ensuremath{2.31^{+0.14}_{-0.05}}\xspace}
\newcommand{\egijRPone}{\ensuremath{1.403^{+0.079}_{-0.043}}\xspace}
\newcommand{\hadfRPone}{\ensuremath{3.773^{+0.088}_{-0.048}}\xspace}
\newcommand{\hidfRPone}{\ensuremath{2.67^{+0.11}_{-0.05}}\xspace}
\newcommand{\bebbRPone}{\ensuremath{1.66^{+0.13}_{-0.05}}\xspace}
\newcommand{\ecdiRPone}{\ensuremath{7.46^{+0.11}_{-0.05}}\xspace}
\newcommand{\gedgRPone}{\ensuremath{1.39^{+0.10}_{-0.05}}\xspace}
\newcommand{\gedgRPtwo}{\ensuremath{2.41^{+0.15}_{-0.06}}\xspace}
\newcommand{\gedgRPthree}{\ensuremath{2.31^{+0.21}_{-0.07}}\xspace}
\newcommand{\jgbhRPone}{\ensuremath{2.13^{+0.11}_{-0.05}}\xspace}
\newcommand{\ghijRPone}{\ensuremath{4.74^{+0.13}_{-0.06}}\xspace}
\newcommand{\fdecRPone}{\ensuremath{2.69^{+0.21}_{-0.09}}\xspace}
\newcommand{\deffRPone}{\ensuremath{3.9^{+0.2}_{-0.1}}\xspace}
\newcommand{\eiadRPone}{\ensuremath{1.92^{+0.11}_{-0.05}}\xspace}
\newcommand{\ifgjRPone}{\ensuremath{7.16^{+0.12}_{-0.06}}\xspace}
\newcommand{\cffcRPone}{\ensuremath{2.461^{+0.079}_{-0.045}}\xspace}
\newcommand{\jjhiMPone}{\ensuremath{5.7\pm2.6}\xspace}
\newcommand{\fcabMPone}{\ensuremath{11\pm7}\xspace}
\newcommand{\eidfMPone}{\ensuremath{19.3\pm5.1}\xspace}
\newcommand{\hehhMPone}{\ensuremath{7.4\pm3.3}\xspace}
\newcommand{\iahiMPone}{\ensuremath{56\pm6}\xspace}
\newcommand{\aiggMPone}{\ensuremath{8\pm15}\xspace}
\newcommand{\jgjjMPone}{\ensuremath{551^{+16}_{-17}}\xspace}
\newcommand{\hhedMPone}{\ensuremath{37.6^{+5.3}_{-4.8}}\xspace}
\newcommand{\befbMPtwo}{\ensuremath{7\pm4}\xspace}
\newcommand{\befbMPone}{\ensuremath{2.55\pm0.38}\xspace}

\newcommand{\eaccMPone}{\ensuremath{5.4^{+2.2}_{-2.1}}\xspace}
\newcommand{\dcijMPone}{\ensuremath{287^{+23}_{-22}}\xspace}
\newcommand{\bejgMPone}{\ensuremath{4.6\pm1.5}\xspace}
\newcommand{\bcgdMPtwo}{\ensuremath{7.7\pm2.7}\xspace}
\newcommand{\bcgdMPone}{\ensuremath{6\pm2}\xspace}
\newcommand{\fffdMPone}{\ensuremath{9.2^{+3.8}_{-3.7}}\xspace}
\newcommand{\fffdMPtwo}{\ensuremath{1142^{+53}_{-45}}\xspace}
\newcommand{\dcbjMPone}{\ensuremath{16\pm4}\xspace}
\newcommand{\ddeiMPone}{\ensuremath{5.1^{+4.5}_{-4.4}}\xspace}
\newcommand{\ddeiMPtwo}{\ensuremath{26.0^{+7.8}_{-7.9}}\xspace}
\newcommand{\fdijMPone}{\ensuremath{15.4\pm4.4}\xspace}
\newcommand{\gafeMPone}{\ensuremath{2.4\pm6.5}\xspace}
\newcommand{\bhiiMPone}{\ensuremath{0\pm4}\xspace}
\newcommand{\bbggMPone}{\ensuremath{17\pm3}\xspace}
\newcommand{\dbfaMPone}{\ensuremath{7.06^{+0.71}_{-0.68}}\xspace}
\newcommand{\dbfaMPtwo}{\ensuremath{357\pm11}\xspace}
\newcommand{\dbfaMPthree}{\ensuremath{13.3\pm1.5}\xspace}
\newcommand{\dbfaMPfour}{\ensuremath{394\pm13}\xspace}
\newcommand{\ccdhMPone}{\ensuremath{3.8^{+4.3}_{-4.2}}\xspace}
\newcommand{\eicdMPtwo}{\ensuremath{5.0^{+2.8}_{-2.9}}\xspace}
\newcommand{\eicdMPone}{\ensuremath{8.03^{+0.88}_{-0.85}}\xspace}
\newcommand{\bggeMPone}{\ensuremath{20^{+21}_{-33}}\xspace}
\newcommand{\gcidMPone}{\ensuremath{8\pm9}\xspace}
\newcommand{\baeiMPone}{\ensuremath{8.0\pm5.2}\xspace}
\newcommand{\bjieMPone}{\ensuremath{16.3\pm1.9}\xspace}
\newcommand{\bjieMPtwo}{\ensuremath{5.7^{+2.3}_{-2.4}}\xspace}
\newcommand{\bjieMPthree}{\ensuremath{13.7\pm2.9}\xspace}
\newcommand{\ggacMPtwo}{\ensuremath{1.0^{+5.9}_{-5.8}}\xspace}
\newcommand{\ggacMPone}{\ensuremath{-1.0^{+4.3}_{-4.2}}\xspace}
\newcommand{\egijMPtwo}{\ensuremath{5.0\pm5.5}\xspace}
\newcommand{\egijMPone}{\ensuremath{4.5\pm3.6}\xspace}
\newcommand{\hadfMPone}{\ensuremath{25.2\pm8.4}\xspace}
\newcommand{\hidfMPone}{\ensuremath{15^{+25}_{-26}}\xspace}
\newcommand{\bebbMPone}{\ensuremath{6.1\pm1.5}\xspace}
\newcommand{\ecdiMPone}{\ensuremath{49.0^{+6.5}_{-6.2}}\xspace}
\newcommand{\gedgMPone}{\ensuremath{-0.9^{+5.1}_{-5.2}}\xspace}
\newcommand{\gedgMPtwo}{\ensuremath{5.1^{+5.1}_{-5.2}}\xspace}
\newcommand{\gedgMPthree}{\ensuremath{12.5^{+6.2}_{-6.1}}\xspace}
\newcommand{\jgbhMPone}{\ensuremath{9.4\pm2.2}\xspace}
\newcommand{\ghijMPone}{\ensuremath{30.2^{+3.5}_{-3.6}}\xspace}
\newcommand{\fdecMPone}{\ensuremath{12\pm18}\xspace}
\newcommand{\deffMPone}{\ensuremath{4.4^{+4.8}_{-4.7}}\xspace}
\newcommand{\eiadMPone}{\ensuremath{0.1^{+7.9}_{-7.7}}\xspace}
\newcommand{\ifgjMPone}{\ensuremath{51\pm11}\xspace}
\newcommand{\cffcMPone}{\ensuremath{7.2^{+1.5}_{-1.4}}\xspace}
\newcommand{\jjhiRHOPone}{\ensuremath{2\pm1}\xspace}
\newcommand{\fcabRHOPone}{\ensuremath{0.33^{+0.22}_{-0.21}}\xspace}
\newcommand{\eidfRHOPone}{\ensuremath{11.9^{+4.2}_{-3.5}}\xspace}
\newcommand{\hehhRHOPone}{\ensuremath{4\pm2}\xspace}
\newcommand{\iahiRHOPone}{\ensuremath{0.339^{+0.061}_{-0.052}}\xspace}
\newcommand{\aiggRHOPone}{\ensuremath{0.9\pm1.9}\xspace}
\newcommand{\jgjjRHOPone}{\ensuremath{0.032^{+0.025}_{-0.013}}\xspace}
\newcommand{\hhedRHOPone}{\ensuremath{0.93^{+0.25}_{-0.19}}\xspace}
\newcommand{\befbRHOPtwo}{\ensuremath{5.1^{+3.5}_{-2.9}}\xspace}
\newcommand{\befbRHOPone}{\ensuremath{7.6^{+2.1}_{-1.7}}\xspace}

\newcommand{\eaccRHOPone}{\ensuremath{13.8^{+7.3}_{-5.9}}\xspace}
\newcommand{\dcijRHOPone}{\ensuremath{0.90^{+0.15}_{-0.13}}\xspace}
\newcommand{\bejgRHOPone}{\ensuremath{5\pm2}\xspace}
\newcommand{\bcgdRHOPtwo}{\ensuremath{2.7^{+1.6}_{-1.1}}\xspace}
\newcommand{\bcgdRHOPone}{\ensuremath{6.5^{+3.7}_{-2.5}}\xspace}
\newcommand{\fffdRHOPone}{\ensuremath{2.8^{+1.3}_{-1.2}}\xspace}
\newcommand{\fffdRHOPtwo}{---\xspace}
\newcommand{\dcbjRHOPone}{\ensuremath{3\pm1}\xspace}
\newcommand{\ddeiRHOPone}{\ensuremath{12^{+12}_{-10}}\xspace}
\newcommand{\ddeiRHOPtwo}{\ensuremath{9.4^{+4.2}_{-3.3}}\xspace}
\newcommand{\fdijRHOPone}{\ensuremath{2.24^{+0.77}_{-0.68}}\xspace}
\newcommand{\gafeRHOPone}{\ensuremath{0.9\pm2.4}\xspace}
\newcommand{\bhiiRHOPone}{\ensuremath{0.0\pm1.6}\xspace}
\newcommand{\bbggRHOPone}{\ensuremath{5.7^{+1.4}_{-1.2}}\xspace}
\newcommand{\dbfaRHOPone}{\ensuremath{6.1^{+1.9}_{-1.4}}\xspace}
\newcommand{\dbfaRHOPtwo}{\ensuremath{0.97^{+0.14}_{-0.12}}\xspace}
\newcommand{\dbfaRHOPthree}{\ensuremath{1.44^{+0.42}_{-0.32}}\xspace}
\newcommand{\dbfaRHOPfour}{---\xspace}
\newcommand{\ccdhRHOPone}{\ensuremath{0.3\pm0.4}\xspace}
\newcommand{\eicdRHOPtwo}{\ensuremath{1.10^{+0.67}_{-0.63}}\xspace}
\newcommand{\eicdRHOPone}{\ensuremath{7.3^{+1.7}_{-1.3}}\xspace}
\newcommand{\bggeRHOPone}{\ensuremath{0.8^{+0.9}_{-1.3}}\xspace}
\newcommand{\gcidRHOPone}{\ensuremath{1\pm1}\xspace}
\newcommand{\baeiRHOPone}{\ensuremath{3.4^{+2.4}_{-2.2}}\xspace}
\newcommand{\bjieRHOPone}{\ensuremath{0.6\pm0.1}\xspace}
\newcommand{\bjieRHOPtwo}{\ensuremath{1.03^{+0.48}_{-0.44}}\xspace}
\newcommand{\bjieRHOPthree}{\ensuremath{1.81^{+0.57}_{-0.45}}\xspace}
\newcommand{\ggacRHOPtwo}{\ensuremath{0.8^{+4.5}_{-4.4}}\xspace}
\newcommand{\ggacRHOPone}{\ensuremath{-0.7\pm3.1}\xspace}
\newcommand{\egijRHOPtwo}{\ensuremath{2\pm2}\xspace}
\newcommand{\egijRHOPone}{\ensuremath{7.1^{+6.2}_{-5.6}}\xspace}
\newcommand{\hadfRHOPone}{\ensuremath{2.36^{+0.88}_{-0.82}}\xspace}
\newcommand{\hidfRHOPone}{\ensuremath{3.9^{+6.9}_{-6.8}}\xspace}
\newcommand{\bebbRHOPone}{\ensuremath{7\pm2}\xspace}
\newcommand{\ecdiRHOPone}{\ensuremath{0.64^{+0.13}_{-0.11}}\xspace}
\newcommand{\gedgRHOPone}{\ensuremath{-1.6^{+9.1}_{-9.5}}\xspace}
\newcommand{\gedgRHOPtwo}{\ensuremath{1.7^{+1.8}_{-1.7}}\xspace}
\newcommand{\gedgRHOPthree}{\ensuremath{4.7^{+2.9}_{-2.4}}\xspace}
\newcommand{\jgbhRHOPone}{\ensuremath{4.6^{+1.4}_{-1.2}}\xspace}
\newcommand{\ghijRHOPone}{\ensuremath{1.71^{+0.34}_{-0.29}}\xspace}
\newcommand{\fdecRHOPone}{\ensuremath{2.7^{+4.4}_{-4.1}}\xspace}
\newcommand{\deffRHOPone}{\ensuremath{0.44^{+0.52}_{-0.48}}\xspace}
\newcommand{\eiadRHOPone}{\ensuremath{0.1^{+6.5}_{-6.4}}\xspace}
\newcommand{\ifgjRHOPone}{\ensuremath{0.7\pm0.2}\xspace}
\newcommand{\cffcRHOPone}{\ensuremath{2.28^{+0.63}_{-0.51}}\xspace}
\newcommand{\jjhiTEQone}{\ensuremath{801\pm20}\xspace}
\newcommand{\fcabTEQone}{\ensuremath{809\pm21}\xspace}
\newcommand{\eidfTEQone}{\ensuremath{734\pm19}\xspace}
\newcommand{\hehhTEQone}{\ensuremath{953\pm25}\xspace}
\newcommand{\iahiTEQone}{\ensuremath{973\pm28}\xspace}
\newcommand{\aiggTEQone}{\ensuremath{1720\pm44}\xspace}
\newcommand{\jgjjTEQone}{\ensuremath{1688\pm47}\xspace}
\newcommand{\hhedTEQone}{\ensuremath{1550^{+58}_{-55}}\xspace}
\newcommand{\befbTEQtwo}{\ensuremath{727\pm21}\xspace}
\newcommand{\befbTEQone}{\ensuremath{1763^{+52}_{-51}}\xspace}

\newcommand{\eaccTEQone}{\ensuremath{1019^{+44}_{-45}}\xspace}
\newcommand{\dcijTEQone}{\ensuremath{1108^{+34}_{-33}}\xspace}
\newcommand{\bejgTEQone}{\ensuremath{1257\pm35}\xspace}
\newcommand{\bcgdTEQtwo}{\ensuremath{862^{+48}_{-50}}\xspace}
\newcommand{\bcgdTEQone}{\ensuremath{1189^{+67}_{-69}}\xspace}
\newcommand{\fffdTEQone}{\ensuremath{968\pm25}\xspace}
\newcommand{\fffdTEQtwo}{\ensuremath{186.1^{+6.6}_{-7.2}}\xspace}
\newcommand{\dcbjTEQone}{\ensuremath{1427\pm51}\xspace}
\newcommand{\ddeiTEQone}{\ensuremath{1200^{+37}_{-36}}\xspace}
\newcommand{\ddeiTEQtwo}{\ensuremath{772\pm24}\xspace}
\newcommand{\fdijTEQone}{\ensuremath{805^{+22}_{-23}}\xspace}
\newcommand{\gafeTEQone}{\ensuremath{1001\pm27}\xspace}
\newcommand{\bhiiTEQone}{\ensuremath{761\pm21}\xspace}
\newcommand{\bbggTEQone}{\ensuremath{571\pm17}\xspace}
\newcommand{\dbfaTEQone}{\ensuremath{1992^{+59}_{-58}}\xspace}
\newcommand{\dbfaTEQtwo}{\ensuremath{1146^{+33}_{-34}}\xspace}
\newcommand{\dbfaTEQthree}{\ensuremath{885\pm26}\xspace}
\newcommand{\dbfaTEQfour}{\ensuremath{219.3\pm6.4}\xspace}
\newcommand{\ccdhTEQone}{\ensuremath{925^{+25}_{-24}}\xspace}
\newcommand{\eicdTEQtwo}{\ensuremath{740^{+21}_{-20}}\xspace}
\newcommand{\eicdTEQone}{\ensuremath{2115\pm59}\xspace}
\newcommand{\bggeTEQone}{\ensuremath{1093\pm31}\xspace}
\newcommand{\gcidTEQone}{\ensuremath{917^{+25}_{-24}}\xspace}
\newcommand{\baeiTEQone}{\ensuremath{932\pm25}\xspace}
\newcommand{\bjieTEQone}{\ensuremath{761^{+22}_{-21}}\xspace}
\newcommand{\bjieTEQtwo}{\ensuremath{577\pm16}\xspace}
\newcommand{\bjieTEQthree}{\ensuremath{500\pm14}\xspace}
\newcommand{\ggacTEQtwo}{\ensuremath{495\pm12}\xspace}
\newcommand{\ggacTEQone}{\ensuremath{666\pm16}\xspace}
\newcommand{\egijTEQtwo}{\ensuremath{877\pm24}\xspace}
\newcommand{\egijTEQone}{\ensuremath{1203^{+34}_{-33}}\xspace}
\newcommand{\hadfTEQone}{\ensuremath{649\pm18}\xspace}
\newcommand{\hidfTEQone}{\ensuremath{769^{+25}_{-26}}\xspace}
\newcommand{\bebbTEQone}{\ensuremath{967^{+24}_{-23}}\xspace}
\newcommand{\ecdiTEQone}{\ensuremath{736\pm22}\xspace}
\newcommand{\gedgTEQone}{\ensuremath{949\pm27}\xspace}
\newcommand{\gedgTEQtwo}{\ensuremath{839\pm24}\xspace}
\newcommand{\gedgTEQthree}{\ensuremath{646\pm18}\xspace}
\newcommand{\jgbhTEQone}{\ensuremath{672\pm17}\xspace}
\newcommand{\ghijTEQone}{\ensuremath{836\pm24}\xspace}
\newcommand{\fdecTEQone}{\ensuremath{949\pm26}\xspace}
\newcommand{\deffTEQone}{\ensuremath{874^{+26}_{-25}}\xspace}
\newcommand{\eiadTEQone}{\ensuremath{1325\pm38}\xspace}
\newcommand{\ifgjTEQone}{\ensuremath{716\pm26}\xspace}
\newcommand{\cffcTEQone}{\ensuremath{281^{+14}_{-12}}\xspace}
\newcommand{\jjhiFLUXone}{\ensuremath{98\pm10}\xspace}
\newcommand{\fcabFLUXone}{\ensuremath{101^{+11}_{-10}}\xspace}
\newcommand{\eidfFLUXone}{\ensuremath{68.9^{+7.4}_{-6.9}}\xspace}
\newcommand{\hehhFLUXone}{\ensuremath{195^{+22}_{-20}}\xspace}
\newcommand{\iahiFLUXone}{\ensuremath{212^{+26}_{-23}}\xspace}
\newcommand{\aiggFLUXone}{\ensuremath{2100\pm200}\xspace}
\newcommand{\jgjjFLUXone}{\ensuremath{1930^{+220}_{-210}}\xspace}
\newcommand{\hhedFLUXone}{\ensuremath{1370^{+220}_{-180}}\xspace}
\newcommand{\befbFLUXtwo}{\ensuremath{66^{+8}_{-7}}\xspace}
\newcommand{\befbFLUXone}{\ensuremath{2290^{+280}_{-260}}\xspace}

\newcommand{\eaccFLUXone}{\ensuremath{255^{+47}_{-42}}\xspace}
\newcommand{\dcijFLUXone}{\ensuremath{357^{+46}_{-41}}\xspace}
\newcommand{\bejgFLUXone}{\ensuremath{592^{+69}_{-64}}\xspace}
\newcommand{\bcgdFLUXtwo}{\ensuremath{131^{+32}_{-28}}\xspace}
\newcommand{\bcgdFLUXone}{\ensuremath{500\pm100}\xspace}
\newcommand{\fffdFLUXone}{\ensuremath{209^{+23}_{-21}}\xspace}
\newcommand{\fffdFLUXtwo}{\ensuremath{0.285\pm0.042}\xspace}
\newcommand{\dcbjFLUXone}{\ensuremath{980^{+150}_{-130}}\xspace}
\newcommand{\ddeiFLUXone}{\ensuremath{492^{+64}_{-57}}\xspace}
\newcommand{\ddeiFLUXtwo}{\ensuremath{84^{+11}_{-10}}\xspace}
\newcommand{\fdijFLUXone}{\ensuremath{100^{+12}_{-11}}\xspace}
\newcommand{\gafeFLUXone}{\ensuremath{238^{+27}_{-25}}\xspace}
\newcommand{\bhiiFLUXone}{\ensuremath{79.7^{+9.1}_{-8.4}}\xspace}
\newcommand{\bbggFLUXone}{\ensuremath{25.3^{+3.2}_{-2.9}}\xspace}
\newcommand{\dbfaFLUXone}{\ensuremath{3740^{+460}_{-420}}\xspace}
\newcommand{\dbfaFLUXtwo}{\ensuremath{409^{+50}_{-46}}\xspace}
\newcommand{\dbfaFLUXthree}{\ensuremath{145^{+18}_{-16}}\xspace}
\newcommand{\dbfaFLUXfour}{\ensuremath{0.549^{+0.067}_{-0.061}}\xspace}
\newcommand{\ccdhFLUXone}{\ensuremath{174^{+20}_{-18}}\xspace}
\newcommand{\eicdFLUXtwo}{\ensuremath{71.0^{+8.2}_{-7.5}}\xspace}
\newcommand{\eicdFLUXone}{\ensuremath{4740^{+550}_{-510}}\xspace}
\newcommand{\bggeFLUXone}{\ensuremath{339^{+41}_{-37}}\xspace}
\newcommand{\gcidFLUXone}{\ensuremath{168^{+19}_{-17}}\xspace}
\newcommand{\baeiFLUXone}{\ensuremath{179^{+20}_{-19}}\xspace}
\newcommand{\bjieFLUXone}{\ensuremath{79.5^{+9.4}_{-8.5}}\xspace}
\newcommand{\bjieFLUXtwo}{\ensuremath{26.2^{+3.1}_{-2.9}}\xspace}
\newcommand{\bjieFLUXthree}{\ensuremath{14.8^{+1.8}_{-1.6}}\xspace}
\newcommand{\ggacFLUXtwo}{\ensuremath{14.3^{+1.4}_{-1.3}}\xspace}
\newcommand{\ggacFLUXone}{\ensuremath{46.7^{+4.7}_{-4.4}}\xspace}
\newcommand{\egijFLUXtwo}{\ensuremath{140^{+16}_{-15}}\xspace}
\newcommand{\egijFLUXone}{\ensuremath{497^{+58}_{-52}}\xspace}
\newcommand{\hadfFLUXone}{\ensuremath{42.1^{+4.8}_{-4.4}}\xspace}
\newcommand{\hidfFLUXone}{\ensuremath{83^{+11}_{-10}}\xspace}
\newcommand{\bebbFLUXone}{\ensuremath{207^{+21}_{-19}}\xspace}
\newcommand{\ecdiFLUXone}{\ensuremath{69.5^{+8.6}_{-7.9}}\xspace}
\newcommand{\gedgFLUXone}{\ensuremath{193^{+23}_{-21}}\xspace}
\newcommand{\gedgFLUXtwo}{\ensuremath{118^{+14}_{-13}}\xspace}
\newcommand{\gedgFLUXthree}{\ensuremath{41.3^{+4.9}_{-4.5}}\xspace}
\newcommand{\jgbhFLUXone}{\ensuremath{48.3^{+5.2}_{-4.7}}\xspace}
\newcommand{\ghijFLUXone}{\ensuremath{116^{+14}_{-13}}\xspace}
\newcommand{\fdecFLUXone}{\ensuremath{192^{+22}_{-21}}\xspace}
\newcommand{\deffFLUXone}{\ensuremath{138^{+17}_{-15}}\xspace}
\newcommand{\eiadFLUXone}{\ensuremath{730^{+86}_{-79}}\xspace}
\newcommand{\ifgjFLUXone}{\ensuremath{62.5^{+9.6}_{-8.5}}\xspace}
\newcommand{\cffcFLUXone}{\ensuremath{1.48^{+0.31}_{-0.24}}\xspace}

\newcommand{\bggePNAMEone}{K2-98 b\xspace}
\newcommand{\jfjgPNAMEone}{K2-199 b\xspace}
\newcommand{\jfjgPERone}{\ensuremath{3.225286\pm0.000078}\xspace}
\newcommand{\jfjgTCone}{\ensuremath{2457221.9649\pm0.0011}\xspace}
\newcommand{\jfjgRPone}{\ensuremath{1.74^{+0.12}_{-0.05}}\xspace}
\newcommand{\jfjgMPone}{\ensuremath{7.1\pm1.8}\xspace}
\newcommand{\jfjgRHOPone}{\ensuremath{6.5^{+2.5}_{-1.9}}\xspace}
\newcommand{\jfjgTEQone}{\ensuremath{842^{+31}_{-30}}\xspace}
\newcommand{\jfjgFLUXone}{\ensuremath{119^{+18}_{-16}}\xspace}

\newcommand{\jfjgPNAMEtwo}{K2-199 c\xspace}
\newcommand{\jfjgPERtwo}{\ensuremath{7.37442\pm0.00012}\xspace}
\newcommand{\jfjgTCtwo}{\ensuremath{2457222.93118\pm0.00069}\xspace}
\newcommand{\jfjgRPtwo}{\ensuremath{2.67^{+0.12}_{-0.05}}\xspace}
\newcommand{\jfjgMPtwo}{\ensuremath{12.6^{+2.3}_{-2.2}}\xspace}
\newcommand{\jfjgRHOPtwo}{\ensuremath{3.21^{+0.91}_{-0.74}}\xspace}
\newcommand{\jfjgTEQtwo}{\ensuremath{639\pm23}\xspace}
\newcommand{\jfjgFLUXtwo}{\ensuremath{39.6^{+6.1}_{-5.4}}\xspace}

\newcommand{\hbagPNAMEone}{HD 89345 b\xspace}
\newcommand{\hbagPERone}{\ensuremath{11.81469\pm0.00044}\xspace}
\newcommand{\hbagTCone}{\ensuremath{2457913.8041\pm0.0011}\xspace}
\newcommand{\hbagRPone}{\ensuremath{7.20^{+0.42}_{-0.15}}\xspace}
\newcommand{\hbagMPone}{\ensuremath{34.1^{+3.4}_{-3.3}}\xspace}
\newcommand{\hbagRHOPone}{\ensuremath{0.55^{+0.12}_{-0.09}}\xspace}
\newcommand{\hbagTEQone}{\ensuremath{993\pm29}\xspace}
\newcommand{\hbagFLUXone}{\ensuremath{231^{+29}_{-26}}\xspace}

\newcommand{\hagfPNAMEone}{K2-3 b\xspace} 
\newcommand{\hagfPERone}{\ensuremath{10.054626^{+0.000009}_{-0.000010}}\xspace} 
\newcommand{\hagfTCone}{\ensuremath{2456813.41843^{+0.00039}_{-0.00038}}\xspace} 
\newcommand{\hagfRPone}{\ensuremath{2.140 \pm 0.264}\xspace} 
\newcommand{\hagfMPone}{\ensuremath{6.48^{+0.99}_{-0.93}}\xspace} 
\newcommand{\hagfRHOPone}{\ensuremath{3.70^{+1.67}_{-1.08}}\xspace} 
\newcommand{\hagfTEQone}{\ensuremath{463 \pm 39}\xspace} 
\newcommand{\hagfFLUXone}{\ensuremath{11\pm4}\xspace} 
 
\newcommand{\hagfPNAMEtwo}{K2-3 c\xspace} 
\newcommand{\hagfPERtwo}{\ensuremath{24.646582^{+0.000039}_{-0.000039}}\xspace} 
\newcommand{\hagfTCtwo}{\ensuremath{2456812.28013^{+0.00090}_{-0.00095}}\xspace} 
\newcommand{\hagfRPtwo}{\ensuremath{1.72^{+0.23}_{-0.22}}\xspace} 
\newcommand{\hagfMPtwo}{\ensuremath{2.14^{+1.08}_{-1.04}}\xspace} 
\newcommand{\hagfRHOPtwo}{\ensuremath{2.98^{+1.96}_{-1.50}}\xspace} 
\newcommand{\hagfTEQtwo}{\ensuremath{344 \pm 29}\xspace} 
\newcommand{\hagfFLUXtwo}{\ensuremath{3\pm1}\xspace} 
 
\newcommand{\hagfPNAMEthree}{K2-3 d\xspace} 
\newcommand{\hagfPERthree}{\ensuremath{44.556456^{+0.000097}_{-0.000087}}\xspace} 
\newcommand{\hagfTCthree}{\ensuremath{2456826.22347^{+0.00053}_{-0.00052}}\xspace} 
\newcommand{\hagfRPthree}{\ensuremath{1.52^{+0.21}_{-0.20}}\xspace} 
\newcommand{\hagfMPthree}{\ensuremath{< 2.80}\xspace} 
\newcommand{\hagfRHOPthree}{\ensuremath{< 5.62}\xspace} 
\newcommand{\hagfTEQthree}{\ensuremath{282 \pm 24}\xspace} 
\newcommand{\hagfFLUXthree}{\ensuremath{282\pm23}\xspace} 
 
\newcommand{\ihidPNAMEone}{K2-291 b\xspace} 
\newcommand{\ihidPERone}{\ensuremath{2.225177^{+0.000066}_{-0.000068}}\xspace} 
\newcommand{\ihidTCone}{\ensuremath{2457830.06163^{+0.00099}_{-0.00104}}\xspace} 
\newcommand{\ihidRPone}{\ensuremath{1.589^{+0.095}_{-0.072}}\xspace} 
\newcommand{\ihidMPone}{\ensuremath{6.49\pm1.16}\xspace} 
\newcommand{\ihidRHOPone}{\ensuremath{8.84^{+2.50}_{-2.03}}\xspace} 
\newcommand{\ihidTEQone}{\ensuremath{1278\pm30}\xspace} 
\newcommand{\ihidFLUXone}{\ensuremath{633^{+59}_{-56}}\xspace} 
 
\newcommand{\bdiaPNAMEone}{HIP 41378 b\xspace} 
\newcommand{\bdiaPERone}{\ensuremath{15.57208 \pm 0.00002}\xspace} 
\newcommand{\bdiaTCone}{\ensuremath{2457152.2818 \pm 0.0012}\xspace} 
\newcommand{\bdiaRPone}{\ensuremath{2.595 \pm 0.036}\xspace} 
\newcommand{\bdiaMPone}{\ensuremath{6.89 \pm 0.88}\xspace} 
\newcommand{\bdiaRHOPone}{\ensuremath{2.17 \pm 0.28}\xspace} 
\newcommand{\bdiaTEQone}{\ensuremath{959^{+9}_{-5}}\xspace} 
\newcommand{\bdiaFLUXone}{\ensuremath{140^{+5}_{-3}}\xspace} 
 
\newcommand{\bdiaPNAMEtwo}{HIP 41378 c\xspace} 
\newcommand{\bdiaPERtwo}{\ensuremath{31.70603 \pm 0.00006}\xspace} 
\newcommand{\bdiaTCtwo}{\ensuremath{2457163.1609^{+0.0023}_{-0.0027}}\xspace} 
\newcommand{\bdiaRPtwo}{\ensuremath{2.727 \pm 0.060}\xspace} 
\newcommand{\bdiaMPtwo}{\ensuremath{4.4 \pm 1.1}\xspace} 
\newcommand{\bdiaRHOPtwo}{\ensuremath{1.19 \pm 0.30}\xspace} 
\newcommand{\bdiaTEQtwo}{\ensuremath{757^{+7}_{-4}}\xspace} 
\newcommand{\bdiaFLUXtwo}{\ensuremath{54^{+2}_{-1}}\xspace} 
 
\newcommand{\bdiaPNAMEthree}{HIP 41378 d\xspace} 
\newcommand{\bdiaPERthree}{\ensuremath{278.3618 \pm 0.0005}\xspace} 
\newcommand{\bdiaTCthree}{\ensuremath{2457166.2604 \pm 0.0017}\xspace} 
\newcommand{\bdiaRPthree}{\ensuremath{3.54 \pm 0.06}\xspace} 
\newcommand{\bdiaMPthree}{\ensuremath{< 4.6}\xspace} 
\newcommand{\bdiaRHOPthree}{\ensuremath{< 0.56}\xspace} 
\newcommand{\bdiaTEQthree}{\ensuremath{367^{+3}_{-2}}\xspace} 
\newcommand{\bdiaFLUXthree}{\ensuremath{3.01^{+0.11}_{-0.06}}\xspace} 
 
\newcommand{\bdiaPNAMEfour}{HIP 41378 e\xspace} 
\newcommand{\bdiaPERfour}{\ensuremath{369 \pm 10}\xspace} 
\newcommand{\bdiaTCfour}{\ensuremath{2457142.0194 \pm 0.0010}\xspace} 
\newcommand{\bdiaRPfour}{\ensuremath{4.92 \pm 0.09}\xspace} 
\newcommand{\bdiaMPfour}{\ensuremath{12 \pm 5}\xspace} 
\newcommand{\bdiaRHOPfour}{\ensuremath{0.55 \pm 0.23}\xspace} 
\newcommand{\bdiaTEQfour}{\ensuremath{335 \pm 4}\xspace} 
\newcommand{\bdiaFLUXfour}{\ensuremath{2.1 \pm 0.1}\xspace} 
 
\newcommand{\bdiaPNAMEfive}{HIP 41378 f\xspace} 
\newcommand{\bdiaPERfive}{\ensuremath{542.07975 \pm 0.00014}\xspace} 
\newcommand{\bdiaTCfive}{\ensuremath{2457186.91423^{+0.00039}_{-0.00038}}\xspace} 
\newcommand{\bdiaRPfive}{\ensuremath{9.2 \pm 0.1}\xspace} 
\newcommand{\bdiaMPfive}{\ensuremath{12 \pm 3}\xspace} 
\newcommand{\bdiaRHOPfive}{\ensuremath{0.09 \pm 0.02}\xspace} 
\newcommand{\bdiaTEQfive}{\ensuremath{294^{+3}_{-1}}\xspace} 
\newcommand{\bdiaFLUXfive}{\ensuremath{1.24^{+0.06}_{-0.02}}\xspace} 
 
\newcommand{\bdiaPNAMEsix}{HIP 41378 g\xspace} 
\newcommand{\bdiaPERsix}{\ensuremath{62.06 \pm 0.32}\xspace} 
\newcommand{\bdiaTCsix}{---\xspace} 
\newcommand{\bdiaRPsix}{---\xspace} 
\newcommand{\bdiaMPsix}{\ensuremath{7 \pm 1.5}\xspace} 
\newcommand{\bdiaRHOPsix}{---\xspace} 
\newcommand{\bdiaTEQsix}{\ensuremath{605 \pm 4.7}\xspace} 
\newcommand{\bdiaFLUXsix}{\ensuremath{22.3^{+0.8}_{-0.5}}\xspace} 
 
\newcommand{\fdfaPNAMEone}{K2-19 b\xspace} 
\newcommand{\fdfaPERone}{\ensuremath{7.9222 \pm 0.0001}\xspace} 
\newcommand{\fdfaTCone}{\ensuremath{2456860.9023 \pm 0.0002}\xspace} 
\newcommand{\fdfaRPone}{\ensuremath{7.0 \pm 0.2}\xspace} 
\newcommand{\fdfaMPone}{\ensuremath{32.4 \pm 1.7}\xspace} 
\newcommand{\fdfaRHOPone}{\ensuremath{0.52\pm0.05}\xspace} 
\newcommand{\fdfaTEQone}{\ensuremath{779\pm21}\xspace} 
\newcommand{\fdfaFLUXone}{\ensuremath{87\pm9}\xspace} 
 
\newcommand{\fdfaPNAMEtwo}{K2-19 c\xspace} 
\newcommand{\fdfaPERtwo}{\ensuremath{11.8993 \pm 0.0008}\xspace} 
\newcommand{\fdfaTCtwo}{\ensuremath{2456853.0007 \pm 0.0004}\xspace} 
\newcommand{\fdfaRPtwo}{\ensuremath{4.1 \pm 0.2}\xspace} 
\newcommand{\fdfaMPtwo}{\ensuremath{10.8 \pm 0.6}\xspace} 
\newcommand{\fdfaRHOPtwo}{\ensuremath{0.87\pm0.14}\xspace} 
\newcommand{\fdfaTEQtwo}{\ensuremath{679\pm18}\xspace} 
\newcommand{\fdfaFLUXtwo}{\ensuremath{51\pm5}\xspace} 
 
\newcommand{\fdfaPNAMEthree}{K2-19 d\xspace} 
\newcommand{\fdfaPERthree}{\ensuremath{2.5081 \pm 0.0002}\xspace} 
\newcommand{\fdfaTCthree}{\ensuremath{2456854.0726 \pm 0.0018}\xspace} 
\newcommand{\fdfaRPthree}{\ensuremath{1.11 \pm 0.05}\xspace} 
\newcommand{\fdfaMPthree}{\ensuremath{< 10}\xspace} 
\newcommand{\fdfaRHOPthree}{\ensuremath{< 40}\xspace} 
\newcommand{\fdfaTEQthree}{\ensuremath{1142\pm31}\xspace} 
\newcommand{\fdfaFLUXthree}{\ensuremath{404\pm44}\xspace} 
 
\newcommand{\ddigPNAMEone}{HD 3167 b\xspace} 
\newcommand{\ddigPERone}{\ensuremath{0.959641^{+0.000012}_{-0.000011}}\xspace} 
\newcommand{\ddigTCone}{\ensuremath{2457394.37454 \pm 0.00043}\xspace} 
\newcommand{\ddigRPone}{\ensuremath{1.70^{+0.15}_{-0.18}}\xspace} 
\newcommand{\ddigMPone}{\ensuremath{5.02 \pm 0.38}\xspace} 
\newcommand{\ddigRHOPone}{\ensuremath{5.60^{+1.43}_{-2.15}}\xspace} 
\newcommand{\ddigTEQone}{\ensuremath{1608 \pm 56}\xspace} 
\newcommand{\ddigFLUXone}{\ensuremath{1625^{+222}_{-244}}\xspace} 
 
\newcommand{\ddigPNAMEtwo}{HD 3167 c\xspace} 
\newcommand{\ddigPERtwo}{\ensuremath{29.8454 \pm 0.0012}\xspace} 
\newcommand{\ddigTCtwo}{\ensuremath{2457394.9788 \pm 0.0012}\xspace} 
\newcommand{\ddigRPtwo}{\ensuremath{3.01^{+0.28}_{-0.42}}\xspace} 
\newcommand{\ddigMPtwo}{\ensuremath{9.80^{+1.24}_{-1.30}}\xspace} 
\newcommand{\ddigRHOPtwo}{\ensuremath{1.97^{+0.59}_{-0.94}}\xspace} 
\newcommand{\ddigTEQtwo}{\ensuremath{511 \pm 18}\xspace} 
\newcommand{\ddigFLUXtwo}{\ensuremath{16.6^{+2.5}_{-2.3}}\xspace} 
 
\newcommand{\ddigPNAMEthree}{HD 3167 d\xspace} 
\newcommand{\ddigPERthree}{\ensuremath{8.509 \pm 0.045}\xspace} 
\newcommand{\ddigTCthree}{\ensuremath{2457806.07^{+0.50}_{-0.52}}\xspace} 
\newcommand{\ddigRPthree}{---\xspace} 
\newcommand{\ddigMPthree}{\ensuremath{6.90 \pm 0.71}\xspace} 
\newcommand{\ddigRHOPthree}{---\xspace} 
\newcommand{\ddigTEQthree}{\ensuremath{776 \pm 28}\xspace} 
\newcommand{\ddigFLUXthree}{\ensuremath{88.9 \pm 6.2}\xspace} 
 
\newcommand{\bajiPNAMEone}{K2-24 b\xspace} 
\newcommand{\bajiPERone}{\ensuremath{20.88977 \pm 0.00035}\xspace} 
\newcommand{\bajiTCone}{\ensuremath{2456905.88977 \pm 0.0055}\xspace} 
\newcommand{\bajiRPone}{\ensuremath{5.4 \pm 0.2}\xspace} 
\newcommand{\bajiMPone}{\ensuremath{19.0 \pm 2.2}\xspace} 
\newcommand{\bajiRHOPone}{\ensuremath{0.64 \pm 0.12}\xspace} 
\newcommand{\bajiTEQone}{\ensuremath{686 \pm 13}\xspace} 
\newcommand{\bajiFLUXone}{\ensuremath{52 \pm 4}\xspace} 
 
\newcommand{\bajiPNAMEtwo}{K2-24 c\xspace} 
\newcommand{\bajiPERtwo}{\ensuremath{42.3391 \pm 0.0012}\xspace} 
\newcommand{\bajiTCtwo}{\ensuremath{2456915.4485 \pm 0.0079}\xspace} 
\newcommand{\bajiRPtwo}{\ensuremath{7.5 \pm 0.3}\xspace} 
\newcommand{\bajiMPtwo}{\ensuremath{15.4 \pm 1.9}\xspace} 
\newcommand{\bajiRHOPtwo}{\ensuremath{0.20 \pm 0.04}\xspace} 
\newcommand{\bajiTEQtwo}{\ensuremath{542 \pm 10}\xspace} 
\newcommand{\bajiFLUXtwo}{\ensuremath{20 \pm 2}\xspace} 
 
\newcommand{\egbePNAMEone}{K2-55 b\xspace} 
\newcommand{\egbePERone}{\ensuremath{2.849271 \pm 0.000029}\xspace} 
\newcommand{\egbeTCone}{\ensuremath{2456983.4229 \pm 0.0004}\xspace} 
\newcommand{\egbeRPone}{\ensuremath{4.43^{+0.29}_{-0.32}}\xspace} 
\newcommand{\egbeMPone}{\ensuremath{44.0 \pm 5.3}\xspace} 
\newcommand{\egbeRHOPone}{\ensuremath{2.8^{+0.7}_{-0.6}}\xspace} 
\newcommand{\egbeTEQone}{\ensuremath{900}\xspace} 
\newcommand{\egbeFLUXone}{\ensuremath{141.3^{+28.8}_{-23.5}}\xspace} 
 
\newcommand{\gghbPNAMEone}{K2-108 b\xspace} 
\newcommand{\gghbPERone}{\ensuremath{4.73401 \pm 0.00024}\xspace} 
\newcommand{\gghbTCone}{\ensuremath{2457145.0965 \pm 0.0019}\xspace} 
\newcommand{\gghbRPone}{\ensuremath{5.28 \pm 0.54}\xspace} 
\newcommand{\gghbMPone}{\ensuremath{59.4 \pm 4.4}\xspace} 
\newcommand{\gghbRHOPone}{\ensuremath{2.22^{+0.77}_{-0.55}}\xspace} 
\newcommand{\gghbTEQone}{\ensuremath{1446 \pm 48}\xspace} 
\newcommand{\gghbFLUXone}{\ensuremath{762 \pm 100}\xspace} 
 
\newcommand{\hieePNAMEone}{HD 106315 b\xspace} 
\newcommand{\hieePERone}{\ensuremath{9.55288\pm 0.00021}\xspace} 
\newcommand{\hieeTCone}{\ensuremath{2457586.5476^{+0.0024}_{-0.0025}}\xspace} 
\newcommand{\hieeRPone}{\ensuremath{2.40 \pm 0.20}\xspace} 
\newcommand{\hieeMPone}{\ensuremath{10.5\pm 3.1}\xspace} 
\newcommand{\hieeRHOPone}{\ensuremath{4.1^{+1.9}_{-1.4}}\xspace} 
\newcommand{\hieeTEQone}{\ensuremath{1040 \pm 18}\xspace} 
\newcommand{\hieeFLUXone}{\ensuremath{277 \pm 20}\xspace} 
 
\newcommand{\hieePNAMEtwo}{HD 106315 c\xspace} 
\newcommand{\hieePERtwo}{\ensuremath{21.05652\pm 0.00012}\xspace} 
\newcommand{\hieeTCtwo}{\ensuremath{2457569.01767^{+0.00097}_{-0.00096}}\xspace} 
\newcommand{\hieeRPtwo}{\ensuremath{4.379 \pm 0.086}\xspace} 
\newcommand{\hieeMPtwo}{\ensuremath{12.0\pm 3.8}\xspace} 
\newcommand{\hieeRHOPtwo}{\ensuremath{0.78^{+0.26}_{-0.25}}\xspace} 
\newcommand{\hieeTEQtwo}{\ensuremath{799 \pm 14}\xspace} 
\newcommand{\hieeFLUXtwo}{\ensuremath{97 \pm 7}\xspace} 
 
\newcommand{\jifiPNAMEone}{GJ 9827 b\xspace} 
\newcommand{\jifiPERone}{\ensuremath{1.2089765\pm 2.3e-06}\xspace} 
\newcommand{\jifiTCone}{\ensuremath{2457738.82586\pm 0.00026}\xspace} 
\newcommand{\jifiRPone}{\ensuremath{1.529 \pm 0.05787}\xspace} 
\newcommand{\jifiMPone}{\ensuremath{4.87\pm 0.37}\xspace} 
\newcommand{\jifiRHOPone}{\ensuremath{7.47^{+1.1}_{-0.95}}\xspace} 
\newcommand{\jifiTEQone}{\ensuremath{1055\pm21}\xspace} 
\newcommand{\jifiFLUXone}{\ensuremath{294\pm24}\xspace} 
 
\newcommand{\jifiPNAMEtwo}{GJ 9827 c\xspace} 
\newcommand{\jifiPERtwo}{\ensuremath{3.648095\pm 2.4e-05}\xspace} 
\newcommand{\jifiTCtwo}{\ensuremath{2457742.19929^{+0.00072}_{-0.00071}}\xspace} 
\newcommand{\jifiRPtwo}{\ensuremath{1.201 \pm 0.04586}\xspace} 
\newcommand{\jifiMPtwo}{\ensuremath{1.92\pm 0.49}\xspace} 
\newcommand{\jifiRHOPtwo}{\ensuremath{6.1^{+1.8}_{-1.6}}\xspace} 
\newcommand{\jifiTEQtwo}{\ensuremath{730\pm15}\xspace} 
\newcommand{\jifiFLUXtwo}{\ensuremath{67\pm5}\xspace} 
 
\newcommand{\jifiPNAMEthree}{GJ 9827 d\xspace} 
\newcommand{\jifiPERthree}{\ensuremath{6.20183\pm 1e-05}\xspace} 
\newcommand{\jifiTCthree}{\ensuremath{2457740.96114^{+0.00045}_{-0.00044}}\xspace} 
\newcommand{\jifiRPthree}{\ensuremath{1.955 \pm 0.07535}\xspace} 
\newcommand{\jifiMPthree}{\ensuremath{3.42\pm 0.62}\xspace} 
\newcommand{\jifiRHOPthree}{\ensuremath{2.51^{+0.57}_{-0.51}}\xspace} 
\newcommand{\jifiTEQthree}{\ensuremath{612\pm12}\xspace} 
\newcommand{\jifiFLUXthree}{\ensuremath{33\pm3}\xspace} 
 
\newcommand{\ecdcPNAMEone}{WASP-107 b\xspace} 
\newcommand{\ecdcPERone}{\ensuremath{5.7214742 \pm 4.3e-7}\xspace} 
\newcommand{\ecdcTCone}{\ensuremath{2457584.329897 \pm 0.000032}\xspace} 
\newcommand{\ecdcRPone}{\ensuremath{10.55 \pm 0.32}\xspace} 
\newcommand{\ecdcMPone}{\ensuremath{30.6 \pm 1.7}\xspace} 
\newcommand{\ecdcRHOPone}{\ensuremath{0.143^{+0.016}_{-0.014}}\xspace} 
\newcommand{\ecdcTEQone}{\ensuremath{525-820}\xspace} 
\newcommand{\ecdcFLUXone}{\ensuremath{51\pm4}\xspace} 
 
\newcommand{\ecdcPNAMEtwo}{WASP-107 c\xspace} 
\newcommand{\ecdcPERtwo}{\ensuremath{1088_{-16}^{+15}}\xspace} 
\newcommand{\ecdcTCtwo}{\ensuremath{2458521_{-56}^{+65}}\xspace} 
\newcommand{\ecdcRPtwo}{---\xspace} 
\newcommand{\ecdcMPtwo}{\ensuremath{115^{+13}_{-14}}\xspace} 
\newcommand{\ecdcRHOPtwo}{---\xspace} 
\newcommand{\ecdcTEQtwo}{\ensuremath{118 \pm 3}\xspace} 
\newcommand{\ecdcFLUXtwo}{\ensuremath{0.046 \pm 0.004}\xspace} 
 
\newcommand{\hbdaPNAMEone}{K2-85 b\xspace} 
\newcommand{\hbdaPERone}{\ensuremath{0.684538626159 \pm 1.0e-5}\xspace} 
\newcommand{\hbdaTCone}{\ensuremath{2457065.36778 \pm  0.00056}\xspace} 
\newcommand{\hbdaRPone}{\ensuremath{1.4 \pm 0.2}\xspace} 
\newcommand{\hbdaMPone}{\ensuremath{4.0 \pm 1.5}\xspace} 
\newcommand{\hbdaRHOPone}{\ensuremath{7.8 \pm 6.3}\xspace} 
\newcommand{\hbdaTEQone}{\ensuremath{1355 \pm 99}\xspace} 
\newcommand{\hbdaFLUXone}{\ensuremath{800 \pm 234}\xspace} 
 
\newcommand{\jjhiSTNAME}{K2-222\xspace} 
\newcommand{\fcabSTNAME}{K2-236\xspace} 
\newcommand{\eidfSTNAME}{K2-418\xspace} 
\newcommand{\hehhSTNAME}{K2-277\xspace} 
\newcommand{\iahiSTNAME}{K2-261\xspace} 
\newcommand{\aiggSTNAME}{K2-100\xspace} 
\newcommand{\jgjjSTNAME}{K2-31\xspace} 
\newcommand{\hhedSTNAME}{K2-39\xspace} 
\newcommand{\befbSTNAME}{K2-229\xspace} 
 
\newcommand{\eaccSTNAME}{K2-111\xspace} 
\newcommand{\dcijSTNAME}{K2-99\xspace} 
\newcommand{\bejgSTNAME}{K2-265\xspace} 
\newcommand{\bcgdSTNAME}{K2-38\xspace} 
\newcommand{\fffdSTNAME}{K2-73\xspace} 
\newcommand{\dcbjSTNAME}{K2-66\xspace} 
\newcommand{\ddeiSTNAME}{K2-36\xspace} 
\newcommand{\fdijSTNAME}{K2-105\xspace} 
\newcommand{\gafeSTNAME}{K2-214\xspace} 
\newcommand{\bhiiSTNAME}{K2-220\xspace} 
\newcommand{\bbggSTNAME}{K2-110\xspace} 
\newcommand{\dbfaSTNAME}{WASP-47\xspace} 
\newcommand{\ccdhSTNAME}{K2-79\xspace} 
\newcommand{\eicdSTNAME}{K2-106\xspace} 
\newcommand{\bggeSTNAME}{K2-98\xspace} 
\newcommand{\gcidSTNAME}{EPIC 213546283\xspace} 
\newcommand{\baeiSTNAME}{EPIC 245991048\xspace} 
\newcommand{\bjieSTNAME}{K2-32\xspace} 
\newcommand{\ggacSTNAME}{K2-62\xspace} 
\newcommand{\egijSTNAME}{K2-189\xspace} 
\newcommand{\hadfSTNAME}{K2-10\xspace} 
\newcommand{\hidfSTNAME}{K2-245\xspace} 
\newcommand{\bebbSTNAME}{K2-216\xspace} 
\newcommand{\ecdiSTNAME}{K2-280\xspace} 
\newcommand{\gedgSTNAME}{K2-37\xspace} 
\newcommand{\jgbhSTNAME}{K2-180\xspace} 
\newcommand{\ghijSTNAME}{K2-27\xspace} 
\newcommand{\fdecSTNAME}{K2-181\xspace} 
\newcommand{\deffSTNAME}{EPIC 245943455\xspace} 
\newcommand{\eiadSTNAME}{K2-61\xspace} 
\newcommand{\ifgjSTNAME}{K2-121\xspace} 
\newcommand{\cffcSTNAME}{K2-18\xspace}

\newcommand{\jjhiEPICID}{220709978\xspace}

\newcommand{\jjhiNOBSHIRES}{55\xspace} 
\newcommand{\fcabNOBSHIRES}{36\xspace} 
\newcommand{\eidfNOBSHIRES}{22\xspace} 
\newcommand{\hehhNOBSHIRES}{26\xspace} 
\newcommand{\iahiNOBSHIRES}{8\xspace} 
\newcommand{\aiggNOBSHIRES}{33\xspace} 
\newcommand{\jgjjNOBSHIRES}{8\xspace} 
\newcommand{\hhedNOBSHIRES}{45\xspace} 
\newcommand{\befbNOBSHIRES}{24\xspace} 
 
\newcommand{\eaccNOBSHIRES}{54\xspace} 
\newcommand{\dcijNOBSHIRES}{19\xspace} 
\newcommand{\bejgNOBSHIRES}{53\xspace} 
\newcommand{\bcgdNOBSHIRES}{65\xspace} 
\newcommand{\fffdNOBSHIRES}{60\xspace} 
\newcommand{\dcbjNOBSHIRES}{44\xspace} 
\newcommand{\ddeiNOBSHIRES}{46\xspace} 
\newcommand{\fdijNOBSHIRES}{31\xspace} 
\newcommand{\gafeNOBSHIRES}{29\xspace} 
\newcommand{\bhiiNOBSHIRES}{28\xspace} 
\newcommand{\bbggNOBSHIRES}{12\xspace} 
\newcommand{\dbfaNOBSHIRES}{76\xspace} 
\newcommand{\ccdhNOBSHIRES}{62\xspace} 
\newcommand{\eicdNOBSHIRES}{39\xspace} 
\newcommand{\bggeNOBSHIRES}{6\xspace} 
\newcommand{\gcidNOBSHIRES}{12\xspace} 
\newcommand{\baeiNOBSHIRES}{16\xspace} 
\newcommand{\bjieNOBSHIRES}{64\xspace} 
\newcommand{\ggacNOBSHIRES}{20\xspace} 
\newcommand{\egijNOBSHIRES}{17\xspace} 
\newcommand{\hadfNOBSHIRES}{22\xspace} 
\newcommand{\hidfNOBSHIRES}{7\xspace} 
\newcommand{\bebbNOBSHIRES}{31\xspace} 
\newcommand{\ecdiNOBSHIRES}{16\xspace} 
\newcommand{\gedgNOBSHIRES}{19\xspace} 
\newcommand{\jgbhNOBSHIRES}{26\xspace} 
 
\newcommand{\fdecNOBSHIRES}{10\xspace} 
\newcommand{\deffNOBSHIRES}{9\xspace} 
\newcommand{\eiadNOBSHIRES}{7\xspace} 
\newcommand{\ifgjNOBSHIRES}{18\xspace} 
\newcommand{\cffcNOBSHIRES}{21\xspace} 
\newcommand{\jjhiNOBSAPF}{32\xspace} 
\newcommand{\fcabNOBSAPF}{2\xspace}

\newcommand{\iahiNOBSAPF}{4\xspace}

\newcommand{\jfjgSTNAME}{K2-199\xspace}

\newcommand{\jfjgNOBSHIRES}{45\xspace}

\newcommand{\hbagSTNAME}{HD 89345\xspace}

\newcommand{\hagfSTNAME}{K2-3\xspace}

\newcommand{\ihidSTNAME}{K2-291\xspace}

\newcommand{\bdiaSTNAME}{HIP 41378\xspace}

\newcommand{\fdfaSTNAME}{K2-19\xspace}

\newcommand{\ddigSTNAME}{HD 3167\xspace} 
 
\newcommand{\ddigEPICID}{220383386\xspace}

\newcommand{\bajiSTNAME}{K2-24\xspace}

\newcommand{\egbeSTNAME}{K2-55\xspace}

\newcommand{\gghbSTNAME}{K2-108\xspace}

\newcommand{\hieeSTNAME}{HD 106315\xspace} 
\newcommand{\hieeSTNAMEKTWO}{K2-109\xspace} 
\newcommand{\hieeEPICID}{201437844\xspace}

\newcommand{\hieeVMAG}{\ensuremath{8.951 \pm 0.018}\xspace}

\newcommand{\jifiSTNAME}{GJ 9827\xspace}

\newcommand{\ecdcSTNAME}{WASP-107\xspace}

\newcommand{\hbdaSTNAME}{K2-85\xspace} 
 
\newcommand{\hbdaEPICID}{210707130\xspace}

\newcommand{\jjhiTEFF}{\ensuremath{5961 \pm 100}\xspace} 
\newcommand{\fcabTEFF}{\ensuremath{6019 \pm 100}\xspace} 
\newcommand{\eidfTEFF}{\ensuremath{5839 \pm 100}\xspace} 
\newcommand{\hehhTEFF}{\ensuremath{5705 \pm 100}\xspace} 
\newcommand{\iahiTEFF}{\ensuremath{5478 \pm 100}\xspace} 
\newcommand{\aiggTEFF}{\ensuremath{6044 \pm 100}\xspace} 
\newcommand{\jgjjTEFF}{\ensuremath{5340 \pm 100}\xspace} 
\newcommand{\hhedTEFF}{\ensuremath{4915 \pm 100}\xspace} 
\newcommand{\befbTEFF}{\ensuremath{5163 \pm 100}\xspace} 
 
\newcommand{\eaccTEFF}{\ensuremath{5832 \pm 100}\xspace} 
\newcommand{\dcijTEFF}{\ensuremath{6053 \pm 100}\xspace} 
\newcommand{\bejgTEFF}{\ensuremath{5435 \pm 100}\xspace} 
\newcommand{\bcgdTEFF}{\ensuremath{5679 \pm 100}\xspace} 
\newcommand{\fffdTEFF}{\ensuremath{5867 \pm 100}\xspace} 
\newcommand{\dcbjTEFF}{\ensuremath{5865 \pm 100}\xspace} 
\newcommand{\ddeiTEFF}{\ensuremath{4836 \pm 100}\xspace} 
\newcommand{\fdijTEFF}{\ensuremath{5373 \pm 100}\xspace} 
\newcommand{\gafeTEFF}{\ensuremath{5875 \pm 100}\xspace} 
\newcommand{\bhiiTEFF}{\ensuremath{5612 \pm 100}\xspace} 
\newcommand{\bbggTEFF}{\ensuremath{4868 \pm 100}\xspace} 
\newcommand{\dbfaTEFF}{\ensuremath{5476 \pm 100}\xspace} 
\newcommand{\ccdhTEFF}{\ensuremath{5853 \pm 100}\xspace} 
\newcommand{\eicdTEFF}{\ensuremath{5496 \pm 100}\xspace} 
\newcommand{\bggeTEFF}{\ensuremath{6103 \pm 100}\xspace} 
\newcommand{\gcidTEFF}{\ensuremath{5685 \pm 100}\xspace} 
\newcommand{\baeiTEFF}{\ensuremath{5773 \pm 100}\xspace} 
\newcommand{\bjieTEFF}{\ensuremath{5274 \pm 100}\xspace} 
\newcommand{\ggacTEFF}{\ensuremath{4455 \pm 70}\xspace} 
\newcommand{\egijTEFF}{\ensuremath{5442 \pm 100}\xspace} 
\newcommand{\hadfTEFF}{\ensuremath{5533 \pm 100}\xspace} 
\newcommand{\hidfTEFF}{\ensuremath{5942 \pm 100}\xspace} 
\newcommand{\bebbTEFF}{\ensuremath{4495 \pm 70}\xspace} 
\newcommand{\ecdiTEFF}{\ensuremath{5741 \pm 100}\xspace} 
\newcommand{\gedgTEFF}{\ensuremath{5352 \pm 100}\xspace} 
\newcommand{\jgbhTEFF}{\ensuremath{5166 \pm 100}\xspace} 
\newcommand{\ghijTEFF}{\ensuremath{5246 \pm 100}\xspace} 
\newcommand{\fdecTEFF}{\ensuremath{5528 \pm 100}\xspace} 
\newcommand{\deffTEFF}{\ensuremath{5310 \pm 100}\xspace} 
\newcommand{\eiadTEFF}{\ensuremath{5748 \pm 100}\xspace} 
\newcommand{\ifgjTEFF}{\ensuremath{4526 \pm 110}\xspace} 
\newcommand{\cffcTEFF}{\ensuremath{3449 \pm 70}\xspace} 
\newcommand{\jjhiLOGG}{\ensuremath{4.330 \pm 0.080}\xspace} 
\newcommand{\fcabLOGG}{\ensuremath{4.180 \pm 0.080}\xspace} 
\newcommand{\eidfLOGG}{\ensuremath{4.420 \pm 0.080}\xspace} 
\newcommand{\hehhLOGG}{\ensuremath{4.450 \pm 0.080}\xspace} 
\newcommand{\iahiLOGG}{\ensuremath{4.020 \pm 0.080}\xspace} 
\newcommand{\aiggLOGG}{\ensuremath{4.400 \pm 0.080}\xspace} 
\newcommand{\jgjjLOGG}{\ensuremath{4.450 \pm 0.083}\xspace} 
\newcommand{\hhedLOGG}{\ensuremath{3.580 \pm 0.080}\xspace} 
\newcommand{\befbLOGG}{\ensuremath{4.530 \pm 0.080}\xspace} 
 
\newcommand{\eaccLOGG}{\ensuremath{4.430 \pm 0.080}\xspace} 
\newcommand{\dcijLOGG}{\ensuremath{3.900 \pm 0.080}\xspace} 
\newcommand{\bejgLOGG}{\ensuremath{4.470 \pm 0.080}\xspace} 
\newcommand{\bcgdLOGG}{\ensuremath{4.320 \pm 0.083}\xspace} 
\newcommand{\fffdLOGG}{\ensuremath{4.390 \pm 0.081}\xspace} 
\newcommand{\dcbjLOGG}{\ensuremath{3.990 \pm 0.080}\xspace} 
\newcommand{\ddeiLOGG}{\ensuremath{4.550 \pm 0.084}\xspace} 
\newcommand{\fdijLOGG}{\ensuremath{4.450 \pm 0.085}\xspace} 
\newcommand{\gafeLOGG}{\ensuremath{4.280 \pm 0.081}\xspace} 
\newcommand{\bhiiLOGG}{\ensuremath{4.450 \pm 0.080}\xspace} 
\newcommand{\bbggLOGG}{\ensuremath{4.530 \pm 0.080}\xspace} 
\newcommand{\dbfaLOGG}{\ensuremath{4.270 \pm 0.080}\xspace} 
\newcommand{\ccdhLOGG}{\ensuremath{4.180 \pm 0.081}\xspace} 
\newcommand{\eicdLOGG}{\ensuremath{4.420 \pm 0.080}\xspace} 
\newcommand{\bggeLOGG}{\ensuremath{4.120 \pm 0.081}\xspace} 
\newcommand{\gcidLOGG}{\ensuremath{4.23 \pm 0.10}\xspace} 
\newcommand{\baeiLOGG}{\ensuremath{4.31 \pm 0.10}\xspace} 
\newcommand{\bjieLOGG}{\ensuremath{4.490 \pm 0.083}\xspace} 
\newcommand{\ggacLOGG}{\ensuremath{4.57 \pm 0.20}\xspace} 
\newcommand{\egijLOGG}{\ensuremath{4.510 \pm 0.085}\xspace} 
\newcommand{\hadfLOGG}{\ensuremath{4.470 \pm 0.085}\xspace} 
\newcommand{\hidfLOGG}{\ensuremath{4.580 \pm 0.081}\xspace} 
\newcommand{\bebbLOGG}{\ensuremath{4.60 \pm 0.20}\xspace} 
\newcommand{\ecdiLOGG}{\ensuremath{4.14 \pm 0.10}\xspace} 
\newcommand{\gedgLOGG}{\ensuremath{4.530 \pm 0.083}\xspace} 
\newcommand{\jgbhLOGG}{\ensuremath{4.630 \pm 0.081}\xspace} 
\newcommand{\ghijLOGG}{\ensuremath{4.480 \pm 0.081}\xspace} 
\newcommand{\fdecLOGG}{\ensuremath{4.350 \pm 0.085}\xspace} 
\newcommand{\deffLOGG}{\ensuremath{4.420 \pm 0.081}\xspace} 
\newcommand{\eiadLOGG}{\ensuremath{4.380 \pm 0.085}\xspace} 
\newcommand{\ifgjLOGG}{\ensuremath{4.63 \pm 0.20}\xspace} 
\newcommand{\cffcLOGG}{\ensuremath{4.60 \pm 0.20}\xspace} 
\newcommand{\jjhiFEH}{\ensuremath{-0.250 \pm 0.044}\xspace} 
\newcommand{\fcabFEH}{\ensuremath{0.140 \pm 0.044}\xspace} 
\newcommand{\eidfFEH}{\ensuremath{-0.110 \pm 0.044}\xspace} 
\newcommand{\hehhFEH}{\ensuremath{0.050 \pm 0.044}\xspace} 
\newcommand{\iahiFEH}{\ensuremath{0.350 \pm 0.043}\xspace} 
\newcommand{\aiggFEH}{\ensuremath{0.300 \pm 0.044}\xspace} 
\newcommand{\jgjjFEH}{\ensuremath{0.180 \pm 0.046}\xspace} 
\newcommand{\hhedFEH}{\ensuremath{0.430 \pm 0.043}\xspace} 
\newcommand{\befbFEH}{\ensuremath{0.040 \pm 0.043}\xspace} 
 
\newcommand{\eaccFEH}{\ensuremath{-0.440 \pm 0.044}\xspace} 
\newcommand{\dcijFEH}{\ensuremath{0.240 \pm 0.044}\xspace} 
\newcommand{\bejgFEH}{\ensuremath{0.050 \pm 0.043}\xspace} 
\newcommand{\bcgdFEH}{\ensuremath{0.230 \pm 0.046}\xspace} 
\newcommand{\fffdFEH}{\ensuremath{0.040 \pm 0.044}\xspace} 
\newcommand{\dcbjFEH}{\ensuremath{-0.050 \pm 0.044}\xspace} 
\newcommand{\ddeiFEH}{\ensuremath{0.000 \pm 0.046}\xspace} 
\newcommand{\fdijFEH}{\ensuremath{0.220 \pm 0.047}\xspace} 
\newcommand{\gafeFEH}{\ensuremath{0.040 \pm 0.044}\xspace} 
\newcommand{\bhiiFEH}{\ensuremath{-0.050 \pm 0.044}\xspace} 
\newcommand{\bbggFEH}{\ensuremath{-0.270 \pm 0.044}\xspace} 
\newcommand{\dbfaFEH}{\ensuremath{0.370 \pm 0.044}\xspace} 
\newcommand{\ccdhFEH}{\ensuremath{-0.000 \pm 0.044}\xspace} 
\newcommand{\eicdFEH}{\ensuremath{0.060 \pm 0.044}\xspace} 
\newcommand{\bggeFEH}{\ensuremath{-0.060 \pm 0.044}\xspace} 
\newcommand{\gcidFEH}{\ensuremath{-0.135 \pm 0.060}\xspace} 
\newcommand{\baeiFEH}{\ensuremath{0.038 \pm 0.060}\xspace} 
\newcommand{\bjieFEH}{\ensuremath{-0.030 \pm 0.046}\xspace} 
\newcommand{\ggacFEH}{\ensuremath{-0.10 \pm 0.12}\xspace} 
\newcommand{\egijFEH}{\ensuremath{-0.100 \pm 0.047}\xspace} 
\newcommand{\hadfFEH}{\ensuremath{-0.070 \pm 0.047}\xspace} 
\newcommand{\hidfFEH}{\ensuremath{-0.450 \pm 0.045}\xspace} 
\newcommand{\bebbFEH}{\ensuremath{0.08 \pm 0.12}\xspace} 
\newcommand{\ecdiFEH}{\ensuremath{0.353 \pm 0.060}\xspace} 
\newcommand{\gedgFEH}{\ensuremath{-0.080 \pm 0.046}\xspace} 
\newcommand{\jgbhFEH}{\ensuremath{-0.710 \pm 0.045}\xspace} 
\newcommand{\ghijFEH}{\ensuremath{0.120 \pm 0.045}\xspace} 
\newcommand{\fdecFEH}{\ensuremath{0.180 \pm 0.047}\xspace} 
\newcommand{\deffFEH}{\ensuremath{0.240 \pm 0.045}\xspace} 
\newcommand{\eiadFEH}{\ensuremath{0.030 \pm 0.048}\xspace} 
\newcommand{\ifgjFEH}{\ensuremath{0.040 \pm 0.080}\xspace} 
\newcommand{\cffcFEH}{\ensuremath{0.00 \pm 0.12}\xspace} 
\newcommand{\jjhiSVAL}{0.1407\xspace} 
\newcommand{\fcabSVAL}{0.1337\xspace} 
\newcommand{\eidfSVAL}{0.1616\xspace} 
\newcommand{\hehhSVAL}{0.186\xspace} 
\newcommand{\iahiSVAL}{0.1382\xspace} 
\newcommand{\aiggSVAL}{0.2772\xspace} 
\newcommand{\jgjjSVAL}{0.2936\xspace} 
\newcommand{\hhedSVAL}{0.173\xspace} 
\newcommand{\befbSVAL}{0.3516\xspace} 
 
\newcommand{\eaccSVAL}{0.1411\xspace} 
\newcommand{\dcijSVAL}{0.119\xspace} 
\newcommand{\bejgSVAL}{0.196\xspace} 
\newcommand{\bcgdSVAL}{0.1435\xspace} 
\newcommand{\fffdSVAL}{0.1586\xspace} 
\newcommand{\dcbjSVAL}{0.1276\xspace} 
\newcommand{\ddeiSVAL}{0.4628\xspace} 
\newcommand{\fdijSVAL}{0.2097\xspace} 
\newcommand{\gafeSVAL}{0.1432\xspace} 
\newcommand{\bhiiSVAL}{0.1391\xspace} 
\newcommand{\bbggSVAL}{0.1761\xspace} 
\newcommand{\dbfaSVAL}{0.1325\xspace} 
\newcommand{\ccdhSVAL}{0.1414\xspace} 
\newcommand{\eicdSVAL}{0.1419\xspace} 
\newcommand{\bggeSVAL}{0.1335\xspace} 
\newcommand{\gcidSVAL}{0.1761\xspace} 
\newcommand{\baeiSVAL}{0.1432\xspace} 
\newcommand{\bjieSVAL}{0.1567\xspace} 
\newcommand{\ggacSVAL}{0.2961\xspace} 
\newcommand{\egijSVAL}{0.1663\xspace} 
\newcommand{\hadfSVAL}{0.1481\xspace} 
\newcommand{\hidfSVAL}{0.1387\xspace} 
\newcommand{\bebbSVAL}{0.612\xspace} 
\newcommand{\ecdiSVAL}{0.127\xspace} 
\newcommand{\gedgSVAL}{0.1787\xspace} 
\newcommand{\jgbhSVAL}{0.2003\xspace} 
\newcommand{\ghijSVAL}{0.1627\xspace} 
\newcommand{\fdecSVAL}{0.183\xspace} 
\newcommand{\deffSVAL}{0.1538\xspace} 
\newcommand{\eiadSVAL}{0.1433\xspace} 
\newcommand{\ifgjSVAL}{0.728\xspace} 
\newcommand{\cffcSVAL}{0.5806\xspace} 
\newcommand{\jjhiRPHK}{$-$5.128\xspace} 
\newcommand{\fcabRPHK}{$-$5.193\xspace} 
\newcommand{\eidfRPHK}{$-$4.984\xspace} 
\newcommand{\hehhRPHK}{$-$4.875\xspace} 
\newcommand{\iahiRPHK}{$-$5.16\xspace} 
\newcommand{\aiggRPHK}{$-$4.526\xspace} 
\newcommand{\jgjjRPHK}{$-$4.647\xspace} 
\newcommand{\hhedRPHK}{$-$5.06\xspace} 
\newcommand{\befbRPHK}{$-$4.603\xspace} 
 
\newcommand{\eaccRPHK}{$-$5.133\xspace} 
\newcommand{\dcijRPHK}{$-$5.393\xspace} 
\newcommand{\bejgRPHK}{$-$4.878\xspace} 
\newcommand{\bcgdRPHK}{$-$5.12\xspace} 
\newcommand{\fffdRPHK}{$-$4.999\xspace} 
\newcommand{\dcbjRPHK}{$-$5.267\xspace} 
\newcommand{\ddeiRPHK}{$-$4.594\xspace} 
\newcommand{\fdijRPHK}{$-$4.843\xspace} 
\newcommand{\gafeRPHK}{$-$5.112\xspace} 
\newcommand{\bhiiRPHK}{$-$5.155\xspace} 
\newcommand{\bbggRPHK}{$-$5.063\xspace} 
\newcommand{\dbfaRPHK}{$-$5.201\xspace} 
\newcommand{\ccdhRPHK}{$-$5.129\xspace} 
\newcommand{\eicdRPHK}{$-$5.135\xspace} 
\newcommand{\bggeRPHK}{$-$5.189\xspace} 
\newcommand{\gcidRPHK}{$-$4.923\xspace} 
\newcommand{\baeiRPHK}{$-$5.119\xspace} 
\newcommand{\bjieRPHK}{$-$5.061\xspace} 
\newcommand{\ggacRPHK}{$-$4.972\xspace} 
\newcommand{\egijRPHK}{$-$4.999\xspace} 
\newcommand{\hadfRPHK}{$-$5.093\xspace} 
\newcommand{\hidfRPHK}{$-$5.148\xspace} 
\newcommand{\bebbRPHK}{$-$4.627\xspace} 
\newcommand{\ecdiRPHK}{$-$5.273\xspace} 
\newcommand{\gedgRPHK}{$-$4.957\xspace} 
\newcommand{\jgbhRPHK}{$-$4.916\xspace} 
\newcommand{\ghijRPHK}{$-$5.037\xspace} 
\newcommand{\fdecRPHK}{$-$4.914\xspace} 
\newcommand{\deffRPHK}{$-$5.073\xspace} 
\newcommand{\eiadRPHK}{$-$5.119\xspace} 
\newcommand{\ifgjRPHK}{$-$4.534\xspace} 
\newcommand{\cffcRPHK}{$-$5.183\xspace} 
\newcommand{\jjhiVSINI}{\ensuremath{1.6 \pm 1.0}\xspace} 
\newcommand{\fcabVSINI}{\ensuremath{0.1 \pm 1.0}\xspace} 
\newcommand{\eidfVSINI}{\ensuremath{1.6 \pm 1.0}\xspace} 
\newcommand{\hehhVSINI}{\ensuremath{2.3 \pm 1.0}\xspace} 
\newcommand{\iahiVSINI}{\ensuremath{3.0 \pm 1.0}\xspace} 
\newcommand{\aiggVSINI}{\ensuremath{13.3 \pm 1.0}\xspace} 
\newcommand{\jgjjVSINI}{\ensuremath{2.8 \pm 1.0}\xspace} 
\newcommand{\hhedVSINI}{\ensuremath{0.1 \pm 1.0}\xspace} 
\newcommand{\befbVSINI}{\ensuremath{2.6 \pm 1.0}\xspace} 
 
\newcommand{\eaccVSINI}{\ensuremath{0.1 \pm 1.0}\xspace} 
\newcommand{\dcijVSINI}{\ensuremath{9.4 \pm 1.0}\xspace} 
\newcommand{\bejgVSINI}{\ensuremath{1.8 \pm 1.0}\xspace} 
\newcommand{\bcgdVSINI}{\ensuremath{1.5 \pm 1.0}\xspace} 
\newcommand{\fffdVSINI}{\ensuremath{2.0 \pm 1.0}\xspace} 
\newcommand{\dcbjVSINI}{\ensuremath{4.0 \pm 1.0}\xspace} 
\newcommand{\ddeiVSINI}{\ensuremath{2.7 \pm 1.0}\xspace} 
\newcommand{\fdijVSINI}{\ensuremath{2.7 \pm 1.0}\xspace} 
\newcommand{\gafeVSINI}{\ensuremath{1.9 \pm 1.0}\xspace} 
\newcommand{\bhiiVSINI}{\ensuremath{1.1 \pm 1.0}\xspace} 
\newcommand{\bbggVSINI}{\ensuremath{0.1 \pm 1.0}\xspace} 
\newcommand{\dbfaVSINI}{\ensuremath{2.2 \pm 1.0}\xspace} 
\newcommand{\ccdhVSINI}{\ensuremath{1.9 \pm 1.0}\xspace} 
\newcommand{\eicdVSINI}{\ensuremath{1.7 \pm 1.0}\xspace} 
\newcommand{\bggeVSINI}{\ensuremath{5.6 \pm 1.0}\xspace} 
\newcommand{\gcidVSINI}{\ensuremath{0.1 \pm 1.0}\xspace} 
\newcommand{\baeiVSINI}{\ensuremath{1.8 \pm 1.0}\xspace} 
\newcommand{\bjieVSINI}{\ensuremath{0.7 \pm 1.0}\xspace} 
\newcommand{\ggacVSINI}{---\xspace} 
\newcommand{\egijVSINI}{\ensuremath{1.9 \pm 1.0}\xspace} 
\newcommand{\hadfVSINI}{\ensuremath{1.5 \pm 1.0}\xspace} 
\newcommand{\hidfVSINI}{\ensuremath{0.1 \pm 1.0}\xspace} 
\newcommand{\bebbVSINI}{---\xspace} 
\newcommand{\ecdiVSINI}{\ensuremath{2.5 \pm 1.0}\xspace} 
\newcommand{\gedgVSINI}{\ensuremath{1.3 \pm 1.0}\xspace} 
\newcommand{\jgbhVSINI}{\ensuremath{0.1 \pm 1.0}\xspace} 
\newcommand{\ghijVSINI}{\ensuremath{1.6 \pm 1.0}\xspace} 
\newcommand{\fdecVSINI}{\ensuremath{2.1 \pm 1.0}\xspace} 
\newcommand{\deffVSINI}{\ensuremath{1.9 \pm 1.0}\xspace} 
\newcommand{\eiadVSINI}{\ensuremath{1.6 \pm 1.0}\xspace} 
\newcommand{\ifgjVSINI}{---\xspace} 
\newcommand{\cffcVSINI}{---\xspace} 
\newcommand{\jjhiMSTAR}{\ensuremath{0.92 \pm 0.04}\xspace} 
\newcommand{\fcabMSTAR}{\ensuremath{1.17 \pm 0.07}\xspace} 
\newcommand{\eidfMSTAR}{\ensuremath{0.95 \pm 0.04}\xspace} 
\newcommand{\hehhMSTAR}{\ensuremath{0.98 \pm 0.04}\xspace} 
\newcommand{\iahiMSTAR}{\ensuremath{1.09 \pm 0.07}\xspace} 
\newcommand{\aiggMSTAR}{\ensuremath{1.21 \pm 0.04}\xspace} 
\newcommand{\jgjjMSTAR}{\ensuremath{0.92 \pm 0.04}\xspace} 
\newcommand{\hhedMSTAR}{\ensuremath{1.38 \pm 0.16}\xspace} 
\newcommand{\befbMSTAR}{\ensuremath{0.84 \pm 0.03}\xspace} 
 
\newcommand{\eaccMSTAR}{\ensuremath{0.83 \pm 0.03}\xspace} 
\newcommand{\dcijMSTAR}{\ensuremath{1.48 \pm 0.12}\xspace} 
\newcommand{\bejgMSTAR}{\ensuremath{0.90 \pm 0.04}\xspace} 
\newcommand{\bcgdMSTAR}{\ensuremath{1.04 \pm 0.05}\xspace} 
\newcommand{\fffdMSTAR}{\ensuremath{1.02 \pm 0.04}\xspace} 
\newcommand{\dcbjMSTAR}{\ensuremath{1.06 \pm 0.07}\xspace} 
\newcommand{\ddeiMSTAR}{\ensuremath{0.76 \pm 0.03}\xspace} 
\newcommand{\fdijMSTAR}{\ensuremath{0.94 \pm 0.04}\xspace} 
\newcommand{\gafeMSTAR}{\ensuremath{1.03 \pm 0.05}\xspace} 
\newcommand{\bhiiMSTAR}{\ensuremath{0.91 \pm 0.04}\xspace} 
\newcommand{\bbggMSTAR}{\ensuremath{0.70 \pm 0.03}\xspace} 
\newcommand{\dbfaMSTAR}{\ensuremath{1.01 \pm 0.05}\xspace} 
\newcommand{\ccdhMSTAR}{\ensuremath{1.01 \pm 0.05}\xspace} 
\newcommand{\eicdMSTAR}{\ensuremath{0.91 \pm 0.04}\xspace} 
\newcommand{\bggeMSTAR}{\ensuremath{1.11^{+0.08}_{-0.06} }\xspace} 
\newcommand{\gcidMSTAR}{\ensuremath{0.89 \pm 0.04}\xspace} 
\newcommand{\baeiMSTAR}{\ensuremath{0.99 \pm 0.05}\xspace} 
\newcommand{\bjieMSTAR}{\ensuremath{0.84 \pm 0.03}\xspace} 
\newcommand{\ggacMSTAR}{\ensuremath{0.67 \pm 0.03}\xspace} 
\newcommand{\egijMSTAR}{\ensuremath{0.85 \pm 0.03}\xspace} 
\newcommand{\hadfMSTAR}{\ensuremath{0.88 \pm 0.04}\xspace} 
\newcommand{\hidfMSTAR}{\ensuremath{0.88 \pm 0.03}\xspace} 
\newcommand{\bebbMSTAR}{\ensuremath{0.72 \pm 0.03}\xspace} 
\newcommand{\ecdiMSTAR}{\ensuremath{1.17 \pm 0.09}\xspace} 
\newcommand{\gedgMSTAR}{\ensuremath{0.84 \pm 0.03}\xspace} 
\newcommand{\jgbhMSTAR}{\ensuremath{0.67 \pm 0.02}\xspace} 
\newcommand{\ghijMSTAR}{\ensuremath{0.88 \pm 0.03}\xspace} 
\newcommand{\fdecMSTAR}{\ensuremath{0.96 \pm 0.04}\xspace} 
\newcommand{\deffMSTAR}{\ensuremath{0.93 \pm 0.04}\xspace} 
\newcommand{\eiadMSTAR}{\ensuremath{0.97 \pm 0.04}\xspace} 
\newcommand{\ifgjMSTAR}{\ensuremath{0.71 \pm 0.03}\xspace} 
\newcommand{\cffcMSTAR}{\ensuremath{0.32 \pm 0.06}\xspace} 
\newcommand{\jjhiRSTAR}{\ensuremath{1.093 \pm 0.038}\xspace} 
\newcommand{\fcabRSTAR}{\ensuremath{1.387 \pm 0.049}\xspace} 
\newcommand{\eidfRSTAR}{\ensuremath{0.998 \pm 0.036}\xspace} 
\newcommand{\hehhRSTAR}{\ensuremath{0.952 \pm 0.035}\xspace} 
\newcommand{\iahiRSTAR}{\ensuremath{1.679 \pm 0.065}\xspace} 
\newcommand{\aiggRSTAR}{\ensuremath{1.227 \pm 0.045}\xspace} 
\newcommand{\jgjjRSTAR}{\ensuremath{1.138 \pm 0.044}\xspace} 
\newcommand{\hhedRSTAR}{\ensuremath{3.08 \pm 0.14}\xspace} 
\newcommand{\befbRSTAR}{\ensuremath{0.774 \pm 0.032}\xspace} 
 
\newcommand{\eaccRSTAR}{\ensuremath{0.884 \pm 0.063}\xspace} 
\newcommand{\dcijRSTAR}{\ensuremath{2.66 \pm 0.11}\xspace} 
\newcommand{\bejgRSTAR}{\ensuremath{0.924 \pm 0.036}\xspace} 
\newcommand{\bcgdRSTAR}{\ensuremath{1.131^{+0.123}_{-0.095} }\xspace} 
\newcommand{\fffdRSTAR}{\ensuremath{1.058 \pm 0.039}\xspace} 
\newcommand{\dcbjRSTAR}{\ensuremath{1.79 \pm 0.10}\xspace} 
\newcommand{\ddeiRSTAR}{\ensuremath{0.717 \pm 0.031}\xspace} 
\newcommand{\fdijRSTAR}{\ensuremath{0.905 \pm 0.035}\xspace} 
\newcommand{\gafeRSTAR}{\ensuremath{1.236 \pm 0.047}\xspace} 
\newcommand{\bhiiRSTAR}{\ensuremath{1.026 \pm 0.040}\xspace} 
\newcommand{\bbggRSTAR}{\ensuremath{0.710 \pm 0.030}\xspace} 
\newcommand{\dbfaRSTAR}{\ensuremath{1.144 \pm 0.049}\xspace} 
\newcommand{\ccdhRSTAR}{\ensuremath{1.247 \pm 0.045}\xspace} 
\newcommand{\eicdRSTAR}{\ensuremath{0.995 \pm 0.039}\xspace} 
\newcommand{\bggeRSTAR}{\ensuremath{1.565 \pm 0.065}\xspace} 
\newcommand{\gcidRSTAR}{\ensuremath{1.154 \pm 0.044}\xspace} 
\newcommand{\baeiRSTAR}{\ensuremath{1.094 \pm 0.041}\xspace} 
\newcommand{\bjieRSTAR}{\ensuremath{0.855 \pm 0.034}\xspace} 
\newcommand{\ggacRSTAR}{\ensuremath{0.696 \pm 0.023}\xspace} 
\newcommand{\egijRSTAR}{\ensuremath{0.879 \pm 0.034}\xspace} 
\newcommand{\hadfRSTAR}{\ensuremath{0.956 \pm 0.037}\xspace} 
\newcommand{\hidfRSTAR}{\ensuremath{0.840^{+0.047}_{-0.034} }\xspace} 
\newcommand{\bebbRSTAR}{\ensuremath{0.699 \pm 0.023}\xspace} 
\newcommand{\ecdiRSTAR}{\ensuremath{1.279 \pm 0.052}\xspace} 
\newcommand{\gedgRSTAR}{\ensuremath{0.809 \pm 0.032}\xspace} 
\newcommand{\jgbhRSTAR}{\ensuremath{0.638 \pm 0.020}\xspace} 
\newcommand{\ghijRSTAR}{\ensuremath{0.876 \pm 0.036}\xspace} 
\newcommand{\fdecRSTAR}{\ensuremath{1.060 \pm 0.042}\xspace} 
\newcommand{\deffRSTAR}{\ensuremath{0.910 \pm 0.038}\xspace} 
\newcommand{\eiadRSTAR}{\ensuremath{0.995 \pm 0.042}\xspace} 
\newcommand{\ifgjRSTAR}{\ensuremath{0.675 \pm 0.033}\xspace} 
\newcommand{\cffcRSTAR}{\ensuremath{0.468 \pm 0.019}\xspace} 
\newcommand{\jfjgTEFF}{\ensuremath{4507 \pm 110}\xspace} 
\newcommand{\jfjgLOGG}{\ensuremath{4.6 \pm 0.2}\xspace} 
\newcommand{\jfjgFEH}{\ensuremath{-0.04 \pm 0.08}\xspace} 
\newcommand{\jfjgSVAL}{0.47\xspace} 
\newcommand{\jfjgRPHK}{$-$4.735\xspace} 
\newcommand{\jfjgVSINI}{---\xspace} 
\newcommand{\jfjgMSTAR}{\ensuremath{0.69 \pm 0.03}\xspace} 
\newcommand{\jfjgRSTAR}{\ensuremath{0.68 \pm 0.03}\xspace} 
\newcommand{\hbagTEFF}{\ensuremath{5472 \pm 110}\xspace} 
\newcommand{\hbagLOGG}{\ensuremath{3.70 \pm 0.10}\xspace} 
\newcommand{\hbagFEH}{\ensuremath{0.44 \pm 0.09}\xspace} 
\newcommand{\hbagSVAL}{0.1334{\xspace}}
\newcommand{\hbagRPHK}{\ensuremath{$-$5.199}\xspace} 
\newcommand{\hbagVSINI}{\ensuremath{2.62 \pm 1.0}\xspace} 
\newcommand{\hbagMSTAR}{\ensuremath{1.17 \pm 0.03}\xspace} 
\newcommand{\hbagRSTAR}{\ensuremath{1.95 \pm 0.18}\xspace} 
\newcommand{\hagfTEFF}{\ensuremath{3896 \pm 189}\xspace} 
\newcommand{\hagfLOGG}{\ensuremath{4.734 \pm 0.062}\xspace} 
\newcommand{\hagfFEH}{\ensuremath{-0.32 \pm 0.13}\xspace} 
\newcommand{\hagfSVAL}{---\xspace} 
\newcommand{\hagfRPHK}{---\xspace} 
\newcommand{\hagfVSINI}{---\xspace} 
\newcommand{\hagfMSTAR}{\ensuremath{0.601 \pm 0.089}\xspace} 
\newcommand{\hagfRSTAR}{\ensuremath{0.561 \pm 0.068}\xspace} 
\newcommand{\ihidTEFF}{\ensuremath{5520 \pm 60}\xspace} 
\newcommand{\ihidLOGG}{\ensuremath{4.50 \pm 0.05}\xspace} 
\newcommand{\ihidFEH}{\ensuremath{0.08 \pm 0.04}\xspace} 
\newcommand{\ihidSVAL}{---\xspace} 
\newcommand{\ihidRPHK}{---\xspace} 
\newcommand{\ihidVSINI}{\ensuremath{< 2}\xspace} 
\newcommand{\ihidMSTAR}{\ensuremath{0.934\pm0.038}\xspace} 
\newcommand{\ihidRSTAR}{\ensuremath{0.899^{+0.035}_{-0.033}}\xspace} 
\newcommand{\bdiaTEFF}{\ensuremath{6320^{+60}_{-30}}\xspace} 
\newcommand{\bdiaLOGG}{\ensuremath{4.294 \pm 0.006}\xspace} 
\newcommand{\bdiaFEH}{\ensuremath{-0.10 \pm 0.07}\xspace} 
\newcommand{\bdiaSVAL}{---\xspace} 
\newcommand{\bdiaRPHK}{\ensuremath{-4.78 \pm 0.03}\xspace} 
\newcommand{\bdiaVSINI}{\ensuremath{5.6 \pm 0.5}\xspace} 
\newcommand{\bdiaMSTAR}{\ensuremath{1.16 \pm 0.04}\xspace} 
\newcommand{\bdiaRSTAR}{\ensuremath{1.273 \pm 0.015}\xspace} 
\newcommand{\fdfaTEFF}{\ensuremath{5322 \pm 100}\xspace} 
\newcommand{\fdfaLOGG}{\ensuremath{4.51 \pm 0.08}\xspace} 
\newcommand{\fdfaFEH}{\ensuremath{0.06 \pm 0.05}\xspace} 
\newcommand{\fdfaSVAL}{---\xspace} 
\newcommand{\fdfaRPHK}{---\xspace} 
\newcommand{\fdfaVSINI}{\ensuremath{< 2}\xspace} 
\newcommand{\fdfaMSTAR}{\ensuremath{0.88 \pm 0.03}\xspace} 
\newcommand{\fdfaRSTAR}{\ensuremath{0.82 \pm 0.03}\xspace} 
\newcommand{\ddigTEFF}{\ensuremath{5261 \pm 60}\xspace} 
\newcommand{\ddigLOGG}{\ensuremath{4.47 \pm 0.05}\xspace} 
\newcommand{\ddigFEH}{\ensuremath{0.04 \pm 0.05}\xspace} 
\newcommand{\ddigSVAL}{---\xspace} 
\newcommand{\ddigRPHK}{\ensuremath{-5.04}\xspace} 
\newcommand{\ddigVSINI}{\ensuremath{1.17 \pm 1.1}\xspace} 
\newcommand{\ddigMSTAR}{\ensuremath{0.866 \pm 0.033}\xspace} 
\newcommand{\ddigRSTAR}{\ensuremath{0.872 \pm 0.057}\xspace} 
\newcommand{\bajiTEFF}{\ensuremath{5625 \pm 60}\xspace} 
\newcommand{\bajiLOGG}{\ensuremath{4.29 \pm 0.05}\xspace} 
\newcommand{\bajiFEH}{\ensuremath{0.34 \pm 0.04}\xspace} 
\newcommand{\bajiSVAL}{0.1269\xspace} 
\newcommand{\bajiRPHK}{\ensuremath{-5.2439}\xspace} 
\newcommand{\bajiVSINI}{---\xspace} 
\newcommand{\bajiMSTAR}{\ensuremath{1.07 \pm 0.06}\xspace} 
\newcommand{\bajiRSTAR}{\ensuremath{1.16 \pm 0.04}\xspace} 
\newcommand{\egbeTEFF}{\ensuremath{4300^{+107}_{-100}}\xspace} 
\newcommand{\egbeLOGG}{\ensuremath{4.566 \pm 0.036}\xspace} 
\newcommand{\egbeFEH}{\ensuremath{0.376 \pm 0.095}\xspace} 
\newcommand{\egbeSVAL}{0.78\xspace} 
\newcommand{\egbeRPHK}{\ensuremath{-4.7339}\xspace} 
\newcommand{\egbeVSINI}{\ensuremath{2.2 \pm 1.0}\xspace} 
\newcommand{\egbeMSTAR}{\ensuremath{0.688 \pm 0.069}\xspace} 
\newcommand{\egbeRSTAR}{\ensuremath{0.715^{+0.043}_{-0.040}}\xspace} 
\newcommand{\gghbTEFF}{\ensuremath{5474 \pm 60}\xspace} 
\newcommand{\gghbLOGG}{\ensuremath{3.99 \pm 0.05}\xspace} 
\newcommand{\gghbFEH}{\ensuremath{0.33 \pm 0.04}\xspace} 
\newcommand{\gghbSVAL}{0.1225\xspace} 
\newcommand{\gghbRPHK}{\ensuremath{-5.2928}\xspace} 
\newcommand{\gghbVSINI}{\ensuremath{2.83 \pm 1.0}\xspace} 
\newcommand{\gghbMSTAR}{\ensuremath{1.121^{+0.065}_{-0.053}}\xspace} 
\newcommand{\gghbRSTAR}{\ensuremath{1.75 \pm 0.14}\xspace} 
\newcommand{\hieeTEFF}{\ensuremath{6364 \pm 87}\xspace} 
\newcommand{\hieeLOGG}{\ensuremath{4.291 \pm 0.025}\xspace} 
\newcommand{\hieeFEH}{\ensuremath{-0.22 \pm 0.09}\xspace} 
\newcommand{\hieeSVAL}{0.14\xspace} 
\newcommand{\hieeRPHK}{---\xspace} 
\newcommand{\hieeVSINI}{\ensuremath{13.2 \pm 1.0}\xspace} 
\newcommand{\hieeMSTAR}{\ensuremath{1.154 \pm 0.042}\xspace} 
\newcommand{\hieeRSTAR}{\ensuremath{1.269 \pm 0.024}\xspace} 
\newcommand{\jifiTEFF}{\ensuremath{4294 \pm 52}\xspace} 
\newcommand{\jifiLOGG}{\ensuremath{4.682 \pm 0.021}\xspace} 
\newcommand{\jifiFEH}{\ensuremath{-0.26 \pm 0.08}\xspace} 
\newcommand{\jifiSVAL}{0.6446\xspace} 
\newcommand{\jifiRPHK}{---\xspace} 
\newcommand{\jifiVSINI}{\ensuremath{2.43 \pm 1.0}\xspace} 
\newcommand{\jifiMSTAR}{\ensuremath{0.593 \pm 0.018}\xspace} 
\newcommand{\jifiRSTAR}{\ensuremath{0.579 \pm 0.018}\xspace} 
\newcommand{\ecdcTEFF}{\ensuremath{4425 \pm 70}\xspace} 
\newcommand{\ecdcLOGG}{\ensuremath{4.633 \pm 0.012}\xspace} 
\newcommand{\ecdcFEH}{\ensuremath{0.02 \pm 0.09}\xspace} 
\newcommand{\ecdcSVAL}{---\xspace} 
\newcommand{\ecdcRPHK}{---\xspace} 
\newcommand{\ecdcVSINI}{\ensuremath{< 2}\xspace} 
\newcommand{\ecdcMSTAR}{\ensuremath{0.683^{+0.017}_{-0.016}}\xspace} 
\newcommand{\ecdcRSTAR}{\ensuremath{0.67 \pm 0.02}\xspace} 
\newcommand{\hbdaTEFF}{\ensuremath{4232 \pm 70}\xspace} 
\newcommand{\hbdaLOGG}{\ensuremath{4.44 \pm 0.10}\xspace} 
\newcommand{\hbdaFEH}{\ensuremath{0.14 \pm 0.09}\xspace} 
\newcommand{\hbdaSVAL}{0.7276\xspace} 
\newcommand{\hbdaRPHK}{\ensuremath{-4.4024}\xspace} 
\newcommand{\hbdaVSINI}{\ensuremath{3 \pm 1}\xspace} 
\newcommand{\hbdaMSTAR}{\ensuremath{0.70 \pm 0.02}\xspace} 
\newcommand{\hbdaRSTAR}{\ensuremath{0.71 \pm 0.10}\xspace} 
\newcommand{\gtthreesig}{51}
\newcommand{\gtfivesig}{32}
\newcommand{\Ntransit}{81}
\newcommand{\Nnontrans}{5}
\newcommand{\Npl}{86}
\newcommand{\Nstars}{55}

\shortauthors{Howard {et~al.}}
\shorttitle{Planet Masses, Radii, and Orbits from NASA's K2 Mission$^1$}

\begin{document}
\pagenumbering{arabic}


 \title{Planet Masses, Radii, and Orbits from NASA's K2 Mission}
\author{Andrew W.\ Howard}    
\affiliation{Department of Astronomy, California Institute of Technology, Pasadena, California, USA}

\author{Evan Sinukoff}    
\affiliation{Department of Astronomy, California Institute of Technology, Pasadena, California, USA}
\affiliation{Institute for Astronomy, University of Hawaii, 2680 Woodlawn Drive, Honolulu, HI, USA} 

\author[0000-0002-3199-2888]{Sarah Blunt}    
\affiliation{Department of Astronomy, California Institute of Technology, Pasadena, California, USA}
\affiliation{Center for Interdisciplinary Exploration and Research in Astrophysics (CIERA) and Department of Physics and Astronomy, Northwestern University, Evanston, IL 60208, USA} 
\affiliation{Department of Astronomy \& Astrophysics, University of California, Santa Cruz, CA, USA} 
\affiliation{Center for Astrophysics \textbar\ Harvard \& Smithsonian, 60 Garden Street, Cambridge, MA 02138, USA}

\author{Erik A.\ Petigura}    
\affiliation{Department of Physics \& Astronomy, University of California Los Angeles, Los Angeles, CA, USA} 

\author{Ian J.\ M. Crossfield}    
\affiliation{Department of Physics and Astronomy, University of Kansas, Lawrence, KS, USA}

\author{Howard Isaacson}    
\affiliation{Astronomy Department, University of California, Berkeley, CA, USA} 

\author{Molly Kosiarek}    
\affiliation{Department of Astronomy \& Astrophysics, University of California, Santa Cruz, CA, USA} 

\author{Ryan A.\ Rubenzahl}    
\affiliation{Department of Astronomy, California Institute of Technology, Pasadena, California, USA}

\author{John M.\ Brewer}    
\affiliation{Department of Physics and Astronomy, San Francisco State University, 1600 Holloway Ave., San Francisco, CA 94132, USA} 

\author{Benjamin J.\ Fulton}    
\affiliation{Department of Astronomy, California Institute of Technology, Pasadena, California, USA}
\affiliation{IPAC - NASA Exoplanet Science Institute, California Institute of Technology, 770 S. Wilson Ave, Pasadena, CA, USA} 

\author{Courtney D.\ Dressing}    
\affiliation{Astronomy Department, University of California, Berkeley, CA, USA} 

\author{Lea A.\ Hirsch}    
\affiliation{Department of Chemical \& Physical Sciences, University of Toronto Mississauga, 3359 Mississauga Road, Mississauga, Ontario L5L 1C6, Canada} 

\author{Heather Knutson}    
\affiliation{Geological and Planetary Sciences, California Institute of Technology, Pasadena, CA, USA} 

\author[0000-0002-4881-3620]{John H.\ Livingston}    
\affiliation{Astrobiology Center, 2-21-1 Osawa, Mitaka, Tokyo 181-8588, Japan} 
\affiliation{National Astronomical Observatory of Japan, 2-21-1 Osawa, Mitaka, Tokyo 181-8588, Japan}
\affiliation{Astronomical Science Program, The Graduate University for Advanced Studies, SOKENDAI, 2-21-1 Osawa, Mitaka, Tokyo 181-8588, Japan}

\author{Sean M.\ Mills}    
\affiliation{Department of Astronomy, California Institute of Technology, Pasadena, California, USA}

\author[0000-0001-8127-5775]{Arpita Roy}    
\affiliation{Space Telescope Science Institute, 3700 San Martin Drive, Baltimore, MD 21218, USA} 
\affiliation{Astrophysics \& Space Institute, Schmidt Sciences, New York, NY 10011, USA}

\author[0000-0002-3725-3058]{Lauren M.\ Weiss}    
\affiliation{Department of Physics and Astronomy, University of Notre Dame, Notre Dame, IN 46556, USA} 

\author{Bjorn Benneke}    
\affiliation{Institut de Recherche sur les Exoplan{\`e}tes, Universit{\'e} de Montr{\'e}al, Montr{\'e}al, QC, Canada}

\author{David R.\ Ciardi}    
\affiliation{IPAC - NASA Exoplanet Science Institute, California Institute of Technology, 770 S. Wilson Ave, Pasadena, CA, USA} 

\author{Jessie L.\ Christiansen}    
\affiliation{IPAC - NASA Exoplanet Science Institute, California Institute of Technology, 770 S. Wilson Ave, Pasadena, CA, USA} 

\author[0000-0001-9662-3496]{William D.\ Cochran}    
\affiliation{McDonald Observatory and Center for Planetary Systems Habitability, The University of Texas, Austin, TX, USA}

\author{Justin R.\ Crepp}    
\affiliation{Department of Physics and Astronomy, University of Notre Dame, Notre Dame, IN 46556, USA} 

\author{Erica Gonzales}    
\affiliation{Department of Astronomy \& Astrophysics, University of California, Santa Cruz, CA, USA}

\author{Brad M.\ S.\ Hansen}    
\affiliation{Department of Physics \& Astronomy, University of California Los Angeles, Los Angeles, CA, USA} 

\author{Kevin Hardegree-Ullman}    
\affiliation{IPAC - NASA Exoplanet Science Institute, California Institute of Technology, 770 S. Wilson Ave, Pasadena, CA, USA} 
\affiliation{Department of Physics and Astronomy, University of Toledo, Toledo, OH, 43606, USA}
\affiliation{Steward Observatory, The University of Arizona, Tucson, AZ 85721, USA}

\author{Steve B.\ Howell}    
\affiliation{NASA Ames Research Center, Moffett Field, CA, USA}

\author{S\'ebastien L\'epine}    
\affiliation{Department of Physics \& Astronomy, Georgia State University, 25 Park Place NE 605, Atlanta, GA, USA}

\author{Arturo O.\ Martinez}    
\affiliation{NASA Ames Research Center, Moffett Field, CA, USA}
\affiliation{Bay Area Environmental Research Institute, Moffett Field, CA, USA}

\author{Leslie A.\ Rogers}    
\affiliation{Department of Astronomy \& Astrophysics, University of Chicago, 5640 S. Ellis Ave, Chicago, IL, USA}

\author{Joshua E.\ Schlieder}    
\affiliation{NASA Goddard Space Flight Center, 8800 Greenbelt Road, Greenbelt, MD, USA}

\author{Michael Werner}    
\affiliation{Jet Propulsion Laboratory, California Institute of Technology, 4800 Oak Grove Drive, Pasadena, CA, USA}


\author{Alex S.\ Polanski}    
\affiliation{Department of Physics and Astronomy, University of Kansas, Lawrence, KS, USA}
\affiliation{Lowell Observatory, 1400 W. Mars Hill Rd. Flagstaff, AZ. 86001. USA}

\author{Isabel Angelo}    
\affiliation{Department of Physics \& Astronomy, University of California Los Angeles, Los Angeles, CA, USA} 

\author{Corey Beard}    
\affiliation{Department of Physics \& Astronomy, The University of California, Irvine, Irvine, CA 92697, USA}
\affiliation{NASA FINESST Fellow}

\author{Aida Behmard}    
\affiliation{Department of Astrophysics, American Museum of Natural History, 200 Central Park West, Manhattan, NY, USA}

\author{Luke G.\ Bouma}    
\affiliation{Department of Astronomy, California Institute of Technology, Pasadena, California, USA}

\author{Casey L.\ Brinkman}    
\affiliation{Institute for Astronomy, University of Hawaii, 2680 Woodlawn Drive, Honolulu, HI, USA} 

\author{Ashley Chontos}    
\affiliation{Department of Astrophysical Sciences, Princeton University, 4 Ivy Lane, Princeton, NJ 08544, USA}

\author{Fei Dai}    
\affiliation{Institute for Astronomy, University of Hawaii, 2680 Woodlawn Drive, Honolulu, HI, USA} 

\author{Paul A.\ Dalba}    
\affiliation{Department of Astronomy \& Astrophysics, University of California, Santa Cruz, CA, USA} 

\author[0000-0002-8965-3969]{Steven Giacalone}    
\affiliation{Department of Astronomy, California Institute of Technology, Pasadena, California, USA}

\author{Samuel K.\ Grunblatt}    
\affiliation{Department of Physics and Astronomy, University of Alabama, Box 870324, Tuscaloosa, AL 35487, USA}

\author{Michelle L.\ Hill}    
\affiliation{Department of Earth and Planetary Sciences, University of California, Riverside, CA 92521, USA}

\author{Stephen R.\ Kane}    
\affiliation{Department of Earth and Planetary Sciences, University of California, Riverside, CA 92521, USA}

\author{Jack Lubin}    
\affiliation{Department of Physics \& Astronomy, University of California Los Angeles, Los Angeles, CA, USA} 

\author{Andrew W.\ Mayo}    
\affiliation{Centre for Star and Planet Formation, Natural History Museum of Denmark \& Niels Bohr Institute, University of Copenhagen, Oster Voldgade 5-7, DK-1350 Copenhagen K., Denmark}

\author{Teo Mocnik}    
\affiliation{Gemini Observatory/NSF NOIRLab, Hilo, HI 96720, USA}

\author{Joseph M.\ Akana Murphy}    
\affiliation{Department of Astronomy \& Astrophysics, University of California, Santa Cruz, CA, USA} 

\author{Malena Rice}    
\affiliation{Department of Astronomy, Yale University, 219 Prospect Street, New Haven, CT 06511, USA}

\author{Lee J.\ Rosenthal}    
\affiliation{Institute for Astronomy, University of Hawaii, 2680 Woodlawn Drive, Honolulu, HI, USA} 


\author{Dakotah Tyler}    
\affiliation{Department of Physics \& Astronomy, University of California Los Angeles, Los Angeles, CA, USA} 

\author{Judah Van Zandt}    
\affiliation{Department of Physics \& Astronomy, University of California Los Angeles, Los Angeles, CA, USA} 

\author{Samuel W.\ Yee}    
\affiliation{Center for Astrophysics \textbar\ Harvard \& Smithsonian, 60 Garden Street, Cambridge, MA 02138, USA}

\begin{abstract}
We report the masses, sizes, and orbital properties of \Npl{} planets orbiting \Nstars{} stars observed by NASA's K2 Mission with follow-up Doppler measurements by the HIRES spectrometer at the W.\,M.\ Keck Observatory 
and the Automated Planet Finder at Lick Observatory.
Eighty-one of the planets were discovered from their transits in the K2 photometry, while 
five 
were found based on subsequent Doppler measurements of transiting planet host stars.  The sizes of the transiting planets range from Earth-size to larger than Jupiter (1--3 \rearth is typical), while the orbital periods range from less than a day to a few months.
For \gtfivesig{} of the planets, the Doppler signal was detected with significance greater than 5-$\sigma$ (\gtthreesig{} were detected with $>$3-$\sigma$ significance).
An important characteristic of this catalog is the use of uniform analysis procedures to determine stellar and planetary properties.  This includes the transit search and fitting procedures applied to the K2 photometry, the Doppler fitting techniques applied to the radial velocities, and the spectral modeling to determine bulk stellar parameters.  Such a uniform treatment will make the catalog useful for statistical studies of the masses, densities, and system architectures of exoplanetary systems.  This work also serves as a data release for all previously unpublished RVs and associated stellar activity indicators obtained by our team for these systems, along with derived stellar and planet parameters.  
\end{abstract}

\section{Introduction}

The Kepler mission operated for four years and discovered 4000$+$ planets whose occurrence patterns and individual characteristics transformed exoplanet science.  Analysis of the Kepler time series photometry demonstrated that small planets vastly outnumber large ones \citep[e.g.,][]{Howard2012,Petigura2013}.  Results from \Kepler\ showed that systems of multiple planets orbiting within 1\,au  \citep[e.g.,][]{Lissauer2011} are a common planetary system architecture despite their dissimilarity with the solar system.  The mission also identified new planet types including circumbinary planets \citep{Doyle2011} and low-density ``super-puffs'' \citep{Masuda2014}.
The population of close-in, small planets have proved particularly interesting.  Variously called ``super-Earths'' and ``sub-Neptunes'', these planets were known from prior Doppler discoveries \citep{Howard2010,Mayor2011}.  However, details of their densities and planetary system architectures were not studied in detail until \Kepler\ data were available.  These planets are commonly in multiplanet systems with low mutual inclinations \citep[e.g.,][]{Fang2012} and low orbital eccentricities \citep[][]{Xie2016,Mills2019}.\footnote{See \cite{Borucki2017} and a special issue of New Astronomy Reviews \citep{Lissauer2018} for reviews of the \Kepler\ mission and its major results.} 

Measurements of the masses and radii of transiting planets provide information about their bulk densities and constrain their bulk compositions.  During the \Kepler\ mission, \cite{Marcy2014} measured the masses of 42 planets orbiting 22 stars.  Subsequent analyses  \citep{Weiss2014,Rogers2015,Wolfgang2015} demonstrated that planets in the size range spanning Earth to Neptune have a range of densities.  High densities for planets smaller than $\sim$1.6\,\rearth\ suggest that these planets are mostly composed of rock and iron (like Earth).  The lower densities of the larger planets mean that these planets are (in part) composed of lower-density material \citep[e.g.,][]{Lopez2014}.  However, the gradient from large to small planets is not smooth and uniformly populated in this domain.  \cite{Fulton2017} showed that low-density sub-Neptunes are nearly distinct from high-density super-Earths.    The ``radius valley'' separating these two types of planets \citep{VanEylen2018b,Fulton2018b} has been interpreted as evidence for photoevaporation by XUV photons of tenuous planetary envelopes of hydrogen and helium \citep{Owen2017}. Core-powered mass loss is an alternative theory in which the power for atmosphere loss comes from the interior of the planet \citep{Ginzburg2018,Gupta2019}.
Understanding the transition from low- to high-density planets with decreasing planet size has been an area of significant study \citep[e.g.,][]{Wolfgang2015,Howe2015,Owen2016,Lehmer2017,Sheng2018,Ning2018,Kanodia2019,Luque2022} and is a motivation for the work described in this paper.

The K2 mission \citep{Howell2014} followed the \Kepler\ mission using the same spacecraft, telescope, and photometer. K2 operated during 2014--2018 with twenty pointings along the ecliptic, each lasting about 80 days.  Each of these ``Campaigns'' produced time-series photometry for bright stars in the $\sim$100 square degree fields.   A primary goal of the K2 mission was to broaden the sample of transiting planets, particularly those orbiting bright stars, for which follow-up measurements are more feasible. In this sense, the K2 project represented a transition from the deep, narrow (and top-down-driven) \Kepler survey to the broad, shallower, and community-driven {\em TESS} survey \citep{Ricker2016,Guerrero2021}.  

Our research collaboration used K2 to observe magnitude-limited samples of stars for most of the K2 fields (typically 5000--7000 GKM dwarf stars per field) through a proposal-driven process.  With these observations, we planned to recreate many aspects of the original \Kepler\ planet search based on new catalogs of planets.  We undertook significant observational programs to validate and characterize the new planets and to measure their masses using the HIRES spectrometer on the Keck I telescope and the Levy spectrometer on the Automated Planet Finder (APF) telescope.  This paper reports the Doppler measurements and planet masses for our K2 study.  Subsequent follow-up papers 
provide statistical analyses of the ensemble planet properties from this catalog with comparisons to theoretical models of bulk planet composition.

The plan for this paper is as follows. Sec.\ \ref{sec:k2_project} provides a description of our K2 project and its goals, as well as the transiting planet search and target selection.  Our spectroscopic observations are described in Sec.\ \ref{sec:stellar_spectroscopy} and our modeling of the spectra to determine stellar properties and planet masses are described in Sec.\ \ref{sec:modeling}.  Details of individual systems and their Keplerian models are provided in Sec.\ \ref{sec:indiv_sys} and Sec.\ \ref{sec:discussion} is a concluding discussion.

\section{K2 Planet Search}
\label{sec:k2_project}

We conducted a search for planets orbiting stars observed by K2 with several goals.  First, we sought to increase the number of \textit{bright} stars with discovered transiting planets.  Such planets are more favorable for follow-up measurements to determine planet masses.  We sought to measure the masses of $\sim$20 K2 planets with sizes between those of Earth and Neptune, roughly doubling the number of planets in that size range with precisely measured masses and radii.  The sample at that time was dominated by \Kepler\ planets and was sufficient to  detect the transition from rocky to gas-dominated (by volume) planets near 1.6\,\rearth. However, the data were too sparse to disentangle other important effects that shape the size-density distribution.  At that time, all of the known small planets with densities of $\gtrsim$5\,g\,cm$^{-3}$ were in close orbits with high equilibrium temperatures.  Their high bulk densities may be a reflection of photoevaporation and atmospheric loss rather than the intrinsic composition distribution of small planets \citep{Lopez2012,Owen2013}.  To disentangle these effects, we sought to preferentially observe K2 planets with equilibrium temperatures and sizes that were not explored in the \Kepler\ sample due to the lack of bright systems. 

Our search was also motivated by a desire to find bright targets for atmospheric transmission spectroscopy studies by the James Webb Space Telescope ({\em JWST}) and the Hubble Space Telescope ({\em HST}).  Mass determinations are crucial to these studies because a planet's surface gravity affects the size of the atmospheric signatures through the atmospheric scale height \citep{Batalha2019}.  Here, we sought to expand the domain of characterized atmospheres by emphasizing planets that are smaller and cooler than hot Jupiters, which were (and remain) the most well-studied to date \citep[e.g.,][]{Sing2016,ERS2023}.

A third significant motivation was to facilitate measurements of detailed patterns of planet occurrence in regions of the Milky Way outside of the Kepler Field.  The goal is to measure variations in planet occurrence between K2 fields to establish whether our detailed measurements from Kepler can be applied directly to the solar neighborhood and to determine how much of the variation in planet occurrence correlates with stellar properties (e.g., metallicity, stellar multiplicity).  This work has been explored in detail in the `Scaling K2' project \citep{Hardegree-Ullman2020,Zink2020,Zink2020b,Zink2021,Christiansen2022,Zink2023,Christiansen2023}.

\begin{figure}
\vspace*{0.1in}
\includegraphics[width=0.48\textwidth]{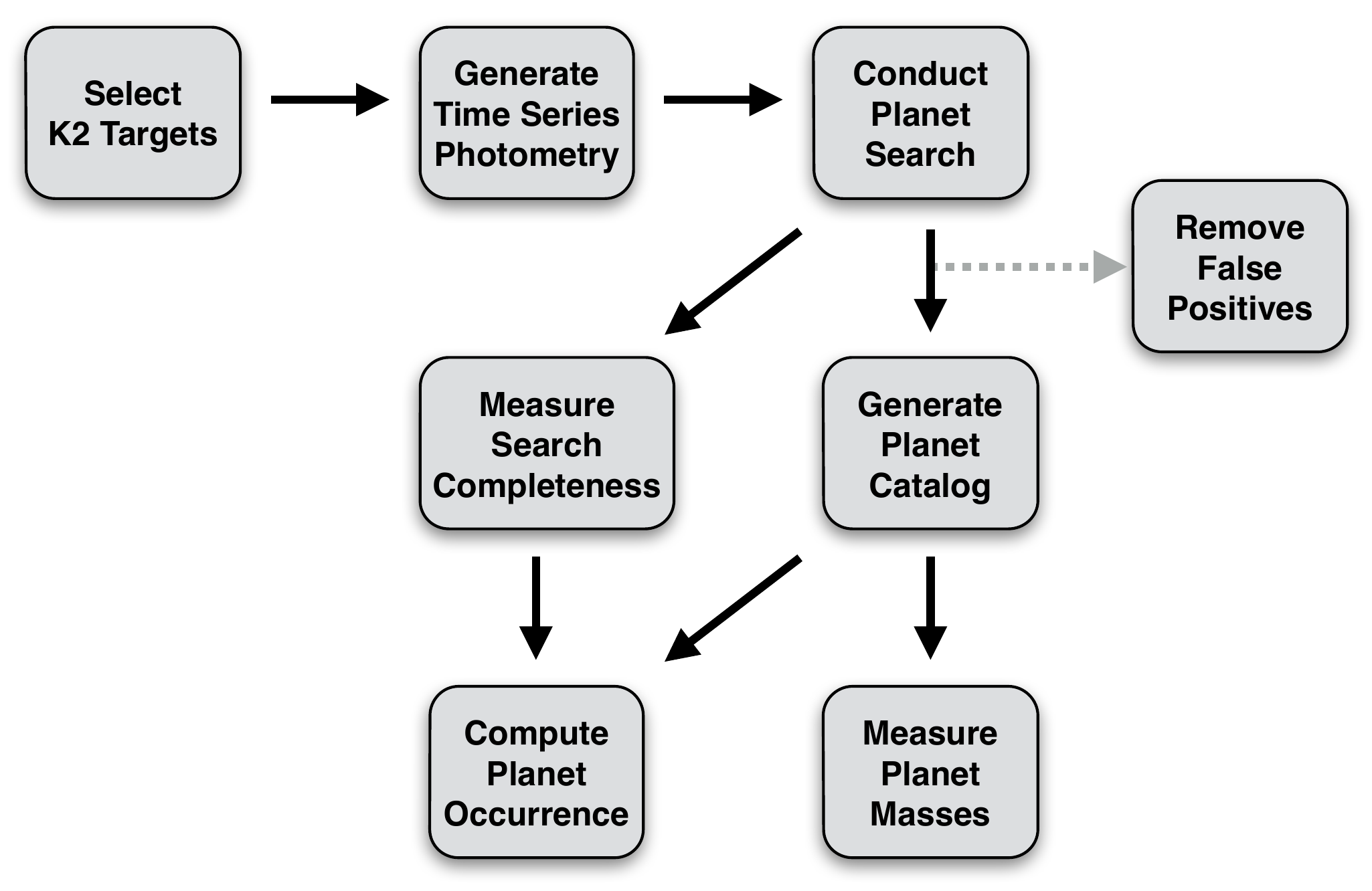}
\caption{Flow chart of our K2 planet search.  This paper describes elements of this process that culminate in the measurement of planet masses; measuring search completeness and computing planet occurrence will be described in subsequent papers.  See Sec.\ \ref{sec:k2_project} for details.}
\label{fig:plan}
\end{figure}

Fig.\ \ref{fig:plan} is a flow chart of the elements of our K2 planet search.  This paper describes target selection (Sec.\ \ref{sec:target_selection}), generation of time series photometry and transiting planet search (Sec.\ \ref{sec:transiting planet search}), generation of a planet catalog (Sec.\ \ref{sec:selected_stars_planets}), planet validation (Sec.\ \ref{sec:validation}), and planet mass measurements (Sec.\ \ref{sec:stellar_spectroscopy}--\ref{sec:indiv_sys}).  Search completeness and planet occurrence based on the catalog in this paper are not addressed in this paper and are subjects for future work.

\subsection{K2 Target Selection}
\label{sec:target_selection}

Our K2 search began with sets of proposals to observe FGK dwarfs and M dwarfs.  We selected $\sim$10,000 FGK dwarfs per K2 Field from the TESS Dwarf Catalog \citep[TDC;][]{Ricker2015}.  The TDC consists of 3 million F5--M5 objects selected from 2MASS and cross-matched with the NOMAD, Tycho-2, Hipparcos, APASS, and UCAC4 catalogs to obtain photometric colors, proper motions, and parallaxes. Giant stars were removed based on reduced proper motion vs.\ $J-H$ color \citep[see][]{CollierCameron2007}. M giants were excluded using JHK color-color cuts \citep{Bessell1988}.  The proposals for FGK dwarfs were limited to \teff\ = 3900--7000\,K and had magnitude cuts of $V \leq 14$.  (The precise cut varied by a few tenths of a magnitude to obtain $\sim$10$^4$ stars per field.) Many of these bright stars were proposed by other teams for a variety of science projects.  The end result was that $\sim$4000--10,000 FGK stars matching the above description were observed by K2 per field.  

We supplemented the FGK dwarf proposals with a series of K2 proposals to observe M dwarfs drawn from the SUPERBLINK proper motion database \citep{lepine:2005}.  Targets were selected using reduced proper-motion and optical/NIR color-cuts and SED fitting to capture most nearby M dwarfs while minimizing contamination from distant giants.  We estimated spectral types using several color-color relations: Eq.~2 of \cite{rodriguez:2013}, Table~5 of \cite{kraus:2007}, and Table~6 of \cite{pecaut:2013}. We then converted the resulting averaged spectral types to stellar radii following \cite{boyajian:2012b}. Late-type targets were prioritized by comparing a nominal Earth-sized transit depth to K2's photometric precision, requiring an expected transit S/N\,$\gtrsim8$.  Furthermore, we required Kepler magnitudes of $<$\,16.5~mag to allow feasible spectroscopic follow-up.

\subsection{Transiting Planet Search}
\label{sec:transiting planet search}

{\referee Our transit search methodology used the \texttt{TERRA}
  pipeline \citep{Petigura2015phd} to identify candidates in our
  custom-detrended \texttt{k2phot} photometry.  Our resulting planet
  and candidate catalogs have been presented in several previous
  papers
  \citep{Sinukoff2016,Crossfield2016,Dressing2017,Petigura2018,Livingston2018,Yu2018,Crossfield2018},
  and we refer the reader to these for details. Once candidates were
  identified,} we fit the light curves for all of the K2 systems using
{\tt k2phot}\footnote{\url{https://github.com/petigura/k2phot}}
\citep{Petigura2015, Aigrain2016} and {\tt everest}
\citep{Luger2016}. We performed the analysis using a methodology
described in detail in \citet{Crossfield2016} and
\citet{Livingston2018}.  In summary, we performed a Markov Chain Monte
Carlo (MCMC) exploration of the parameter space using {\tt emcee}
\citet{Foreman-Mackey2013} and {\tt batman} \citep{Kreidberg2015} to
fit the transit parameters. We first removed any long-term trends in
the photometry and isolated the individual planet transits in 3$\times
T_{14}$ segments centered on the mid-transit times to reduce the
computation expense ($T_{14}$ is the transit duration from first to
last contact); in multi-planet systems we fit the transits for each
planet separately.

The free parameters in the transit model are: orbital period $P_\mathrm{orb}$, mid-transit time $T_0$, scaled planet radius $R_p/R_\star$, scaled semi-major axis $a/R_\star$, impact parameter $b\equiv a\cos i/R_\star$, and quadratic limb-darkening coefficients ($q_1$ and $q_2$) under the transformation of \citet{Kipping2013}. We additionally included the logarithm of the Gaussian errors (log\,$\sigma$) and a constant out-of-transit baseline offset to allow for variation in the normalization of the light curve. We included Gaussian priors on the limb darkening coefficients informed from a Monte Carlo sampling of an interpolated grid of theoretical limb darkening coefficients \citep{Claret2012}. This allows for propagation of uncertainties in host star effective temperature \teff, surface gravity \logg, and metallicity \feh.

We performed a preliminary non-linear least squares fit using {\tt lmfit} \citep{Newville2014}. We initialized 100 ``walkers'' around the least squares solution and ran the MCMC for 5000 steps, discarding the first 3000 steps as ``burn-in'' before inspecting the chains and posteriors for convergence. We then calculated the autocorrelation time using the {\tt acor} package in Python to check that we had a sufficient number of independent samples. The median and 68\% credible intervals of the marginalized posterior distribution for the planet radius and orbital period are reported in Table~\ref{tb:planet_props}.   {\referee Most of our targets have also subsequently been observed at lower precision by {\em TESS}, but we leave a joint analysis of {\em K2} and {\em TESS} photometry and RVs for future analysis.}


\subsection{Selected Stars and Planets}
\label{sec:selected_stars_planets}

The targets selected for observation in our Keck/HIRES radial velocity
program were taken from our {\em K2} transit light curve transit
search efforts described above. {\referee We selected our targets
  manually for the list of all {\em K2} targets after accounting for
  stellar properties (brightness, activity, rotation), planet
  properties (inferred mass, predicted RV semi-amplitude, proximity of
  the orbital and stellar rotation periods), and follow-up and vetting
  observations (high-dispersion spectroscopy, high-resolution imaging,
  and {\em K2} photometric diagnostics) --- as well as our attempt to
  produce an eventual mass catalog that would serve a broad range of
  science cases.}

\begin{figure}
\includegraphics[width=0.45\textwidth, trim=0 {.33\textwidth} {.45\textwidth} 0, clip]{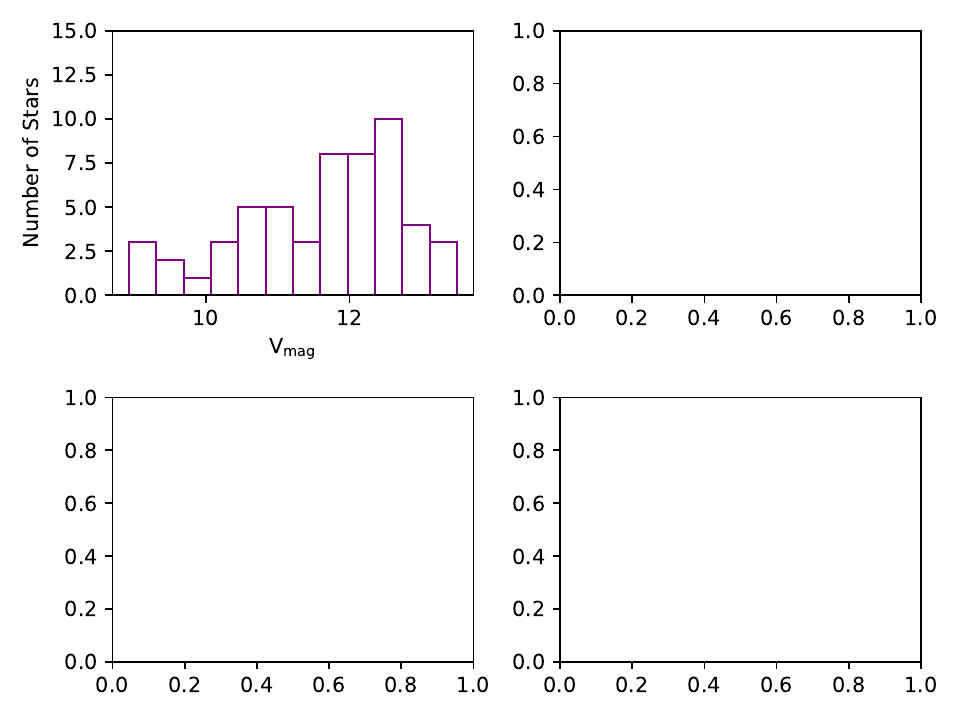}
\caption{Distribution of planet host star brightness ($V$).}
\label{fig:vmag_hist}
\end{figure}

\begin{figure*}
\includegraphics[width=\textwidth]{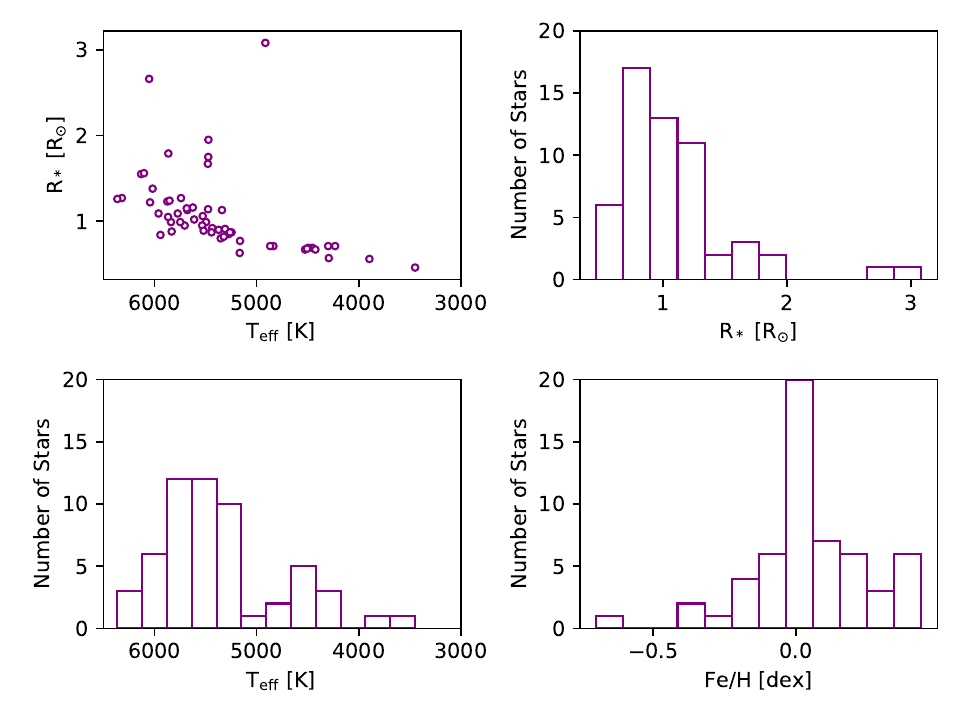}
\caption{Properties of planet host stars showing distributions of stellar temperature, radius, and iron abundance.}
\label{fig:stellar_hist}
\end{figure*}

Table~\ref{tb:star_pars} lists the stars observed for our HIRES program; the table notes the stars' names, EPIC and TIC numbers, the number of RV observations $N_\mathrm{obs}$ from HIRES, the APF, and the literature used in our analysis; and the number of planets ($N_p$) in each system. Fig.~\ref{fig:vmag_hist} shows that our stars span $V$ magnitudes of roughly 9 to 13.5; fainter on average than many modern {\em TESS} targets \citep{Guerrero2021}, but significantly brighter than the average {\em Kepler} host star sample. The overall bulk properties of the target stars are listed in Table~\ref{tb:star_props}.  Fig.~\ref{fig:stellar_hist} shows that our sample focuses predominately on stars with roughly Sun-like properties, while also including a smaller number of evolved stars and cooler dwarfs.

Finally, Table~\ref{tb:planet_props} lists the derived properties of the planets orbiting our target stars.  This includes orbital period, mid-transit time, radius, mass, density, instellation flux, and equilibrium temperature as derived from our {\em K2} transit light curve and Keck/HIRES radial velocity analyses.

\begin{deluxetable*}{lccccccccccc}
\tablecaption{Host star Parameters\label{tb:star_pars}}
\tabletypesize{\scriptsize}
\tablehead{\colhead{Name} & \colhead{Other } & \colhead{EPIC No.} & \colhead{TIC No.} & \colhead{Field} & \colhead{RA} & \colhead{Dec.} & \colhead{$V$ (mag)} & \multicolumn{3}{c}{$N_{\rm obs}$} & \colhead{$n_p$}\\
\cline{9-11} 
&Name&&&&&&& HIRES & APF & Literature&
}
\decimals
\startdata
HD 3167 & K2-96 & 220383386 & 318707987 & 8 & 00:34:58 & $+$04:22:53.3 &                8.941 $\pm$ 0.015 & 60 & 116 & 76 & 2 \\
HIP 41378 & --- & 211311380 & 366443426 & 5/18 & 08:26:28 & +10:04:49.3 &          8.93 $\pm$ 0.02 & 218 & 0 & 389 & 5 \\
HD 106315 & K2-109 & 201437844 & 56815340 & 10 & 12:13:53 & $-$00:23:36.5 &        8.95 $\pm$ 0.02 & 352 & 125 & 155 & 2 \\
HD 89345 & K2-234 & 248777106 & 281731203 & 14 & 10:18:41 & +10:07:44.0 &          9.38 $\pm$ 0.03 & 12 & 21 & 66 & 1 \\
K2-222 & --- & 220709978 & 257774438 & 8 & 01:05:51 & +11:45:12.3 &                  9.54 $\pm$ 0.03 & 55 & 32 & 0 & 1 \\
K2-291 & --- & 247418783 & 27039476 & 13 & 05:05:47 & +21:32:55 &                     10.01 $\pm$ 0.03 & 50 & 0 & 25 & 1 \\
K2-236 & --- & 211945201 & 243244680 & 5 & 09:06:18 & +19:24:08.1 &                  10.15 $\pm$ 0.06 & 36 & 2 & 19 & 1 \\
K2-418 & --- & 229004835 & 94924542 & 10 & 12:25:57 & $-$01:24:16.5 &        10.23 $\pm$ 0.04 & 22 & 0 & 0 & 1 \\
K2-277 & --- & 212357477 & 404421005 & 6 & 13:28:04 & $-$15:56:16.2 &                10.36 $\pm$ 0.05 & 26 & 0 & 0 & 1 \\
GJ 9827 & K2-135 & 246389858 & 301289516 & 12 & 23:27:05 & $-$01:17:10.6 &                10.51 $\pm$ 0.069 & 92 & 0 & 142 & 3 \\
K2-261 & --- & 201498078 & 281731203 & 14 & 10:52:08 & +00:29:36.1 &                  10.61 $\pm$ 0.06 & 8 & 4 & 0 & 1 \\
K2-100 & --- & 211990866 & 307733361 & 5 & 08:38:24 & +20:06:21.8 &                  10.65 $\pm$ 0.09 & 33 & 0 & 0 & 1 \\
K2-31 & --- & 204129699 & 50171060 & 2 & 16:21:46 & $-$23:32:52.3 &                10.8 $\pm$ 0.07 & 8 & 0 & 9 & 1 \\
K2-39 & --- & 206247743 & 250977648 & 3 & 22:33:28 & $-$09:01:22.0 &                10.83 $\pm$ 0.07 & 45 & 0 & 30 & 1 \\
K2-229 & --- & 228801451 & 98720809 & 10 & 12:27:30 & $-$06:43:18.7 &                10.98 $\pm$ 0.08 & 24 & 0 & 115 & 2 \\
K2-111 & --- & 210894022 & 14227229 & 4 & 03:59:34 & +21:17:55.3 &                  11.14 $\pm$ 0.04 & 54 & 0 & 18 & 1 \\
K2-99 & --- & 212803289 & 176966903 & 6 & 13:55:06 & $-$05:26:32.9 &                11.15 $\pm$ 0.1 & 19 & 0 & 33 & 1 \\
K2-265 & --- & 206011496 & 146364192 & 3 & 22:48:08 & $-$14:29:40.9 &                11.19 $\pm$ 0.1 & 53 & 0 & 0 & 1 \\
K2-24 & --- & 203771098 & 68048686 & 2 & 16:10:18 & $-$24:59:25.0 &                11.28 $\pm$ 0.1 & 63 & 0 & 0 & 2 \\
K2-38 & --- & 204221263 & 12666215 & 2 & 16:00:08 & $-$23:11:21.4 &                11.34 $\pm$ 0.11 & 65 & 0 & 0 & 2 \\
K2-73 & --- & 206245553 & 38354061 & 3 & 22:20:06 & $-$09:03:21.9 &                11.35 $\pm$ 0.15 & 60 & 0 & 0 & 2 \\
WASP-107 & --- & 228724232 & 429302040 & 10 & 12:33:33 & $-$10:08:46.2 &        11.47 $\pm$ 0.2 & 50 & 0 & 31 & 1 \\
K2-66 & --- & 206153219 & 50183101 & 3 & 22:06:06 & $-$10:42:41.6 &                11.71 $\pm$ 0.19 & 44 & 0 & 0 & 1 \\
K2-36 & --- & 201713348 & 363445121 & 1 & 11:17:48 & +03:51:59.0 &                  11.73 $\pm$ 0.23 & 46 & 0 & 0 & 2 \\
K2-105 & --- & 211525389 & 6892385 & 5 & 08:21:41 & +13:29:51.1 &                          11.75 $\pm$ 0.2 & 31 & 0 & 0 & 1 \\
K2-214 & --- & 220376054 & 344657681 & 8 & 00:59:30 & +04:13:40.1 &                  11.8 $\pm$ 0.21 & 29 & 0 & 0 & 1 \\
K2-220 & --- & 220621788 & 266012991 & 8 & 00:51:05 & +09:31:00.5 &                  11.89 $\pm$ 0.02 & 28 & 0 & 0 & 1 \\
K2-110 & --- & 212521166 & 287333762 & 6 & 13:49:24 & $-$12:17:04.2 &                11.91 $\pm$ 0.07 & 12 & 0 & 27 & 1 \\
WASP-47 & K2-23 & 206103150 & 102264230 & 3 & 22:04:49 & $-$12:01:08.0 &                11.99 $\pm$ 0.01 & 76 & 0 & 143 & 4 \\
K2-79 & --- & 210402237 & 435339558 & 4 & 03:41:01 & +13:31:09.7 &                  12.07 $\pm$ 0.06 & 62 & 0 & 0 & 1 \\
K2-106 & --- & 220674823 & 266015990 & 8 & 00:52:19 & +10:47:40.9 &                  12.1 $\pm$ 0.21 & 39 & 0 & 53 & 2 \\
K2-98 & --- & 211391664 & 366410512 & 5 & 08:25:57 & +11:30:40.1 &                  12.17 $\pm$ 0.03 & 6 & 0 & 12 & 1 \\
K2-3 & --- & 201367065 & 173103335 & 1 & 11:29:20 & $-$01:27:17.2 &                12.17 $\pm$ 0.01 & 74 & 0 & 360 & 3 \\
EPIC 213546283 & --- & 213546283 & 2670610 & 7 & 19:17:30 & $-$29:02:57.1 &                12.21 $\pm$ 0.3 & 12 & 0 & 0 & 1 \\
K2-199 & --- & 212779596 & 2621213 & 6 & 13:55:36 & $-$06:08:10.1 &                        12.29 $\pm$ 0.02 & 45 & 0 & 0 & 2 \\
EPIC 245991048 & --- & 245991048 & 9030096 & 12 & 23:42:31 & $-$09:42:48.8 &                12.3 $\pm$ 0.25 & 16 & 0 & 0 & 1 \\
K2-32 & --- & 205071984 & 437444661 & 2 & 16:49:42 & $-$19:32:34.2 &                12.31 $\pm$ 0.03 & 64 & 0 & 0 & 3 \\
K2-108 & --- & 211736671 & 27635334 & 5 & 08:13:32 & $+$16:25:11.0 &                12.33 $\pm$ 0.01 & 20 & 0 & 0 & 1 \\
K2-62 & --- & 206096602 & 434094657 & 3 & 22:17:27 & $-$12:11:15.0 &                12.4 $\pm$ 0.04 & 20 & 0 & 0 & 1 \\
K2-189 & --- & 212394689 & 422349881 & 6 & 13:34:29 & $-$15:02:10.9 &                12.4 $\pm$ 0.21 & 17 & 0 & 0 & 1 \\
K2-10 & --- & 201577035 & 363573185 & 1 & 11:28:29 & +01:41:26.3 &                  12.42 $\pm$ 0.02 & 22 & 0 & 25 & 1 \\
EPIC 201357835 & --- & 201357835 & 147677251 & 10 & 12:20:44 & $-$01:35:17.9 &        12.44 $\pm$ 0.03 & 7 & 0 & 0 & 1 \\
K2-216 & --- & 220481411 & 418761354 & 8 & 00:45:55 & +06:20:49.1 &                  12.48 $\pm$ 0.05 & 31 & 0 & 29 & 1 \\
K2-280 & --- & 216494238 & 119605900 & 7 & 19:26:23 & $-$22:14:51.6 &                12.54 $\pm$ 0.04 & 16 & 0 & 0 & 1 \\
K2-37 & --- & 203826436 & 68504570 & 2 & 16:13:48 & $-$24:47:13.4 &                12.57 $\pm$ 0.03 & 19 & 0 & 0 & 3 \\
K2-180 & --- & 211319617 & 366411016 & 5 & 08:25:51 & +10:14:49.1 &                  12.6 $\pm$ 0.02 & 26 & 0 & 0 & 1 \\
K2-27 & --- & 201546283 & 363548415 & 1 & 11:26:04 & +01:13:50.7 &                  12.64 $\pm$ 0.02 & 15 & 0 & 31 & 1 \\
K2-181 & --- & 211355342 & 366528389 & 5 & 08:30:13 & +10:54:37.1 &                  12.75 $\pm$ 0.03 & 10 & 0 & 0 & 1 \\
EPIC 245943455 & --- & 245943455 & 49735922 & 12 & 23:30:51 & $-$11:04:38.1 &        12.82 $\pm$ 0.03 & 9 & 0 & 0 & 1 \\
K2-85 & --- & 210707130 & 14160842 & 4 & 03:57:52 & 18:27:55.0 &                   12.8 $\pm$ 0.5 & 21 & 0 & 0 & 1 \\
K2-61 & --- & 206044803 & 402314147 & 3 & 22:38:42 & $-$13:33:36.1 &                12.99 $\pm$ 0.02 & 7 & 0 & 0 & 1 \\
K2-121 & --- & 211818569 & 7059054 & 5 & 08:27:45 & +17:34:45.8 &                          13.32 $\pm$ 0.03 & 18 & 0 & 0 & 1 \\
K2-18 & --- & 201912552 & 388804061 & 1 & 11:30:15 & +07:35:18.2 &                  13.5 $\pm$ 0.05 & 21 & 0 & 133 & 2 \\
K2-55 & --- & 205924614 & 2028887614 & 3 & 22:15:00 & $-$17:15:02.6 &                13.55 $\pm$ 0.02 & 12 & 0 & 0 & 1 \\
K2-19 & --- & 201505350 & 281885301 & 1 & 11:39:50 & +00:36:12.9 &                  13 $\pm$ 0.01 & 51 & 0 & 0 & 3 \\

\enddata
\end{deluxetable*}

\begin{deluxetable*}{lcccccccc}
\tablecaption{Host star Properties\label{tb:star_props}}
\tabletypesize{\scriptsize}
\tablehead{Name & \teff & \logg & \feh & \shk & \lrphk & \vsini & \Mstar & \Rstar \\ 
& (K) & (dex) & (dex) & & & (\kms) & (\msun) & (\rsun) }
\startdata
\ddigSTNAME & \ddigTEFF & \ddigLOGG & \ddigFEH & \ddigSVAL & \ddigRPHK & \ddigVSINI & \ddigMSTAR & \ddigRSTAR \\
\bdiaSTNAME & \bdiaTEFF & \bdiaLOGG & \bdiaFEH & \bdiaSVAL & \bdiaRPHK & \bdiaVSINI & \bdiaMSTAR & \bdiaRSTAR \\
\hieeSTNAME & \hieeTEFF & \hieeLOGG & \hieeFEH & \hieeSVAL & \hieeRPHK & \hieeVSINI & \hieeMSTAR & \hieeRSTAR \\
\hbagSTNAME & \hbagTEFF & \hbagLOGG & \hbagFEH & \hbagSVAL & \hbagRPHK & \hbagVSINI & \hbagMSTAR & \hbagRSTAR \\
\jjhiSTNAME & \jjhiTEFF & \jjhiLOGG & \jjhiFEH & \jjhiSVAL & \jjhiRPHK & \jjhiVSINI & \jjhiMSTAR & \jjhiRSTAR \\
\ihidSTNAME & \ihidTEFF & \ihidLOGG & \ihidFEH & \ihidSVAL & \ihidRPHK & \ihidVSINI & \ihidMSTAR & \ihidRSTAR \\
\fcabSTNAME & \fcabTEFF & \fcabLOGG & \fcabFEH & \fcabSVAL & \fcabRPHK & \fcabVSINI & \fcabMSTAR & \fcabRSTAR \\
\eidfSTNAME & \eidfTEFF & \eidfLOGG & \eidfFEH & \eidfSVAL & \eidfRPHK & \eidfVSINI & \eidfMSTAR & \eidfRSTAR \\
\hehhSTNAME & \hehhTEFF & \hehhLOGG & \hehhFEH & \hehhSVAL & \hehhRPHK & \hehhVSINI & \hehhMSTAR & \hehhRSTAR \\
\jifiSTNAME & \jifiTEFF & \jifiLOGG & \jifiFEH & \jifiSVAL & \jifiRPHK & \jifiVSINI & \jifiMSTAR & \jifiRSTAR \\
\iahiSTNAME & \iahiTEFF & \iahiLOGG & \iahiFEH & \iahiSVAL & \iahiRPHK & \iahiVSINI & \iahiMSTAR & \iahiRSTAR \\
\aiggSTNAME & \aiggTEFF & \aiggLOGG & \aiggFEH & \aiggSVAL & \aiggRPHK & \aiggVSINI & \aiggMSTAR & \aiggRSTAR \\
\jgjjSTNAME & \jgjjTEFF & \jgjjLOGG & \jgjjFEH & \jgjjSVAL & \jgjjRPHK & \jgjjVSINI & \jgjjMSTAR & \jgjjRSTAR \\
\hhedSTNAME & \hhedTEFF & \hhedLOGG & \hhedFEH & \hhedSVAL & \hhedRPHK & \hhedVSINI & \hhedMSTAR & \hhedRSTAR \\
\befbSTNAME & \befbTEFF & \befbLOGG & \befbFEH & \befbSVAL & \befbRPHK & \befbVSINI & \befbMSTAR & \befbRSTAR \\
\eaccSTNAME & \eaccTEFF & \eaccLOGG & \eaccFEH & \eaccSVAL & \eaccRPHK & \eaccVSINI & \eaccMSTAR & \eaccRSTAR \\
\dcijSTNAME & \dcijTEFF & \dcijLOGG & \dcijFEH & \dcijSVAL & \dcijRPHK & \dcijVSINI & \dcijMSTAR & \dcijRSTAR \\
\bejgSTNAME & \bejgTEFF & \bejgLOGG & \bejgFEH & \bejgSVAL & \bejgRPHK & \bejgVSINI & \bejgMSTAR & \bejgRSTAR \\
\bajiSTNAME & \bajiTEFF & \bajiLOGG & \bajiFEH & \bajiSVAL & \bajiRPHK & \bajiVSINI & \bajiMSTAR & \bajiRSTAR \\
\bcgdSTNAME & \bcgdTEFF & \bcgdLOGG & \bcgdFEH & \bcgdSVAL & \bcgdRPHK & \bcgdVSINI & \bcgdMSTAR & \bcgdRSTAR \\
\fffdSTNAME & \fffdTEFF & \fffdLOGG & \fffdFEH & \fffdSVAL & \fffdRPHK & \fffdVSINI & \fffdMSTAR & \fffdRSTAR \\
\ecdcSTNAME & \ecdcTEFF & \ecdcLOGG & \ecdcFEH & \ecdcSVAL & \ecdcRPHK & \ecdcVSINI & \ecdcMSTAR & \ecdcRSTAR \\
\dcbjSTNAME & \dcbjTEFF & \dcbjLOGG & \dcbjFEH & \dcbjSVAL & \dcbjRPHK & \dcbjVSINI & \dcbjMSTAR & \dcbjRSTAR \\
\ddeiSTNAME & \ddeiTEFF & \ddeiLOGG & \ddeiFEH & \ddeiSVAL & \ddeiRPHK & \ddeiVSINI & \ddeiMSTAR & \ddeiRSTAR \\
\fdijSTNAME & \fdijTEFF & \fdijLOGG & \fdijFEH & \fdijSVAL & \fdijRPHK & \fdijVSINI & \fdijMSTAR & \fdijRSTAR \\
\gafeSTNAME & \gafeTEFF & \gafeLOGG & \gafeFEH & \gafeSVAL & \gafeRPHK & \gafeVSINI & \gafeMSTAR & \gafeRSTAR \\
\bhiiSTNAME & \bhiiTEFF & \bhiiLOGG & \bhiiFEH & \bhiiSVAL & \bhiiRPHK & \bhiiVSINI & \bhiiMSTAR & \bhiiRSTAR \\
\bbggSTNAME & \bbggTEFF & \bbggLOGG & \bbggFEH & \bbggSVAL & \bbggRPHK & \bbggVSINI & \bbggMSTAR & \bbggRSTAR \\
\dbfaSTNAME & \dbfaTEFF & \dbfaLOGG & \dbfaFEH & \dbfaSVAL & \dbfaRPHK & \dbfaVSINI & \dbfaMSTAR & \dbfaRSTAR \\
\ccdhSTNAME & \ccdhTEFF & \ccdhLOGG & \ccdhFEH & \ccdhSVAL & \ccdhRPHK & \ccdhVSINI & \ccdhMSTAR & \ccdhRSTAR \\
\eicdSTNAME & \eicdTEFF & \eicdLOGG & \eicdFEH & \eicdSVAL & \eicdRPHK & \eicdVSINI & \eicdMSTAR & \eicdRSTAR \\
\bggeSTNAME & \bggeTEFF & \bggeLOGG & \bggeFEH & \bggeSVAL & \bggeRPHK & \bggeVSINI & \bggeMSTAR & \bggeRSTAR \\
\hagfSTNAME & \hagfTEFF & \hagfLOGG & \hagfFEH & \hagfSVAL & \hagfRPHK & \hagfVSINI & \hagfMSTAR & \hagfRSTAR \\
\gcidSTNAME & \gcidTEFF & \gcidLOGG & \gcidFEH & \gcidSVAL & \gcidRPHK & \gcidVSINI & \gcidMSTAR & \gcidRSTAR \\
\jfjgSTNAME & \jfjgTEFF & \jfjgLOGG & \jfjgFEH & \jfjgSVAL & \jfjgRPHK & \jfjgVSINI & \jfjgMSTAR & \jfjgRSTAR \\
\baeiSTNAME & \baeiTEFF & \baeiLOGG & \baeiFEH & \baeiSVAL & \baeiRPHK & \baeiVSINI & \baeiMSTAR & \baeiRSTAR \\
\bjieSTNAME & \bjieTEFF & \bjieLOGG & \bjieFEH & \bjieSVAL & \bjieRPHK & \bjieVSINI & \bjieMSTAR & \bjieRSTAR \\
\gghbSTNAME & \gghbTEFF & \gghbLOGG & \gghbFEH & \gghbSVAL & \gghbRPHK & \gghbVSINI & \gghbMSTAR & \gghbRSTAR \\
\ggacSTNAME & \ggacTEFF & \ggacLOGG & \ggacFEH & \ggacSVAL & \ggacRPHK & \ggacVSINI & \ggacMSTAR & \ggacRSTAR \\
\egijSTNAME & \egijTEFF & \egijLOGG & \egijFEH & \egijSVAL & \egijRPHK & \egijVSINI & \egijMSTAR & \egijRSTAR \\
\hadfSTNAME & \hadfTEFF & \hadfLOGG & \hadfFEH & \hadfSVAL & \hadfRPHK & \hadfVSINI & \hadfMSTAR & \hadfRSTAR \\
\hidfSTNAME & \hidfTEFF & \hidfLOGG & \hidfFEH & \hidfSVAL & \hidfRPHK & \hidfVSINI & \hidfMSTAR & \hidfRSTAR \\
\bebbSTNAME & \bebbTEFF & \bebbLOGG & \bebbFEH & \bebbSVAL & \bebbRPHK & \bebbVSINI & \bebbMSTAR & \bebbRSTAR \\
\ecdiSTNAME & \ecdiTEFF & \ecdiLOGG & \ecdiFEH & \ecdiSVAL & \ecdiRPHK & \ecdiVSINI & \ecdiMSTAR & \ecdiRSTAR \\
\gedgSTNAME & \gedgTEFF & \gedgLOGG & \gedgFEH & \gedgSVAL & \gedgRPHK & \gedgVSINI & \gedgMSTAR & \gedgRSTAR \\
\jgbhSTNAME & \jgbhTEFF & \jgbhLOGG & \jgbhFEH & \jgbhSVAL & \jgbhRPHK & \jgbhVSINI & \jgbhMSTAR & \jgbhRSTAR \\
\ghijSTNAME & \ghijTEFF & \ghijLOGG & \ghijFEH & \ghijSVAL & \ghijRPHK & \ghijVSINI & \ghijMSTAR & \ghijRSTAR \\
\fdecSTNAME & \fdecTEFF & \fdecLOGG & \fdecFEH & \fdecSVAL & \fdecRPHK & \fdecVSINI & \fdecMSTAR & \fdecRSTAR \\
\deffSTNAME & \deffTEFF & \deffLOGG & \deffFEH & \deffSVAL & \deffRPHK & \deffVSINI & \deffMSTAR & \deffRSTAR \\
\hbdaSTNAME & \hbdaTEFF & \hbdaLOGG & \hbdaFEH & \hbdaSVAL & \hbdaRPHK & \hbdaVSINI & \hbdaMSTAR & \hbdaRSTAR \\
\eiadSTNAME & \eiadTEFF & \eiadLOGG & \eiadFEH & \eiadSVAL & \eiadRPHK & \eiadVSINI & \eiadMSTAR & \eiadRSTAR \\
\ifgjSTNAME & \ifgjTEFF & \ifgjLOGG & \ifgjFEH & \ifgjSVAL & \ifgjRPHK & \ifgjVSINI & \ifgjMSTAR & \ifgjRSTAR \\
\cffcSTNAME & \cffcTEFF & \cffcLOGG & \cffcFEH & \cffcSVAL & \cffcRPHK & \cffcVSINI & \cffcMSTAR & \cffcRSTAR \\
\egbeSTNAME & \egbeTEFF & \egbeLOGG & \egbeFEH & \egbeSVAL & \egbeRPHK & \egbeVSINI & \egbeMSTAR & \egbeRSTAR \\
\fdfaSTNAME & \fdfaTEFF & \fdfaLOGG & \fdfaFEH & \fdfaSVAL & \fdfaRPHK & \fdfaVSINI & \fdfaMSTAR & \fdfaRSTAR \\

\enddata
\end{deluxetable*}

\begin{deluxetable*}{lccccccc}
\tablecaption{Planet Properties\label{tb:planet_props}}
\tabletypesize{\scriptsize}
\tablehead{Name & Orbital Period & Transit Time & Radius & Mass & Density & $T_{\rm eq}$ & Flux \\
 & (days) & (BJD) & (\rearth) & (\mearth) & (\gmc) & (K) & ($F_\earth$)}
\startdata
\ddigPNAMEone & \ddigPERone & \ddigTCone & \ddigRPone & \ddigMPone & \ddigRHOPone & \ddigTEQone & \ddigFLUXone \\
\ddigPNAMEtwo & \ddigPERtwo & \ddigTCtwo & \ddigRPtwo & \ddigMPtwo & \ddigRHOPtwo & \ddigTEQtwo & \ddigFLUXtwo \\
\ddigPNAMEthree & \ddigPERthree & \ddigTCthree & \ddigRPthree & \ddigMPthree & \ddigRHOPthree & \ddigTEQthree & \ddigFLUXthree \\
\bdiaPNAMEone & \bdiaPERone & \bdiaTCone & \bdiaRPone & \bdiaMPone & \bdiaRHOPone & \bdiaTEQone & \bdiaFLUXone \\
\bdiaPNAMEtwo & \bdiaPERtwo & \bdiaTCtwo & \bdiaRPtwo & \bdiaMPtwo & \bdiaRHOPtwo & \bdiaTEQtwo & \bdiaFLUXtwo \\
\bdiaPNAMEthree & \bdiaPERthree & \bdiaTCthree & \bdiaRPthree & \bdiaMPthree & \bdiaRHOPthree & \bdiaTEQthree & \bdiaFLUXthree \\
\bdiaPNAMEfour & \bdiaPERfour & \bdiaTCfour & \bdiaRPfour & \bdiaMPfour & \bdiaRHOPfour & \bdiaTEQfour & \bdiaFLUXfour \\
\bdiaPNAMEfive & \bdiaPERfive & \bdiaTCfive & \bdiaRPfive & \bdiaMPfive & \bdiaRHOPfive & \bdiaTEQfive & \bdiaFLUXfive \\
\bdiaPNAMEsix & \bdiaPERsix & \bdiaTCsix & \bdiaRPsix & \bdiaMPsix & \bdiaRHOPsix & \bdiaTEQsix & \bdiaFLUXsix \\
\hieePNAMEone & \hieePERone & \hieeTCone & \hieeRPone & \hieeMPone & \hieeRHOPone & \hieeTEQone & \hieeFLUXone \\
\hieePNAMEtwo & \hieePERtwo & \hieeTCtwo & \hieeRPtwo & \hieeMPtwo & \hieeRHOPtwo & \hieeTEQtwo & \hieeFLUXtwo \\
\hbagPNAMEone & \hbagPERone & \hbagTCone & \hbagRPone & \hbagMPone & \hbagRHOPone & \hbagTEQone & \hbagFLUXone \\
\jjhiPNAMEone & \jjhiPERone & \jjhiTCone & \jjhiRPone & \jjhiMPone & \jjhiRHOPone & \jjhiTEQone & \jjhiFLUXone \\
\ihidPNAMEone & \ihidPERone & \ihidTCone & \ihidRPone & \ihidMPone & \ihidRHOPone & \ihidTEQone & \ihidFLUXone \\
\fcabPNAMEone & \fcabPERone & \fcabTCone & \fcabRPone & \fcabMPone & \fcabRHOPone & \fcabTEQone & \fcabFLUXone \\
\eidfPNAMEone & \eidfPERone & \eidfTCone & \eidfRPone & \eidfMPone & \eidfRHOPone & \eidfTEQone & \eidfFLUXone \\
\hehhPNAMEone & \hehhPERone & \hehhTCone & \hehhRPone & \hehhMPone & \hehhRHOPone & \hehhTEQone & \hehhFLUXone \\
\jifiPNAMEone & \jifiPERone & \jifiTCone & \jifiRPone & \jifiMPone & \jifiRHOPone & \jifiTEQone & \jifiFLUXone \\
\jifiPNAMEtwo & \jifiPERtwo & \jifiTCtwo & \jifiRPtwo & \jifiMPtwo & \jifiRHOPtwo & \jifiTEQtwo & \jifiFLUXtwo \\
\jifiPNAMEthree & \jifiPERthree & \jifiTCthree & \jifiRPthree & \jifiMPthree & \jifiRHOPthree & \jifiTEQthree & \jifiFLUXthree \\
\iahiPNAMEone & \iahiPERone & \iahiTCone & \iahiRPone & \iahiMPone & \iahiRHOPone & \iahiTEQone & \iahiFLUXone \\
\aiggPNAMEone & \aiggPERone & \aiggTCone & \aiggRPone & \aiggMPone & \aiggRHOPone & \aiggTEQone & \aiggFLUXone \\
\jgjjPNAMEone & \jgjjPERone & \jgjjTCone & \jgjjRPone & \jgjjMPone & \jgjjRHOPone & \jgjjTEQone & \jgjjFLUXone \\
\hhedPNAMEone & \hhedPERone & \hhedTCone & \hhedRPone & \hhedMPone & \hhedRHOPone & \hhedTEQone & \hhedFLUXone \\
\befbPNAMEone & \befbPERone & \befbTCone & \befbRPone & \befbMPone & \befbRHOPone & \befbTEQone & \befbFLUXone \\
\befbPNAMEtwo & \befbPERtwo & \befbTCtwo & \befbRPtwo & \befbMPtwo & \befbRHOPtwo & \befbTEQtwo & \befbFLUXtwo \\
\eaccPNAMEone & \eaccPERone & \eaccTCone & \eaccRPone & \eaccMPone & \eaccRHOPone & \eaccTEQone & \eaccFLUXone \\
\dcijPNAMEone & \dcijPERone & \dcijTCone & \dcijRPone & \dcijMPone & \dcijRHOPone & \dcijTEQone & \dcijFLUXone \\
\bejgPNAMEone & \bejgPERone & \bejgTCone & \bejgRPone & \bejgMPone & \bejgRHOPone & \bejgTEQone & \bejgFLUXone \\
\bajiPNAMEone & \bajiPERone & \bajiTCone & \bajiRPone & \bajiMPone & \bajiRHOPone & \bajiTEQone & \bajiFLUXone \\
\bajiPNAMEtwo & \bajiPERtwo & \bajiTCtwo & \bajiRPtwo & \bajiMPtwo & \bajiRHOPtwo & \bajiTEQtwo & \bajiFLUXtwo \\
\bcgdPNAMEone & \bcgdPERone & \bcgdTCone & \bcgdRPone & \bcgdMPone & \bcgdRHOPone & \bcgdTEQone & \bcgdFLUXone \\
\bcgdPNAMEtwo & \bcgdPERtwo & \bcgdTCtwo & \bcgdRPtwo & \bcgdMPtwo & \bcgdRHOPtwo & \bcgdTEQtwo & \bcgdFLUXtwo \\
\fffdPNAMEone & \fffdPERone & \fffdTCone & \fffdRPone & \fffdMPone & \fffdRHOPone & \fffdTEQone & \fffdFLUXone \\
\fffdPNAMEtwo & \fffdPERtwo & \fffdTCtwo & \fffdRPtwo & \fffdMPtwo & \fffdRHOPtwo & \fffdTEQtwo & \fffdFLUXtwo \\
\ecdcPNAMEone & \ecdcPERone & \ecdcTCone & \ecdcRPone & \ecdcMPone & \ecdcRHOPone & \ecdcTEQone & \ecdcFLUXone \\
\ecdcPNAMEtwo & \ecdcPERtwo & \ecdcTCtwo & \ecdcRPtwo & \ecdcMPtwo & \ecdcRHOPtwo & \ecdcTEQtwo & \ecdcFLUXtwo \\
\dcbjPNAMEone & \dcbjPERone & \dcbjTCone & \dcbjRPone & \dcbjMPone & \dcbjRHOPone & \dcbjTEQone & \dcbjFLUXone \\
\ddeiPNAMEtwo & \ddeiPERtwo & \ddeiTCtwo & \ddeiRPtwo & \ddeiMPtwo & \ddeiRHOPtwo & \ddeiTEQtwo & \ddeiFLUXtwo \\
\ddeiPNAMEone & \ddeiPERone & \ddeiTCone & \ddeiRPone & \ddeiMPone & \ddeiRHOPone & \ddeiTEQone & \ddeiFLUXone \\
\fdijPNAMEone & \fdijPERone & \fdijTCone & \fdijRPone & \fdijMPone & \fdijRHOPone & \fdijTEQone & \fdijFLUXone \\
\gafePNAMEone & \gafePERone & \gafeTCone & \gafeRPone & \gafeMPone & \gafeRHOPone & \gafeTEQone & \gafeFLUXone \\
\bhiiPNAMEone & \bhiiPERone & \bhiiTCone & \bhiiRPone & \bhiiMPone & \bhiiRHOPone & \bhiiTEQone & \bhiiFLUXone \\
\bbggPNAMEone & \bbggPERone & \bbggTCone & \bbggRPone & \bbggMPone & \bbggRHOPone & \bbggTEQone & \bbggFLUXone \\
\dbfaPNAMEtwo & \dbfaPERtwo & \dbfaTCtwo & \dbfaRPtwo & \dbfaMPtwo & \dbfaRHOPtwo & \dbfaTEQtwo & \dbfaFLUXtwo \\
\dbfaPNAMEfour & \dbfaPERfour & \dbfaTCfour & \dbfaRPfour & \dbfaMPfour & \dbfaRHOPfour & \dbfaTEQfour & \dbfaFLUXfour \\
\dbfaPNAMEthree & \dbfaPERthree & \dbfaTCthree & \dbfaRPthree & \dbfaMPthree & \dbfaRHOPthree & \dbfaTEQthree & \dbfaFLUXthree \\
\dbfaPNAMEone & \dbfaPERone & \dbfaTCone & \dbfaRPone & \dbfaMPone & \dbfaRHOPone & \dbfaTEQone & \dbfaFLUXone \\
\ccdhPNAMEone & \ccdhPERone & \ccdhTCone & \ccdhRPone & \ccdhMPone & \ccdhRHOPone & \ccdhTEQone & \ccdhFLUXone \\
\eicdPNAMEone & \eicdPERone & \eicdTCone & \eicdRPone & \eicdMPone & \eicdRHOPone & \eicdTEQone & \eicdFLUXone \\
\eicdPNAMEtwo & \eicdPERtwo & \eicdTCtwo & \eicdRPtwo & \eicdMPtwo & \eicdRHOPtwo & \eicdTEQtwo & \eicdFLUXtwo \\
\bggePNAMEone & \bggePERone & \bggeTCone & \bggeRPone & \bggeMPone & \bggeRHOPone & \bggeTEQone & \bggeFLUXone \\
\hagfPNAMEone & \hagfPERone & \hagfTCone & \hagfRPone & \hagfMPone & \hagfRHOPone & \hagfTEQone & \hagfFLUXone \\
\hagfPNAMEtwo & \hagfPERtwo & \hagfTCtwo & \hagfRPtwo & \hagfMPtwo & \hagfRHOPtwo & \hagfTEQtwo & \hagfFLUXtwo \\
\hagfPNAMEthree & \hagfPERthree & \hagfTCthree & \hagfRPthree & \hagfMPthree & \hagfRHOPthree & \hagfTEQthree & \hagfFLUXthree \\
\gcidPNAMEone & \gcidPERone & \gcidTCone & \gcidRPone & \gcidMPone & \gcidRHOPone & \gcidTEQone & \gcidFLUXone \\
\jfjgPNAMEone & \jfjgPERone & \jfjgTCone & \jfjgRPone & \jfjgMPone & \jfjgRHOPone & \jfjgTEQone & \jfjgFLUXone \\
\jfjgPNAMEtwo & \jfjgPERtwo & \jfjgTCtwo & \jfjgRPtwo & \jfjgMPtwo & \jfjgRHOPtwo & \jfjgTEQtwo & \jfjgFLUXtwo \\
\baeiPNAMEone & \baeiPERone & \baeiTCone & \baeiRPone & \baeiMPone & \baeiRHOPone & \baeiTEQone & \baeiFLUXone \\
\bjiePNAMEone & \bjiePERone & \bjieTCone & \bjieRPone & \bjieMPone & \bjieRHOPone & \bjieTEQone & \bjieFLUXone \\
\bjiePNAMEtwo & \bjiePERtwo & \bjieTCtwo & \bjieRPtwo & \bjieMPtwo & \bjieRHOPtwo & \bjieTEQtwo & \bjieFLUXtwo \\
\bjiePNAMEthree & \bjiePERthree & \bjieTCthree & \bjieRPthree & \bjieMPthree & \bjieRHOPthree & \bjieTEQthree & \bjieFLUXthree \\
\gghbPNAMEone & \gghbPERone & \gghbTCone & \gghbRPone & \gghbMPone & \gghbRHOPone & \gghbTEQone & \gghbFLUXone \\
\ggacPNAMEone & \ggacPERone & \ggacTCone & \ggacRPone & \ggacMPone & \ggacRHOPone & \ggacTEQone & \ggacFLUXone \\
\ggacPNAMEtwo & \ggacPERtwo & \ggacTCtwo & \ggacRPtwo & \ggacMPtwo & \ggacRHOPtwo & \ggacTEQtwo & \ggacFLUXtwo \\
\egijPNAMEone & \egijPERone & \egijTCone & \egijRPone & \egijMPone & \egijRHOPone & \egijTEQone & \egijFLUXone \\
\egijPNAMEtwo & \egijPERtwo & \egijTCtwo & \egijRPtwo & \egijMPtwo & \egijRHOPtwo & \egijTEQtwo & \egijFLUXtwo \\
\hadfPNAMEone & \hadfPERone & \hadfTCone & \hadfRPone & \hadfMPone & \hadfRHOPone & \hadfTEQone & \hadfFLUXone \\
\hidfPNAMEone & \hidfPERone & \hidfTCone & \hidfRPone & \hidfMPone & \hidfRHOPone & \hidfTEQone & \hidfFLUXone \\
\bebbPNAMEone & \bebbPERone & \bebbTCone & \bebbRPone & \bebbMPone & \bebbRHOPone & \bebbTEQone & \bebbFLUXone \\
\ecdiPNAMEone & \ecdiPERone & \ecdiTCone & \ecdiRPone & \ecdiMPone & \ecdiRHOPone & \ecdiTEQone & \ecdiFLUXone \\
\gedgPNAMEone & \gedgPERone & \gedgTCone & \gedgRPone & \gedgMPone & \gedgRHOPone & \gedgTEQone & \gedgFLUXone \\
\gedgPNAMEtwo & \gedgPERtwo & \gedgTCtwo & \gedgRPtwo & \gedgMPtwo & \gedgRHOPtwo & \gedgTEQtwo & \gedgFLUXtwo \\
\gedgPNAMEthree & \gedgPERthree & \gedgTCthree & \gedgRPthree & \gedgMPthree & \gedgRHOPthree & \gedgTEQthree & \gedgFLUXthree \\
\jgbhPNAMEone & \jgbhPERone & \jgbhTCone & \jgbhRPone & \jgbhMPone & \jgbhRHOPone & \jgbhTEQone & \jgbhFLUXone \\
\ghijPNAMEone & \ghijPERone & \ghijTCone & \ghijRPone & \ghijMPone & \ghijRHOPone & \ghijTEQone & \ghijFLUXone \\
\fdecPNAMEone & \fdecPERone & \fdecTCone & \fdecRPone & \fdecMPone & \fdecRHOPone & \fdecTEQone & \fdecFLUXone \\
\deffPNAMEone & \deffPERone & \deffTCone & \deffRPone & \deffMPone & \deffRHOPone & \deffTEQone & \deffFLUXone \\
\hbdaPNAMEone & \hbdaPERone & \hbdaTCone & \hbdaRPone & \hbdaMPone & \hbdaRHOPone & \hbdaTEQone & \hbdaFLUXone \\
\eiadPNAMEone & \eiadPERone & \eiadTCone & \eiadRPone & \eiadMPone & \eiadRHOPone & \eiadTEQone & \eiadFLUXone \\
\ifgjPNAMEone & \ifgjPERone & \ifgjTCone & \ifgjRPone & \ifgjMPone & \ifgjRHOPone & \ifgjTEQone & \ifgjFLUXone \\
\cffcPNAMEone & \cffcPERone & \cffcTCone & \cffcRPone & \cffcMPone & \cffcRHOPone & \cffcTEQone & \cffcFLUXone \\
\egbePNAMEone & \egbePERone & \egbeTCone & \egbeRPone & \egbeMPone & \egbeRHOPone & \egbeTEQone & \egbeFLUXone \\
\fdfaPNAMEone & \fdfaPERone & \fdfaTCone & \fdfaRPone & \fdfaMPone & \fdfaRHOPone & \fdfaTEQone & \fdfaFLUXone \\
\fdfaPNAMEtwo & \fdfaPERtwo & \fdfaTCtwo & \fdfaRPtwo & \fdfaMPtwo & \fdfaRHOPtwo & \fdfaTEQtwo & \fdfaFLUXtwo \\
\fdfaPNAMEthree & \fdfaPERthree & \fdfaTCthree & \fdfaRPthree & \fdfaMPthree & \fdfaRHOPthree & \fdfaTEQthree & \fdfaFLUXthree \\
\enddata
\end{deluxetable*}

\begin{figure*}
\includegraphics[width=\textwidth]{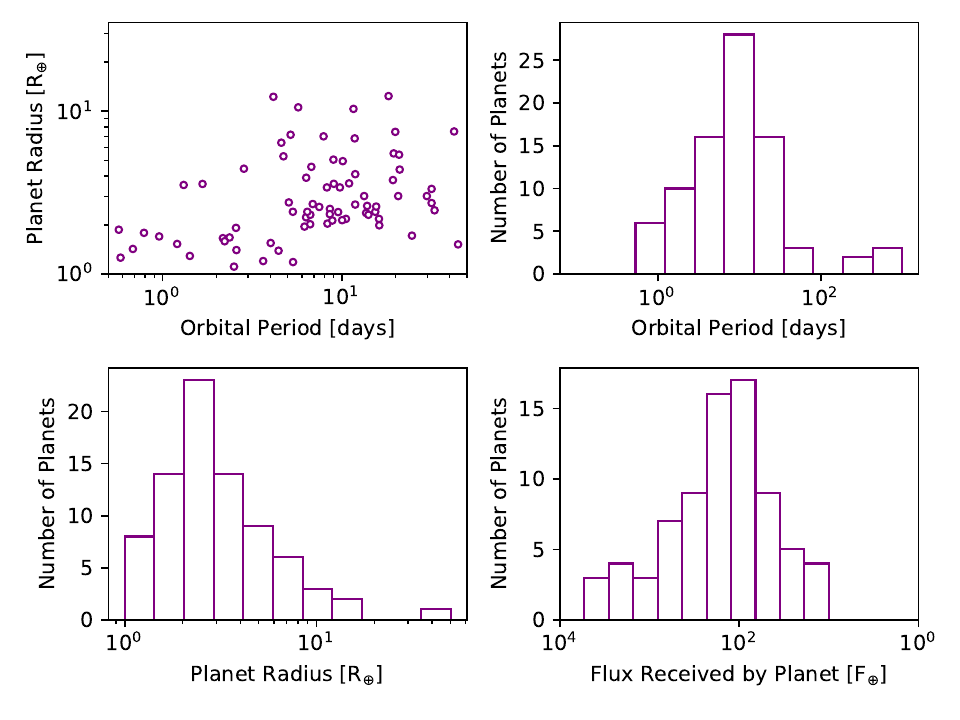}
\caption{Distributions of planet size, orbital period, and flux received for the planets described in this paper. 
}
\label{fig:planet_hist}
\end{figure*}

\subsection{Planet Validation}
\label{sec:validation}

Most of the planets for which we made mass measurements have been validated using statistical techniques \citep{Montet2015, Crossfield2016, Livingston2018, Mayo2018} or through mass measurements.  These systems can be identified in the tables below by their `K2 names' (e.g., `K2-3') or names from other common catalogs (HD, HIP, GJ).  The subsections of Sec.\ \ref{sec:indiv_sys} provides references for the papers that validate each planet.  A handful of planets in this paper were not validated prior to this work and we adopt `EPIC names' for these (e.g., EPIC 213546283). 

\section{Stellar Spectroscopy}
\label{sec:stellar_spectroscopy}

\subsection{Observations}

We used the HIRES spectrometer \citep{Vogt1994} on the Keck I telescope at the W.\ M.\ Keck Observatory to measure optical spectra of all targets presented in this paper.  These spectra span 3640--7990 \AA{} and most have a resolution of $R$ = 60,000.  We used the spectra to determine properties of host stars and for precise Doppler measurements to measure planets (see Sec.\ \ref{sec:doppler}).  Our HIRES observations followed standard  procedures of the California Planet Search \citep[CPS;][]{Howard2010_CPS}.  

The K2 targets in this paper span a range of $V$-band magnitudes. We therefore adopted different spectral signal-to-noise targets for each star using the exposure meter on HIRES.  These exposure meter levels ranged from 250,000 counts for the brightest stars (corresponding to SNR = 200 per reduced pixel on blaze near 550\,nm) to 30,000 counts for the faintest stars (SNR = 75).  Note that the exposure meter counts are in units of flux received by the photo-multiplier tube in approximately $V$-band (in arbitrary units), which are proportional to the number of photoelectrons recorded by HIRES over an appropriate wavelength range.  

For a few bright stars that are noted in Sec.\ \ref{sec:indiv_sys}, we recorded three exposures per night.  We adopted this strategy to help mitigate stellar jitter on approximately hour-long timescales due to granulation on the star's surface or to improve the Doppler precision for stars with high \vsini.  The survey was designed so that each star would be observed on at least 15 nights to provide a consistent floor of Doppler detectability.  In practice, we met this threshold for 42/\Nstars\ stars.  Some stars were observed more frequently because of sky conditions or because of interest in noteworthy systems (e.g., HD 3167).  The median number of nights with an observation is 30 and 14 stars have observations of 50 on more nights.

For a few of the stars that are particularly bright, we also obtained spectra and corresponding RVs from the Levy Spectrometer on the 2.4\,m Automated Planet Finder (APF) telescope \citep{Vogt2014} with a spectral resolution of $R$ = 100,000 and a similar wavelength coverage.  Our APF observing procedures are described in \cite{Fulton2015}.  Systems with APF observations are noted in Sec.\ \ref{sec:indiv_sys}.

\subsection{Doppler Measurements}
\label{sec:doppler}

We determined precise, relative radial velocities (RVs) for Keck/HIRES and APF/Levy spectra using the iodine cell method.  With this technique, the stars are observed with a glass cell of gaseous iodine mounted in front of the spectrometer entrance slit.  Thousands of narrow iodine lines in the wavelength range 5000--6200\,\AA\ are imprinted onto the stellar spectra.  These molecular absorption lines provide a robust wavelength reference against which Doppler shifts are measured and strongly constrain the shape of the spectrometer instrumental profile at the time of each observation \citep{Marcy1992,Valenti1995}.

Our procedure for determining RVs is descended from \cite{Butler1996} and is part of the standard procedures of the California Planet Search \citep{Howard2010_CPS}.  We measured the Doppler shift of each star-times-iodine spectrum by forward-modeling $\sim$700 spectral chunks, each 2\,\AA\ wide. The ingredients of the forward model in each chunk are a de-convolved spectrum of the target star, a high-resolution/high-SNR spectrum of the iodine absorption cell, a description of the instrumental profile, a description of wavelength solution, and the Doppler shift.  We measured stellar activity from the Mt.\ Wilson \shk\ and \lrphk\ values from the HIRES spectra using the procedures in \cite{Isaacson2010}.

We also included RV measurements from other facilities that are in the published literature, where available for particular K2 systems.  Table \ref{tab:rvs} provides a comprehensive list of HIRES, APF, and literature RVs and \shk\ values used in this paper.

\begin{deluxetable*}{lccccc}
\tablecaption{Radial Velocity Measurements\label{tab:rvs}}
\tablehead{Name & Time & RV & Uncertainty & $S_\mathrm{HK}$ \tablenotemark{a} & Instrument \\ 
 &  (BJD) & (\ms) & (\ms) & & }
\startdata
HD 3167 & 2457633.77119 & 0.76 & 1.38 & --- & APF \\
HD 3167 & 2457600.98655 & $-$1.74 & 1.09 & --- & APF \\
HD 3167 & 2457600.85553 & $-$0.53 & 1.36 & --- & APF \\
HD 3167 & 2457599.99676 & 6.02 & 1.23 & --- & APF \\
HD 3167 & 2457599.97510 & $-$1.77 & 1.34 & --- & APF \\

\enddata
\tablenotetext{a}{Uncertainties on $S_\mathrm{HK}$ are 0.002 for all Keck-HIRES measurements.}
\tablecomments{This table will be available in its entirety in machine-readable form in the published paper.  A portion is shown here for guidance regarding its form and content. RV measurements from literature papers that we use in our analyses are also given in this table. The references for each of these points are given in the relevant subsection of Section \ref{sec:indiv_sys}.}
\end{deluxetable*}

\begin{deluxetable}{lr}
\tablecaption{RV Instrument Summary\label{tab:inst}}
\tablehead{Name & $N_\mathrm{RV}$}
\startdata
 HIRES  &  1988  \\ 
 APF  &  346  \\ 
 HARPS  &  475  \\ 
 HARPS-N  &  196  \\ 
 FIES  &  93  \\ 
 CORALIE  &  76  \\ 
 CARMENES  &  58  \\ 
 PFS  &  94  \\ 
 PARAS  &  19  \\ 
 CORPOST  &  6  \\ 
 {\referee McDonald 2.7m}  &  6  \\ 
 FIES-POST  &  4  \\ 
 HDS  &  3  \\ 
\enddata
\end{deluxetable}

\section{Modeling}
\label{sec:modeling}

\subsection{Stellar Characterization}

We determined stellar parameters for each system from a spectroscopic analysis from a HIRES spectrum of the host star. For stars hotter than 4700 K, we measured $T_{\rm eff}$, log $g$, v sin$i$ and [Fe/H] using Spectroscopy Made Easy \citep[SME,][]{Brewer2015}. We followed \citet{Brewer2018} to estimate uncertainties as a function of spectral SNR and subsequently inflated our uncertainty on $T_{\rm eff}$ to 100 K to account for systematic differences between different spectroscopic modeling tools. For some of these hot stars, the HIRES spectrum had previously been analyzed with SME by \citet{Brewer2016} and \citet{Brewer2018}; for these we adopt the measurements shown in those catalogs. For three hot stars (EPIC 213546283, EPIC 216494238, EPIC 245991048) we used SpecMatch \citep{Petigura2015} instead which has been shown to have good agreement with SME \citep{Petigura2017b}. 

For stars cooler than 4700 K, we used SpecMatch-Emp \citep{Yee2017} as the increase in molecular lines limits the reliability of stellar characterization pipelines such as SME and SpecMatch. SpecMatch-EMP compares the HIRES spectra with a library of well-characterized stars to derive $T_{\rm eff}$, log $g$, v sin$i$, and [Fe/H]. 

Afterwards, we derived the stellar mass and radius using isoclassify \citep{huber2017}, see \citet{Fulton2018} for a detailed description of our procedures. All of the stellar properties are listed in Table~\ref{tb:star_props}.

\subsection{Doppler Time Series Modeling}

\subsubsection{Keplerian Models}

We analyzed the radial velocities for each system using RadVel, an open source Python package for fitting Keplerian orbits to radial velocity data sets~\citep{Fulton2018}. The model for the radial velocity ($v_r$) of the star is given by a sum over contributions from its orbiting planets:
\begin{eqnarray}\label{eq:rv}
    v_r(t) &=& \sum_k^{N_{pl}} v_{r,k}(t) + \gamma + \dot{\gamma}(t-t_0) + \ddot{\gamma}(t-t_0)^2, \nonumber \\
    v_{r,k}(t) &=& K_k\left[\cos(\nu_k+\omega_k) + e_k\cos\omega_k\right],
\end{eqnarray}
where the RV semi-amplitude $K$, orbital eccentricity $e$, and argument of periastron $\omega$ are free parameters, $t_0$ is a reference epoch, and $N_{pl}$ is the number of planets in the model. In some models we also included $\dot{\gamma}$ and/or $\ddot{\gamma}$ as free parameters, which model a linear and quadratic RV trend, respectively. A nonzero $\dot{\gamma}$ or $\ddot{\gamma}$  may indicate a distant planet which has only partially completed a full orbit during the observing baseline. The offset term $\gamma$ is unique for each instrument, so for datasets compiled from multiple instruments we assigned a different $\gamma$ term for each one. Finally, the true anomaly $\nu$ is determined from the time of observation, a reference time (e.g., the time of conjunction $t_c$), the orbital eccentricity, and the planet's orbital period $P$ from Kepler's equation, which RadVel solves numerically~\citep[see][]{Fulton2018}.

\subsubsection{Gaussian Process Modeling of Activity}
\label{sec:gp_modeling}

For systems with \lrphk $>-4.9$ and more than 30 available RV observations, we adopted a Gaussian Process (GP) noise model to account for the effects of stellar activity on the data (see, e.g., \citealt{Kosiarek2019b}, \citealt{Lopez-Morales2016}, \citealt{Grunblatt2015}, \citealt{Haywood2014}).
{\referee The choice of \lrphk $> -4.9$ was made to apply activity modeling to systems where the expected activity amplitude is detectable for HIRES observations which have systematic and photon-limited errors of $\sim$2~m~s$^{-1}$ for the brightest stars and a little worse for fainter stars.  This corresponds to a cutoff of roughly more active than the Sun, which varies between \lrphk = $-4.9$ in active periods to \lrphk = $-4.95$ in quiet parts of the cycle.}
The basic approach was to characterize stellar activity variations by measuring the covariance properties of photometry and then apply that model to the RVs.  \cite{Kosiarek2020} demonstrated good agreement between hyperparameter posteriors for  simultaneous photometry and radial velocities of the Sun. 

Our GP model for activity uses a quasi-periodic kernel in all cases; see \citealt{Grunblatt2015} for a definition of this kernel and a description of each parameter. An element in the covariance matrix kernel is given by

\begin{equation}\label{eq:gp}
    C_{ij} = \eta_1^2 \exp{\bigg\{-\frac{|t_i-t_j|^2}{\eta_2^2} -\frac{\sin^2(\pi|t_i-t_j|/\eta_3)}{2\eta_4^2}\bigg\}}.
\end{equation} This kernel is physically motivated, with a free parameter representing the stellar rotation period ($\eta_3$), an exponential free parameter representing the characteristic evolution time of an activity region ($\eta_2$), a scale parameter representing the relative weights of the periodic and exponential decay components ($\eta_4$), and an amplitude ($\eta_1$).

For each system that meets the requirements for a GP fit, we initially trained the GP on non-detrended Everest photometry (rebinned to one point every 15 hours for computational tractability) to constrain the hyperparameters $\eta_3$, $\eta_4$, and $\eta_2$, following \cite{Grunblatt2015}. We imposed a Gaussian prior on $\eta_4$ centered at 0.5 and with $\sigma=0.05$, following \cite{Lopez-Morales2016}, and uniform priors between 0 and the time range of the photometry (ending date minus beginning date) on the timescale parameters $\eta_3$ and $\eta_2$. We used \texttt{george} \citep{Foreman-Mackey2015a} and \texttt{emcee} \citep{Foreman-Mackey2013} to produce posterior distributions of the photometric dataset modeled with GP regression. The posteriors on $\eta_3$, $\eta_4$, and $\eta_2$ were then used as priors on the same parameters in an orbit fit with GP regression using the same kernel, but with independent amplitude ($\eta_1$) parameters for each telescope contributing RV observations to the dataset. We use \texttt{radvel} \citep{Fulton2017} to compute RV and GP posteriors. When performing a GP fit, we also held eccentricity for all planets fixed at 0 unless otherwise specified. 

Other approaches to GP regression for RV orbit-fitting are explored in the literature (see, e.g. \citealt{Rajpaul2015}, \citealt{Jones2017}). We chose to train the GPs presented in this work on photometry primarily because of the availability of complete, consistently derived photometric datasets.

\subsubsection{MCMC Analyses}
\label{sec:rv_mcmc}

The procedure for fitting the above model to the observed RV data within RadVel is as follows. We first performed a maximum-a-posteriori (MAP) fit using Powell’s method (\citealt{Powell1964}, as implemented in \texttt{scipy.optimize.minimize}, \citealt{Virtanen2020}). We then used the resulting solution to seed a Markov chain Monte Carlo (MCMC, as implemented in \texttt{emcee}, \citealt{Foreman-Mackey2013}) to estimate the full posterior distribution. We ran eight independent ensembles in parallel, each containing 50 walkers. We checked that the MCMC was ``burned-in'' by computing the Gelman-Rubin statistic~\citep[G-R,][]{Gelman2003} and ensured that G-R $< 1.03$ for all free parameters. We saved up to 10,000 of the MCMC samples after this point or until we determined the chains were ``well-mixed'' (G-R $< 1.01$) and that the number of independent samples~\citep[$T_z$ statistic,][]{Ford2006} is $> 1000$ for all free parameters for at least five consecutive checks. Once the MCMC was complete, a second MAP fit was run starting at the median values determined by the MCMC posteriors. This is the final MAP solution quoted alongside the posterior quantiles estimated from the MCMC chains.

We invoked a physical prior on the orbital eccentricity $e \in [0, 0.99)$. The RV semiamplitude is allowed to be negative, which while nonphysical eliminates the statistical bias that a $K > 0$ prior would introduce and cause us to overestimate the masses of planets detected at lower significance~\citep{LucySweeney1971}. For all transiting planets we adopted fixed values for the orbital period $P$ and time of conjunction/time of mid-transit $t_c$. In all other cases we used uninformative priors.

\subsubsection{Model Comparison}
\label{sec:rv_model_comparison}

For each system we explored more complex models by re-running the fitting procedure with additional free parameters, namely including trends ($\gamma,\,\dot{\gamma}$) and eccentric orbits $(e,\,\omega)$. For eccentric models we fit using the basis $(\sqrt{e}\cos\omega,\, \sqrt{e}\sin\omega)$ to improve the MCMC sampling efficiency and convergence~\citep{Fulton2018}. For stars with high activity we also tested models that included a Gaussian Process (see Section~\ref{sec:gp_modeling} for more details). We selected between models based on the AIC statistic which estimates a goodness of fit while also penalizing models with more free parameters~\citep{Akaike1974}. We favored models with lower AIC values (better, simpler fit) to those with higher values (worse or overly-complex fit).

\subsubsection{Searches for Additional Planets}
\label{sec:rv_additional_planet_search}


For systems with more than 40 RVs, we conducted a search for additional planets using a periodogram analysis.  See Fig.\ \ref{fig:rvs_epic220709978_resid} for an example.  In that figure, the black line shows the normalized difference in $\chi^2$ for the adopted model (based on two transiting planets in this case) compared to a model with one additional planet, as a function of the orbital period of the additional planet.  Model parameters for all planets not otherwise constrained were allowed to vary with each trial period for the additional planet, which was assumed to be in a circular orbit.  The periodogram power $\Delta\chi^2/\chi^2$ was computed using the 2DKLS formalism of \cite{Otoole2009}.  The threshold for detection of additional planets using the RVs was a 1\% false alarm probability using the empirical method in \cite{Howard2016}.

\section{Discussion}
\label{sec:discussion}



\begin{figure*}
\plotone{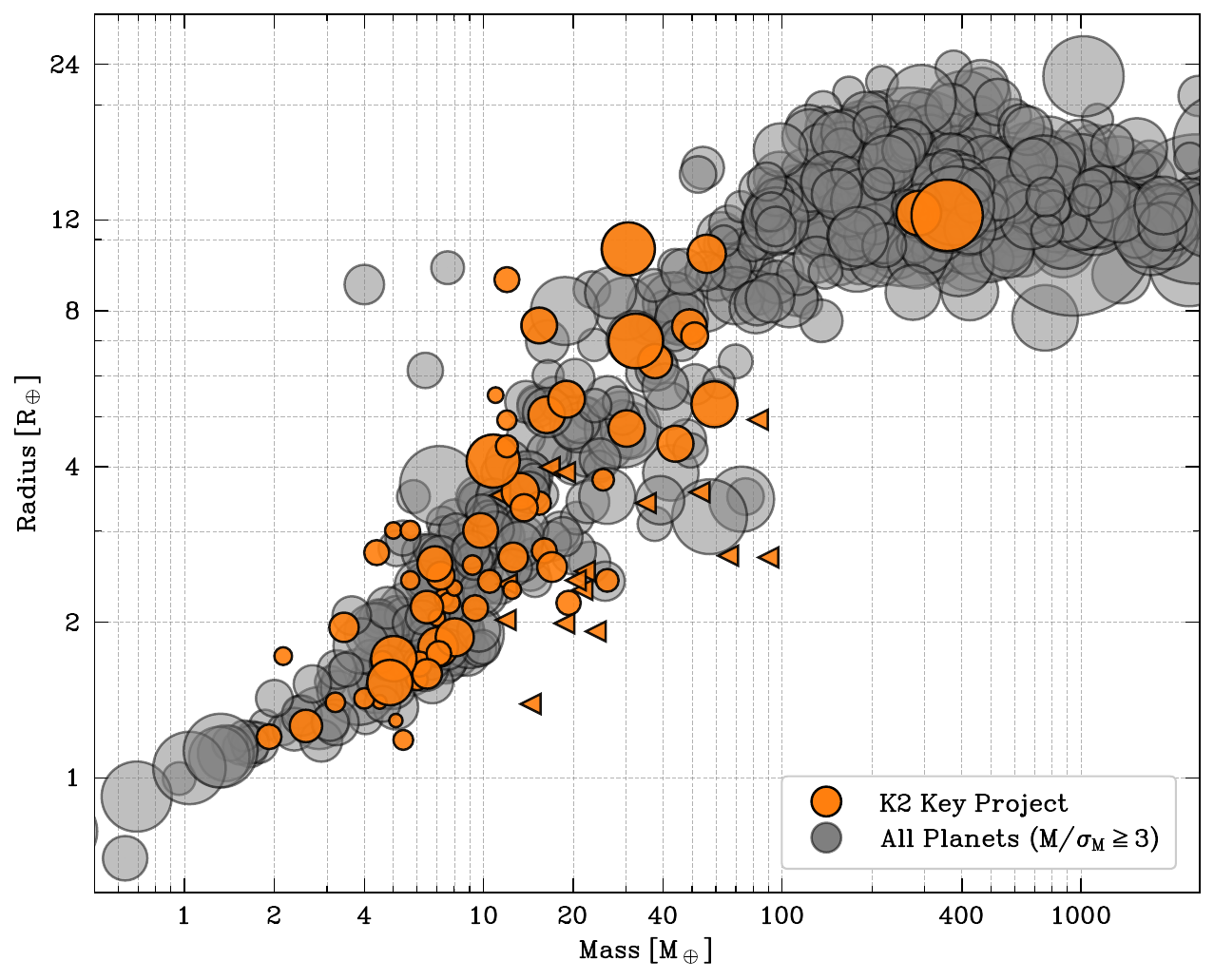}
\caption{Mass-radius diagram for our full sample (orange circles, with triangles indicating 3$\sigma$ upper mass limits) in the context of all known exoplanets with a $\ge$3$\sigma$ mass measurements (NASA Exoplanet Archive, 2023/09/24). Point sizes are scaled so that more precise mass measurements have larger marker sizes. The sub-Neptune portion of this diagram is shown in Fig.~\ref{fig:MR2}.   }
\label{fig:MR1}
\end{figure*}

\begin{figure}
\plotone{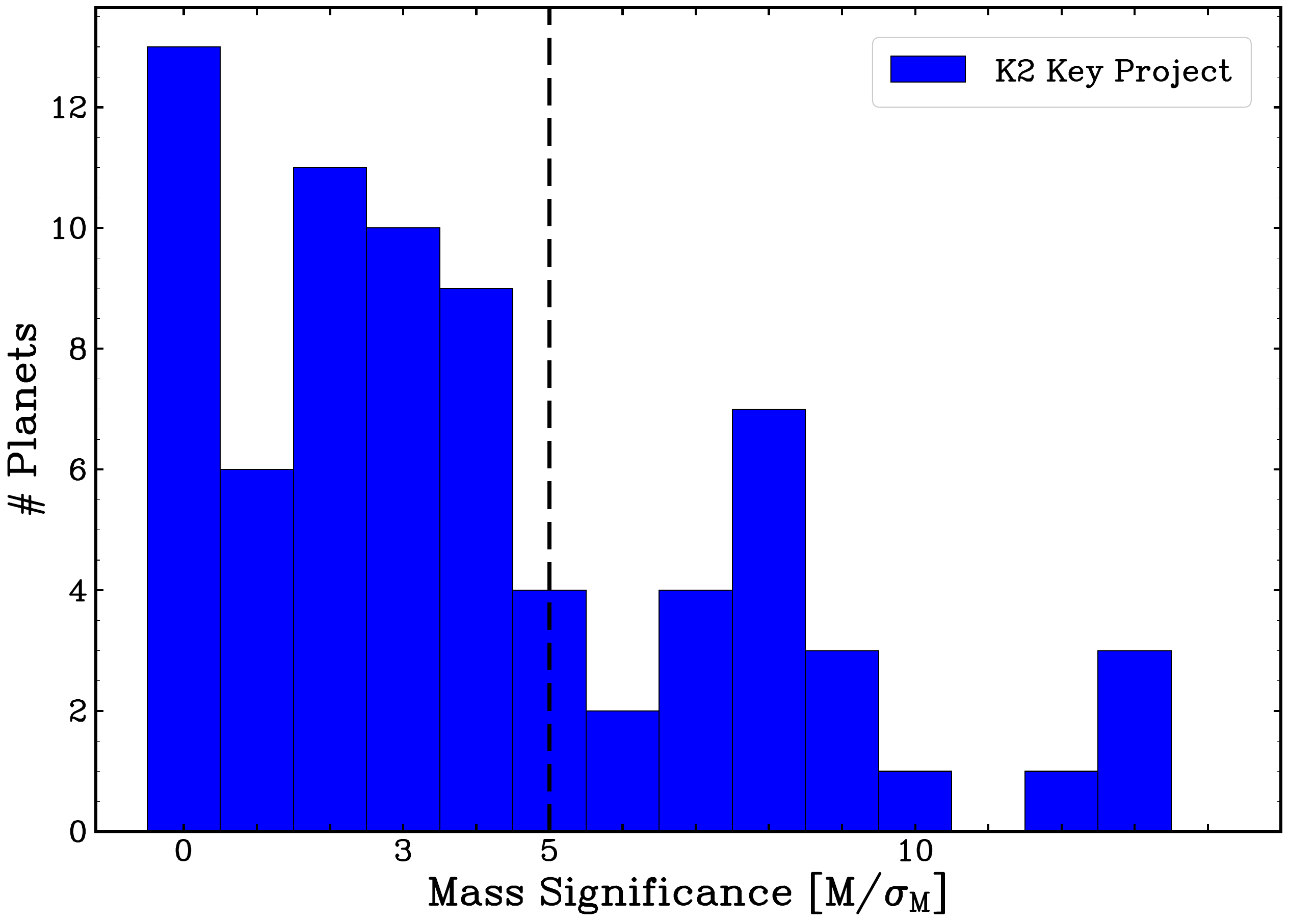}
\caption{Mass significance ($M_p/\sigma_{M_p}$) for all planets in our sample.  \gtfivesig{} planet masses are measured at $\ge$5$\sigma$ (vertical dashed line), while \gtthreesig{} are measured at $\ge$3$\sigma$.  }
\label{fig:mass_sig}
\end{figure}

\begin{figure*}
\plotone{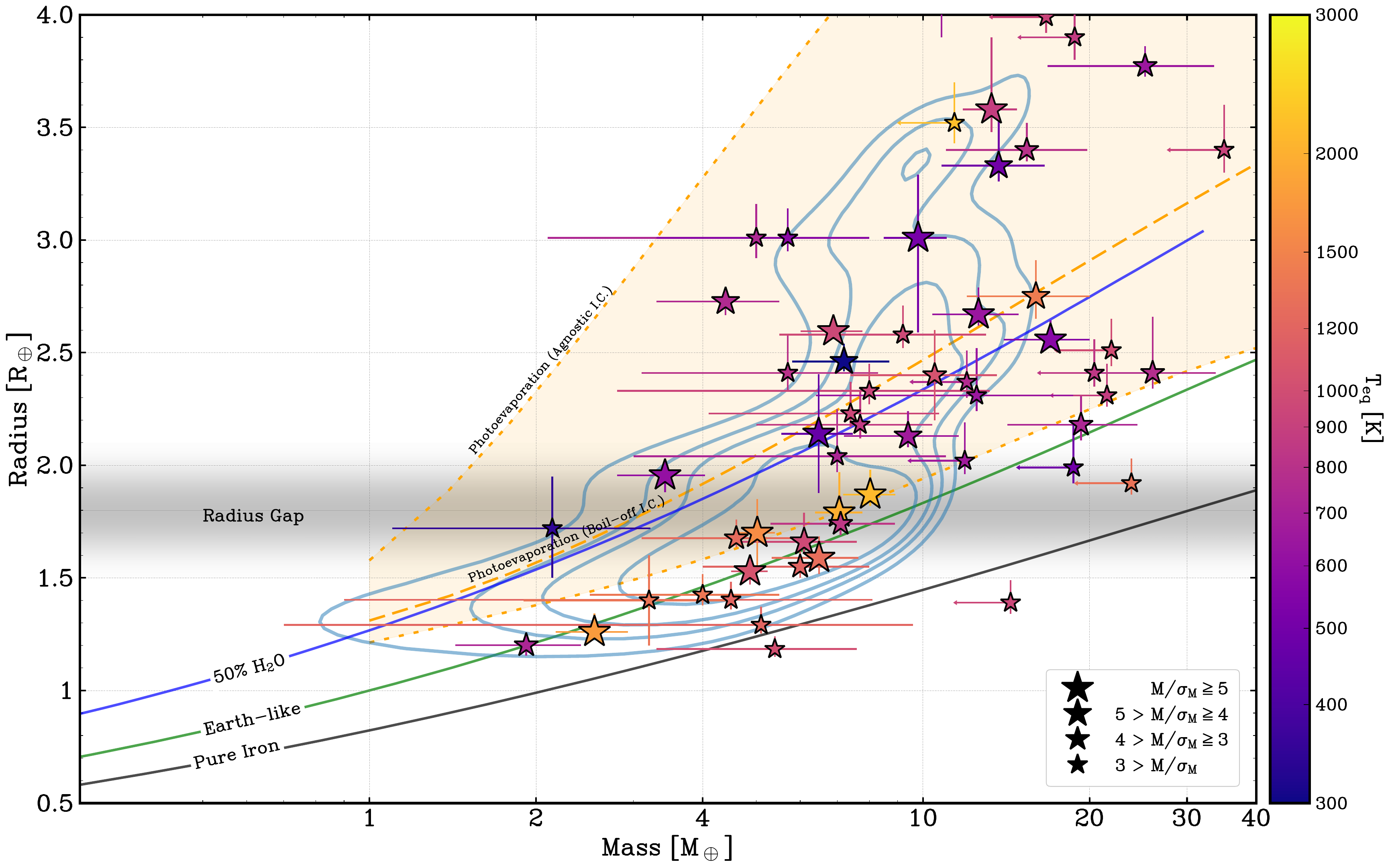}
\caption{Our measured masses and radii for sub-Neptune-size planets.  Points are color-coded based on the equilibrium temperatures of the planets, with marker size scaled so that more precise mass measurements have larger marker sizes; the lowest precision measurements are plotted as 3$\sigma$ upper limits.  The underlying blue contours represent a KDE analysis of all known exoplanets around FGK stars (NASA Exoplanet Archive, 2023/09/24).  We also show composition tracks for three different bulk compositions \citep[solid curves, from][]{Zeng2019} as well as the photoevaporation limits (dashed/dotted curves) from \cite{Rogers2023}.}
\label{fig:MR2}
\end{figure*}

\subsection{Summary of Results}
Figure~\ref{fig:MR1} shows the full distribution of planet masses and radii measured for all \Ntransit{} transiting planets in our sample.   These range from  small, low-mass super-Earths, such as GJ~9827c (Sec.~\ref{sec:GJ9827}) and K2-229b (Sec.~\ref{sec:K2229}), to massive hot Jupiters, such as K2-99b (Sec.~\ref{sec:K299}) and WASP-47b (Sec.~\ref{sec:WASP47}). The masses of many of these planets are tightly constrained: from our RV analysis, \gtfivesig{} planet masses are measured at $\ge$5$\sigma$ (vertical dashed line), while \gtthreesig{} are measured at $\ge$3$\sigma$ (see Fig.~\ref{fig:mass_sig}).

By design, our sample mainly consists of sub-Neptune size planets.  Of the \Ntransit{} transiting planets in our sample with well-determined radii, 66 have $R_p < 4 R_\oplus$. Fig.~\ref{fig:MR2} shows the masses and radii of these small planets in the context of the known exoplanet population along with a Kernel Density Estimate (KDE) map, and reveals that our targets are generally representative of the sub-Neptune exoplanet population.  

{\referee Our final sample also includes five non-transiting planets
  that were thus not identified from the initial K2 photometry.  These
  are HD 3167d (Sec.~\ref{sec:hd3167}), HIP 41378g
  (Sec.~\ref{sec:hip41378}), K2-73c (Sec.~\ref{sec:k2_73}), WASP-47c
  (Sec.~\ref{sec:WASP47}), and WASP-107c (Sec.~\ref{sec:wasp_107}).
  Some systems also show evidence for likely long-term accelerations
  indicating additional, long-period bodies in the systems
  \citep[cf.][]{bonomo:2023}; these systems will be the subject of a
  follow-up paper (Rhem et al., in prep.).}

\subsection{Future Work}
Large samples of planet masses and radii offer numerous opportunities for follow-up studies \citep{Teske2021,Polanski2024}.  For example, one can attempt to measure the diversity of planet core mass and bulk composition using interior structure models. Such efforts can attempt to measure the intrinsic spread of core compositions (and how many planet cores are approximately Earth-like), as well as to seek correlations between stellar properties such as mass or [Fe/H] with estimated planet core mass.

Such data sets can also be used to examine how planet density varies with planet mass or radius \citep[e.g.][]{Luque2022}, and how the super-Earth/mini-Neptune `radius valley' \citep{Fulton2017} manifests in planet mass or density. They can also be used to empirically determine parametric, probabalistic `mass-radius relations' that demonstrate how planet mass can be predicted from planet radius \citep[e.g.,][]{Wolfgang2015,Ning2018}.  

In addition, RV observations of these systems over long time baselines can reveal the presence of massive, long-period companions on wide orbits. Since the RV observations presented here were mostly acquired from $\sim$2015--2018, just a few observations in the present day could reveal long-term accelerations or `trends.' The  detection of such RV trends can reveal new substellar companions suitable for high-contrast characterization \citep{Crepp2014}, and systematic analyses can constrain planet formation and migration via planetary system architectures \citep{Bonomo2023,VanZandt2023}.  Though one initial study has already been done in this direction using K2 targets \citep{Bonomo2023}, the  target overlap is  surprisingly small and so more work could still be done in this arena.

\subsection{Conclusions}
In summary, we have measured the masses, radii, and stellar \&\ orbital properties of \Npl{} planets orbiting \Nstars{}\ {\referee stars}. All these targets were identified using custom transit-search analysis of data from NASA's four-year {\em K2} mission, and masses were measured using Keck/HIRES radial velocity spectroscopy.  Our sample includes:
\begin{itemize}
    \item \Nstars{} targeted planetary systems, mostly around Sun-like stars but spanning a range of stellar parameters (Tables~\ref{tb:star_pars} and~\ref{tb:star_props}, and Figs.~\ref{fig:stellar_hist} and~\ref{fig:vmag_hist});
    \item  \Ntransit{} transiting planets with mass measurements (or constraints) from our analysis (Table~\ref{tb:planet_props}, and Figs.~\ref{fig:planet_hist} and~\ref{fig:MR1});
    \item \Nnontrans{} non-transiting planets identified through our RV observations;
    \item \gtfivesig{} mass measurements at $\ge$5$\sigma$ significance (\gtthreesig{} at $>$3$\sigma$; Fig.~\ref{fig:mass_sig});
    \item 66 planets with sizes $< 4 R_\oplus$ (Fig.~\ref{fig:MR2}).
\end{itemize}

{\referee Our analysis presents newly-measured planet masses for a
  large sample of {\em K2} systems.  In particular, we present mass
  measurements of $\ge$3$\sigma$ significance for four systems that
  previously lacked such measurements: K2-10 (Sec.~\ref{sec:k2_10}),
  K2-55 (Sec.~\ref{sec:k2_55}), K2-105 (Sec.~\ref{sec:k2_105}), and
  K2-121 (Sec.~\ref{sec:k2_121}). We also present other measurements
    that improve on the literature precision, but remain at the
    $2--3\sigma$ level, for three systems: K2-73
    (Sec.~\ref{sec:k2_73}), K2-85 (Sec.~\ref{sec:k2_85}), and K2-277
    (Sec.~\ref{sec:k2_277}). Finally, we present mass upper limits
    ($<$2$\sigma$ mass precision), and thus conclusively rule out
    eclipsing-binary false-positive scenarios, for 12 systems that
    were candidates (or statistically validated as planets) but that
    lacked previous mass measurements: K2-37, K2-61, K2-62, K2-181,
    K2-189, K2-214, K2-220, EPIC 201357835, EPIC 202089657, EPIC
    213546283, EPIC 245943455, and EPIC 245991048 (see individual
    sections in Appendix~\ref{sec:indiv_sys}).}

Together with ongoing RV follow-up of transiting exoplanets discovered by {\em TESS}  \citep[][{\referee Armstrong et al., in review}]{Teske2021,Chontos2022,Crossfield2025}, our survey of {\em K2} planets continues to reveal new worlds suitable for future study.  For example, multiple planets in our sample have already been observed by {\em HST} \citep[GJ~9827, HD~3167, HD~106315, HIP~41378, K2-3][]{Roy2023,Evans2020,Kreidberg2022,DiamondLowe2022} or are being observed by {\em JWST} (WASP-47e, GJ~9827d, HD~106315c).  Between atmospheric spectroscopy,  subsequent (long-term) RV monitoring,  dynamical studies, and improved stellar parameters, our sample of planetary systems should enable intriguing new studies for many years to come.

\acknowledgments{
The K2 planet search described here was conducted over several years with contributions from dozens of astronomers. 
The target stars were selected and characterized by AOM, AWH, BB, BJF, EAP, ES, IJMC, JHL, JLC, JMB, and SL. 
Observing proposals were written by AWH, BB, BMSH, CDD, EAP, ES, HK, IJMC, JES, JLC, LAR, and TPG. 
The radial velocity observing program with HIRES was planned and coordinated by AWH, HI, and ES.  HIRES observations were conducted by IA, CB, AB, SB, LGB, CLB, AC, IJMC, FD, PAD, BJF, SG, SKG, MLH, LAH, AWH, HI, MK, JAL, AWM, SMM, TM, JMAM, ASP, EAP, MR, LJR, RAR, ES, DT, JVZ, and LMW (those who observed K2 targets with HIRES for 10$+$ nights).
High-resolution imaging and associated target vetting were done by DRC, EG, IJMC, JES, JRC, KHU, LAH, MW, and SBH. Analyses of the K2 light curves were conducted by JHL.
Analyses of radial velocity measurements were conducted by SB, RAR, ES, and MK, and AWH. AWH, SB, RR, MK, IJMC, and ES wrote this paper.  

The authors thank: the anonymous referee for comments that improved the quality of the paper; Dr.\ Thomas Greene for useful discussions during this project; and Drs.\ Johanna Teske and Michael Endl for help observing some of our targets.

The scope and extent of this project were enabled by NASA's Key Strategic Mission Support program that provided 40 nights at the W.\,M.\ Keck Observatory. We gratefully acknowledge their support throughout that program.  
We are grateful to the time assignment committees of NASA, the California Institute of Technology, the University of Hawai'i, and the University of California for their generous allocations of observing time. 
We thank additional observers who contributed to the RV measurements from the W.\,M.\ Keck Observatory.
AWH acknowledges support for our K2 team through a NASA Astrophysics Data Analysis Program grant. 
AWH and IJMC acknowledge support from the K2 Guest Observer Program.  LMW acknowledges support from the NASA Exoplanet Research Program (grant no. 80NSSC23K0269).
Some of the data presented in this paper were obtained from the Mikulski Archive for Space Telescopes (MAST). STScI is operated by the Association of Universities for Research in Astronomy, Inc., under NASA contract NAS5-26555. Support for MAST for non-HST data is provided by the NASA Office of Space Science via grant NNX09AF08G and by other grants and contracts. This research has also used the NASA Exoplanet Archive, operated by the California Institute of Technology, under contract with the National Aeronautics and Space Administration under the Exoplanet Exploration Program. This research has used the NASA/IPAC Infrared Science Archive, operated by the Jet Propulsion Laboratory, California Institute of Technology, under contract with the National Aeronautics and Space Administration. 
This research has used the NASA Exoplanet Follow-Up Observation Program website, operated by the California Institute of Technology, under contract with the National Aeronautics and Space Administration under the Exoplanet Exploration Program. 
Finally, we recognize and acknowledge the cultural role that the summit of Maunakea has within the indigenous Hawaiian community. We are grateful to conduct observations from this mountain.}

\facility{Keck:I (HIRES), Keck:II (NIRC2), Kepler, Automated Planet Finder (Levy)}

\software{RadVel \citep{Fulton2018}, emcee \citep{Foreman-Mackey2013}, isochrones \citep{Morton2015-isochrones}, acor (\url{https://github.com/dfm/acor}), Specmatch-Emp \citep{Yee2017}, isoclassify, \citep{huber2017} \citep{Petigura2015}, everest \citep{Luger2016}}

\bibliographystyle{aasjournal}
\bibliography{k2cat}

\begin{thebibliography}{}
\expandafter\ifx\csname natexlab\endcsname\relax\def\natexlab#1{#1}\fi
\providecommand{\url}[1]{\href{#1}{#1}}
\providecommand{\dodoi}[1]{doi:~\href{http://doi.org/#1}{\nolinkurl{#1}}}
\providecommand{\doeprint}[1]{\href{http://ascl.net/#1}{\nolinkurl{http://ascl.net/#1}}}
\providecommand{\doarXiv}[1]{\href{https://arxiv.org/abs/#1}{\nolinkurl{https://arxiv.org/abs/#1}}}

\bibitem[{{Adams} {et~al.}(2016){Adams}, {Jackson}, \& {Endl}}]{Adams2016}
{Adams}, E.~R., {Jackson}, B., \& {Endl}, M. 2016, \aj, 152, 47,
  \dodoi{10.3847/0004-6256/152/2/47}

\bibitem[{{Adams} {et~al.}(2017){Adams}, {Jackson}, {Endl}, {Cochran},
  {MacQueen}, {Duev}, {Jensen-Clem}, {Salama}, {Ziegler}, {Baranec},
  {Kulkarni}, {Law}, \& {Riddle}}]{Adams2017}
{Adams}, E.~R., {Jackson}, B., {Endl}, M., {et~al.} 2017, \aj, 153, 82,
  \dodoi{10.3847/1538-3881/153/2/82}

\bibitem[{{Aigrain} {et~al.}(2016){Aigrain}, {Parviainen}, \&
  {Pope}}]{Aigrain2016}
{Aigrain}, S., {Parviainen}, H., \& {Pope}, B.~J.~S. 2016, \mnras, 459, 2408,
  \dodoi{10.1093/mnras/stw706}

\bibitem[{{Akaike}(1974)}]{Akaike1974}
{Akaike}, H. 1974, IEEE Transactions on Automatic Control, 19, 716,
  \dodoi{10.1109/TAC.1974.1100705}

\bibitem[{{Akana Murphy} {et~al.}(2021){Akana Murphy}, {Kosiarek}, {Batalha},
  {Gonzales}, {Isaacson}, {Petigura}, {Weiss}, {Grunblatt}, {Ciardi}, {Fulton},
  {Hirsch}, {Behmard}, \& {Rosenthal}}]{Murphy2021}
{Akana Murphy}, J.~M., {Kosiarek}, M.~R., {Batalha}, N.~M., {et~al.} 2021, \aj,
  162, 294, \dodoi{10.3847/1538-3881/ac2830}

\bibitem[{Akinsanmi {et~al.}(2020)Akinsanmi, Santos, Faria, Oshagh, Barros,
  Santerne, \& Charnoz}]{Akinsanmi2020}
Akinsanmi, B., Santos, N.~C., Faria, J.~P., {et~al.} 2020, A\&A, 635, L8,
  \dodoi{10.1051/0004-6361/202037618}

\bibitem[{{Alam} {et~al.}(2022){Alam}, {Kirk}, {Dressing}, {L{\'o}pez-Morales},
  {Ohno}, {Gao}, {Akinsanmi}, {Santerne}, {Grouffal}, {Adibekyan}, {Barros},
  {Buchhave}, {Crossfield}, {Dai}, {Deleuil}, {Giacalone}, {Lillo-Box},
  {Marley}, {Mayo}, {Mortier}, {Santos}, {Sousa}, {Turtelboom}, {Wheatley}, \&
  {Vanderburg}}]{Alam2022}
{Alam}, M.~K., {Kirk}, J., {Dressing}, C.~D., {et~al.} 2022, \apjl, 927, L5,
  \dodoi{10.3847/2041-8213/ac559d}

\bibitem[{{Almenara} {et~al.}(2016){Almenara}, {D{\'{\i}}az}, {Bonfils}, \&
  {Udry}}]{Almenara2016}
{Almenara}, J.~M., {D{\'{\i}}az}, R.~F., {Bonfils}, X., \& {Udry}, S. 2016,
  \aap, 595, L5, \dodoi{10.1051/0004-6361/201629770}

\bibitem[{{Almenara} {et~al.}(2015){Almenara}, {Astudillo-Defru}, {Bonfils},
  {Forveille}, {Santerne}, {Albrecht}, {Barros}, {Bouchy}, {Delfosse},
  {Demangeon}, {Diaz}, {H{\'e}brard}, {Mayor}, {Neves}, {Rojo}, {Santos}, \&
  {W{\"u}nsche}}]{Almenara2015}
{Almenara}, J.~M., {Astudillo-Defru}, N., {Bonfils}, X., {et~al.} 2015, \aap,
  581, L7, \dodoi{10.1051/0004-6361/201525918}

\bibitem[{{Anderson} {et~al.}(2017){Anderson}, {Collier Cameron}, {Delrez},
  {Doyle}, {Gillon}, {Hellier}, {Jehin}, {Lendl}, {Maxted}, {Madhusudhan},
  {Pepe}, {Pollacco}, {Queloz}, {S{\'e}gransan}, {Smalley}, {Smith}, {Triaud},
  {Turner}, {Udry}, \& {West}}]{Anderson2017}
{Anderson}, D.~R., {Collier Cameron}, A., {Delrez}, L., {et~al.} 2017, \aap,
  604, A110, \dodoi{10.1051/0004-6361/201730439}

\bibitem[{{Armstrong} {et~al.}(2015){Armstrong}, {Santerne}, {Veras}, {Barros},
  {Demangeon}, {Lillo-Box}, {McCormac}, {Osborn}, {Tsantaki}, {Almenara},
  {Barrado}, {Boisse}, {Bonomo}, {Brown}, {Bruno}, {Rey Cerda}, {Courcol},
  {Deleuil}, {D{\'{\i}}az}, {Doyle}, {H{\'e}brard}, {Kirk}, {Lam}, {Pollacco},
  {Rajpurohit}, {Spake}, \& {Walker}}]{Armstrong2015}
{Armstrong}, D.~J., {Santerne}, A., {Veras}, D., {et~al.} 2015, \aap, 582, A33,
  \dodoi{10.1051/0004-6361/201526008}

\bibitem[{{Barrag{\'a}n} {et~al.}(2016){Barrag{\'a}n}, {Grziwa}, {Gandolfi},
  {Fridlund}, {Endl}, {Deeg}, {Cagigal}, {Lanza}, {Prada Moroni}, {Smith},
  {Korth}, {Bedell}, {Cabrera}, {Cochran}, {Cusano}, {Csizmadia},
  {Eigm{\"u}ller}, {Erikson}, {Guenther}, {Hatzes}, {Nespral}, {P{\"a}tzold},
  {Prieto-Arranz}, \& {Rauer}}]{Barragan2016}
{Barrag{\'a}n}, O., {Grziwa}, S., {Gandolfi}, D., {et~al.} 2016, \aj, 152, 193,
  \dodoi{10.3847/0004-6256/152/6/193}

\bibitem[{{Barrag{\'a}n} {et~al.}(2019){Barrag{\'a}n}, Aigrain, Kubyshkina,
  Gandolfi, Livingston, Fridlund, Fossati, Korth, Parviainen, Malavolta, Palle,
  Deeg, Nowak, Rajpaul, Zicher, Antoniciello, Narita, Albrecht, Bedin, Cabrera,
  Cochran, de Leon, Eigmüller, Fukui, Granata, Grziwa, Guenther, Hatzes,
  Kusakabe, Latham, Libralato, Luque, Monta{\~n}{\'e}s-Rodr{\'i}guez, Murgas,
  Nardiello, Pagano, Piotto, Persson, Redfield, \& Tamura}]{Barragan2019}
{Barrag{\'a}n}, O., Aigrain, S., Kubyshkina, D., {et~al.} 2019, Monthly Notices
  of the Royal Astronomical Society, 490, 698, \dodoi{10.1093/mnras/stz2569}

\bibitem[{{Barros} {et~al.}(2016){Barros}, {Demangeon}, \&
  {Deleuil}}]{Barros2016}
{Barros}, S.~C.~C., {Demangeon}, O., \& {Deleuil}, M. 2016, \aap, 594, A100,
  \dodoi{10.1051/0004-6361/201628902}

\bibitem[{{Barros} {et~al.}(2015){Barros}, {Almenara}, {Demangeon}, {Tsantaki},
  {Santerne}, {Armstrong}, {Barrado}, {Brown}, {Deleuil}, {Lillo-Box},
  {Osborn}, {Pollacco}, {Abe}, {Andre}, {Bendjoya}, {Boisse}, {Bonomo},
  {Bouchy}, {Bruno}, {Cerda}, {Courcol}, {D{\'{\i}}az}, {H{\'e}brard}, {Kirk},
  {Lachuri{\'e}}, {Lam}, {Martinez}, {McCormac}, {Moutou}, {Rajpurohit},
  {Rivet}, {Spake}, {Suarez}, {Toublanc}, \& {Walker}}]{Barros2015}
{Barros}, S.~C.~C., {Almenara}, J.~M., {Demangeon}, O., {et~al.} 2015, \mnras,
  454, 4267, \dodoi{10.1093/mnras/stv2271}

\bibitem[{{Barros} {et~al.}(2017){Barros}, {Gosselin}, {Lillo-Box}, {Bayliss},
  {Delgado Mena}, {Brugger}, {Santerne}, {Armstrong}, {Adibekyan}, {Armstrong},
  {Barrado}, {Bento}, {Boisse}, {Bonomo}, {Bouchy}, {Brown}, {Cochran},
  {Collier Cameron}, {Deleuil}, {Demangeon}, {D{\'{\i}}az}, {Doyle},
  {Dumusque}, {Ehrenreich}, {Espinoza}, {Faedi}, {Faria}, {Figueira}, {Foxell},
  {H{\'e}brard}, {Hojjatpanah}, {Jackman}, {Lendl}, {Ligi}, {Lovis}, {Melo},
  {Mousis}, {Neal}, {Osborn}, {Pollacco}, {Santos}, {Sefako}, {Shporer},
  {Sousa}, {Triaud}, {Udry}, {Vigan}, \& {Wyttenbach}}]{Barros2017}
{Barros}, S.~C.~C., {Gosselin}, H., {Lillo-Box}, J., {et~al.} 2017, \aap, 608,
  A25, \dodoi{10.1051/0004-6361/201731276}

\bibitem[{{Batalha} {et~al.}(2019){Batalha}, {Lewis}, {Fortney}, {Batalha},
  {Kempton}, {Lewis}, \& {Line}}]{Batalha2019}
{Batalha}, N.~E., {Lewis}, T., {Fortney}, J.~J., {et~al.} 2019, \apjl, 885,
  L25, \dodoi{10.3847/2041-8213/ab4909}

\bibitem[{{Becker} {et~al.}(2015){Becker}, {Vanderburg}, {Adams}, {Rappaport},
  \& {Schwengeler}}]{Becker2015}
{Becker}, J.~C., {Vanderburg}, A., {Adams}, F.~C., {Rappaport}, S.~A., \&
  {Schwengeler}, H.~M. 2015, \apjl, 812, L18,
  \dodoi{10.1088/2041-8205/812/2/L18}

\bibitem[{Becker {et~al.}(2018)Becker, Vanderburg, Rodriguez, Omohundro, Adams,
  Stassun, Yao, Hartman, Pepper, Bakos, Barentsen, Beatty, Bhatti, Chontos,
  Cameron, Hellier, Huber, James, Kuhn, Lund, Pollacco, Siverd, Stevens,
  de~Miranda~Cardoso, \& West}]{Becker2018}
Becker, J.~C., Vanderburg, A., Rodriguez, J.~E., {et~al.} 2018, The
  Astronomical Journal, 157, 19, \dodoi{10.3847/1538-3881/aaf0a2}

\bibitem[{{Beichman} {et~al.}(2016){Beichman}, {Livingston}, {Werner},
  {Gorjian}, {Krick}, {Deck}, {Knutson}, {Wong}, {Petigura}, {Christiansen},
  {Ciardi}, {Greene}, {Schlieder}, {Line}, {Crossfield}, {Howard}, \&
  {Sinukoff}}]{Beichman2016}
{Beichman}, C., {Livingston}, J., {Werner}, M., {et~al.} 2016, \apj, 822, 39,
  \dodoi{10.3847/0004-637X/822/1/39}

\bibitem[{{Benneke} {et~al.}(2017){Benneke}, {Werner}, {Petigura}, {Knutson},
  {Dressing}, {Crossfield}, {Schlieder}, {Livingston}, {Beichman},
  {Christiansen}, {Krick}, {Gorjian}, {Howard}, {Sinukoff}, {Ciardi}, \&
  {Akeson}}]{Benneke2017}
{Benneke}, B., {Werner}, M., {Petigura}, E., {et~al.} 2017, \apj, 834, 187,
  \dodoi{10.3847/1538-4357/834/2/187}

\bibitem[{{Benneke} {et~al.}(2019){Benneke}, {Wong}, {Piaulet}, {Knutson},
  {Lothringer}, {Morley}, {Crossfield}, {Gao}, {Greene}, {Dressing},
  {Dragomir}, {Howard}, {McCullough}, {Kempton}, {Fortney}, \&
  {Fraine}}]{Benneke2019}
{Benneke}, B., {Wong}, I., {Piaulet}, C., {et~al.} 2019, \apjl, 887, L14,
  \dodoi{10.3847/2041-8213/ab59dc}

\bibitem[{Berardo {et~al.}(2019)Berardo, Crossfield, Werner, Petigura,
  Christiansen, Ciardi, Dressing, Fulton, Gorjian, Greene, Hardegree-Ullman,
  Kane, Livingston, Morales, \& Schlieder}]{Berardo2019}
Berardo, D., Crossfield, I. J.~M., Werner, M., {et~al.} 2019, The Astronomical
  Journal, 157, 185, \dodoi{10.3847/1538-3881/ab100c}

\bibitem[{{Bessell} \& {Brett}(1988)}]{Bessell1988}
{Bessell}, M.~S., \& {Brett}, J.~M. 1988, \pasp, 100, 1134,
  \dodoi{10.1086/132281}

\bibitem[{{Bonomo} {et~al.}(2023{\natexlab{a}}){Bonomo}, {Dumusque}, {Massa},
  {Mortier}, {Bongiolatti}, {Malavolta}, {Sozzetti}, {Buchhave}, {Damasso},
  {Haywood}, {Morbidelli}, {Latham}, {Molinari}, {Pepe}, {Poretti}, {Udry},
  {Affer}, {Boschin}, {Charbonneau}, {Cosentino}, {Cretignier}, {Ghedina},
  {Lega}, {L{\'o}pez-Morales}, {Margini}, {Mart{\'\i}nez Fiorenzano}, {Mayor},
  {Micela}, {Pedani}, {Pinamonti}, {Rice}, {Sasselov}, {Tronsgaard}, \&
  {Vanderburg}}]{bonomo:2023}
{Bonomo}, A.~S., {Dumusque}, X., {Massa}, A., {et~al.} 2023{\natexlab{a}},
  \aap, 677, A33, \dodoi{10.1051/0004-6361/202346211}

\bibitem[{{Bonomo} {et~al.}(2023{\natexlab{b}}){Bonomo}, {Dumusque}, {Massa},
  {Mortier}, {Bongiolatti}, {Malavolta}, {Sozzetti}, {Buchhave}, {Damasso},
  {Haywood}, {Morbidelli}, {Latham}, {Molinari}, {Pepe}, {Poretti}, {Udry},
  {Affer}, {Boschin}, {Charbonneau}, {Cosentino}, {Cretignier}, {Ghedina},
  {Lega}, {L{\'o}pez-Morales}, {Margini}, {Mart{\'\i}nez Fiorenzano}, {Mayor},
  {Micela}, {Pedani}, {Pinamonti}, {Rice}, {Sasselov}, {Tronsgaard}, \&
  {Vanderburg}}]{Bonomo2023}
---. 2023{\natexlab{b}}, \aap, 677, A33, \dodoi{10.1051/0004-6361/202346211}

\bibitem[{Borsato {et~al.}(2019)Borsato, Malavolta, Piotto, Buchhave, Mortier,
  Rice, Cameron, Coffinet, Sozzetti, Charbonneau, Cosentino, Dumusque,
  Figueira, Latham, Lopez-Morales, Mayor, Micela, Molinari, Pepe, Phillips,
  Poretti, Udry, \& Watson}]{Borsato2017}
Borsato, L., Malavolta, L., Piotto, G., {et~al.} 2019, Monthly Notices of the
  Royal Astronomical Society, 484, 3233, \dodoi{10.1093/mnras/stz181}

\bibitem[{{Borucki}(2017)}]{Borucki2017}
{Borucki}, W.~J. 2017, Proceedings of the American Philosophical Society, 161,
  38

\bibitem[{{Bourrier} {et~al.}(2022){Bourrier}, {Deline}, {Krenn}, {Egger},
  {Petit}, {Malavolta}, {Cretignier}, {Billot}, {Broeg}, {Flor{\'e}n},
  {Queloz}, {Alibert}, {Bonfanti}, {Bonomo}, {Delisle}, {Demangeon}, {Demory},
  {Dumusque}, {Ehrenreich}, {Haywood}, {Howell}, {Lendl}, {Mortier}, {Nigro},
  {Salmon}, {Sousa}, {Wilson}, {Adibekyan}, {Alonso}, {Anglada}, {B{\'a}rczy},
  {Barrado y Navascues}, {Barros}, {Baumjohann}, {Beck}, {Benz}, {Biondi},
  {Bonfils}, {Brandeker}, {Cabrera}, {Charnoz}, {Csizmadia}, {Collier Cameron},
  {Damasso}, {Davies}, {Deleuil}, {Delrez}, {Di Fabrizio}, {Erikson},
  {Fortier}, {Fossati}, {Fridlund}, {Gandolfi}, {Gillon}, {G{\"u}del}, {Heng},
  {Hoyer}, {Isaak}, {Kiss}, {Laskar}, {Lecavelier des Etangs}, {Lorenzi},
  {Lovis}, {Magrin}, {Massa}, {Maxted}, {Nascimbeni}, {Olofsson}, {Ottensamer},
  {Pagano}, {Pall{\'e}}, {Peter}, {Piotto}, {Pollacco}, {Ragazzoni}, {Rando},
  {Rauer}, {Ribas}, {Santos}, {Scandariato}, {S{\'e}gransan}, {Simon}, {Smith},
  {Steller}, {Szab{\'o}}, {Thomas}, {Udry}, {Van Grootel}, {Verrecchia},
  {Walton}, {Beck}, {Buder}, {Ratti}, {Ulmer}, \& {Viotto}}]{Bourrier2022}
{Bourrier}, V., {Deline}, A., {Krenn}, A., {et~al.} 2022, \aap, 668, A31,
  \dodoi{10.1051/0004-6361/202243778}

\bibitem[{{Bourrier} {et~al.}(2023){Bourrier}, {Attia}, {Mallonn}, {Marret},
  {Lendl}, {Konig}, {Krenn}, {Cretignier}, {Allart}, {Henry}, {Bryant},
  {Leleu}, {Nielsen}, {Hebrard}, {Hara}, {Ehrenreich}, {Seidel}, {dos Santos},
  {Lovis}, {Bayliss}, {Cegla}, {Dumusque}, {Boisse}, {Boucher}, {Bouchy},
  {Pepe}, {Lavie}, {Rey Cerda}, {S{\'e}gransan}, {Udry}, \&
  {Vrignaud}}]{Bourrier2023}
{Bourrier}, V., {Attia}, O., {Mallonn}, M., {et~al.} 2023, \aap, 669, A63,
  \dodoi{10.1051/0004-6361/202245004}

\bibitem[{{Boyajian} {et~al.}(2012){Boyajian}, {von Braun}, {van Belle},
  {McAlister}, {ten Brummelaar}, {Kane}, {Muirhead}, {Jones}, {White},
  {Schaefer}, {Ciardi}, {Henry}, {L{\'o}pez-Morales}, {Ridgway}, {Gies}, {Jao},
  {Rojas-Ayala}, {Parks}, {Sturmann}, {Sturmann}, {Turner}, {Farrington},
  {Goldfinger}, \& {Berger}}]{boyajian:2012b}
{Boyajian}, T.~S., {von Braun}, K., {van Belle}, G., {et~al.} 2012, \apj, 757,
  112, \dodoi{10.1088/0004-637X/757/2/112}

\bibitem[{{Brahm} {et~al.}(2019){Brahm}, {Espinoza}, {Rabus}, {Jord{\'a}n},
  {D{\'\i}az}, {Rojas}, {Vu{\v{c}}kovi{\'c}}, {Zapata}, {Cort{\'e}s}, {Drass},
  {Jenkins}, {Lachaume}, {Pantoja}, {Sarkis}, {Soto}, {Vasquez}, {Henning}, \&
  {Jones}}]{Brahm2019}
{Brahm}, R., {Espinoza}, N., {Rabus}, M., {et~al.} 2019, \mnras, 483, 1970,
  \dodoi{10.1093/mnras/sty3230}

\bibitem[{{Brewer} \& {Fischer}(2018)}]{Brewer2018}
{Brewer}, J.~M., \& {Fischer}, D.~A. 2018, \apjs, 237, 38,
  \dodoi{10.3847/1538-4365/aad501}

\bibitem[{{Brewer} {et~al.}(2015){Brewer}, {Fischer}, {Basu}, {Valenti}, \&
  {Piskunov}}]{Brewer2015}
{Brewer}, J.~M., {Fischer}, D.~A., {Basu}, S., {Valenti}, J.~A., \& {Piskunov},
  N. 2015, \apj, 805, 126, \dodoi{10.1088/0004-637X/805/2/126}

\bibitem[{{Brewer} {et~al.}(2016){Brewer}, {Fischer}, {Valenti}, \&
  {Piskunov}}]{Brewer2016}
{Brewer}, J.~M., {Fischer}, D.~A., {Valenti}, J.~A., \& {Piskunov}, N. 2016,
  \apjs, 225, 32, \dodoi{10.3847/0067-0049/225/2/32}

\bibitem[{{Bryant} \& {Bayliss}(2022)}]{Bryant2022}
{Bryant}, E.~M., \& {Bayliss}, D. 2022, \aj, 163, 197,
  \dodoi{10.3847/1538-3881/ac58ff}

\bibitem[{{Butler} {et~al.}(1996){Butler}, {Marcy}, {Williams}, {McCarthy},
  {Dosanjh}, \& {Vogt}}]{Butler1996}
{Butler}, R.~P., {Marcy}, G.~W., {Williams}, E., {et~al.} 1996, \pasp, 108,
  500, \dodoi{10.1086/133755}

\bibitem[{{Chakraborty} {et~al.}(2018){Chakraborty}, {Roy}, {Sharma},
  {Mahadevan}, {Chaturvedi}, {Prasad}, \& {Anandarao}}]{Chakraborty2018}
{Chakraborty}, A., {Roy}, A., {Sharma}, R., {et~al.} 2018, ArXiv e-prints.
\newblock \doarXiv{1805.03466}

\bibitem[{{Chontos} {et~al.}(2022){Chontos}, {Murphy}, {MacDougall},
  {Fetherolf}, {Van Zandt}, {Rubenzahl}, {Beard}, {Huber}, {Batalha},
  {Crossfield}, {Dressing}, {Fulton}, {Howard}, {Isaacson}, {Kane}, {Petigura},
  {Robertson}, {Roy}, {Weiss}, {Behmard}, {Dai}, {Dalba}, {Giacalone}, {Hill},
  {Lubin}, {Mayo}, {Mo{\v{c}}nik}, {Polanski}, {Rosenthal}, {Scarsdale},
  {Turtelboom}, {Ricker}, {Vanderspek}, {Latham}, {Seager}, {Winn}, {Jenkins},
  {Quinn}, {Guerrero}, {Collins}, {Ciardi}, {Shporer}, {Goeke}, {Levine},
  {Ting}, {Bieryla}, {Collins}, {Kielkopf}, {Barkaoui}, {Benni},
  {Esparza-Borges}, {Conti}, {Hooton}, {Kagetani}, {Laloum}, {Marino},
  {Massey}, {Murgas}, {Papini}, {Schwarz}, {Srdoc}, {Stockdale}, {Wang},
  {Wittrock}, \& {Zou}}]{Chontos2022}
{Chontos}, A., {Murphy}, J. M.~A., {MacDougall}, M.~G., {et~al.} 2022, \aj,
  163, 297, \dodoi{10.3847/1538-3881/ac6266}

\bibitem[{{Christiansen} {et~al.}(2017){Christiansen}, {Vanderburg}, {Burt},
  {Fulton}, {Batygin}, {Benneke}, {Brewer}, {Charbonneau}, {Ciardi}, {Collier
  Cameron}, {Coughlin}, {Crossfield}, {Dressing}, {Greene}, {Howard}, {Latham},
  {Molinari}, {Mortier}, {Mullally}, {Pepe}, {Rice}, {Sinukoff}, {Sozzetti},
  {Thompson}, {Udry}, {Vogt}, {Barman}, {Batalha}, {Bouchy}, {Buchhave},
  {Butler}, {Cosentino}, {Dupuy}, {Ehrenreich}, {Fiorenzano}, {Hansen},
  {Henning}, {Hirsch}, {Holden}, {Isaacson}, {Johnson}, {Knutson}, {Kosiarek},
  {L{\'o}pez-Morales}, {Lovis}, {Malavolta}, {Mayor}, {Micela}, {Motalebi},
  {Petigura}, {Phillips}, {Piotto}, {Rogers}, {Sasselov}, {Schlieder},
  {S{\'e}gransan}, {Watson}, \& {Weiss}}]{Christiansen2017}
{Christiansen}, J.~L., {Vanderburg}, A., {Burt}, J., {et~al.} 2017, \aj, 154,
  122, \dodoi{10.3847/1538-3881/aa832d}

\bibitem[{{Christiansen} {et~al.}(2022){Christiansen}, {Bhure}, {Zink},
  {Hardegree-Ullman}, {Adkins}, {Hedges}, {Morton}, {Bieryla}, {Ciardi},
  {Cochran}, {Dressing}, {Everett}, {Isaacson}, {Livingston}, {Ziegler},
  {Berlind}, {Calkins}, {Esquerdo}, {Latham}, {Endl}, {MacQueen}, {Fulton},
  {Hirsch}, {Howard}, {Weiss}, {Allen}, {Berberyann}, {Ciardi}, {Dunlavy},
  {Glassford}, {Dai}, {Hirano}, {Tamura}, {Beichman}, {Gonzales}, {Schlieder},
  {Barclay}, {Crossfield}, {Gilbert}, {Matthews}, {Giacalone}, \&
  {Petigura}}]{Christiansen2022}
{Christiansen}, J.~L., {Bhure}, S., {Zink}, J.~K., {et~al.} 2022, \aj, 163,
  244, \dodoi{10.3847/1538-3881/ac5c4c}

\bibitem[{{Christiansen} {et~al.}(2023){Christiansen}, {Zink},
  {Hardegree-Ullman}, {Fernandes}, {Hopkins}, {Rebull}, {Boley}, {Bergsten}, \&
  {Bhure}}]{Christiansen2023}
{Christiansen}, J.~L., {Zink}, J.~K., {Hardegree-Ullman}, K.~K., {et~al.} 2023,
  \aj, 166, 248, \dodoi{10.3847/1538-3881/acf9f9}

\bibitem[{{Claret} {et~al.}(2012){Claret}, {Hauschildt}, \&
  {Witte}}]{Claret2012}
{Claret}, A., {Hauschildt}, P.~H., \& {Witte}, S. 2012, VizieR Online Data
  Catalog, J/A+A/546/A14

\bibitem[{{Cloutier} {et~al.}(2017){Cloutier}, {Astudillo-Defru}, {Doyon},
  {Bonfils}, {Almenara}, {Benneke}, {Bouchy}, {Delfosse}, {Ehrenreich},
  {Forveille}, {Lovis}, {Mayor}, {Menou}, {Murgas}, {Pepe}, {Rowe}, {Santos},
  {Udry}, \& {W{\"u}nsche}}]{Cloutier2017}
{Cloutier}, R., {Astudillo-Defru}, N., {Doyon}, R., {et~al.} 2017, \aap, 608,
  A35, \dodoi{10.1051/0004-6361/201731558}

\bibitem[{{Cloutier} {et~al.}(2019){Cloutier}, {Astudillo-Defru}, {Doyon},
  {Bonfils}, {Almenara}, {Bouchy}, {Delfosse}, {Forveille}, {Lovis}, {Mayor},
  {Menou}, {Murgas}, {Pepe}, {Santos}, {Udry}, \& {W{\"u}nsche}}]{Cloutier2019}
---. 2019, \aap, 621, A49, \dodoi{10.1051/0004-6361/201833995}

\bibitem[{{Collier Cameron} {et~al.}(2007){Collier Cameron}, {Wilson}, {West},
  {Hebb}, {Wang}, {Aigrain}, {Bouchy}, {Christian}, {Clarkson}, {Enoch},
  {Esposito}, {Guenther}, {Haswell}, {H{\'e}brard}, {Hellier}, {Horne},
  {Irwin}, {Kane}, {Loeillet}, {Lister}, {Maxted}, {Mayor}, {Moutou}, {Parley},
  {Pollacco}, {Pont}, {Queloz}, {Ryans}, {Skillen}, {Street}, {Udry}, \&
  {Wheatley}}]{CollierCameron2007}
{Collier Cameron}, A., {Wilson}, D.~M., {West}, R.~G., {et~al.} 2007, \mnras,
  380, 1230, \dodoi{10.1111/j.1365-2966.2007.12195.x}

\bibitem[{{Crepp} {et~al.}(2014){Crepp}, {Johnson}, {Howard}, {Marcy},
  {Brewer}, {Fischer}, {Wright}, \& {Isaacson}}]{Crepp2014}
{Crepp}, J.~R., {Johnson}, J.~A., {Howard}, A.~W., {et~al.} 2014, \apj, 781,
  29, \dodoi{10.1088/0004-637X/781/1/29}

\bibitem[{{Crossfield} {et~al.}(2015){Crossfield}, {Petigura}, {Schlieder},
  {Howard}, {Fulton}, {Aller}, {Ciardi}, {L{\'e}pine}, {Barclay}, {de Pater},
  {de Kleer}, {Quintana}, {Christiansen}, {Schlafly}, {Kaltenegger}, {Crepp},
  {Henning}, {Obermeier}, {Deacon}, {Weiss}, {Isaacson}, {Hansen}, {Liu},
  {Greene}, {Howell}, {Barman}, \& {Mordasini}}]{Crossfield2015}
{Crossfield}, I.~J.~M., {Petigura}, E., {Schlieder}, J.~E., {et~al.} 2015,
  \apj, 804, 10, \dodoi{10.1088/0004-637X/804/1/10}

\bibitem[{{Crossfield} {et~al.}(2016){Crossfield}, {Ciardi}, {Petigura},
  {Sinukoff}, {Schlieder}, {Howard}, {Beichman}, {Isaacson}, {Dressing},
  {Christiansen}, {Fulton}, {L{\'e}pine}, {Weiss}, {Hirsch}, {Livingston},
  {Baranec}, {Law}, {Riddle}, {Ziegler}, {Howell}, {Horch}, {Everett}, {Teske},
  {Martinez}, {Obermeier}, {Benneke}, {Scott}, {Deacon}, {Aller}, {Hansen},
  {Mancini}, {Ciceri}, {Brahm}, {Jord{\'a}n}, {Knutson}, {Henning}, {Bonnefoy},
  {Liu}, {Crepp}, {Lothringer}, {Hinz}, {Bailey}, {Skemer}, \&
  {Defrere}}]{Crossfield2016}
{Crossfield}, I.~J.~M., {Ciardi}, D.~R., {Petigura}, E.~A., {et~al.} 2016,
  \apjs, 226, 7, \dodoi{10.3847/0067-0049/226/1/7}

\bibitem[{{Crossfield} {et~al.}(2017){Crossfield}, {Ciardi}, {Isaacson},
  {Howard}, {Petigura}, {Weiss}, {Fulton}, {Sinukoff}, {Schlieder}, {Mawet},
  {Ruane}, {de Pater}, {de Kleer}, {Davies}, {Christiansen}, {Dressing},
  {Hirsch}, {Benneke}, {Crepp}, {Kosiarek}, {Livingston}, {Gonzales},
  {Beichman}, \& {Knutson}}]{Crossfield2017}
{Crossfield}, I.~J.~M., {Ciardi}, D.~R., {Isaacson}, H., {et~al.} 2017, \aj,
  153, 255, \dodoi{10.3847/1538-3881/aa6e01}

\bibitem[{{Crossfield} {et~al.}(2018){Crossfield}, {Guerrero}, {David},
  {Quinn}, {Feinstein}, {Huang}, {Yu}, {Collins}, {Fulton}, {Benneke},
  {Peterson}, {Bieryla}, {Schlieder}, {Kosiarek}, {Bristow}, {Newton},
  {Bedell}, {Latham}, {Christiansen}, {Esquerdo}, {Berlind}, {Calkins},
  {Shporer}, {Burt}, {Ballard}, {Rodriguez}, {Mehrle}, {Dressing},
  {Livingston}, {Petigura}, {Seager}, {Dittmann}, {Berardo}, {Sha}, {Essack},
  {Zhan}, {Owens}, {Kain}, {Isaacson}, {Ciardi}, {Gonzales}, {Howard}, \&
  {Cardoso}}]{Crossfield2018}
{Crossfield}, I. J.~M., {Guerrero}, N., {David}, T., {et~al.} 2018, \apjs, 239,
  5, \dodoi{10.3847/1538-4365/aae155}

\bibitem[{{Crossfield} {et~al.}(2025){Crossfield}, {Polanski}, {Robertson},
  {Murphy}, {Turtelboom}, {Luque}, {Beatty}, {Daylan}, {Isaacson}, {Brande},
  {Kreidberg}, {Batalha}, {Huber}, {Rhem}, {Dressing}, {Kane}, {Bossett},
  {Gagnebin}, {Kroft}, {Premnath}, {Rogers}, {Collins}, {Latham}, {Watkins},
  {Ciardi}, {Howell}, {Savel}, {Berlind}, {Calkins}, {Esquerdo}, {Mink},
  {Clark}, {Lund}, {Matson}, {Everett}, {Schlieder}, {Matthews}, {Giacalone},
  {Barclay}, {Zambelli}, {Plavchan}, {Ellingson}, {Bowen}, {Srdoc}, {McLeod},
  {Schwarz}, {Barkaoui}, {Kamler}, {Murgas}, {Palle}, {Narita}, {Fukui},
  {Relles}, {Bieryla}, {Girardin}, {Massey}, {Stockdale}, {Lewin}, {Papini},
  {Guerra}, {Conti}, {Yal{\c{c}}inkaya}, {Ba{\c{s}}t{\"u}rk}, \&
  {Mourad}}]{Crossfield2025}
{Crossfield}, I. J.~M., {Polanski}, A.~S., {Robertson}, P., {et~al.} 2025, \aj,
  169, 89, \dodoi{10.3847/1538-3881/ad9aa6}

\bibitem[{{Dai} {et~al.}(2019){Dai}, {Masuda}, {Winn}, \& {Zeng}}]{Dai2019}
{Dai}, F., {Masuda}, K., {Winn}, J.~N., \& {Zeng}, L. 2019, \apj, 883, 79,
  \dodoi{10.3847/1538-4357/ab3a3b}

\bibitem[{{Dai} \& {Winn}(2017)}]{Dai2017}
{Dai}, F., \& {Winn}, J.~N. 2017, \aj, 153, 205,
  \dodoi{10.3847/1538-3881/aa65d1}

\bibitem[{{Dai} {et~al.}(2015){Dai}, {Winn}, {Arriagada}, {Butler}, {Crane},
  {Johnson}, {Shectman}, {Teske}, {Thompson}, {Vanderburg}, \&
  {Wittenmyer}}]{Dai2015}
{Dai}, F., {Winn}, J.~N., {Arriagada}, P., {et~al.} 2015, \apjl, 813, L9,
  \dodoi{10.1088/2041-8205/813/1/L9}

\bibitem[{{Dai} {et~al.}(2016){Dai}, {Winn}, {Albrecht}, {Arriagada},
  {Bieryla}, {Butler}, {Crane}, {Hirano}, {Johnson}, {Kiilerich}, {Latham},
  {Narita}, {Nowak}, {Palle}, {Ribas}, {Rogers}, {Sanchis-Ojeda}, {Shectman},
  {Teske}, {Thompson}, {Van Eylen}, {Vanderburg}, {Wittenmyer}, \&
  {Yu}}]{Dai2016}
{Dai}, F., {Winn}, J.~N., {Albrecht}, S., {et~al.} 2016, \apj, 823, 115,
  \dodoi{10.3847/0004-637X/823/2/115}

\bibitem[{{Dalal} {et~al.}(2019){Dalal}, {H{\'e}brard}, {Lecavelier des
  {\'E}tangs}, {Petit}, {Bourrier}, {Laskar}, {K{\"o}nig}, \&
  {Correia}}]{Dalal2019}
{Dalal}, S., {H{\'e}brard}, G., {Lecavelier des {\'E}tangs}, A., {et~al.} 2019,
  \aap, 631, A28, \dodoi{10.1051/0004-6361/201935944}

\bibitem[{{Damasso} {et~al.}(2018){Damasso}, {Bonomo}, {Astudillo-Defru},
  {Bonfils}, {Malavolta}, {Sozzetti}, {Lopez}, {Zeng}, {Haywood}, {Irwin},
  {Mortier}, {Vanderburg}, {Maldonado}, {Lanza}, {Affer}, {Almenara},
  {Benatti}, {Biazzo}, {Bignamini}, {Borsa}, {Bouchy}, {Buchhave}, {Cameron},
  {Carleo}, {Charbonneau}, {Claudi}, {Cosentino}, {Covino}, {Delfosse},
  {Desidera}, {Di Fabrizio}, {Dressing}, {Esposito}, {Fares}, {Figueira},
  {Fiorenzano}, {Forveille}, {Giacobbe}, {Gonz{\'a}lez-{\'A}lvarez}, {Gratton},
  {Harutyunyan}, {Johnson}, {Latham}, {Leto}, {Lopez-Morales}, {Lovis},
  {Maggio}, {Mancini}, {Masiero}, {Mayor}, {Micela}, {Molinari}, {Motalebi},
  {Murgas}, {Nascimbeni}, {Pagano}, {Pepe}, {Phillips}, {Piotto}, {Poretti},
  {Rainer}, {Rice}, {Santos}, {Sasselov}, {Scandariato}, {S{\'e}gransan},
  {Smareglia}, {Udry}, {Watson}, \& {W{\"u}nsche}}]{Damasso2018}
{Damasso}, M., {Bonomo}, A.~S., {Astudillo-Defru}, N., {et~al.} 2018, ArXiv
  e-prints.
\newblock \doarXiv{1802.08320}

\bibitem[{{Damasso, M.} {et~al.}(2019){Damasso, M.}, {Zeng, L.}, {Malavolta,
  L.}, {Mayo, A.}, {Sozzetti, A.}, {Mortier, A.}, {Buchhave, L. A.},
  {Vanderburg, A.}, {Lopez-Morales, M.}, {Bonomo, A. S.}, {Cameron, A. C.},
  {Coffinet, A.}, {Figueira, P.}, {Latham, D. W.}, {Mayor, M.}, {Molinari, E.},
  {Pepe, F.}, {Phillips, D. F.}, {Poretti, E.}, {Rice, K.}, {Udry, S.}, \&
  {Watson, C. A.}}]{Damasso2019}
{Damasso, M.}, {Zeng, L.}, {Malavolta, L.}, {et~al.} 2019, A\&A, 624, A38,
  \dodoi{10.1051/0004-6361/201834671}

\bibitem[{{Dattilo} {et~al.}(2019){Dattilo}, {Vanderburg}, {Shallue}, {Mayo},
  {Berlind}, {Bieryla}, {Calkins}, {Esquerdo}, {Everett}, {Howell}, {Latham},
  {Scott}, \& {Yu}}]{Dattilo2019}
{Dattilo}, A., {Vanderburg}, A., {Shallue}, C.~J., {et~al.} 2019, \aj, 157,
  169, \dodoi{10.3847/1538-3881/ab0e12}

\bibitem[{{Diamond-Lowe} {et~al.}(2022){Diamond-Lowe}, {Kreidberg}, {Harman},
  {Kempton}, {Rogers}, {Joyce}, {Eastman}, {King}, {Kopparapu}, {Youngblood},
  {Kosiarek}, {Livingston}, {Hardegree-Ullman}, \&
  {Crossfield}}]{DiamondLowe2022}
{Diamond-Lowe}, H., {Kreidberg}, L., {Harman}, C.~E., {et~al.} 2022, \aj, 164,
  172, \dodoi{10.3847/1538-3881/ac7807}

\bibitem[{{dos Santos} {et~al.}(2020){dos Santos}, {Ehrenreich}, {Bourrier},
  {Astudillo-Defru}, {Bonfils}, {Forget}, {Lovis}, {Pepe}, \&
  {Udry}}]{dosSantos2020}
{dos Santos}, L.~A., {Ehrenreich}, D., {Bourrier}, V., {et~al.} 2020, \aap,
  634, L4, \dodoi{10.1051/0004-6361/201937327}

\bibitem[{{Doyle} {et~al.}(2011){Doyle}, {Carter}, {Fabrycky}, {Slawson},
  {Howell}, {Winn}, {Orosz}, {P{\v{r}}sa}, {Welsh}, {Quinn}, {Latham},
  {Torres}, {Buchhave}, {Marcy}, {Fortney}, {Shporer}, {Ford}, {Lissauer},
  {Ragozzine}, {Rucker}, {Batalha}, {Jenkins}, {Borucki}, {Koch}, {Middour},
  {Hall}, {McCauliff}, {Fanelli}, {Quintana}, {Holman}, {Caldwell}, {Still},
  {Stefanik}, {Brown}, {Esquerdo}, {Tang}, {Furesz}, {Geary}, {Berlind},
  {Calkins}, {Short}, {Steffen}, {Sasselov}, {Dunham}, {Cochran}, {Boss},
  {Haas}, {Buzasi}, \& {Fischer}}]{Doyle2011}
{Doyle}, L.~R., {Carter}, J.~A., {Fabrycky}, D.~C., {et~al.} 2011, Science,
  333, 1602, \dodoi{10.1126/science.1210923}

\bibitem[{Dressing {et~al.}(2017)Dressing, Newton, Schlieder, Charbonneau,
  Knutson, Vanderburg, \& Sinukoff}]{Dressing2017}
Dressing, C.~D., Newton, E.~R., Schlieder, J.~E., {et~al.} 2017, The
  Astrophysical Journal, 836, 167, \dodoi{10.3847/1538-4357/836/2/167}

\bibitem[{{Dressing} {et~al.}(2018){Dressing}, {Sinukoff}, {Fulton}, {Lopez},
  {Beichman}, {Howard}, {Knutson}, {Werner}, {Benneke}, {Crossfield},
  {Isaacson}, {Krick}, {Gorjian}, {Livingston}, {Petigura}, {Schlieder},
  {Akeson}, {Batygin}, {Christiansen}, {Ciardi}, {Crepp}, {Gonzales},
  {Hardegree-Ullman}, {Hirsch}, {Kosiarek}, \& {Weiss}}]{Dressing2018}
{Dressing}, C.~D., {Sinukoff}, E., {Fulton}, B.~J., {et~al.} 2018, \aj, 156,
  70, \dodoi{10.3847/1538-3881/aacf99}

\bibitem[{{Dressing} {et~al.}(2019){Dressing}, {Hardegree-Ullman}, {Schlieder},
  {Newton}, {Vand erburg}, {Feinstein}, {Duvvuri}, {Arnold}, {Bristow},
  {Thackeray}, {Schwab Abrahams}, {Ciardi}, {Crossfield}, {Yu}, {Martinez},
  {Christiansen}, {Crepp}, \& {Isaacson}}]{Dressing2019}
{Dressing}, C.~D., {Hardegree-Ullman}, K., {Schlieder}, J.~E., {et~al.} 2019,
  \aj, 158, 87, \dodoi{10.3847/1538-3881/ab2895}

\bibitem[{{Fang} \& {Margot}(2012)}]{Fang2012}
{Fang}, J., \& {Margot}, J.-L. 2012, \apj, 761, 92,
  \dodoi{10.1088/0004-637X/761/2/92}

\bibitem[{{Ford}(2006)}]{Ford2006}
{Ford}, E.~B. 2006, \apj, 642, 505, \dodoi{10.1086/500802}

\bibitem[{{Foreman-Mackey}(2015)}]{Foreman-Mackey2015a}
{Foreman-Mackey}, D. 2015, {George: Gaussian Process regression}.
\newblock \doeprint{1511.015}

\bibitem[{{Foreman-Mackey} {et~al.}(2013){Foreman-Mackey}, {Hogg}, {Lang}, \&
  {Goodman}}]{Foreman-Mackey2013}
{Foreman-Mackey}, D., {Hogg}, D.~W., {Lang}, D., \& {Goodman}, J. 2013, \pasp,
  125, 306, \dodoi{10.1086/670067}

\bibitem[{{Foreman-Mackey} {et~al.}(2015){Foreman-Mackey}, {Montet}, {Hogg},
  {Morton}, {Wang}, \& {Sch{\"o}lkopf}}]{Foreman-Mackey2015}
{Foreman-Mackey}, D., {Montet}, B.~T., {Hogg}, D.~W., {et~al.} 2015, \apj, 806,
  215, \dodoi{10.1088/0004-637X/806/2/215}

\bibitem[{{Fortney} {et~al.}(2007){Fortney}, {Marley}, \&
  {Barnes}}]{Fortney2007}
{Fortney}, J.~J., {Marley}, M.~S., \& {Barnes}, J.~W. 2007, \apj, 659, 1661,
  \dodoi{10.1086/512120}

\bibitem[{{Fridlund} {et~al.}(2017){Fridlund}, {Gaidos}, {Barrag{\'a}n},
  {Persson}, {Gandolfi}, {Cabrera}, {Hirano}, {Kuzuhara}, {Csizmadia}, {Nowak},
  {Endl}, {Grziwa}, {Korth}, {Pfaff}, {Bitsch}, {Johansen}, {Mustill},
  {Davies}, {Deeg}, {Palle}, {Cochran}, {Eigm{\"u}ller}, {Erikson}, {Guenther},
  {Hatzes}, {Kiilerich}, {Kudo}, {MacQueen}, {Narita}, {Nespral},
  {P{\"a}tzold}, {Prieto-Arranz}, {Rauer}, \& {Van Eylen}}]{Fridlund2017}
{Fridlund}, M., {Gaidos}, E., {Barrag{\'a}n}, O., {et~al.} 2017, \aap, 604,
  A16, \dodoi{10.1051/0004-6361/201730822}

\bibitem[{{Fukui} {et~al.}(2016){Fukui}, {Livingston}, {Narita}, {Hirano},
  {Onitsuka}, {Ryu}, \& {Kusakabe}}]{Fukui2016}
{Fukui}, A., {Livingston}, J., {Narita}, N., {et~al.} 2016, \aj, 152, 171,
  \dodoi{10.3847/0004-6256/152/6/171}

\bibitem[{{Fulton} \& {Petigura}(2018)}]{Fulton2018b}
{Fulton}, B.~J., \& {Petigura}, E.~A. 2018, \aj, 156, 264,
  \dodoi{10.3847/1538-3881/aae828}

\bibitem[{{Fulton} {et~al.}(2018){Fulton}, {Petigura}, {Blunt}, \&
  {Sinukoff}}]{Fulton2018}
{Fulton}, B.~J., {Petigura}, E.~A., {Blunt}, S., \& {Sinukoff}, E. 2018, \pasp,
  130, 044504, \dodoi{10.1088/1538-3873/aaaaa8}

\bibitem[{{Fulton} {et~al.}(2015){Fulton}, {Weiss}, {Sinukoff}, {Isaacson},
  {Howard}, {Marcy}, {Henry}, {Holden}, \& {Kibrick}}]{Fulton2015}
{Fulton}, B.~J., {Weiss}, L.~M., {Sinukoff}, E., {et~al.} 2015, \apj, 805, 175,
  \dodoi{10.1088/0004-637X/805/2/175}

\bibitem[{{Fulton} {et~al.}(2017){Fulton}, {Petigura}, {Howard}, {Isaacson},
  {Marcy}, {Cargile}, {Hebb}, {Weiss}, {Johnson}, {Morton}, {Sinukoff},
  {Crossfield}, \& {Hirsch}}]{Fulton2017}
{Fulton}, B.~J., {Petigura}, E.~A., {Howard}, A.~W., {et~al.} 2017, \aj, 154,
  109, \dodoi{10.3847/1538-3881/aa80eb}

\bibitem[{{Gaia Collaboration} {et~al.}(2018){Gaia Collaboration}, {Brown},
  {Vallenari}, {Prusti}, {de Bruijne}, {Babusiaux}, {Bailer-Jones}, {Biermann},
  {Evans}, {Eyer}, \& et~al.}]{gaia:2018}
{Gaia Collaboration}, {Brown}, A.~G.~A., {Vallenari}, A., {et~al.} 2018, \aap,
  616, A1, \dodoi{10.1051/0004-6361/201833051}

\bibitem[{{Gandolfi} {et~al.}(2017){Gandolfi}, {Barrag{\'a}n}, {Hatzes},
  {Fridlund}, {Fossati}, {Donati}, {Johnson}, {Nowak}, {Prieto-Arranz},
  {Albrecht}, {Dai}, {Deeg}, {Endl}, {Grziwa}, {Hjorth}, {Korth}, {Nespral},
  {Saario}, {Smith}, {Antoniciello}, {Alarcon}, {Bedell}, {Blay}, {Brems},
  {Cabrera}, {Csizmadia}, {Cusano}, {Cochran}, {Eigm{\"u}ller}, {Erikson},
  {Gonz{\'a}lez Hern{\'a}ndez}, {Guenther}, {Hirano}, {Su{\'a}rez
  Mascare{\~n}o}, {Narita}, {Palle}, {Parviainen}, {P{\"a}tzold}, {Persson},
  {Rauer}, {Saviane}, {Schmidtobreick}, {Van Eylen}, {Winn}, \&
  {Zakhozhay}}]{Gandolfi2017}
{Gandolfi}, D., {Barrag{\'a}n}, O., {Hatzes}, A.~P., {et~al.} 2017, \aj, 154,
  123, \dodoi{10.3847/1538-3881/aa832a}

\bibitem[{Gelman {et~al.}(2003)Gelman, Carlin, Stern, \& Rubin}]{Gelman2003}
Gelman, A., Carlin, J.~B., Stern, H.~S., \& Rubin, D.~B. 2003, Bayesian Data
  Analysis, 2nd edn. (Chapman and Hall)

\bibitem[{{Ginzburg} {et~al.}(2018){Ginzburg}, {Schlichting}, \&
  {Sari}}]{Ginzburg2018}
{Ginzburg}, S., {Schlichting}, H.~E., \& {Sari}, R. 2018, \mnras, 476, 759,
  \dodoi{10.1093/mnras/sty290}

\bibitem[{{Grunblatt} {et~al.}(2015){Grunblatt}, {Howard}, \&
  {Haywood}}]{Grunblatt2015}
{Grunblatt}, S.~K., {Howard}, A.~W., \& {Haywood}, R.~D. 2015, \apj, 808, 127,
  \dodoi{10.1088/0004-637X/808/2/127}

\bibitem[{{Grziwa} {et~al.}(2016){Grziwa}, {Gandolfi}, {Csizmadia}, {Fridlund},
  {Parviainen}, {Deeg}, {Cabrera}, {Djupvik}, {Albrecht}, {Palle},
  {P{\"a}tzold}, {B{\'e}jar}, {Prieto-Arranz}, {Eigm{\"u}ller}, {Erikson},
  {Fynbo}, {Guenther}, {Hatzes}, {Kiilerich}, {Korth}, {Kuutma},
  {Monta{\~n}{\'e}s-Rodr{\'{\i}}guez}, {Nespral}, {Nowak}, {Rauer}, {Saario},
  {Sebastian}, \& {Slumstrup}}]{Grziwa2016}
{Grziwa}, S., {Gandolfi}, D., {Csizmadia}, S., {et~al.} 2016, \aj, 152, 132,
  \dodoi{10.3847/0004-6256/152/5/132}

\bibitem[{{Guenther} {et~al.}(2024){Guenther}, {Goffo}, {Sebastian}, {Smith},
  {Persson}, {Fridlund}, {Gandolfi}, \& {Korth}}]{Guenther2024}
{Guenther}, E.~W., {Goffo}, E., {Sebastian}, D., {et~al.} 2024, arXiv e-prints,
  arXiv:2402.09322, \dodoi{10.48550/arXiv.2402.09322}

\bibitem[{{Guenther} {et~al.}(2017){Guenther}, {Barrag{\'a}n}, {Dai},
  {Gandolfi}, {Hirano}, {Fridlund}, {Fossati}, {Chau}, {Helled}, {Korth},
  {Prieto-Arranz}, {Nespral}, {Antoniciello}, {Deeg}, {Hjorth}, {Grziwa},
  {Albrecht}, {Hatzes}, {Rauer}, {Csizmadia}, {Smith}, {Cabrera}, {Narita},
  {Arriagada}, {Burt}, {Butler}, {Cochran}, {Crane}, {Eigm{\"u}ller},
  {Erikson}, {Johnson}, {Kiilerich}, {Kubyshkina}, {Palle}, {Persson},
  {P{\"a}tzold}, {Sabotta}, {Sato}, {Shectman}, {Teske}, {Thompson}, {Van
  Eylen}, {Nowak}, {Vanderburg}, {Winn}, \& {Wittenmyer}}]{Guenther2017}
{Guenther}, E.~W., {Barrag{\'a}n}, O., {Dai}, F., {et~al.} 2017, \aap, 608,
  A93, \dodoi{10.1051/0004-6361/201730885}

\bibitem[{{Guerrero} {et~al.}(2021){Guerrero}, {Seager}, {Huang}, {Vanderburg},
  {Garcia Soto}, {Mireles}, {Hesse}, {Fong}, {Glidden}, {Shporer}, {Latham},
  {Collins}, {Quinn}, {Burt}, {Dragomir}, {Crossfield}, {Vanderspek},
  {Fausnaugh}, {Burke}, {Ricker}, {Daylan}, {Essack}, {G{\"u}nther}, {Osborn},
  {Pepper}, {Rowden}, {Sha}, {Villanueva}, {Yahalomi}, {Yu}, {Ballard},
  {Batalha}, {Berardo}, {Chontos}, {Dittmann}, {Esquerdo}, {Mikal-Evans},
  {Jayaraman}, {Krishnamurthy}, {Louie}, {Mehrle}, {Niraula}, {Rackham},
  {Rodriguez}, {Rowden}, {Sousa-Silva}, {Watanabe}, {Wong}, {Zhan},
  {Zivanovic}, {Christiansen}, {Ciardi}, {Swain}, {Lund}, {Mullally},
  {Fleming}, {Rodriguez}, {Boyd}, {Quintana}, {Barclay}, {Col{\'o}n},
  {Rinehart}, {Schlieder}, {Clampin}, {Jenkins}, {Twicken}, {Caldwell},
  {Coughlin}, {Henze}, {Lissauer}, {Morris}, {Rose}, {Smith}, {Tenenbaum},
  {Ting}, {Wohler}, {Bakos}, {Bean}, {Berta-Thompson}, {Bieryla}, {Bouma},
  {Buchhave}, {Butler}, {Charbonneau}, {Doty}, {Ge}, {Holman}, {Howard},
  {Kaltenegger}, {Kane}, {Kjeldsen}, {Kreidberg}, {Lin}, {Minsky}, {Narita},
  {Paegert}, {P{\'a}l}, {Palle}, {Sasselov}, {Spencer}, {Sozzetti}, {Stassun},
  {Torres}, {Udry}, \& {Winn}}]{Guerrero2021}
{Guerrero}, N.~M., {Seager}, S., {Huang}, C.~X., {et~al.} 2021, \apjs, 254, 39,
  \dodoi{10.3847/1538-4365/abefe1}

\bibitem[{{Guilluy} {et~al.}(2023){Guilluy}, {Bourrier}, {Jaziri}, {Dethier},
  {Mounzer}, {Giacobbe}, {Attia}, {Allart}, {Bonomo}, {Dos Santos}, {Rainer},
  \& {Sozzetti}}]{Guilluy2023}
{Guilluy}, G., {Bourrier}, V., {Jaziri}, Y., {et~al.} 2023, \aap, 676, A130,
  \dodoi{10.1051/0004-6361/202346419}

\bibitem[{{Gupta} \& {Schlichting}(2019)}]{Gupta2019}
{Gupta}, A., \& {Schlichting}, H.~E. 2019, \mnras, 487, 24,
  \dodoi{10.1093/mnras/stz1230}

\bibitem[{{Hardegree-Ullman} {et~al.}(2020){Hardegree-Ullman}, {Zink},
  {Christiansen}, {Dressing}, {Ciardi}, \& {Schlieder}}]{Hardegree-Ullman2020}
{Hardegree-Ullman}, K.~K., {Zink}, J.~K., {Christiansen}, J.~L., {et~al.} 2020,
  \apjs, 247, 28, \dodoi{10.3847/1538-4365/ab7230}

\bibitem[{{Haywood} {et~al.}(2014){Haywood}, {Collier Cameron}, {Queloz},
  {Barros}, {Deleuil}, {Fares}, {Gillon}, {Lanza}, {Lovis}, {Moutou}, {Pepe},
  {Pollacco}, {Santerne}, {S{\'e}gransan}, \& {Unruh}}]{Haywood2014}
{Haywood}, R.~D., {Collier Cameron}, A., {Queloz}, D., {et~al.} 2014, \mnras,
  443, 2517, \dodoi{10.1093/mnras/stu1320}

\bibitem[{{Hejazi} {et~al.}(2023){Hejazi}, {Crossfield}, {Nordlander},
  {Mansfield}, {Souto}, {Marfil}, {Coria}, {Brande}, {Polanski}, {Hand}, \&
  {Wienke}}]{Hejazi2023}
{Hejazi}, N., {Crossfield}, I. J.~M., {Nordlander}, T., {et~al.} 2023, \apj,
  949, 79, \dodoi{10.3847/1538-4357/accb97}

\bibitem[{{Heller} {et~al.}(2019){Heller}, {Rodenbeck, Kai}, \& {Hippke,
  Michael}}]{Heller2019}
{Heller}, R., {Rodenbeck, Kai}, \& {Hippke, Michael}. 2019, A\&A, 625, A31,
  \dodoi{10.1051/0004-6361/201935276}

\bibitem[{{Hellier} {et~al.}(2012){Hellier}, {Anderson}, {Collier Cameron},
  {Doyle}, {Fumel}, {Gillon}, {Jehin}, {Lendl}, {Maxted}, {Pepe}, {Pollacco},
  {Queloz}, {S{\'e}gransan}, {Smalley}, {Smith}, {Southworth}, {Triaud},
  {Udry}, \& {West}}]{Hellier2012}
{Hellier}, C., {Anderson}, D.~R., {Collier Cameron}, A., {et~al.} 2012, \mnras,
  426, 739, \dodoi{10.1111/j.1365-2966.2012.21780.x}

\bibitem[{{Howard} \& {Fulton}(2016)}]{Howard2016}
{Howard}, A.~W., \& {Fulton}, B.~J. 2016, Publications of the Astronomical
  Society of the Pacific, 128, 114401, \dodoi{10.1088/1538-3873/128/969/114401}

\bibitem[{{Howard} {et~al.}(2010{\natexlab{a}}){Howard}, {Marcy}, {Johnson},
  {Fischer}, {Wright}, {Isaacson}, {Valenti}, {Anderson}, {Lin}, \&
  {Ida}}]{Howard2010}
{Howard}, A.~W., {Marcy}, G.~W., {Johnson}, J.~A., {et~al.} 2010{\natexlab{a}},
  Science, 330, 653, \dodoi{10.1126/science.1194854}

\bibitem[{{Howard} {et~al.}(2010{\natexlab{b}}){Howard}, {Johnson}, {Marcy},
  {Fischer}, {Wright}, {Bernat}, {Henry}, {Peek}, {Isaacson}, {Apps}, {Endl},
  {Cochran}, {Valenti}, {Anderson}, \& {Piskunov}}]{Howard2010_CPS}
{Howard}, A.~W., {Johnson}, J.~A., {Marcy}, G.~W., {et~al.} 2010{\natexlab{b}},
  \apj, 721, 1467, \dodoi{10.1088/0004-637X/721/2/1467}

\bibitem[{{Howard} {et~al.}(2012){Howard}, {Marcy}, {Bryson}, {Jenkins},
  {Rowe}, {Batalha}, {Borucki}, {Koch}, {Dunham}, {Gautier}, {Van Cleve},
  {Cochran}, {Latham}, {Lissauer}, {Torres}, {Brown}, {Gilliland}, {Buchhave},
  {Caldwell}, {Christensen-Dalsgaard}, {Ciardi}, {Fressin}, {Haas}, {Howell},
  {Kjeldsen}, {Seager}, {Rogers}, {Sasselov}, {Steffen}, {Basri},
  {Charbonneau}, {Christiansen}, {Clarke}, {Dupree}, {Fabrycky}, {Fischer},
  {Ford}, {Fortney}, {Tarter}, {Girouard}, {Holman}, {Johnson}, {Klaus},
  {Machalek}, {Moorhead}, {Morehead}, {Ragozzine}, {Tenenbaum}, {Twicken},
  {Quinn}, {Isaacson}, {Shporer}, {Lucas}, {Walkowicz}, {Welsh}, {Boss},
  {Devore}, {Gould}, {Smith}, {Morris}, {Prsa}, {Morton}, {Still}, {Thompson},
  {Mullally}, {Endl}, \& {MacQueen}}]{Howard2012}
{Howard}, A.~W., {Marcy}, G.~W., {Bryson}, S.~T., {et~al.} 2012, \apjs, 201,
  15, \dodoi{10.1088/0067-0049/201/2/15}

\bibitem[{{Howe} \& {Burrows}(2015)}]{Howe2015}
{Howe}, A.~R., \& {Burrows}, A. 2015, \apj, 808, 150,
  \dodoi{10.1088/0004-637X/808/2/150}

\bibitem[{{Howell} {et~al.}(2014){Howell}, {Sobeck}, {Haas}, {Still},
  {Barclay}, {Mullally}, {Troeltzsch}, {Aigrain}, {Bryson}, {Caldwell},
  {Chaplin}, {Cochran}, {Huber}, {Marcy}, {Miglio}, {Najita}, {Smith},
  {Twicken}, \& {Fortney}}]{Howell2014}
{Howell}, S.~B., {Sobeck}, C., {Haas}, M., {et~al.} 2014, \pasp, 126, 398,
  \dodoi{10.1086/676406}

\bibitem[{{Huber}(2017)}]{huber2017}
{Huber}, D. 2017, {Isoclassify: V1.2}, v1.2,  Zenodo,
  \dodoi{10.5281/zenodo.573372}

\bibitem[{{Isaacson} \& {Fischer}(2010)}]{Isaacson2010}
{Isaacson}, H., \& {Fischer}, D. 2010, \apj, 725, 875,
  \dodoi{10.1088/0004-637X/725/1/875}

\bibitem[{{Jin} \& {Mordasini}(2018)}]{Sheng2018}
{Jin}, S., \& {Mordasini}, C. 2018, \apj, 853, 163,
  \dodoi{10.3847/1538-4357/aa9f1e}

\bibitem[{{Johnson} {et~al.}(2018){Johnson}, {Dai}, {Justesen}, {Gandolfi},
  {Hatzes}, {Nowak}, {Endl}, {Cochran}, {Hidalgo}, {Watanabe}, {Parviainen},
  {Hirano}, {Villanueva}, {Prieto-Arranz}, {Narita}, {Palle}, {Guenther},
  {Barrag{\'a}n}, {Trifonov}, {Niraula}, {MacQueen}, {Cabrera}, {Csizmadia},
  {Eigm{\"u}ller}, {Grziwa}, {Korth}, {P{\"a}tzold}, {Smith}, {Albrecht},
  {Alonso}, {Deeg}, {Erikson}, {Esposito}, {Fridlund}, {Fukui}, {Kusakabe},
  {Kuzuhara}, {Livingston}, {Monta{\~n}es Rodriguez}, {Nespral}, {Persson},
  {Purismo}, {Raimundo}, {Rauer}, {Ribas}, {Tamura}, {Van Eylen}, \&
  {Winn}}]{Johnson2018}
{Johnson}, M.~C., {Dai}, F., {Justesen}, A.~B., {et~al.} 2018, \mnras, 481,
  596, \dodoi{10.1093/mnras/sty2238}

\bibitem[{{Jones} {et~al.}(2017){Jones}, {Stenning}, {Ford}, {Wolpert},
  {Loredo}, {Gilbertson}, \& {Dumusque}}]{Jones2017}
{Jones}, D.~E., {Stenning}, D.~C., {Ford}, E.~B., {et~al.} 2017, arXiv
  e-prints, arXiv:1711.01318.
\newblock \doarXiv{1711.01318}

\bibitem[{{JWST Transiting Exoplanet Community Early Release Science Team}
  {et~al.}(2023){JWST Transiting Exoplanet Community Early Release Science
  Team}, {Ahrer}, {Alderson}, {Batalha}, {Batalha}, {Bean}, {Beatty}, {Bell},
  {Benneke}, {Berta-Thompson}, {Carter}, {Crossfield}, {Espinoza}, {Feinstein},
  {Fortney}, {Gibson}, {Goyal}, {Kempton}, {Kirk}, {Kreidberg},
  {L{\'o}pez-Morales}, {Line}, {Lothringer}, {Moran}, {Mukherjee}, {Ohno},
  {Parmentier}, {Piaulet}, {Rustamkulov}, {Schlawin}, {Sing}, {Stevenson},
  {Wakeford}, {Allen}, {Birkmann}, {Brande}, {Crouzet}, {Cubillos}, {Damiano},
  {D{\'e}sert}, {Gao}, {Harrington}, {Hu}, {Kendrew}, {Knutson}, {Lagage},
  {Leconte}, {Lendl}, {MacDonald}, {May}, {Miguel}, {Molaverdikhani}, {Moses},
  {Murray}, {Nehring}, {Nikolov}, {Petit dit de la Roche}, {Radica}, {Roy},
  {Stassun}, {Taylor}, {Waalkes}, {Wachiraphan}, {Welbanks}, {Wheatley},
  {Aggarwal}, {Alam}, {Banerjee}, {Barstow}, {Blecic}, {Casewell}, {Changeat},
  {Chubb}, {Col{\'o}n}, {Coulombe}, {Daylan}, {de Val-Borro}, {Decin}, {Dos
  Santos}, {Flagg}, {France}, {Fu}, {Garc{\'\i}a Mu{\~n}oz}, {Gizis},
  {Glidden}, {Grant}, {Heng}, {Henning}, {Hong}, {Inglis}, {Iro}, {Kataria},
  {Komacek}, {Krick}, {Lee}, {Lewis}, {Lillo-Box}, {Lustig-Yaeger}, {Mancini},
  {Mandell}, {Mansfield}, {Marley}, {Mikal-Evans}, {Morello}, {Nixon}, {Ortiz
  Ceballos}, {Piette}, {Powell}, {Rackham}, {Ramos-Rosado}, {Rauscher},
  {Redfield}, {Rogers}, {Roman}, {Roudier}, {Scarsdale}, {Shkolnik},
  {Southworth}, {Spake}, {Steinrueck}, {Tan}, {Teske}, {Tremblin}, {Tsai},
  {Tucker}, {Turner}, {Valenti}, {Venot}, {Waldmann}, {Wallack}, {Zhang}, \&
  {Zieba}}]{ERS2023}
{JWST Transiting Exoplanet Community Early Release Science Team}, {Ahrer},
  E.-M., {Alderson}, L., {et~al.} 2023, \nat, 614, 649,
  \dodoi{10.1038/s41586-022-05269-w}

\bibitem[{{Kane} {et~al.}(2020){Kane}, {Fetherolf}, \& {Hill}}]{Kane2020}
{Kane}, S.~R., {Fetherolf}, T., \& {Hill}, M.~L. 2020, \aj, 159, 176,
  \dodoi{10.3847/1538-3881/ab7818}

\bibitem[{{Kanodia} {et~al.}(2019){Kanodia}, {Wolfgang}, {Stefansson}, {Ning},
  \& {Mahadevan}}]{Kanodia2019}
{Kanodia}, S., {Wolfgang}, A., {Stefansson}, G.~K., {Ning}, B., \& {Mahadevan},
  S. 2019, \apj, 882, 38, \dodoi{10.3847/1538-4357/ab334c}

\bibitem[{{Kipping}(2013)}]{Kipping2013}
{Kipping}, D.~M. 2013, \mnras, 435, 2152, \dodoi{10.1093/mnras/stt1435}

\bibitem[{{Korth} {et~al.}(2019){Korth}, {Csizmadia}, {Gandolfi}, {Fridlund},
  {P{\"a}tzold}, {Hirano}, {Livingston}, {Persson}, {Deeg}, {Justesen},
  {Barrag{\'a}n}, {Grziwa}, {Endl}, {Tronsgaard}, {Dai}, {Cochran}, {Albrecht},
  {Alonso}, {Cabrera}, {Cauley}, {Cusano}, {Eigm{\"u}ller}, {Erikson},
  {Esposito}, {Guenther}, {Hatzes}, {Hidalgo}, {Kuzuhara}, {Monta{\~n}es},
  {Napolitano}, {Narita}, {Niraula}, {Nespral}, {Nowak}, {Palle}, {Petrillo},
  {Redfield}, {Prieto-Arranz}, {Rauer}, {Smith}, {Tortora}, {Van Eylen}, \&
  {Winn}}]{Korth2019}
{Korth}, J., {Csizmadia}, S., {Gandolfi}, D., {et~al.} 2019, \mnras, 482, 1807,
  \dodoi{10.1093/mnras/sty2760}

\bibitem[{{Kosiarek} \& {Crossfield}(2020)}]{Kosiarek2020}
{Kosiarek}, M.~R., \& {Crossfield}, I. J.~M. 2020, \aj, 159, 271,
  \dodoi{10.3847/1538-3881/ab8d3a}

\bibitem[{{Kosiarek} {et~al.}(2019{\natexlab{a}}){Kosiarek}, {Blunt},
  {L{\'o}pez-Morales}, {Crossfield}, {Sinukoff}, {Petigura}, {Gonzales},
  {Poretti}, {Malavolta}, {Howard}, {Isaacson}, {Haywood}, {Ciardi}, {Bristow},
  {Collier Cameron}, {Charbonneau}, {Dressing}, {Figueira}, {Fulton}, {Hardee},
  {Hirsch}, {Latham}, {Mortier}, {Nava}, {Schlieder}, {Vanderburg}, {Weiss},
  {Bonomo}, {Bouchy}, {Buchhave}, {Coffinet}, {Damasso}, {Dumusque}, {Lovis},
  {Mayor}, {Micela}, {Molinari}, {Pepe}, {Phillips}, {Piotto}, {Rice},
  {Sasselov}, {S{\'e}gransan}, {Sozzetti}, {Udry}, \& {Watson}}]{Kosiarek2019b}
{Kosiarek}, M.~R., {Blunt}, S., {L{\'o}pez-Morales}, M., {et~al.}
  2019{\natexlab{a}}, \aj, 157, 116, \dodoi{10.3847/1538-3881/aafe83}

\bibitem[{{Kosiarek} {et~al.}(2019{\natexlab{b}}){Kosiarek}, {Crossfield},
  {Hardegree-Ullman}, {Livingston}, {Benneke}, {Henry}, {Howard}, {Berardo},
  {Blunt}, {Fulton}, {Hirsch}, {Howard}, {Isaacson}, {Petigura}, {Sinukoff},
  {Weiss}, {Bonfils}, {Dressing}, {Knutson}, {Schlieder}, {Werner}, {Gorjian},
  {Krick}, {Morales}, {Astudillo-Defru}, {Almenara}, {Delfosse}, {Forveille},
  {Lovis}, {Mayor}, {Murgas}, {Pepe}, {Santos}, {Udry}, {Corbett}, {Fors},
  {Law}, {Ratzloff}, \& {del Ser}}]{Kosiarek2019}
{Kosiarek}, M.~R., {Crossfield}, I. J.~M., {Hardegree-Ullman}, K.~K., {et~al.}
  2019{\natexlab{b}}, \aj, 157, 97, \dodoi{10.3847/1538-3881/aaf79c}

\bibitem[{{Kosiarek} {et~al.}(2021){Kosiarek}, {Berardo}, {Crossfield},
  {Laguna}, {Piaulet}, {Akana Murphy}, {Howell}, {Henry}, {Isaacson}, {Fulton},
  {Weiss}, {Petigura}, {Behmard}, {Hirsch}, {Teske}, {Burt}, {Mills},
  {Chontos}, {Mo{\v{c}}nik}, {Howard}, {Werner}, {Livingston}, {Krick},
  {Beichman}, {Gorjian}, {Kreidberg}, {Morley}, {Christiansen}, {Morales},
  {Scott}, {Crane}, {Wang}, {Shectman}, {Rosenthal}, {Grunblatt}, {Rubenzahl},
  {Dalba}, {Giacalone}, {Villanueva}, {Liu}, {Dai}, {Hill}, {Rice}, {Kane}, \&
  {Mayo}}]{Kosiarek2020b}
{Kosiarek}, M.~R., {Berardo}, D.~A., {Crossfield}, I. J.~M., {et~al.} 2021,
  \aj, 161, 47, \dodoi{10.3847/1538-3881/abca39}

\bibitem[{{Kostov} {et~al.}(2019){Kostov}, {Mullally}, {Quintana}, {Coughlin},
  {Mullally}, {Barclay}, {Col{\'o}n}, {Schlieder}, {Barentsen}, \&
  {Burke}}]{Kostov2019}
{Kostov}, V.~B., {Mullally}, S.~E., {Quintana}, E.~V., {et~al.} 2019, \aj, 157,
  124, \dodoi{10.3847/1538-3881/ab0110}

\bibitem[{{Kraus} \& {Hillenbrand}(2007{\natexlab{a}})}]{kraus:2007}
{Kraus}, A.~L., \& {Hillenbrand}, L.~A. 2007{\natexlab{a}}, \aj, 134, 2340,
  \dodoi{10.1086/522831}

\bibitem[{{Kraus} \& {Hillenbrand}(2007{\natexlab{b}})}]{Kraus2007}
---. 2007{\natexlab{b}}, \aj, 134, 2340, \dodoi{10.1086/522831}

\bibitem[{{Kreidberg}(2015)}]{Kreidberg2015}
{Kreidberg}, L. 2015, \pasp, 127, 1161, \dodoi{10.1086/683602}

\bibitem[{{Kreidberg} {et~al.}(2018){Kreidberg}, {Line}, {Thorngren}, {Morley},
  \& {Stevenson}}]{Kreidberg2018}
{Kreidberg}, L., {Line}, M.~R., {Thorngren}, D., {Morley}, C.~V., \&
  {Stevenson}, K.~B. 2018, \apjl, 858, L6, \dodoi{10.3847/2041-8213/aabfce}

\bibitem[{{Kreidberg} {et~al.}(2022){Kreidberg}, {Molli{\`e}re}, {Crossfield},
  {Thorngren}, {Kawashima}, {Morley}, {Benneke}, {Mikal-Evans}, {Berardo},
  {Kosiarek}, {Gorjian}, {Ciardi}, {Christiansen}, {Dragomir}, {Dressing},
  {Fortney}, {Fulton}, {Greene}, {Hardegree-Ullman}, {Howard}, {Howell},
  {Isaacson}, {Krick}, {Livingston}, {Lothringer}, {Morales}, {Petigura},
  {Rodriguez}, {Schlieder}, \& {Weiss}}]{Kreidberg2022}
{Kreidberg}, L., {Molli{\`e}re}, P., {Crossfield}, I. J.~M., {et~al.} 2022,
  \aj, 164, 124, \dodoi{10.3847/1538-3881/ac85be}

\bibitem[{{Kruse} {et~al.}(2019){Kruse}, {Agol}, {Luger}, \&
  {Foreman-Mackey}}]{Kruse2019}
{Kruse}, E., {Agol}, E., {Luger}, R., \& {Foreman-Mackey}, D. 2019, \apjs, 244,
  11, \dodoi{10.3847/1538-4365/ab346b}

\bibitem[{{Lam} {et~al.}(2018){Lam}, {Santerne}, {Sousa}, {Vigan}, {Armstrong,
  D. J.}, {Barros, S. C. C.}, {Brugger, B.}, {Adibekyan, V.}, {Almenara,
  J.-M.}, {Delgado Mena, E.}, {Dumusque, X.}, {Barrado, D.}, {Bayliss, D.},
  {Bonomo, A. S.}, {Bouchy, F.}, {Brown, D. J. A.}, {Ciardi, D.}, {Deleuil,
  M.}, {Demangeon, O.}, {Faedi, F.}, {Foxell, E.}, {Jackman, J. A. G.}, {King,
  G. W.}, {Kirk, J.}, {Ligi, R.}, {Lillo-Box, J.}, {Lopez, T.}, {Lovis, C.},
  {Louden, T.}, {Nielsen, L. D.}, {McCormac, J.}, {Mousis, O.}, {Osborn, H.
  P.}, {Pollacco, D.}, {Santos, N. C.}, {Udry, S.}, \& {Wheatley, P.
  J.}}]{Lam2018}
{Lam}, K. W.~F., {Santerne}, A., {Sousa}, S.~G., {et~al.} 2018, A\&A, 620, A77,
  \dodoi{10.1051/0004-6361/201834073}

\bibitem[{{Lehmer} \& {Catling}(2017)}]{Lehmer2017}
{Lehmer}, O.~R., \& {Catling}, D.~C. 2017, \apj, 845, 130,
  \dodoi{10.3847/1538-4357/aa8137}

\bibitem[{{L{\'e}pine} \& {Shara}(2005)}]{lepine:2005}
{L{\'e}pine}, S., \& {Shara}, M.~M. 2005, \aj, 129, 1483,
  \dodoi{10.1086/427854}

\bibitem[{{Lillo-Box} {et~al.}(2020){Lillo-Box}, {Lopez}, {Santerne},
  {Nielsen}, {Barros}, {Deleuil}, {Acu{\~n}a}, {Mousis}, {Sousa}, {Adibekyan},
  {Armstrong}, {Barrado}, {Bayliss}, {Brown}, {Demangeon}, {Dumusque},
  {Figueira}, {Hojjatpanah}, {Osborn}, {Santos}, \& {Udry}}]{LilloBox2020}
{Lillo-Box}, J., {Lopez}, T.~A., {Santerne}, A., {et~al.} 2020, \aap, 640, A48,
  \dodoi{10.1051/0004-6361/202037896}

\bibitem[{{Lissauer} \& {Eisberg}(2018)}]{Lissauer2018}
{Lissauer}, J.~J., \& {Eisberg}, J. 2018, \nar, 83, 1,
  \dodoi{10.1016/j.newar.2019.04.002}

\bibitem[{{Lissauer} {et~al.}(2011){Lissauer}, {Ragozzine}, {Fabrycky},
  {Steffen}, {Ford}, {Jenkins}, {Shporer}, {Holman}, {Rowe}, {Quintana},
  {Batalha}, {Borucki}, {Bryson}, {Caldwell}, {Carter}, {Ciardi}, {Dunham},
  {Fortney}, {Gautier}, {Howell}, {Koch}, {Latham}, {Marcy}, {Morehead}, \&
  {Sasselov}}]{Lissauer2011}
{Lissauer}, J.~J., {Ragozzine}, D., {Fabrycky}, D.~C., {et~al.} 2011, \apjs,
  197, 8, \dodoi{10.1088/0067-0049/197/1/8}

\bibitem[{Livingston {et~al.}(2018)Livingston, Crossfield, Petigura, Gonzales,
  Ciardi, Beichman, Christiansen, Dressing, Henning, Howard, Isaacson, Fulton,
  Kosiarek, Schlieder, Sinukoff, \& Tamura}]{Livingston2018}
Livingston, J.~H., Crossfield, I. J.~M., Petigura, E.~A., {et~al.} 2018, The
  Astronomical Journal, 156, 277, \dodoi{10.3847/1538-3881/aae778}

\bibitem[{{Livingston} {et~al.}(2018){Livingston}, {Endl}, {Dai}, {Cochran},
  {Barragan}, {Gandolfi}, {Hirano}, {Grziwa}, {Smith}, {Albrecht}, {Cabrera},
  {Csizmadia}, {de Leon}, {Deeg}, {Eigm{\"u}ller}, {Erikson}, {Everett},
  {Fridlund}, {Fukui}, {Guenther}, {Hatzes}, {Howell}, {Korth}, {Narita},
  {Nespral}, {Nowak}, {Palle}, {P{\"a}tzold}, {Persson}, {Prieto-Arranz},
  {Rauer}, {Tamura}, {Van Eylen}, \& {Winn}}]{Livingston2018b}
{Livingston}, J.~H., {Endl}, M., {Dai}, F., {et~al.} 2018, \aj, 156, 78,
  \dodoi{10.3847/1538-3881/aaccde}

\bibitem[{{Lopez} \& {Fortney}(2014)}]{Lopez2014}
{Lopez}, E.~D., \& {Fortney}, J.~J. 2014, \apj, 792, 1,
  \dodoi{10.1088/0004-637X/792/1/1}

\bibitem[{{Lopez} {et~al.}(2012){Lopez}, {Fortney}, \& {Miller}}]{Lopez2012}
{Lopez}, E.~D., {Fortney}, J.~J., \& {Miller}, N. 2012, \apj, 761, 59,
  \dodoi{10.1088/0004-637X/761/1/59}

\bibitem[{{L{\'o}pez-Morales} {et~al.}(2016){L{\'o}pez-Morales}, {Haywood},
  {Coughlin}, {Zeng}, {Buchhave}, {Giles}, {Affer}, {Bonomo}, {Charbonneau},
  {Collier Cameron}, {Consentino}, {Dressing}, {Dumusque}, {Figueira},
  {Fiorenzano}, {Harutyunyan}, {Johnson}, {Latham}, {Lopez}, {Lovis},
  {Malavolta}, {Mayor}, {Micela}, {Molinari}, {Mortier}, {Motalebi},
  {Nascimbeni}, {Pepe}, {Phillips}, {Piotto}, {Pollacco}, {Queloz}, {Rice},
  {Sasselov}, {Segransan}, {Sozzetti}, {Udry}, {Vanderburg}, \&
  {Watson}}]{Lopez-Morales2016}
{L{\'o}pez-Morales}, M., {Haywood}, R.~D., {Coughlin}, J.~L., {et~al.} 2016,
  \aj, 152, 204, \dodoi{10.3847/0004-6256/152/6/204}

\bibitem[{{Lucy} \& {Sweeney}(1971)}]{LucySweeney1971}
{Lucy}, L.~B., \& {Sweeney}, M.~A. 1971, \aj, 76, 544, \dodoi{10.1086/111159}

\bibitem[{{Luger} {et~al.}(2016){Luger}, {Agol}, {Kruse}, {Barnes}, {Becker},
  {Foreman-Mackey}, \& {Deming}}]{Luger2016}
{Luger}, R., {Agol}, E., {Kruse}, E., {et~al.} 2016, \aj, 152, 100,
  \dodoi{10.3847/0004-6256/152/4/100}

\bibitem[{{Luque} \& {Pall{\'e}}(2022)}]{Luque2022}
{Luque}, R., \& {Pall{\'e}}, E. 2022, Science, 377, 1211,
  \dodoi{10.1126/science.abl7164}

\bibitem[{{Madhusudhan} {et~al.}(2020){Madhusudhan}, {Nixon}, {Welbanks},
  {Piette}, \& {Booth}}]{Madhusudhan2020}
{Madhusudhan}, N., {Nixon}, M.~C., {Welbanks}, L., {Piette}, A. A.~A., \&
  {Booth}, R.~A. 2020, \apjl, 891, L7, \dodoi{10.3847/2041-8213/ab7229}

\bibitem[{{Madhusudhan} {et~al.}(2023){Madhusudhan}, {Sarkar}, {Constantinou},
  {Holmberg}, {Piette}, \& {Moses}}]{Madhusudhan2023}
{Madhusudhan}, N., {Sarkar}, S., {Constantinou}, S., {et~al.} 2023, \apjl, 956,
  L13, \dodoi{10.3847/2041-8213/acf577}

\bibitem[{Malavolta {et~al.}(2017)Malavolta, Borsato, Granata, Piotto, Lopez,
  Vanderburg, Figueira, Mortier, Nascimbeni, Affer, Bonomo, Bouchy, Buchhave,
  Charbonneau, Cameron, Cosentino, Dressing, Dumusque, Fiorenzano, Harutyunyan,
  Haywood, Johnson, Latham, Lopez-Morales, Lovis, Mayor, Micela, Molinari,
  Motalebi, Pepe, Phillips, Pollacco, Queloz, Rice, Sasselov, S{\'{e}}gransan,
  Sozzetti, Udry, \& Watson}]{Malavolta2017}
Malavolta, L., Borsato, L., Granata, V., {et~al.} 2017, The Astronomical
  Journal, 153, 224, \dodoi{10.3847/1538-3881/aa6897}

\bibitem[{{Mann} {et~al.}(2017){Mann}, {Gaidos}, {Vanderburg}, {Rizzuto},
  {Ansdell}, {Medina}, {Mace}, {Kraus}, \& {Sokal}}]{Mann2017a}
{Mann}, A.~W., {Gaidos}, E., {Vanderburg}, A., {et~al.} 2017, \aj, 153, 64,
  \dodoi{10.1088/1361-6528/aa5276}

\bibitem[{{Marcy} \& {Butler}(1992)}]{Marcy1992}
{Marcy}, G.~W., \& {Butler}, R.~P. 1992, \pasp, 104, 270,
  \dodoi{10.1086/132989}

\bibitem[{{Marcy} {et~al.}(2014){Marcy}, {Isaacson}, {Howard}, {Rowe},
  {Jenkins}, {Bryson}, {Latham}, {Howell}, {Gautier}, {Batalha}, {Rogers},
  {Ciardi}, {Fischer}, {Gilliland}, {Kjeldsen}, {Christensen-Dalsgaard},
  {Huber}, {Chaplin}, {Basu}, {Buchhave}, {Quinn}, {Borucki}, {Koch}, {Hunter},
  {Caldwell}, {Van Cleve}, {Kolbl}, {Weiss}, {Petigura}, {Seager}, {Morton},
  {Johnson}, {Ballard}, {Burke}, {Cochran}, {Endl}, {MacQueen}, {Everett},
  {Lissauer}, {Ford}, {Torres}, {Fressin}, {Brown}, {Steffen}, {Charbonneau},
  {Basri}, {Sasselov}, {Winn}, {Sanchis-Ojeda}, {Christiansen}, {Adams},
  {Henze}, {Dupree}, {Fabrycky}, {Fortney}, {Tarter}, {Holman}, {Tenenbaum},
  {Shporer}, {Lucas}, {Welsh}, {Orosz}, {Bedding}, {Campante}, {Davies},
  {Elsworth}, {Handberg}, {Hekker}, {Karoff}, {Kawaler}, {Lund}, {Lundkvist},
  {Metcalfe}, {Miglio}, {Silva Aguirre}, {Stello}, {White}, {Boss}, {Devore},
  {Gould}, {Prsa}, {Agol}, {Barclay}, {Coughlin}, {Brugamyer}, {Mullally},
  {Quintana}, {Still}, {Thompson}, {Morrison}, {Twicken}, {D{\'e}sert},
  {Carter}, {Crepp}, {H{\'e}brard}, {Santerne}, {Moutou}, {Sobeck}, {Hudgins},
  {Haas}, {Robertson}, {Lillo-Box}, \& {Barrado}}]{Marcy2014}
{Marcy}, G.~W., {Isaacson}, H., {Howard}, A.~W., {et~al.} 2014, \apjs, 210, 20,
  \dodoi{10.1088/0067-0049/210/2/20}

\bibitem[{{Martinez} {et~al.}(2017){Martinez}, {Crossfield}, {Schlieder},
  {Dressing}, {Obermeier}, {Livingston}, {Ciceri}, {Peacock}, {Beichman},
  {L{\'e}pine}, {Aller}, {Chance}, {Petigura}, {Howard}, \&
  {Werner}}]{Martinez2017}
{Martinez}, A.~O., {Crossfield}, I.~J.~M., {Schlieder}, J.~E., {et~al.} 2017,
  \apj, 837, 72, \dodoi{10.3847/1538-4357/aa56c7}

\bibitem[{{Masuda}(2014)}]{Masuda2014}
{Masuda}, K. 2014, \apj, 783, 53, \dodoi{10.1088/0004-637X/783/1/53}

\bibitem[{{Mayo} {et~al.}(2018){Mayo}, {Vanderburg}, {Latham}, {Bieryla},
  {Morton}, {Buchhave}, {Dressing}, {Beichman}, {Berlind}, {Calkins}, {Ciardi},
  {Crossfield}, {Esquerdo}, {Everett}, {Gonzales}, {Hirsch}, {Horch}, {Howard},
  {Howell}, {Livingston}, {Patel}, {Petigura}, {Schlieder}, {Scott}, {Schumer},
  {Sinukoff}, {Teske}, \& {Winters}}]{Mayo2018}
{Mayo}, A.~W., {Vanderburg}, A., {Latham}, D.~W., {et~al.} 2018, \aj, 155, 136,
  \dodoi{10.3847/1538-3881/aaadff}

\bibitem[{{Mayor} {et~al.}(2011){Mayor}, {Marmier}, {Lovis}, {Udry},
  {S{\'e}gransan}, {Pepe}, {Benz}, {Bertaux}, {Bouchy}, {Dumusque}, {Lo Curto},
  {Mordasini}, {Queloz}, \& {Santos}}]{Mayor2011}
{Mayor}, M., {Marmier}, M., {Lovis}, C., {et~al.} 2011, arXiv e-prints,
  arXiv:1109.2497.
\newblock \doarXiv{1109.2497}

\bibitem[{{Mikal-Evans} {et~al.}(2021){Mikal-Evans}, {Crossfield}, {Benneke},
  {Kreidberg}, {Moses}, {Morley}, {Thorngren}, {Molli{\`e}re},
  {Hardegree-Ullman}, {Brewer}, {Christiansen}, {Ciardi}, {Dragomir},
  {Dressing}, {Fortney}, {Gorjian}, {Greene}, {Hirsch}, {Howard}, {Howell},
  {Isaacson}, {Kosiarek}, {Krick}, {Livingston}, {Lothringer}, {Morales},
  {Petigura}, {Schlieder}, \& {Werner}}]{Evans2020}
{Mikal-Evans}, T., {Crossfield}, I. J.~M., {Benneke}, B., {et~al.} 2021, \aj,
  161, 18, \dodoi{10.3847/1538-3881/abc874}

\bibitem[{{Mills} {et~al.}(2019){Mills}, {Howard}, {Petigura}, {Fulton},
  {Isaacson}, \& {Weiss}}]{Mills2019}
{Mills}, S.~M., {Howard}, A.~W., {Petigura}, E.~A., {et~al.} 2019, \aj, 157,
  198, \dodoi{10.3847/1538-3881/ab1009}

\bibitem[{{Montet} {et~al.}(2015){Montet}, {Morton}, {Foreman-Mackey},
  {Johnson}, {Hogg}, {Bowler}, {Latham}, {Bieryla}, \& {Mann}}]{Montet2015}
{Montet}, B.~T., {Morton}, T.~D., {Foreman-Mackey}, D., {et~al.} 2015, \apj,
  809, 25, \dodoi{10.1088/0004-637X/809/1/25}

\bibitem[{{Mortier} {et~al.}(2020){Mortier}, {Zapatero Osorio}, {Malavolta},
  {Alibert}, {Rice}, {Lillo-Box}, {Vanderburg}, {Oshagh}, {Buchhave},
  {Adibekyan}, {Delgado Mena}, {Lopez-Morales}, {Charbonneau}, {Sousa},
  {Lovis}, {Affer}, {Allende Prieto}, {Barros}, {Benatti}, {Bonomo}, {Boschin},
  {Bouchy}, {Cabral}, {Collier Cameron}, {Cosentino}, {Cristiani}, {Demangeon},
  {Di Marcantonio}, {D'Odorico}, {Dumusque}, {Ehrenreich}, {Figueira},
  {Fiorenzano}, {Ghedina}, {Gonz{\'a}lez Hern{\'a}ndez}, {Haldemann},
  {Harutyunyan}, {Haywood}, {Latham}, {Lavie}, {Lo Curto}, {Maldonado},
  {Manescau}, {Martins}, {Mayor}, {M{\'e}gevand}, {Mehner}, {Micela}, {Molaro},
  {Molinari}, {Nunes}, {Pepe}, {Palle}, {Phillips}, {Piotto}, {Pinamonti},
  {Poretti}, {Riva}, {Rebolo}, {Santos}, {Sasselov}, {Sozzetti}, {Su{\'a}rez
  Mascare{\~n}o}, {Udry}, {West}, {Watson}, \& {Wilson}}]{Mortier2020}
{Mortier}, A., {Zapatero Osorio}, M.~R., {Malavolta}, L., {et~al.} 2020,
  \mnras, 499, 5004, \dodoi{10.1093/mnras/staa3144}

\bibitem[{{Morton}(2015)}]{Morton2015-isochrones}
{Morton}, T.~D. 2015, {isochrones: Stellar model grid package}, Astrophysics
  Source Code Library.
\newblock \doeprint{1503.010}

\bibitem[{{Narita} {et~al.}(2015){Narita}, {Hirano}, {Fukui}, {Hori},
  {Sanchis-Ojeda}, {Winn}, {Ryu}, {Kusakabe}, {Kudo}, {Onitsuka}, {Delrez},
  {Gillon}, {Jehin}, {McCormac}, {Holman}, {Izumiura}, {Takeda}, {Tamura}, \&
  {Yanagisawa}}]{Narita2015}
{Narita}, N., {Hirano}, T., {Fukui}, A., {et~al.} 2015, \apj, 815, 47,
  \dodoi{10.1088/0004-637X/815/1/47}

\bibitem[{{Narita} {et~al.}(2017){Narita}, {Hirano}, {Fukui}, {Hori}, {Dai},
  {Yu}, {Livingston}, {Ryu}, {Nowak}, {Kuzuhara}, {Sato}, {Takeda}, {Albrecht},
  {Kudo}, {Kusakabe}, {Palle}, {Ribas}, {Tamura}, {Van Eylen}, \&
  {Winn}}]{Narita2017}
---. 2017, \pasj, 69, 29, \dodoi{10.1093/pasj/psx002}

\bibitem[{{Nava} {et~al.}(2022){Nava}, {L{\'o}pez-Morales}, {Mortier}, {Zeng},
  {Giles}, {Bieryla}, {Vanderburg}, {Buchhave}, {Poretti}, {Saar}, {Dumusque},
  {Latham}, {Charbonneau}, {Damasso}, {Bonomo}, {Lovis}, {Collier Cameron},
  {Eastman}, {Sozzetti}, {Cosentino}, {Pedani}, {Pepe}, {Molinari}, {Sasselov},
  {Mayor}, {Stalport}, {Malavolta}, {Rice}, {Watson}, {Martinez Fiorenzano}, \&
  {Di Fabrizio}}]{Nava2022}
{Nava}, C., {L{\'o}pez-Morales}, M., {Mortier}, A., {et~al.} 2022, \aj, 163,
  41, \dodoi{10.3847/1538-3881/ac3141}

\bibitem[{{Nespral} {et~al.}(2017){Nespral}, {Gandolfi}, {Deeg}, {Borsato},
  {Fridlund}, {Barrag{\'a}n}, {Alonso}, {Grziwa}, {Korth}, {Albrecht},
  {Cabrera}, {Csizmadia}, {Nowak}, {Kuutma}, {Saario}, {Eigm{\"u}ller},
  {Erikson}, {Guenther}, {Hatzes}, {Monta{\~n}{\'e}s Rodr{\'{\i}}guez},
  {Palle}, {P{\"a}tzold}, {Prieto-Arranz}, {Rauer}, \&
  {Sebastian}}]{Nespral2017}
{Nespral}, D., {Gandolfi}, D., {Deeg}, H.~J., {et~al.} 2017, \aap, 601, A128,
  \dodoi{10.1051/0004-6361/201628639}

\bibitem[{{Neveu-VanMalle} {et~al.}(2016){Neveu-VanMalle}, {Queloz},
  {Anderson}, {Brown}, {Collier Cameron}, {Delrez}, {D{\'{\i}}az}, {Gillon},
  {Hellier}, {Jehin}, {Lister}, {Pepe}, {Rojo}, {S{\'e}gransan}, {Triaud},
  {Turner}, \& {Udry}}]{Neveu-VanMalle2016}
{Neveu-VanMalle}, M., {Queloz}, D., {Anderson}, D.~R., {et~al.} 2016, \aap,
  586, A93, \dodoi{10.1051/0004-6361/201526965}

\bibitem[{Newville {et~al.}(2014)Newville, Stensitzki, Allen, \&
  Ingargiola}]{Newville2014}
Newville, M., Stensitzki, T., Allen, D.~B., \& Ingargiola, A. 2014, {LMFIT:
  Non-Linear Least-Square Minimization and Curve-Fitting for Python¶},
  \dodoi{10.5281/zenodo.11813}

\bibitem[{{Ning} {et~al.}(2018){Ning}, {Wolfgang}, \& {Ghosh}}]{Ning2018}
{Ning}, B., {Wolfgang}, A., \& {Ghosh}, S. 2018, \apj, 869, 5,
  \dodoi{10.3847/1538-4357/aaeb31}

\bibitem[{{Niraula} {et~al.}(2017){Niraula}, {Redfield}, {Dai}, {Barrag{\'a}n},
  {Gandolfi}, {Cauley}, {Hirano}, {Korth}, {Smith}, {Prieto-Arranz}, {Grziwa},
  {Fridlund}, {Persson}, {Justesen}, {Winn}, {Albrecht}, {Cochran},
  {Csizmadia}, {Duvvuri}, {Endl}, {Hatzes}, {Livingston}, {Narita}, {Nespral},
  {Nowak}, {P{\"a}tzold}, {Palle}, \& {Van Eylen}}]{Niraula2017}
{Niraula}, P., {Redfield}, S., {Dai}, F., {et~al.} 2017, \aj, 154, 266,
  \dodoi{10.3847/1538-3881/aa957c}

\bibitem[{{Nowak} {et~al.}(2020){Nowak}, {Palle}, {Gandolfi}, {Deeg}, {Hirano},
  {Barrag{\'a}n}, {Kuzuhara}, {Dai}, {Luque}, {Persson}, {Fridlund}, {Johnson},
  {Korth}, {Livingston}, {Grziwa}, {Mathur}, {Hatzes}, {Prieto-Arranz},
  {Nespral}, {Hidalgo}, {Hjorth}, {Albrecht}, {Van Eylen}, {Lam}, {Cochran},
  {Esposito}, {Csizmadia}, {Guenther}, {Kabath}, {Blay}, {Brahm}, {Jord{\'a}n},
  {Espinoza}, {Rojas}, {Casasayas Barris}, {Rodler}, {Alonso Sobrino},
  {Cabrera}, {Carleo}, {Chaushev}, {de Leon}, {Eigm{\"u}ller}, {Endl},
  {Erikson}, {Fukui}, {Georgieva}, {Gonz{\'a}lez-Cuesta}, {Knudstrup}, {Lund},
  {Monta{\~n}es Rodr{\'\i}guez}, {Murgas}, {Narita}, {Niraula}, {P{\"a}tzold},
  {Rauer}, {Redfield}, {Ribas}, {Skarka}, {Smith}, \& {Subjak}}]{Nowak2020}
{Nowak}, G., {Palle}, E., {Gandolfi}, D., {et~al.} 2020, \mnras, 497, 4423,
  \dodoi{10.1093/mnras/staa2077}

\bibitem[{{Osborn} {et~al.}(2017){Osborn}, {Santerne}, {Barros}, {Santos},
  {Dumusque}, {Malavolta}, {Armstrong}, {Hojjatpanah}, {Demangeon},
  {Adibekyan}, {Almenara}, {Barrado}, {Bayliss}, {Boisse}, {Bouchy}, {Brown},
  {Cameron}, {Charbonneau}, {Deleuil}, {Delgado Mena}, {D{\'{\i}}az},
  {H{\'e}brard}, {Kirk}, {King}, {Lam}, {Latham}, {Lillo-Box}, {Louden},
  {Lovis}, {Marmier}, {McCormac}, {Molinari}, {Pepe}, {Pollacco}, {Sousa},
  {Udry}, \& {Walker}}]{Osborn2017}
{Osborn}, H.~P., {Santerne}, A., {Barros}, S.~C.~C., {et~al.} 2017, \aap, 604,
  A19, \dodoi{10.1051/0004-6361/201628932}

\bibitem[{{O'Toole} {et~al.}(2009){O'Toole}, {Tinney}, {Jones}, {Butler},
  {Marcy}, {Carter}, \& {Bailey}}]{Otoole2009}
{O'Toole}, S.~J., {Tinney}, C.~G., {Jones}, H.~R.~A., {et~al.} 2009, \mnras,
  392, 641, \dodoi{10.1111/j.1365-2966.2008.14051.x}

\bibitem[{{Owen} \& {Wu}(2013)}]{Owen2013}
{Owen}, J.~E., \& {Wu}, Y. 2013, \apj, 775, 105,
  \dodoi{10.1088/0004-637X/775/2/105}

\bibitem[{{Owen} \& {Wu}(2016)}]{Owen2016}
---. 2016, \apj, 817, 107, \dodoi{10.3847/0004-637X/817/2/107}

\bibitem[{{Owen} \& {Wu}(2017)}]{Owen2017}
---. 2017, \apj, 847, 29, \dodoi{10.3847/1538-4357/aa890a}

\bibitem[{{Passegger} {et~al.}(2024){Passegger}, {Su{\'a}rez Mascare{\~n}o},
  {Allart}, {Gonz{\'a}lez Hern{\'a}ndez}, {Lovis}, {Lavie}, {Silva},
  {M{\"u}ller}, {Tabernero}, {Cristiani}, {Pepe}, {Rebolo}, {Santos},
  {Adibekyan}, {Alibert}, {Allende Prieto}, {Barros}, {Bouchy},
  {Castro-Gonz{\'a}lez}, {D'Odorico}, {Dumusque}, {Di Marcantonio},
  {Ehrenreich}, {Figueira}, {G{\'e}nova Santos}, {Lo Curto}, {Martins},
  {Mehner}, {Micela}, {Molaro}, {Nari}, {Nunes}, {Pall{\'e}}, {Poretti},
  {Rodrigues}, {Sousa}, {Sozzetti}, {Udry}, \& {Zapatero
  Osorio}}]{Passegger2024}
{Passegger}, V.~M., {Su{\'a}rez Mascare{\~n}o}, A., {Allart}, R., {et~al.}
  2024, arXiv e-prints, arXiv:2401.06276, \dodoi{10.48550/arXiv.2401.06276}

\bibitem[{{Pecaut} \& {Mamajek}(2013)}]{pecaut:2013}
{Pecaut}, M.~J., \& {Mamajek}, E.~E. 2013, \apjs, 208, 9,
  \dodoi{10.1088/0067-0049/208/1/9}

\bibitem[{{Persson} {et~al.}(2018){Persson}, {Fridlund}, {Barrag{\'a}n}, {Dai},
  {Gandolfi}, {Hatzes}, {Hirano}, {Grziwa}, {Korth}, {Prieto-Arranz},
  {Fossati}, {Van Eylen}, {Justesen}, {Livingston}, {Kubyshkina}, {Deeg},
  {Guenther}, {Nowak}, {Eigm{\"u}ller}, {Csizmadia}, {Smith}, {Erikson},
  {Alonso Sobrino}, {Cochran}, {Endl}, {Esposito}, {Fukui}, {Heeren},
  {Hidalgo}, {Hjorth}, {Kuzuhara}, {Narita}, {Nespral}, {Palle}, {P{\"a}tzold},
  {Rauer}, {Rodler}, \& {Winn}}]{Persson2018}
{Persson}, C.~M., {Fridlund}, M., {Barrag{\'a}n}, O., {et~al.} 2018, ArXiv
  e-prints.
\newblock \doarXiv{1805.04774}

\bibitem[{{Petigura}(2015)}]{Petigura2015phd}
{Petigura}, E.~A. 2015, ArXiv e-prints.
\newblock \doarXiv{1510.03902}

\bibitem[{{Petigura} {et~al.}(2013){Petigura}, {Howard}, \&
  {Marcy}}]{Petigura2013}
{Petigura}, E.~A., {Howard}, A.~W., \& {Marcy}, G.~W. 2013, Proceedings of the
  National Academy of Science, 110, 19273, \dodoi{10.1073/pnas.1319909110}

\bibitem[{{Petigura} {et~al.}(2015){Petigura}, {Schlieder}, {Crossfield},
  {Howard}, {Deck}, {Ciardi}, {Sinukoff}, {Allers}, {Best}, {Liu}, {Beichman},
  {Isaacson}, {Hansen}, \& {L{\'e}pine}}]{Petigura2015}
{Petigura}, E.~A., {Schlieder}, J.~E., {Crossfield}, I. J.~M., {et~al.} 2015,
  \apj, 811, 102, \dodoi{10.1088/0004-637X/811/2/102}

\bibitem[{{Petigura} {et~al.}(2016){Petigura}, {Howard}, {Lopez}, {Deck},
  {Fulton}, {Crossfield}, {Ciardi}, {Chiang}, {Lee}, {Isaacson}, {Beichman},
  {Hansen}, {Schlieder}, \& {Sinukoff}}]{Petigura2016}
{Petigura}, E.~A., {Howard}, A.~W., {Lopez}, E.~D., {et~al.} 2016, \apj, 818,
  36, \dodoi{10.3847/0004-637X/818/1/36}

\bibitem[{{Petigura} {et~al.}(2017{\natexlab{a}}){Petigura}, {Howard}, {Marcy},
  {Johnson}, {Isaacson}, {Cargile}, {Hebb}, {Fulton}, {Weiss}, {Morton},
  {Winn}, {Rogers}, {Sinukoff}, {Hirsch}, \& {Crossfield}}]{Petigura2017b}
{Petigura}, E.~A., {Howard}, A.~W., {Marcy}, G.~W., {et~al.}
  2017{\natexlab{a}}, \aj, 154, 107, \dodoi{10.3847/1538-3881/aa80de}

\bibitem[{{Petigura} {et~al.}(2017{\natexlab{b}}){Petigura}, {Sinukoff},
  {Lopez}, {Crossfield}, {Howard}, {Brewer}, {Fulton}, {Isaacson}, {Ciardi},
  {Howell}, {Everett}, {Horch}, {Hirsch}, {Weiss}, \&
  {Schlieder}}]{Petigura2017}
{Petigura}, E.~A., {Sinukoff}, E., {Lopez}, E.~D., {et~al.} 2017{\natexlab{b}},
  \aj, 153, 142, \dodoi{10.3847/1538-3881/aa5ea5}

\bibitem[{{Petigura} {et~al.}(2018{\natexlab{a}}){Petigura}, {Crossfield},
  {Isaacson}, {Beichman}, {Christiansen}, {Dressing}, {Fulton}, {Howard},
  {Kosiarek}, {L{\'e}pine}, {Schlieder}, {Sinukoff}, \& {Yee}}]{Petigura2018}
{Petigura}, E.~A., {Crossfield}, I.~J.~M., {Isaacson}, H., {et~al.}
  2018{\natexlab{a}}, \aj, 155, 21, \dodoi{10.3847/1538-3881/aa9b83}

\bibitem[{{Petigura} {et~al.}(2018{\natexlab{b}}){Petigura}, {Benneke},
  {Batygin}, {Fulton}, {Werner}, {Krick}, {Gorjian}, {Sinukoff}, {Deck},
  {Mills}, \& {Deming}}]{Petigura2018b}
{Petigura}, E.~A., {Benneke}, B., {Batygin}, K., {et~al.} 2018{\natexlab{b}},
  \aj, 156, 89, \dodoi{10.3847/1538-3881/aaceac}

\bibitem[{{Petigura} {et~al.}(2020){Petigura}, {Livingston}, {Batygin},
  {Mills}, {Werner}, {Isaacson}, {Fulton}, {Howard}, {Weiss}, {Espinoza},
  {Jontof-Hutter}, {Shporer}, {Bayliss}, \& {Barros}}]{Petigura2019}
{Petigura}, E.~A., {Livingston}, J., {Batygin}, K., {et~al.} 2020, \aj, 159, 2,
  \dodoi{10.3847/1538-3881/ab5220}

\bibitem[{Piaulet {et~al.}(2021)Piaulet, Benneke, Rubenzahl, Howard, Lee,
  Thorngren, Angus, Peterson, Schlieder, Werner, Kreidberg, Jaouni, Crossfield,
  Ciardi, Petigura, Livingston, Dressing, Fulton, Beichman, Christiansen,
  Gorjian, Hardegree-Ullman, Krick, \& Sinukoff}]{Piaulet2021}
Piaulet, C., Benneke, B., Rubenzahl, R.~A., {et~al.} 2021, \apj, 161, 70,
  \dodoi{10.3847/1538-3881/abcd3c}

\bibitem[{{Polanski} {et~al.}(2024){Polanski}, {Lubin}, {Beard}, {Akana
  Murphy}, {Rubenzahl}, {Hill}, {Crossfield}, {Chontos}, {Robertson},
  {Isaacson}, {Kane}, {Ciardi}, {Batalha}, {Dressing}, {Fulton}, {Howard},
  {Huber}, {Petigura}, {Weiss}, {Angelo}, {Behmard}, {Blunt}, {Brinkman},
  {Dai}, {Dalba}, {Fetherolf}, {Giacalone}, {Hirsch}, {Holcomb}, {Kosiarek},
  {Mayo}, {MacDougall}, {Mo{\v{c}}nik}, {Pidhorodetska}, {Rice}, {Rosenthal},
  {Scarsdale}, {Turtelboom}, {Tyler}, {Van Zandt}, {Yee}, {Coria}, {Dulz},
  {Hartman}, {Householder}, {Lange}, {Langford}, {Louden}, {Siegel}, {Gilbert},
  {Gonzales}, {Schlieder}, {Boyle}, {Christiansen}, {Clark}, {Fernandes},
  {Lund}, {Savel}, {Gill}, {Beichman}, {Matson}, {Matthews}, {Furlan},
  {Howell}, {Scott}, {Everett}, {Livingston}, {Ershova}, {Cheryasov},
  {Safonov}, {Lillo-Box}, {Barrado}, \& {Morales-Calder{\'o}n}}]{Polanski2024}
{Polanski}, A.~S., {Lubin}, J., {Beard}, C., {et~al.} 2024, \apjs, 272, 32,
  \dodoi{10.3847/1538-4365/ad4484}

\bibitem[{{Pope} {et~al.}(2016){Pope}, {Parviainen}, \& {Aigrain}}]{Pope2016}
{Pope}, B.~J.~S., {Parviainen}, H., \& {Aigrain}, S. 2016, \mnras, 461, 3399,
  \dodoi{10.1093/mnras/stw1373}

\bibitem[{Powell(1964)}]{Powell1964}
Powell, M. J.~D. 1964, The Computer Journal, 7, 155,
  \dodoi{10.1093/comjnl/7.2.155}

\bibitem[{{Prieto-Arranz} {et~al.}(2018){Prieto-Arranz}, {Palle}, {Gandolfi},
  {Barrag{\'a}n}, {Guenther}, {Dai}, {Fridlund}, {Hirano}, {Livingston},
  {Luque}, {Niraula}, {Persson}, {Redfield}, {Albrecht}, {Alonso},
  {Antoniciello}, {Cabrera}, {Cochran}, {Csizmadia}, {Deeg}, {Eigm{\"u}ller},
  {Endl}, {Erikson}, {Everett}, {Fukui}, {Grziwa}, {Hatzes}, {Hidalgo},
  {Hjorth}, {Korth}, {Lorenzo-Oliveira}, {Murgas}, {Narita}, {Nespral},
  {Nowak}, {P{\"a}tzold}, {Monta{\~n}ez Rodr{\'\i}guez}, {Rauer}, {Ribas},
  {Smith}, {Trifonov}, {Van Eylen}, \& {Winn}}]{PrietoArranz2018}
{Prieto-Arranz}, J., {Palle}, E., {Gandolfi}, D., {et~al.} 2018, \aap, 618,
  A116, \dodoi{10.1051/0004-6361/201832872}

\bibitem[{{Radica} {et~al.}(2022){Radica}, {Artigau}, {Lafreni{\'e}re},
  {Cadieux}, {Cook}, {Doyon}, {Amado}, {Caballero}, {Henning}, {Quirrenbach},
  {Reiners}, \& {Ribas}}]{Radica2022}
{Radica}, M., {Artigau}, {\'E}., {Lafreni{\'e}re}, D., {et~al.} 2022, \mnras,
  517, 5050, \dodoi{10.1093/mnras/stac3024}

\bibitem[{{Rajpaul} {et~al.}(2015){Rajpaul}, {Aigrain}, {Osborne}, {Reece}, \&
  {Roberts}}]{Rajpaul2015}
{Rajpaul}, V., {Aigrain}, S., {Osborne}, M.~A., {Reece}, S., \& {Roberts}, S.
  2015, \mnras, 452, 2269, \dodoi{10.1093/mnras/stv1428}

\bibitem[{{Rice} {et~al.}(2019){Rice}, {Malavolta}, {Mayo}, {Mortier},
  {Buchhave}, {Affer}, {Vanderburg}, {Lopez-Morales}, {Poretti}, {Zeng},
  {Collier Cameron}, {Damasso}, {Coffinet}, {Latham}, {Bonomo}, {Bouchy},
  {Charbonneau}, {Dumusque}, {Figueira}, {Martinez Fiorenzano}, {Haywood},
  {Johnson}, {Lopez}, {Lovis}, {Mayor}, {Micela}, {Molinari}, {Nascimbeni},
  {Nava}, {Pepe}, {Phillips}, {Piotto}, {Sasselov}, {S{\'e}gransan},
  {Sozzetti}, {Udry}, \& {Watson}}]{Rice2019}
{Rice}, K., {Malavolta}, L., {Mayo}, A., {et~al.} 2019, \mnras, 484, 3731,
  \dodoi{10.1093/mnras/stz130}

\bibitem[{{Ricker} {et~al.}(2015){Ricker}, {Winn}, {Vanderspek}, {Latham},
  {Bakos}, {Bean}, {Berta-Thompson}, {Brown}, {Buchhave}, {Butler}, {Butler},
  {Chaplin}, {Charbonneau}, {Christensen-Dalsgaard}, {Clampin}, {Deming},
  {Doty}, {De Lee}, {Dressing}, {Dunham}, {Endl}, {Fressin}, {Ge}, {Henning},
  {Holman}, {Howard}, {Ida}, {Jenkins}, {Jernigan}, {Johnson}, {Kaltenegger},
  {Kawai}, {Kjeldsen}, {Laughlin}, {Levine}, {Lin}, {Lissauer}, {MacQueen},
  {Marcy}, {McCullough}, {Morton}, {Narita}, {Paegert}, {Palle}, {Pepe},
  {Pepper}, {Quirrenbach}, {Rinehart}, {Sasselov}, {Sato}, {Seager},
  {Sozzetti}, {Stassun}, {Sullivan}, {Szentgyorgyi}, {Torres}, {Udry}, \&
  {Villasenor}}]{Ricker2015}
{Ricker}, G.~R., {Winn}, J.~N., {Vanderspek}, R., {et~al.} 2015, Journal of
  Astronomical Telescopes, Instruments, and Systems, 1, 014003,
  \dodoi{10.1117/1.JATIS.1.1.014003}

\bibitem[{{Ricker} {et~al.}(2016){Ricker}, {Vanderspek}, {Winn}, {Seager},
  {Berta-Thompson}, {Levine}, {Villasenor}, {Latham}, {Charbonneau}, {Holman},
  {Johnson}, {Sasselov}, {Szentgyorgyi}, {Torres}, {Bakos}, {Brown},
  {Christensen-Dalsgaard}, {Kjeldsen}, {Clampin}, {Rinehart}, {Deming}, {Doty},
  {Dunham}, {Ida}, {Kawai}, {Sato}, {Jenkins}, {Lissauer}, {Jernigan},
  {Kaltenegger}, {Laughlin}, {Lin}, {McCullough}, {Narita}, {Pepper},
  {Stassun}, \& {Udry}}]{Ricker2016}
{Ricker}, G.~R., {Vanderspek}, R., {Winn}, J., {et~al.} 2016, in Society of
  Photo-Optical Instrumentation Engineers (SPIE) Conference Series, Vol. 9904,
  Space Telescopes and Instrumentation 2016: Optical, Infrared, and Millimeter
  Wave, ed. H.~A. {MacEwen}, G.~G. {Fazio}, M.~{Lystrup}, N.~{Batalha},
  N.~{Siegler}, \& E.~C. {Tong}, 99042B, \dodoi{10.1117/12.2232071}

\bibitem[{{Rizzuto} {et~al.}(2017){Rizzuto}, {Mann}, {Vanderburg}, {Kraus}, \&
  {Covey}}]{Rizzuto2017}
{Rizzuto}, A.~C., {Mann}, A.~W., {Vanderburg}, A., {Kraus}, A.~L., \& {Covey},
  K.~R. 2017, \aj, 154, 224, \dodoi{10.3847/1538-3881/aa9070}

\bibitem[{{Rodriguez} {et~al.}(2013){Rodriguez}, {Zuckerman}, {Kastner},
  {Bessell}, {Faherty}, \& {Murphy}}]{rodriguez:2013}
{Rodriguez}, D.~R., {Zuckerman}, B., {Kastner}, J.~H., {et~al.} 2013, \apj,
  774, 101, \dodoi{10.1088/0004-637X/774/2/101}

\bibitem[{{Rodriguez} {et~al.}(2018){Rodriguez}, {Vanderburg}, {Eastman},
  {Mann}, {Crossfield}, {Ciardi}, {Latham}, \& {Quinn}}]{Rodriguez2018}
{Rodriguez}, J.~E., {Vanderburg}, A., {Eastman}, J.~D., {et~al.} 2018, \aj,
  155, 72, \dodoi{10.3847/1538-3881/aaa292}

\bibitem[{{Rodriguez} {et~al.}(2017){Rodriguez}, {Zhou}, {Vanderburg},
  {Eastman}, {Kreidberg}, {Cargile}, {Bieryla}, {Latham}, {Irwin}, {Mayo},
  {Calkins}, {Esquerdo}, \& {Mink}}]{Rodriguez2017}
{Rodriguez}, J.~E., {Zhou}, G., {Vanderburg}, A., {et~al.} 2017, \aj, 153, 256,
  \dodoi{10.3847/1538-3881/aa6dfb}

\bibitem[{{Rodr{\'\i}guez Mart{\'\i}nez} {et~al.}(2023){Rodr{\'\i}guez
  Mart{\'\i}nez}, {Gaudi}, {Schulze}, {Acu{\~n}a}, {Kolecki}, {Johnson},
  {Asnodkar}, {Boley}, {Deleuil}, {Mousis}, {Panero}, \& {Wang}}]{Martinez2023}
{Rodr{\'\i}guez Mart{\'\i}nez}, R., {Gaudi}, B.~S., {Schulze}, J.~G., {et~al.}
  2023, \aj, 165, 97, \dodoi{10.3847/1538-3881/acb04b}

\bibitem[{{Rogers} {et~al.}(2023){Rogers}, {Owen}, \&
  {Schlichting}}]{Rogers2023}
{Rogers}, J.~G., {Owen}, J.~E., \& {Schlichting}, H.~E. 2023, arXiv e-prints,
  arXiv:2311.12295, \dodoi{10.48550/arXiv.2311.12295}

\bibitem[{{Rogers}(2015)}]{Rogers2015}
{Rogers}, L.~A. 2015, \apj, 801, 41, \dodoi{10.1088/0004-637X/801/1/41}

\bibitem[{{Roy} {et~al.}(2023){Roy}, {Benneke}, {Piaulet}, {Gully-Santiago},
  {Crossfield}, {Morley}, {Kreidberg}, {Mikal-Evans}, {Brande}, {Delisle},
  {Greene}, {Hardegree-Ullman}, {Barman}, {Christiansen}, {Dragomir},
  {Fortney}, {Howard}, {Kosiarek}, \& {Lothringer}}]{Roy2023}
{Roy}, P.-A., {Benneke}, B., {Piaulet}, C., {et~al.} 2023, \apjl, 954, L52,
  \dodoi{10.3847/2041-8213/acebf0}

\bibitem[{Rubenzahl {et~al.}(2021)Rubenzahl, Dai, Howard, Chontos, Giacalone,
  Lubin, Rosenthal, Isaacson, Batalha, Crossfield, Dressing, Fulton, Huber,
  Kane, Petigura, Robertson, Roy, Weiss, Beard, Hill, Mayo, Mocnik, Murphy, \&
  Scarsdale}]{Rubenzahl2021}
Rubenzahl, R.~A., Dai, F., Howard, A.~W., {et~al.} 2021, \apj, 161, 119,
  \dodoi{10.3847/1538-3881/abd177}

\bibitem[{{Sanchis-Ojeda} {et~al.}(2015){Sanchis-Ojeda}, {Winn}, {Dai},
  {Howard}, {Isaacson}, {Marcy}, {Petigura}, {Sinukoff}, {Weiss}, {Albrecht},
  {Hirano}, \& {Rogers}}]{Sanchis-Ojeda2015}
{Sanchis-Ojeda}, R., {Winn}, J.~N., {Dai}, F., {et~al.} 2015, \apjl, 812, L11,
  \dodoi{10.1088/2041-8205/812/1/L11}

\bibitem[{{Santerne} {et~al.}(2018){Santerne}, {Brugger}, {Armstrong},
  {Adibekyan}, {Lillo-Box}, {Gosselin}, {Aguichine}, {Almenara}, {Barrado},
  {Barros}, {Bayliss}, {Boisse}, {Bonomo}, {Bouchy}, {Brown}, {Deleuil},
  {Delgado Mena}, {Demangeon}, {D{\'{\i}}az}, {Doyle}, {Dumusque}, {Faedi},
  {Faria}, {Figueira}, {Foxell}, {Giles}, {H{\'e}brard}, {Hojjatpanah},
  {Hobson}, {Jackman}, {King}, {Kirk}, {Lam}, {Ligi}, {Lovis}, {Louden},
  {McCormac}, {Mousis}, {Neal}, {Osborn}, {Pepe}, {Pollacco}, {Santos},
  {Sousa}, {Udry}, \& {Vigan}}]{Santerne2018}
{Santerne}, A., {Brugger}, B., {Armstrong}, D.~J., {et~al.} 2018, Nature
  Astronomy, \dodoi{10.1038/s41550-018-0420-5UNREFEREED:}

\bibitem[{{Santerne} {et~al.}(2019){Santerne}, {Malavolta}, {Kosiarek}, {Dai},
  {Dressing}, {Dumusque}, {Hara}, {Lopez}, {Mortier}, {Vanderburg},
  {Adibekyan}, {Armstrong}, {Barrado}, {Barros}, {Bayliss}, {Berardo},
  {Boisse}, {Bonomo}, {Bouchy}, {Brown}, {Buchhave}, {Butler}, {Collier
  Cameron}, {Cosentino}, {Crane}, {Crossfield}, {Damasso}, {Deleuil}, {Delgado
  Mena}, {Demangeon}, {D{\'\i}az}, {Donati}, {Figueira}, {Fulton}, {Ghedina},
  {Harutyunyan}, {H{\'e}brard}, {Hirsch}, {Hojjatpanah}, {Howard}, {Isaacson},
  {Latham}, {Lillo-Box}, {L{\'o}pez-Morales}, {Lovis}, {Martinez Fiorenzano},
  {Molinari}, {Mousis}, {Moutou}, {Nava}, {Nielsen}, {Osborn}, {Petigura},
  {Phillips}, {Pollacco}, {Poretti}, {Rice}, {Santos}, {S{\'e}gransan},
  {Shectman}, {Sinukoff}, {Sousa}, {Sozzetti}, {Teske}, {Udry}, {Vigan},
  {Wang}, {Watson}, {Weiss}, {Wheatley}, \& {Winn}}]{Santerne2019}
{Santerne}, A., {Malavolta}, L., {Kosiarek}, M.~R., {et~al.} 2019, arXiv
  e-prints, arXiv:1911.07355, \dodoi{10.48550/arXiv.1911.07355}

\bibitem[{{Sarkis} {et~al.}(2018){Sarkis}, {Henning}, {K{\"u}rster},
  {Trifonov}, {Zechmeister}, {Tal-Or}, {Anglada-Escud{\'e}}, {Hatzes},
  {Lafarga}, {Dreizler}, {Ribas}, {Caballero}, {Reiners}, {Mallonn}, {Morales},
  {Kaminski}, {Aceituno}, {Amado}, {B{\'e}jar}, {Hagen}, {Jeffers},
  {Quirrenbach}, {Launhardt}, {Marvin}, \& {Montes}}]{Sarkis2018}
{Sarkis}, P., {Henning}, T., {K{\"u}rster}, M., {et~al.} 2018, \aj, 155, 257,
  \dodoi{10.3847/1538-3881/aac108}

\bibitem[{{Schmitt} {et~al.}(2016){Schmitt}, {Tokovinin}, {Wang}, {Fischer},
  {Kristiansen}, {LaCourse}, {Gagliano}, {Tan}, {Schwengeler}, {Omohundro},
  {Venner}, {Terentev}, {Schmitt}, {Jacobs}, {Winarski}, {Sejpka}, {Jek},
  {Boyajian}, {Brewer}, {Ishikawa}, {Lintott}, {Lynn}, {Schawinski}, {Schwamb},
  \& {Weiksnar}}]{Schmitt2016}
{Schmitt}, J.~R., {Tokovinin}, A., {Wang}, J., {et~al.} 2016, \aj, 151, 159,
  \dodoi{10.3847/0004-6256/151/6/159}

\bibitem[{{Sing} {et~al.}(2016){Sing}, {Fortney}, {Nikolov}, {Wakeford},
  {Kataria}, {Evans}, {Aigrain}, {Ballester}, {Burrows}, {Deming},
  {D{\'e}sert}, {Gibson}, {Henry}, {Huitson}, {Knutson}, {Lecavelier Des
  Etangs}, {Pont}, {Showman}, {Vidal-Madjar}, {Williamson}, \&
  {Wilson}}]{Sing2016}
{Sing}, D.~K., {Fortney}, J.~J., {Nikolov}, N., {et~al.} 2016, \nat, 529, 59,
  \dodoi{10.1038/nature16068}

\bibitem[{{Sinukoff} {et~al.}(2016){Sinukoff}, {Howard}, {Petigura},
  {Schlieder}, {Crossfield}, {Ciardi}, {Fulton}, {Isaacson}, {Aller},
  {Baranec}, {Beichman}, {Hansen}, {Knutson}, {Law}, {Liu}, {Riddle}, \&
  {Dressing}}]{Sinukoff2016}
{Sinukoff}, E., {Howard}, A.~W., {Petigura}, E.~A., {et~al.} 2016, \apj, 827,
  78, \dodoi{10.3847/0004-637X/827/1/78}

\bibitem[{{Sinukoff} {et~al.}(2017{\natexlab{a}}){Sinukoff}, {Howard},
  {Petigura}, {Fulton}, {Crossfield}, {Isaacson}, {Gonzales}, {Crepp},
  {Brewer}, {Hirsch}, {Weiss}, {Ciardi}, {Schlieder}, {Benneke},
  {Christiansen}, {Dressing}, {Hansen}, {Knutson}, {Kosiarek}, {Livingston},
  {Greene}, {Rogers}, \& {L{\'e}pine}}]{Sinukoff2017b}
---. 2017{\natexlab{a}}, \aj, 153, 271, \dodoi{10.3847/1538-3881/aa725f}

\bibitem[{{Sinukoff} {et~al.}(2017{\natexlab{b}}){Sinukoff}, {Howard},
  {Petigura}, {Fulton}, {Isaacson}, {Weiss}, {Brewer}, {Hansen}, {Hirsch},
  {Christiansen}, {Crepp}, {Crossfield}, {Schlieder}, {Ciardi}, {Beichman},
  {Knutson}, {Benneke}, {Dressing}, {Livingston}, {Deck}, {L{\'e}pine}, \&
  {Rogers}}]{Sinukoff2017a}
---. 2017{\natexlab{b}}, \aj, 153, 70, \dodoi{10.3847/1538-3881/153/2/70}

\bibitem[{{Smith} {et~al.}(2017){Smith}, {Gandolfi}, {Barrag{\'a}n}, {Bowler},
  {Csizmadia}, {Endl}, {Fridlund}, {Grziwa}, {Guenther}, {Hatzes}, {Nowak},
  {Albrecht}, {Alonso}, {Cabrera}, {Cochran}, {Deeg}, {Cusano},
  {Eigm{\"u}ller}, {Erikson}, {Hidalgo}, {Hirano}, {Johnson}, {Korth}, {Mann},
  {Narita}, {Nespral}, {Palle}, {P{\"a}tzold}, {Prieto-Arranz}, {Rauer},
  {Ribas}, {Tingley}, \& {Wolthoff}}]{Smith2017}
{Smith}, A.~M.~S., {Gandolfi}, D., {Barrag{\'a}n}, O., {et~al.} 2017, \mnras,
  464, 2708, \dodoi{10.1093/mnras/stw2487}

\bibitem[{{Smith} {et~al.}(2022){Smith}, {Breton}, {Csizmadia}, {Dai},
  {Gandolfi}, {Garc{\'\i}a}, {Howard}, {Isaacson}, {Korth}, {Lam}, {Mathur},
  {Nowak}, {P{\'e}rez Hern{\'a}ndez}, {Persson}, {Albrecht}, {Barrag{\'a}n},
  {Cabrera}, {Cochran}, {Deeg}, {Fridlund}, {Georgieva}, {Goffo}, {Guenther},
  {Hatzes}, {Kabath}, {Livingston}, {Luque}, {Palle}, {Redfield}, {Rodler},
  {Serrano}, \& {Van Eylen}}]{Smith2022}
{Smith}, A.~M.~S., {Breton}, S.~N., {Csizmadia}, S., {et~al.} 2022, \mnras,
  510, 5035, \dodoi{10.1093/mnras/stab3497}

\bibitem[{{Spake} {et~al.}(2018){Spake}, {Sing}, {Evans}, {Oklop{\v{c}}i{\'c}},
  {}, {Bourrier}, {Kreidberg}, {Rackham}, {Irwin}, {Ehrenreich}, {Wyttenbach},
  {Wakeford}, {Zhou}, {Chubb}, {Nikolov}, {Goyal}, {Henry}, {Williamson},
  {Blumenthal}, {Anderson}, {Hellier}, {Charbonneau}, {Udry}, \&
  {Madhusudhan}}]{Spake2018}
{Spake}, J.~J., {Sing}, D.~K., {Evans}, T.~M., {et~al.} 2018, \nat, 557, 68,
  \dodoi{10.1038/s41586-018-0067-5}

\bibitem[{{Teske} {et~al.}(2021){Teske}, {Wang}, {Wolfgang}, {Gan},
  {Plotnykov}, {Armstrong}, {Butler}, {Cale}, {Crane}, {Howard}, {Jensen},
  {Law}, {Shectman}, {Plavchan}, {Valencia}, {Vanderburg}, {Ricker},
  {Vanderspek}, {Latham}, {Seager}, {Winn}, {Jenkins}, {Adibekyan}, {Barrado},
  {Barros}, {Benkhaldoun}, {Brown}, {Bryant}, {Burt}, {Caldwell},
  {Charbonneau}, {Cloutier}, {Collins}, {Collins}, {Colon}, {Conti},
  {Demangeon}, {Eastman}, {Elmufti}, {Feng}, {Flowers}, {Guerrero},
  {Hojjatpanah}, {Irwin}, {Isopi}, {Lillo-Box}, {Mallia}, {Massey}, {Mori},
  {Mullally}, {Narita}, {Nishiumi}, {Osborn}, {Paegert}, {de Leon}, {Quinn},
  {Reefe}, {Schwarz}, {Shporer}, {Soubkiou}, {Sousa}, {Stockdale}, {Str{\o}m},
  {Tan}, {Tang}, {Tenenbaum}, {Wheatley}, {Wittrock}, {Yahalomi}, \&
  {Zohrabi}}]{Teske2021}
{Teske}, J., {Wang}, S.~X., {Wolfgang}, A., {et~al.} 2021, \apjs, 256, 33,
  \dodoi{10.3847/1538-4365/ac0f0a}

\bibitem[{{Teske} {et~al.}(2018){Teske}, {Wang}, {Wolfgang}, {Dai}, {Shectman},
  {Butler}, {Crane}, \& {Thompson}}]{Teske2018}
{Teske}, J.~K., {Wang}, S., {Wolfgang}, A., {et~al.} 2018, \aj, 155, 148,
  \dodoi{10.3847/1538-3881/aaab56}

\bibitem[{{Toledo-Padr{\'o}n} {et~al.}(2020){Toledo-Padr{\'o}n}, {Lovis},
  {Su{\'a}rez Mascare{\~n}o}, {Barros}, {Gonz{\'a}lez Hern{\'a}ndez},
  {Sozzetti}, {Bouchy}, {Zapatero Osorio}, {Rebolo}, {Cristiani}, {Pepe},
  {Santos}, {Sousa}, {Tabernero}, {Lillo-Box}, {Bossini}, {Adibekyan},
  {Allart}, {Damasso}, {D'Odorico}, {Figueira}, {Lavie}, {Lo Curto}, {Mehner},
  {Micela}, {Modigliani}, {Nunes}, {Pall{\'e}}, {Abreu}, {Affolter}, {Alibert},
  {Aliverti}, {Allende Prieto}, {Alves}, {Amate}, {Avila}, {Baldini}, {Bandy},
  {Benatti}, {Benz}, {Bianco}, {Broeg}, {Cabral}, {Calderone}, {Cirami},
  {Coelho}, {Conconi}, {Coretti}, {Cumani}, {Cupani}, {Deiries}, {Dekker},
  {Delabre}, {Demangeon}, {Di Marcantonio}, {Ehrenreich}, {Fragoso}, {Genolet},
  {Genoni}, {G{\'e}nova Santos}, {Hughes}, {Iwert}, {Knudstrup}, {Landoni},
  {Lizon}, {Maire}, {Manescau}, {Martins}, {M{\'e}gevand}, {Molaro},
  {Monteiro}, {Monteiro}, {Moschetti}, {Mueller}, {Oggioni}, {Oliveira},
  {Oshagh}, {Pariani}, {Pasquini}, {Poretti}, {Rasilla}, {Redaelli}, {Riva},
  {Santana Tschudi}, {Santin}, {Santos}, {Segovia}, {Sosnowska}, {Span{\`o}},
  {Tenegi}, {Udry}, {Zanutta}, \& {Zerbi}}]{Toledo-Padron2020}
{Toledo-Padr{\'o}n}, B., {Lovis}, C., {Su{\'a}rez Mascare{\~n}o}, A., {et~al.}
  2020, \aap, 641, A92, \dodoi{10.1051/0004-6361/202038187}

\bibitem[{{Tsiaras} {et~al.}(2019){Tsiaras}, {Waldmann}, {Tinetti}, {Tennyson},
  \& {Yurchenko}}]{Tsiaras2019}
{Tsiaras}, A., {Waldmann}, I.~P., {Tinetti}, G., {Tennyson}, J., \&
  {Yurchenko}, S.~N. 2019, Nature Astronomy, 3, 1086,
  \dodoi{10.1038/s41550-019-0878-9}

\bibitem[{{Valenti} {et~al.}(1995){Valenti}, {Butler}, \&
  {Marcy}}]{Valenti1995}
{Valenti}, J.~A., {Butler}, R.~P., \& {Marcy}, G.~W. 1995, \pasp, 107, 966,
  \dodoi{10.1086/133645}

\bibitem[{{Van Eylen} {et~al.}(2018{\natexlab{a}}){Van Eylen}, {Agentoft},
  {Lundkvist}, {Kjeldsen}, {Owen}, {Fulton}, {Petigura}, \&
  {Snellen}}]{VanEylen2018b}
{Van Eylen}, V., {Agentoft}, C., {Lundkvist}, M.~S., {et~al.}
  2018{\natexlab{a}}, \mnras, 479, 4786, \dodoi{10.1093/mnras/sty1783}

\bibitem[{{Van Eylen} {et~al.}(2016{\natexlab{a}}){Van Eylen}, {Albrecht},
  {Gandolfi}, {Dai}, {Winn}, {Hirano}, {Narita}, {Bruntt}, {Prieto-Arranz},
  {B{\'e}jar}, {Nowak}, {Lund}, {Palle}, {Ribas}, {Sanchis-Ojeda}, {Yu},
  {Arriagada}, {Butler}, {Crane}, {Handberg}, {Deeg}, {Jessen-Hansen},
  {Johnson}, {Nespral}, {Rogers}, {Ryu}, {Shectman}, {Shrotriya}, {Slumstrup},
  {Takeda}, {Teske}, {Thompson}, {Vanderburg}, \& {Wittenmyer}}]{vanEylen2016b}
{Van Eylen}, V., {Albrecht}, S., {Gandolfi}, D., {et~al.} 2016{\natexlab{a}},
  \aj, 152, 143, \dodoi{10.3847/0004-6256/152/5/143}

\bibitem[{{Van Eylen} {et~al.}(2016{\natexlab{b}}){Van Eylen}, {Nowak},
  {Albrecht}, {Palle}, {Ribas}, {Bruntt}, {Perger}, {Gandolfi}, {Hirano},
  {Sanchis-Ojeda}, {Kiilerich}, {Prieto-Arranz}, {Badenas}, {Dai}, {Deeg},
  {Guenther}, {Monta{\~n}{\'e}s-Rodr{\'{\i}}guez}, {Narita}, {B{\'e}jar},
  {Shrotriya}, {Winn}, \& {Sebastian}}]{vanEylen2016a}
{Van Eylen}, V., {Nowak}, G., {Albrecht}, S., {et~al.} 2016{\natexlab{b}},
  \apj, 820, 56, \dodoi{10.3847/0004-637X/820/1/56}

\bibitem[{{Van Eylen} {et~al.}(2018{\natexlab{b}}){Van Eylen}, {Dai}, {Mathur},
  {Gandolfi}, {Albrecht}, {Fridlund}, {Garc{\'\i}a}, {Guenther}, {Hjorth},
  {Justesen}, {Livingston}, {Lund}, {P{\'e}rez Hern{\'a}ndez}, {Prieto-Arranz},
  {Regulo}, {Bugnet}, {Everett}, {Hirano}, {Nespral}, {Nowak}, {Palle}, {Silva
  Aguirre}, {Trifonov}, {Winn}, {Barrag{\'a}n}, {Beck}, {Chaplin}, {Cochran},
  {Csizmadia}, {Deeg}, {Endl}, {Heeren}, {Grziwa}, {Hatzes}, {Hidalgo},
  {Korth}, {Mathis}, {Monta{\~n}es Rodriguez}, {Narita}, {Patzold}, {Persson},
  {Rodler}, \& {Smith}}]{VanEylen2018}
{Van Eylen}, V., {Dai}, F., {Mathur}, S., {et~al.} 2018{\natexlab{b}}, \mnras,
  478, 4866, \dodoi{10.1093/mnras/sty1390}

\bibitem[{{Van Zandt} {et~al.}(2023){Van Zandt}, {Petigura}, {MacDougall},
  {Gilbert}, {Lubin}, {Barclay}, {Batalha}, {Crossfield}, {Dressing}, {Fulton},
  {Howard}, {Huber}, {Isaacson}, {Kane}, {Robertson}, {Roy}, {Weiss},
  {Behmard}, {Beard}, {Chontos}, {Dai}, {Dalba}, {Fetherolf}, {Giacalone},
  {Henze}, {Hill}, {Hirsch}, {Holcomb}, {Howell}, {Jenkins}, {Latham}, {Mayo},
  {Mireles}, {Mo{\v{c}}nik}, {Murphy}, {Pidhorodetska}, {Polanski}, {Ricker},
  {Rosenthal}, {Rubenzahl}, {Seager}, {Scarsdale}, {Turtelboom}, {Vanderspek},
  \& {Winn}}]{VanZandt2023}
{Van Zandt}, J., {Petigura}, E.~A., {MacDougall}, M., {et~al.} 2023, \aj, 165,
  60, \dodoi{10.3847/1538-3881/aca6ef}

\bibitem[{{Vanderburg} {et~al.}(2016{\natexlab{a}}){Vanderburg}, {Bieryla},
  {Duev}, {Jensen-Clem}, {Latham}, {Mayo}, {Baranec}, {Berlind}, {Kulkarni},
  {Law}, {Nieberding}, {Riddle}, \& {Salama}}]{Vanderburg2016-3167}
{Vanderburg}, A., {Bieryla}, A., {Duev}, D.~A., {et~al.} 2016{\natexlab{a}},
  \apjl, 829, L9, \dodoi{10.3847/2041-8205/829/1/L9}

\bibitem[{{Vanderburg} {et~al.}(2016{\natexlab{b}}){Vanderburg}, {Latham},
  {Buchhave}, {Bieryla}, {Berlind}, {Calkins}, {Esquerdo}, {Welsh}, \&
  {Johnson}}]{Vanderburg2016-catalog}
{Vanderburg}, A., {Latham}, D.~W., {Buchhave}, L.~A., {et~al.}
  2016{\natexlab{b}}, \apjs, 222, 14, \dodoi{10.3847/0067-0049/222/1/14}

\bibitem[{Vanderburg {et~al.}(2016)Vanderburg, Becker, Kristiansen, Bieryla,
  Duev, Jensen-Clem, Morton, Latham, Adams, Baranec, Berlind, Calkins,
  Esquerdo, Kulkarni, Law, Riddle, Salama, \&
  Schmitt}]{Vanderburg2016-hip41378}
Vanderburg, A., Becker, J.~C., Kristiansen, M.~H., {et~al.} 2016, The
  Astrophysical Journal, 827, L10, \dodoi{10.3847/2041-8205/827/1/l10}

\bibitem[{{Vanderburg} {et~al.}(2017){Vanderburg}, {Becker}, {Buchhave},
  {Mortier}, {Lopez}, {Malavolta}, {Haywood}, {Latham}, {Charbonneau},
  {L{\'o}pez-Morales}, {Adams}, {Bonomo}, {Bouchy}, {Collier Cameron},
  {Cosentino}, {Di Fabrizio}, {Dumusque}, {Fiorenzano}, {Harutyunyan},
  {Johnson}, {Lorenzi}, {Lovis}, {Mayor}, {Micela}, {Molinari}, {Pedani},
  {Pepe}, {Piotto}, {Phillips}, {Rice}, {Sasselov}, {S{\'e}gransan},
  {Sozzetti}, {Udry}, \& {Watson}}]{Vanderburg2017}
{Vanderburg}, A., {Becker}, J.~C., {Buchhave}, L.~A., {et~al.} 2017, \aj, 154,
  237, \dodoi{10.3847/1538-3881/aa918b}

\bibitem[{Virtanen {et~al.}(2020)Virtanen, Gommers, Oliphant, Haberland, Reddy,
  Cournapeau, Burovski, Peterson, Weckesser, Bright, {van der Walt}, Brett,
  Wilson, Millman, Mayorov, Nelson, Jones, Kern, Larson, Carey, Polat, Feng,
  Moore, {VanderPlas}, Laxalde, Perktold, Cimrman, Henriksen, Quintero, Harris,
  Archibald, Ribeiro, Pedregosa, {van Mulbregt}, \& {SciPy 1.0
  Contributors}}]{Virtanen2020}
Virtanen, P., Gommers, R., Oliphant, T.~E., {et~al.} 2020, Nature Methods, 17,
  261, \dodoi{10.1038/s41592-019-0686-2}

\bibitem[{{Vogt} {et~al.}(1994){Vogt}, {Allen}, {Bigelow}, {Bresee}, {Brown},
  {Cantrall}, {Conrad}, {Couture}, {Delaney}, {Epps}, {Hilyard}, {Hilyard},
  {Horn}, {Jern}, {Kanto}, {Keane}, {Kibrick}, {Lewis}, {Osborne},
  {Pardeilhan}, {Pfister}, {Ricketts}, {Robinson}, {Stover}, {Tucker}, {Ward},
  \& {Wei}}]{Vogt1994}
{Vogt}, S.~S., {Allen}, S.~L., {Bigelow}, B.~C., {et~al.} 1994, Society of
  Photo-Optical Instrumentation Engineers (SPIE) Conference Series, Vol. 2198,
  {HIRES: the high-resolution echelle spectrometer on the Keck 10-m Telescope}
  ({Crawford}, David L. and {Craine}, Eric R.), 362, \dodoi{10.1117/12.176725}

\bibitem[{{Vogt} {et~al.}(2014){Vogt}, {Radovan}, {Kibrick}, {Butler},
  {Alcott}, {Allen}, {Arriagada}, {Bolte}, {Burt}, {Cabak}, {Chloros},
  {Cowley}, {Deich}, {Dupraw}, {Earthman}, {Epps}, {Faber}, {Fischer}, {Gates},
  {Hilyard}, {Holden}, {Johnston}, {Keiser}, {Kanto}, {Katsuki}, {Laiterman},
  {Lanclos}, {Laughlin}, {Lewis}, {Lockwood}, {Lynam}, {Marcy}, {McLean},
  {Miller}, {Misch}, {Peck}, {Pfister}, {Phillips}, {Rivera}, {Sand ford},
  {Saylor}, {Stover}, {Thompson}, {Walp}, {Ward}, {Wareham}, {Wei}, \&
  {Wright}}]{Vogt2014}
{Vogt}, S.~S., {Radovan}, M., {Kibrick}, R., {et~al.} 2014, \pasp, 126, 359,
  \dodoi{10.1086/676120}

\bibitem[{{Weiss} \& {Marcy}(2014)}]{Weiss2014}
{Weiss}, L.~M., \& {Marcy}, G.~W. 2014, \apjl, 783, L6,
  \dodoi{10.1088/2041-8205/783/1/L6}

\bibitem[{{Weiss} {et~al.}(2017){Weiss}, {Deck}, {Sinukoff}, {Petigura},
  {Agol}, {Lee}, {Becker}, {Howard}, {Isaacson}, {Crossfield}, {Fulton},
  {Hirsch}, \& {Benneke}}]{Weiss2017}
{Weiss}, L.~M., {Deck}, K.~M., {Sinukoff}, E., {et~al.} 2017, \aj, 153, 265,
  \dodoi{10.3847/1538-3881/aa6c29}

\bibitem[{{Wittenmyer} {et~al.}(2018){Wittenmyer}, {Sharma}, {Stello}, {Buder},
  {Kos}, {Asplund}, {Duong}, {Lin}, {Lind}, {Ness}, {Zwitter}, {Horner},
  {Clark}, {Kane}, {Huber}, {Bland-Hawthorn}, {Casey}, {De Silva}, {D'Orazi},
  {Freeman}, {Martell}, {Simpson}, {Zucker}, {Anguiano}, {Casagrande},
  {Esdaile}, {Hon}, {Ireland}, {Kafle}, {Khanna}, {Marshall}, {Saddon},
  {Traven}, \& {Wright}}]{Wittenmyer2018}
{Wittenmyer}, R.~A., {Sharma}, S., {Stello}, D., {et~al.} 2018, \aj, 155, 84,
  \dodoi{10.3847/1538-3881/aaa3e4}

\bibitem[{{Wittenmyer} {et~al.}(2020){Wittenmyer}, {Clark}, {Sharma}, {Stello},
  {Horner}, {Kane}, {Stevens}, {Wright}, {Spina}, {{\v{C}}otar}, {Asplund},
  {Bland-Hawthorn}, {Buder}, {Casey}, {De Silva}, {D'Orazi}, {Freeman}, {Kos},
  {Lewis}, {Lin}, {Lind}, {Martell}, {Simpson}, {Zucker}, \&
  {Zwitter}}]{Wittenmyer2020}
{Wittenmyer}, R.~A., {Clark}, J.~T., {Sharma}, S., {et~al.} 2020, \mnras, 496,
  851, \dodoi{10.1093/mnras/staa1528}

\bibitem[{{Wolfgang} \& {Lopez}(2015)}]{Wolfgang2015}
{Wolfgang}, A., \& {Lopez}, E. 2015, \apj, 806, 183,
  \dodoi{10.1088/0004-637X/806/2/183}

\bibitem[{{Xie} {et~al.}(2016){Xie}, {Dong}, {Zhu}, {Huber}, {Zheng}, {De Cat},
  {Fu}, {Liu}, {Luo}, {Wu}, {Zhang}, {Zhang}, {Zhou}, {Cao}, {Hou}, {Wang}, \&
  {Zhang}}]{Xie2016}
{Xie}, J.-W., {Dong}, S., {Zhu}, Z., {et~al.} 2016, Proceedings of the National
  Academy of Science, 113, 11431, \dodoi{10.1073/pnas.1604692113}

\bibitem[{{Yee} {et~al.}(2017){Yee}, {Petigura}, \& {von Braun}}]{Yee2017}
{Yee}, S.~W., {Petigura}, E.~A., \& {von Braun}, K. 2017, \apj, 836, 77,
  \dodoi{10.3847/1538-4357/836/1/77}

\bibitem[{{Yu} {et~al.}(2018){Yu}, {Crossfield}, {Schlieder}, {Kosiarek},
  {Feinstein}, {Livingston}, {Howard}, {Benneke}, {Petigura}, {Bristow},
  {Christiansen}, {Ciardi}, {Crepp}, {Dressing}, {Fulton}, {Gonzales},
  {Hardegree-Ullman}, {Henning}, {Isaacson}, {L{\'e}pine}, {Martinez},
  {Morales}, \& {Sinukoff}}]{Yu2018}
{Yu}, L., {Crossfield}, I. J.~M., {Schlieder}, J.~E., {et~al.} 2018, \aj, 156,
  22, \dodoi{10.3847/1538-3881/aac6e6}

\bibitem[{{Zeng} {et~al.}(2019){Zeng}, {Jacobsen}, {Sasselov}, {Petaev},
  {Vanderburg}, {Lopez-Morales}, {Perez-Mercader}, {Mattsson}, {Li}, {Heising},
  {Bonomo}, {Damasso}, {Berger}, {Cao}, {Levi}, \& {Wordsworth}}]{Zeng2019}
{Zeng}, L., {Jacobsen}, S.~B., {Sasselov}, D.~D., {et~al.} 2019, Proceedings of
  the National Academy of Science, 116, 9723, \dodoi{10.1073/pnas.1812905116}

\bibitem[{{Zink} {et~al.}(2020{\natexlab{a}}){Zink}, {Hardegree-Ullman},
  {Christiansen}, {Dressing}, {Crossfield}, {Petigura}, {Schlieder}, \&
  {Ciardi}}]{Zink2020}
{Zink}, J.~K., {Hardegree-Ullman}, K.~K., {Christiansen}, J.~L., {et~al.}
  2020{\natexlab{a}}, \aj, 159, 154, \dodoi{10.3847/1538-3881/ab7448}

\bibitem[{{Zink} {et~al.}(2020{\natexlab{b}}){Zink}, {Hardegree-Ullman},
  {Christiansen}, {Petigura}, {Dressing}, {Schlieder}, {Ciardi}, \&
  {Crossfield}}]{Zink2020b}
---. 2020{\natexlab{b}}, \aj, 160, 94, \dodoi{10.3847/1538-3881/aba123}

\bibitem[{{Zink} {et~al.}(2021){Zink}, {Hardegree-Ullman}, {Christiansen},
  {Bhure}, {Adkins}, {Petigura}, {Dressing}, {Crossfield}, \&
  {Schlieder}}]{Zink2021}
---. 2021, \aj, 162, 259, \dodoi{10.3847/1538-3881/ac2309}

\bibitem[{{Zink} {et~al.}(2023){Zink}, {Hardegree-Ullman}, {Christiansen},
  {Petigura}, {Boley}, {Bhure}, {Rice}, {Yee}, {Isaacson}, {Fernandes},
  {Howard}, {Blunt}, {Lubin}, {Chontos}, {Pidhorodetska}, \&
  {MacDougall}}]{Zink2023}
---. 2023, \aj, 165, 262, \dodoi{10.3847/1538-3881/acd24c}

\end{thebibliography}

\appendix
\section{Individual Sections}
\label{sec:indiv_sys}


This appendix describes each planetary system considered in this paper and our analysis of its photometry and RVs.  Individual subsections provide details of each analysis, tables of system parameters, and figures with results. The entire set of planetary properties is listed in Table \ref{tb:planet_props} and is plotted in Fig.\ \ref{fig:planet_hist}, as well as in Figs.~\ref{fig:MR1} and~\ref{fig:MR2}.

\subsection{HD 3167} 
\label{sec:hd3167}

\ddigSTNAME (K2-96, EPIC \ddigEPICID) is a bright late G dwarf from Campaign 8.  \cite{Vanderburg2016-3167} discovered two transiting planets with radii of 1.6 \rearth (planet b) and 2.9 \rearth (planet c) and orbital periods of 0.96 days and 29.8 days, respectively.  \cite{Christiansen2017} found masses of $5.02 \pm 0.38$ \mearth and $9.80_{-1.24}^{+1.30}$ \mearth for the two planets using RVs from HIRES (60 measurements), APF (116), and HARPS-N (76).  They also discovered a third nontransiting planet with a period of 8.5 days and a mass of $6.90 \pm 0.71$ \mearth. Subsequently, \cite{Gandolfi2017} characterized by the system using new RVs from FIES (37 measurements), HARPS (45), and HARPS-N (32).  They did not detect planet d, but found masses of $5.69 \pm 0.44$ \mearth and $8.33_{-1.85}^{+1.79}$ \mearth for planets b and c, respectively. A later analysis including CHEOPS transits and additional RVs reported a second nontransiting planet, with masses (or minimum masses) of $4.7\pm 0.3$,  $10.7 \pm 0.8$,   $5.0 \pm 0.5$, and  $9.7  \pm 1.2 M_\oplus$ for planets b, c, d, and e, respectively \citep{Bourrier2022}, largely consistent with previous results.

Planet c was revealed to be in a near-polar orbit via Rossiter-McLaughlin measurements at HARPS-N \citep{Dalal2019}, and H$_2$O vapor was detected in its atmosphere by {\em HST} transmission spectroscopy \citep{Evans2020}.

We adopt the stellar and planet parameters from \cite{Christiansen2017}. 
The full set of adopted parameters are listed in Tables \ref{tb:star_pars}, \ref{tb:star_props} and \ref{tb:planet_props}.

\subsection{HD 106315} 


\hieeSTNAME (\hieeSTNAMEKTWO, EPIC \hieeEPICID) is a bright ($V = \hieeVMAG$)  F5 dwarf observed in Campaign 10.  \cite{Crossfield2017} and \cite{Rodriguez2017} concurrently announced the discovery of two small transiting planets with orbital periods of 9.5 days (planet b) and 21 days (planet c) and sizes of 2.4 \rearth and 4.3 \rearth.  The star is rapidly rotating with \vsini = \hieeVSINI \kms, consistent with its early spectral type, making the system a challenging target for Doppler measurements. The two planets orbiting \hieeSTNAME were also noted in the \cite{Mayo2018} catalog.

\cite{Barros2017} collected 130 RVs from HARPS from which they measured a mass of $12.6 \pm 3.2$ \mearth and  density of $4.7 \pm 1.7$ \gmc for planet b and a mass of $15.2 \pm 3.7$ \mearth and a density of $1.01 \pm 0.29$ \gmc for planet c.  

\citet{Kosiarek2020b} followed with an analysis based on 352 HIRES observations and 25 PFS observations, in addition to the HARPS RVs. They additionally updated the stellar parameters from Gaia DR2 \citep{gaia:2018} and propagated these forward to calculate updated planet radii. They found that a circular 2-planet fit without a Gaussian process best fits the data based on the AIC statistic, resulting in planet masses of $10.5\pm3.1$ \mearth for planet b and $12.0\pm3.8$ \mearth for planet c. \hieeSTNAME\ b is a super-Earth with a small volatile mass that increases its radius and 
\hieeSTNAME\ c is a Neptune-sized planet consistent with having a rocky core and a 10\% H/He envelope.  A tentative hint of H$_2$O was seen in planet c's atmosphere via {\em HST}/WFC3 transmission spectroscpoy \citep{Kreidberg2022}, and the planet is being targeted with {\em JWST} MIRI/LRS spectroscopy in GO-2950.
For our analysis, we adopt the \citet{Kosiarek2020b} solution.  See Tables \ref{tb:star_pars} and \ref{tb:star_props} for stellar properties and Table \ref{tb:planet_props} for planet parameters.

\subsection{K2-85} 
\label{sec:k2_85}

\hbdaSTNAME (EPIC \hbdaEPICID) is a late K dwarf with elevated chromospheric activity (\lrphk = $-4.67$) from Campaign 4 that hosts an ultra-short period planet ($P$ = 0.7 days) with a radius of 1.4 \rearth.
See Tables \ref{tb:star_pars} and \ref{tb:star_props} for stellar properties and Table \ref{tb:planet_props} for precise planet parameters.   \cite{Dressing2017} used near-infrared spectroscopy to measure a stellar radius of $0.68 \pm 0.03$\,\rsun.  \cite{Barros2016} designated the object as a planet candidate. 
It is validated and appears in the \cite{Crossfield2016}, \cite{Adams2016}, and \cite{Mayo2018} catalogs.
Our HIRES measurements reveal a planet mass of $4.0\pm1.5$ \mearth, as listed in Table \ref{tb:planet_props}.



\begin{figure*}
\epsscale{1.0}
\plotone{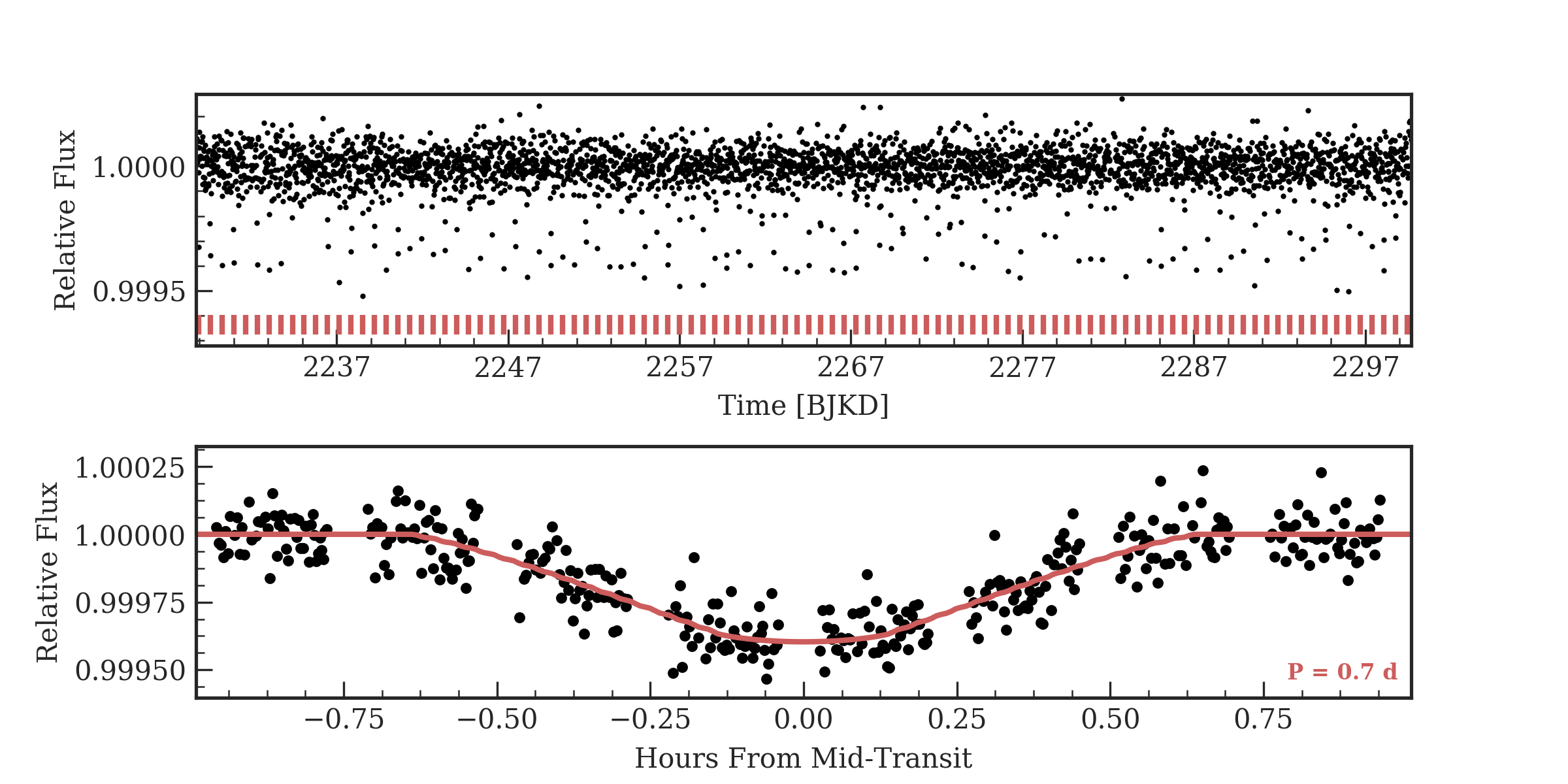}
\caption{Time series (top) and phase-folded (bottom) light curve for the planet orbiting \hbdaSTNAME.  Plot formatting is the same as in Fig.\ \ref{fig:lc_epic220709978}.}
\label{fig:lc_epic210707130}
\end{figure*}

\begin{figure}
\epsscale{1.0}
\plotone{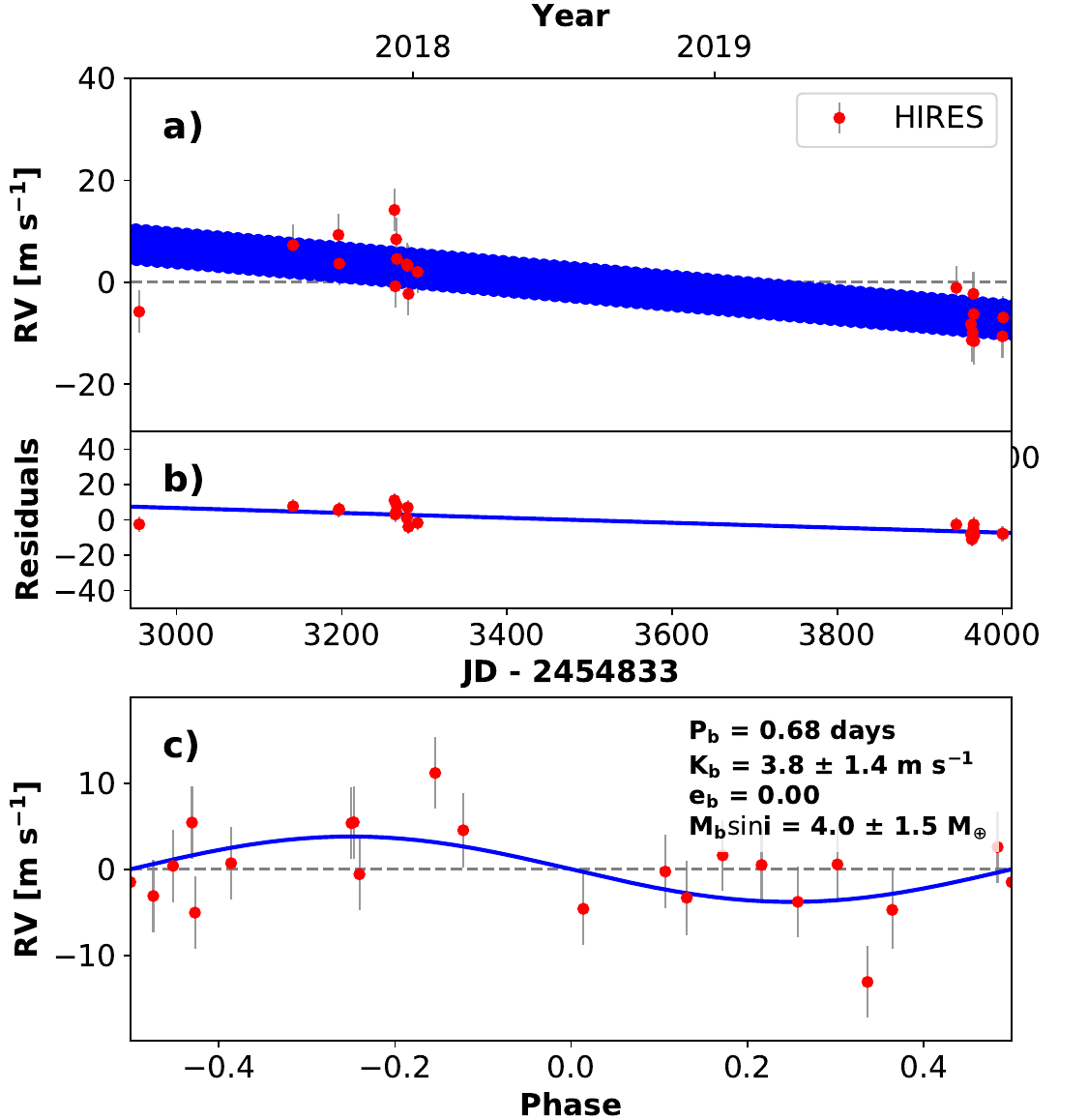}
\caption{RVs and Keplerian model for \hbdaSTNAME. Symbols, lines, and annotations are similar to those in Fig.\ \ref{fig:rvs_epic220709978}.}
\label{fig:rvs_k2-85}
\end{figure}

\subsection{K2-222} 

\begin{figure*}
\epsscale{1.0}
\plotone{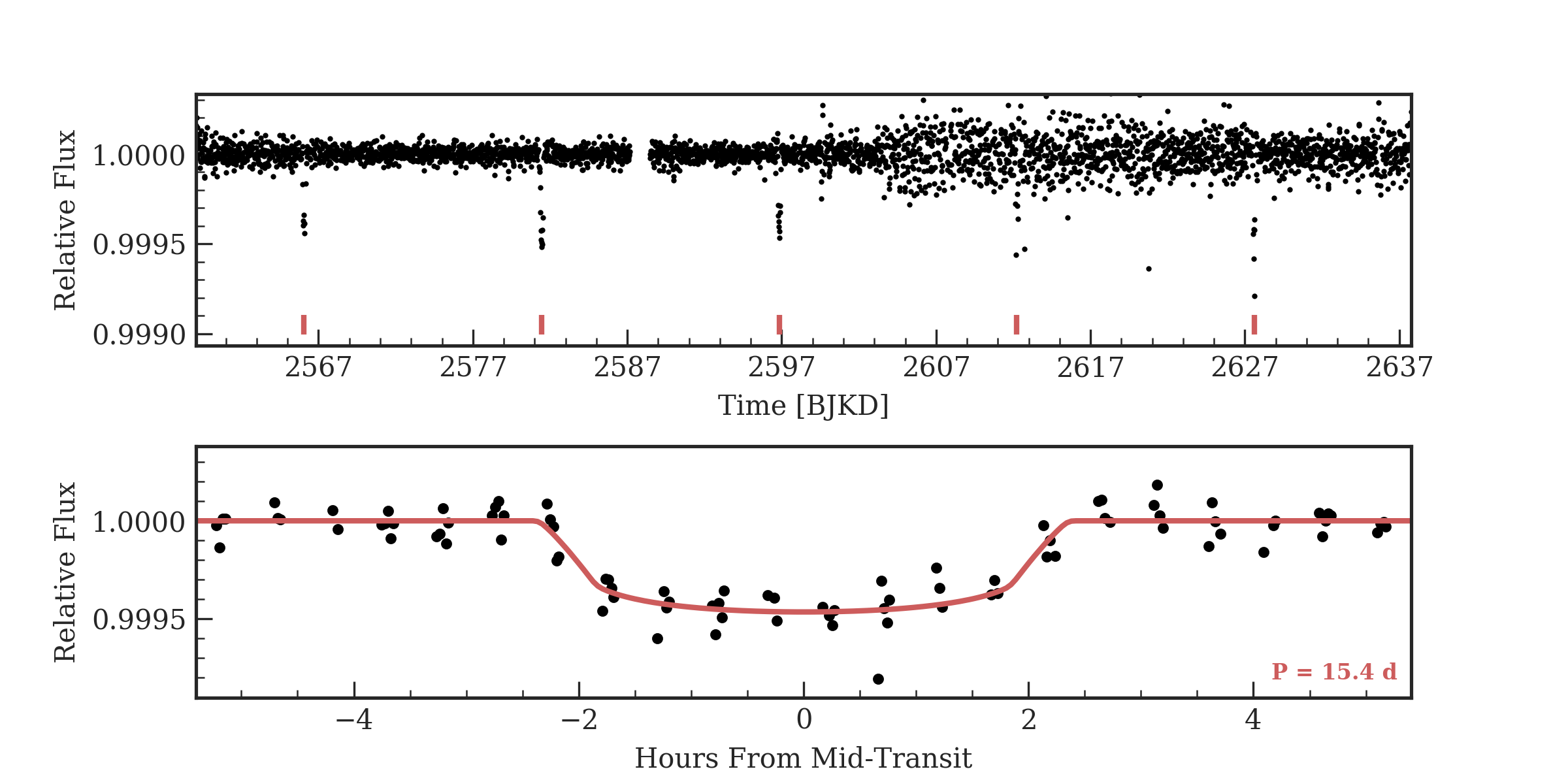}
\caption{De-trended, time series (top) and phase-folded (middle and bottom) light curve for \jjhiSTNAME.  Transit epochs are marked by vertical ticks in the time series, and the maximum a posteriori transit model is shown as a colored line in the phase-folded panel.  Gray points are not included in the light curve fit.  Different color ticks indicate the planets in the time series, and there are separate phase-folded panels for each planet in the middle and lower rows.  In the top panel, time is expressed in units of days with BKJD = BJD$_\mathrm{TBD}$ $-$ 2,545,833.}
\label{fig:lc_epic220709978}
\end{figure*}

\jjhiSTNAME (EPIC \jjhiEPICID) is a G0 dwarf observed by K2 in Campaign 8 with one detected transiting planet with an orbital period of 15 days and a size of 2.4 \rearth.  See Tables \ref{tb:star_pars}  and \ref{tb:star_props} for stellar properties and Table \ref{tb:planet_props} for precise planet parameters.
\cite{Petigura2018} lists this object as a planet candidates and  \cite{Mayo2018} validated it as a planet.  Our fit of the EVEREST light curve of the K2 photometry for \jjhiSTNAME is shown in Fig.\ \ref{fig:lc_epic220709978}.

\begin{figure}
\epsscale{1.0}
\plotone{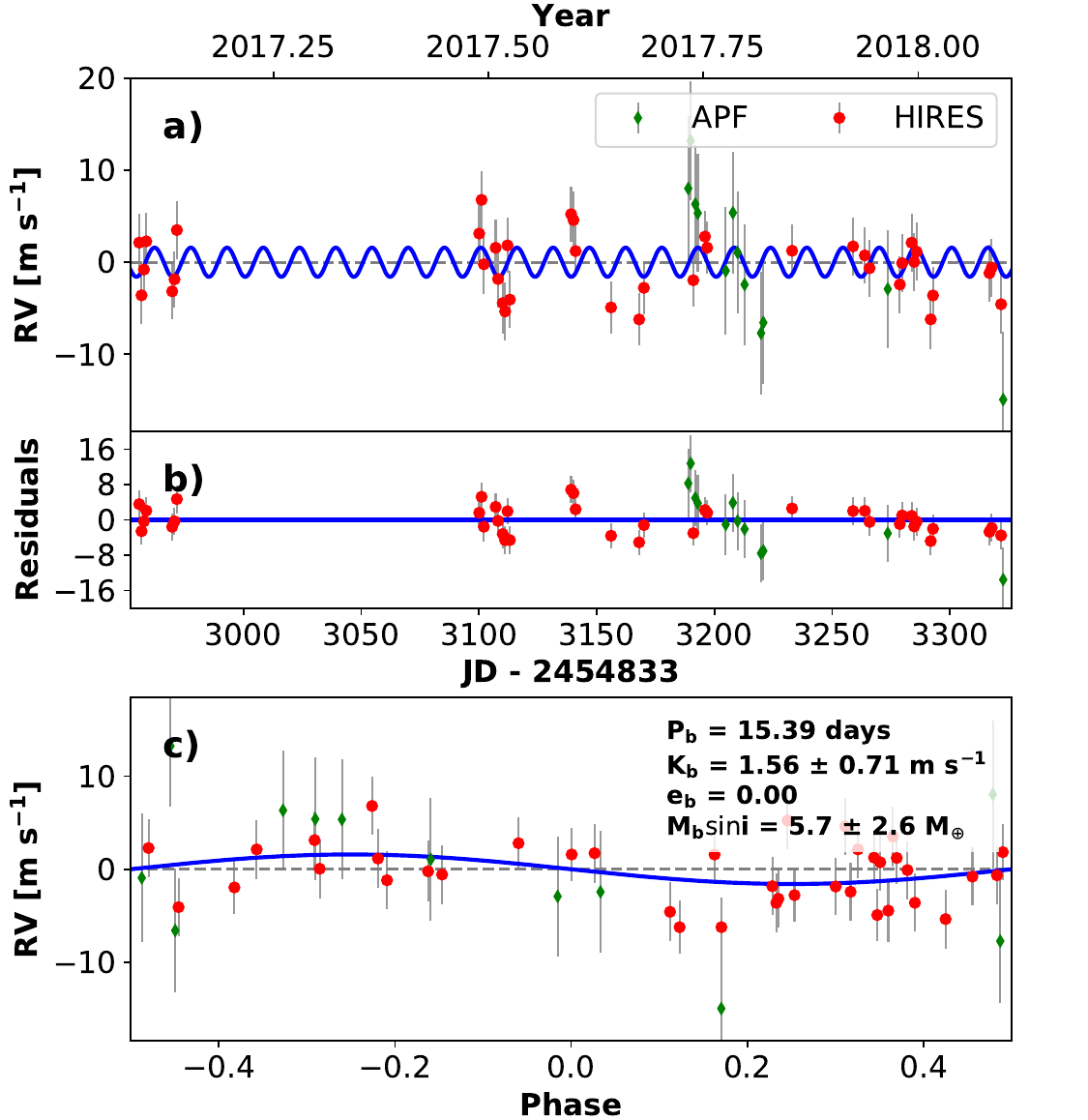}
\caption{RVs and Keplerian model for \jjhiSTNAME.  Panel a shows the time series RVs from HIRES and HARPS with the best-fit Keplerian model in blue.  The residuals to this model are shown in panel b.  Panels c and d show the same RVs phased to the orbital periods of the planets with annotations indicating parameters of the model.  This paper contains many similar plots for other systems, each showing the time-series RVs and residuals on the top and one panel each for the phased RVs for each planet.  Data from the instrument (or segment of data from an instrument needing a separate zero point in the analysis) are labeled with separate symbols.}
\label{fig:rvs_epic220709978}
\end{figure}

\begin{figure}
\epsscale{1.0}
\plotone{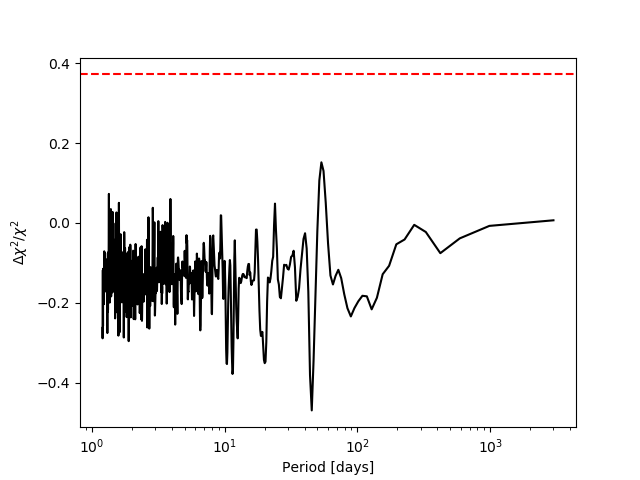}
\caption{Periodogram search of the RVs for a third planet orbiting \jjhiSTNAME.  The black line shows the normalized difference in $\chi^2$ for the adopted two-planet model (Table \ref{tab:epic220709978}) compared to a model with one additional planet, as a function of the orbital period of the additional planet.  The parameters not listed as fixed in Table \ref{tab:epic220709978} were allowed to vary with each trial period for the third planet, which is assumed to be in a circular orbit.  The periodogram power $\Delta\chi^2/\chi^2$ was calculated using the 2DKLS formalism of \cite{Otoole2009}.  The dashed red line denotes the 1\% false alarm probability level computed using the empirical method in \cite{Howard2016}.  In this case, we did not find a compelling period for a prospective third planet.}
\label{fig:rvs_epic220709978_resid}
\end{figure}

We acquired \jjhiNOBSHIRES RVs of \jjhiSTNAME with HIRES (exposure meter setting of 250,000) and \jjhiNOBSAPF with the APF.  We modeled the system as a single planet in a circular orbit, with the orbital period and phase fixed to the transit ephemeris.  
Our model did not include any explicit priors, and RVs within one night from a particular telescope were binned.    
We considered more complicated models with orbital eccentricity and/or a linear RV trend, but rejected these because of insufficient evidence based on the AICc statistic.
The results of this analysis are listed in Table \ref{tab:epic220709978} and the best-fit model is shown in Fig. \ref{fig:rvs_epic220709978}. 
Our analysis finds a low jitter for the HIRES RVs and a high value for the APF data (see Table \ref{tab:epic220709978}), suggesting underestimated errors on the latter.  Using the method in Sec.\ \ref{sec:rv_additional_planet_search}, we also searched the RVs for additional planets, but we found no compelling signals (Fig.\ \ref{fig:rvs_epic220709978_resid}).
Although the star is bright, we only detected the planet's signal at 2.4-$\sigma$.
Subsequent analysis by other teams securely measured the planet to be 8--9\,$M_\oplus$ \citep{Nava2022,Bonomo2023}.


\begin{deluxetable}{lrrr}
\tablecolumns{4}
\tabletypesize{\scriptsize}
\tablecaption{ K2-222 System Parameters \label{tab:epic220709978}}
\tablehead{
  \colhead{Parameter} & 
  \colhead{Credible} & 
  \colhead{Maximum} & 
  \colhead{Units} \\
  \colhead{} & 
  \colhead{Interval} & 
  \colhead{Likelihood} & 
  \colhead{} }
\startdata
\sidehead{\bf{RV Analysis -- MCMC Step Parameters}} 
$P_{b}$ & $\equiv15.3863$ & $\equiv15.3863$ & days \\ 
$T_{\mathrm{conj},b}$ & $\equiv2457399.0652$ & $\equiv2457399.0652$ & BJD$_\mathrm{TBD}$ \\ 
$\sqrt{e}\cos{\omega}_{b}$ & $\equiv0.0$ & $\equiv0.0$ &  \\ 
$\sqrt{e}\sin{\omega}_{b}$ & $\equiv0.0$ & $\equiv0.0$ &  \\ 
$K_{b}$ & $1.6^{+0.7}_{-0.7}$ & $1.6$ & m s$^{-1}$ \\ 
$\gamma_{\rm HIRES}$ & $-1.22^{+0.54}_{-0.55}$ & $-1.21$ & m s$-1$ \\ 
$\gamma_{\rm APF}$ & $1.2\pm2.2$ & $1.2$ & m s$-1$ \\ 
$\dot{\gamma}$ & $\equiv0.0$ & $\equiv0.0$ & m s$^{-1}$ day$^{-1}$ \\ 
$\ddot{\gamma}$ & $\equiv0.0$ & $\equiv0.0$ & m s$^{-1}$ day$^{-2}$ \\ 
$\sigma_{\rm HIRES}$ & $2.83^{+0.62}_{-0.24}$ & $2.68$ & $\rm m\ s^{-1}$ \\ 
$\sigma_{\rm APF}$ & $4^{+2}_{-13}$ & $6$ & $\rm m\ s^{-1}$ \\ 
\hline
\sidehead{\bf{Orbital \& Physical Parameters}}
$P_{b}$ & $\equiv15.3863$ & $\equiv15.3863$ & days \\
$T_{\mathrm{conj},b}$ & $\equiv2457399.0652$ & $\equiv2457399.0652$ & BJD$_\mathrm{TBD}$ \\
$e_{b}$ & $\equiv0.0$ & $\equiv0.0$ &  \\
$\omega_{b}$ & $\equiv0.0$ & $\equiv0.0$ & radians \\
$K_{b}$ & $1.6^{+0.7}_{-0.7}$ & $1.6$ & m s$^{-1}$ \\
$M_b$ & $5.7^{+2.5}_{-2.7}$ & $5.8$ & $M_\earth$ \\
$R_b/R_*$ & $0.0202^{+0.0013}_{-0.0008}$ & $0.0203$ &  \\
$\rho_b$ & $2\pm1$ & $2$ & g cm$^{-3}$ \\
$R_b$ & $2.41^{+0.15}_{-0.09}$ & $2.42$ & $R_\earth$ \\
\hline
\hline
\sidehead{\bf{Priors}}
Parameter & Prior \\
\hline
\\None \\
\enddata
\end{deluxetable}

\subsection{K2-291} 


K2-291 (EPIC 247418783) is a G0 dwarf star observed in Campaign 13 that hosts one super-Earth transiting planet (R$_p$=1.5 \rearth) in a short-period orbit ($P$ = 2.23 d).  The host star has elevated activity with \lrphk = -4.79 dex.


K2-291b was discovered, confirmed, and characterized by \cite{Kosiarek2019b}. They discovered the planet in the K2 Campaign 13 data and measured the mass using a combination of HIRES and HARPS-N RVs. \cite{Kosiarek2019b} 
used a Gaussian process to characterize stellar activity and found that a circular, 1-planet model with a GP characterizes the data best, resulting in a mass of M$_b$=6.5 \mearth.
We adopt their solution; see see Tables \ref{tb:star_pars} and \ref{tb:star_props} for stellar properties and Table \ref{tb:planet_props} for the planet parameters.

\subsection{K2-236} 

\begin{figure*}
\epsscale{1.0}
\plotone{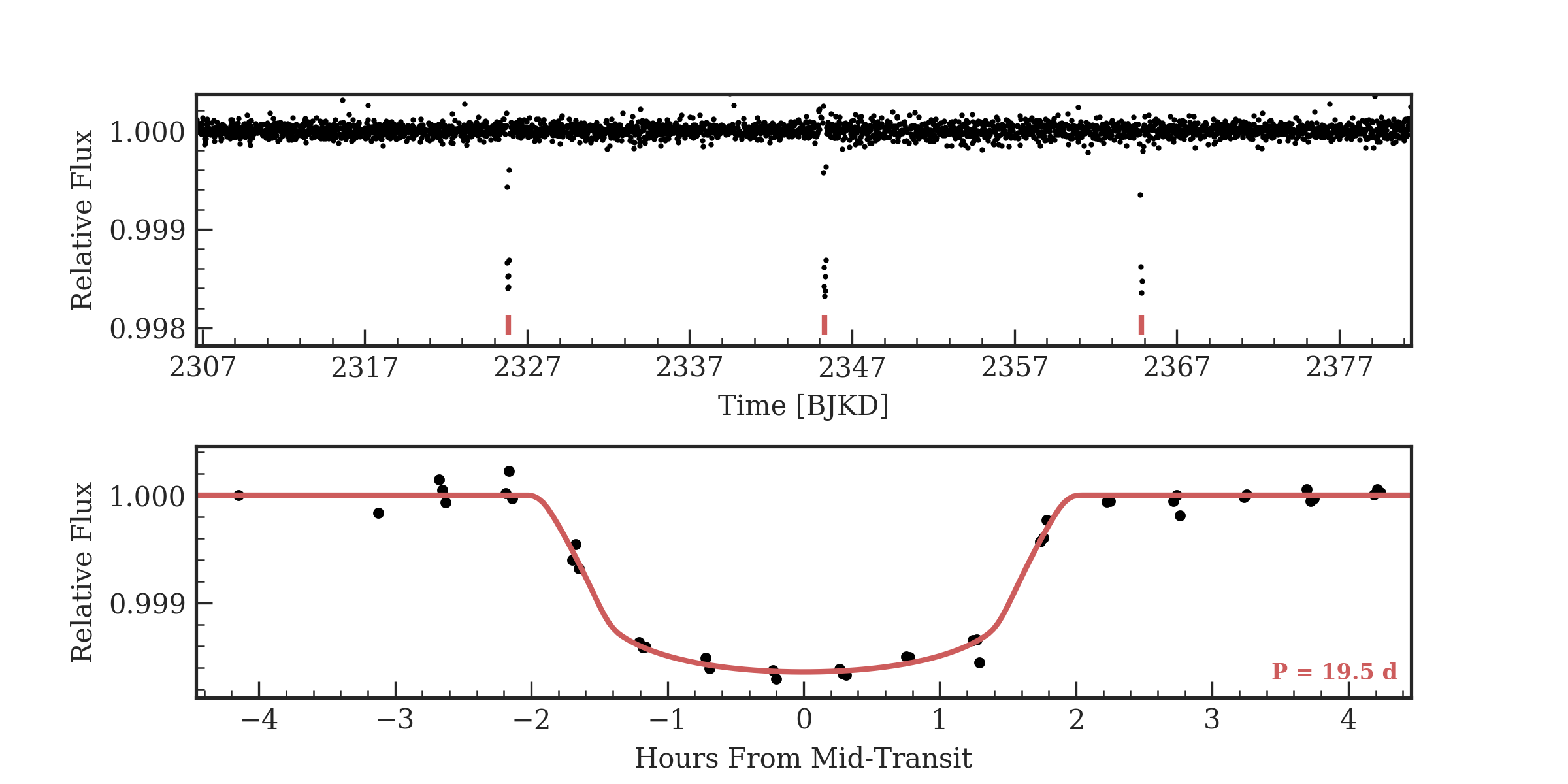}
\caption{Time series (top) and phase-folded (bottom) light curve for the planet orbiting \fcabSTNAME.  Plot formatting is the same as in Fig.\ \ref{fig:lc_epic220709978}.}
\label{fig:lc_epic211945201}
\end{figure*}

\fcabSTNAME is a nearly solar-type star from Campaign 5 with slightly elevated \teff, \feh, and \Rstar compared to the Sun.  The star was re-observed in Campaign 16, although we only analyzed the photometry from Field 5.  We detected one transiting planet with a size of 5.7 \rearth and an orbital period of 19.5 days.
See Tables \ref{tb:star_pars}  and \ref{tb:star_props} for stellar properties and Table \ref{tb:planet_props} for precise planet parameters.
The planet is noted in the catalogs by \cite{Barros2016}, \cite{Petigura2018}, and \cite{Mayo2018} as a candidate.  Our fit of the EVEREST light curve of the K2 photometry for \fcabSTNAME is shown in Fig.\ \ref{fig:lc_epic211945201}.

\cite{Chakraborty2018} studied this system using the PARAS spectrometer and found a mass of  $27^{+14}_{-13}$ \mearth and a density of $0.65^{+0.34}_{-0.30}$ \gmc based on 19 RVs.  They also validated the planet using VESPA, finding a false positive probability of 2\%.

\begin{figure}
\epsscale{1.0}
\plotone{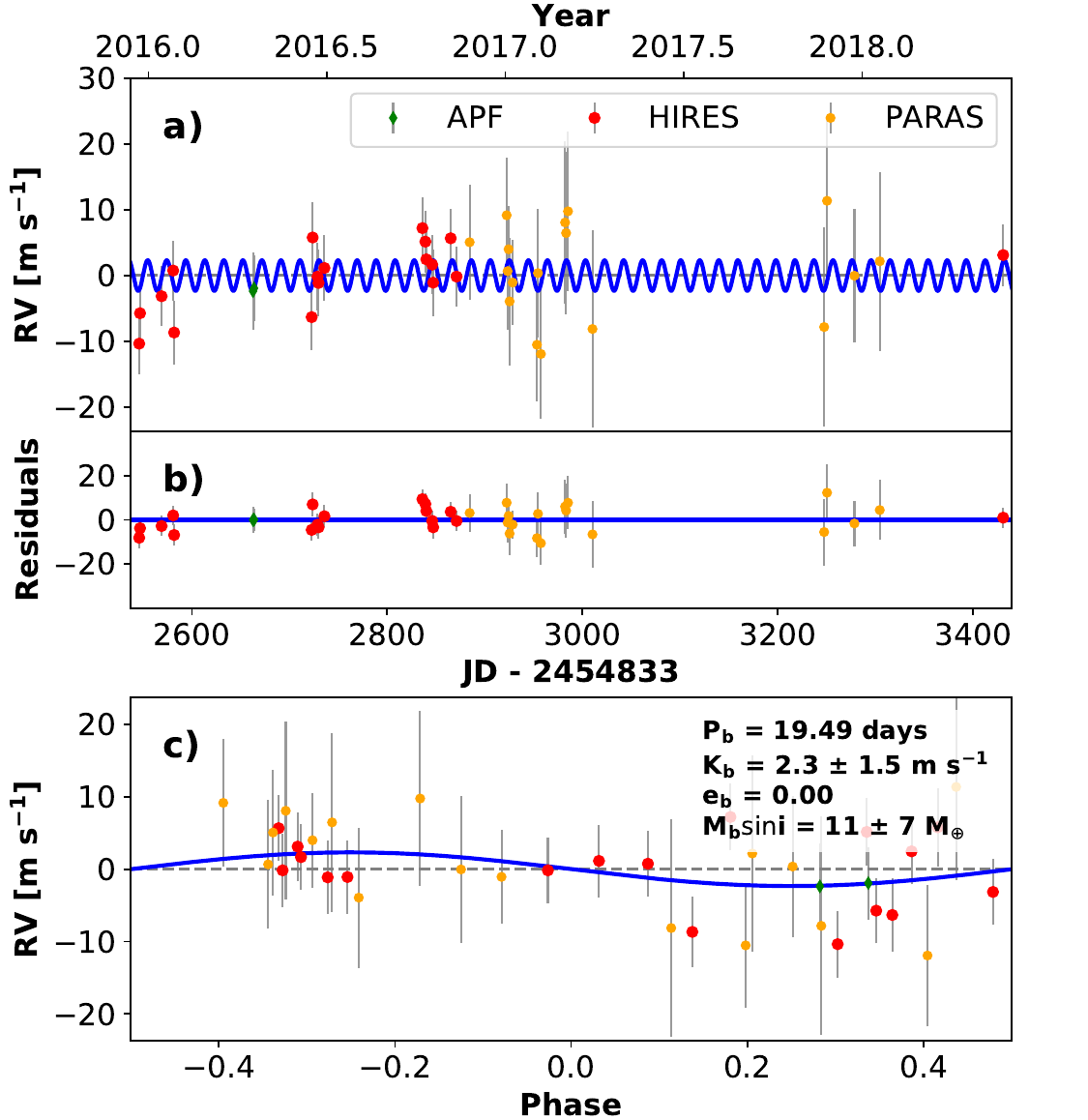}
\caption{RVs and Keplerian model for \fcabSTNAME.  Symbols, lines, and annotations are similar to those in Fig.\ \ref{fig:rvs_epic220709978}.}
\label{fig:rvs_epic211945201}
\end{figure}

We acquired \fcabNOBSHIRES RVs of \fcabSTNAME with HIRES, typically with an exposure meter setting of 60,000 counts.  We also acquired \fcabNOBSAPF RVs from the APF.  
We modeled the system as a single planet in a circular orbit with the orbital period and phase fixed to the transit ephemeris.  RVs from PARAS, HIRES, and APF were included, with binning applied per night and per telescope.  The results of this analysis are listed in Table \ref{tab:epic211945201} and the best-fit model is shown in Fig.\ \ref{fig:rvs_epic211945201}.   We considered more complicated models with eccentricity and/or an RV slope, but rejected these because of insufficient evidence using the AICc statistic.  Our combined analysis of PARAS, HIRES, and APF data finds a lower mass and density for the planet than \cite{Chakraborty2018}, although neither analysis detects the planet signal with high confidence.  We conclude that \fcabPNAMEone is a sub-Saturn-size planet with low density.


\begin{deluxetable}{lrrr}
\tablecolumns{4}
\tabletypesize{\scriptsize}
\tablecaption{ K2-236 System Parameters \label{tab:epic211945201}}
\tablehead{
  \colhead{Parameter} & 
  \colhead{Credible} & 
  \colhead{Maximum} & 
  \colhead{Units} \\
  \colhead{} & 
  \colhead{Interval} & 
  \colhead{Likelihood} & 
  \colhead{} }
\startdata
\sidehead{\bf{RV Analysis -- MCMC Step Parameters}} 
$P_{b}$ & $\equiv19.491$ & $\equiv19.491$ & days \\ 
$T_{\mathrm{conj},b}$ & $\equiv2457158.8266$ & $\equiv2457158.8266$ & BJD$_\mathrm{TBD}$ \\ 
$\sqrt{e}\cos{\omega}_{b}$ & $\equiv0.0$ & $\equiv0.0$ &  \\ 
$\sqrt{e}\sin{\omega}_{b}$ & $\equiv0.0$ & $\equiv0.0$ &  \\ 
$K_{b}$ & $2.3^{+1.6}_{-1.4}$ & $2.3$ & m s$^{-1}$ \\ 
$\gamma_{\rm PARAS}$ & $1260.7^{+2.6}_{-2.5}$ & $1260.6$ & m s$-1$ \\ 
$\gamma_{\rm HIRES}$ & $-1.0\pm1.3$ & $-1.0$ & m s$-1$ \\ 
$\gamma_{\rm APF}$ & $0^{+7100}_{-5200}$ & $0$ & m s$-1$ \\ 
$\dot{\gamma}$ & $\equiv0.0$ & $\equiv0.0$ & m s$^{-1}$ day$^{-1}$ \\ 
$\ddot{\gamma}$ & $\equiv0.0$ & $\equiv0.0$ & m s$^{-1}$ day$^{-2}$ \\ 
$\sigma_{\rm PARAS}$ & $-0.0^{+3.3}_{-3.5}$ & $-0.0$ & $\rm m\ s^{-1}$ \\ 
$\sigma_{\rm HIRES}$ & $5^{+1.8}_{-0.5}$ & $4.4$ & $\rm m\ s^{-1}$ \\ 
$\sigma_{\rm APF}$ & $0^{+22000}_{-11000}$ & $0$ & $\rm m\ s^{-1}$ \\ 
\hline
\sidehead{\bf{Orbital \& Physical Parameters}}
$P_{b}$ & $\equiv19.491$ & $\equiv19.491$ & days \\
$T_{\mathrm{conj},b}$ & $\equiv2457158.8266$ & $\equiv2457158.8266$ & BJD$_\mathrm{TBD}$ \\
$e_{b}$ & $\equiv0.0$ & $\equiv0.0$ &  \\
$\omega_{b}$ & $\equiv0.0$ & $\equiv0.0$ & radians \\
$K_{b}$ & $2.3^{+1.6}_{-1.4}$ & $2.3$ & m s$^{-1}$ \\
$M_b$ & $11^{+7.2}_{-6.7}$ & $10.8$ & $M_\earth$ \\
$R_b/R_*$ & $0.0375^{+0.0004}_{-0.0015}$ & $0.0384$ &  \\
$\rho_b$ & $0.35^{+0.29}_{-0.18}$ & $0.30$ & g cm$^{-3}$ \\
$R_b$ & $\equiv5.4972$ & $\equiv5.4972$ & $R_\earth$ \\
\hline
\hline
\sidehead{\bf{Priors}}
Parameter & Prior \\
\hline
\\None \\
\enddata
\end{deluxetable}

\subsection{K2-418 (EPIC 229004835)} 
\label{sec:epic229004835}

\begin{figure*}
\epsscale{1.0}
\plotone{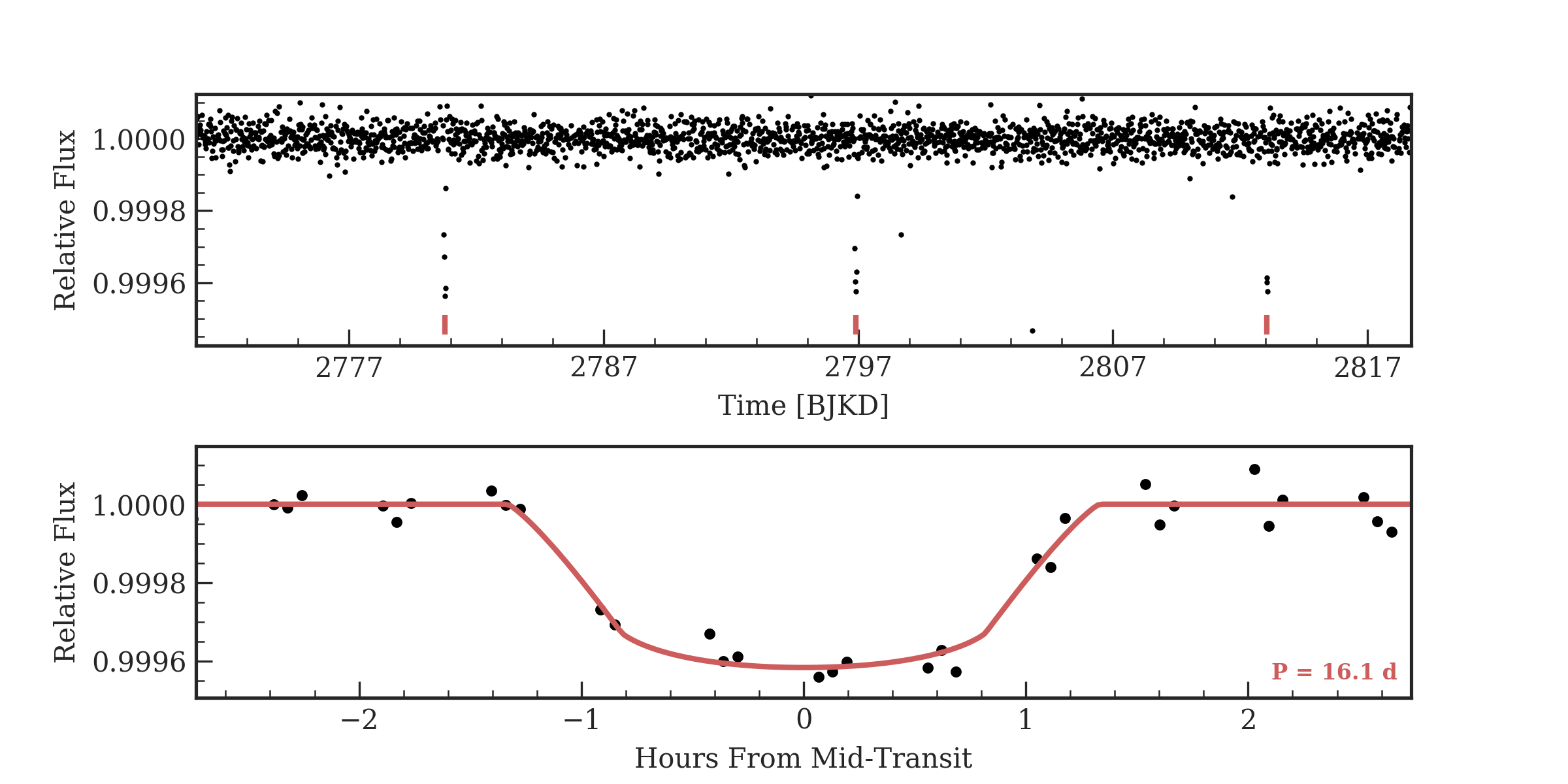}
\caption{Time series (top) and phase-folded (bottom) light curve for the planet orbiting \eidfSTNAME.  Plot formatting is the same as in Fig.\ \ref{fig:lc_epic220709978}.}
\label{fig:lc_epic229004835}
\end{figure*}

\eidfSTNAME {\referee (EPIC-229004835)} is a solar-type star observed in Campaign 10 that has one transiting planet with an orbital period of 16 days and a size of 2.1 \rearth.
The object was identified by our pipeline \citep{Livingston2018} and in \cite{Mayo2018} as a planet candidate.  See Tables \ref{tb:star_pars} and \ref{tb:star_props} for stellar properties and Table \ref{tb:planet_props} for precise planet parameters.    Our fit of the EVEREST light curve of the K2 photometry for \eidfSTNAME is shown in Fig.\ \ref{fig:lc_epic229004835}. Our observations described below are sufficient to validate the planet candidate according to the criteria listed in Section~\ref{sec:validation}.

\begin{figure}
\epsscale{1.0}
\plotone{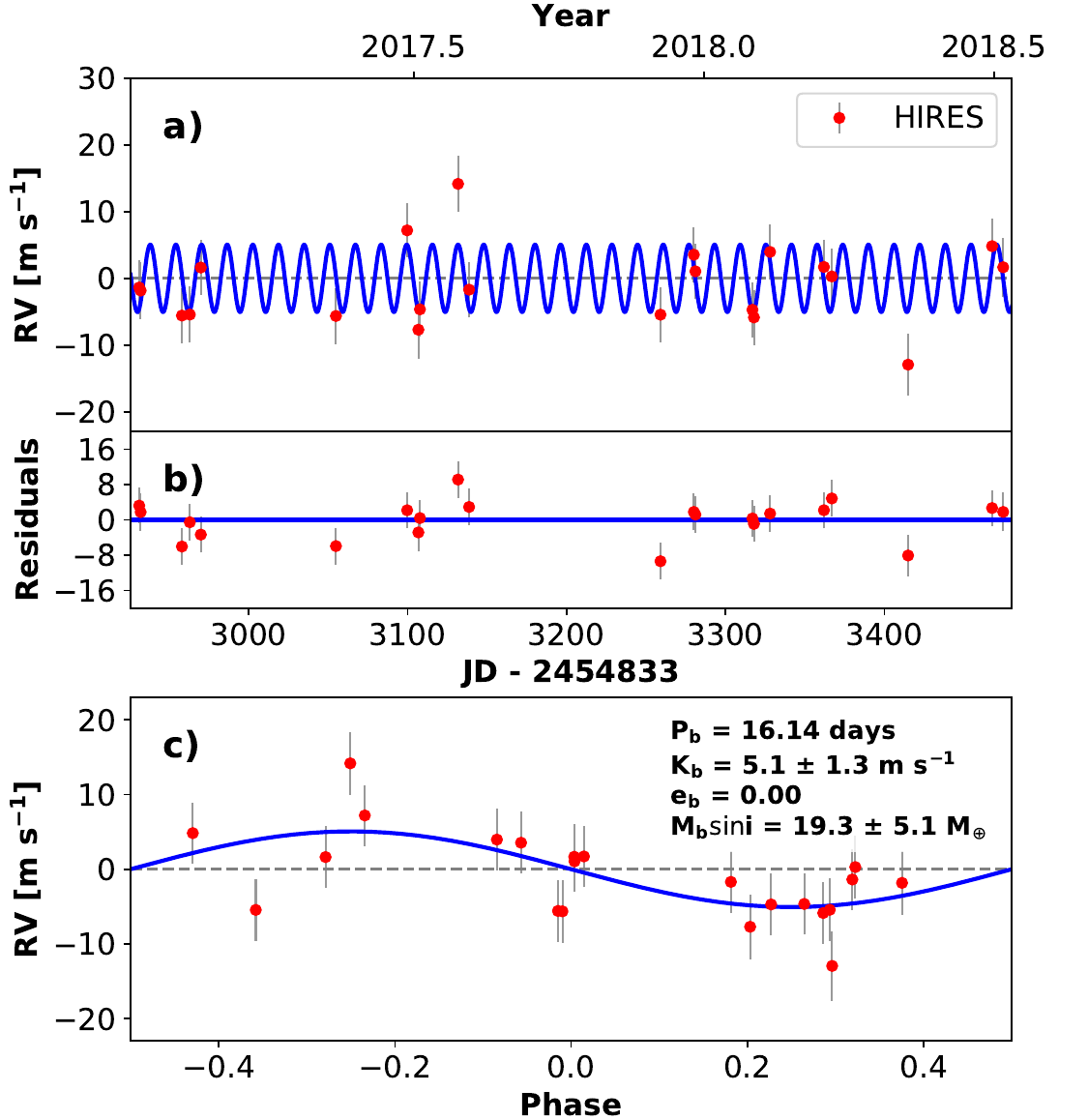}
\caption{RVs and Keplerian model for \eidfSTNAME.  Symbols, lines, and annotations are similar to those in Fig.\ \ref{fig:rvs_epic220709978}.}
\label{fig:rvs_epic229004835}
\end{figure}

We acquired \eidfNOBSHIRES RVs of \eidfSTNAME with HIRES, typically with an exposure meter setting of 250,000.  
We modeled the system as a single planet in a circular orbit, with no additional priors.  
The results of this analysis are listed in Table \ref{tab:epic229004835} and the best fit model is shown in Fig.\ \ref{fig:rvs_epic229004835}. 
We considered more complicated models, but found insufficient evidence to justify inclusion of orbital eccentricity or a linear RV trend based on the AICc statistic. 

The size of {\referee K2-418 b} (\eidfRPone \rearth)  places it on the sub-Neptune side of the radius valleyradius valley \citep{Fulton2017}.  However, the density we measured of \eidfRHOPone \gmc suggests a rocky composition. This may be driven in part by sparse phase coverage and a possible outlier RV at upper quadrature (see Fig.\ \ref{fig:rvs_epic229004835}).  The star has \lrphk = \eidfRPHK so an outlier RV cannot be explained by elevated stellar activity.  Subsequent observations and analysis reveal a mass of $10.4\pm1.5 M_\oplus$ \citep{Bonomo2023}, broadly consistent with, but more precise than our measurement.  



\begin{deluxetable}{lrrr}
\tablecolumns{4}
\tabletypesize{\scriptsize}
\tablecaption{ EPIC 229004835 System Parameters \label{tab:epic229004835}}
\tablehead{
  \colhead{Parameter} & 
  \colhead{Credible} & 
  \colhead{Maximum} & 
  \colhead{Units} \\
  \colhead{} & 
  \colhead{Interval} & 
  \colhead{Likelihood} & 
  \colhead{} }
\startdata
\sidehead{\bf{RV Analysis -- MCMC Step Parameters}} 
$P_{b}$ & $\equiv16.1388$ & $\equiv16.1388$ & days \\ 
$T_{\mathrm{conj},b}$ & $\equiv2457613.7661$ & $\equiv2457613.7661$ & BJD$_\mathrm{TBD}$ \\ 
$\sqrt{e}\cos{\omega}_{b}$ & $\equiv0.0$ & $\equiv0.0$ &  \\ 
$\sqrt{e}\sin{\omega}_{b}$ & $\equiv0.0$ & $\equiv0.0$ &  \\ 
$K_{b}$ & $5.1^{+1.3}_{-1.4}$ & $5.1$ & m s$^{-1}$ \\ 
$\gamma_{\rm HIRES}$ & $-1\pm1.0$ & $-1.0$ & m s$-1$ \\ 
$\dot{\gamma}$ & $\equiv0.0$ & $\equiv0.0$ & m s$^{-1}$ day$^{-1}$ \\ 
$\ddot{\gamma}$ & $\equiv0.0$ & $\equiv0.0$ & m s$^{-1}$ day$^{-2}$ \\ 
$\sigma_{\rm HIRES}$ & $4.27^{+1.30}_{-0.30}$ & $3.80$ & $\rm m\ s^{-1}$ \\ 
\hline
\sidehead{\bf{Orbital \& Physical Parameters}}
$P_{b}$ & $\equiv16.1388$ & $\equiv16.1388$ & days \\
$T_{\mathrm{conj},b}$ & $\equiv2457613.7661$ & $\equiv2457613.7661$ & BJD$_\mathrm{TBD}$ \\
$e_{b}$ & $\equiv0.0$ & $\equiv0.0$ &  \\
$\omega_{b}$ & $\equiv0.0$ & $\equiv0.0$ & radians \\
$K_{b}$ & $5.1^{+1.3}_{-1.4}$ & $5.1$ & m s$^{-1}$ \\
$M_b$ & $19.3\pm5.0$ & $19.0$ & $M_\earth$ \\
$R_b/R_*$ & $0.0190^{+0.0016}_{-0.0002}$ & $0.0185$ &  \\
$\rho_b$ & $9.9^{+0.1}_{-6.1}$ & $13.0$ & g cm$^{-3}$ \\
$R_b$ & $\equiv2.1832$ & $\equiv2.1832$ & $R_\earth$ \\
\hline
\hline
\sidehead{\bf{Priors}}
Parameter & Prior \\
\hline
\\None \\
\enddata
\end{deluxetable}

\subsection{K2-277} 
\label{sec:k2_277}

\begin{figure*}
\epsscale{1.0}
\plotone{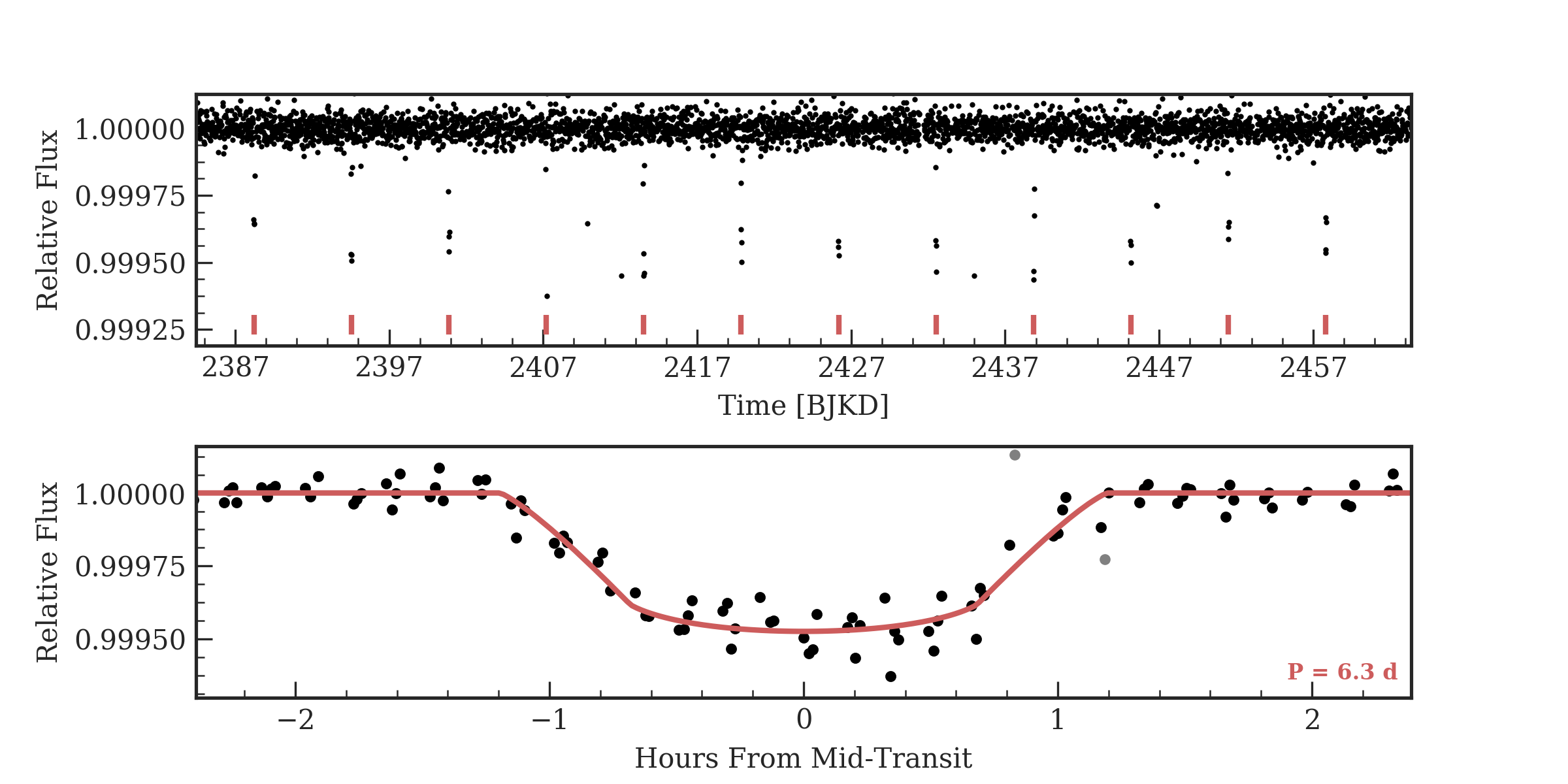}
\caption{Time series (top) and phase-folded (bottom) light curve for the planet orbiting \hehhSTNAME.  Plot formatting is the same as in Fig.\ \ref{fig:lc_epic220709978}.}
\label{fig:lc_epic212357477}
\end{figure*}

\hehhSTNAME is a solar-type star from Campaign 6 with one transiting planet with a radius of 2.1 \rearth and an orbital period of 6.3 days.
The planet was listed as a candidate in \cite{Petigura2018} and was subsequently validated by \cite{Livingston2018}.  It was not listed in \cite{Mayo2018} even though Campaign 6 was covered in that catalog.
See Tables \ref{tb:star_pars}  and \ref{tb:star_props} for stellar properties and Table \ref{tb:planet_props} for precise planet parameters.  Our fit of the EVEREST light curve of the K2 photometry for \hehhSTNAME is shown in Fig.\ \ref{fig:lc_epic212357477}.

\begin{figure}
\epsscale{1.0}
\plotone{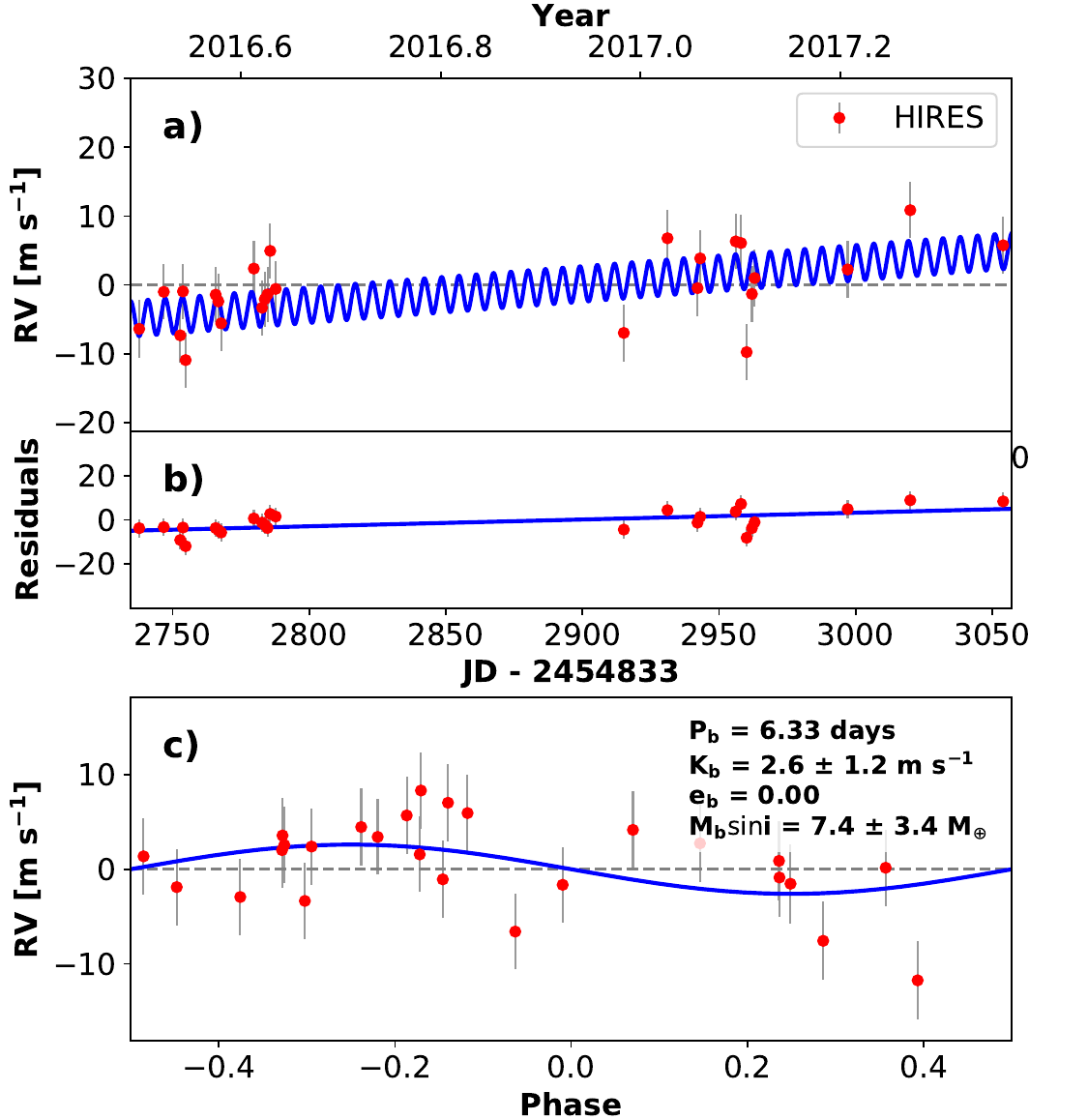}
\caption{RVs and Keplerian model for \hehhSTNAME.  Symbols, lines, and annotations are similar to those in Fig.\ \ref{fig:rvs_epic220709978}.}
\label{fig:rvs_epic212357477}
\end{figure}

We acquired \hehhNOBSHIRES 
RVs of \hehhSTNAME with HIRES, typically with an exposure meter setting of 250,000 counts.  
We modeled the system as a single planet in a circular orbit with the orbital period and phase fixed to the values from the transit ephemeris.  The results of this analysis are listed in Table \ref{tab:epic212357477} and the best fit model is shown in Fig. \ref{fig:rvs_epic212357477}. 
Note that our model includes a linear RV trend that is strongly favored compared to a model without a trend based on the AICc statistic.  We rejected models with an eccentric orbit based on similar AICc comparisons.
\hehhPNAMEone is detected with only $\sim$2-$\sigma$ significance based on the HIRES RVs. As a result, this sub-Neptune's bulk density of \hehhRHOPone \gmc is consistent with a rocky composition or a significant gas envelope.

\begin{deluxetable}{lrrr}
\tablecolumns{4}
\tabletypesize{\scriptsize}
\tablecaption{ K2-277 System Parameters \label{tab:epic212357477}}
\tablehead{
  \colhead{Parameter} & 
  \colhead{Credible} & 
  \colhead{Maximum} & 
  \colhead{Units} \\
  \colhead{} & 
  \colhead{Interval} & 
  \colhead{Likelihood} & 
  \colhead{} }
\startdata
\sidehead{\bf{RV Analysis -- MCMC Step Parameters}} 
$P_{b}$ & $\equiv6.3268$ & $\equiv6.3268$ & days \\ 
$T_{\mathrm{conj},b}$ & $\equiv2457221.2301$ & $\equiv2457221.2301$ & BJD$_\mathrm{TBD}$ \\ 
$\sqrt{e}\cos{\omega}_{b}$ & $\equiv0.0$ & $\equiv0.0$ &  \\ 
$\sqrt{e}\sin{\omega}_{b}$ & $\equiv0.0$ & $\equiv0.0$ &  \\ 
$K_{b}$ & $2.6\pm1.2$ & $2.6$ & m s$^{-1}$ \\ 
$\gamma_{\rm HIRES}$ & $-1.57^{+0.95}_{-0.97}$ & $-1.56$ & m s$-1$ \\ 
$\dot{\gamma}$ & $0.0308^{+0.0088}_{-0.0087}$ & $0.0307$ & m s$^{-1}$ day$^{-1}$ \\ 
$\ddot{\gamma}$ & $\equiv0.0$ & $\equiv0.0$ & m s$^{-1}$ day$^{-2}$ \\ 
$\sigma_{\rm HIRES}$ & $4.24^{+1.20}_{-0.20}$ & $3.80$ & $\rm m\ s^{-1}$ \\ 
\hline
\sidehead{\bf{Orbital \& Physical Parameters}}
$P_{b}$ & $\equiv6.3268$ & $\equiv6.3268$ & days \\
$T_{\mathrm{conj},b}$ & $\equiv2457221.2301$ & $\equiv2457221.2301$ & BJD$_\mathrm{TBD}$ \\
$e_{b}$ & $\equiv0.0$ & $\equiv0.0$ &  \\
$\omega_{b}$ & $\equiv0.0$ & $\equiv0.0$ & radians \\
$K_{b}$ & $2.6\pm1.2$ & $2.6$ & m s$^{-1}$ \\
$M_b$ & $7.4^{+3.4}_{-3.3}$ & $7.4$ & $M_\earth$ \\
$R_b/R_*$ & $0.0202^{+0.0017}_{-0.0001}$ & $0.0198$ &  \\
$\rho_b$ & $3.5^{+0.5}_{-2.8}$ & $4.7$ & g cm$^{-3}$ \\
$R_b$ & $\equiv2.2324$ & $\equiv2.2324$ & $R_\earth$ \\
\hline
\hline
\sidehead{\bf{Priors}}
Parameter & Prior \\
\hline
\\None \\
\enddata
\tablecomments{}{
Reference epoch for $\dot{\gamma}$,$\ddot{\gamma}$: 2457728.9046480004 
}
\end{deluxetable}

\subsection{GJ 9827} 
\label{sec:GJ9827}
GJ9827 is a relatively bright ($V$ = 10.4) K6V star in Campaign 12.  
\cite{Niraula2017} and \cite{Rodriguez2018} discovered three transiting planets with sizes and orbital periods of 1.8, 1.4, and 2.1 \rearth, and 1.21, 3.65, and 6.20 days, respectively.  

Over the last few years, this system has had considerable follow-up. \cite{Teske2018} measured the mass of planet b ($M_b\approx 8$ \mearth) and the upper limits on planets c and d ($M_c<2.5$ \mearth, $M_d<$5.6 \mearth) based on RVs from PFS. \cite{PrietoArranz2018} added RVs from HARPS and HARPS-N to determine the masses of all three planets ($M_b=3.74\pm0.50$ \mearth, $M_c=1.47 \pm 0.59$ \mearth, and $M_d=2.38\pm0.71$ \mearth. \cite{Rice2019} refined these measurements with additional HARPS-N data and a Gaussian process fit informed from the K2 light curve ($M_b=4.91\pm0.49$ \mearth and $M_d =4.04 \pm 0.84$ \mearth). \cite{Passegger2024} recently reported an analysis also consistent with these previous studies.

\citet{Kosiarek2020b} followed with 92 HIRES measurements and performed an analysis that included all previously published data. They additionally updated the stellar parameters from Gaia DR2 information and propagated these to update the planet radii. We adopt their solution; see see Tables \ref{tb:star_pars} and \ref{tb:star_props} for stellar properties and Table \ref{tb:planet_props} for the planet parameters. They find a circular 3-planet fit that includes a Gaussian process informed from the $S_{\rm HK}$ best fits the data based on the AICc statistic, resulting in planet masses of $M_b=4.87\pm0.37$ \mearth, $M_c=1.92\pm0.49$ \mearth, and $M_d=3.42\pm0.62$ \mearth. \jifiSTNAME\ b and c are both rocky super-Earths consistent with an Earth-like composition. \jifiSTNAME\ d is a super-Earth that requires a small amount of volatiles to explain its radius.  This volatile envelope was confirmed by the detection of H$_2$O in planet d's atmosphere through {\em HST}/WFC3 transmission spectroscopy \citep{Roy2023}.
\subsection{K2-261} 




\iahiSTNAME is a G7 star in Field 14 with one transiting planet with a radius of 9.7\,\rearth and an orbital period of 11.6 days.  The planet was discovered and characterized independently by \citet{Johnson2018} and \citet{Brahm2019}. \cite{Johnson2018} found K2-261b to be a warm Saturn with a mass of $70.9\pm9.9$\,\mearth in an eccentric orbit ($e=0.39\pm0.15$) based on RVs from FIES, HARPS-N, and HARPS. \cite{Brahm2019} found a lower mass for this eccentric planet ($M_b=56.9\pm6.7$\,\mearth, $e=0.42\pm0.03$) based on RVs from Coralie, FEROS, and HARPS.

\begin{figure*}
\epsscale{1.0}
\plotone{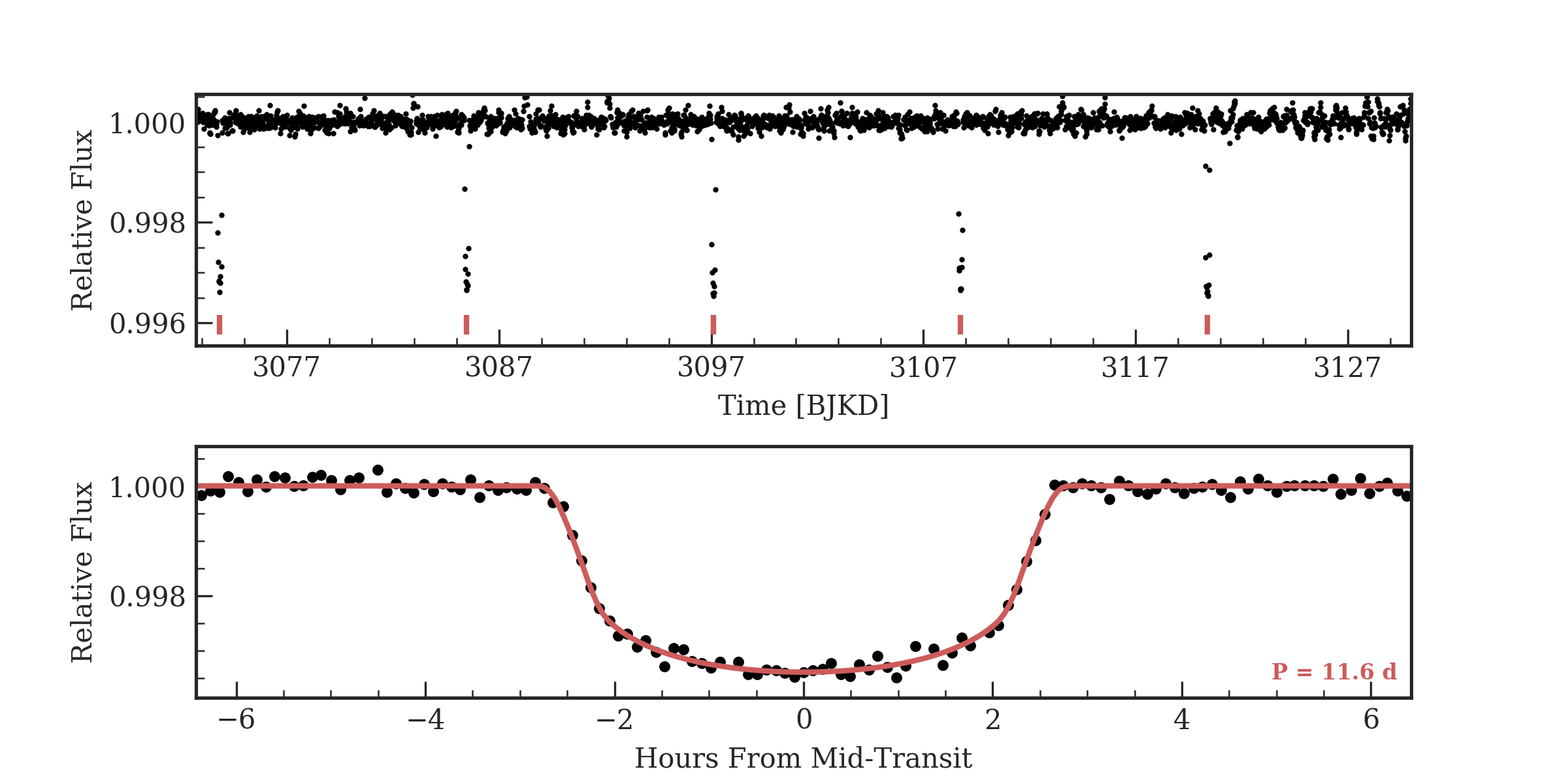}
\caption{Time series (top) and phase-folded (bottom) light curve for the planet orbiting \iahiSTNAME.  Plot formatting is the same as in Fig.\ \ref{fig:lc_epic220709978}.}
\label{fig:lc_epic201498078}
\end{figure*}

\begin{figure}
\epsscale{1.0}
\plotone{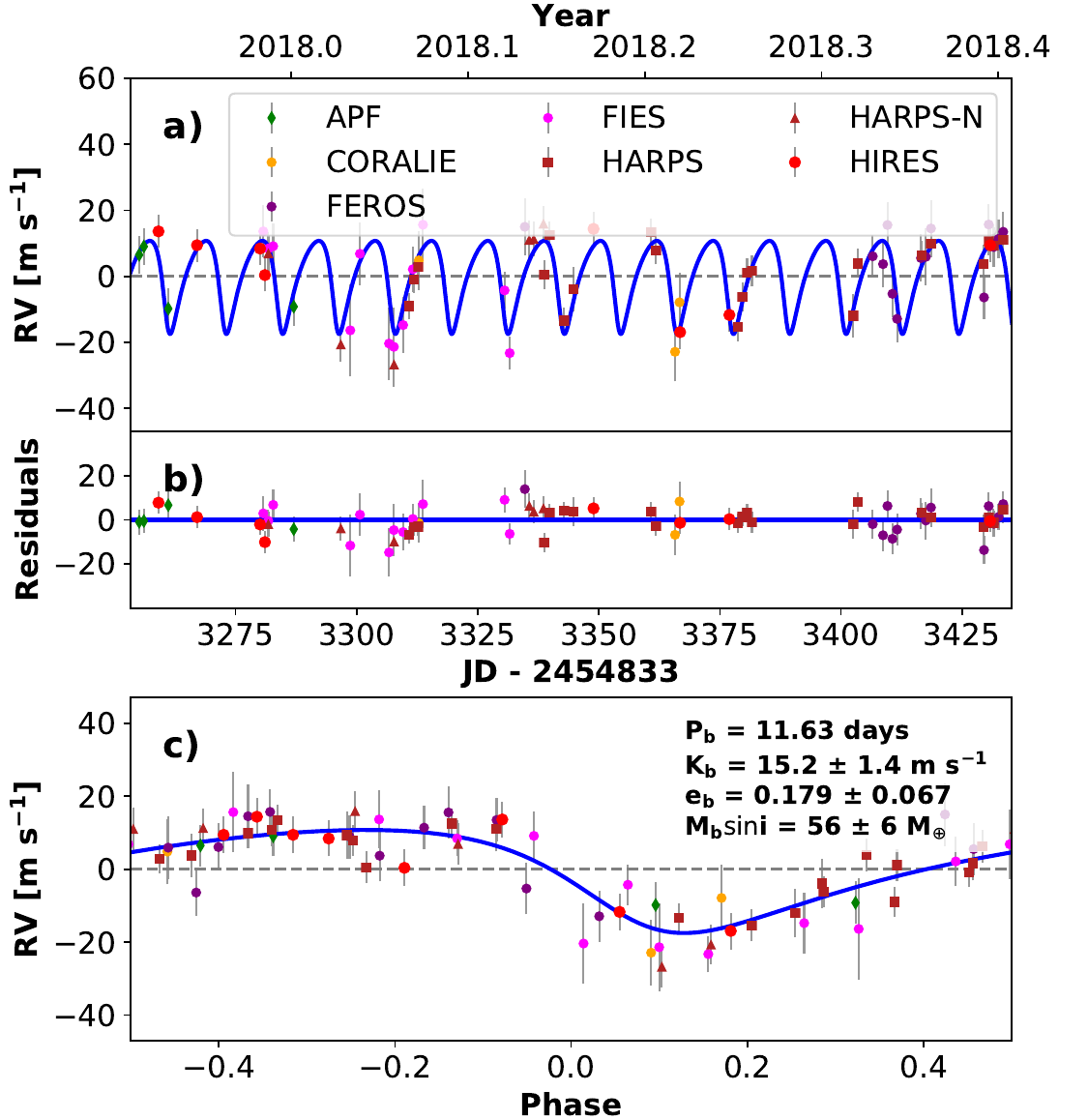}
\caption{RVs and Keplerian model for \iahiSTNAME.  Symbols, lines, and annotations are similar to those in Fig.\ \ref{fig:rvs_epic220709978}.}
\label{fig:rvs_epic201498078}
\end{figure}

We acquired \iahiNOBSHIRES RVs of \iahiSTNAME with HIRES, typically with an exposure meter setting of 50,000, and \iahiNOBSAPF RVs with the APF.  
We fit all of the data with a model of a single planet in an eccentric orbit with the orbital period and phase fixed to the transit ephemeris.  We find a mass of \iahiMPone\,\mearth, which is consistent with previous results, and a lower eccentricity of $e = 0.179 \pm 0.067$ based on the global analysis.  An eccentric orbit is clearly favored over a circular one in a model comparison test ($\Delta$AICc = 24.24), and a linear trend is not warranted based on an AICc comparison.  The results of this analysis are listed in Table 
\ref{tab:epic201498078} and the best fit model is shown in Fig.\ \ref{fig:rvs_epic201498078}.  \iahiPNAMEone is a low-density warm-Saturn planet.
See Tables \ref{tb:star_pars} and \ref{tb:star_props} for stellar properties and Table \ref{tb:planet_props} for precise planet parameters.

\begin{deluxetable}{lrrr}
\tablecolumns{4}
\tabletypesize{\scriptsize}
\tablecaption{ K2-261 System Parameters \label{tab:epic201498078}}
\tablehead{
  \colhead{Parameter} & 
  \colhead{Credible} & 
  \colhead{Maximum} & 
  \colhead{Units} \\
  \colhead{} & 
  \colhead{Interval} & 
  \colhead{Likelihood} & 
  \colhead{} }
\startdata
\sidehead{\bf{RV Analysis -- MCMC Step Parameters}} 
$P_{b}$ & $\equiv11.634$ & $\equiv11.634$ & days \\ 
$T_{\mathrm{conj},b}$ & $\equiv2457906.8406$ & $\equiv2457906.8406$ & BJD$_\mathrm{TBD}$ \\ 
$\sqrt{e}\cos{\omega}_{b}$ & $-0.33^{+0.11}_{-0.08}$ & $-0.34$ &  \\ 
$\sqrt{e}\sin{\omega}_{b}$ & $0.22^{+0.17}_{-0.23}$ & $0.30$ &  \\ 
$K_{b}$ & $15.2^{+1.6}_{-1.5}$ & $14.7$ & m s$^{-1}$ \\ 
$\gamma_{\rm HIRES}$ & $-7.0^{+2.6}_{-2.7}$ & $-7.0$ & m s$-1$ \\ 
$\gamma_{\rm HARPS-N}$ & $3334\pm4$ & $3333.5$ &  \\ 
$\gamma_{\rm HARPS}$ & $3341.5^{+1.4}_{-1.5}$ & $3341.6$ & m s$-1$ \\
\enddata
\end{deluxetable}

\subsection{K2-100} 



\aiggSTNAME is a relatively young dwarf star with \teff = \aiggTEFF K.  Evidence for youth includes elevated stellar activity (\lrphk = \aiggRPHK), membership in the Praesepe cluster \citep[e.g.,][]{Kraus2007} that has an age of $\sim$600 Myr, $\sim$1\%  spot modulation in K2 photometry with a rotational period of 4.3 days \citep{Mann2017a}, and \vsini = \aiggVSINI \kms.
The transiting planet has an orbital period of 1.7 days and a radius of 3.7 \rearth, apparently inflated for the short orbital period.  
See Tables \ref{tb:star_pars}  and \ref{tb:star_props} for stellar properties and Table \ref{tb:planet_props} for precise planet parameters.

\begin{figure*}
\epsscale{1.0}
\plotone{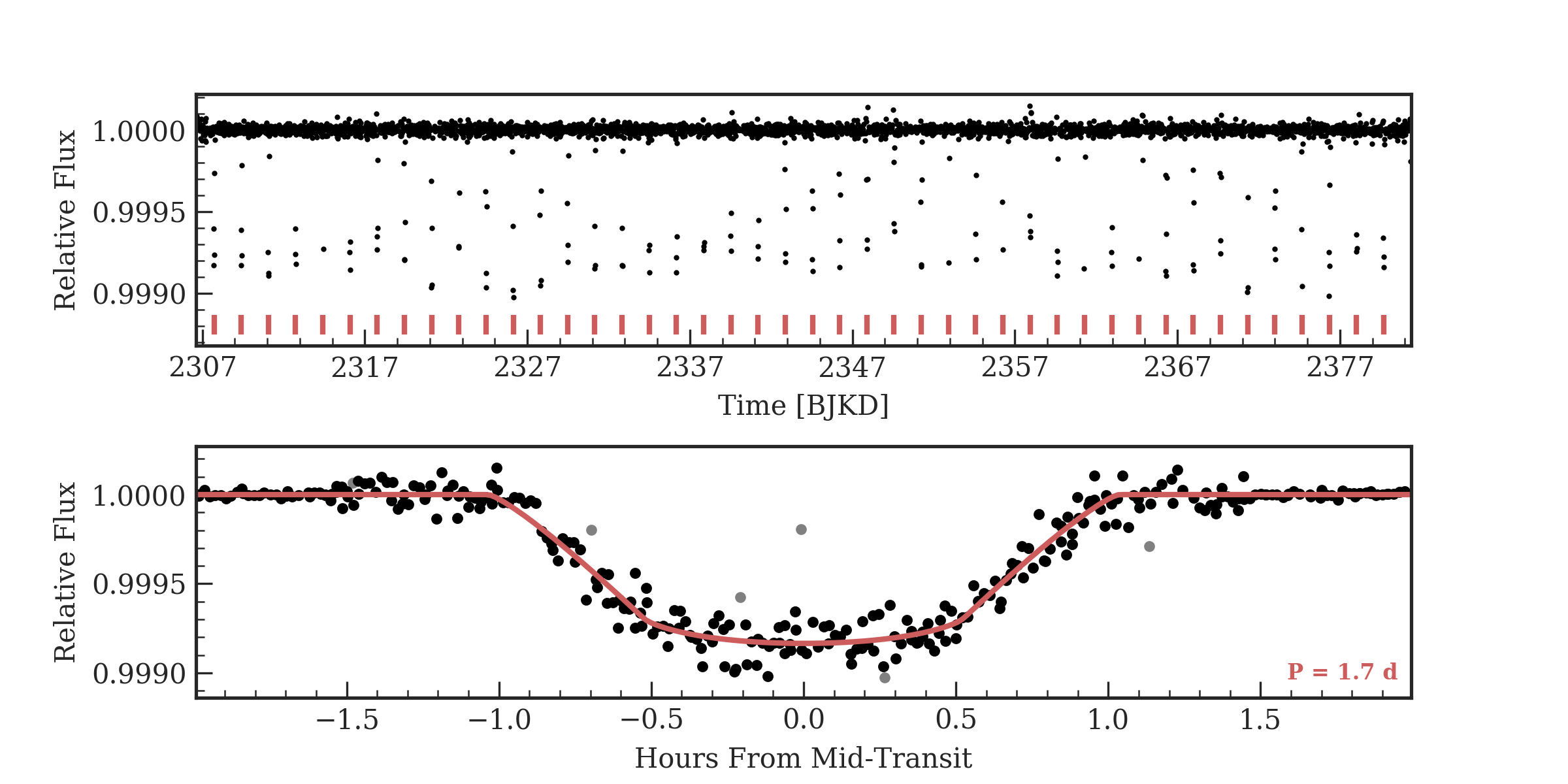}
\caption{Time series (top) and phase-folded (bottom) light curve for the planet orbiting \aiggSTNAME.  Plot formatting is the same as in Fig.\ \ref{fig:lc_epic220709978}.}
\label{fig:lc_epic211990866}
\end{figure*}

\aiggPNAMEone was noted the catalog of seven planets in Praesepe by \cite{Mann2017a}, who also validated the planet with a false positive analysis.  The system was noted in the uniform search for planets in clusters by \cite{Rizzuto2017}.  \cite{Pope2016} and \cite{Petigura2018} list \aiggPNAMEone as a planet candidate, although these papers did not attempt statistical validation.  Our fit of the EVEREST light curve of the K2 photometry for \aiggSTNAME is shown in Fig.\ \ref{fig:lc_epic211990866}.

\cite{Barragan2019} gathered 78 RVs of this star using HARPS.  Their analysis of the RVs using a Gaussian process model trained on activity indices gives a planet mass of $10.6 \pm 3.0$ \mearth\ and a density of $2.04^{+0.66}_{-0.61}$\,\gmc.

We acquired \aiggNOBSHIRES RVs of \aiggSTNAME with HIRES, typically with an exposure meter setting of 125,000.  Due to the high \vsini, we collected three exposures per night separated by an hour or more each (when possible).  We also adopted a strategy of observing the star on consecutive nights that partially freezes out the stellar activity on rotational timescales, giving greater sensitivity to the planet's signal on orbital timescales.

\begin{figure}
\epsscale{1.0}
\plotone{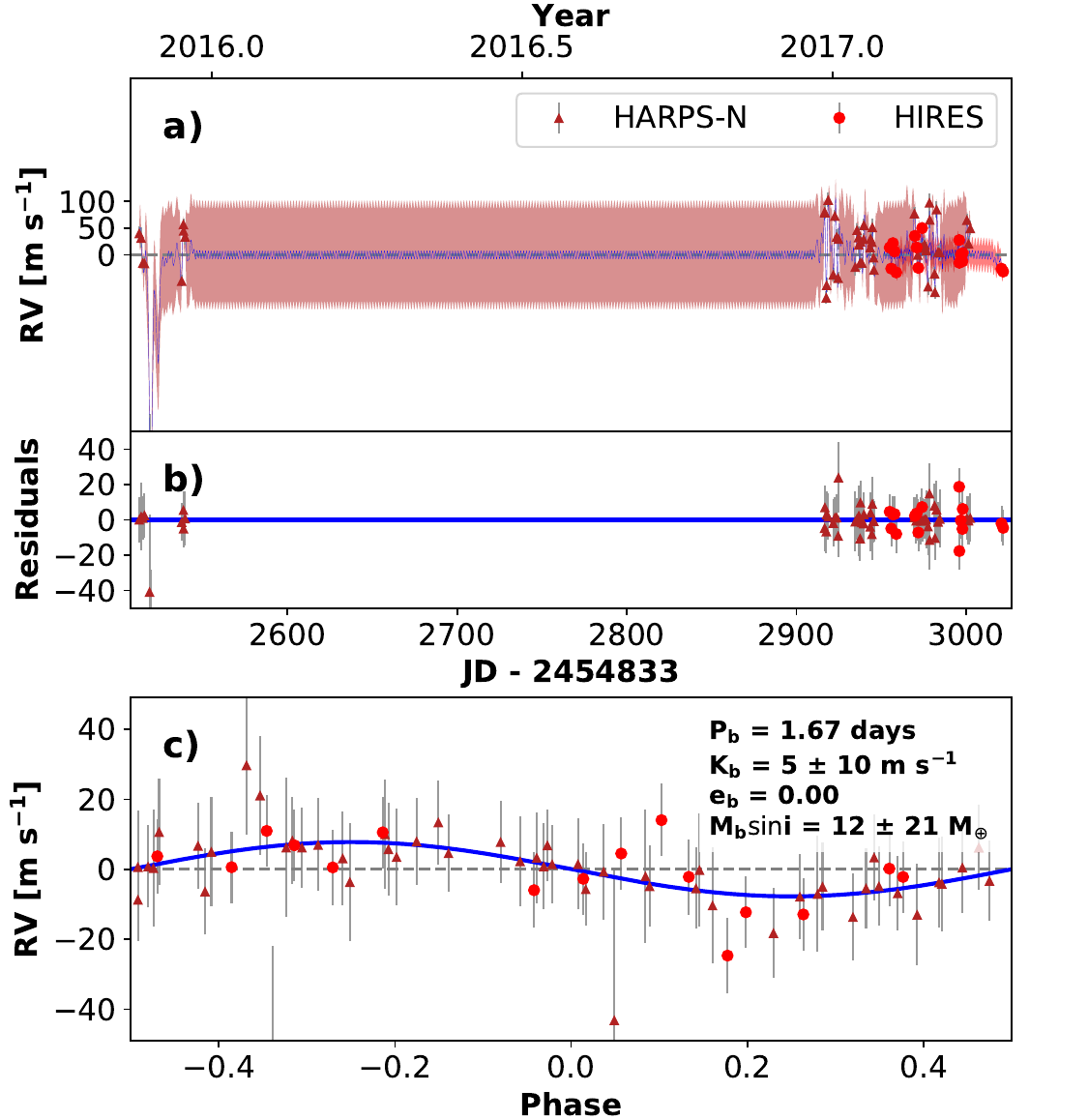}
\caption{RVs and Keplerian model for \aiggSTNAME.  Symbols, lines, and annotations are similar to those in Fig.\ \ref{fig:rvs_epic220709978}.}
\label{fig:rvs_epic211990866}
\end{figure}

We modeled the system as a single planet in a circular orbit with the orbital period and phase fixed to the transit ephemeris.  We included our HIRES RVs and the HARPS measurements from \cite{Barragan2019}.  HIRES RVs were binned within one night.  The results are listed in Table \ref{tab:epic211990866} and the best-fit model is shown in Fig. \ref{fig:rvs_epic211990866}.  We considered more complicated models with free eccentricity and/or a linear RV trend, but rejected those based on model comparison using the AICc statistic.  \aiggSTNAME is a good candidate for a Gaussian process model due to the high activity (\lrphk = \aiggRPHK). Combining our HIRES RVs with observations from HARPS-N satisfies our criteria for attempting a Gaussian process model. However, the photometry for this system is not particularly constraining, and hence we adopted an untrained GP.

The high stellar flux received by \aiggPNAMEone (\aiggFLUXone \fearth), combined with its approximately Neptune size and relative youth, suggests that it is inflated.  We do not detect the Doppler signal from \aiggPNAMEone, likely due to the high stellar jitter (see Table \ref{tab:epic211990866}), but we rule out a high density for this planet. Our constraint on the Doppler semiamplitude is consistent with the value $K_b = 10.6 \pm 3.0$ \ms\ obtained by \cite{Barragan2019}, who used a GP trained on activity indices to obtain a detection.  We elected not to train the GP on activity indices for consistency with our analysis of other systems in this paper. \aiggSTNAME may be a good target for IR Doppler spectrometers that may be less affected by spot-related activity. Such measurements would still need to contend with the reduced information content of stellar spectra given the elevated \vsini.





\begin{deluxetable}{lrrr}
\tablecolumns{4}
\tabletypesize{\scriptsize}
\tablecaption{ K2-100 System Parameters \label{tab:epic211990866}}
\tablehead{
  \colhead{Parameter} & 
  \colhead{Credible} & 
  \colhead{Maximum} & 
  \colhead{Units} \\
  \colhead{} & 
  \colhead{Interval} & 
  \colhead{Likelihood} & 
  \colhead{} }
\startdata
\sidehead{\bf{RV Analysis -- MCMC Step Parameters}} 
$P_{b}$ & $\equiv1.6739$ & $\equiv1.6739$ & days \\ 
$T_{\mathrm{conj},b}$ & $\equiv2457140.7197$ & $\equiv2457140.7197$ & BJD$_\mathrm{TBD}$ \\ 
$\sqrt{e}\cos{\omega}_{b}$ & $\equiv0.0$ & $\equiv0.0$ &  \\ 
$\sqrt{e}\sin{\omega}_{b}$ & $\equiv0.0$ & $\equiv0.0$ &  \\ 
$K_{b}$ & $3.7\pm7.3$ & $8.0$ & m s$^{-1}$ \\ 
$\gamma_{\rm HIRES}$ & $-1^{+20}_{-22}$ & $-1$ & m s$-1$ \\ 
$\gamma_{\rm HARPS-N}$ & $34342^{+65}_{-69}$ & $34380$ & m s$-1$ \\ 
$\dot{\gamma}$ & $\equiv0.0$ & $\equiv0.0$ & m s$^{-1}$ day$^{-1}$ \\
\enddata
\end{deluxetable}

\subsection{K2-31} 

\begin{figure*}
\epsscale{1.0}
\plotone{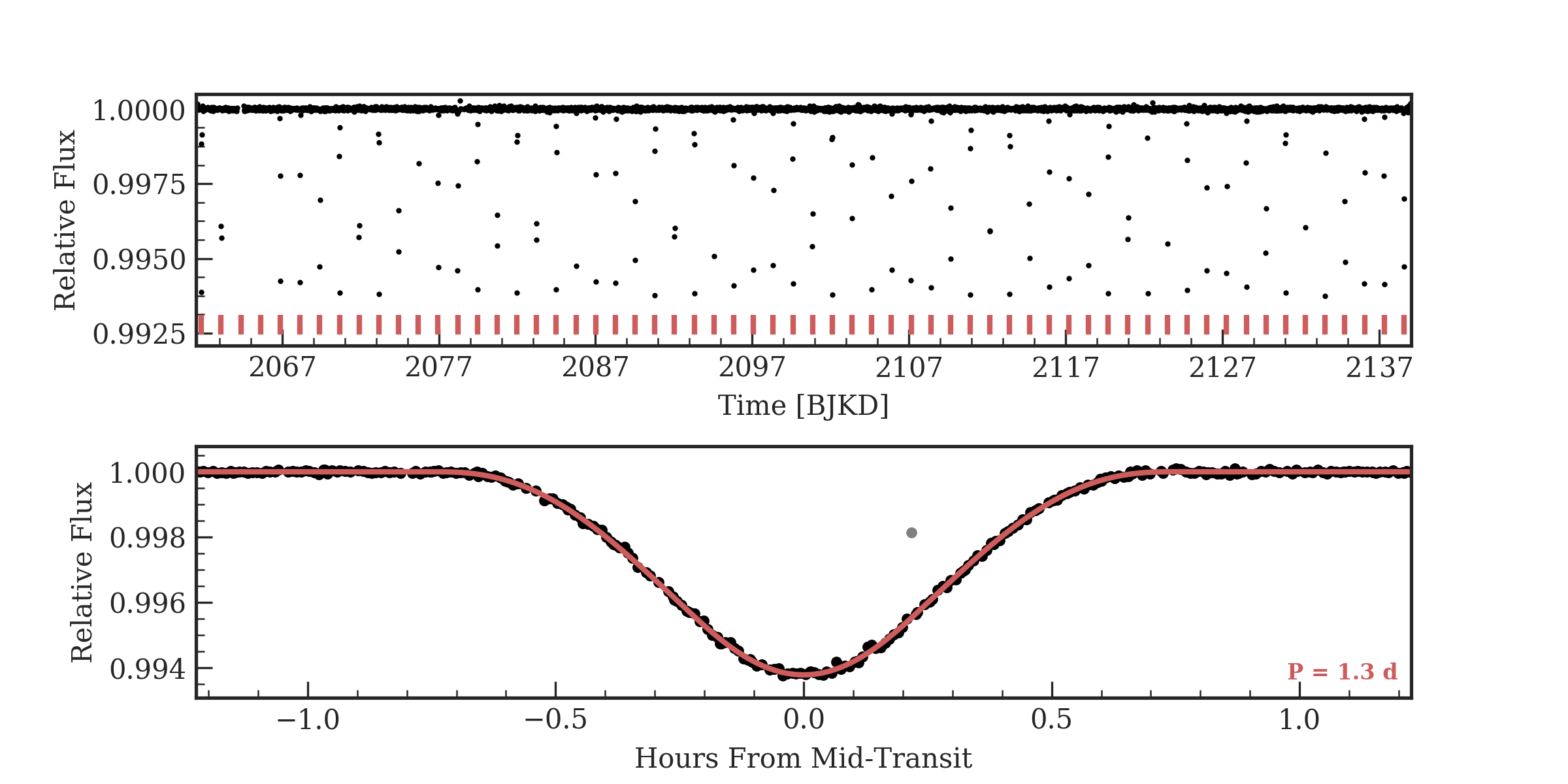}
\caption{Time series (top) and phase-folded (bottom) light curve for the planet orbiting \jgjjSTNAME.  Plot formatting is the same as in Fig.\ \ref{fig:lc_epic220709978}.}
\label{fig:lc_epic204129699}
\end{figure*}

\jgjjSTNAME is a late G dwarf with a close-in, transiting sub-Saturn size planet (\Rp = \jgjjRPone \rearth, $P$ = 1.3 days).
See Tables \ref{tb:star_pars}  and \ref{tb:star_props} for stellar properties and Table \ref{tb:planet_props} for precise planet parameters determined by our analysis.
The planet was noted the catalogs of \cite{Barros2016}, \cite{Crossfield2016}, \cite{Vanderburg2016-catalog}, and  \cite{Schmitt2016}, and has been validated.  Our fit of the EVEREST light curve of the K2 photometry for \jgjjSTNAME is shown in Fig.\ \ref{fig:lc_epic204129699}.  The V shape of the transit is due to the high impact parameter (grazing transit) and the short K2 sampling rate (30 min) compared to the transit duration (1 hr).


\cite{Grziwa2016} characterized the system using 6 FIES RVs and 3 HARPS RVs and measured a planet mass of $563.8 \pm 25.1$ \mearth.  Due to the grazing transit ($b$ = 0.90--1.05), its measured \Rp =  8.0--15.7 \rearth is highly uncertain. 
\cite{Dai2016} added seven RVs from PFS RVs and 10 from TRES.  They modeled all available RVs and found a mass of $564.2^{+9.5}_{-8.3}$ \mearth with $e < 0.027$ (95\% confidence).

\begin{figure}
\epsscale{1.0}
\plotone{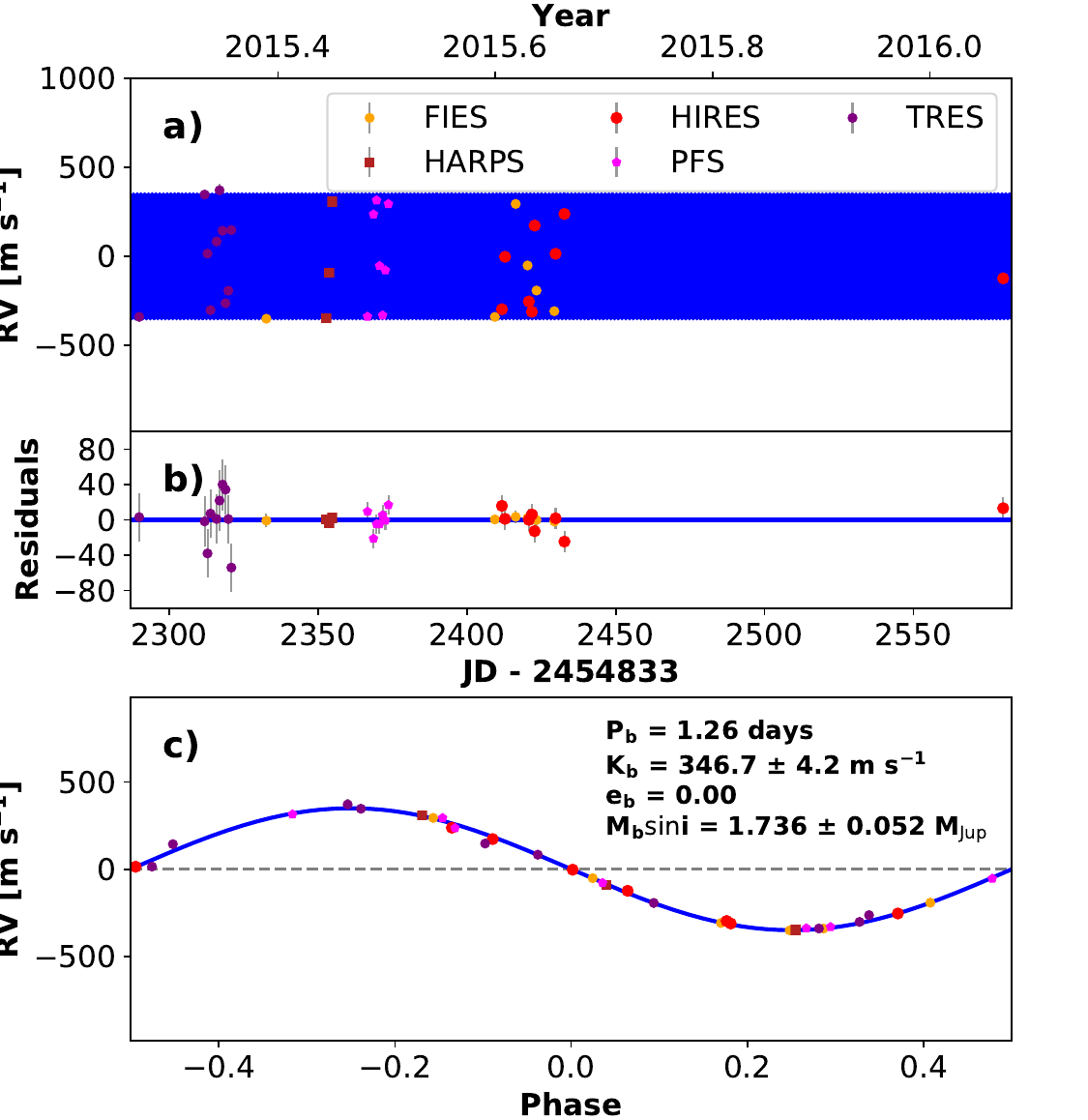}
\caption{RVs and Keplerian model for \jgjjSTNAME.  Symbols, lines, and annotations are similar to those in Fig.\ \ref{fig:rvs_epic220709978}.}
\label{fig:rvs_k2-31}
\end{figure}

We acquired \jgjjNOBSHIRES RVs of \jgjjSTNAME with HIRES, typically with an exposure meter setting of 50,000 counts.  The star is outside of our target sample and was mostly observed in poor conditions because it is bright and the Doppler signal was expected to be large.  
The star has somewhat elevated stellar activity (\lrphk = \jgjjRPHK).
We modeled the system as a single planet in a circular orbit with the orbital period and phase fixed to the transit ephemeris, without additional priors. 
The results are listed in Table \ref{tab:epic204129699} and the best-fit model is shown in Fig.\ \ref{fig:rvs_k2-31}. We find a mass of 551$^{+16}_{-17}$ \mearth, consistent with previous studies. 
We considered more complicated models but found insufficient evidence to justify inclusion of orbital eccentricity or a linear RV trend based on the AICc statistic. 


\begin{deluxetable}{lrrr}
\tablecolumns{4}
\tabletypesize{\scriptsize}
\tablecaption{ K2-31 System Parameters \label{tab:epic204129699}}
\tablehead{
  \colhead{Parameter} & 
  \colhead{Credible} & 
  \colhead{Maximum} & 
  \colhead{Units} \\
  \colhead{} & 
  \colhead{Interval} & 
  \colhead{Likelihood} & 
  \colhead{} }
\startdata
\sidehead{\bf{RV Analysis -- MCMC Step Parameters}} 
$P_{b}$ & $\equiv1.2578$ & $\equiv1.2578$ & days \\ 
$T_{\mathrm{conj},b}$ & $\equiv2456893.599$ & $\equiv2456893.599$ & BJD$_\mathrm{TBD}$ \\ 
$\sqrt{e}\cos{\omega}_{b}$ & $\equiv0.0$ & $\equiv0.0$ &  \\ 
$\sqrt{e}\sin{\omega}_{b}$ & $\equiv0.0$ & $\equiv0.0$ &  \\ 
$K_{b}$ & $346.7^{+3.7}_{-4.8}$ & $349.7$ & m s$^{-1}$ \\ 
$\gamma_{\rm TRES}$ & $-147\pm10$ & $-147$ & m s$-1$ \\ 
$\gamma_{\rm PFS}$ & $53.4^{+5.5}_{-5.7}$ & $53.6$ & m s$-1$ \\ 
$\gamma_{\rm HIRES}$ & $69\pm5$ & $69.9$ & m s$-1$ \\
\enddata
\end{deluxetable}

\subsection{K2-39} 


\hhedSTNAME is a metal-rich (\feh = \hhedFEH dex) K-type subgiant with one transiting planet with a radius of 6.5 \rearth and an orbital period of 4.6 days.
The planet is validated and appears in the \cite{Crossfield2016}, \cite{Vanderburg2016-catalog}, and \cite{Schmitt2016} catalogs.  
See Tables \ref{tb:star_pars}  and \ref{tb:star_props} for stellar properties and Table \ref{tb:planet_props} for precise planet parameters.  

\begin{figure*}
\epsscale{1.0}
\plotone{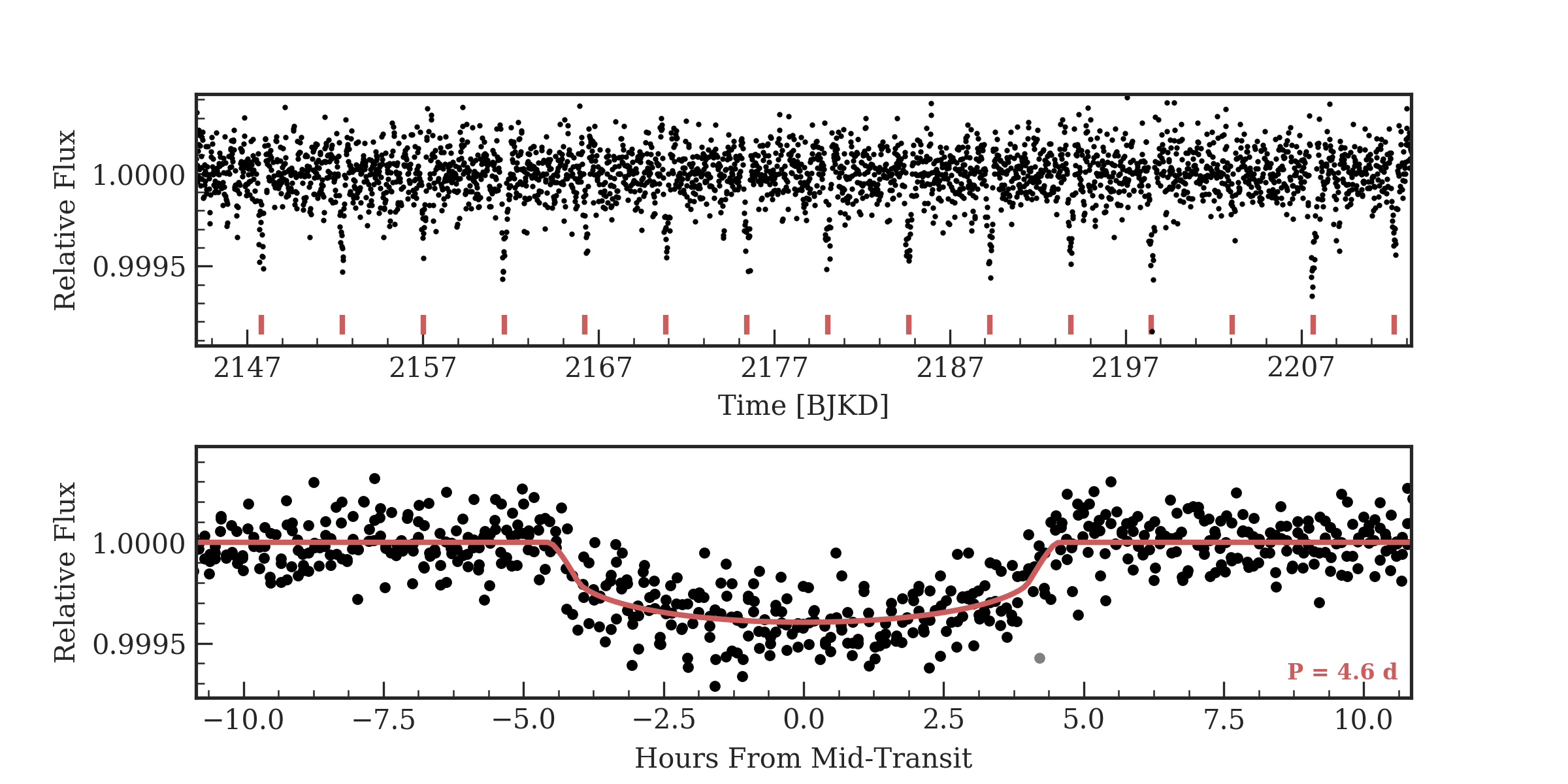}
\caption{Time series (top) and phase-folded (bottom) light curve for the planet orbiting \hhedSTNAME.  Plot formatting is the same as in Fig.\ \ref{fig:lc_epic220709978}.  Photometry was generated using K2phot, instead of EVEREST as with other stars.}
\label{fig:lc_epic206247743}
\end{figure*}

\cite{vanEylen2016b} measured a mass of $50.3^{+9.7}_{-9.4}$ \mearth for \hhedPNAMEone using RVs from HARPS (7 measurements), FIES (17), and PFS (6).  Their model also includes a long-term RV trend with curvature.  \cite{Petigura2017} also measured a mass of $30.9 \pm 4.6$ \mearth, using data from \cite{vanEylen2016b} and 42 new HIRES RVs. 

We attempted to reduce the data using EVEREST, but found unusually high photometric noise. Instead, we adopted a light curve of the K2 photometry produced using K2phot.  The photometry is shown in Fig.\ \ref{fig:lc_epic206247743}. \cite{Petigura2017} noted the discrepancy in \rprstar between the adopted values from \cite{Crossfield2016} and \cite{vanEylen2016b}, reexamined the photometry using multiple pipelines, and adopted a value of $0.0179 \pm 0.0013$ (closer to the \cite{vanEylen2016b} result). Our adopted value of \rprstar = $0.0180 \pm 0.0008$ is consistent with this.

\begin{figure}
\epsscale{1.0}
\plotone{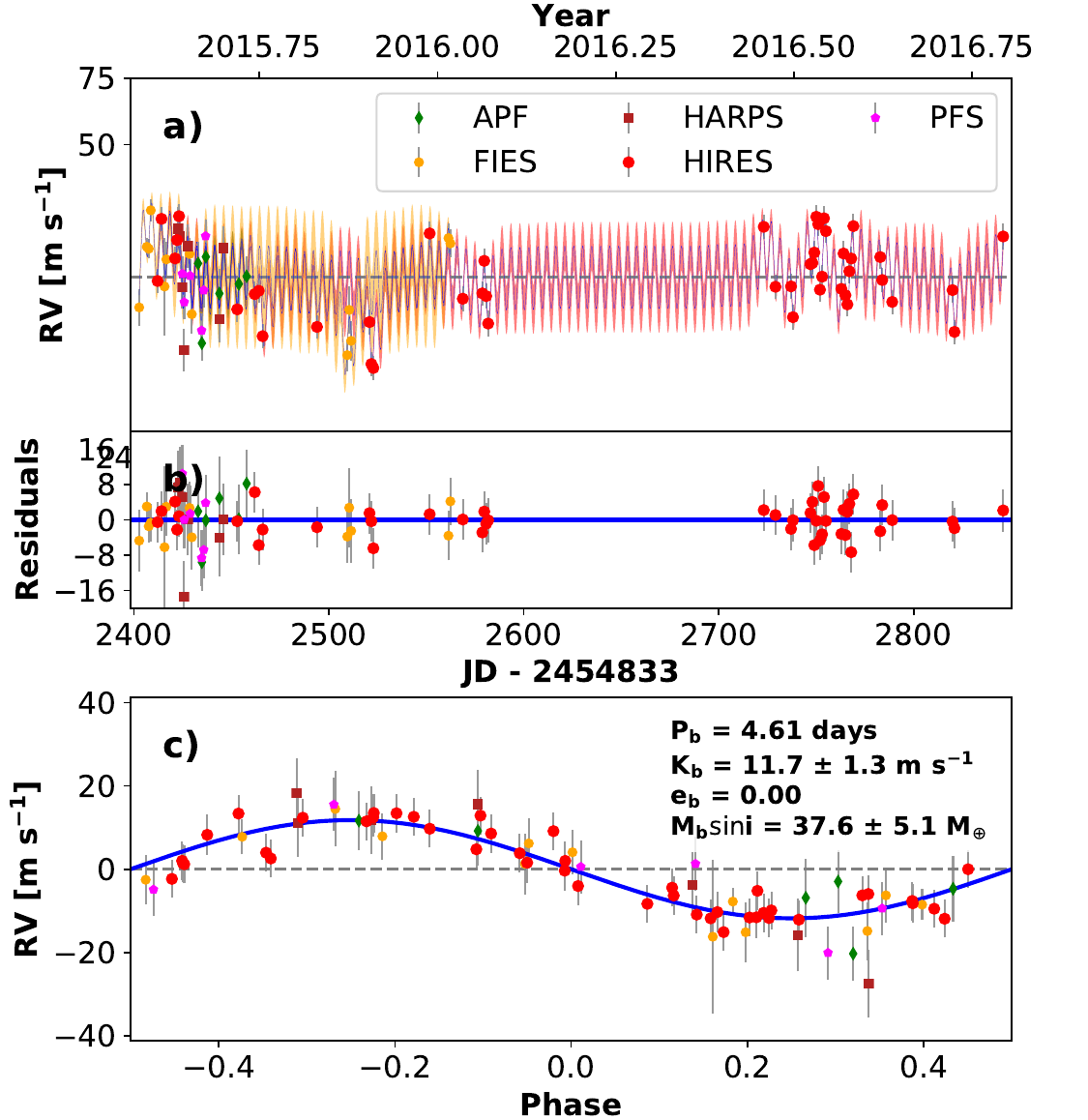}
\caption{RVs and Keplerian model for \hhedSTNAME. Symbols, lines, and annotations are similar to those in Fig.\ \ref{fig:rvs_epic220709978}.}
\label{fig:rvs_k2-39}
\end{figure}

We acquired a total of \hhedNOBSHIRES RVs of \hhedSTNAME with HIRES, typically with an exposure meter setting of 80,000.  
We modeled the system as a single planet in a circular orbit with the orbital period and phase from the transit ephemerides. Our model does not include a linear trend as in \cite{vanEylen2016b}, or orbital eccentricity, based on model comparison using the AICc statistic. Since this system meets our $N_{obs}$ and activity thresholds (Sec.\ \ref{sec:gp_modeling}), we include a GP trained on the photometry to model the stellar variability.
The results of this analysis are listed in Table \ref{tab:epic206247743} and the best-fit model is shown in Fig.\ \ref{fig:rvs_k2-39}.
\hhedPNAMEone is giant planet with a radius and density similar to Saturn's.

\begin{figure}
\epsscale{1.0}
\plotone{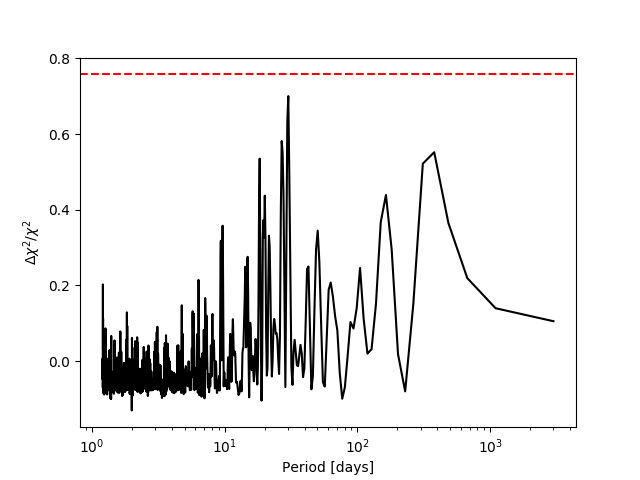}
\caption{Periodogram search of the RVs showing no evidence for a second planet orbiting \hhedSTNAME.  Lines and annotations are similar to those in Fig.\ \ref{fig:rvs_epic220709978_resid}.}
\label{fig:rvs_epic206247743_resid}
\end{figure}

We searched for additional planets in the system using Keplerian models without a Gaussian process. First, we searched for two-planet solutions with circular orbits, as shown in Fig.\ \ref{fig:rvs_epic206247743_resid}. No candidate period was identified that exceeded the 1\% false alarm threshold. We do see excess power in a series of peaks near 30 days and at $\sim$1 year, both of which correspond to patterns in our observing cadence. We performed a deeper search by fitting for the most significant period in Fig.\ \ref{fig:rvs_epic206247743_resid} at 24 days and a third trial period. This periodogram (not shown) did not reveal any convincing candidate periods.


\begin{deluxetable}{lrrr}
\tablecolumns{4}
\tabletypesize{\scriptsize}
\tablecaption{ K2-39 System Parameters \label{tab:epic206247743}}
\tablehead{
  \colhead{Parameter} & 
  \colhead{Credible} & 
  \colhead{Maximum} & 
  \colhead{Units} \\
  \colhead{} & 
  \colhead{Interval} & 
  \colhead{Likelihood} & 
  \colhead{} }
\startdata
\sidehead{\bf{RV Analysis -- MCMC Step Parameters}} 
$P_{b}$ & $\equiv4.6055$ & $\equiv4.6055$ & days \\ 
$T_{\mathrm{conj},b}$ & $\equiv2456985.4268$ & $\equiv2456985.4268$ & BJD$_\mathrm{TBD}$ \\ 
$\sqrt{e}\cos{\omega}_{b}$ & $\equiv0.0$ & $\equiv0.0$ &  \\ 
$\sqrt{e}\sin{\omega}_{b}$ & $\equiv0.0$ & $\equiv0.0$ &  \\ 
$K_{b}$ & $11.7\pm1.3$ & $11.8$ & m s$^{-1}$ \\ 
$\gamma_{\rm PFS}$ & $-1^{+4.2}_{-3.9}$ & $-1.0$ & m s$-1$ \\ 
$\gamma_{\rm HIRES}$ & $-1.4^{+2.8}_{-3.4}$ & $-1.1$ & m s$-1$ \\ 
$\gamma_{\rm HARPS}$ & $24490^{+4.9}_{-4.7}$ & $24489.7$ & m s$-1$ \\
\enddata
\end{deluxetable}

\subsection{K2-229} 
\label{sec:K2229}


\befbSTNAME is a K0 dwarf with two transiting planets that have sizes 1.2 \rearth and 2.0 \rearth and orbital periods of 14 hr and 8 days.  See Tables \ref{tb:star_pars} and \ref{tb:star_props} for stellar properties and Table \ref{tb:planet_props} for precise planet parameters.

The planets were first noted in \cite{Mayo2018}, where they were validated.  \cite{Santerne2018} followed up with 104 high-cadence RVs from HARPS that provided precise mass estimates.  Their model included a GP constrained to the rotational period of $18.1 \pm 0.3$ days.  They found masses of $2.59 \pm 0.43$ \mearth and $< 21.3$ \mearth (95\% confidence), with corresponding densities of $8.9 \pm 2.1$ \gmc and $< 12.8$ \gmc, for planets b and c respectively.  They concluded that planet b has a 30\%/70\% mass fraction for rock/iron composition, i.e., closer to Mercury than Earth.  The \cite{Santerne2018} analysis also included a nontransiting planet with $P = 31.0 \pm 0.1$ days and a mass of $< 25.1$ \mearth.  


\begin{figure*}
\epsscale{1.0}
\plotone{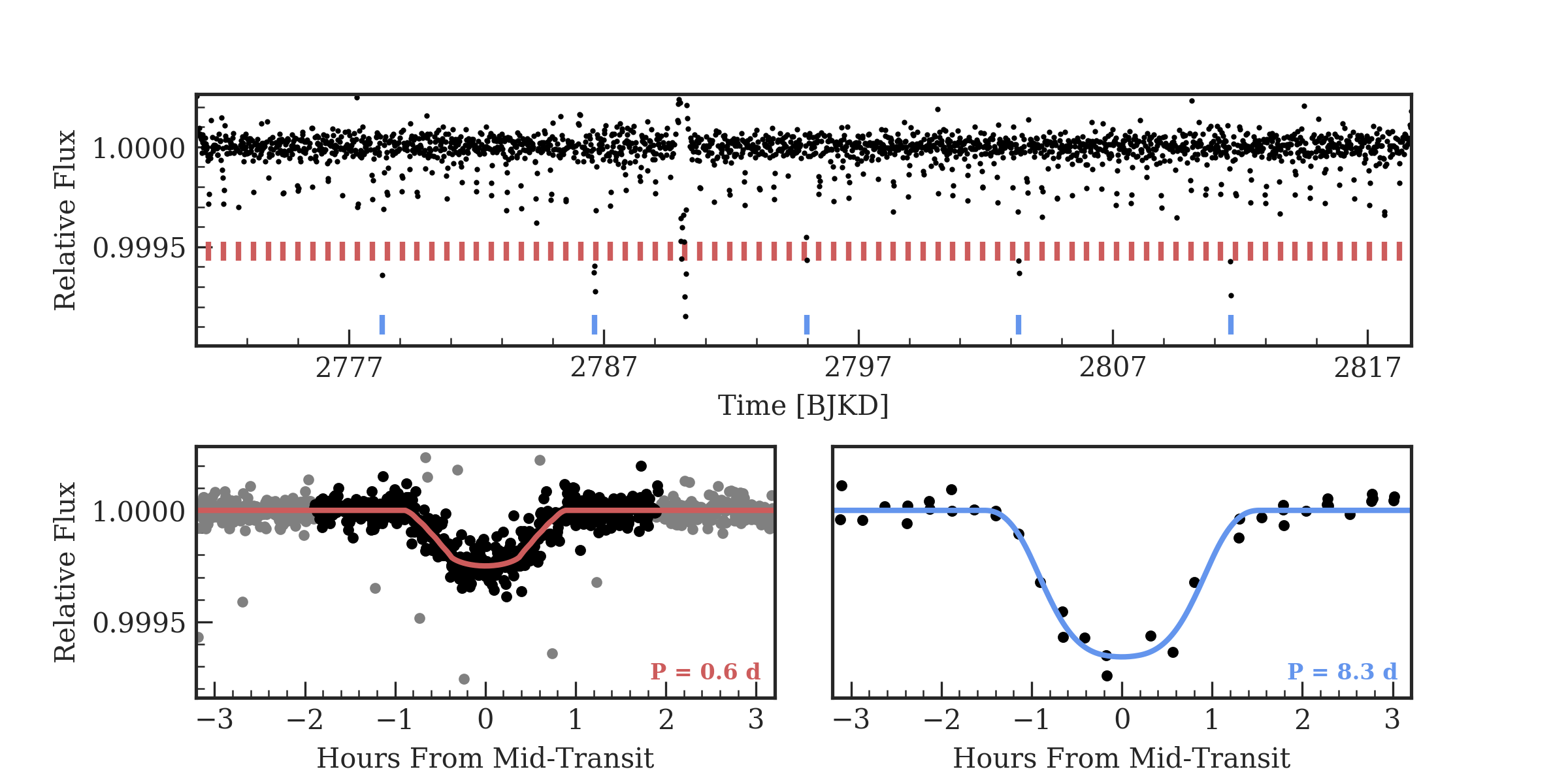}
\caption{Time series (top) and phase-folded (bottom) light curve for the planet orbiting \befbSTNAME.  Plot formatting is the same as in Fig.\ \ref{fig:lc_epic220709978}.}
\label{fig:lc_epic228801451}
\end{figure*}

\begin{figure}
\epsscale{1.0}
\plotone{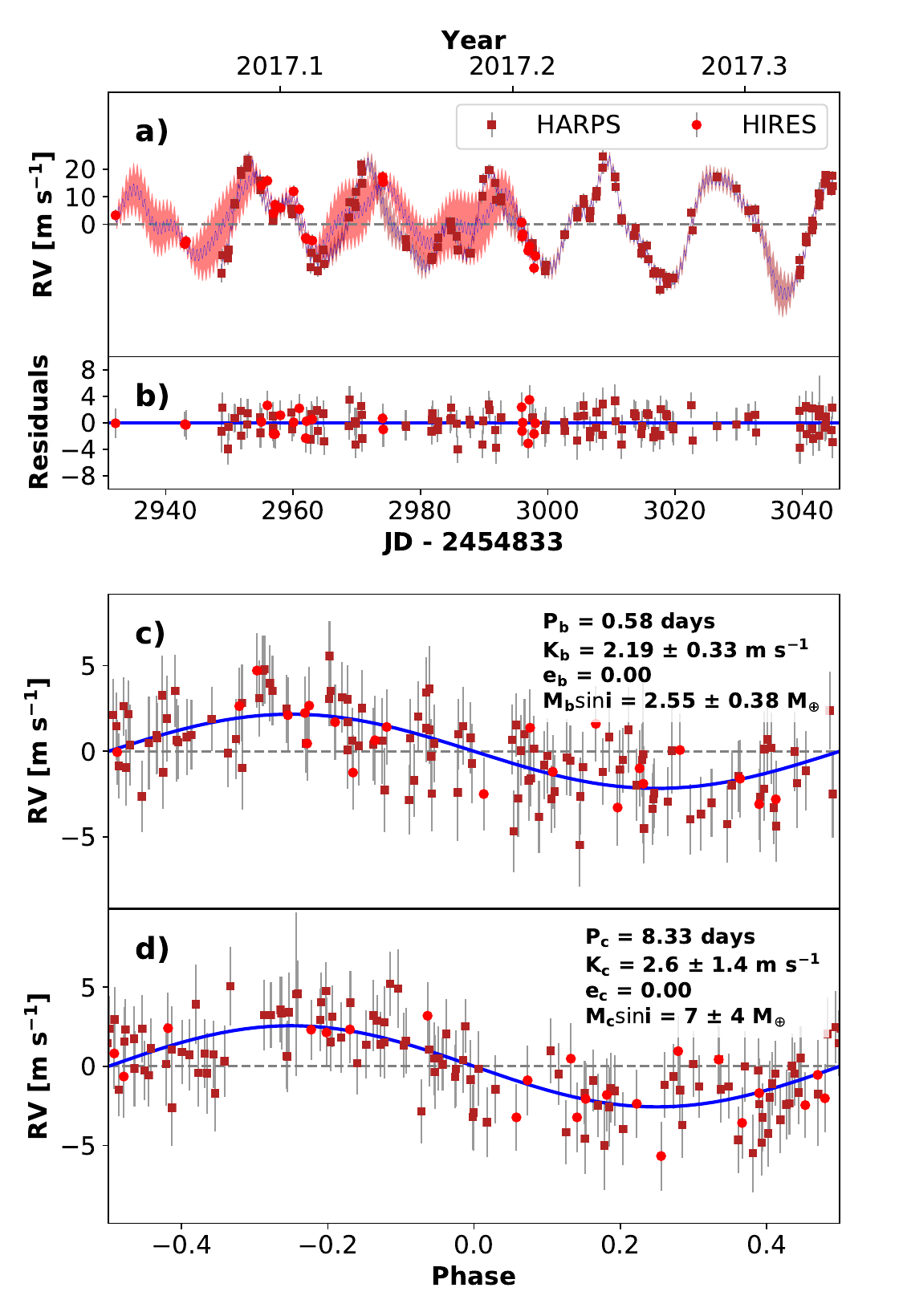}
\caption{RVs and Keplerian model for \befbSTNAME.  Symbols, lines, and annotations are similar to those in Fig.\ \ref{fig:rvs_epic220709978}.}
\label{fig:rvs_k2-229}
\end{figure}

Our fit of the EVEREST light curve of the K2 photometry for \befbSTNAME is shown in Fig.\ \ref{fig:lc_epic228801451}.
We acquired \befbNOBSHIRES RVs of \befbSTNAME with HIRES, typically with an exposure meter setting of 125,000 counts.  
The host star has significant activity, with \lrphk = \befbRPHK.
We modeled the system as a two-planet fit, including a Gaussian process to account for the stellar activity. We considered adding additional parameters to the model including an RV trend, planet eccentricities, and a third non-transiting planet, but rejected those based on model comparison using the AICc statistic.  The results of this analysis are listed in Table \ref{tab:epic228801451} and the best-fit model is shown in Fig. \ref{fig:rvs_k2-229}.


The primary result of \cite{Santerne2018} is that \befbPNAMEone has a high density ($8.9 \pm 2.1$ \gmc), which implies a large iron fraction (70\%).  Density depends on \Rp, \rprstar, \Rstar, and \Mp.
Our mass measurement is consistent with that from \citet{Santerne2018}; however we measure a larger \rprstar resulting in a smaller density ($7.5^{+2.05}_{-1.61}$ \gmc), requiring a smaller iron fraction (0--30\%) based on the \cite{Fortney2007} mass-radius models. 



\begin{deluxetable}{lrrr}
\tablecolumns{4}
\tabletypesize{\scriptsize}
\tablecaption{ K2-229 System Parameters \label{tab:epic228801451}}
\tablehead{
  \colhead{Parameter} & 
  \colhead{Credible} & 
  \colhead{Maximum} & 
  \colhead{Units} \\
  \colhead{} & 
  \colhead{Interval} & 
  \colhead{Likelihood} & 
  \colhead{} }
\startdata
\sidehead{\bf{RV Analysis -- MCMC Step Parameters}} 
$P_{b}$ & $\equiv0.5843$ & $\equiv0.5843$ & days \\ 
$T_{\mathrm{conj},b}$ & $\equiv2457605.0861$ & $\equiv2457605.0861$ & BJD$_\mathrm{TBD}$ \\ 
$\sqrt{e}\cos{\omega}_{b}$ & $\equiv0.0$ & $\equiv0.0$ &  \\ 
$\sqrt{e}\sin{\omega}_{b}$ & $\equiv0.0$ & $\equiv0.0$ &  \\ 
$K_{b}$ & $2.19\pm0.30$ & $2.20$ & m s$^{-1}$ \\ 
$P_{c}$ & $\equiv8.3262$ & $\equiv8.3262$ & days \\ 
$T_{\mathrm{conj},c}$ & $\equiv2457611.3227$ & $\equiv2457611.3227$ & BJD$_\mathrm{TBD}$ \\ 
$\sqrt{e}\cos{\omega}_{c}$ & $\equiv0.0$ & $\equiv0.0$ &  \\
\enddata
\end{deluxetable}

\subsection{K2-111} 



\eaccSTNAME is a solar-type star from Field 4 with low metallicity (\feh = \eaccFEH).  It has one transiting planet with a radius of 1.3 \rearth and an orbital period of 5.3 days.  See Tables \ref{tb:star_pars}  and \ref{tb:star_props} for stellar properties and Table \ref{tb:planet_props} for precise planet parameters.

\begin{figure*}
\epsscale{1.0}
\plotone{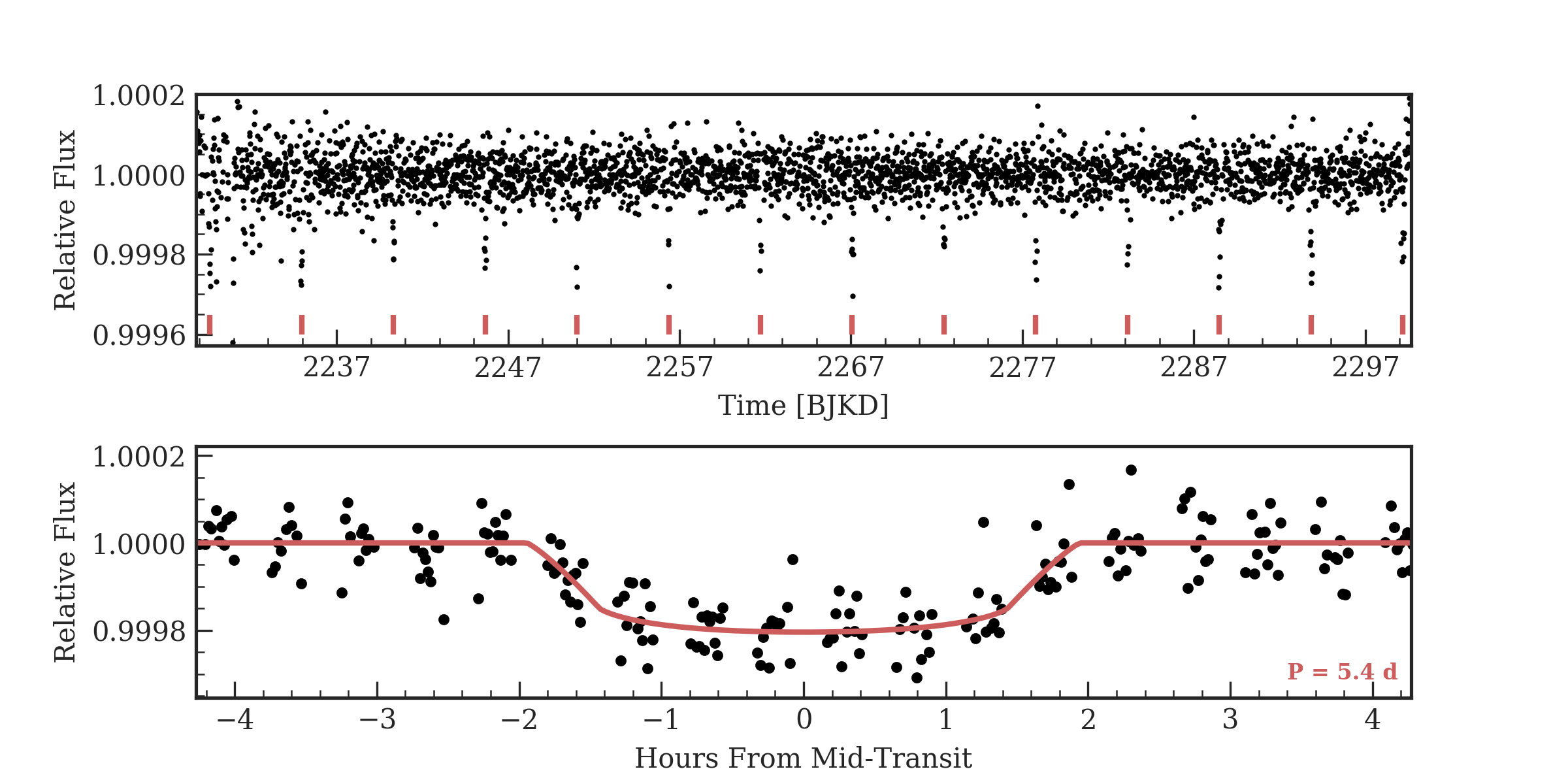}
\caption{Time series (top) and phase-folded (bottom) light curve for the planet orbiting \eaccSTNAME.  Plot formatting is the same as in Fig.\ \ref{fig:lc_epic220709978}.}
\label{fig:lc_epic210894022}
\end{figure*}

The planet was initially listed as a false positive in \cite{Crossfield2016}.   \cite{Fridlund2017} validated the system and measured a mass of $8.6 \pm 3.9$ \mearth using six FIES and 12 HARPS-N RVs.  They also determined that the star is in the background of the Hyades cluster, at four times the distance.  A subsequent analysis with more data reveal a lower and more precise mass of $5.6 \pm 0.7 M_\oplus$ and a second, non-transiting planet \cite{Mortier2020,Bonomo2023}. Our fit of the EVEREST light curve of the K2 photometry for \eaccSTNAME is shown in Fig.\ \ref{fig:lc_epic210894022}. We find a radius that is consistent with, but more precise than, the value from \cite{Mayo2018}.

\begin{figure}
\epsscale{1.0}
\plotone{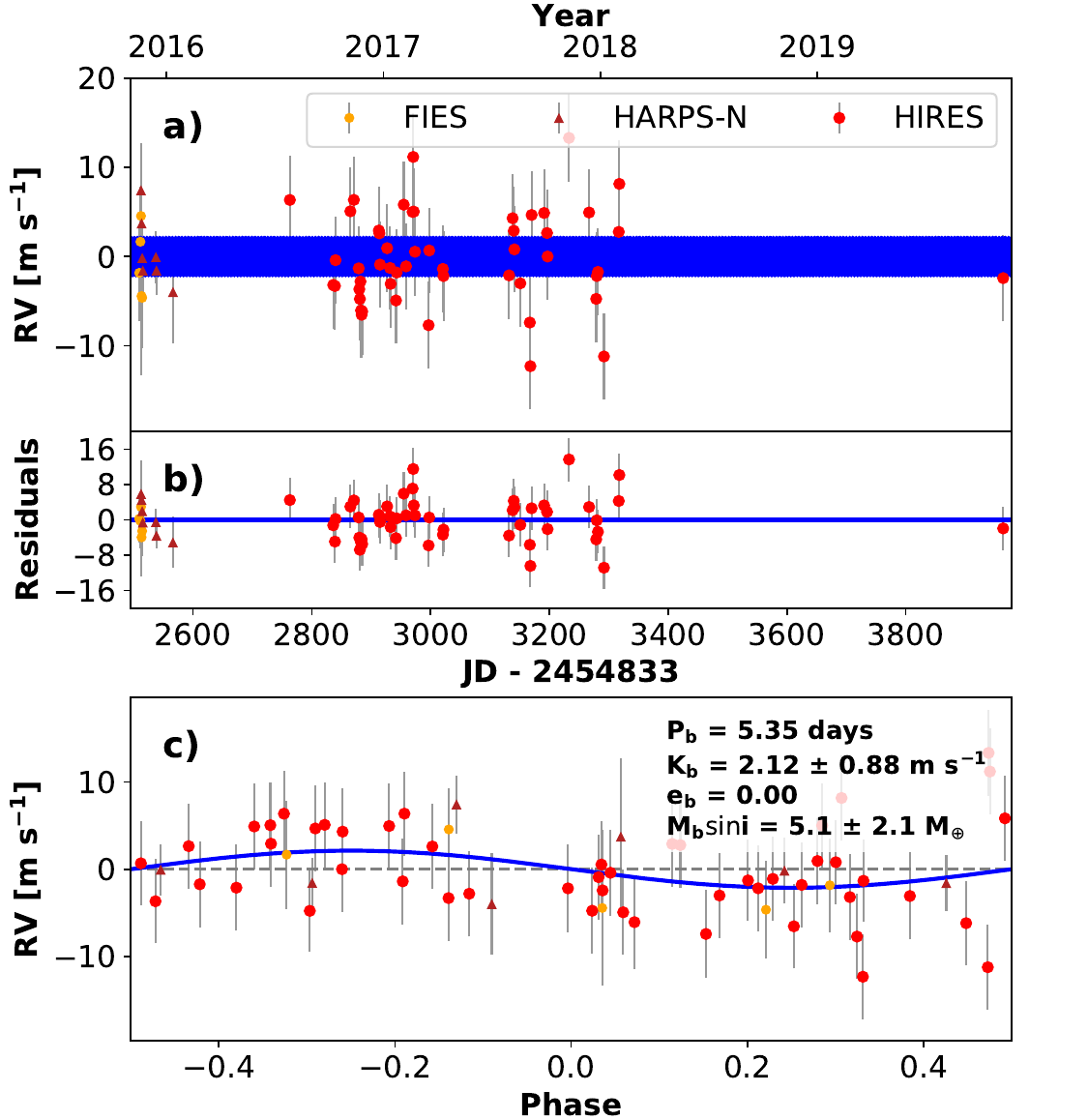}
\caption{RVs and Keplerian model for \eaccSTNAME.  Symbols, lines, and annotations are similar to those in Fig.\ \ref{fig:rvs_epic220709978}.}
\label{fig:rvs_k2-111}
\end{figure}

We acquired \eaccNOBSHIRES RVs of \eaccSTNAME with HIRES, typically with an exposure meter setting of 125,000.  
Using the RVs from HIRES, HARPS-N, and FIES, we modeled the system as a single planet with the orbital period and phase fixed to the transit ephemeris.  The results of this analysis are listed in Table \ref{tab:epic210894022}
and the best-fit model is shown in Fig.\ \ref{fig:rvs_k2-111}. 
\cite{Fridlund2017} noted a trend in the residual RVs to their one-planet fit. We found weak evidence for this trend in the HIRES RVs; however, a model with a trend yielded a posterior for $\dot{\gamma}$ consistent with zero. We thus adopted the simpler model with no trend. We also ruled out a eccentric orbit based on an AICc comparison. Our mass of \eaccMPone \mearth is consistent with the literature values  \citep{Fridlund2017,Mortier2020,Bonomo2023}. Our improved mass and radius estimates yield a high density (\eaccRHOPone \gmc) consistent with a rocky composition, more confidently placing \eaccPNAMEone in the super-Earth category.

\begin{deluxetable}{lrrr}
\tablecolumns{4}
\tabletypesize{\scriptsize}
\tablecaption{ K2-111 System Parameters \label{tab:epic210894022}}
\tablehead{
  \colhead{Parameter} & 
  \colhead{Credible} & 
  \colhead{Maximum} & 
  \colhead{Units} \\
  \colhead{} & 
  \colhead{Interval} & 
  \colhead{Likelihood} & 
  \colhead{} }
\startdata
\sidehead{\bf{RV Analysis -- MCMC Step Parameters}} 
$P_{b}$ & $\equiv5.3523$ & $\equiv5.3523$ & days \\ 
$T_{\mathrm{conj},b}$ & $\equiv2457067.9628$ & $\equiv2457067.9628$ & BJD$_\mathrm{TBD}$ \\ 
$\sqrt{e}\cos{\omega}_{b}$ & $\equiv0.0$ & $\equiv0.0$ &  \\ 
$\sqrt{e}\sin{\omega}_{b}$ & $\equiv0.0$ & $\equiv0.0$ &  \\ 
$K_{b}$ & $2.23\pm0.89$ & $2.10$ & m s$^{-1}$ \\ 
$\gamma_{\rm HIRES}$ & $-1.24^{+0.70}_{-0.71}$ & $-1.19$ & m s$-1$ \\ 
$\gamma_{\rm HARPS-N}$ & $-1.1^{+1.8}_{-1.7}$ & $-1.5$ & m s$-1$ \\ 
$\gamma_{\rm FIES}$ & $\equiv-16397.5644$ & $\equiv-16397.5644$ & m s$-1$ \\ 
$\dot{\gamma}$ & $\equiv0.0$ & $\equiv0.0$ & m s$^{-1}$ day$^{-1}$ \\ 
$\ddot{\gamma}$ & $\equiv0.0$ & $\equiv0.0$ & m s$^{-1}$ day$^{-2}$ \\ 
$\sigma_{\rm HIRES}$ & $4.66^{+0.61}_{-0.52}$ & $4.61$ & $\rm m\ s^{-1}$ \\ 
$\sigma_{\rm HARPS-N}$ & $14.1^{+1.4}_{-1.2}$ & $14.0$ & $\rm m\ s^{-1}$ \\ 
$\sigma_{\rm FIES}$ & $\equiv0.0$ & $\equiv0.0$ & $\rm m\ s^{-1}$ \\ 
\hline
\sidehead{\bf{Orbital \& Physical Parameters}}
$P_{b}$ & $\equiv5.3523$ & $\equiv5.3523$ & days \\
$T_{\mathrm{conj},b}$ & $\equiv2457067.9628$ & $\equiv2457067.9628$ & BJD$_\mathrm{TBD}$ \\
$e_{b}$ & $\equiv0.0$ & $\equiv0.0$ &  \\
$\omega_{b}$ & $\equiv0.0$ & $\equiv0.0$ & radians \\
$K_{b}$ & $2.23\pm0.89$ & $2.10$ & m s$^{-1}$ \\
$M_b$ & $5.4\pm2.2$ & $5.4$ & $M_\earth$ \\
$\rho_b$ & $17.3^{+7.5}_{-7.1}$ & $12.0$ & g cm$^{-3}$ \\
\hline
\hline
\sidehead{\bf{Priors}}
Parameter & Prior \\
\hline
$\sigma_{\rm HIRES}$ & $\mathcal{U}(0,1e+100)$ \\
$\sigma_{\rm HARPS-N}$ & $\mathcal{U}(0,1e+100)$ \\
$\sigma_{\rm FIES}$ & $\mathcal{U}(0,1e+100)$ \\
\enddata
\end{deluxetable}

\begin{figure}
\epsscale{1.0}
\plotone{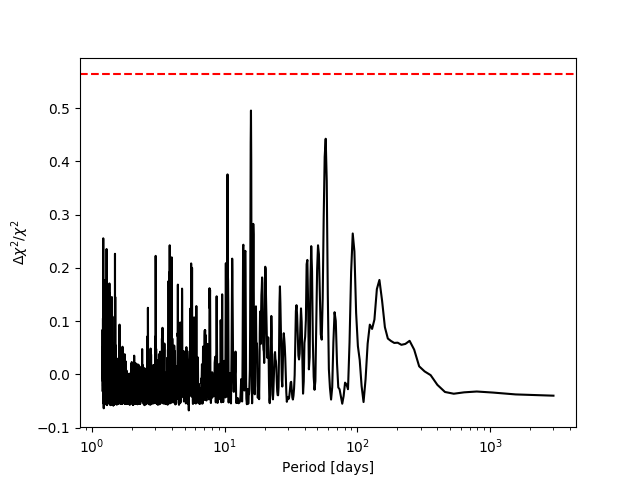}
\caption{Periodogram search of the RVs showing no evidence for a second planet orbiting \eaccSTNAME.  Lines and annotations are similar to those in Fig.\ \ref{fig:rvs_epic220709978_resid}.}
\label{fig:rvs_epic210894022_resid}
\end{figure}
\subsection{K2-99} 
\label{sec:K299}




\dcijSTNAME is subgiant from Campaign 6 and 17 that hosts one transiting sub-Saturn-size planet with an orbital period of 18 days.  See Tables \ref{tb:star_pars}  and \ref{tb:star_props} for stellar properties and Table \ref{tb:planet_props} for precise planet parameters.
\cite{Smith2017} first noted the planet and measured a mass of $308 \pm 29$ \mearth based on 14 FIES RVs, 5 HARPS-N RVs, 6 {\referee McDonald 2.7m} RVs, and 8 HARPS RVs.  Their model included a free eccentricity ($e = 0.19 \pm 0.04$) and linear RV trend ($-2.12 \pm 0.04$ \msy).  The planet is also listed in the catalogs by \cite{Pope2016}, \cite{Crossfield2018}, \cite{Petigura2018}, and \cite{Mayo2018}. A more recent RV analysis updates the sizes of the host star and planet to 2.55\,$R_\odot$ and 1.06\,$R_\mathrm{Jup}$, respectively, and identifies an outer planet c with a 522-day period and mass of 8.4\,$M_\mathrm{Jup}$ \citep{Smith2022}.

\begin{figure*}
\epsscale{1.0}
\plotone{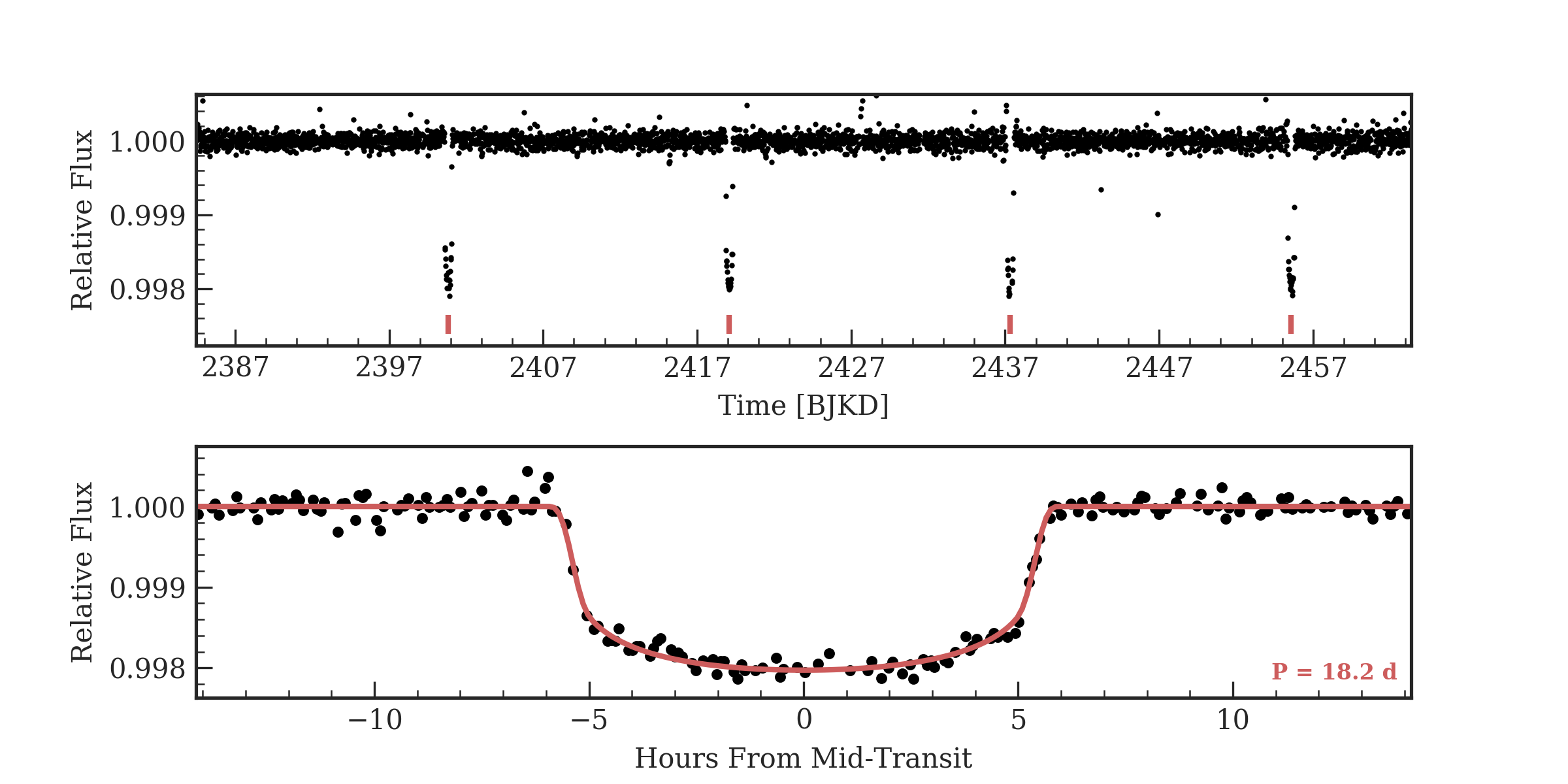}
\caption{Time series (top) and phase-folded (bottom) light curve for the planet orbiting \dcijSTNAME.  Plot formatting is the same as in Fig.\ \ref{fig:lc_epic220709978}.}
\label{fig:lc_epic212803289}
\end{figure*}

\begin{figure}
\epsscale{1.0}
\plotone{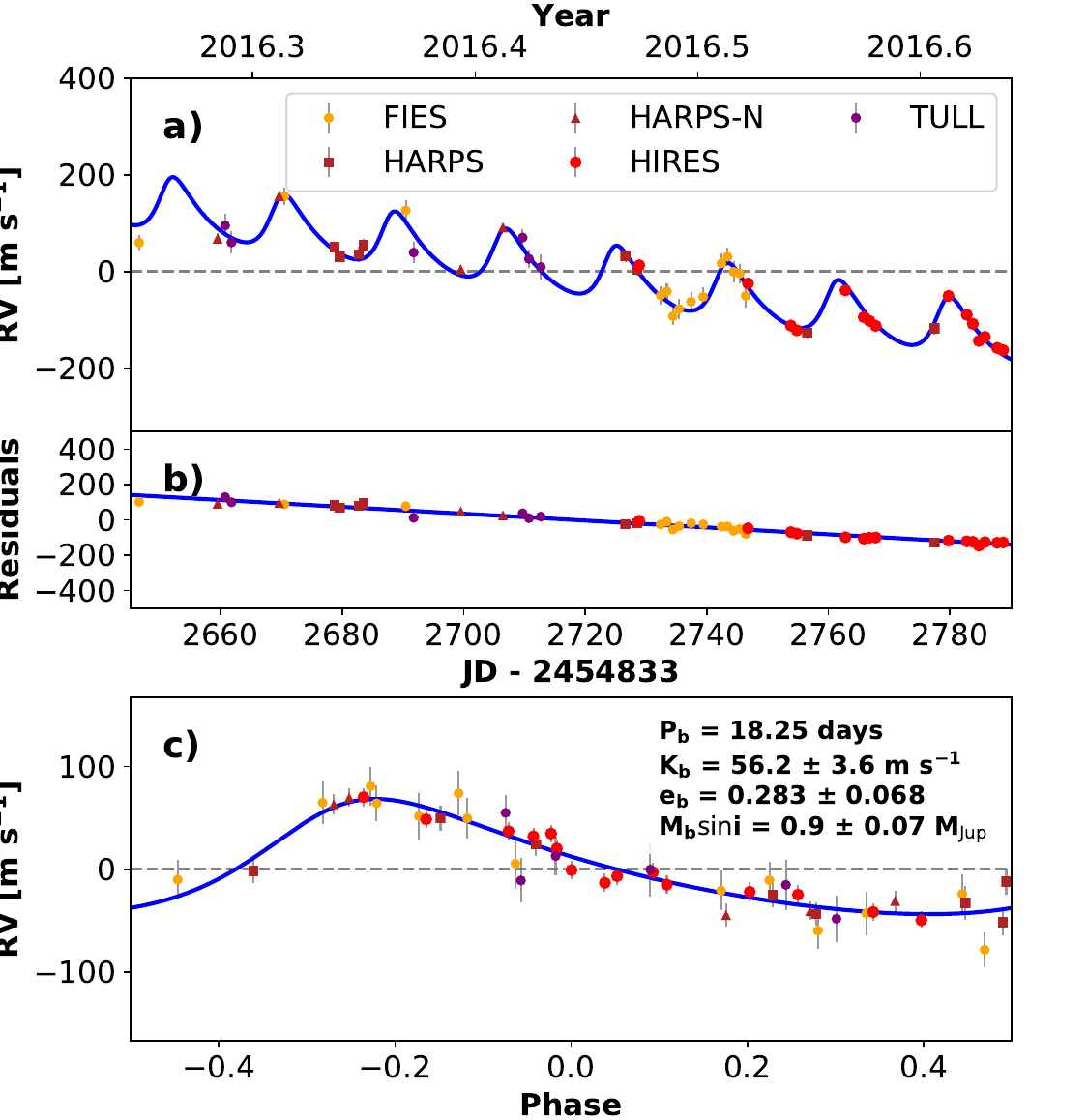}
\caption{RVs and Keplerian model for \dcijSTNAME.  Symbols, lines, and annotations are similar to those in Fig.\ \ref{fig:rvs_epic220709978}.}
\label{fig:rvs_k2-99}
\end{figure}

Our fit of the EVEREST light curve of the K2 photometry for \dcijSTNAME is shown in Fig.\ \ref{fig:lc_epic212803289}.  We acquired \dcijNOBSHIRES RVs of \dcijSTNAME with HIRES, typically with an exposure meter setting of 60,000.  We modeled the system as a single planet with free eccentricity and a linear RV trend.  The results of this analysis are listed in Table \ref{tab:epic212803289}
and the best-fit model is shown in Fig.\ \ref{fig:rvs_k2-99}.  The next-best model had a circular orbit and a linear trend and \dAICc of 17 compared to the selected model. \dcijPNAMEone is a giant planet in a moderately eccentric orbit.


\begin{deluxetable}{lrrr}
\tablecolumns{4}
\tabletypesize{\scriptsize}
\tablecaption{ K2-99 System Parameters \label{tab:epic212803289}}
\tablehead{
  \colhead{Parameter} & 
  \colhead{Credible} & 
  \colhead{Maximum} & 
  \colhead{Units} \\
  \colhead{} & 
  \colhead{Interval} & 
  \colhead{Likelihood} & 
  \colhead{} }
\startdata
\sidehead{\bf{RV Analysis -- MCMC Step Parameters}} 
$P_{b}$ & $\equiv18.249$ & $\equiv18.249$ & days \\ 
$T_{\mathrm{conj},b}$ & $\equiv2457233.8255$ & $\equiv2457233.8255$ & BJD$_\mathrm{TBD}$ \\ 
$\sqrt{e}\cos{\omega}_{b}$ & $0.391^{+0.046}_{-0.053}$ & $0.405$ &  \\ 
$\sqrt{e}\sin{\omega}_{b}$ & $-0.36^{+0.13}_{-0.11}$ & $-0.36$ &  \\ 
$K_{b}$ & $56.2^{+3.7}_{-3.6}$ & $56.0$ & m s$^{-1}$ \\ 
$\gamma_{\rm TULL}$ & $408^{+12}_{-13}$ & $407$ & m s$-1$ \\ 
$\gamma_{\rm HIRES}$ & $91.2\pm7.6$ & $91.0$ & m s$-1$ \\ 
$\gamma_{\rm HARPS-N}$ & $-2649^{+9}_{-10}$ & $-2649.5$ & m s$-1$ \\
\enddata
\tablecomments{}{
Reference epoch for $\dot{\gamma}$,$\ddot{\gamma}$: 2457550.6875529997 
}
\end{deluxetable}

\subsection{K2-265} 

\begin{figure*}
\epsscale{1.0}
\plotone{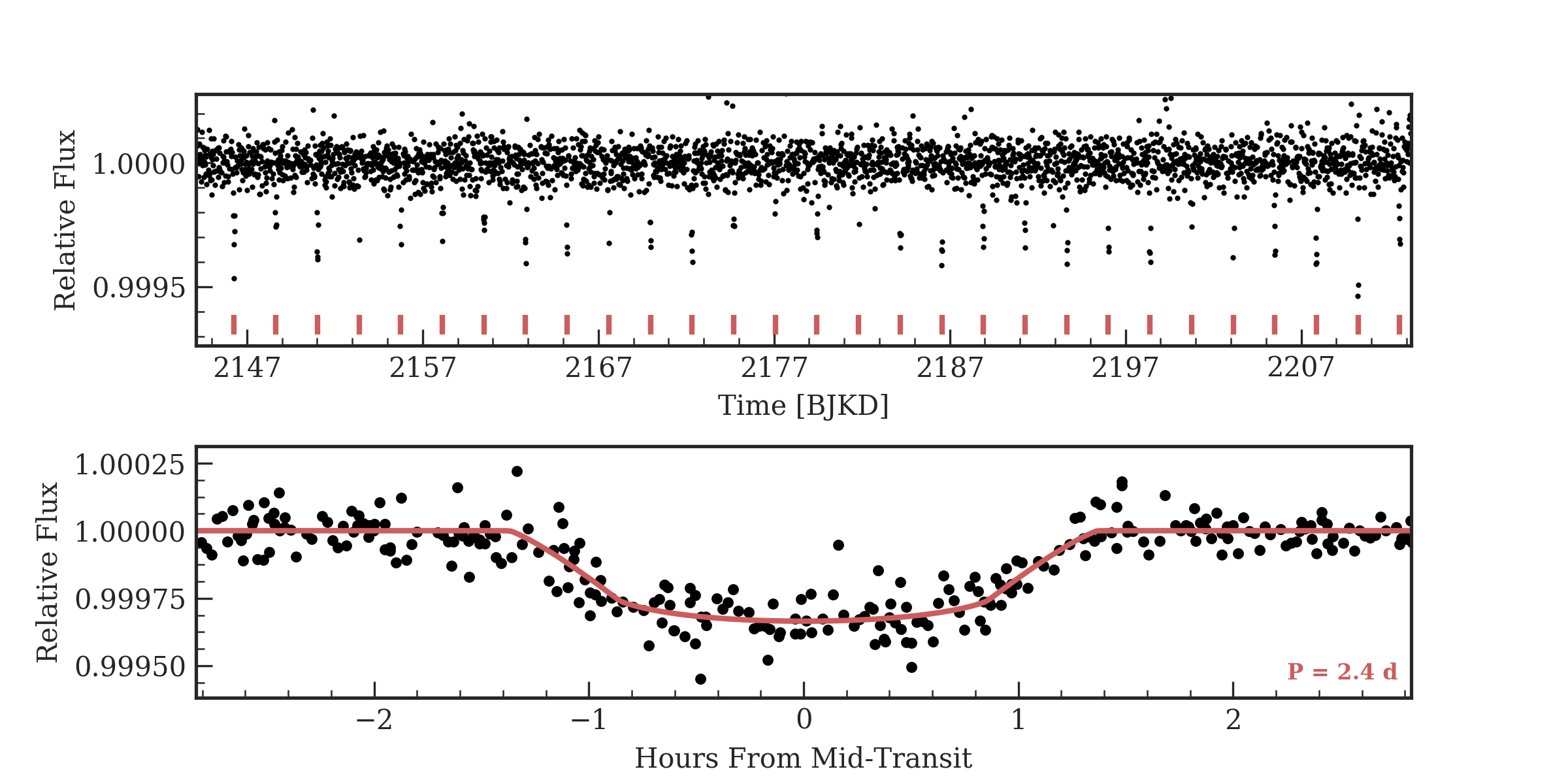}
\caption{Time series (top) and phase-folded (bottom) light curve for the planet orbiting \bejgSTNAME.  Plot formatting is the same as in Fig.\ \ref{fig:lc_epic220709978}.}
\label{fig:lc_epic206011496}
\end{figure*}

\bejgSTNAME is a late G dwarf with one transiting planet with a radius of 1.7 \rearth and an orbital period of 2.4 days.  See Tables \ref{tb:star_pars}  and \ref{tb:star_props} for stellar properties and Table \ref{tb:planet_props} for planet parameters.  The planet was listed as a candidate in \cite{Crossfield2016}, who noted an imaged companion with a separation of $\sim$1\arcsec and $\Delta$K = 2.8 mag.  The transit duration is consistent with the planet orbiting the brighter star.  \cite{Vanderburg2016-catalog} and \cite{Mayo2018} also list the object as a planet candidate.  
\cite{Lam2018} validated the system using K2 photometry and measured a mass of $6.54 \pm 0.84$ \mearth and a density of $7.1 \pm 1.8$ \gmc based on 153 HARPS RVs. 
Our fit of the EVEREST light curve of the K2 photometry for \bejgSTNAME is shown in Fig.\ \ref{fig:lc_epic206011496}.

We acquired \bejgNOBSHIRES RVs of \bejgSTNAME with HIRES, typically with an exposure meter setting of 125,000.  
With \lrphk$=$\bejgRPHK and \bejgNOBSHIRES\ RVs from HIRES, \bejgSTNAME is an excellent candidate for GP regression. As described in Sec.\ \ref{sec:gp_modeling}, we trained the GP hyperparameters on non-detrended Everest photometry before using computing RV orbit posteriors. For comparison, we also performed an RV orbit fit using an untrained GP. The trained and the untrained GP models produce semi-amplitudes values for planet b that are consistent within 1$\sigma$ ($2.4\pm0.8$ and $2.1\pm0.8$, respectively), but the trained GP produces a median value $15\%$ higher. The median value of $K_b$ returned by the trained GP fit is identical to that produced by a non-GP fit, but the uncertainty on $K_b$ is almost halved.  Note that the apparent correlated noise features in the RVs near 2017.5 and 2018.0 are also seen in the $S_{HK}$ time series and are modeled by the GP.  This system illustrates the power of GP regression to improve the precision of orbit fitting analyses. 

\begin{deluxetable}{lrrr}
\tablecolumns{4}
\tabletypesize{\scriptsize}
\tablecaption{ K2-265 System Parameters \label{tab:epic206011496}}
\tablehead{
  \colhead{Parameter} & 
  \colhead{Credible} & 
  \colhead{Maximum} & 
  \colhead{Units} \\
  \colhead{} & 
  \colhead{Interval} & 
  \colhead{Likelihood} & 
  \colhead{} }
\startdata
\sidehead{\bf{RV Analysis -- MCMC Step Parameters}} 
$P_{b}$ & $\equiv2.3691$ & $\equiv2.3691$ & days \\ 
$T_{\mathrm{conj},b}$ & $\equiv2456981.645$ & $\equiv2456981.645$ & BJD$_\mathrm{TBD}$ \\ 
$\sqrt{e}\cos{\omega}_{b}$ & $\equiv0.0$ & $\equiv0.0$ &  \\ 
$\sqrt{e}\sin{\omega}_{b}$ & $\equiv0.0$ & $\equiv0.0$ &  \\ 
$K_{b}$ & $2.38^{+0.70}_{-0.80}$ & $2.40$ & m s$^{-1}$ \\ 
$\gamma_{\rm HIRES}$ & $-2.0^{+1.2}_{-1.5}$ & $-1.9$ & m s$-1$ \\ 
$\dot{\gamma}$ & $\equiv0.0$ & $\equiv0.0$ & m s$^{-1}$ day$^{-1}$ \\ 
$\ddot{\gamma}$ & $\equiv0.0$ & $\equiv0.0$ & m s$^{-1}$ day$^{-2}$ \\
\enddata
\end{deluxetable}

\begin{figure}
\epsscale{1.0}
\plotone{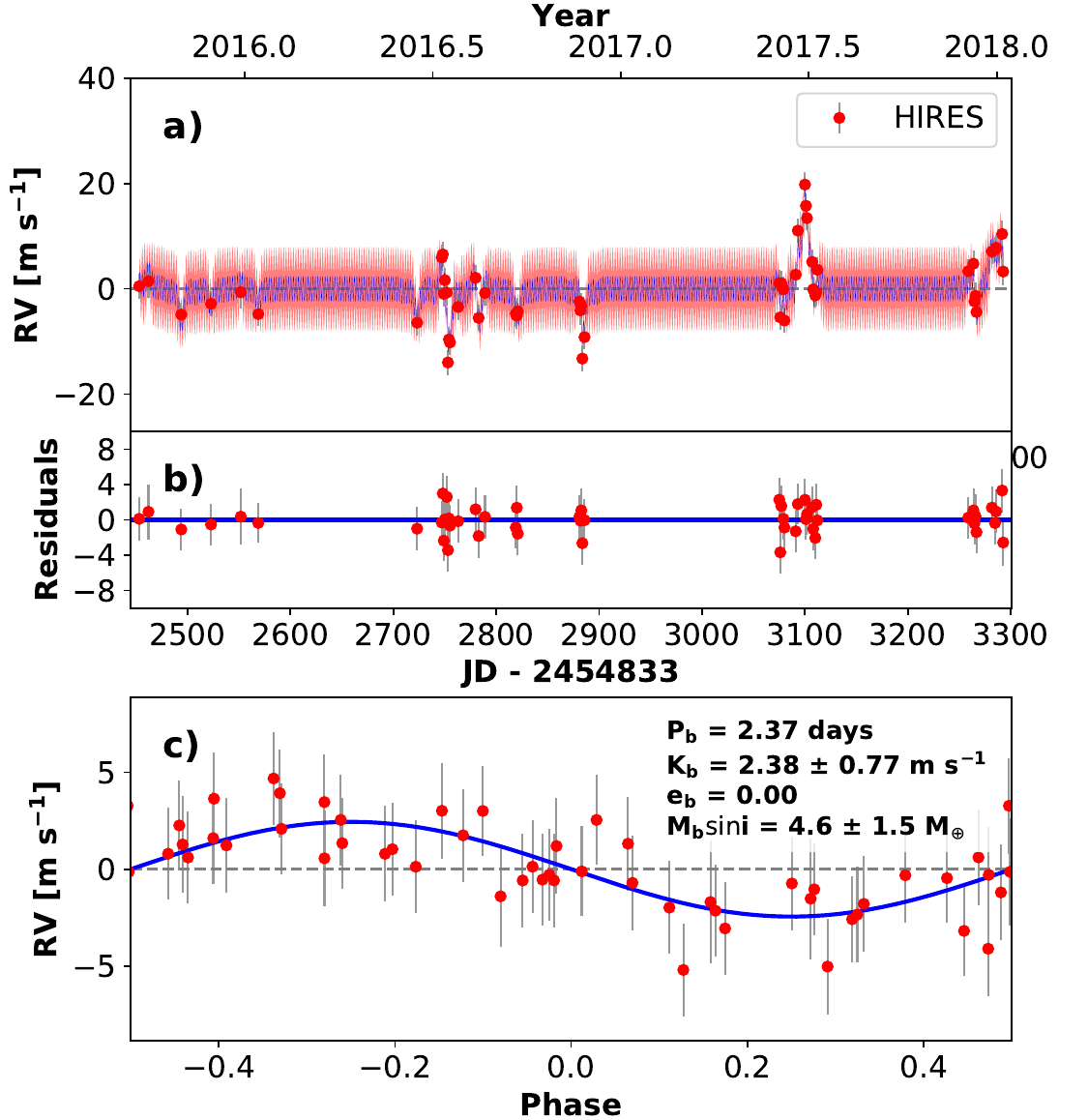}
\caption{RVs and Keplerian model for \bejgSTNAME.
Symbols, lines, and annotations are similar to those in Fig.\ \ref{fig:rvs_epic220709978}.}
\label{fig:rvs_epic206011496}
\end{figure}

\begin{figure}
\epsscale{1.0}
\plotone{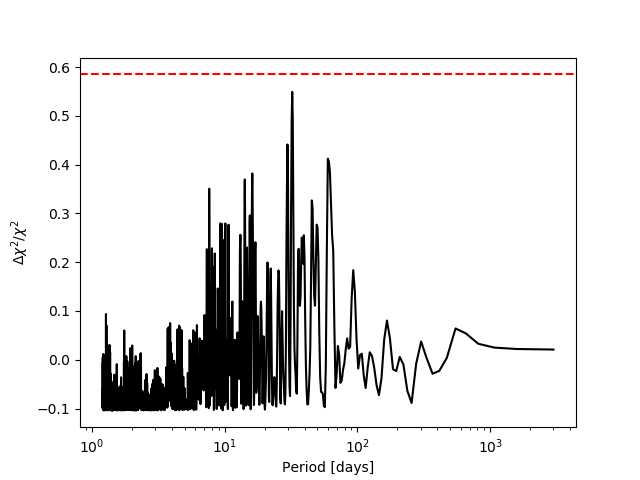}
\caption{Periodogram search of the RVs showing no evidence for a second planet orbiting \bejgSTNAME.  The period with the most significant periodogram peak is at 32 d.  Lines and annotations are similar to those in Fig.\ \ref{fig:rvs_epic220709978_resid}.}
\label{fig:rvs_epic206011496_resid}
\end{figure}
\subsection{K2-24} 


\bajiSTNAME is a metal-rich G dwarf with two transiting planets that have sizes 5.4\,\rearth\ and 7.5\,\rearth, between that of Uranus and Saturn.  The orbital periods (20.9 days and 42.4 days) are within 1\% of a 2:1 ratio.  The system is noted in the catalogs of \cite{Crossfield2016}, \cite{Vanderburg2016-catalog}, \cite{Sinukoff2016}, \cite{Wittenmyer2018}.

\cite{Petigura2016} validated the planets and measured masses of $21.0 \pm 5.4$\,\mearth and (planet b) $27.0 \pm 6.9$\,\mearth (planet c) based on 32 RVs from HIRES.  The bulk densities of $0.63 \pm 0.25$\,\gmc and $0.31 \pm 0.12$\,\gmc, respectively, are low for planets of that size, and modeling suggested that thick envelopes of H/He to needed to account for the masses and radii.  \cite{Dai2016} measured 16 PFS RVs and 10 HARPS RVs.  Their model of the PFS, HARPS, and HIRES RVs gave masses of $19.8^{+4.5}_{-4.4}$\,\mearth and $26.0^{+5.8}_{-6.1}$\,\mearth, respectively.
\cite{Petigura2018b} reported on additional HIRES RVs taken this project (63 RVs total).  Their analysis included transit times from K2 photometry as well as four transit epochs from Spitzer.  The additional RVs over a longer baseline revealed a non-transiting planet (K2-24d) with an orbital period of 428 days.  Modeling the RVs and TTVs simultaneously gave much tighter constraints for the planet masses ($19.0^{+2.2}_{-2.1}$\,\mearth and $15.4^{+1.9}_{-1.8}$\,\mearth, respectively) and eccentricities ($e \sim 0.08$ for both planets).  Interestingly, K2-24b is 20\% less massive than K2-24c despite being 40\% larger.
We adopt the planet parameters from \cite{Petigura2018b}.  
The full set of adopted parameters (stellar and planetary) is listed in Tables \ref{tb:star_pars}, \ref{tb:star_props} and \ref{tb:planet_props}.

\subsection{K2-38} 


\bcgdSTNAME is a metal-rich, solar-type star in Field 2 with two detected transiting planets.  The planets have sizes 1.6 \rearth and 2.4 \rearth and periods of 4 days and 10 days, respectively.
See Tables \ref{tb:star_pars}  and \ref{tb:star_props} for stellar properties and Table \ref{tb:planet_props} for precise planet parameters.

The planets orbiting \bcgdSTNAME were first identified and validated in \cite{Sinukoff2016}, who measured preliminary masses using 14 RVs from HIRES.  
They modeled the system as two planets in circular orbits and found masses of $12.0 \pm 2.9$ \mearth and $9.9 \pm 4.6$ \mearth, corresponding to densities of $17.5^{+8.5}_{-6.2}$ \gmc and $3.6^{+2.7}_{-1.9}$ \gmc, respectively.  Their model also included a linear trend of $-37 \pm 11$ \msy; this model was favored over a flat model with $\Delta$ BIC = 5.6.
This analysis suggests that the inner planet has an implausibly high bulk density, while the outer planet was detected with $\sim$2-$\sigma$ significance and a density intermediate between that of rocky and gas-dominated planets.  
The two planets also appear in catalogs by \cite{Crossfield2016} and \cite{Wittenmyer2018}. More recently, \cite{Bonomo2023} report planet masses of $7.7\pm 1.2 M_\oplus$ and $7.4 \pm 1.4 M_\oplus$ for planets b and c, respectively; consistent with the values we report below.

\begin{figure*}
\epsscale{1.0}
\plotone{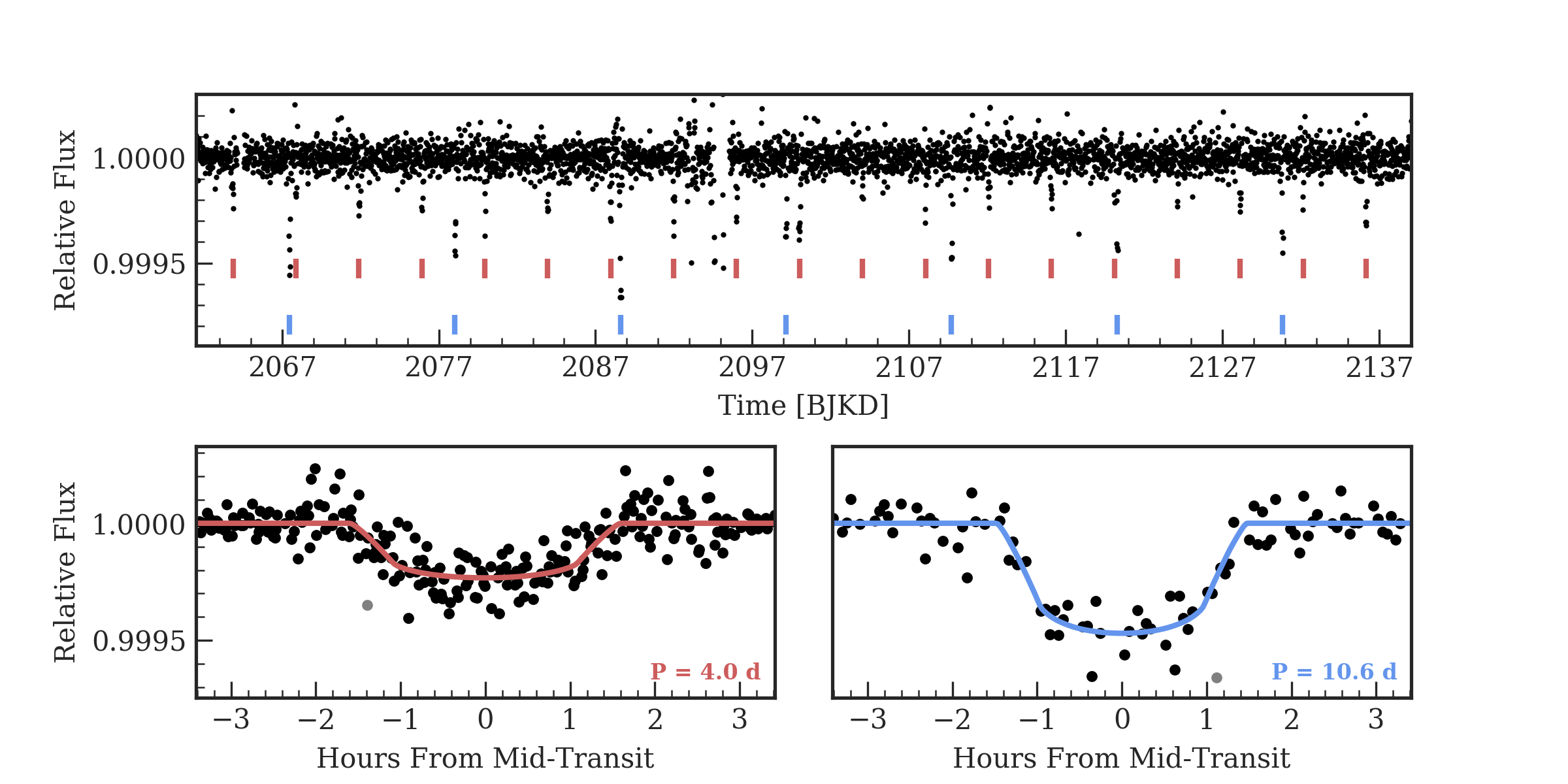}
\caption{Time series (top) and phase-folded (bottom) light curve for the planet orbiting \bcgdSTNAME.  Plot formatting is the same as in Fig.\ \ref{fig:lc_epic220709978}.}
\label{fig:lc_epic204221263}
\end{figure*}

Our fit of the EVEREST light curve of the K2 photometry for \bcgdSTNAME is shown in Fig.\ \ref{fig:lc_epic204221263}.  
We acquired \bcgdNOBSHIRES RVs of \bcgdSTNAME with HIRES (including the 14 RVs from \cite{Sinukoff2016}), typically with an exposure meter setting of 125,000 counts.  We also acquired three template exposures using the B3 decker on HIRES and an exposure meter setting of 250,000 to correct for a spurious 1 yr signal present in the RV time series generated using the first two templates.  The amplitude of this signal ($\sim$2--4 \ms) varied depending on the template used.  The signal is nearly absent in the final template (from which the RVs in Table \ref{tab:rvs} were computed).  It appears that poor selection of the B stars (used for template deconvolution) is responsible for the systematic errors in the previous reductions.

\begin{figure}
\epsscale{1.0}
\plotone{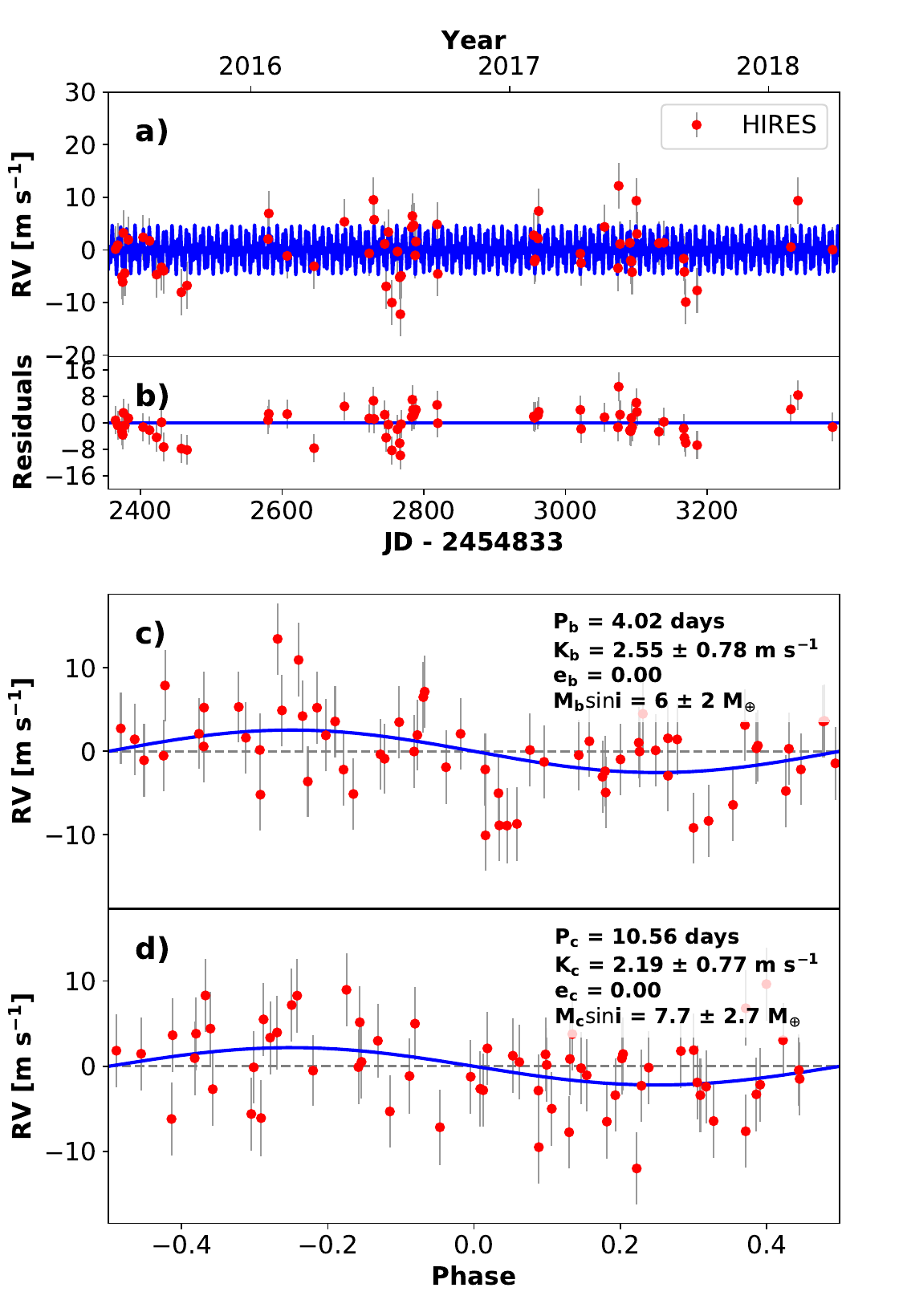}
\caption{RVs and Keplerian model for \bcgdSTNAME.  Symbols, lines, and annotations are similar to those in Fig.\ \ref{fig:rvs_epic220709978}.}
\label{fig:rvs_epic204221263}
\end{figure}

\begin{figure}
\epsscale{1.0}
\plotone{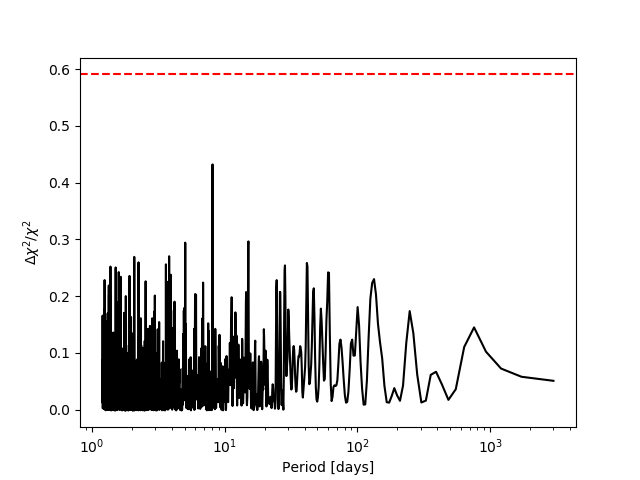}
\caption{Periodogram search of the RVs showing no evidence for a second planet orbiting \bcgdSTNAME.  Lines and annotations are similar to those in Fig.\ \ref{fig:rvs_epic220709978_resid}.}
\label{fig:rvs_epic204221263_resid}
\end{figure}

We modeled the system as two planets in circular orbits with periods and phases fixed to the values from the K2 ephemerides. We found that an eccentric fit is significantly favored (from an AICc comparison), however the best-fit eccentricity is at unphysically high values ($e_b \sim 0.8$) for the short period and compact nature of the two planets. As a result, we adopt fits with circular orbis. The results of this analysis are listed in Table \ref{tab:epic204221263} and the best-fit model is shown in Fig. \ref{fig:rvs_epic204221263}. We found no evidence for an additional planet based on a periodogram search of the RVs (Fig.\ \ref{fig:rvs_epic204221263_resid}).

\begin{deluxetable}{lrrr}
\tablecolumns{4}
\tabletypesize{\scriptsize}
\tablecaption{ K2-38 System Parameters \label{tab:epic204221263}}
\tablehead{
  \colhead{Parameter} & 
  \colhead{Credible} & 
  \colhead{Maximum} & 
  \colhead{Units} \\
  \colhead{} & 
  \colhead{Interval} & 
  \colhead{Likelihood} & 
  \colhead{} }
\startdata
\sidehead{\bf{RV Analysis -- MCMC Step Parameters}} 
$P_{b}$ & $\equiv4.0167$ & $\equiv4.0167$ & days \\ 
$T_{\mathrm{conj},b}$ & $\equiv2456896.8707$ & $\equiv2456896.8707$ & BJD$_\mathrm{TBD}$ \\ 
$\sqrt{e}\cos{\omega}_{b}$ & $\equiv0.0$ & $\equiv0.0$ &  \\ 
$\sqrt{e}\sin{\omega}_{b}$ & $\equiv0.0$ & $\equiv0.0$ &  \\ 
$K_{b}$ & $2.55\pm0.78$ & $2.55$ & m s$^{-1}$ \\ 
$P_{c}$ & $\equiv10.561$ & $\equiv10.561$ & days \\ 
$T_{\mathrm{conj},c}$ & $\equiv2456900.4751$ & $\equiv2456900.4751$ & BJD$_\mathrm{TBD}$ \\ 
$\sqrt{e}\cos{\omega}_{c}$ & $\equiv0.0$ & $\equiv0.0$ &  \\
\enddata
\end{deluxetable}

We confirm with significantly higher precision the determinations in \cite{Sinukoff2016} that \bcgdPNAMEone is a dense (\bcgdRHOPone \gmc), rocky planet, while the density of \bcgdPNAMEtwo (\bcgdRHOPtwo \gmc) is intermediate between that of a solid planet and one dominated by gas.  

During the preparations of this paper, \cite{Toledo-Padron2020} published a Doppler analysis of 43 new ESPRESSO RVs and the 14 HIRES RVs from \cite{Sinukoff2016}.  They found masses of $7.3 \pm 1.1$ \mearth and $8.3 \pm 1.3$ \mearth and densities of 
$11.0^{+4.1}_{-2.8}$ \gmc and $3.8^{+1.8}_{-1.1}$ \gmc, respectively for the two planets.  We did not perform a global analysis that includes the ESPRESSO data.

\subsection{K2-73}  
\label{sec:k2_73}

\fffdSTNAME is a solar-type star in Field 3 with one transiting planet with a size of 2.3 \rearth and an orbital period of 7.5 days.
See Tables \ref{tb:star_pars}  and \ref{tb:star_props} for stellar properties and Table \ref{tb:planet_props} for precise planet parameters.
The planet was validated in the \cite{Crossfield2016} catalog and also noted in the catalogs of \cite{Vanderburg2016-catalog}, \cite{Schmitt2016}, \cite{Barros2016}, \cite{Mayo2018}, and \cite{Wittenmyer2018}.

\begin{figure*}
\epsscale{1.0}
\plotone{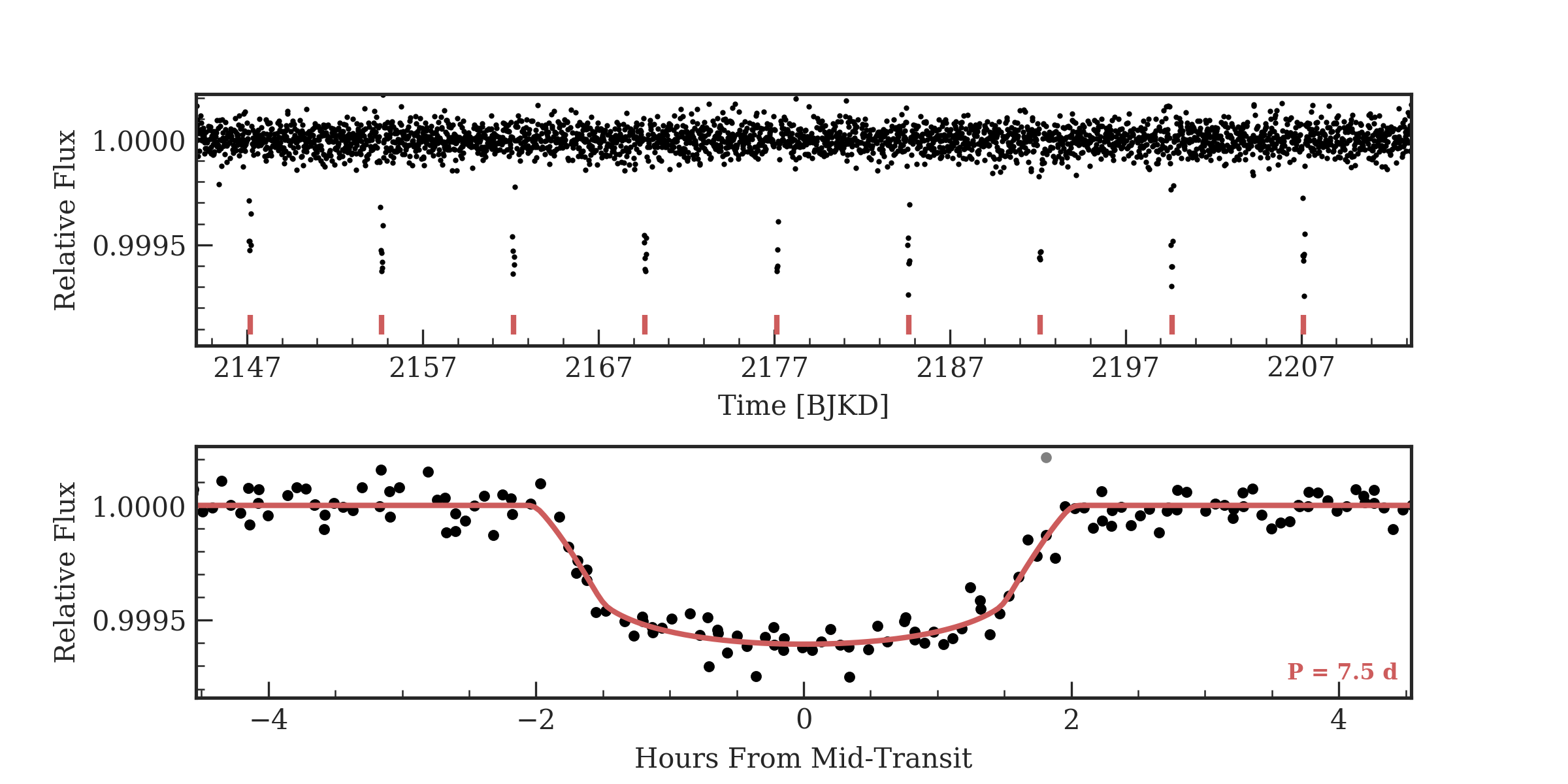}
\caption{Time series (top) and phase-folded (bottom) light curve for the planet orbiting \fffdSTNAME.  Plot formatting is the same as in Fig.\ \ref{fig:lc_epic220709978}.}
\label{fig:lc_epic206245553}
\end{figure*}

\begin{figure}
\epsscale{1.0}
\plotone{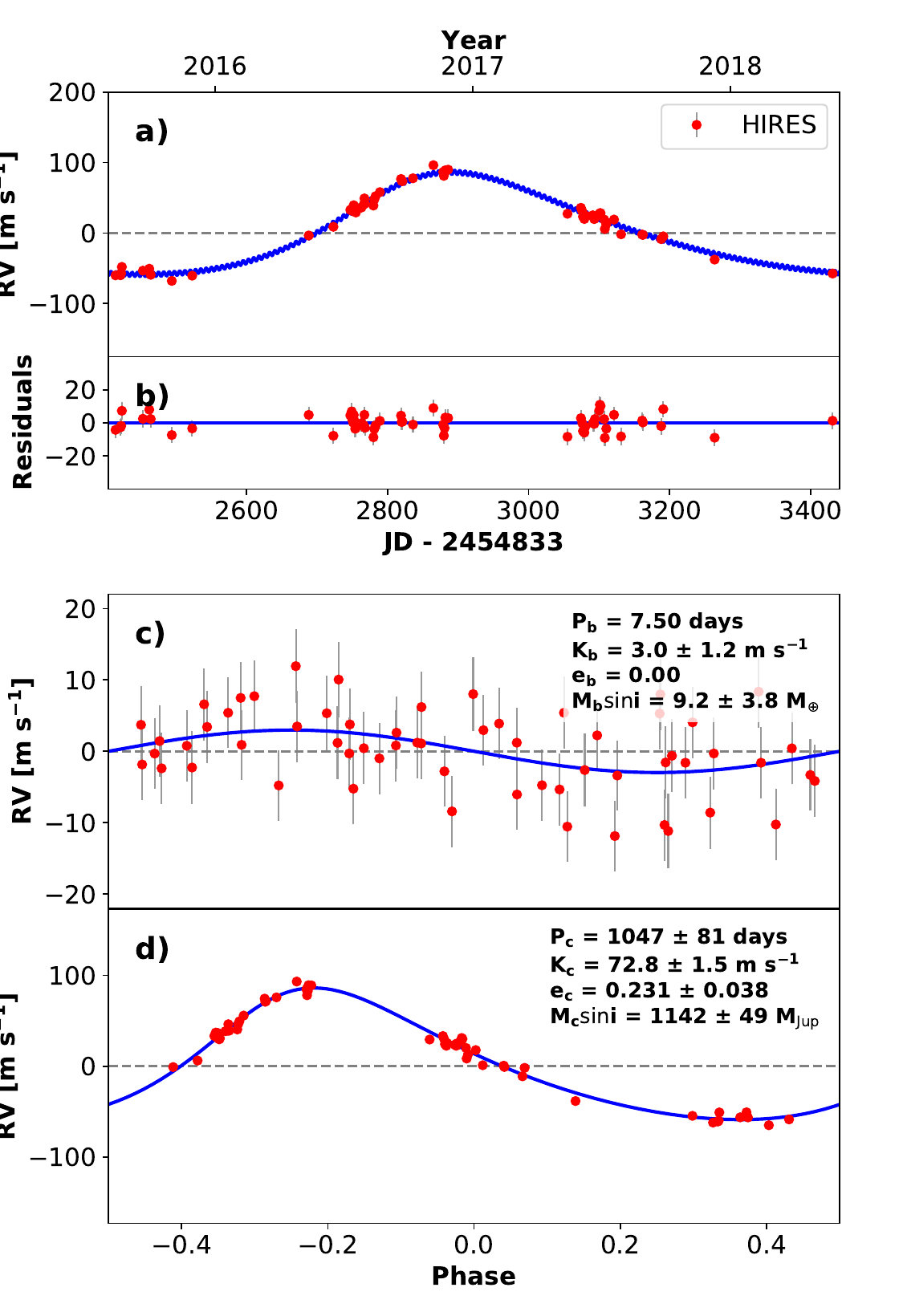}
\caption{RVs and Keplerian model for \fffdSTNAME.  
Symbols, lines, and annotations are similar to those in Fig.\ \ref{fig:rvs_epic220709978}.}
\label{fig:rvs_k2-73}
\end{figure}

Our fit of the EVEREST light curve of the K2 photometry for \fffdSTNAME is shown in Fig.\ \ref{fig:lc_epic206245553}.  We acquired \fffdNOBSHIRES RVs of \fffdSTNAME with HIRES, typically with an exposure meter setting of 80,000 counts.  The time series RVs are dominated by a large amplitude signal with a timescale of longer than one year.  We modeled the system as a transiting planet in a circular orbit with the orbital period and phase fixed to the transit ephemeris and a second, non-transiting planet with a long orbital period.  
The results of this analysis are listed in Table \ref{tab:epic206245553} and the best-fit model is shown in Fig.\ \ref{fig:rvs_k2-73}.
\fffdPNAMEone is a short-period sub-Neptune with a density intermediate between those expected for gas-dominated and rocky and rocky planets ($\rho_b=2.8^{+1.3}_{-1.2}$ g cm$^{-3}$).  
\fffdSTNAME\ c is a giant planet with a 2.7-year orbital perioda and \msini = \fffdMPtwo \mearth.
We also find a significant eccentricity for planet c (see Table Fig.\ \ref{fig:rvs_k2-73}).
Planet c has a predicted astrometric motion of 24 $\mu$as and should be detectable by Gaia.  Such a detection would measure the orbital inclination and determine if that planet is coplanar with the inner transiting planet b.


\begin{deluxetable}{lrrr}
\tablecolumns{4}
\tabletypesize{\scriptsize}
\tablecaption{ K2-73 System Parameters \label{tab:epic206245553}}
\tablehead{
  \colhead{Parameter} & 
  \colhead{Credible} & 
  \colhead{Maximum} & 
  \colhead{Units} \\
  \colhead{} & 
  \colhead{Interval} & 
  \colhead{Likelihood} & 
  \colhead{} }
\startdata
\sidehead{\bf{RV Analysis -- MCMC Step Parameters}} 
$P_{b}$ & $\equiv7.4956$ & $\equiv7.4956$ & days \\ 
$T_{\mathrm{conj},b}$ & $\equiv2456987.6725$ & $\equiv2456987.6725$ & BJD$_\mathrm{TBD}$ \\ 
$\sqrt{e}\cos{\omega}_{b}$ & $\equiv0.0$ & $\equiv0.0$ &  \\ 
$\sqrt{e}\sin{\omega}_{b}$ & $\equiv0.0$ & $\equiv0.0$ &  \\ 
$K_{b}$ & $3.0\pm1.2$ & $3.0$ & m s$^{-1}$ \\ 
$P_{c}$ & $1000\pm100$ & $1000$ & days \\ 
$T_{\mathrm{conj},c}$ & $2456905^{+62}_{-99}$ & $2456905$ & BJD$_\mathrm{TBD}$ \\ 
$\sqrt{e}\cos{\omega}_{c}$ & $0.397^{+0.065}_{-0.061}$ & $0.397$ &  \\
\enddata
\end{deluxetable}

\begin{figure}
\epsscale{1.0}
\plotone{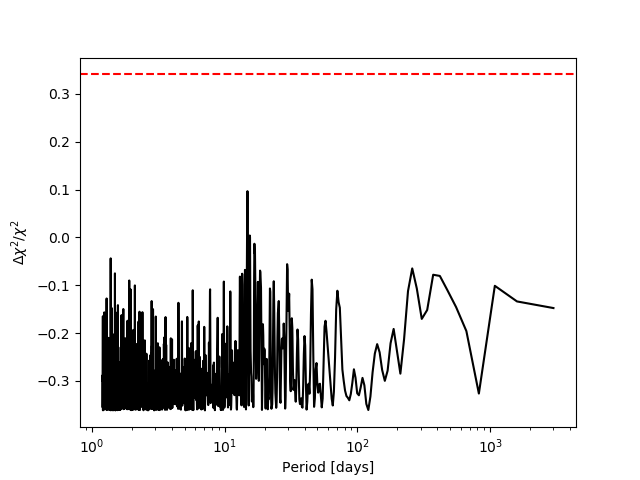}
\caption{Periodogram search of the RVs showing no evidence for a second planet orbiting \fffdSTNAME.  Lines and annotations are similar to those in Fig.\ \ref{fig:rvs_epic220709978_resid}.}
\label{fig:rvs_206245553_resid}
\end{figure}
\subsection{WASP-107} 
\label{sec:wasp_107}


\ecdcSTNAME was known to host a very low-density, short-period planet prior to observations by K2 in Campaign 10.
\cite{Anderson2017} discovered \ecdcPNAMEone prior to K2 observations based on photometry from the WASP survey.  They followed up the planet with 32 RVs from CORALIE and measured a mass of $38 \pm 3$ \mearth.  With a mass only 2.2 times that of Neptune (0.40 times that of Saturn), but a radius 0.94 times that of Jupiter, WASP-107b is among the lowest density gas giant planets.  It is a compelling target for transit spectroscopy because of the deep transit depth and large scale height of the planetary atmosphere.  The host star is a late K dwarf with \teff = \ecdcTEFF K. WASP-107b also resides in a highly misaligned orbit, as predicted by spot-crossing anomalies~\citep{Dai2017} and later confirmed by Rossiter-McLaughlin measurements~\citep{Rubenzahl2021}. This may be due to the dynamical influence of the outer companion as discussed in \citet{Rubenzahl2021}. The planet is notable as the first exoplanet with a detected Helium outflow in the metastable 1083~nm line \citep{Spake2018} and is an excellent target for transmission spectroscopy with {\em HST} \citep{Kreidberg2018} and JWST. Our adopted stellar parameters are in Tables \ref{tb:star_pars}, \ref{tb:star_props} and \ref{tb:planet_props}.  Unusually for a cool host star, a detailed set of elemental abundances has also been measured \citep{Hejazi2023}.

\begin{figure*}
\epsscale{1.0}
\plotone{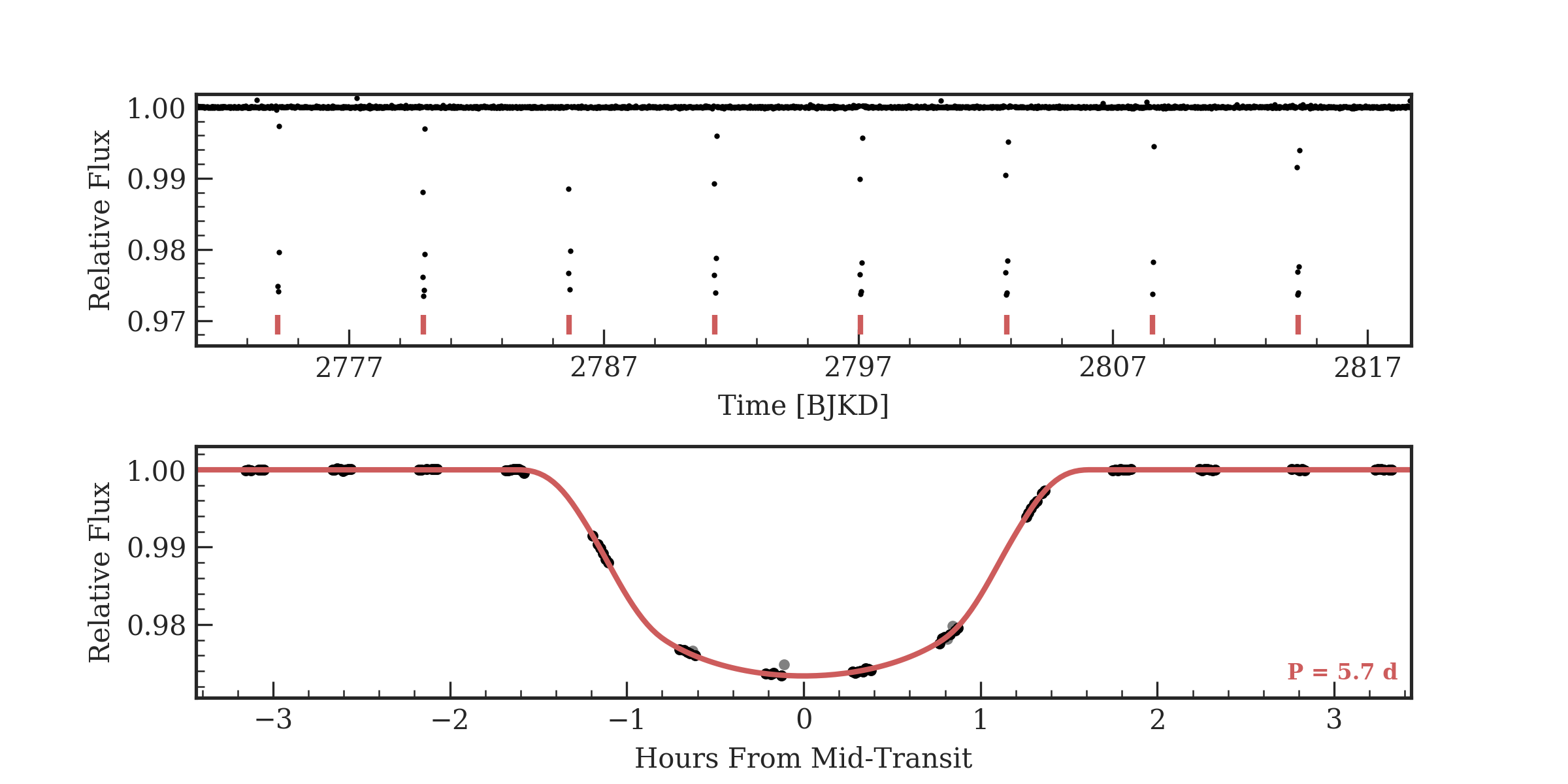}
\caption{Time series (top) and phase-folded (bottom) light curve for the planet orbiting \ecdcSTNAME.  Plot formatting is the same as in Fig.\ \ref{fig:lc_epic220709978}.}
\label{fig:lc_epic228724232}
\end{figure*}

We adopt the Keplerian solution of \citet{Piaulet2021}, who jointly fit our HIRES observations with the RVs from \cite{Anderson2017} and found a smaller and more precise mass for planet b ($30.5 \pm 1.7$ \mearth). The HIRES RVs show a significant long-period signature which \citet{Piaulet2021} found is best modeled by a second planet with $M_c\sin i = 0.36~M_J$ on a wide $\sim 1090$~day eccentric $e_c = 0.28$ orbit. An AICc comparison between circular and eccentric fits for planet b showed a significant preference for the circular model, consistent with the small eccentricity ($0.06 \pm 0.04$) reported by \citet{Piaulet2021} and the expected bias for $e$ given a prior bounded below at 0.



\subsection{K2-66} 


\dcbjSTNAME is a slightly evolved solar-temperature star from Campaign 3 with one transiting planet with a size of 2.4 \rearth and an orbital period of 5 days.
See Tables \ref{tb:star_pars}  and \ref{tb:star_props} for stellar properties and Table \ref{tb:planet_props} for precise planet parameters.
\dcbjSTNAME was first reported in \cite{Vanderburg2016-catalog} and was later statistically validated by \cite{Crossfield2016}. 
\cite{Sinukoff2017b} measured the mass of $21.3 \pm 3.6$ \mearth using 38 HIRES RVs.  We provide an update to that measurement here.  
  
\begin{figure*}
\epsscale{1.0}
\plotone{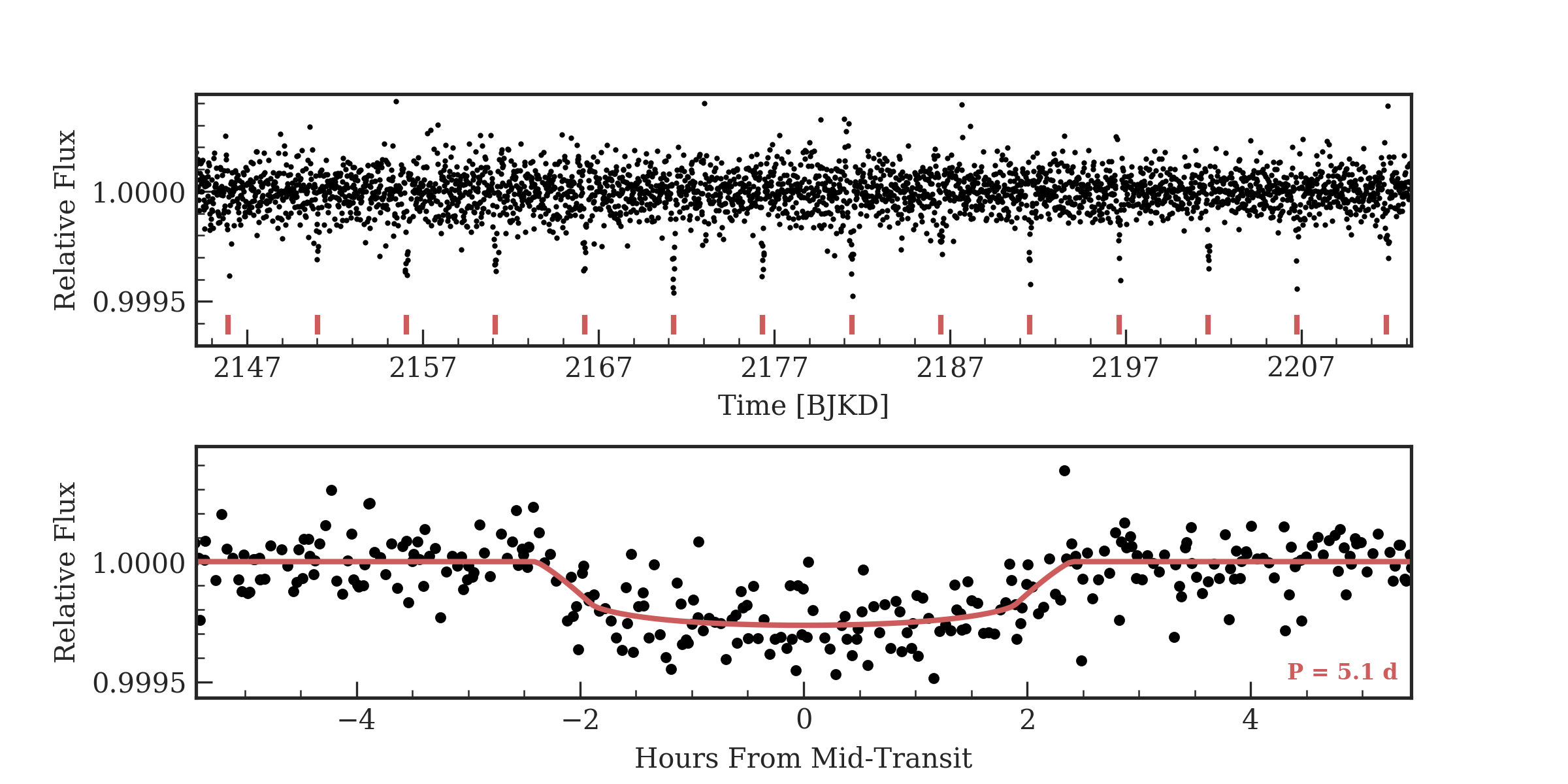}
\caption{Time series (top) and phase-folded (bottom) light curve for the planet orbiting \dcbjSTNAME.  Plot formatting is the same as in Fig.\ \ref{fig:lc_epic220709978}.}
\label{fig:lc_epic206153219}
\end{figure*}

\begin{figure}
\epsscale{1.0}
\plotone{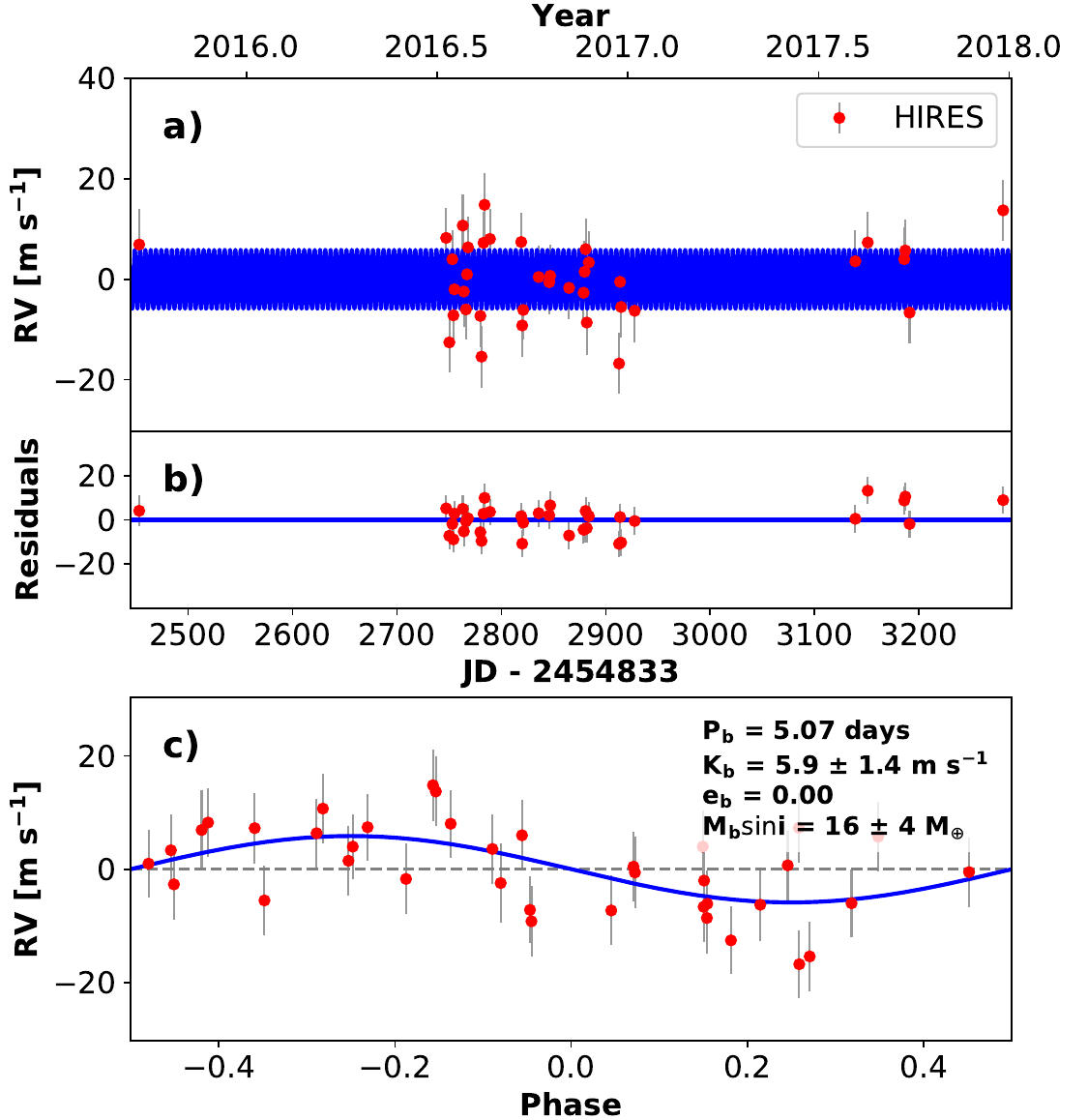}
\caption{RVs and Keplerian model for \dcbjSTNAME. Symbols, lines, and annotations are similar to those in Fig.\ \ref{fig:rvs_epic220709978}.}
\label{fig:rvs_epic206153219}
\end{figure}

Our fit of the EVEREST light curve of the K2 photometry for \dcbjSTNAME is shown in Fig.\ \ref{fig:lc_epic206153219}.  We acquired \dcbjNOBSHIRES RVs with HIRES (38 of which were reported in \citet{Sinukoff2017b}), typically with an exposure meter setting of 50,000 counts.  
We modeled the system as a single planet in a circular orbit with orbital period and phase fixed to transit ephemeris.  The results of this analysis are listed in Table \ref{tab:epic206153219} and the best-fit model is shown in Fig.\ \ref{fig:rvs_epic206153219}.  We considered other models including ones with an eccentric orbit and/or a linear RV trend.  The model with a linear trend had a slightly lower AICc statistic, but the difference (\dAICc = 1.7) was insufficient to justify the additional complexity of the model.  \dcbjPNAMEone is a massive sub-Neptune (\dcbjMPone \mearth).


\begin{deluxetable}{lrrr}
\tablecolumns{4}
\tabletypesize{\scriptsize}
\tablecaption{ K2-66 System Parameters \label{tab:epic206153219}}
\tablehead{
  \colhead{Parameter} & 
  \colhead{Credible} & 
  \colhead{Maximum} & 
  \colhead{Units} \\
  \colhead{} & 
  \colhead{Interval} & 
  \colhead{Likelihood} & 
  \colhead{} }
\startdata
\sidehead{\bf{RV Analysis -- MCMC Step Parameters}} 
$P_{b}$ & $\equiv5.0694$ & $\equiv5.0694$ & days \\ 
$T_{\mathrm{conj},b}$ & $\equiv2456984.0075$ & $\equiv2456984.0075$ & BJD$_\mathrm{TBD}$ \\ 
$\sqrt{e}\cos{\omega}_{b}$ & $\equiv0.0$ & $\equiv0.0$ &  \\ 
$\sqrt{e}\sin{\omega}_{b}$ & $\equiv0.0$ & $\equiv0.0$ &  \\ 
$K_{b}$ & $5.9^{+1.3}_{-1.4}$ & $5.9$ & m s$^{-1}$ \\ 
$\gamma_{\rm HIRES}$ & $-1.6\pm1.0$ & $-2.0$ & m s$-1$ \\ 
$\dot{\gamma}$ & $\equiv0.0$ & $\equiv0.0$ & m s$^{-1}$ day$^{-1}$ \\ 
$\ddot{\gamma}$ & $\equiv0.0$ & $\equiv0.0$ & m s$^{-1}$ day$^{-2}$ \\ 
$\sigma_{\rm HIRES}$ & $5.89^{+1.30}_{-0.40}$ & $5.60$ & $\rm m\ s^{-1}$ \\ 
\hline
\sidehead{\bf{Orbital \& Physical Parameters}}
$P_{b}$ & $\equiv5.0694$ & $\equiv5.0694$ & days \\
$T_{\mathrm{conj},b}$ & $\equiv2456984.0075$ & $\equiv2456984.0075$ & BJD$_\mathrm{TBD}$ \\
$e_{b}$ & $\equiv0.0$ & $\equiv0.0$ &  \\
$\omega_{b}$ & $\equiv0.0$ & $\equiv0.0$ & radians \\
$K_{b}$ & $5.9^{+1.3}_{-1.4}$ & $5.9$ & m s$^{-1}$ \\
$M_b$ & $16^{+3.8}_{-4.1}$ & $16.5$ & $M_\earth$ \\
$R_b/R_*$ & $0.01515^{+0.00110}_{-0.00030}$ & $0.01490$ &  \\
$\rho_b$ & $4.2^{+1.8}_{-0.7}$ & $3.6$ & g cm$^{-3}$ \\
$R_b$ & $\equiv2.7545$ & $\equiv2.7545$ & $R_\earth$ \\
\hline
\hline
\sidehead{\bf{Priors}}
Parameter & Prior \\
\hline
\\None \\
\enddata
\end{deluxetable}

\subsection{K2-36} 

\begin{figure*}
\epsscale{1.0}
\plotone{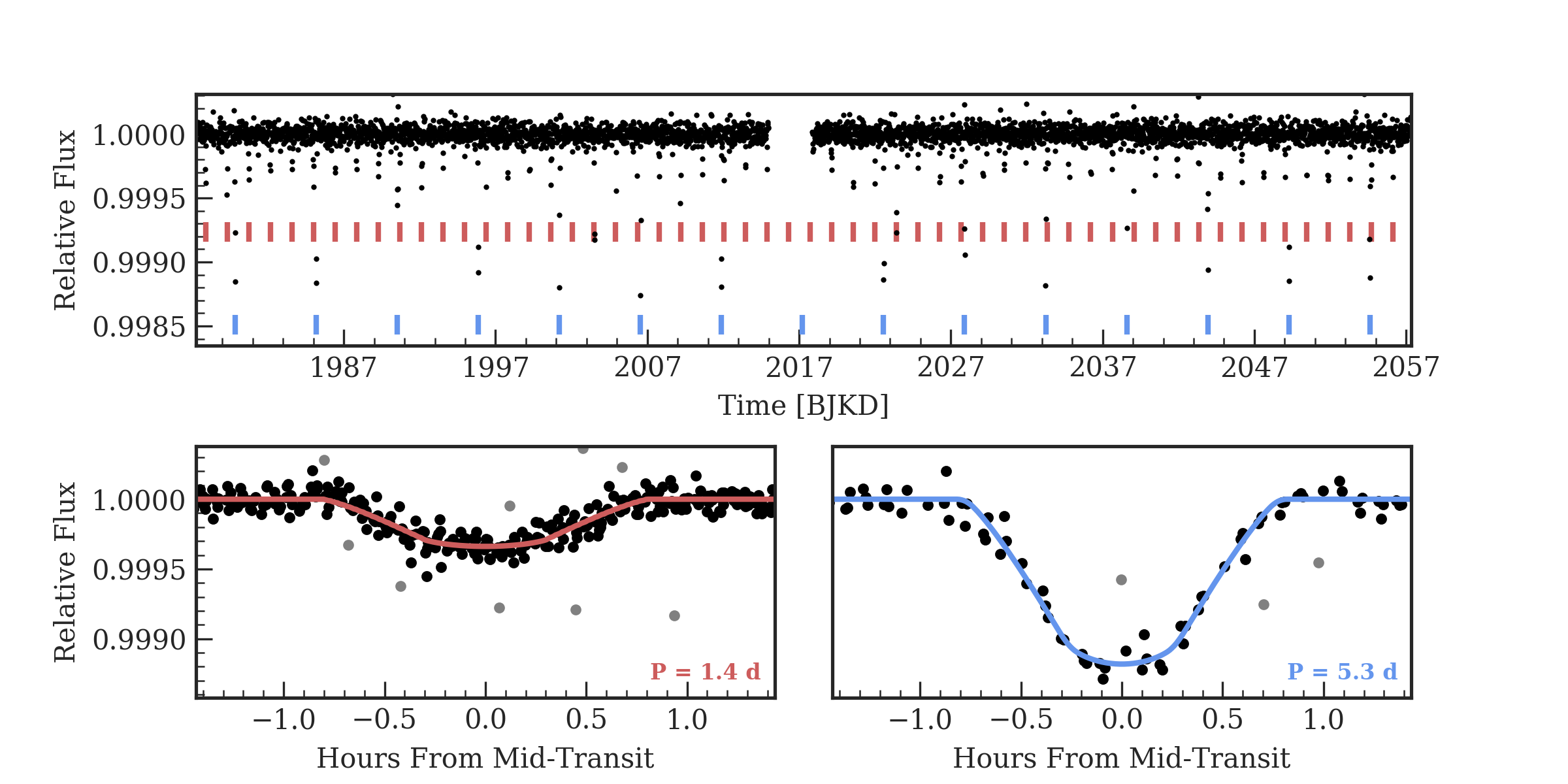}
\caption{Time series (top) and phase-folded (bottom) light curve for the planet orbiting \ddeiSTNAME.  Plot formatting is the same as in Fig.\ \ref{fig:lc_epic220709978}.}
\label{fig:lc_epic201713348}
\end{figure*}

\ddeiSTNAME is a late-K dwarf with two transiting planets that have sizes of 1.4 \rearth and 2.6 \rearth and orbital periods of 1.4 days and 5.3 days.
See Tables \ref{tb:star_pars}  and \ref{tb:star_props} for stellar properties and Table \ref{tb:planet_props} for precise planet parameters.
The planets were statistically validated in \cite{Crossfield2016} and also noted in \cite{Vanderburg2016-catalog} and \cite{Sinukoff2016}.  Our fit of the EVEREST light curve of the K2 photometry for \ddeiSTNAME is shown in Fig.\ \ref{fig:lc_epic201713348}. \cite{Damasso2019} characterized the system using 81 RVs obtained with HARPS-N and found masses of $3.9 \pm 1.1$~\mearth and $7.8 \pm 2.3$~\mearth for planets b and c, respectively, consistent with the values subsequently reported by \cite{Bonomo2023} of $4.3 \pm1.4 M_\oplus$ and $7.9 \pm 2.8 M_\oplus$, respectively.

We acquired \ddeiNOBSHIRES RVs of \ddeiSTNAME with HIRES, typically with an exposure meter setting of 80,000 counts.  The star is moderately active with \lrphk = \ddeiRPHK, which, combined with our $N_\mathrm{obs,HIRES}=$\ddeiNOBSHIRES, satisfies our prerequisites for a GP regression analysis. 
As described in Sec.\ \ref{sec:gp_modeling}, we trained the GP hyperparameters on non-detrended Everest photometry before using computing RV orbit posteriors. For comparison purposes, we also performed an RV orbit fit using an untrained GP. The trained hyperparameters are clearly peaked and constrained to a portion of the parameter space allowed by the priors, whereas the untrained hyperparameter posteriors extend over the entire allowable parameter space. Without training, $\lambda$ is peaked at $\approx2$ days, between the orbital period values of the two planets, and near half the value of the period of planet b. With training, however, the values of both timescale parameters are significantly larger than either planet period. For both planets, a trained GP reduces the uncertainties on semiamplitudes by more than $50\%$ and favors higher median semiamplitude values, as compared with a non-GP model. This highlights not only a case where using GP regression cuts down on orbital parameter uncertainty, but also one where GP training provides a well-motivated prior keeping the timescale parameters away from the planet orbital periods. 

\begin{figure}
\epsscale{1.0}
\plotone{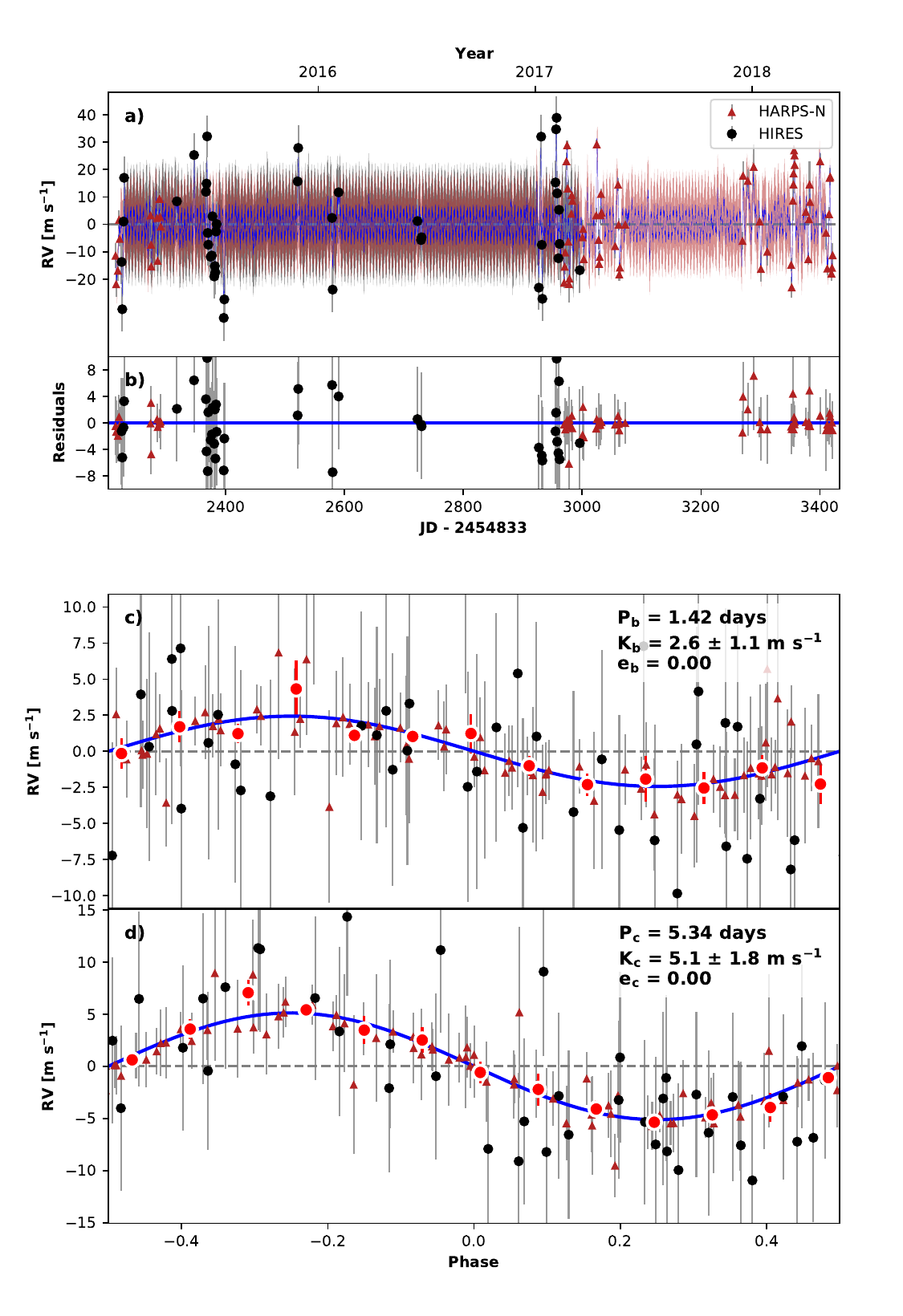}
\caption{RVs and Keplerian model for \ddeiSTNAME. Symbols, lines, and annotations are similar to those in Fig.\ \ref{fig:rvs_epic220709978}.}
\label{fig:rvs_epic201713348}
\end{figure}

Our GP analysis, combined with the higher-precision HIRES RVs, yields larger mass determinations for both planets than previous work \citep{Damasso2019,Bonomo2023}. A model with a circular orbit and no trend is preferred based on an AICc comparison. The results of this analysis are listed in Table \ref{tab:epic201713348} 
and the best-fit model is shown in Fig.\ \ref{fig:rvs_epic201713348}.
\ddeiPNAMEone is a short-period super-Earth with a high (but not precisely determined) density (\ddeiRHOPone \gmc) suggesting a rocky composition. The larger mass for \ddeiPNAMEtwo makes it more like typical sub-Neptunes, as its 2.2 \rearth radius suggests an appreciable H/He envelope. 

\begin{deluxetable}{lrrr}
\tablecolumns{4}
\tabletypesize{\scriptsize}
\tablecaption{ K2-36 System Parameters \label{tab:epic201713348}}
\tablehead{
  \colhead{Parameter} & 
  \colhead{Credible} & 
  \colhead{Maximum} & 
  \colhead{Units} \\
  \colhead{} & 
  \colhead{Interval} & 
  \colhead{Likelihood} & 
  \colhead{} }
\startdata
\sidehead{\bf{RV Analysis -- MCMC Step Parameters}} 
$P_{b}$ & $\equiv1.4226$ & $\equiv1.4226$ & days \\ 
$T_{\mathrm{conj},b}$ & $\equiv2456827.963$ & $\equiv2456827.963$ & BJD$_\mathrm{TBD}$ \\ 
$\sqrt{e}\cos{\omega}_{b}$ & $\equiv0.0$ & $\equiv0.0$ &  \\ 
$\sqrt{e}\sin{\omega}_{b}$ & $\equiv0.0$ & $\equiv0.0$ &  \\ 
$K_{b}$ & $4\pm3$ & $3.3$ & m s$^{-1}$ \\ 
$P_{c}$ & $\equiv5.3408$ & $\equiv5.3408$ & days \\ 
$T_{\mathrm{conj},c}$ & $\equiv2456812.841$ & $\equiv2456812.841$ & BJD$_\mathrm{TBD}$ \\ 
$\sqrt{e}\cos{\omega}_{c}$ & $\equiv0.0$ & $\equiv0.0$ &  \\
\enddata
\end{deluxetable}

\subsection{K2-105} 
\label{sec:k2_105}

\begin{figure*}
\epsscale{1.0}
\plotone{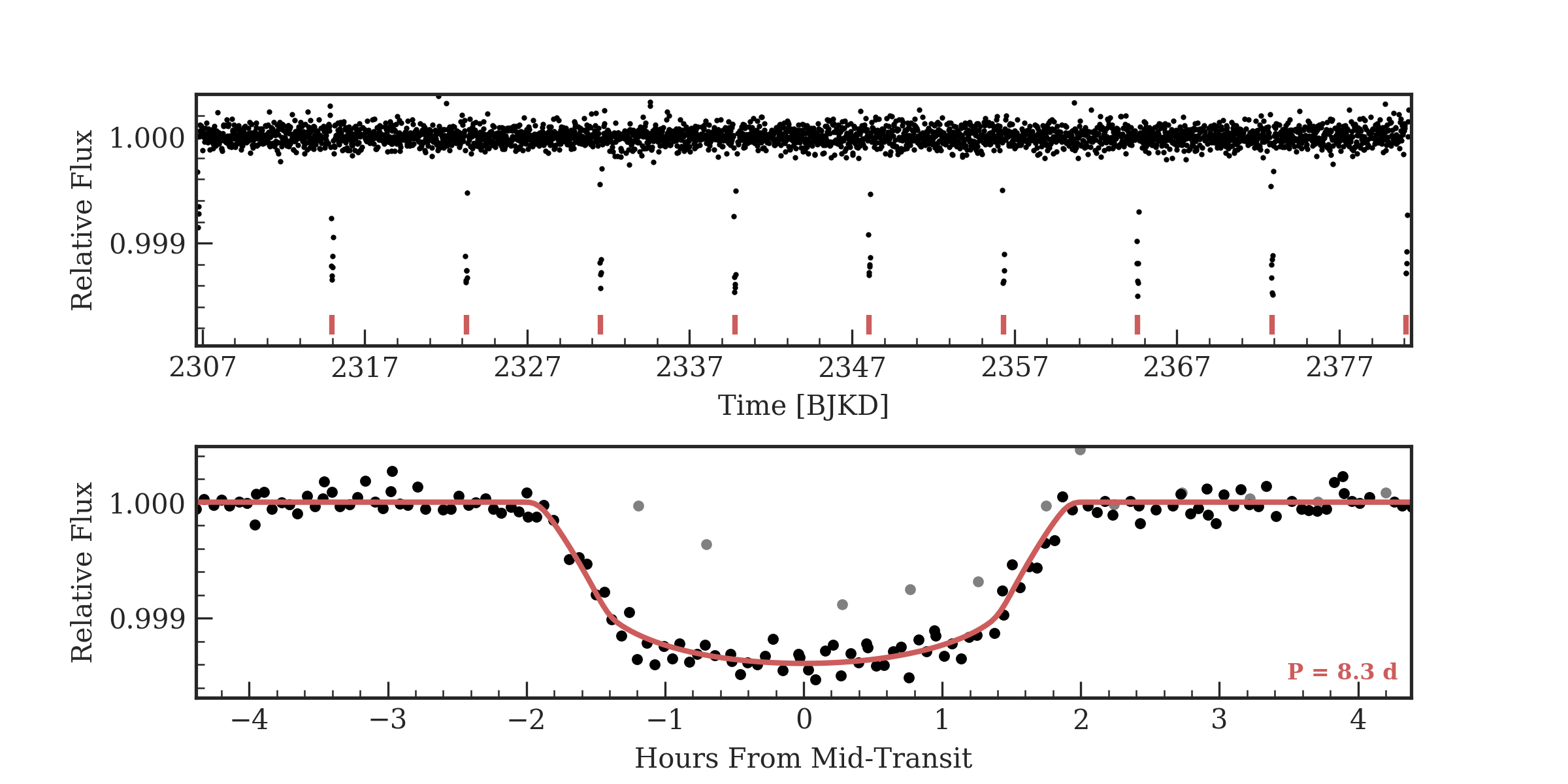}
\caption{Time series (top) and phase-folded (bottom) light curve for the planet orbiting \fdijSTNAME.  Plot formatting is the same as in Fig.\ \ref{fig:lc_epic220709978}.}
\label{fig:lc_epic211525389}
\end{figure*}

\fdijSTNAME is a late G dwarf in Field 5 with one transiting planet with a radius of 3.4 \rearth and an orbital period of 8 days.
See Tables \ref{tb:star_pars}  and \ref{tb:star_props} for stellar properties and Table \ref{tb:planet_props} for precise planet parameters.
The planet was discovered by \cite{Narita2017} who estimated a mass upper limit of 90 \mearth (3-$\sigma$) based on eight RVs from Subaru/HDS gathered in three clusters.  The planet is also noted in the \cite{Petigura2018} and \cite{Mayo2018} catalogs. HARPS-N transit spectroscopy hints at a misaligned orbit \citep{Bourrier2023} and GIANO observations detect no sign of mass loss via the metastable Helium line \citep{Guilluy2023}. Our fit of the EVEREST light curve of the K2 photometry for \fdijSTNAME is shown in Fig.\ \ref{fig:lc_epic211525389}.

\begin{figure}
\epsscale{1.0}
\plotone{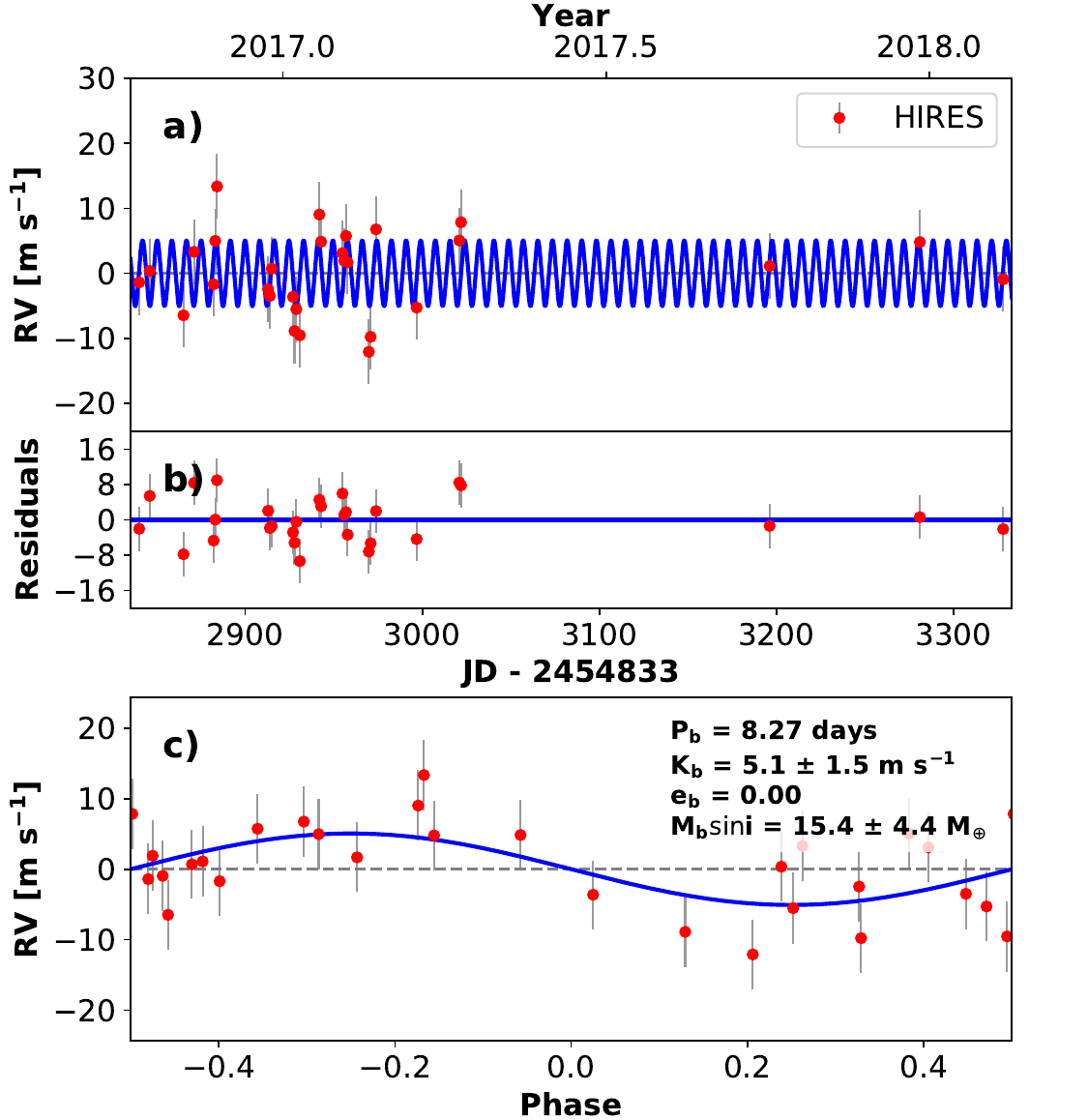}
\caption{RVs and Keplerian model for \fdijSTNAME. Symbols, lines, and annotations are similar to those in Fig.\ \ref{fig:rvs_epic220709978}.}
\label{fig:rvs_k2-105}
\end{figure}

We acquired \fdijNOBSHIRES RVs of \fdijSTNAME with HIRES, typically with an exposure meter setting of 80,000 counts.  We modeled the system as a single planet in a circular orbit with the orbital period and phased fixed to the transit ephemeris.  The results of this analysis are listed in Table \ref{tab:epic211525389} and the best fit model is shown in Fig.\ \ref{fig:rvs_k2-105}.  We considered more complicated models but found insufficient evidence to justify inclusion of orbital eccentricity or a linear RV trend based on the AICc statistic.  \fdijPNAMEone is a short-period planet with a size and density comparable to Neptune's.

\begin{deluxetable}{lrrr}
\tablecolumns{4}
\tabletypesize{\scriptsize}
\tablecaption{ K2-105 System Parameters \label{tab:epic211525389}}
\tablehead{
  \colhead{Parameter} & 
  \colhead{Credible} & 
  \colhead{Maximum} & 
  \colhead{Units} \\
  \colhead{} & 
  \colhead{Interval} & 
  \colhead{Likelihood} & 
  \colhead{} }
\startdata
\sidehead{\bf{RV Analysis -- MCMC Step Parameters}} 
$P_{b}$ & $\equiv8.2673$ & $\equiv8.2673$ & days \\ 
$T_{\mathrm{conj},b}$ & $\equiv2457147.9887$ & $\equiv2457147.9887$ & BJD$_\mathrm{TBD}$ \\ 
$\sqrt{e}\cos{\omega}_{b}$ & $\equiv0.0$ & $\equiv0.0$ &  \\ 
$\sqrt{e}\sin{\omega}_{b}$ & $\equiv0.0$ & $\equiv0.0$ &  \\ 
$K_{b}$ & $5.1^{+1.4}_{-1.5}$ & $5.1$ & m s$^{-1}$ \\ 
$\gamma_{\rm HIRES}$ & $-1\pm1$ & $-1$ & m s$-1$ \\ 
$\dot{\gamma}$ & $\equiv0.0$ & $\equiv0.0$ & m s$^{-1}$ day$^{-1}$ \\ 
$\ddot{\gamma}$ & $\equiv0.0$ & $\equiv0.0$ & m s$^{-1}$ day$^{-2}$ \\ 
$\sigma_{\rm HIRES}$ & $5.1^{+1.2}_{-0.3}$ & $4.8$ & $\rm m\ s^{-1}$ \\ 
\hline
\sidehead{\bf{Orbital \& Physical Parameters}}
$P_{b}$ & $\equiv8.2673$ & $\equiv8.2673$ & days \\
$T_{\mathrm{conj},b}$ & $\equiv2457147.9887$ & $\equiv2457147.9887$ & BJD$_\mathrm{TBD}$ \\
$e_{b}$ & $\equiv0.0$ & $\equiv0.0$ &  \\
$\omega_{b}$ & $\equiv0.0$ & $\equiv0.0$ & radians \\
$K_{b}$ & $5.1^{+1.4}_{-1.5}$ & $5.1$ & m s$^{-1}$ \\
$M_b$ & $15.4^{+4.3}_{-4.5}$ & $15.5$ & $M_\earth$ \\
$R_b/R_*$ & $0.0339^{+0.0003}_{-0.0015}$ & $0.0349$ &  \\
$\rho_b$ & $2.09^{+0.65}_{-0.59}$ & $2.07$ & g cm$^{-3}$ \\
$R_b$ & $3.40\pm0.10$ & $3.40$ & $R_\earth$ \\
\hline
\hline
\sidehead{\bf{Priors}}
Parameter & Prior \\
\hline
\\None \\
\enddata
\end{deluxetable}

\subsection{K2-214} 

\begin{figure*}
\epsscale{1.0}
\plotone{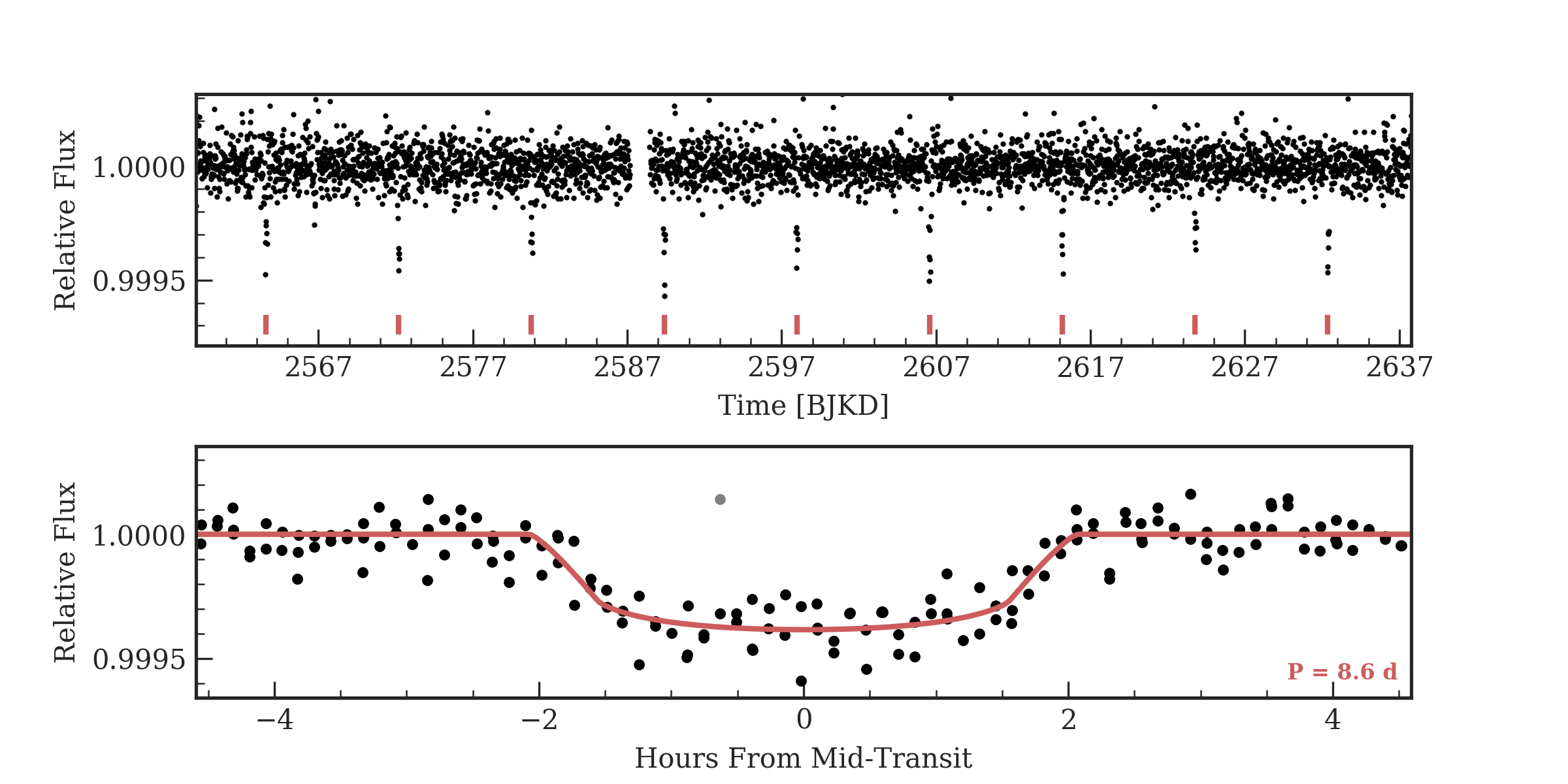}
\caption{Time series (top) and phase-folded (bottom) light curve for the planet orbiting \gafeSTNAME.  Plot formatting is the same as in Fig.\ \ref{fig:lc_epic220709978}.}
\label{fig:lc_epic220376054}
\end{figure*}

\gafeSTNAME is a slightly evolved, solar-temperature star from Campaign 8 with one transiting planet with a 2.5 \rearth radius and an 8.5-day orbital period.
See Tables \ref{tb:star_pars}  and \ref{tb:star_props} for stellar properties and Table \ref{tb:planet_props} for precise planet parameters.
The star is noted in the catalogs of \cite{Petigura2018} and \cite{Mayo2018}, and validated in the latter.
Our fit of the EVEREST light curve of the K2 photometry for \gafeSTNAME is shown in Fig.\ \ref{fig:lc_epic220376054}.  

\begin{figure}
\epsscale{1.0}
\plotone{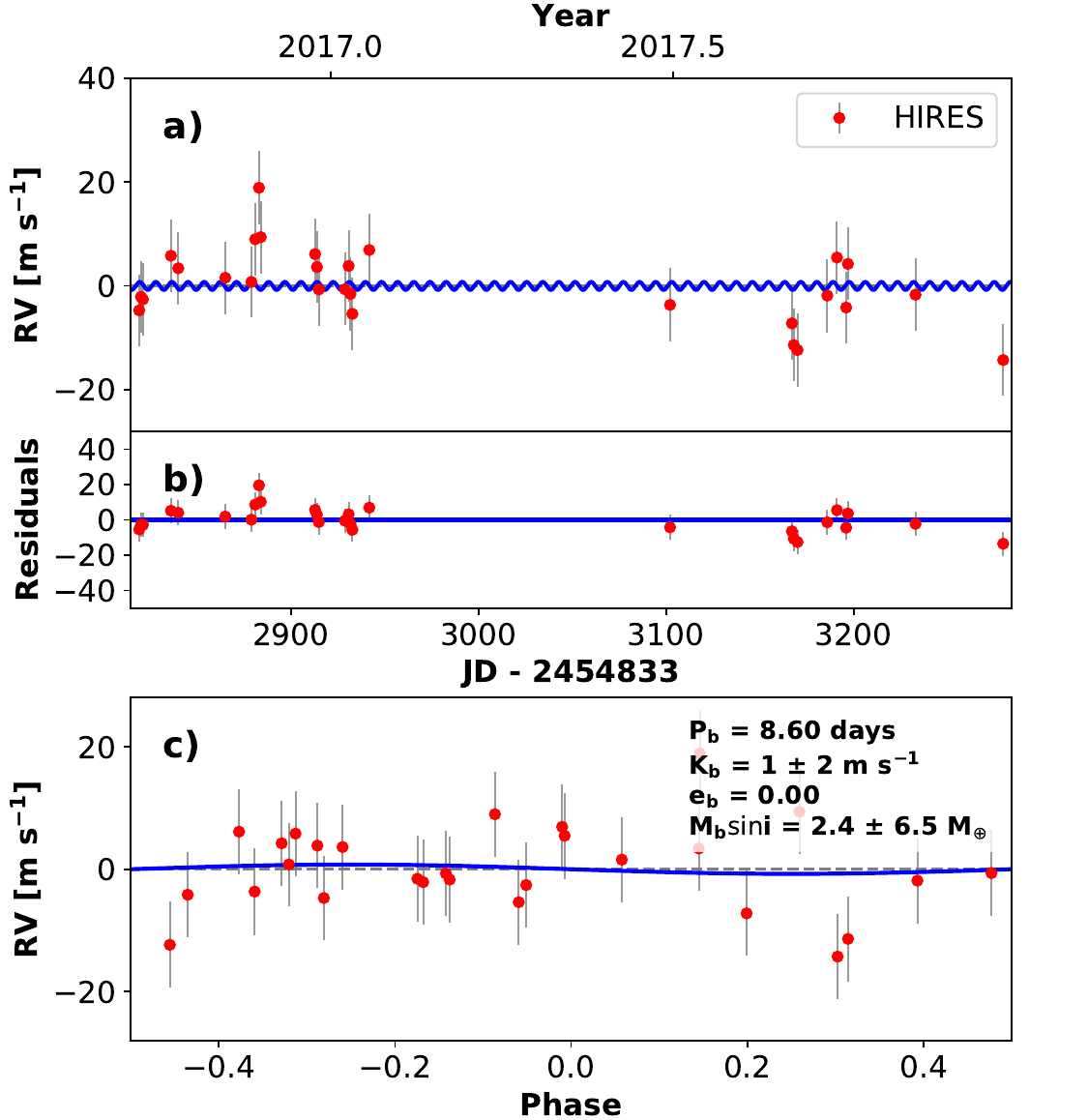}
\caption{RVs and Keplerian model for \gafeSTNAME.  Symbols, lines, and annotations are similar to those in Fig.\ \ref{fig:rvs_epic220709978}.}
\label{fig:rvs_k2-214}
\end{figure}

We acquired \gafeNOBSHIRES RVs of \gafeSTNAME with HIRES, typically with an exposure meter setting of 80,000 counts.  We modeled the system as a single planet in a circular orbit whose period and phase are fixed to the transit ephemeris.  The results of this analysis are listed in Table \ref{tab:epic220376054} and the adopted model is shown in Fig.\  \ref{fig:rvs_k2-214}.
While the preferred model is a circular orbit with a trend, \gafePNAMEone is poorly characterized by our HIRES measurements due to high jitter (see Table \ref{tab:epic220376054}) and the planet is not detected in RVs. As a result, we adopt a circular model with no trend and provide the best-fit semiamplitude as an upper limit. This system does not have enough observations to meet our requirements for a GP analysis, thus more observations are needed to better understand \gafeSTNAME.

\begin{deluxetable}{lrrr}
\tablecolumns{4}
\tabletypesize{\scriptsize}
\tablecaption{ K2-214 System Parameters \label{tab:epic220376054}}
\tablehead{
  \colhead{Parameter} & 
  \colhead{Credible} & 
  \colhead{Maximum} & 
  \colhead{Units} \\
  \colhead{} & 
  \colhead{Interval} & 
  \colhead{Likelihood} & 
  \colhead{} }
\startdata
\sidehead{\bf{RV Analysis -- MCMC Step Parameters}} 
$P_{b}$ & $\equiv8.5966$ & $\equiv8.5966$ & days \\ 
$T_{\mathrm{conj},b}$ & $\equiv2457396.6009$ & $\equiv2457396.6009$ & BJD$_\mathrm{TBD}$ \\ 
$\sqrt{e}\cos{\omega}_{b}$ & $\equiv0.0$ & $\equiv0.0$ &  \\ 
$\sqrt{e}\sin{\omega}_{b}$ & $\equiv0.0$ & $\equiv0.0$ &  \\ 
$K_{b}$ & $1^{+1.7}_{-2.3}$ & $1.0$ & m s$^{-1}$ \\ 
$\gamma_{\rm HIRES}$ & $-0.6\pm1.5$ & $-0.6$ & m s$-1$ \\ 
$\dot{\gamma}$ & $\equiv0.0$ & $\equiv0.0$ & m s$^{-1}$ day$^{-1}$ \\ 
$\ddot{\gamma}$ & $\equiv0.0$ & $\equiv0.0$ & m s$^{-1}$ day$^{-2}$ \\ 
$\sigma_{\rm HIRES}$ & $7.2^{+1.7}_{-0.5}$ & $6.7$ & $\rm m\ s^{-1}$ \\ 
\hline
\sidehead{\bf{Orbital \& Physical Parameters}}
$P_{b}$ & $\equiv8.5966$ & $\equiv8.5966$ & days \\
$T_{\mathrm{conj},b}$ & $\equiv2457396.6009$ & $\equiv2457396.6009$ & BJD$_\mathrm{TBD}$ \\
$e_{b}$ & $\equiv0.0$ & $\equiv0.0$ &  \\
$\omega_{b}$ & $\equiv0.0$ & $\equiv0.0$ & radians \\
$K_{b}$ & $1^{+1.7}_{-2.3}$ & $1.0$ & m s$^{-1}$ \\
$M_b$ & $2.4^{+5.7}_{-7.3}$ & $3.3$ & $M_\earth$ \\
$R_b/R_*$ & $0.018^{+0.002}_{-0.000}$ & $0.018$ &  \\
$\rho_b$ & $0.8^{+1.7}_{-2.7}$ & $1.3$ & g cm$^{-3}$ \\
$R_b$ & $\equiv2.5122$ & $\equiv2.5122$ & $R_\earth$ \\
\hline
\hline
\sidehead{\bf{Priors}}
Parameter & Prior \\
\hline
\\None \\
\enddata
\end{deluxetable}

\subsection{K2-220} 


\bhiiSTNAME is a G dwarf from Campaign 8 with one transiting planet with a 2.3-\rearth radius and a 13-day orbital period.  
See Tables \ref{tb:star_pars} and \ref{tb:star_props} for stellar properties and Table \ref{tb:planet_props} for precise planet parameters.  
The planet was noted in the catalogs by \cite{Petigura2018} and \cite{Mayo2018}, with validation in the latter.

\begin{figure*}
\epsscale{1.0}
\plotone{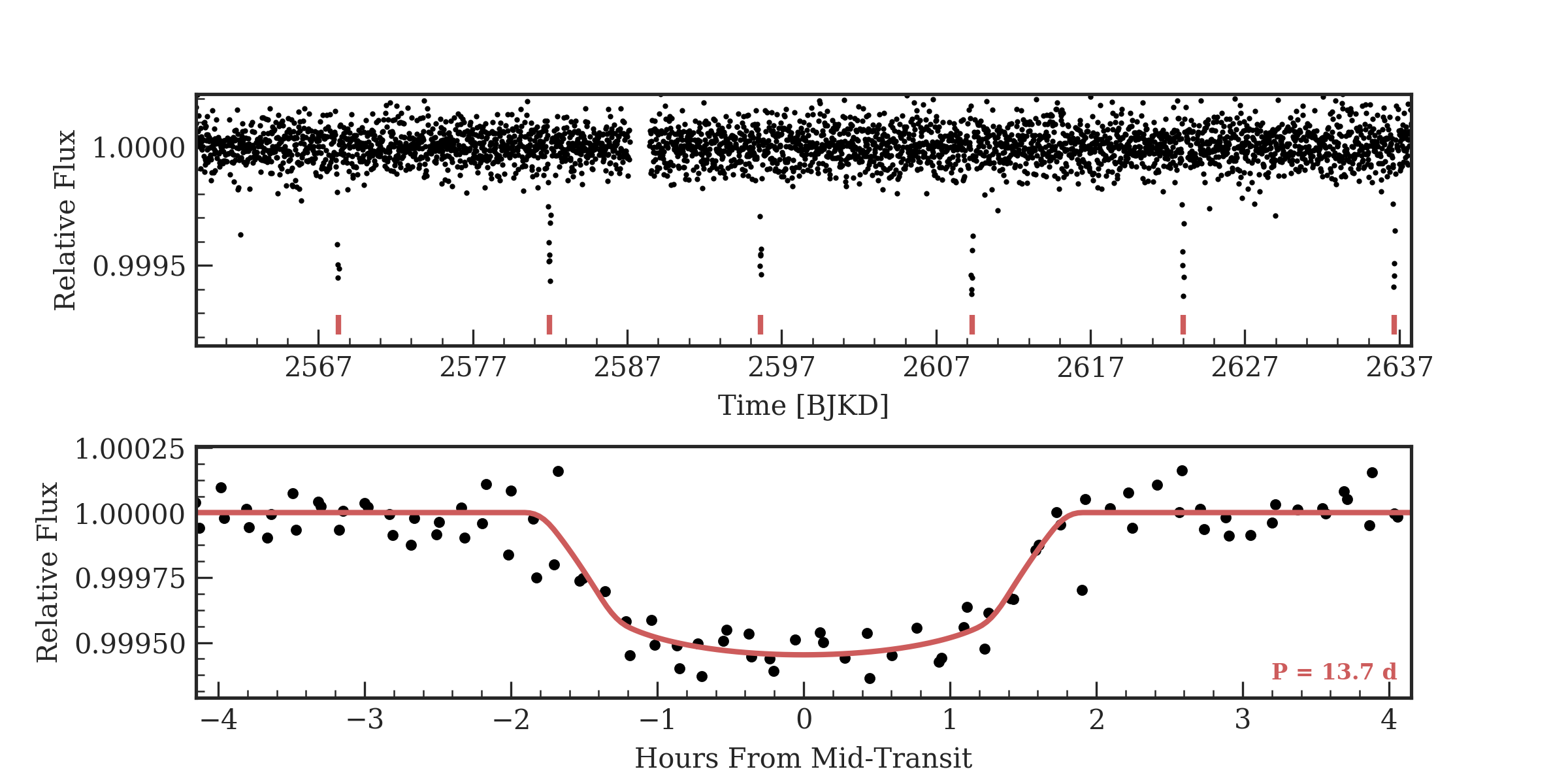}
\caption{Time series (top) and phase-folded (bottom) light curve for the planet orbiting \bhiiSTNAME.  Plot formatting is the same as in Fig.\ \ref{fig:lc_epic220709978}.}
\label{fig:lc_epic220621788}
\end{figure*}

\begin{figure}
\epsscale{1.0}
\plotone{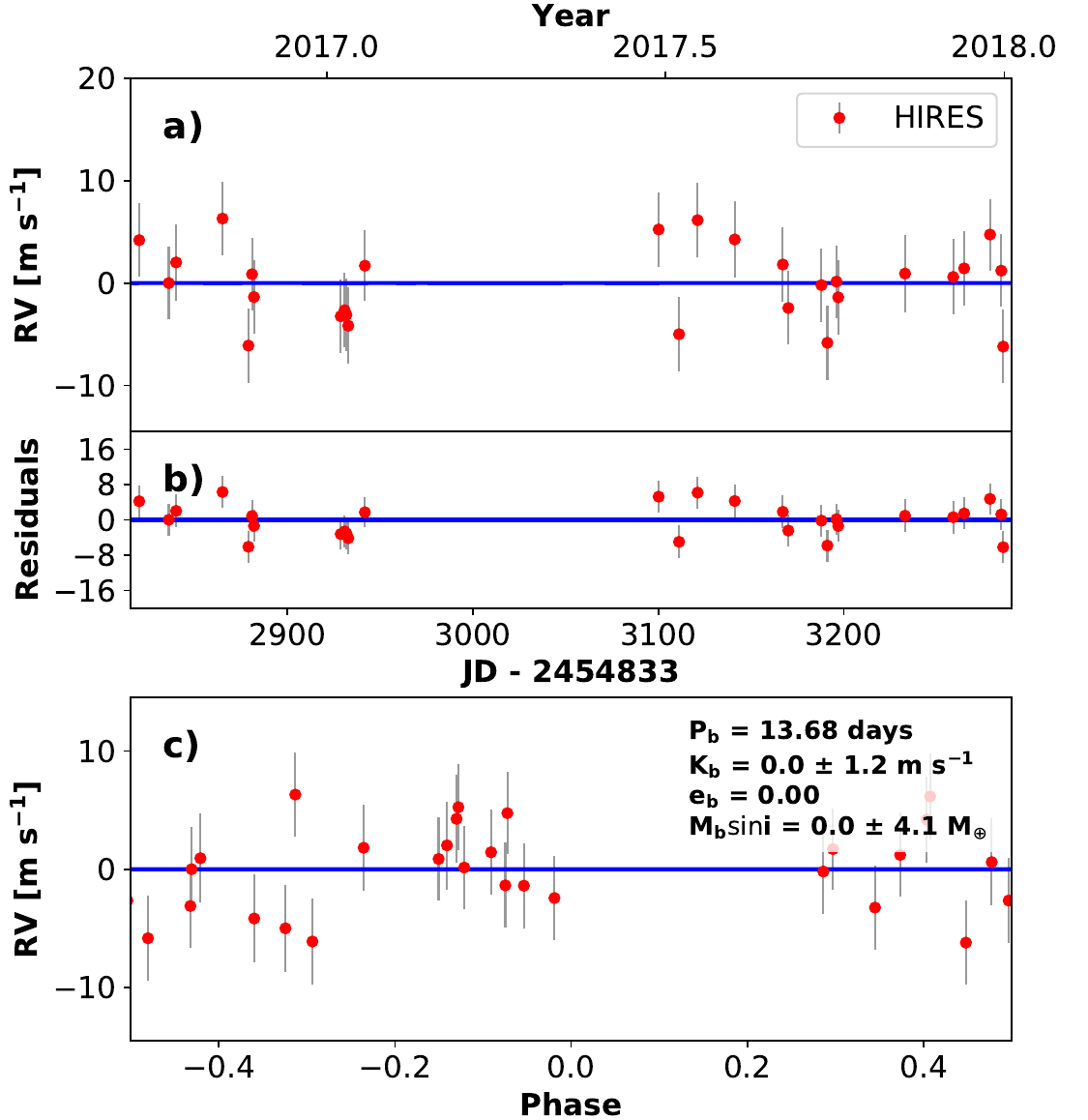}
\caption{RVs and Keplerian model for \bhiiSTNAME. Symbols, lines, and annotations are similar to those in Fig.\ \ref{fig:rvs_epic220709978}.}
\label{fig:rvs_k2-220}
\end{figure}

Our fit of the EVEREST light curve of the K2 photometry for \bhiiSTNAME is shown in Fig.\ \ref{fig:lc_epic220621788}.
We acquired \bhiiNOBSHIRES RVs of \bhiiSTNAME with HIRES, typically with an exposure meter setting of 80,000 counts.  We modeled the system as a single planet in a circular orbit with the period and phase fixed to the transit ephemeris.  The results of this analysis are listed in Table \ref{tab:epic220621788} and the best-fit model is shown in Fig.\ \ref{fig:rvs_k2-220}.  We do not detect the Doppler signal from \bhiiPNAMEone, but we can rule out a rocky composition based on a measured density of \bhiiRHOPone \gmc.  A gas-dominated composition is not surprising for a planet of this size.

\begin{deluxetable}{lrrr}
\tablecolumns{4}
\tabletypesize{\scriptsize}
\tablecaption{ K2-220 System Parameters \label{tab:epic220621788}}
\tablehead{
  \colhead{Parameter} & 
  \colhead{Credible} & 
  \colhead{Maximum} & 
  \colhead{Units} \\
  \colhead{} & 
  \colhead{Interval} & 
  \colhead{Likelihood} & 
  \colhead{} }
\startdata
\sidehead{\bf{RV Analysis -- MCMC Step Parameters}} 
$P_{b}$ & $\equiv13.6819$ & $\equiv13.6819$ & days \\ 
$T_{\mathrm{conj},b}$ & $\equiv2457401.2729$ & $\equiv2457401.2729$ & BJD$_\mathrm{TBD}$ \\ 
$\sqrt{e}\cos{\omega}_{b}$ & $\equiv0.0$ & $\equiv0.0$ &  \\ 
$\sqrt{e}\sin{\omega}_{b}$ & $\equiv0.0$ & $\equiv0.0$ &  \\ 
$K_{b}$ & $0.0\pm1.2$ & $0.0$ & m s$^{-1}$ \\ 
$\gamma_{\rm HIRES}$ & $-0.72^{+0.76}_{-0.79}$ & $-0.72$ & m s$-1$ \\ 
$\dot{\gamma}$ & $\equiv0.0$ & $\equiv0.0$ & m s$^{-1}$ day$^{-1}$ \\ 
$\ddot{\gamma}$ & $\equiv0.0$ & $\equiv0.0$ & m s$^{-1}$ day$^{-2}$ \\ 
$\sigma_{\rm HIRES}$ & $3.4^{+1.0}_{-0.3}$ & $3.1$ & $\rm m\ s^{-1}$ \\ 
\hline
\sidehead{\bf{Orbital \& Physical Parameters}}
$P_{b}$ & $\equiv13.6819$ & $\equiv13.6819$ & days \\
$T_{\mathrm{conj},b}$ & $\equiv2457401.2729$ & $\equiv2457401.2729$ & BJD$_\mathrm{TBD}$ \\
$e_{b}$ & $\equiv0.0$ & $\equiv0.0$ &  \\
$\omega_{b}$ & $\equiv0.0$ & $\equiv0.0$ & radians \\
$K_{b}$ & $0.0\pm1.2$ & $0.0$ & m s$^{-1}$ \\
$M_b$ & $0.0\pm4.0$ & $0.0$ & $M_\earth$ \\
$R_b/R_*$ & $\equiv0.0215$ & $\equiv0.0215$ &  \\
$\rho_b$ & $-0.0\pm1.6$ & $0.0$ & g cm$^{-3}$ \\
$R_b$ & $\equiv2.367$ & $\equiv2.367$ & $R_\earth$ \\
\hline
\hline
\sidehead{\bf{Priors}}
Parameter & Prior \\
\hline
\\None \\
\enddata
\end{deluxetable}

\subsection{K2-110} 


\bbggSTNAME is a late K dwarf from Campaign 6 with one transiting planet with a radius 2.5 \rearth and a period of 13 days.
See Tables \ref{tb:star_pars}  and \ref{tb:star_props} for stellar properties and Table \ref{tb:planet_props} for precise planet parameters.

\bbggPNAMEone was discovered by \cite{Osborn2017} who measured a mass of $16.7 \pm 3.2$ \mearth and a density of $5.2 \pm 1.2$ \gmc based on 17 HARPS RVs and 11 HARPS-N RVs.  The density is unusually high for a sub-Neptune planet.  The planet was also noted in the catalogs by \cite{Mayo2018} and \cite{Petigura2018}. \cite{Bonomo2023} report a mass of $15.9 \pm 2.7 M_\oplus$.

\begin{figure*}
\epsscale{1.0}
\plotone{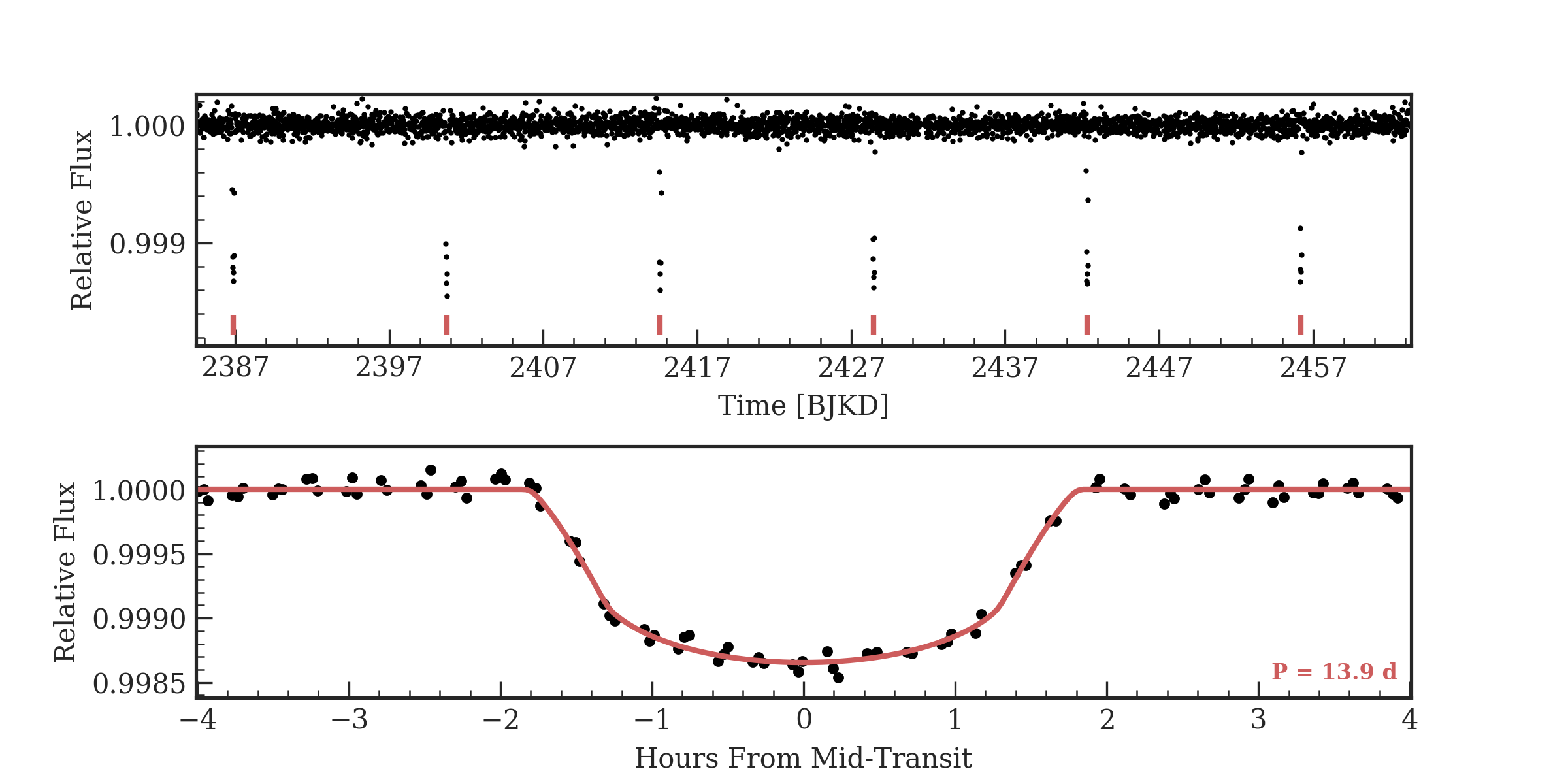}
\caption{Time series (top) and phase-folded (bottom) light curve for the planet orbiting \bbggSTNAME.  Plot formatting is the same as in Fig.\ \ref{fig:lc_epic220709978}.}
\label{fig:lc_epic212521166}
\end{figure*}

\begin{figure}
\epsscale{1.0}
\plotone{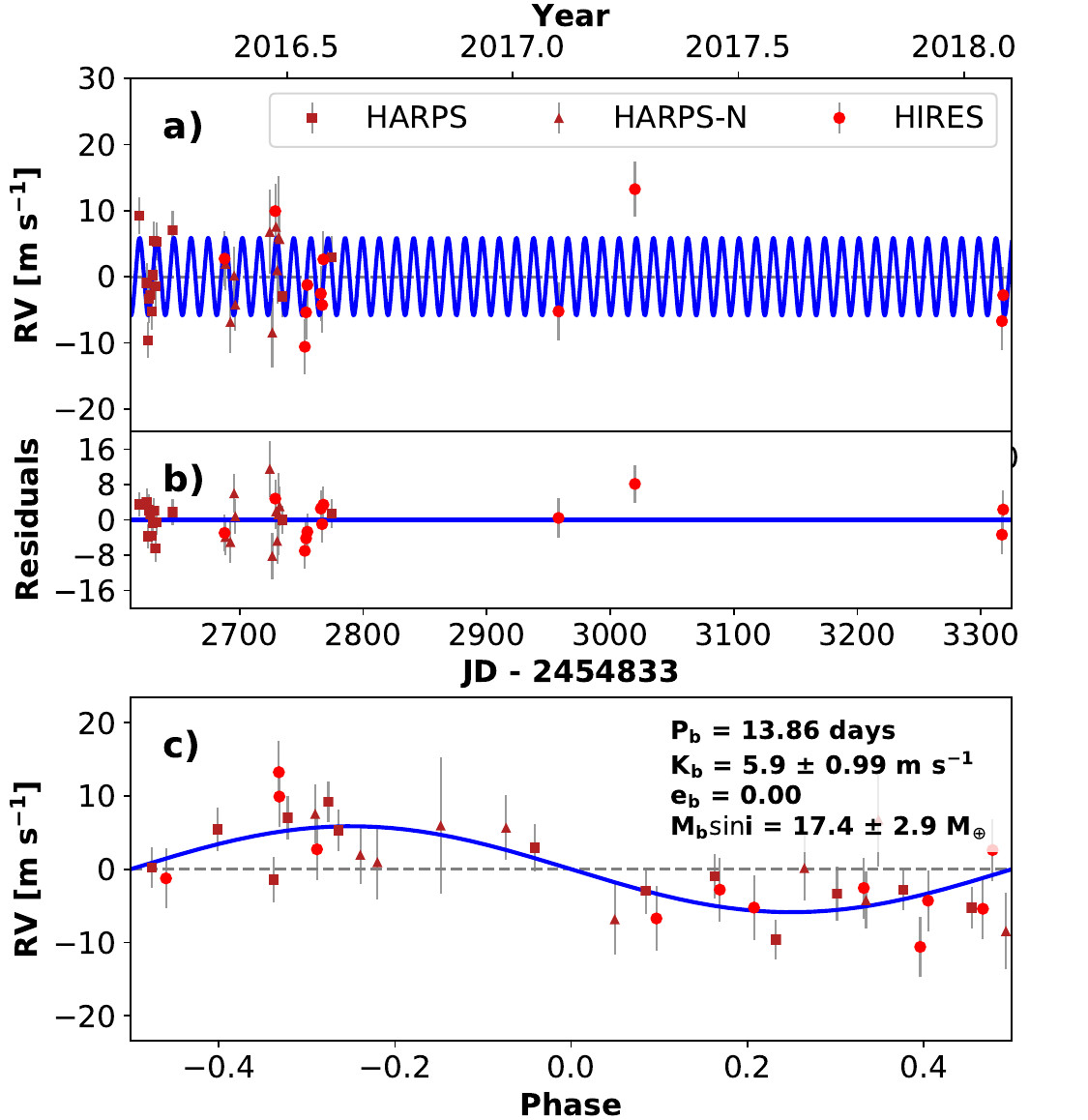}
\caption{RVs and Keplerian model for \bbggSTNAME.  Symbols, lines, and annotations are similar to those in Fig.\ \ref{fig:rvs_epic220709978}.}
\label{fig:rvs_k2-110}
\end{figure}

Our fit of the EVEREST light curve of the K2 photometry for \bbggSTNAME is shown in Fig.\ \ref{fig:lc_epic212521166}.
We acquired \bbggNOBSHIRES RVs of \bbggSTNAME with HIRES, typically with an exposure meter setting of 100,000 counts.  We modeled the HIRES, HARPS-N, and HARPS RVs as a single planet in a circular orbit with the period and phase fixed to the transit ephemeris.  We rejected more complicated models with orbital eccentricity and/or a linear RV trend based on the AICc statistic.  The results of our analysis are listed in Table \ref{tab:epic212521166} and the best fit model is shown in Fig.\ \ref{fig:rvs_k2-110}.  Our combined analysis confirms the previous results \citep{Osborn2017,Bonomo2023}.  \bbggPNAMEone has an unusually high density (\bbggRHOPone \gmc) for its size.

\begin{deluxetable}{lrrr}
\tablecolumns{4}
\tabletypesize{\scriptsize}
\tablecaption{ K2-110 System Parameters \label{tab:epic212521166}}
\tablehead{
  \colhead{Parameter} & 
  \colhead{Credible} & 
  \colhead{Maximum} & 
  \colhead{Units} \\
  \colhead{} & 
  \colhead{Interval} & 
  \colhead{Likelihood} & 
  \colhead{} }
\startdata
\sidehead{\bf{RV Analysis -- MCMC Step Parameters}} 
$P_{b}$ & $\equiv13.8637$ & $\equiv13.8637$ & days \\ 
$T_{\mathrm{conj},b}$ & $\equiv2457233.7386$ & $\equiv2457233.7386$ & BJD$_\mathrm{TBD}$ \\ 
$\sqrt{e}\cos{\omega}_{b}$ & $\equiv0.0$ & $\equiv0.0$ &  \\ 
$\sqrt{e}\sin{\omega}_{b}$ & $\equiv0.0$ & $\equiv0.0$ &  \\ 
$K_{b}$ & $5.80\pm0.99$ & $6.00$ & m s$^{-1}$ \\ 
$\gamma_{\rm HIRES}$ & $-0.5\pm1.4$ & $-0.5$ & m s$-1$ \\ 
$\gamma_{\rm HARPS-N}$ & $-21637.2^{+1.8}_{-1.7}$ & $-21637.3$ &  \\ 
$\gamma_{\rm HARPS}$ & $-21632.83^{+0.95}_{-0.94}$ & $-21632.83$ & m s$-1$ \\
\enddata
\end{deluxetable}

\subsection{WASP-47} 
\label{sec:WASP47}

\dbfaSTNAME has an extensive history that precedes and includes the K2 mission.  The star is a slightly evolved, metal-rich, solar-temperature star.  
See Tables \ref{tb:star_pars}  and \ref{tb:star_props} for stellar properties and Table \ref{tb:planet_props} for the precise planet parameters adopted in this paper.
WASP-47b is a hot Jupiter discovered by \cite{Hellier2012} using ground-based photometry.  They characterized the system using 19 RVs from CORALIE and measured a planet mass of $362 \pm 16$ \mearth.  \cite{Neveu-VanMalle2016} added 26 CORALIE RVs and discovered an outer companion (WASP-47c) with a mass of $359 \pm19$ \mearth and an orbital period of 572 days.

\cite{Becker2015} searched the K2 photometry and discovered two additional transiting planets: WASP-47d with a period of 9 days and a radius of 3.6 \rearth, and WASP-47e with a period of 0.8 days and a radius of 1.8 \rearth.  Note that the letters attached to the planets (b--e) follow their order of discovery, not orbital period order.  \cite{Becker2015} also conducted the first TTV analysis of the system, finding planet masses of $341^{+73}_{-55}$ \mearth (WASP-47b), $< 22$ \mearth (WASP-47e), and $15.2 \pm 7$ \mearth (WASP-47d).  For the K2 mission, the planets orbiting \dbfaSTNAME are included in the \cite{Crossfield2016}, \cite{Vanderburg2016-catalog}, \cite{Adams2016}, \cite{Barros2016}, \cite{Adams2017}, and \cite{Wittenmyer2018} catalogs. 

Spurred by the high multiplicity and unusual system architecture, several teams pursued RV measurements of this system.  First, \cite{Dai2015} reported 27 RVs from PFS.  Their analysis did not include the prior CORALIE RVs and found masses of
$12.2 \pm 3.7$ \mearth (WASP-47e), 
$370 \pm 29$ \mearth (WASP-47b), and 
$10.4 \pm 8.4$ \mearth(WASP-47d). 

\cite{Almenara2016} conducted a dynamical analysis based on transit times from K2 and RVs from \cite{Hellier2012} and \cite{Dai2015} and \cite{Neveu-VanMalle2016}.  They found masses of 
$364 \pm 9$ \mearth (WASP-47b), 
$361^{+80}_{-54}$ \mearth (\msini; WASP-47c), 
$15.7 \pm 1.1$ \mearth, (WASP-47d), and 
$9.1^{+1.8}_{-2.9}$ \mearth (WASP-47e).

\cite{Sinukoff2017a} added 47 HIRES RVs.  Combined with the previous CORALIE and PFS RVs, they measured masses of 
$356 \pm 12$ \mearth (WASP-47b), 
$411 \pm 18$ \mearth (\msini; WASP-47c), 
$12.8 \pm 2.7$ \mearth (WASP-47d), and 
$9.1 \pm 1.2$ \mearth (WASP-47e).  
A dynamical analysis by \cite{Weiss2017} considered all available RVs at the time of \cite{Sinukoff2017a} as well as transit times from K2.  In particular, \cite{Weiss2017} found improved masses for WASP-47d ($13.6 \pm 2.0$ \mearth) compared to RV alone ($12.8 \pm 2.7$ \mearth) or TTVs alone ($16.1 \pm 3.8$ \mearth).

\cite{Vanderburg2017} added 69 HARPS-N RVs and performed a combined analysis with all available RVs.  They found masses of 
$363.1 \pm 7.3$ \mearth (WASP-47b), 
$398.2 \pm 9.3$ \mearth (\msini; WASP-47c),  
$13.1 \pm 1.5$ \mearth (WASP-47d),  
$6.83 \pm 0.66$ \mearth (WASP-47e).  Their analysis was the most precise RV-only analysis to date, which we update slightly here with new HIRES RVs.

\cite{Dai2019} incorporated parallax information from Gaia in a reanalysis of this system, including a Gaussian process to account for correlated noise. (They only list a mass for planet e, which is consistent with \cite{Vanderburg2017})
\cite{Sanchis-Ojeda2015} measured the obliquity of the giant planet, WASP-47b, using RV measurements from HIRES.  They modeled the Rossiter-McLaughlin curve and found a projected obliquity of $\lambda = 0 \pm 24^\circ$, consistent with a spin orbit alignment. Most recently, \cite{Bryant2022} used ESPRESSO RVs to measure planet e's mass to be $6.77 \pm 0.57 M_\oplus$, consistent with our value below.

\cite{Kane2020} analyzed the K2 photometry to look for phase signatures for the planets to constrain their albedos. They determined that WASP-47b is potentially a ``dark'' planet with an albedo of 0.016 and WASP-47e shows early evidence of also having a low albedo.  Planet e is also being targeted by JWST observations in GO-3615.

\begin{figure*}
\epsscale{1.0}
\plotone{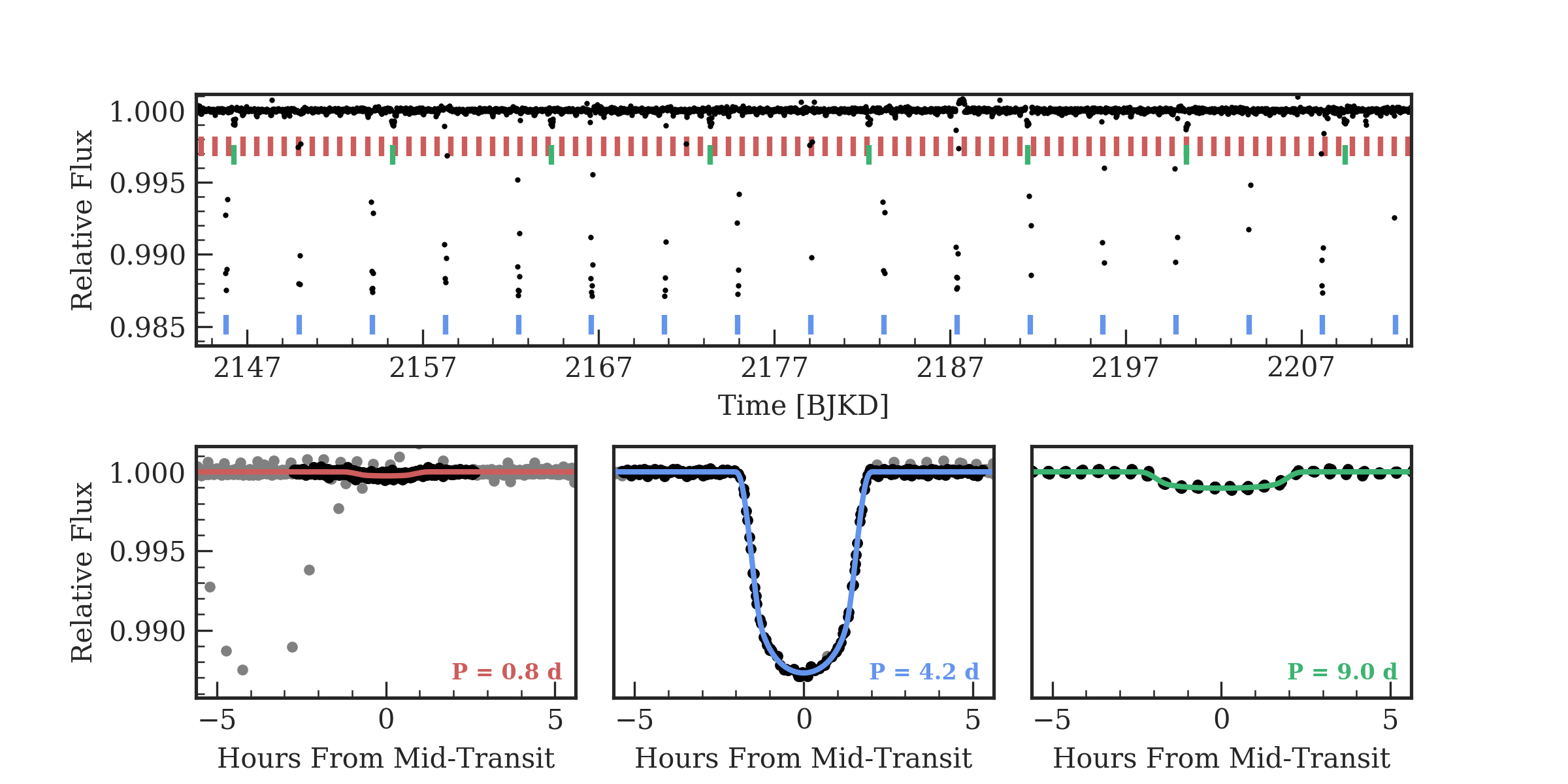}
\caption{Time series (top) and phase-folded (bottom) light curve for the planet orbiting \dbfaSTNAME.  Plot formatting is the same as in Fig.\ \ref{fig:lc_epic220709978}.}
\label{fig:lc_epic206103150}
\end{figure*}

\begin{figure}
\epsscale{0.6}
\plotone{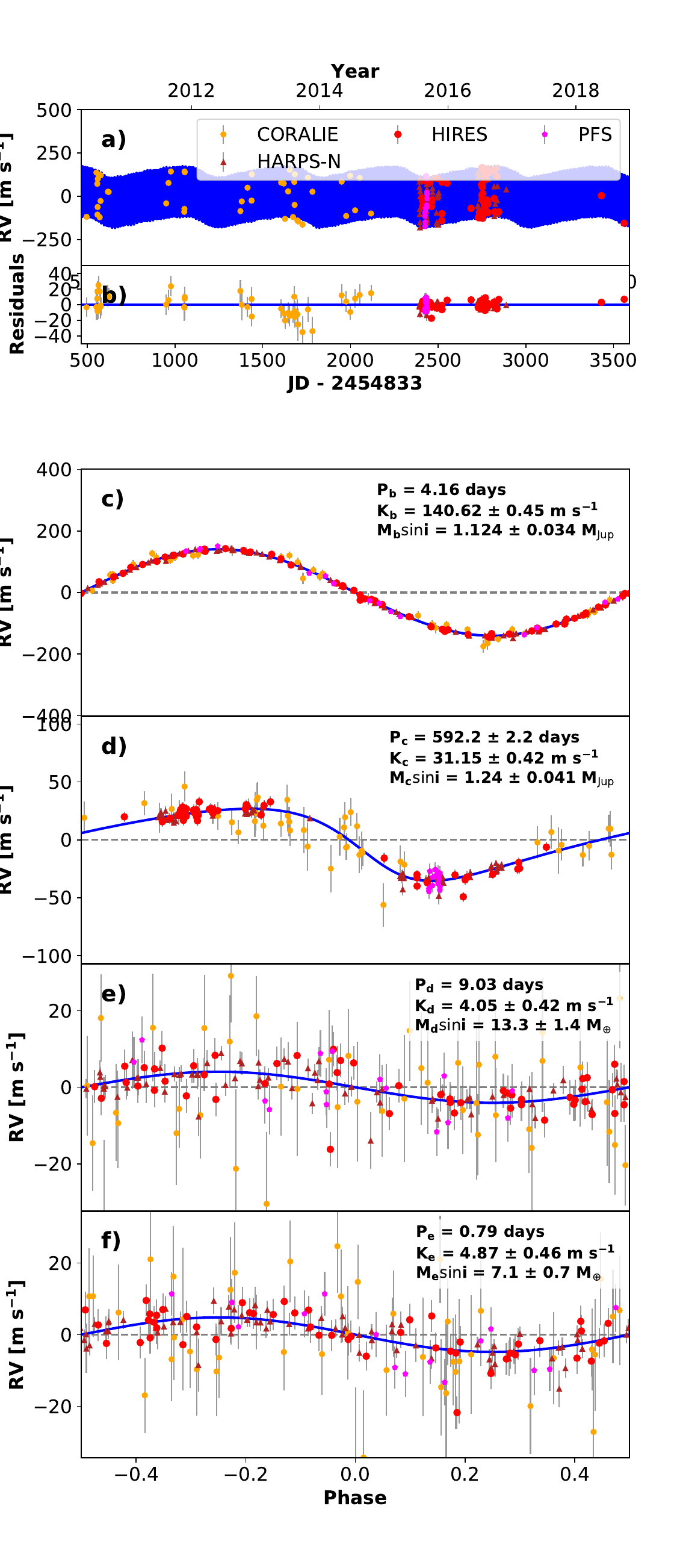}
\caption{RVs and Keplerian model for  \dbfaSTNAME.  Symbols, lines, and annotations are similar to those in Fig.\ \ref{fig:rvs_epic220709978}.}
\label{fig:rvs_wasp-47}
\end{figure}

\begin{figure}
\epsscale{1.0}
\plotone{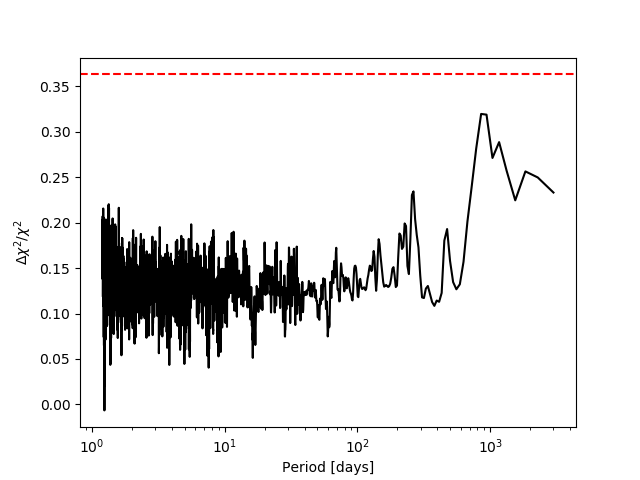}
\caption{Periodogram search of the RVs showing no evidence for a fifth planet orbiting \dbfaSTNAME.  Lines and annotations are similar to those in Fig.\ \ref{fig:rvs_epic220709978_resid}.}
\label{fig:rvs_wasp47_resid}
\end{figure}

Our fit of the EVEREST light curve of the K2 photometry for \dbfaSTNAME is shown in Fig.\ \ref{fig:lc_epic206103150}.  We acquired a total of \dbfaNOBSHIRES RVs of \dbfaSTNAME with HIRES, including those reported in \cite{Sinukoff2017a} and \cite{Sanchis-Ojeda2015}.  These HIRES observations typically had an exposure meter setting of 50,000 counts.  We modeled the system as a four-planet system with the three transiting planets fixed in circular orbits and WASP-47c in an eccentric orbit. 
We searched for additional planets in the RVs and we found no evidence of a fifth planet (Fig.\ \ref{fig:rvs_wasp47_resid}). 
The results of this analysis are listed in Table \ref{tab:wasp47} and the best fit model is shown in Fig.\ \ref{fig:rvs_wasp-47}.
We found planet masses of 
$357 \pm 11$ \mearth (WASP-47b), 
$395 \pm 13$ \mearth (\msini; WASP-47c),  
$12.8 \pm 1.4$ \mearth (WASP-47d),  
$7.38 \pm 0.72$ \mearth (WASP-47e). Our results are consistent with the results from \cite{Vanderburg2017} to within 1$\sigma$. 


\begin{deluxetable}{lrrr}
\tablecolumns{4}
\tabletypesize{\scriptsize}
\tablecaption{ WASP-47 System Parameters \label{tab:wasp47}}
\tablehead{
  \colhead{Parameter} & 
  \colhead{Credible} & 
  \colhead{Maximum} & 
  \colhead{Units} \\
  \colhead{} & 
  \colhead{Interval} & 
  \colhead{Likelihood} & 
  \colhead{} }
\startdata
\sidehead{\bf{RV Analysis -- MCMC Step Parameters}} 
$P_{e}$ & $\equiv0.7896$ & $\equiv0.7896$ & days \\ 
$T_{\mathrm{conj},e}$ & $\equiv2456981.3436$ & $\equiv2456981.3436$ & BJD$_\mathrm{TBD}$ \\ 
$\sqrt{e}\cos{\omega}_{e}$ & $\equiv0.0$ & $\equiv0.0$ &  \\ 
$\sqrt{e}\sin{\omega}_{e}$ & $\equiv0.0$ & $\equiv0.0$ &  \\ 
$K_{e}$ & $4.85^{+0.46}_{-0.45}$ & $4.82$ & m s$^{-1}$ \\ 
$P_{b}$ & $\equiv4.1592$ & $\equiv4.1592$ & days \\ 
$T_{\mathrm{conj},b}$ & $\equiv2456982.9772$ & $\equiv2456982.9772$ & BJD$_\mathrm{TBD}$ \\ 
$\sqrt{e}\cos{\omega}_{b}$ & $\equiv0.0$ & $\equiv0.0$ &  \\
\enddata
\end{deluxetable}

\subsection{K2-79} 


\ccdhSTNAME is a slightly evolved, solar-temperature star in Field 4 with one transiting planet that has a radius of 3.7 \rearth and an orbital period of 11 days. 
See Tables \ref{tb:star_pars}  and \ref{tb:star_props} for stellar properties and Table \ref{tb:planet_props} for precise planet parameters.  \ccdhPNAMEone is in the \cite{Crossfield2016} and \cite{Mayo2018} catalogs, the latter of which validated the planet. The planet's mass has been previously measured to be 9--12~$M_\oplus$ \citep{Nava2022,Bonomo2023}.

\begin{figure*}
\epsscale{1.0}
\plotone{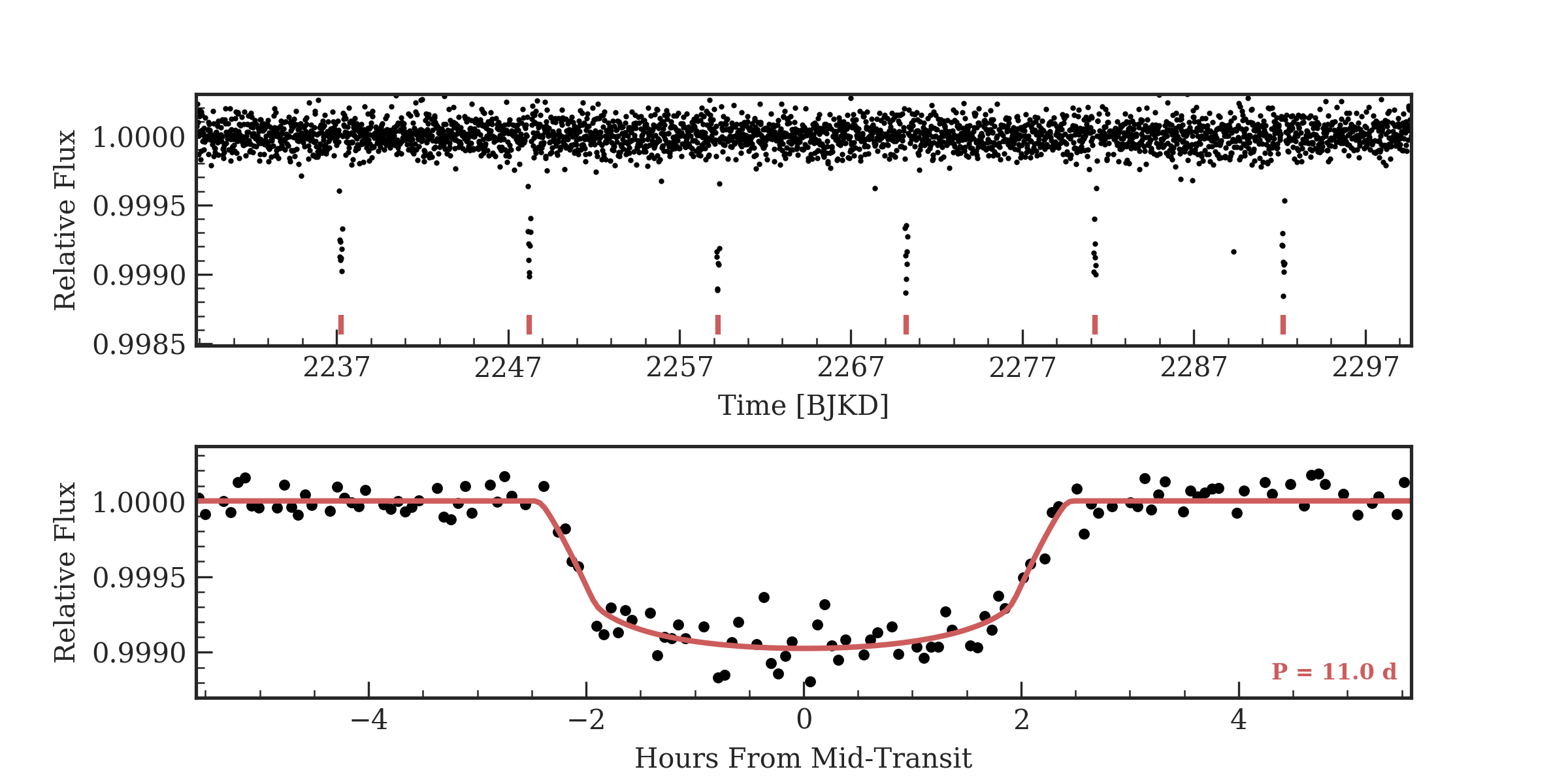}
\caption{Time series (top) and phase-folded (bottom) light curve for the planet orbiting \ccdhSTNAME.  Plot formatting is the same as in Fig.\ \ref{fig:lc_epic220709978}.}
\label{fig:lc_epic210402237}
\end{figure*}

\begin{figure}
\epsscale{1.0}
\plotone{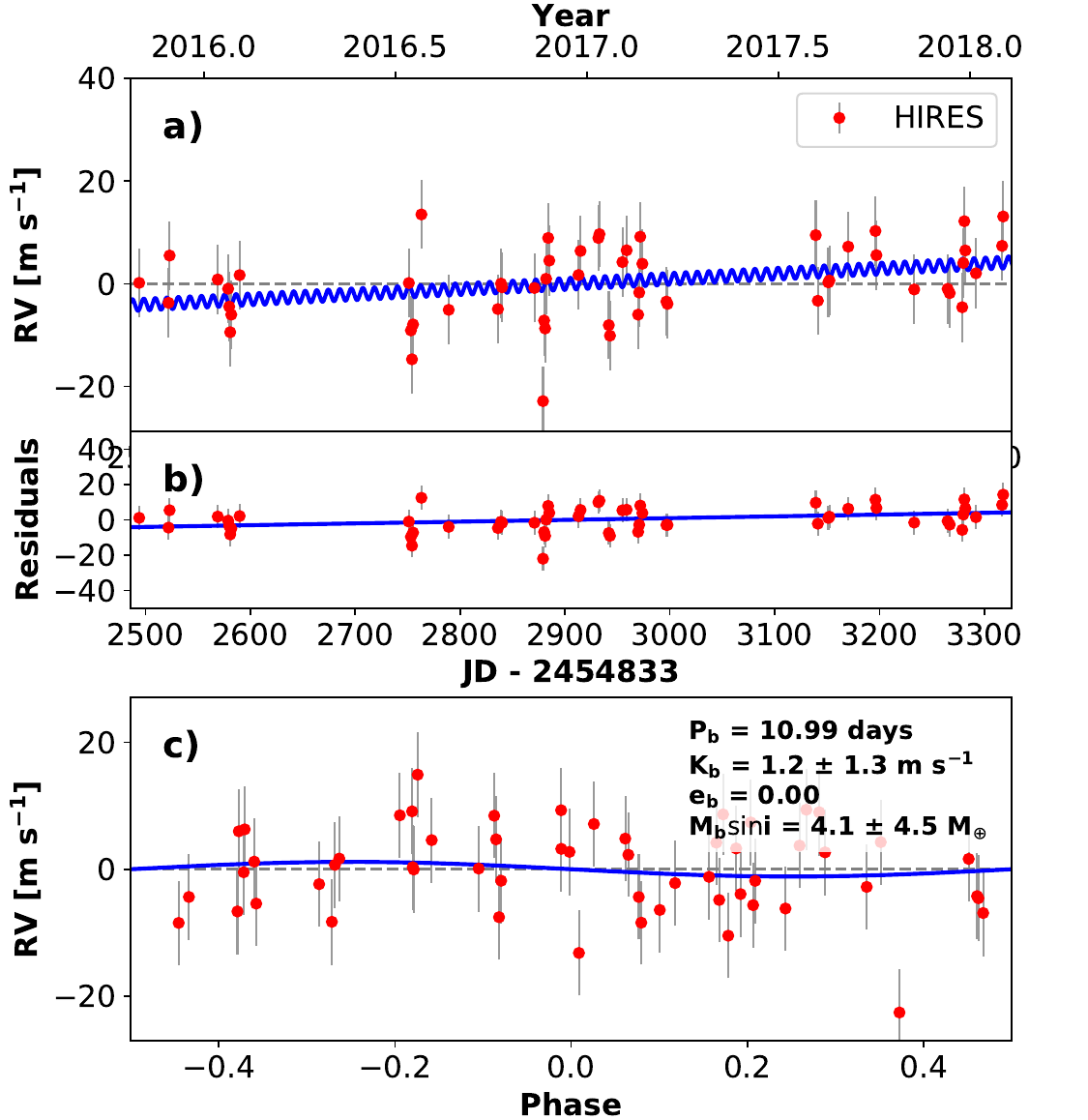}
\caption{RVs and Keplerian model for \ccdhSTNAME.  Symbols, lines, and annotations are similar to those in Fig.\ \ref{fig:rvs_epic220709978}.}
\label{fig:rvs_k2-79}
\end{figure}

Our fit of the EVEREST light curve of the K2 photometry for this star is shown in Fig.\ \ref{fig:lc_epic210402237}.  We acquired \ccdhNOBSHIRES RVs with HIRES, typically with an exposure meter setting of 50,000 counts.  We modeled the system as a single planet in a circular orbit with the orbital period and phase fixed to the transit ephemeris.  Our model also included a linear RV trend which we justified based on \dAICc = 4 compared to a model without a trend.  The results of this analysis are listed in Table \ref{tab:epic210402237} and the best-fit model is shown in Fig.\ \ref{fig:rvs_k2-79}.  \ccdhPNAMEone is a Neptune-sized planet whose Doppler signal we detected with 1-$\sigma$ significance and has a low density (\ccdhRHOPone \gmc).  The measurement of this signal was hampered by an unexpectedly high jitter for a star only modestly evolved and with low activity (\lrphk = \ccdhRPHK). A search for additional planets found no significant signals in a $\Delta$BIC periodogram. Our observations are consistent with the more precise measurements in the literature \citep{Nava2022,Bonomo2023}.


\begin{deluxetable}{lrrr}
\tablecolumns{4}
\tabletypesize{\scriptsize}
\tablecaption{ K2-79 System Parameters \label{tab:epic210402237}}
\tablehead{
  \colhead{Parameter} & 
  \colhead{Credible} & 
  \colhead{Maximum} & 
  \colhead{Units} \\
  \colhead{} & 
  \colhead{Interval} & 
  \colhead{Likelihood} & 
  \colhead{} }
\startdata
\sidehead{\bf{RV Analysis -- MCMC Step Parameters}} 
$P_{b}$ & $\equiv10.995$ & $\equiv10.995$ & days \\ 
$T_{\mathrm{conj},b}$ & $\equiv2457070.2428$ & $\equiv2457070.2428$ & BJD$_\mathrm{TBD}$ \\ 
$\sqrt{e}\cos{\omega}_{b}$ & $\equiv0.0$ & $\equiv0.0$ &  \\ 
$\sqrt{e}\sin{\omega}_{b}$ & $\equiv0.0$ & $\equiv0.0$ &  \\ 
$K_{b}$ & $1.1\pm1.2$ & $1.1$ & m s$^{-1}$ \\ 
$\gamma_{\rm HIRES}$ & $0.8\pm1.3$ & $0.8$ & m s$-1$ \\ 
$\dot{\gamma}$ & $0.0087\pm0.0034$ & $0.0088$ & m s$^{-1}$ day$^{-1}$ \\ 
$\ddot{\gamma}$ & $\equiv0.0$ & $\equiv0.0$ & m s$^{-1}$ day$^{-2}$ \\ 
$\sigma_{\rm HIRES}$ & $6.37^{+0.75}_{-0.65}$ & $6.10$ & $\rm m\ s^{-1}$ \\ 
\hline
\sidehead{\bf{Orbital \& Physical Parameters}}
$P_{b}$ & $\equiv10.995$ & $\equiv10.995$ & days \\
$T_{\mathrm{conj},b}$ & $\equiv2457070.2428$ & $\equiv2457070.2428$ & BJD$_\mathrm{TBD}$ \\
$e_{b}$ & $\equiv0.0$ & $\equiv0.0$ &  \\
$\omega_{b}$ & $\equiv0.0$ & $\equiv0.0$ & radians \\
$K_{b}$ & $1.1\pm1.2$ & $1.1$ & m s$^{-1}$ \\
$M_b$ & $3.8\pm4.3$ & $3.8$ & $M_\earth$ \\
$R_b/R_*$ & $0.0289^{+0.0008}_{-0.0005}$ & $0.0286$ &  \\
$\rho_b$ & $0.33\pm0.37$ & $0.40$ & g cm$^{-3}$ \\
$R_b$ & $3.99^{+0.11}_{-0.07}$ & $3.89$ & $R_\earth$ \\
\hline
\hline
\sidehead{\bf{Priors}}
Parameter & Prior \\
\hline
$jit_\mathrm{HIRES}$ & $\mathcal{U}(0,1e+100)$ \\
\enddata
\tablecomments{}{
Reference epoch for $\dot{\gamma}$,$\ddot{\gamma}$: 2458061.964141 
}
\end{deluxetable}

\begin{figure}
\epsscale{1.0}
\plotone{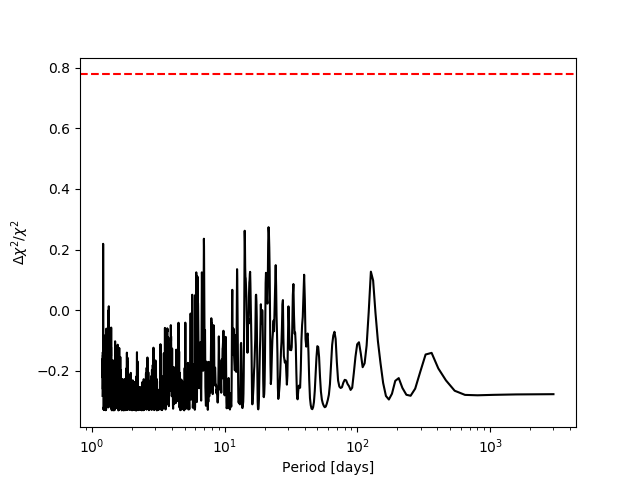}
\caption{Periodogram search of the RVs showing no evidence for a second planet orbiting \ccdhSTNAME.  Lines and annotations are similar to those in Fig.\ \ref{fig:rvs_epic220709978_resid}.}
\label{fig:rvs_epic210402237_resid}
\end{figure}
\subsection{K2-106} 

\begin{figure*}
\epsscale{1.0}
\plotone{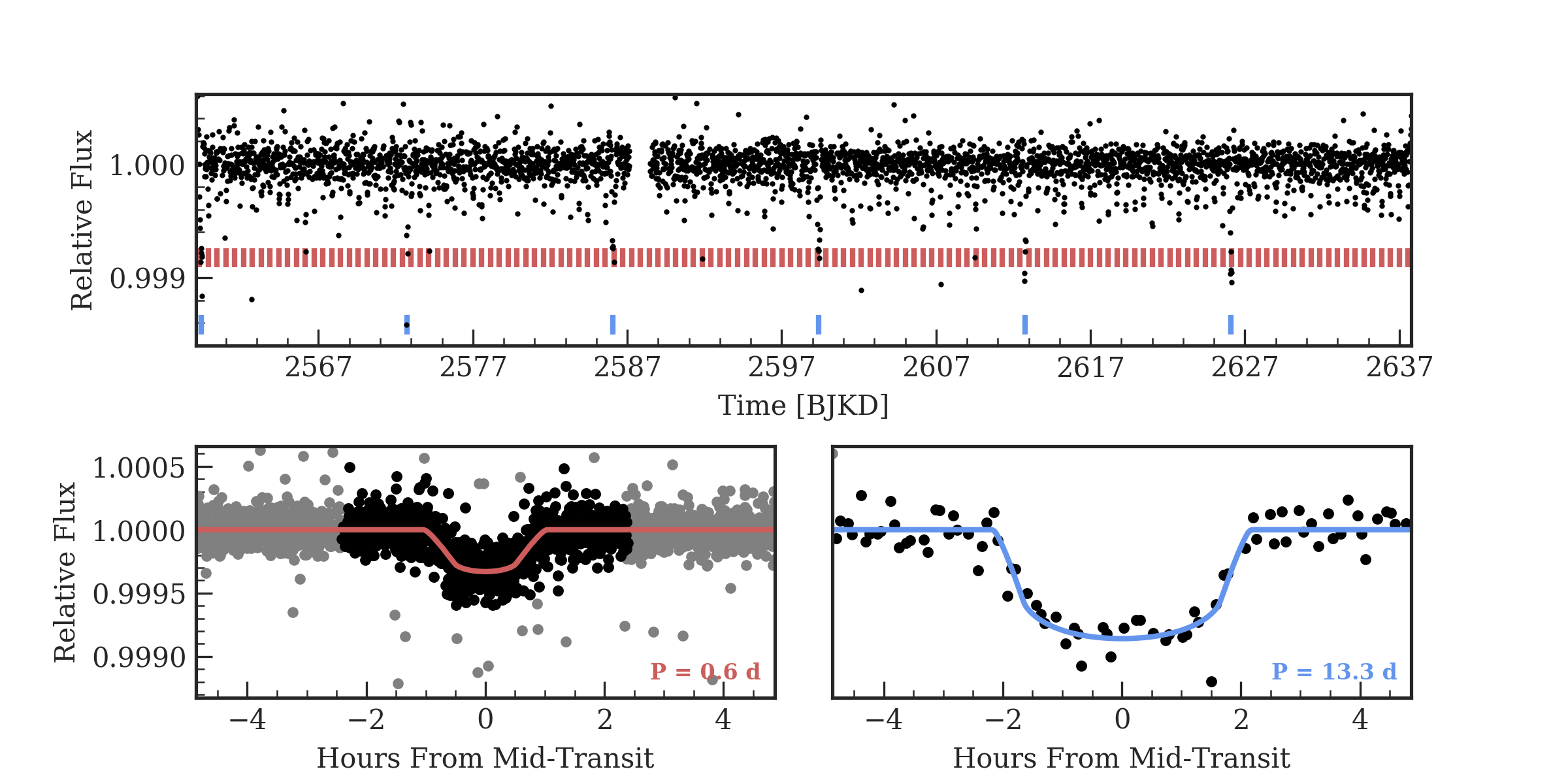}
\caption{Time series (top) and phase-folded (bottom) light curve for the planets orbiting \eicdSTNAME.  Plot formatting is the same as in Fig.\ \ref{fig:lc_epic220709978}.}
\label{fig:lc_epic220674823}
\end{figure*}

\eicdSTNAME is a slightly evolved, solar-temperature star from Campaign 8 with two transiting planets.  The first planet has a radius of 1.8 \rearth with an ultra-short period of 0.8 days.  The second planet has a size of 2.7 \rearth and a period of 13 days. See Tables \ref{tb:star_pars}  and \ref{tb:star_props} for stellar properties and Table \ref{tb:planet_props} for precise planet parameters adopted in this paper.
Our fit of the EVEREST light curve of the K2 photometry for \eicdSTNAME is shown in Fig.\ \ref{fig:lc_epic220674823}.  
The system was first reported as a candidate in \cite{Vanderburg2016-catalog}, and also in the catalogs by \cite{Petigura2018} and \cite{Mayo2018}.  The system was first validated by \cite{Adams2017}.  

\cite{Sinukoff2017b} characterized the system by measuring planet masses of $9.0 \pm 1.6$ \mearth for the USP and $< 24.4$ \mearth (99.7\% confidence) for the outer planet, based on 35 HIRES RVs. A later analysis by \cite{Guenther2017} measured masses of $8.36^{+0.96}_{-0.94}$ \mearth (\eicdPNAMEone) and $5.8^{+3.3}_{-3.0}$ \mearth (\eicdPNAMEtwo) based on RVs from PFS (13), HDS (3), FIES (6), HARPS-N (12),  HARPS (20), as well as the earlier HIRES RVs. \citep{Martinez2023} report masses of $8.5 \pm 1.0 M_\oplus$  and $5.9 \pm 3.3 M_\oplus$,  \cite{Bonomo2023} find $8.2 \pm 0.8 M_\oplus$ and $8.9 \pm 2.4 M_\oplus$, and \cite{Guenther2024} find $7.8 \pm 0.7 M_\oplus$ and $7.3 \pm 2.5 M_\oplus$.

\begin{figure}
\epsscale{1.0}
\plotone{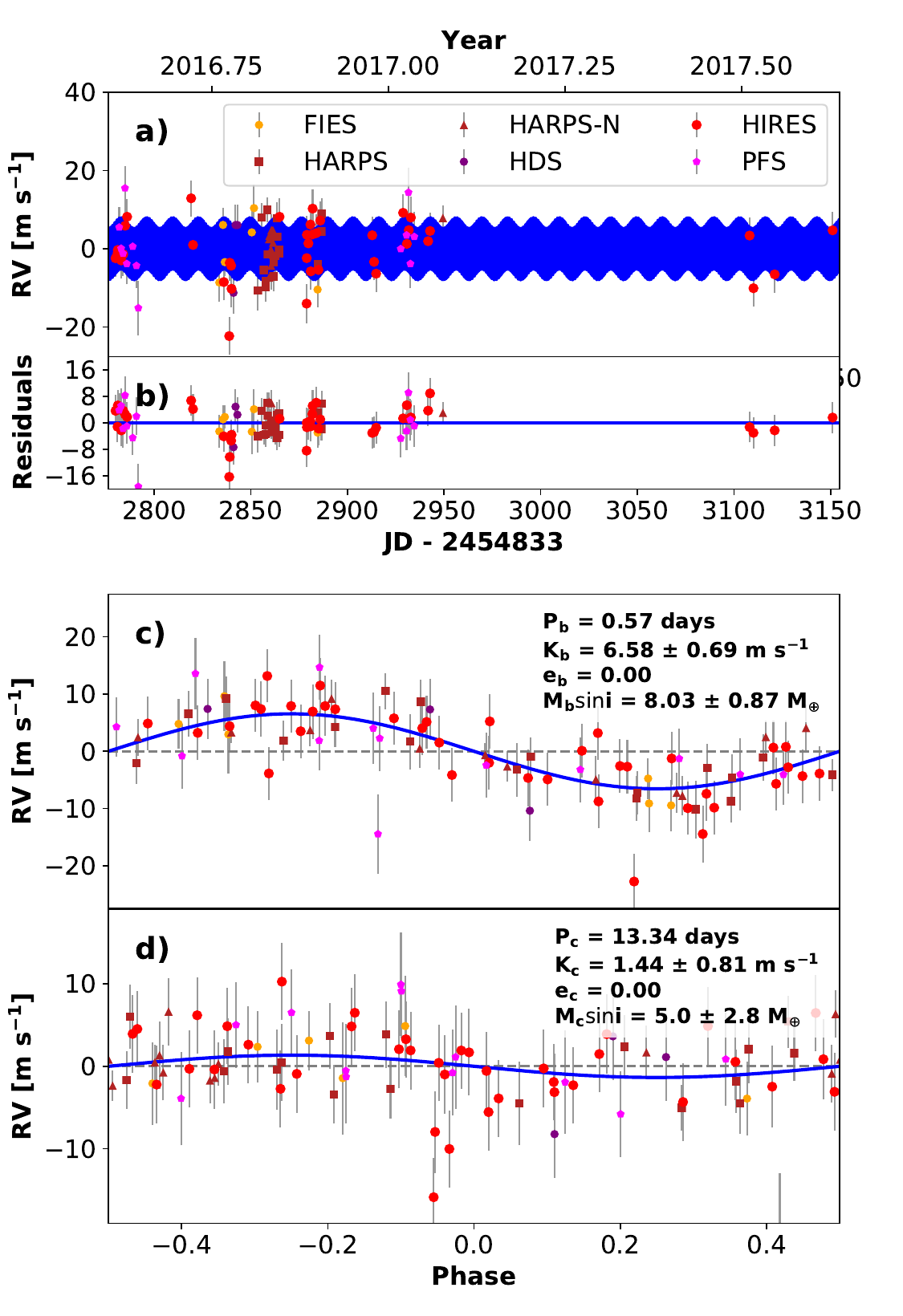}
\caption{RVs and Keplerian model for \eicdSTNAME.  Symbols, lines, and annotations are similar to those in Fig.\ \ref{fig:rvs_epic220709978}.}
\label{fig:rvs_k2-106}
\end{figure}

\begin{figure}
\epsscale{1.0}
\plotone{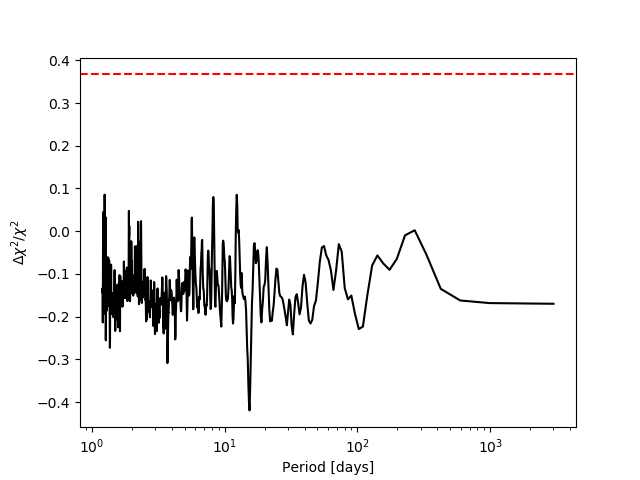}
\caption{Periodogram search of the RVs showing no evidence for a third planet orbiting \eicdSTNAME.  Lines and annotations are similar to those in Fig.\ \ref{fig:rvs_epic220709978_resid}.}
\label{fig:rvs_epic220674823_resid}
\end{figure}

We acquired \eicdNOBSHIRES RVs of \eicdSTNAME with HIRES, including those previously reported in \cite{Sinukoff2017b}.
The observations have a typical exposure meter setting of 80,000 counts.  
We modeled the system as two planets in circular orbits with orbital periods and phases fixed to the transit ephemeris.  We considered more complicated models with free eccentricities and/or a linear RV trend, but rejected those models based on the AIC statistic. Furthermore, we do not find evidence for a third planet in the system (Fig.~\ref{fig:rvs_epic220674823_resid}). We binned each RV dataset into intervals of 1.4 hr.  The results of our analysis are listed in Table \ref{tab:epic220674823} and the best-fit model is shown in Fig.\ \ref{fig:rvs_k2-106}. 
\eicdPNAMEone has a size and density (\eicdRHOPone \gmc) consistent with a rocky super-Earth.  The Doppler signal of the outer planet is only detected with $\sim$1-$\sigma$ significance, but we can rule out a rocky composition based on the density estimate of \eicdRHOPtwo \gmc. Our results are consistent with those of previous (and more recent) studies \citep{Martinez2023,Bonomo2023,Guenther2024}.

\begin{deluxetable}{lrrr}
\tablecolumns{4}
\tabletypesize{\scriptsize}
\tablecaption{ K2-106 System Parameters \label{tab:epic220674823}}
\tablehead{
  \colhead{Parameter} & 
  \colhead{Credible} & 
  \colhead{Maximum} & 
  \colhead{Units} \\
  \colhead{} & 
  \colhead{Interval} & 
  \colhead{Likelihood} & 
  \colhead{} }
\startdata
\sidehead{\bf{RV Analysis -- MCMC Step Parameters}} 
$P_{b}$ & $\equiv0.5713$ & $\equiv0.5713$ & days \\ 
$T_{\mathrm{conj},b}$ & $\equiv2457393.4405$ & $\equiv2457393.4405$ & BJD$_\mathrm{TBD}$ \\ 
$\sqrt{e}\cos{\omega}_{b}$ & $\equiv0.0$ & $\equiv0.0$ &  \\ 
$\sqrt{e}\sin{\omega}_{b}$ & $\equiv0.0$ & $\equiv0.0$ &  \\ 
$K_{b}$ & $6.6\pm0.7$ & $6.5$ & m s$^{-1}$ \\ 
$P_{c}$ & $\equiv13.3392$ & $\equiv13.3392$ & days \\ 
$T_{\mathrm{conj},c}$ & $\equiv2457405.733$ & $\equiv2457405.733$ & BJD$_\mathrm{TBD}$ \\ 
$\sqrt{e}\cos{\omega}_{c}$ & $\equiv0.0$ & $\equiv0.0$ &  \\
\enddata
\end{deluxetable}

\subsection{K2-98} 


\bggeSTNAME is a late-F type star from Campaign 5 with \vsini = \bggeVSINI \kms.  The star hosts one transiting planet with radius of 4.9 \rearth and an orbital period of 10 days.
See Tables \ref{tb:star_pars}  and \ref{tb:star_props} for stellar properties and Table \ref{tb:planet_props} for precise planet parameters.
The planet was discovered by \cite{Barragan2016} who measured a mass using 4 RVs from FIES, 4 from HARPS, and 4 from HARPS-N.  The system is also in the catalogs of \cite{Petigura2018} and \cite{Mayo2018}.

\begin{figure*}
\epsscale{1.0}
\plotone{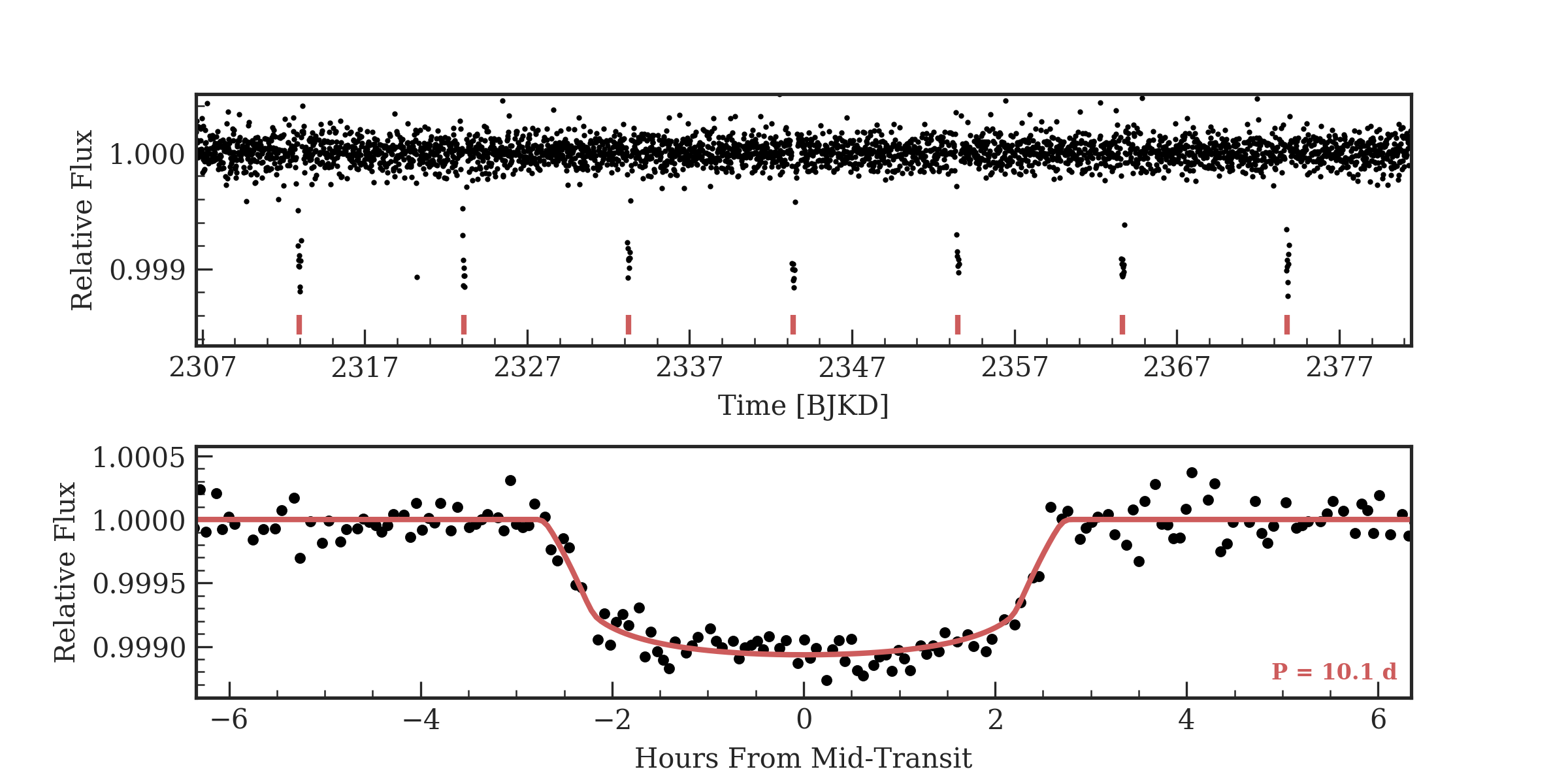}
\caption{Time series (top) and phase-folded (bottom) light curve for the planet orbiting \bggeSTNAME.  Plot formatting is the same as in Fig.\ \ref{fig:lc_epic220709978}.}
\label{fig:lc_epic211391664}
\end{figure*}

\begin{figure}
\epsscale{1.0}
\plotone{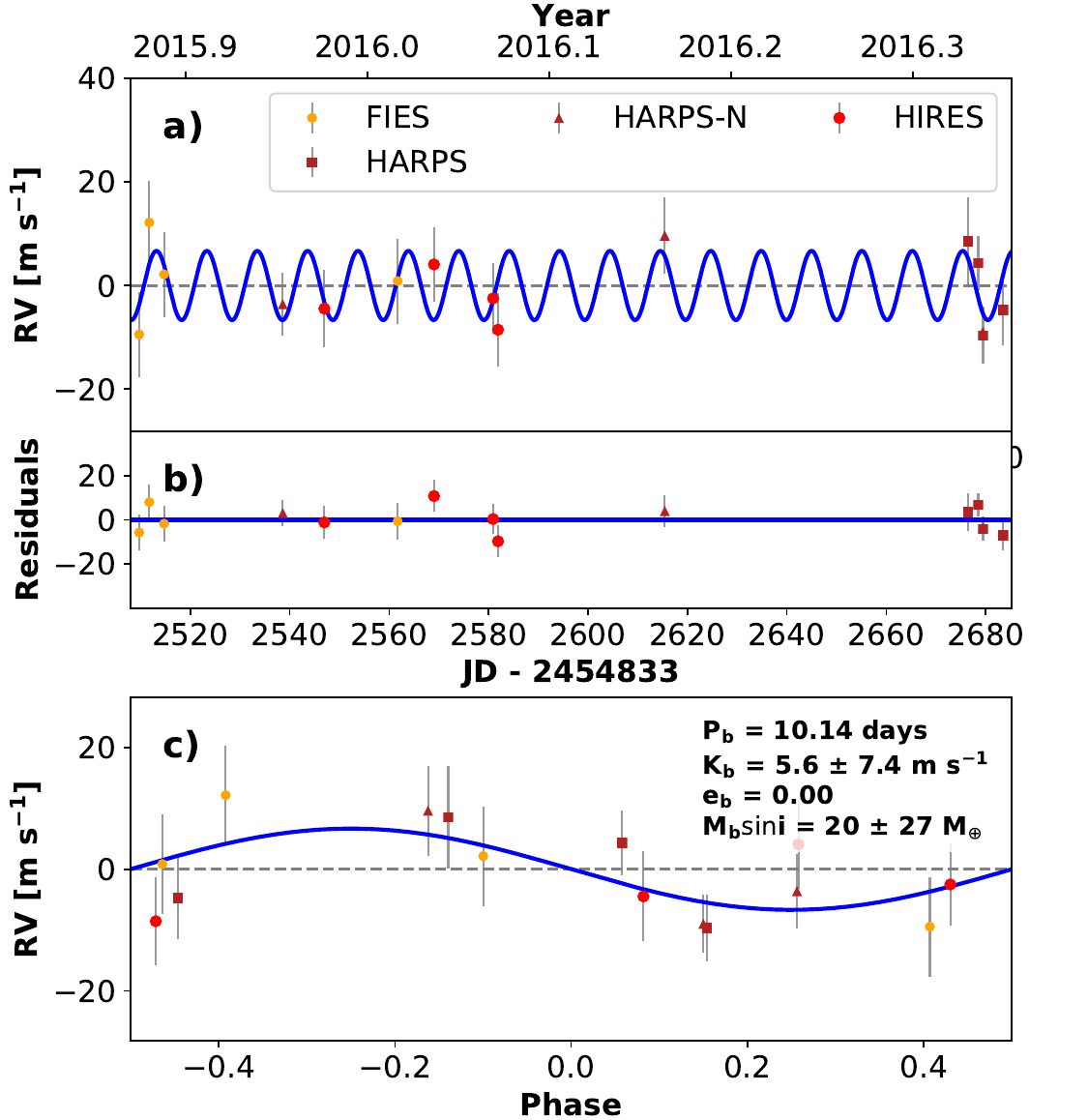}
\caption{RVs and Keplerian model for \bggeSTNAME. Symbols, lines, and annotations are similar to those in Fig.\ \ref{fig:rvs_epic220709978}.}
\label{fig:rvs_k2-98}
\end{figure}

Our fit of the EVEREST light curve of the K2 photometry for \bggeSTNAME is shown in Fig.\ \ref{fig:lc_epic211391664}.  We acquired \bggeNOBSHIRES RVs of \bggeSTNAME spanning 35 days with HIRES, typically with an exposure meter setting of 60,000.  We modeled the RVs from HIRES, FIES, and HARPS as a single planet in a circular orbit with the orbital period and phase fixed to the transit ephemeris.  Observations taken on the same night by the same telescope were binned in our analysis.  The results of this analysis are listed in Table \ref{tab:epic211391664} and the best-fit model is shown in Fig.\ \ref{fig:rvs_k2-98}.  

We note that \cite{Barragan2016} found a statistically significant mass of $32.1 \pm 8.1$ \mearth while our analysis found \bggeMPone \mearth.  The values are consistent, but the \cite{Barragan2016} uncertainty is much smaller.  We attribute this discrepancy mainly to differences in modeling.  Importantly, \cite{Barragan2016} did not include jitter (so far as we can tell) and required $K$ to be $>1$ \ms by including a uniform prior that forced $K$ to be in the range 1--1000 \ms).  We find that jitter is a critical part of the model; failure to include it can result in significantly underestimated uncertainties.  Our model found large and poorly determined values for the jitter of each instrument (except for FIES), in part due to sparse measurements; see Table \ref{tab:epic211391664}.  Since each instrument only contributes 4--\bggeNOBSHIRES RVs to the analysis, we suggest that additional RVs of \bggeSTNAME would refine the mass measurement.

\begin{deluxetable}{lrrr}
\tablecolumns{4}
\tabletypesize{\scriptsize}
\tablecaption{ K2-98 System Parameters \label{tab:epic211391664}}
\tablehead{
  \colhead{Parameter} & 
  \colhead{Credible} & 
  \colhead{Maximum} & 
  \colhead{Units} \\
  \colhead{} & 
  \colhead{Interval} & 
  \colhead{Likelihood} & 
  \colhead{} }
\startdata
\sidehead{\bf{RV Analysis -- MCMC Step Parameters}} 
$P_{b}$ & $\equiv10.1369$ & $\equiv10.1369$ & days \\ 
$T_{\mathrm{conj},b}$ & $\equiv2457145.9796$ & $\equiv2457145.9796$ & BJD$_\mathrm{TBD}$ \\ 
$\sqrt{e}\cos{\omega}_{b}$ & $\equiv0.0$ & $\equiv0.0$ &  \\ 
$\sqrt{e}\sin{\omega}_{b}$ & $\equiv0.0$ & $\equiv0.0$ &  \\ 
$K_{b}$ & $5.6^{+5.8}_{-9.1}$ & $7.0$ & m s$^{-1}$ \\ 
$\gamma_{\rm HIRES}$ & $0.9^{+8.9}_{-5.8}$ & $3.0$ & m s$-1$ \\ 
$\gamma_{\rm HARPS-N}$ & $76741.0^{+8.5}_{-9.4}$ & $76740.3$ & m s$-1$ \\ 
$\gamma_{\rm HARPS}$ & $76746.7^{+5.9}_{-6.7}$ & $76747.2$ & m s$-1$ \\
\enddata
\end{deluxetable}

\subsection{K2-3} 


K2-3 is an M0 dwarf star with 3 transiting planets with sizes of 2.1, 1.7, and 1.5 \rearth and orbital periods of 10.1, 24.6, and 44.6 d.  The planetary system was discovered in \cite{Crossfield2015} as {\em K2}'s first multiplanet system, and is among the best studied {\em K2} systems to date.  It also appears in the catalogs by \cite{Montet2015,Foreman-Mackey2015,Vanderburg2016-catalog,Crossfield2016,Barros2016,Sinukoff2016,Martinez2017,Kruse2019}. \cite{Beichman2016} refined the ephemerides and radii of the three planets with Spitzer transits and \cite{Fukui2016} refined the parameters of K2-3d with a ground-based transit. 

Because the planets are desirable targets for transit spectroscopy, several groups have gathered RVs to measure planet masses.  \citet{Almenara2015} measured 66 RVs from HARPS and determined that the masses of planets b, c, and d were $8.4 \pm 2.1$, $2.1^{+2.1}_{-1.3}$, and $11.1 \pm 3.5$ \mearth, respectively. They cautioned that the RV semiamplitudes of planets c and d are likely affected by stellar activity. 
\citet{Dai2016} measured 31 RVs with PFS on Magellan and modeled the RVs with available HARPS data, giving planet masses of $7.7 \pm 2.0$, $<12.6$, and $11.3^{+5.9}_{-5.8}$ \mearth\ for b, c, and d. 
\citet{Damasso2018} performed a RV analysis on a total of 132 HARPS and 197 HARPS-N measurements, including the Almenara sample. This HARPS analysis found that the masses of planets b and c are 6.6 $\pm$ 1.1 and 3.1$^{+1.3}_{-1.2}$ \mearth, respectively. They estimated the mass of planet d to be 2.7$^{+1.2}_{-0.8}$ \mearth from a suite of injection-recovery tests. \cite{DiamondLowe2022} conduct a reanalysis of the system and measure masses of $5.1 \pm 0.6 M_\oplus$ and $2.7 \pm 0.9 M_\oplus$ for planets b and c, with again non-detection for planet d.

We acquired 50 RVs with HIRES, typically with an exposure meter setting of 80,000. We modeled the HIRES RVs and above HARPS and PFS data with a 3-planet model using a Gaussian process to describe stellar activity. The observations and analysis are described in \cite{Kosiarek2019}, whose results we adopt here. See Tables \ref{tb:star_pars} and \ref{tb:star_props} for stellar properties and Table \ref{tb:planet_props} for precise planet parameters.
\subsection{EPIC 213546283} 

\begin{figure*}
\epsscale{1.0}
\plotone{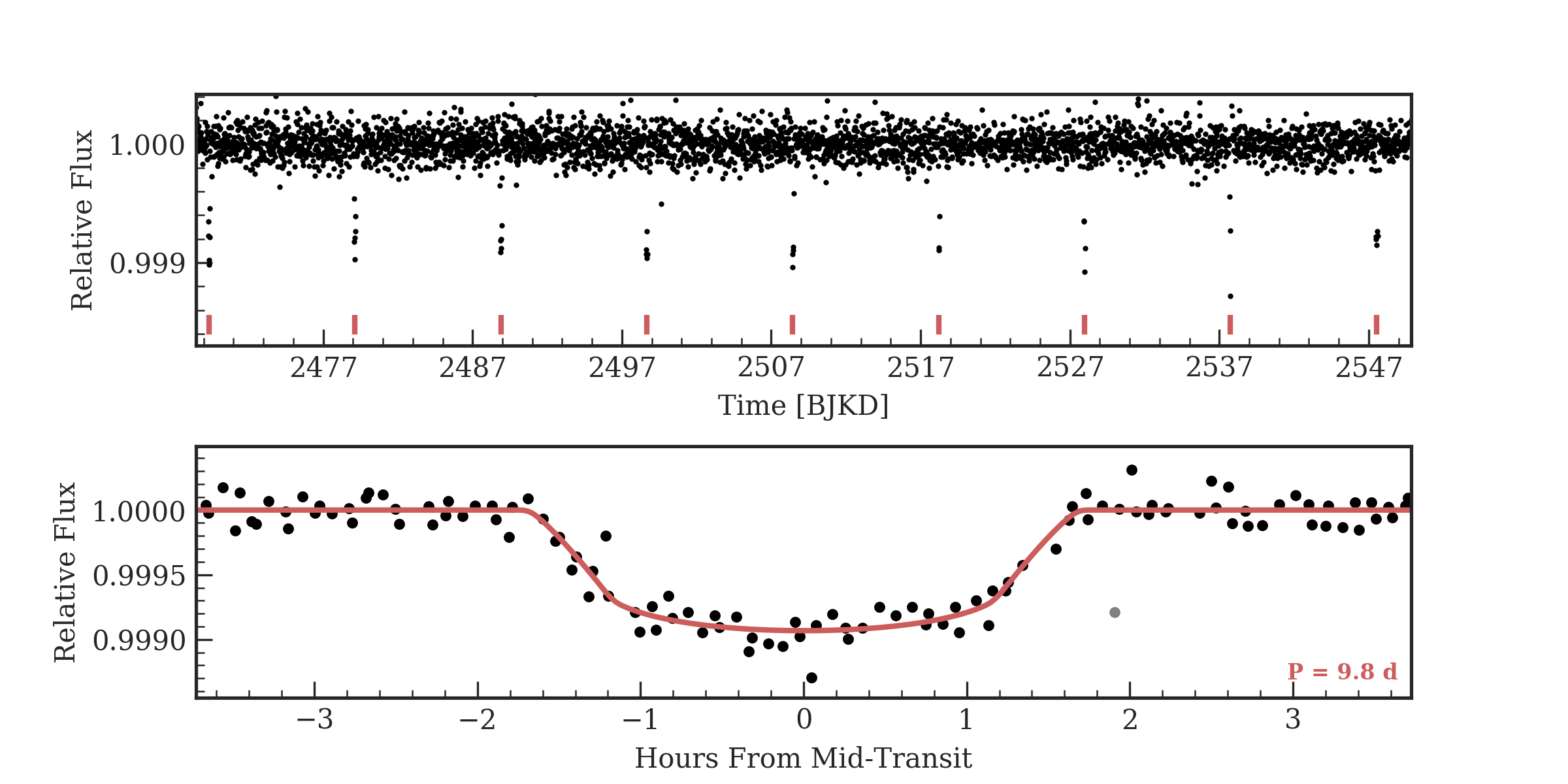}
\caption{Time series (top) and phase-folded (bottom) light curve for the planet orbiting \gcidSTNAME.  Plot formatting is the same as in Fig.\ \ref{fig:lc_epic220709978}.}
\label{fig:lc_epic213546283}
\end{figure*}

\gcidSTNAME is a slightly evolved mid-G star in Field 7 with one transiting planet with a 3.3 \rearth radius in a 10-day orbital period.
See Tables \ref{tb:star_pars} and \ref{tb:star_props} for stellar properties and Table \ref{tb:planet_props} for precise planet parameters.
Our fit of the EVEREST light curve of the K2 photometry for \gcidSTNAME is shown in Fig.\ \ref{fig:lc_epic213546283}.  
The planet is listed in the catalogs of \cite{Petigura2018} and \cite{Mayo2018}, both of which list \gcidPNAMEone as a planet candidate.  \cite{Livingston2018} estimated a false positive probability of 0.034 and designated the system as a planet candidate as well.  Our measurements are not sufficient to {\referee confirm the planet but we do successfully rule out massive eclipsing binary false-positive scenarios, increasing the likelihood that this signal is planetary in origin.}

\begin{figure}
\epsscale{1.0}
\plotone{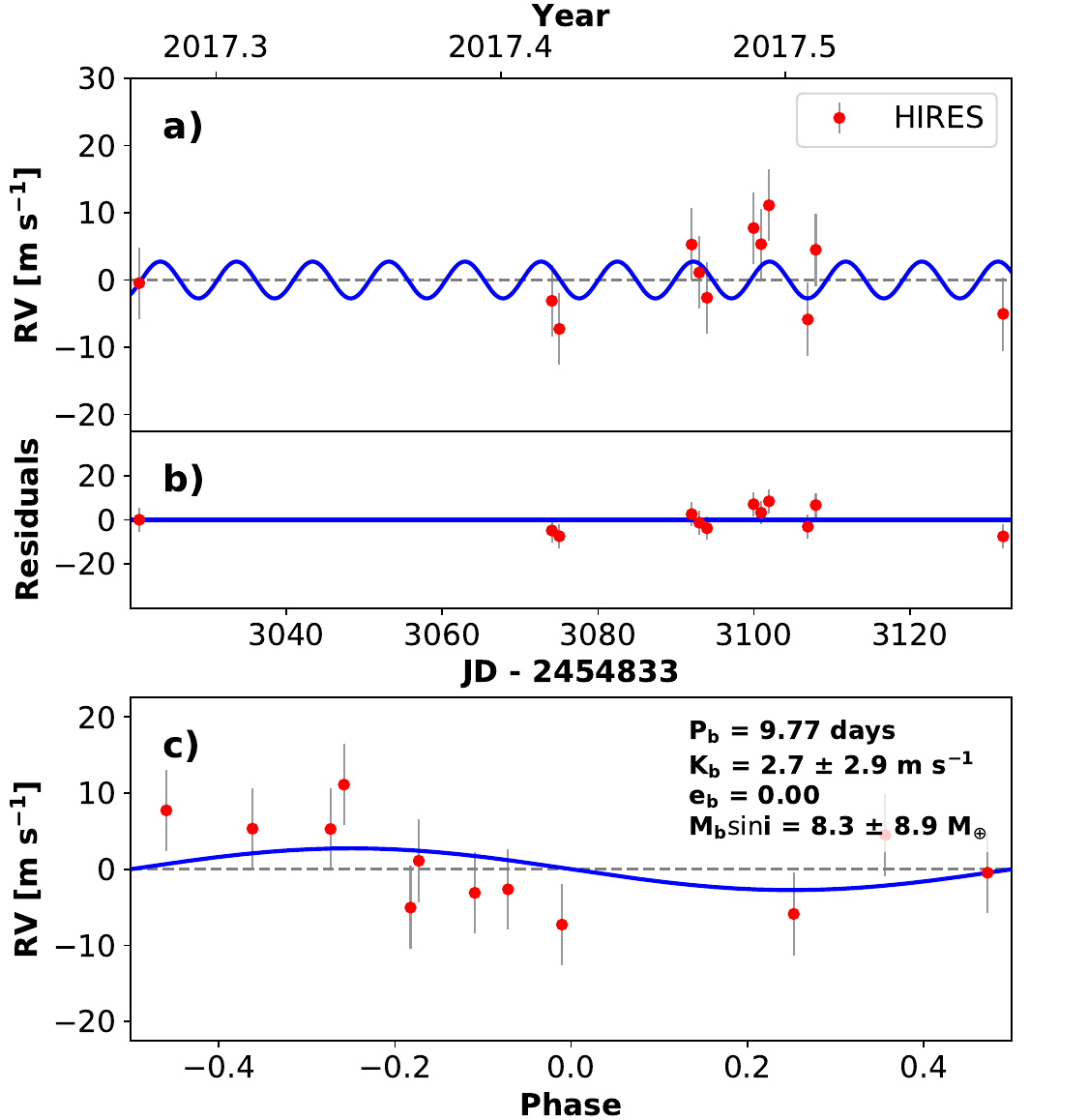}
\caption{RVs and Keplerian model for \gcidSTNAME. Symbols, lines, and annotations are similar to those in Fig.\ \ref{fig:rvs_epic220709978}.}
\label{fig:rvs_epic213546283}
\end{figure}

We acquired \gcidNOBSHIRES RVs of \gcidSTNAME with HIRES, typically with an exposure meter setting of 60,000 counts.  We modeled the system as a single planet in a circular orbit with the orbital period and phase fixed to the transit ephemeris.  The results of this analysis are listed in Table \ref{tab:epic213546283} and the best-fit model is shown in Fig.\ \ref{fig:rvs_epic213546283}.
We detected \gcidPNAMEone with 1-$\sigma$ significance, which gives a density of \gcidRHOPone \gmc, consistent with similarly sized planets Neptune and Uranus.

The stellar rotation period of \gcidSTNAME was readily apparent from quasiperiodic variations in its non-detrended Everest light curve. Although this system did not meet our minimum number of observations requirement, we tried a GP model trained on Everest photometry. This model gave $K$ = $4 \pm 3$ \ms. 

\begin{deluxetable}{lrrr}
\tablecolumns{4}
\tabletypesize{\scriptsize}
\tablecaption{ EPIC 213546283 System Parameters \label{tab:epic213546283}}
\tablehead{
  \colhead{Parameter} & 
  \colhead{Credible} & 
  \colhead{Maximum} & 
  \colhead{Units} \\
  \colhead{} & 
  \colhead{Interval} & 
  \colhead{Likelihood} & 
  \colhead{} }
\startdata
\sidehead{\bf{RV Analysis -- MCMC Step Parameters}} 
$P_{b}$ & $\equiv9.7706$ & $\equiv9.7706$ & days \\ 
$T_{\mathrm{conj},b}$ & $\equiv2457312.1249$ & $\equiv2457312.1249$ & BJD$_\mathrm{TBD}$ \\ 
$\sqrt{e}\cos{\omega}_{b}$ & $\equiv0.0$ & $\equiv0.0$ &  \\ 
$\sqrt{e}\sin{\omega}_{b}$ & $\equiv0.0$ & $\equiv0.0$ &  \\ 
$K_{b}$ & $2.7^{+2.9}_{-2.8}$ & $2.7$ & m s$^{-1}$ \\ 
$\gamma_{\rm HIRES}$ & $-2.0\pm2.1$ & $-2.0$ & m s$-1$ \\ 
$\dot{\gamma}$ & $\equiv0.0$ & $\equiv0.0$ & m s$^{-1}$ day$^{-1}$ \\ 
$\ddot{\gamma}$ & $\equiv0.0$ & $\equiv0.0$ & m s$^{-1}$ day$^{-2}$ \\ 
$\sigma_{\rm HIRES}$ & $6^{+2}_{-1}$ & $4.8$ & $\rm m\ s^{-1}$ \\ 
\hline
\sidehead{\bf{Orbital \& Physical Parameters}}
$P_{b}$ & $\equiv9.7706$ & $\equiv9.7706$ & days \\
$T_{\mathrm{conj},b}$ & $\equiv2457312.1249$ & $\equiv2457312.1249$ & BJD$_\mathrm{TBD}$ \\
$e_{b}$ & $\equiv0.0$ & $\equiv0.0$ &  \\
$\omega_{b}$ & $\equiv0.0$ & $\equiv0.0$ & radians \\
$K_{b}$ & $2.7^{+2.9}_{-2.8}$ & $2.7$ & m s$^{-1}$ \\
$M_b$ & $8\pm9$ & $8.4$ & $M_\earth$ \\
$R_b/R_*$ & $0.0285^{+0.0017}_{-0.0008}$ & $0.0289$ &  \\
$\rho_b$ & $1.1\pm1.2$ & $1.0$ & g cm$^{-3}$ \\
$R_b$ & $3.41^{+0.21}_{-0.10}$ & $3.63$ & $R_\earth$ \\
\hline
\hline
\sidehead{\bf{Priors}}
Parameter & Prior \\
\hline
$\sigma_{\rm HIRES}$ & $\mathcal{U}(0,1e+100)$ \\
\enddata
\end{deluxetable}

\subsection{K2-199} 


\jfjgSTNAME is a K dwarf with two transiting planets with radii 1.8 \rearth and 2.9 \rearth and orbital periods of 3.2 and 7.4 days, respectively.  
See Tables \ref{tb:star_pars}  and \ref{tb:star_props} for stellar properties and Table \ref{tb:planet_props} for precise planet parameters.
The two planets are in the \cite{Petigura2018} and \cite{Mayo2018} catalogs, the latter of which validated them. The planets' masses are reported by \cite{Murphy2021} to be $6.9 \pm 1.8 M_\oplus$ and $12.4 \pm 2.3 M_\oplus$.

\begin{figure*}
\epsscale{1.0}
\plotone{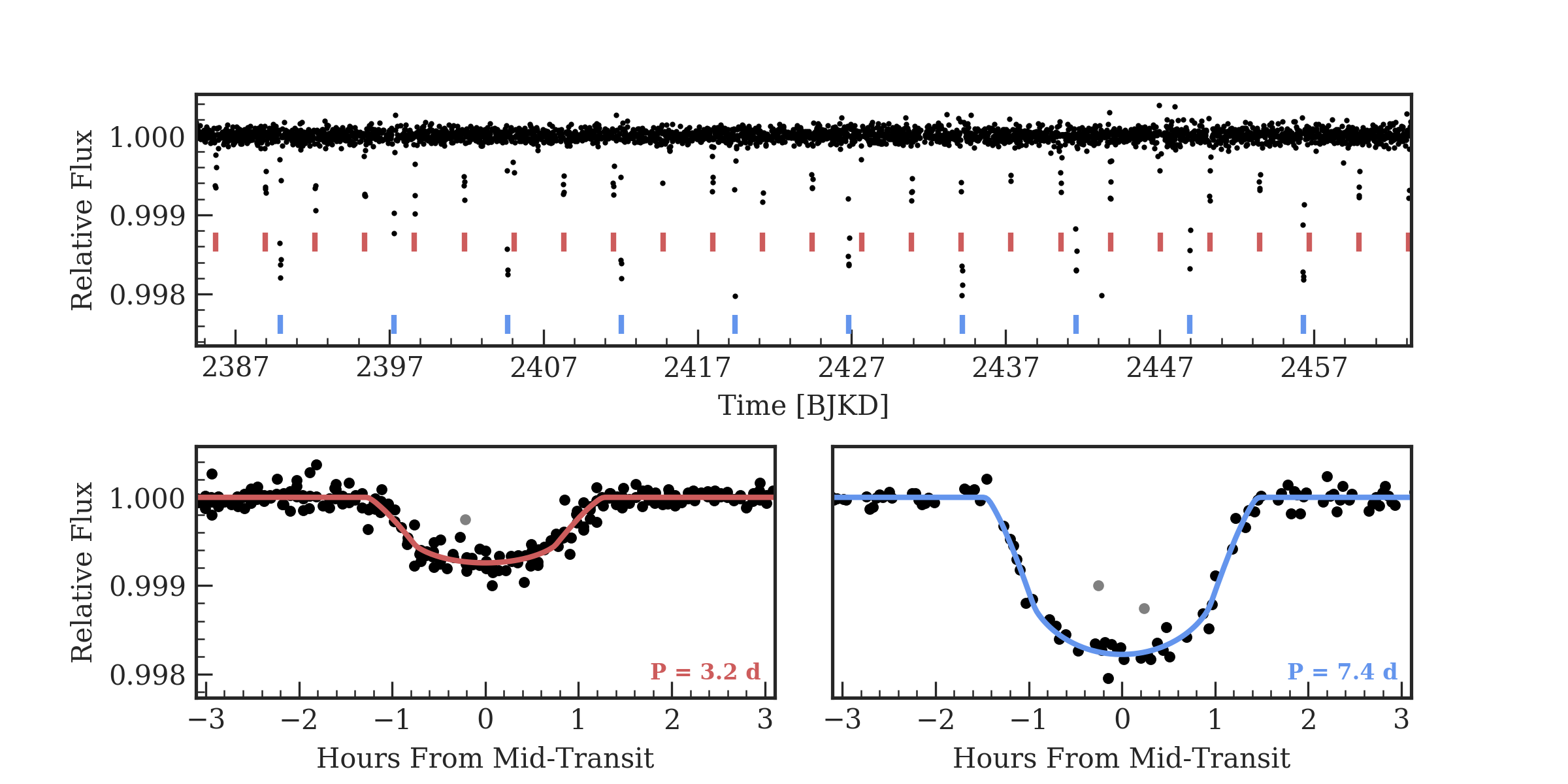}
\caption{Time series (top) and phase-folded (bottom) light curve for the planet orbiting \jfjgSTNAME.  Plot formatting is the same as in Fig.\ \ref{fig:lc_epic220709978}.}
\label{fig:lc_epic212779596}
\end{figure*}

\begin{figure}
\epsscale{1.0}
\plotone{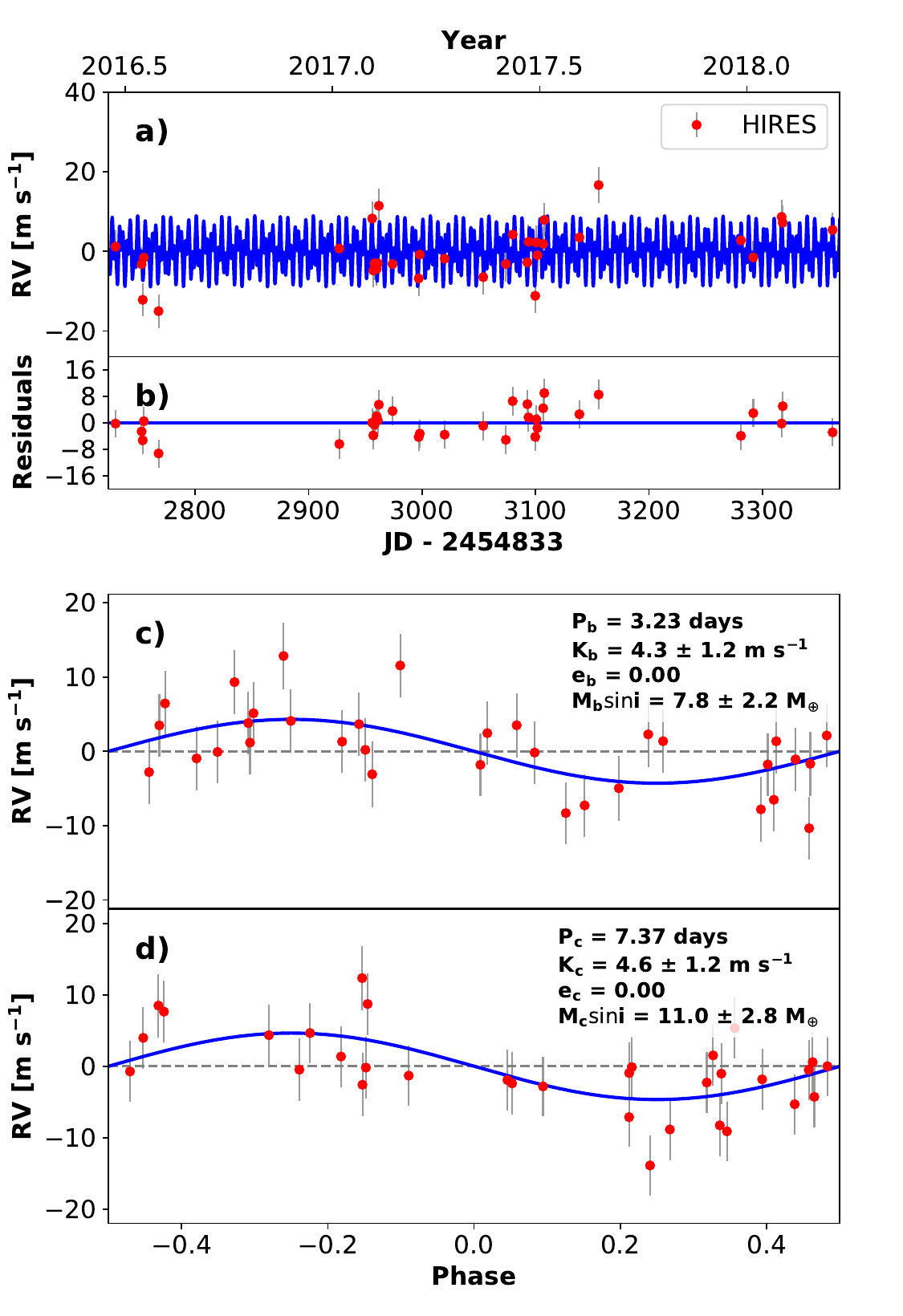}
\caption{RVs and Keplerian model for \jfjgSTNAME. Symbols, lines, and annotations are similar to those in Fig.\ \ref{fig:rvs_epic220709978}.}
\label{fig:rvs_epic212779596}
\end{figure}

Our fit of the EVEREST light curve of the K2 photometry for \jfjgSTNAME is shown in Fig.\ \ref{fig:lc_epic212779596}.  We acquired \jfjgNOBSHIRES RVs of \jfjgSTNAME with HIRES, typically with an exposure meter setting of 60,000.  
We modeled the system as two planets in circular orbits with the orbital periods and phases fixed to the transit ephemerides.  We rejected more complicated models with free eccentricities with/without a linear RV trend using the AICc statistic. A linear trend is slightly preferred (dAICc = 2.5) in the circular model, but because of the low significance we adopt the simpler no-trend model.  The results of our analysis are listed in Table \ref{tab:epic212779596} and the best-fit model is shown in Fig.\ \ref{fig:rvs_epic212779596}.  We detected the Doppler signals of both planets with high significance, consistent with previous measurements \citep{Murphy2021}.  \jfjgPNAMEone is a hot super-Earth with a high density consistent with a rocky composition.  \jfjgPNAMEtwo is a sub-Neptune with an intermediate density. Extending the baseline with more observations would help constrain a possible trend.

\begin{deluxetable}{lrrr}
\tablecolumns{4}
\tabletypesize{\scriptsize}
\tablecaption{ K2-199 System Parameters \label{tab:epic212779596}}
\tablehead{
  \colhead{Parameter} & 
  \colhead{Credible} & 
  \colhead{Maximum} & 
  \colhead{Units} \\
  \colhead{} & 
  \colhead{Interval} & 
  \colhead{Likelihood} & 
  \colhead{} }
\startdata
\sidehead{\bf{RV Analysis -- MCMC Step Parameters}} 
$P_{b}$ & $\equiv3.2253$ & $\equiv3.2253$ & days \\ 
$T_{\mathrm{conj},b}$ & $\equiv2457221.9649$ & $\equiv2457221.9649$ & BJD$_\mathrm{TBD}$ \\ 
$\sqrt{e}\cos{\omega}_{b}$ & $\equiv0.0$ & $\equiv0.0$ &  \\ 
$\sqrt{e}\sin{\omega}_{b}$ & $\equiv0.0$ & $\equiv0.0$ &  \\ 
$K_{b}$ & $4\pm1$ & $4.0$ & m s$^{-1}$ \\ 
$P_{c}$ & $\equiv7.3744$ & $\equiv7.3744$ & days \\ 
$T_{\mathrm{conj},c}$ & $\equiv2457222.9312$ & $\equiv2457222.9312$ & BJD$_\mathrm{TBD}$ \\ 
$\sqrt{e}\cos{\omega}_{c}$ & $\equiv0.0$ & $\equiv0.0$ &  \\
\enddata
\end{deluxetable}

\subsection{EPIC 245991048} 


\baeiSTNAME is a solar-type star in Field 12 with one transiting planet with a radius of 2.2 \rearth and an orbital period of 8 days. The planet does not appear in any catalogs to date, but was detected by our pipeline.  Our observations are insufficient to validate the planet.  See Tables \ref{tb:star_pars}  and \ref{tb:star_props} for stellar properties and Table \ref{tb:planet_props} for precise planet parameters.

\begin{figure*}
\epsscale{1.0}
\plotone{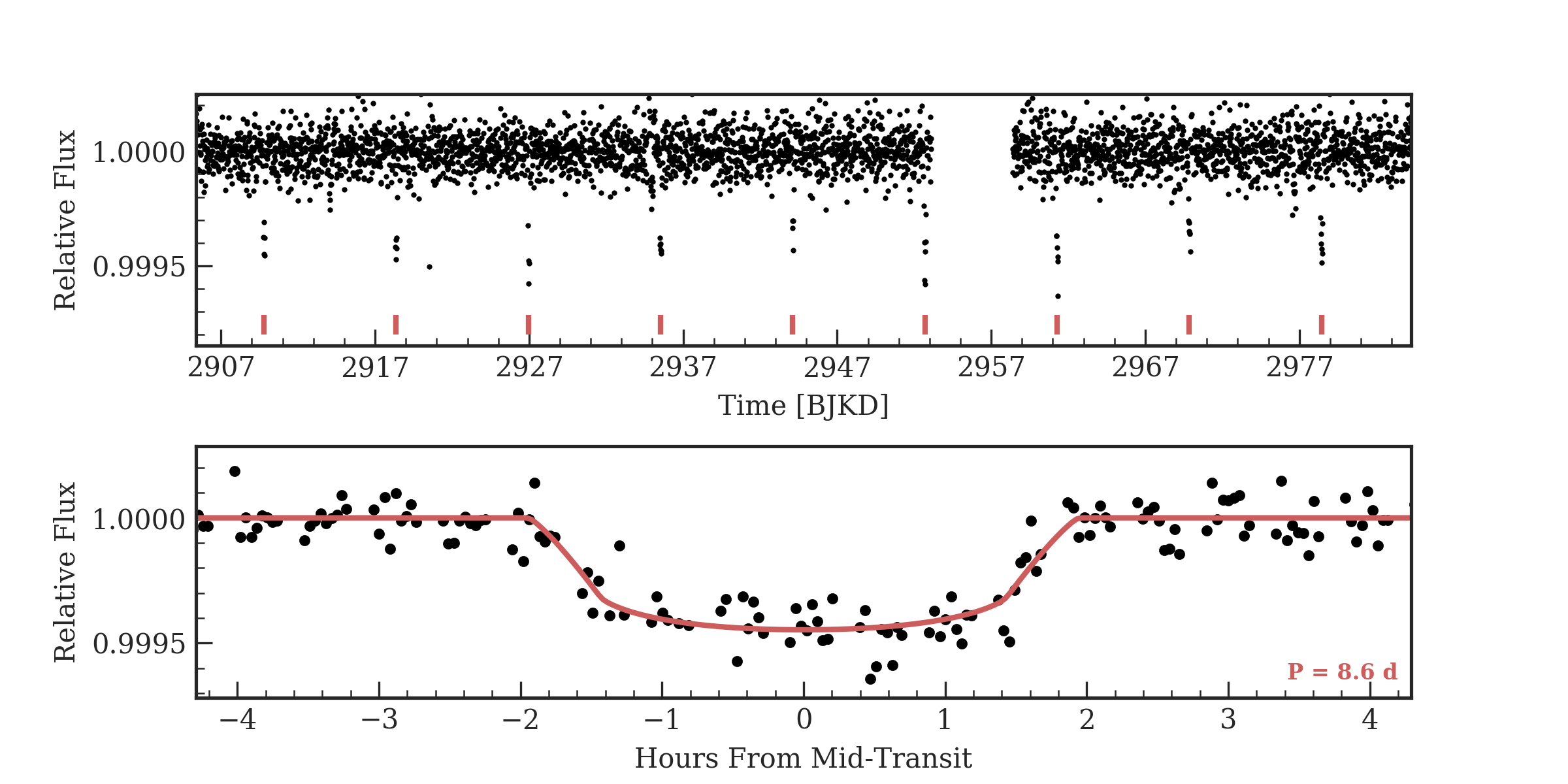}
\caption{Time series (top) and phase-folded (bottom) light curve for the planet orbiting \baeiSTNAME.  Plot formatting is the same as in Fig.\ \ref{fig:lc_epic220709978}.}
\label{fig:lc_epic245991048}
\end{figure*}

\begin{figure}
\epsscale{1.0}
\plotone{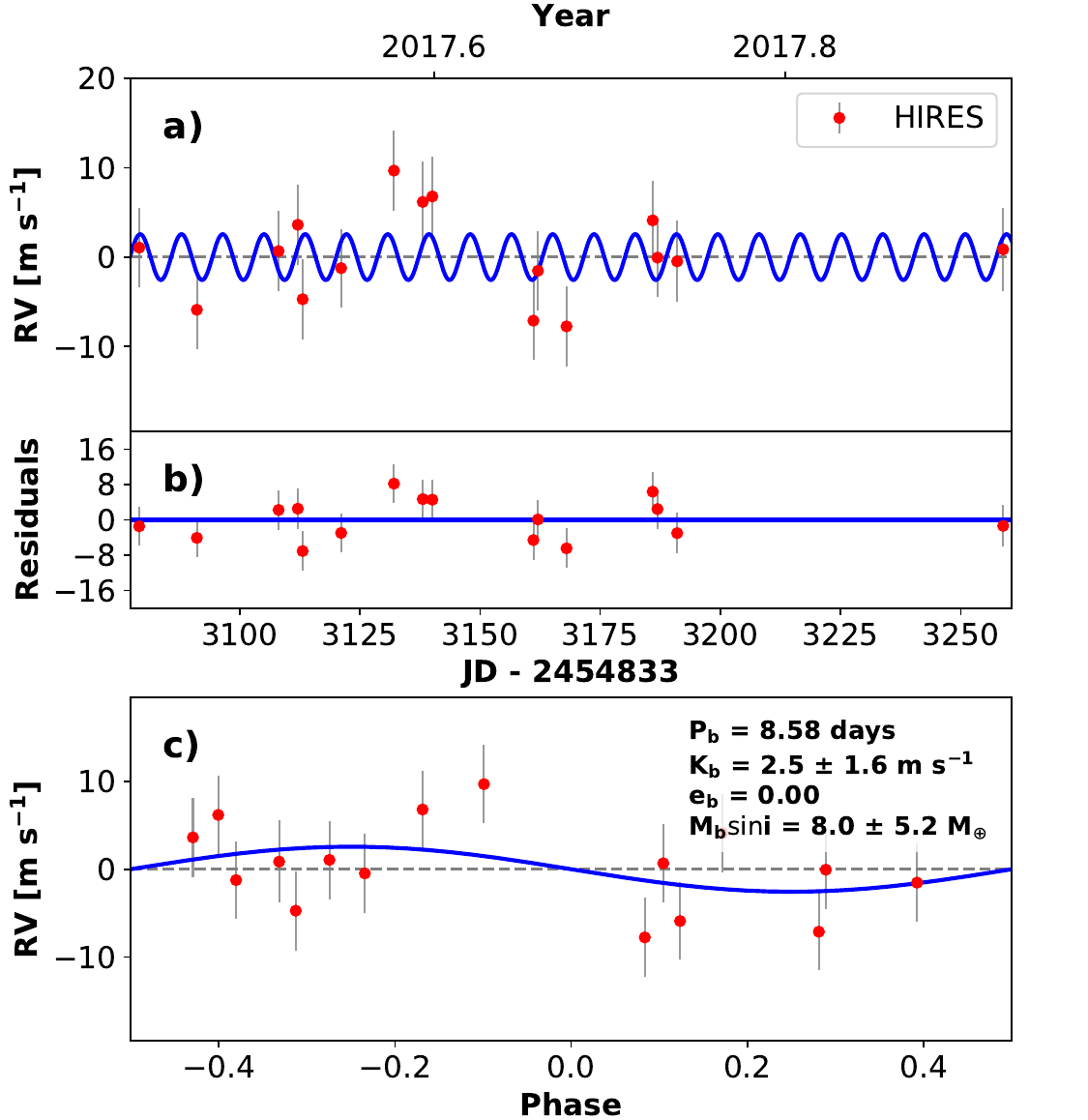}
\caption{RVs and Keplerian model for \baeiSTNAME. Symbols, lines, and annotations are similar to those in Fig.\ \ref{fig:rvs_epic220709978}.}
\label{fig:rvs_epic245991048}
\end{figure}

Our fit of the EVEREST light curve of K2 photometry for \baeiSTNAME is shown in Fig. \ref{fig:lc_epic245991048}.  We acquired \baeiNOBSHIRES RVs of \baeiSTNAME with HIRES, typically with an exposure meter setting of 80,000 counts.  We modeled the system as a single planet in a circular orbit with an orbital period and phase fixed to the transit ephemeris.  We rejected more complicated models with a free eccentricity and a linear RV trend based on the AICc statistic.  The results of this analysis are listed in Table \ref{tab:epic245991048} and the best fit model is shown in Fig.\ \ref{fig:rvs_epic245991048}. 
\baeiPNAMEone appears to be a sub-Neptune with an intermediate bulk density.

We tried a GP fit for this system but adopted a non-GP fit because of the small number of available RVs. A photometry-trained GP found $K_b$ = $2.3 \pm 1.9$ \ms, while an untrained GP found $K_b$ = $1.5 \pm 2.3$ \ms. Note that the median stellar rotation period parameter, $\eta_3$, returned by the trained GP is different from the $\approx50$ day periodic signal seen in the residual plot of Figure \ref{fig:rvs_epic245991048}.

\begin{deluxetable}{lrrr}
\tablecolumns{4}
\tabletypesize{\scriptsize}
\tablecaption{ EPIC 245991048 System Parameters \label{tab:epic245991048}}
\tablehead{
  \colhead{Parameter} & 
  \colhead{Credible} & 
  \colhead{Maximum} & 
  \colhead{Units} \\
  \colhead{} & 
  \colhead{Interval} & 
  \colhead{Likelihood} & 
  \colhead{} }
\startdata
\sidehead{\bf{RV Analysis -- MCMC Step Parameters}} 
$P_{b}$ & $\equiv8.5829$ & $\equiv8.5829$ & days \\ 
$T_{\mathrm{conj},b}$ & $\equiv2457742.7821$ & $\equiv2457742.7821$ & BJD$_\mathrm{TBD}$ \\ 
$\sqrt{e}\cos{\omega}_{b}$ & $\equiv0.0$ & $\equiv0.0$ &  \\ 
$\sqrt{e}\sin{\omega}_{b}$ & $\equiv0.0$ & $\equiv0.0$ &  \\ 
$K_{b}$ & $2.5^{+1.6}_{-1.7}$ & $2.6$ & m s$^{-1}$ \\ 
$\gamma_{\rm HIRES}$ & $-0.9\pm1.3$ & $-0.9$ & m s$-1$ \\ 
$\dot{\gamma}$ & $\equiv0.0$ & $\equiv0.0$ & m s$^{-1}$ day$^{-1}$ \\ 
$\ddot{\gamma}$ & $\equiv0.0$ & $\equiv0.0$ & m s$^{-1}$ day$^{-2}$ \\ 
$\sigma_{\rm HIRES}$ & $4.4^{+1.8}_{-1.3}$ & $4.0$ & $\rm m\ s^{-1}$ \\ 
\hline
\sidehead{\bf{Orbital \& Physical Parameters}}
$P_{b}$ & $\equiv8.5829$ & $\equiv8.5829$ & days \\
$T_{\mathrm{conj},b}$ & $\equiv2457742.7821$ & $\equiv2457742.7821$ & BJD$_\mathrm{TBD}$ \\
$e_{b}$ & $\equiv0.0$ & $\equiv0.0$ &  \\
$\omega_{b}$ & $\equiv0.0$ & $\equiv0.0$ & radians \\
$K_{b}$ & $2.5^{+1.6}_{-1.7}$ & $2.6$ & m s$^{-1}$ \\
$M_b$ & $8.0\pm5.0$ & $8.0$ & $M_\earth$ \\
$R_b/R_*$ & $0.0196^{+0.0013}_{-0.0002}$ & $0.0193$ &  \\
$\rho_b$ & $3.3^{+1.9}_{-2.6}$ & $3.7$ & g cm$^{-3}$ \\
$R_b$ & $2.33^{+0.15}_{-0.03}$ & $2.30$ & $R_\earth$ \\
\hline
\hline
\sidehead{\bf{Priors}}
Parameter & Prior \\
\hline
\\None \\
\enddata
\end{deluxetable}

\subsection{K2-32} 



\bjieSTNAME is a K0 dwarf from Campaign 2 with three transiting planets with orbital periods near the 3:2:1 commensurability.  The planets have sizes 5.1 \rearth, 3 \rearth, and 3.4 \rearth, and orbital periods of 9 days, 21 days, and 32 days, respectively. 
 See Tables \ref{tb:star_pars}  and \ref{tb:star_props} for stellar properties and Table \ref{tb:planet_props} for precise planet parameters.

The planets were listed as candidates in \cite{Vanderburg2016-catalog} and were subsequently confirmed in \cite{Sinukoff2016}. \cite{Dai2016} obtained 43 RVs from HARPS and 6 RVs from PFS, measured a mass for the innermost planet, and obtained upper limits for plances c and d. \cite{Petigura2017} added 31 HIRES observations and found masses of $16.5 \pm 2.7 \mearth$, $< 12.1 \mearth$ (95\% confidence), and $10.3 \pm 4.7 \mearth$ for planets b, c, and d respectively.   After our analysis was complete, \cite{Heller2019} applied a new transit least squares algorithm and detected a fourth transiting planet with a radius of 1 \rearth and an orbital period of 4.35 days, which would make the commensurability chain near 1:2:3:7. Our transit search did not recover this planet. With an expected RV semiamplitude of only $0.4$,\ms, we do not expect it to contribute significantly to the Keplerian model. \cite{LilloBox2020} measure masses of planets b, c, d, and e of $15 \pm 1.8$, $8.1 \pm 2.4$, $6.7  \pm 2.5$, and $2.1 \pm 1.2 M_\oplus$, respectively.

\begin{figure*}
\epsscale{1.0}
\plotone{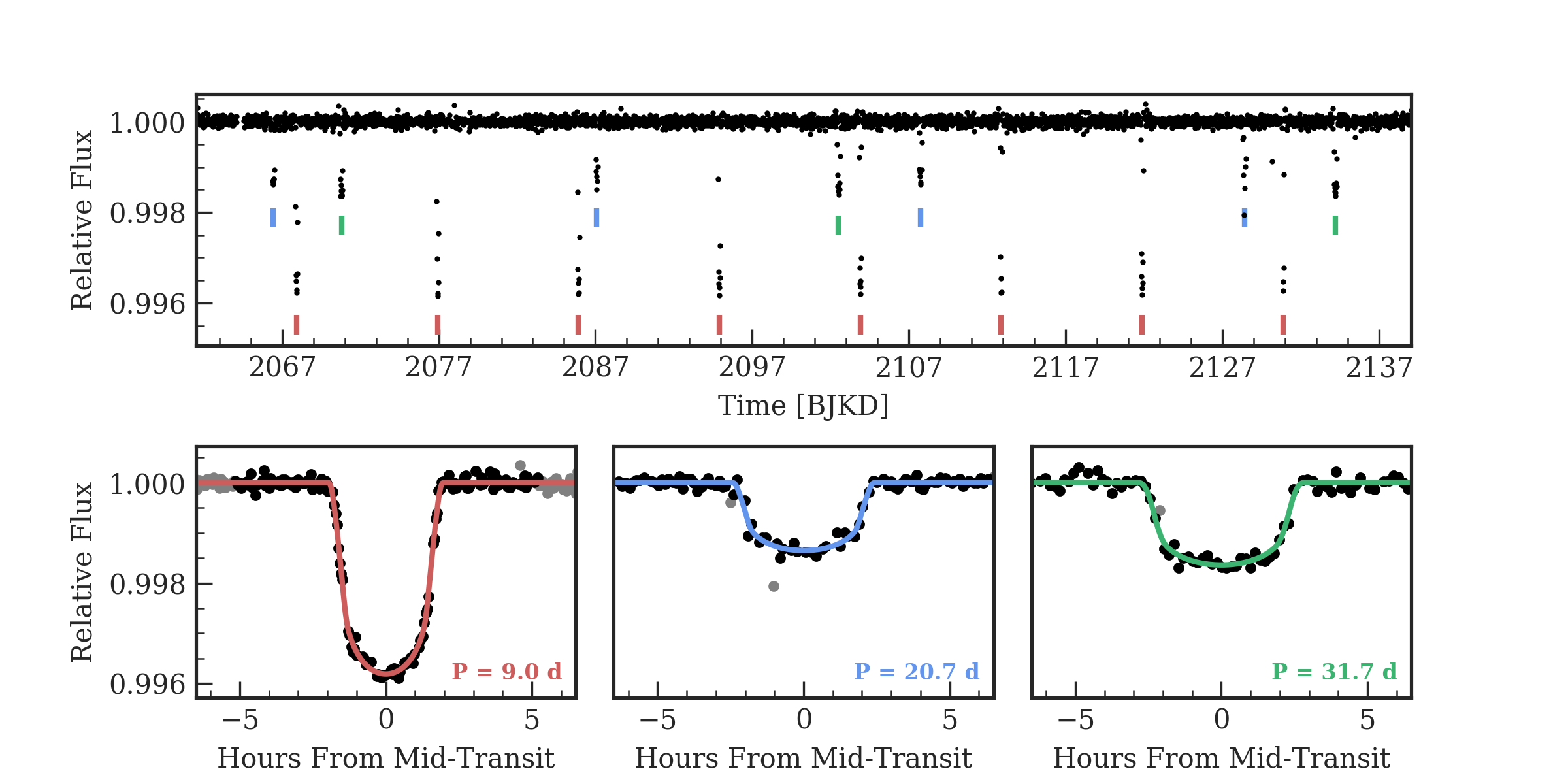}
\caption{Time series (top) and phase-folded (bottom) light curve for the planet orbiting \bjieSTNAME.  Plot formatting is the same as in Fig.\ \ref{fig:lc_epic220709978}.}
\label{fig:lc_epic205071984}
\end{figure*}

Our fit of the EVEREST light curve of the K2 photometry for \bjieSTNAME  is shown in Fig.\ \ref{fig:lc_epic205071984}.

\begin{figure}
\epsscale{1.0}
\plotone{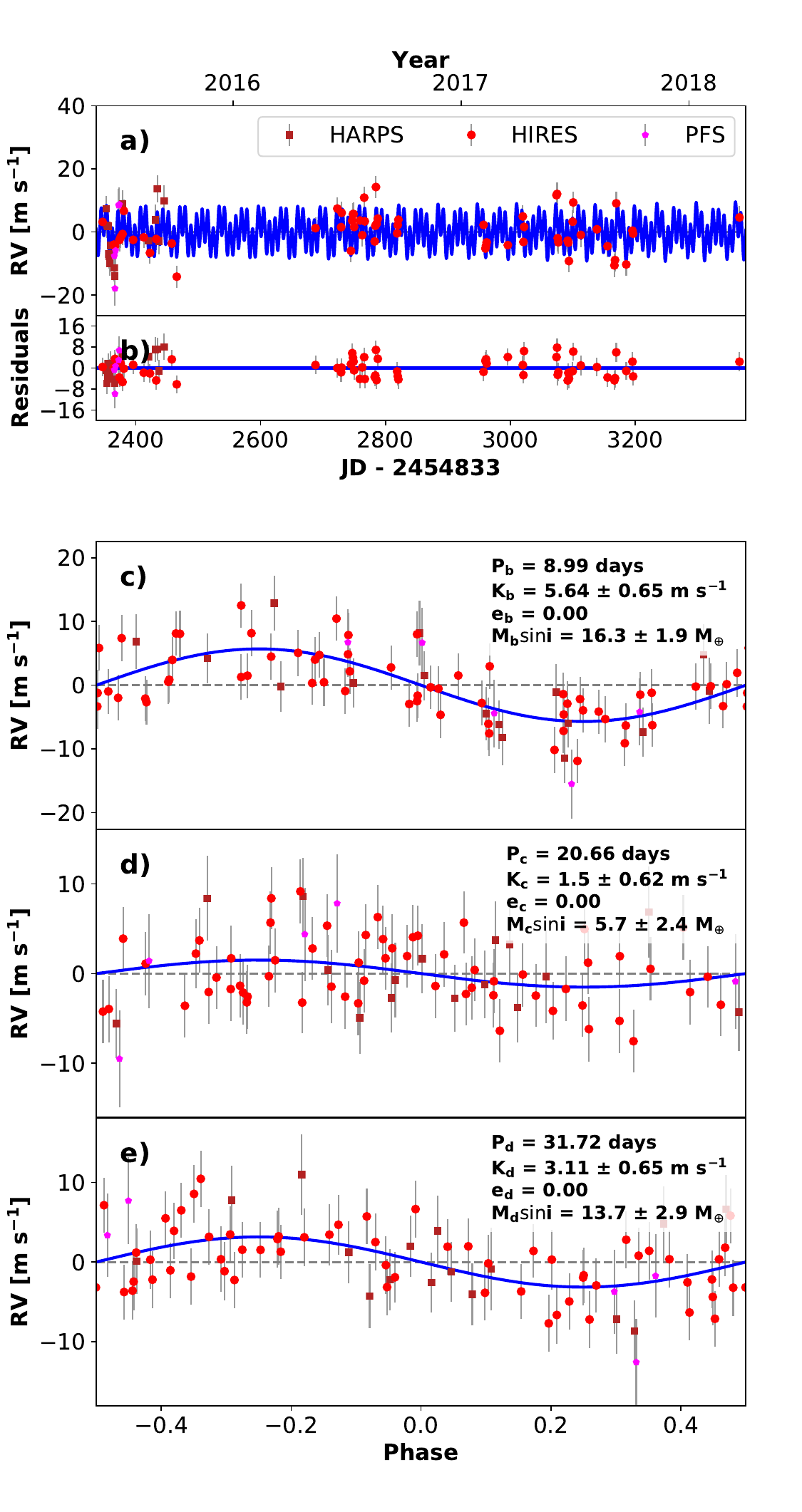}
\caption{RVs and Keplerian model for \bjieSTNAME. Symbols, lines, and annotations are similar to those in Fig.\ \ref{fig:rvs_epic220709978}.}
\label{fig:rvs_epic205071984}
\end{figure}

We acquired \bjieNOBSHIRES RVs of \bjieSTNAME with HIRES, typically with an exposure meter setting of 60,000 counts.  
We modeled the system as three planets (b, c, d) in circular orbits with orbital periods and phases fixed to transit ephemerides and included the HARPS and PFS RVs from \cite{Dai2016} in our analysis.  The results of this analysis are listed in Table \ref{tab:epic205071984}
and the best-fit model is shown in Fig.\ \ref{fig:rvs_epic205071984}. More complex models including eccentric orbits for each planet and a linear trend are strongly disfavored by an AICc comparison. We recovered planets b and d and find masses generally consistent with those in the literature \citep{Petigura2017,LilloBox2020}.  With the additional RVs we also measured a $\sim$2$\sigma$ mass for planet c of \bjieMPtwo\,\mearth. At $\sim$0.4\,\ms, the putative fourth planet at 4.35 days represents a challenge for next-generation EPRV instruments.

\begin{deluxetable}{lrrr}
\tablecolumns{4}
\tabletypesize{\scriptsize}
\tablecaption{ K2-32 System Parameters \label{tab:epic205071984}}
\tablehead{
  \colhead{Parameter} & 
  \colhead{Credible} & 
  \colhead{Maximum} & 
  \colhead{Units} \\
  \colhead{} & 
  \colhead{Interval} & 
  \colhead{Likelihood} & 
  \colhead{} }
\startdata
\sidehead{\bf{RV Analysis -- MCMC Step Parameters}} 
$P_{b}$ & $\equiv8.992$ & $\equiv8.992$ & days \\ 
$T_{\mathrm{conj},b}$ & $\equiv2456909.9188$ & $\equiv2456909.9188$ & BJD$_\mathrm{TBD}$ \\ 
$\sqrt{e}\cos{\omega}_{b}$ & $\equiv0.0$ & $\equiv0.0$ &  \\ 
$\sqrt{e}\sin{\omega}_{b}$ & $\equiv0.0$ & $\equiv0.0$ &  \\ 
$K_{b}$ & $5.64^{+0.65}_{-0.66}$ & $5.70$ & m s$^{-1}$ \\ 
$P_{c}$ & $\equiv20.6609$ & $\equiv20.6609$ & days \\ 
$T_{\mathrm{conj},c}$ & $\equiv2456961.4065$ & $\equiv2456961.4065$ & BJD$_\mathrm{TBD}$ \\ 
$\sqrt{e}\cos{\omega}_{c}$ & $\equiv0.0$ & $\equiv0.0$ &  \\
\enddata
\end{deluxetable}


\subsection{K2-108} 


\gghbSTNAME is mid-G star starting to ascend the subgiant branch in Campaign 5 with one transiting planet with a radius of 5.2 \rearth and an orbital period of 4.7 days.
See Tables \ref{tb:star_pars}  and \ref{tb:star_props} for stellar properties and Table \ref{tb:planet_props} for precise planet parameters.  The planet is listed in the \cite{Petigura2018} and \cite{Mayo2018} catalogs.
\gghbPNAMEone was discovered and confirmed by \cite{Petigura2017}, whose solution we adopt here.  They found a mass of $59.4 \pm 4.4$ \mearth with a linear trend using 20 HIRES RVs.  These RVs were mistakenly omitted from \cite{Petigura2017}, but are listed in Table \ref{tab:rvs}.  




\subsection{K2-62} 


\ggacSTNAME is a late K dwarf in Field 3 with two transiting planets with orbital periods of 6.7 days and 16 days.  Both planets have radii of about 2 \rearth.  See Tables \ref{tb:star_pars}  and \ref{tb:star_props} for stellar properties and Table \ref{tb:planet_props} for precise planet parameters.  The planets were validated by \cite{Crossfield2016} and appear in the \cite{Vanderburg2016-catalog} and \cite{Mayo2018} catalogs.

\begin{figure*}
\epsscale{1.0}
\plotone{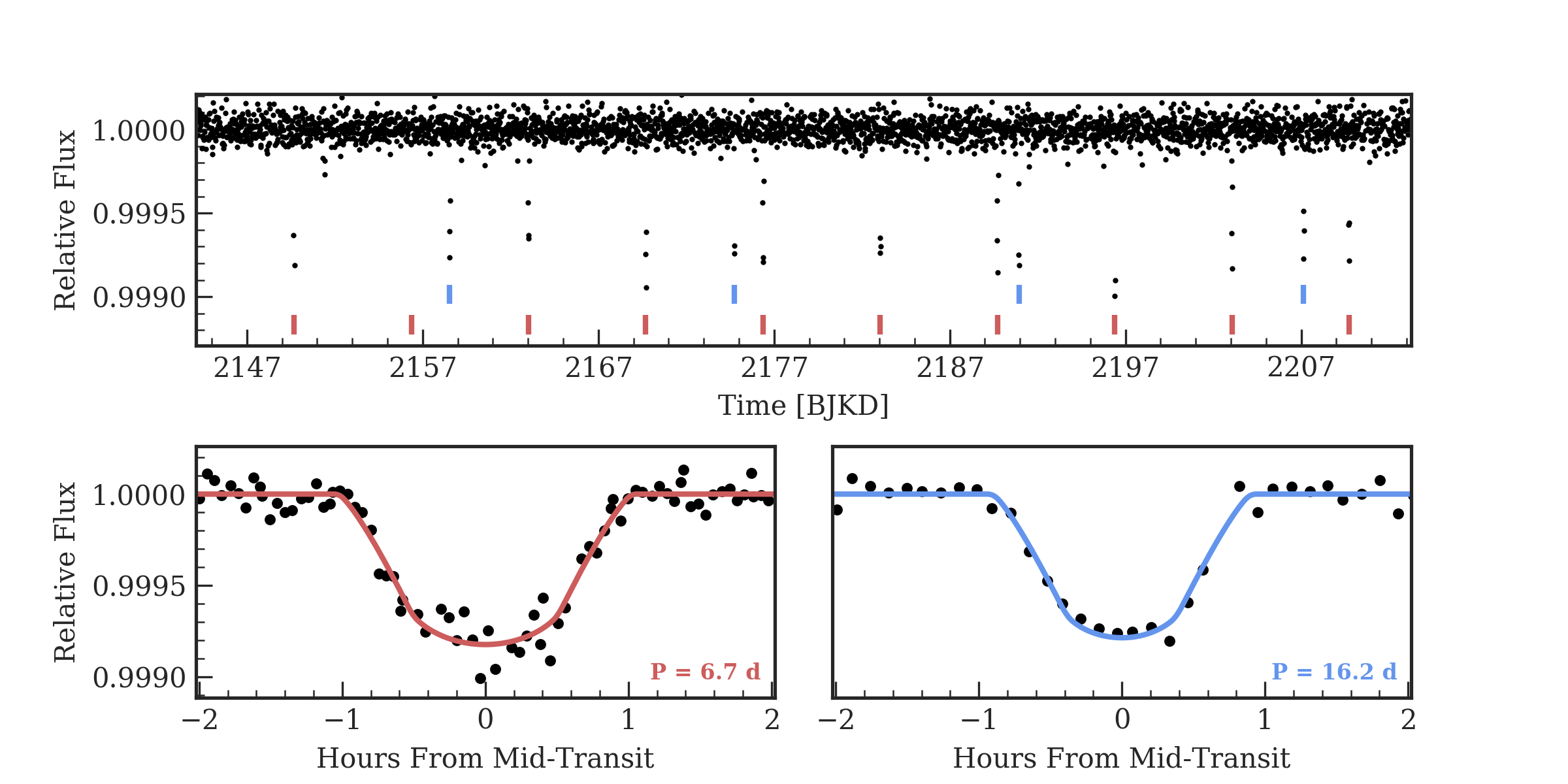}
\caption{Time series (top) and phase-folded (bottom) light curve for the planet orbiting \ggacSTNAME.  Plot formatting is the same as in Fig.\ \ref{fig:lc_epic220709978}.}
\label{fig:lc_epic206096602}
\end{figure*}

\begin{figure}
\epsscale{1.0}
\plotone{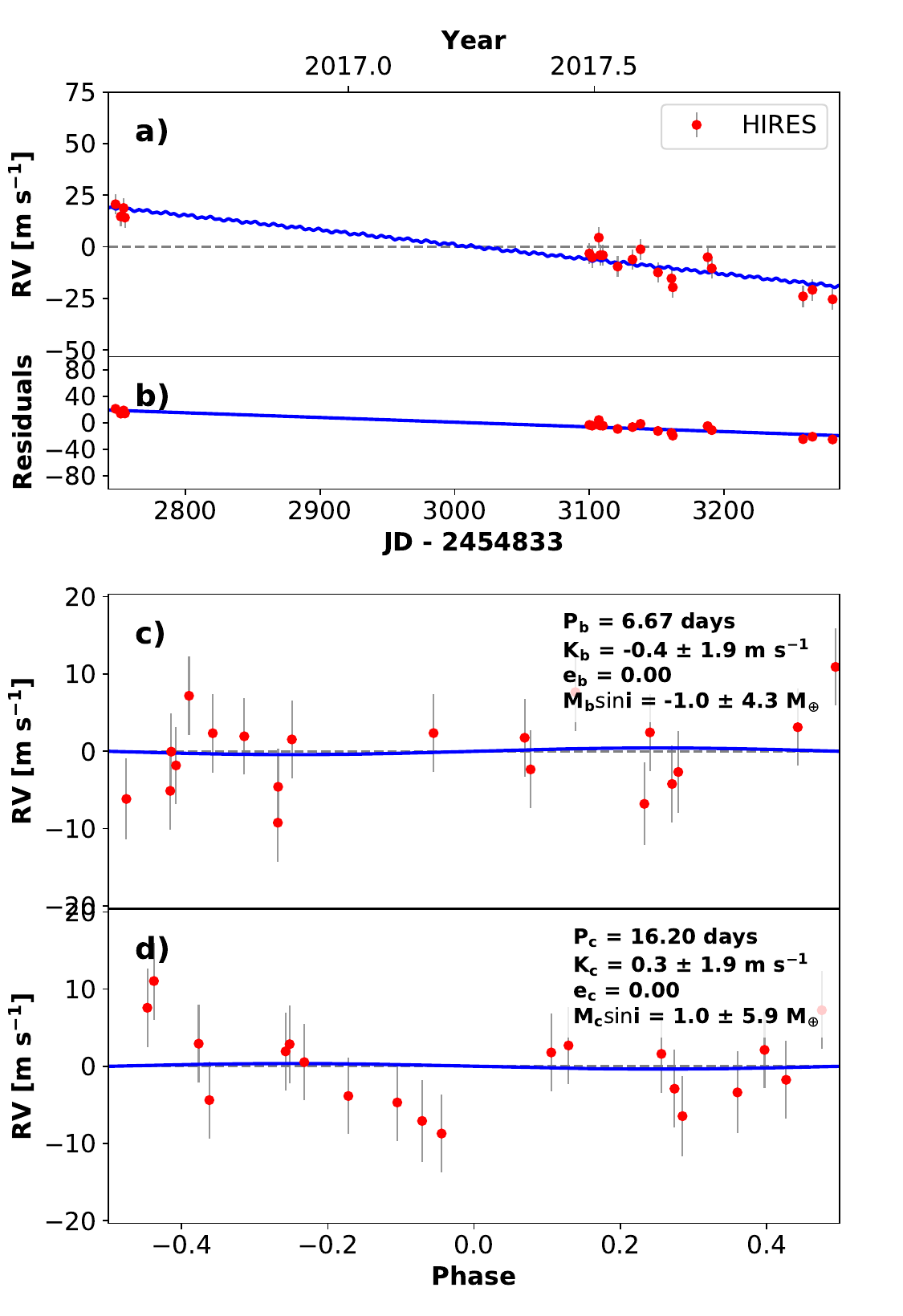}
\caption{RVs and Keplerian model for \ggacSTNAME. Symbols, lines, and annotations are similar to those in Fig.\ \ref{fig:rvs_epic220709978}.}
\label{fig:rvs_k2-62}
\end{figure}

Our fit of the EVEREST light curve of the K2 photometry for \ggacSTNAME is shown in Fig.\ \ref{fig:lc_epic206096602}.  We acquired \ggacNOBSHIRES RVs of \ggacSTNAME with HIRES, typically with an exposure meter setting of 60,000 counts.  We modeled the system as two planets in circular orbits with orbital periods and phases fixed to the transit ephemerides.  We adopted a model with a linear trend based on \dAICc = 32 compared to a flat model.  Similarly, we rejected models with eccentric orbits.  The results of our analysis are listed in Table \ref{tab:epic206096602} and the best fit model is shown in Fig.\ \ref{fig:rvs_k2-62}.
Doppler signals from the two planets are not detected (1-$\sigma$ significance).  The limits on density do not meaningfully constrain the compositions of these two planets.

\begin{deluxetable}{lrrr}
\tablecolumns{4}
\tabletypesize{\scriptsize}
\tablecaption{ K2-62 System Parameters \label{tab:epic206096602}}
\tablehead{
  \colhead{Parameter} & 
  \colhead{Credible} & 
  \colhead{Maximum} & 
  \colhead{Units} \\
  \colhead{} & 
  \colhead{Interval} & 
  \colhead{Likelihood} & 
  \colhead{} }
\startdata
\sidehead{\bf{RV Analysis -- MCMC Step Parameters}} 
$P_{b}$ & $\equiv6.672$ & $\equiv6.672$ & days \\ 
$T_{\mathrm{conj},b}$ & $\equiv2456982.6852$ & $\equiv2456982.6852$ & BJD$_\mathrm{TBD}$ \\ 
$\sqrt{e}\cos{\omega}_{b}$ & $\equiv0.0$ & $\equiv0.0$ &  \\ 
$\sqrt{e}\sin{\omega}_{b}$ & $\equiv0.0$ & $\equiv0.0$ &  \\ 
$K_{b}$ & $-0.4\pm1.9$ & $-0.4$ & m s$^{-1}$ \\ 
$P_{c}$ & $\equiv16.197$ & $\equiv16.197$ & days \\ 
$T_{\mathrm{conj},c}$ & $\equiv2456991.5453$ & $\equiv2456991.5453$ & BJD$_\mathrm{TBD}$ \\ 
$\sqrt{e}\cos{\omega}_{c}$ & $\equiv0.0$ & $\equiv0.0$ &  \\
\enddata
\tablecomments{}{
Reference epoch for $\dot{\gamma}$,$\ddot{\gamma}$: 2457847.4101299997 
}
\end{deluxetable}

\subsection{K2-189} 


\egijSTNAME is a late-G/early-K dwarf from Campaign 6 with two transiting planets with radii of 1.5 and 2.5 \rearth and orbital periods of 2.6 and 6.7 days.
See Tables \ref{tb:star_pars}  and \ref{tb:star_props} for stellar properties and Table \ref{tb:planet_props} for precise planet parameters.
The planets were validated by \cite{Mayo2018,Barros2016} and also appear in \cite{Petigura2018}.

\begin{figure*}
\epsscale{1.0}
\plotone{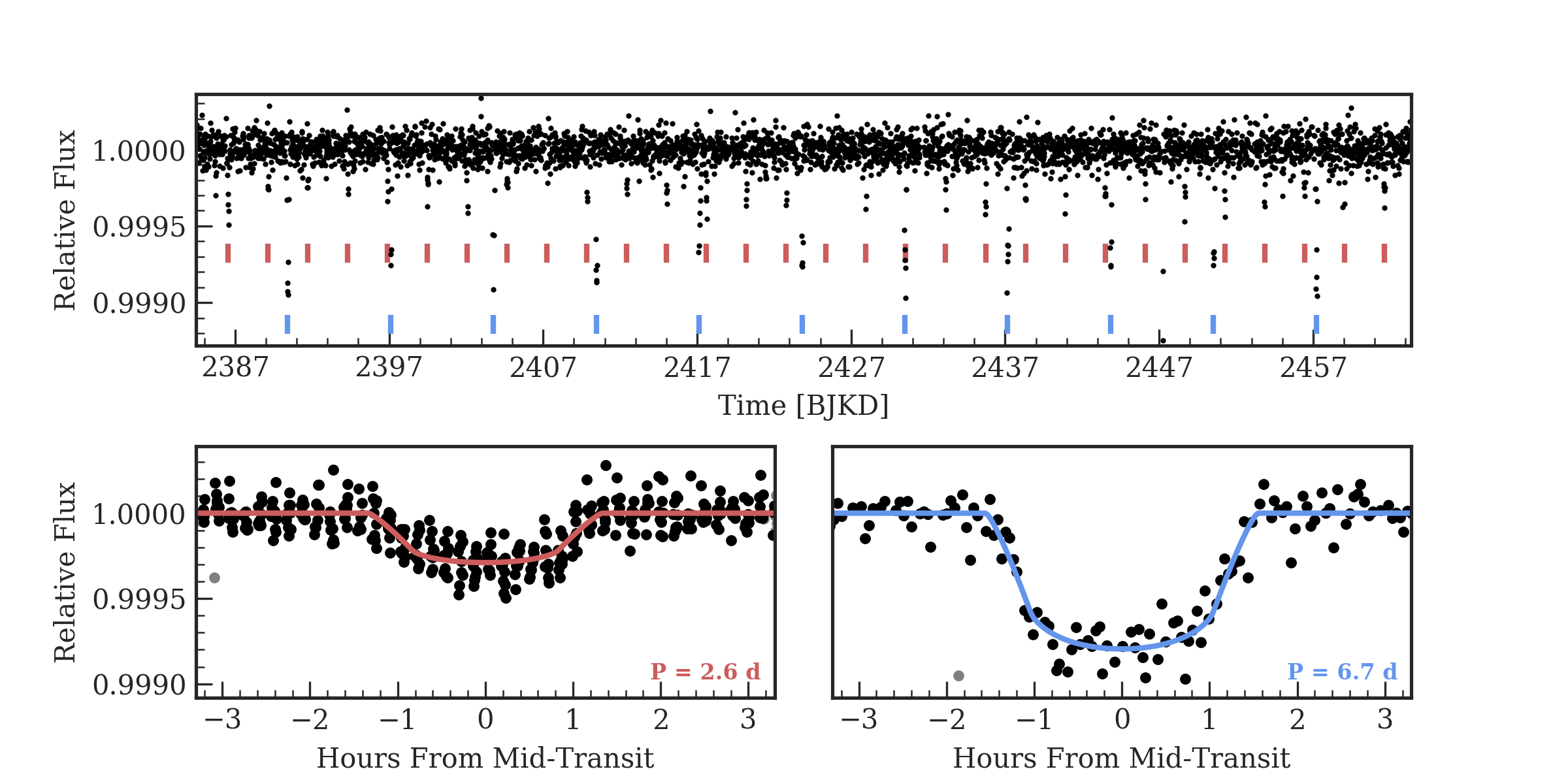}
\caption{Time series (top) and phase-folded (bottom) light curve for the planet orbiting \egijSTNAME.  Plot formatting is the same as in Fig.\ \ref{fig:lc_epic220709978}.}
\label{fig:lc_epic212394689}
\end{figure*}

\begin{figure}
\epsscale{1.0}
\plotone{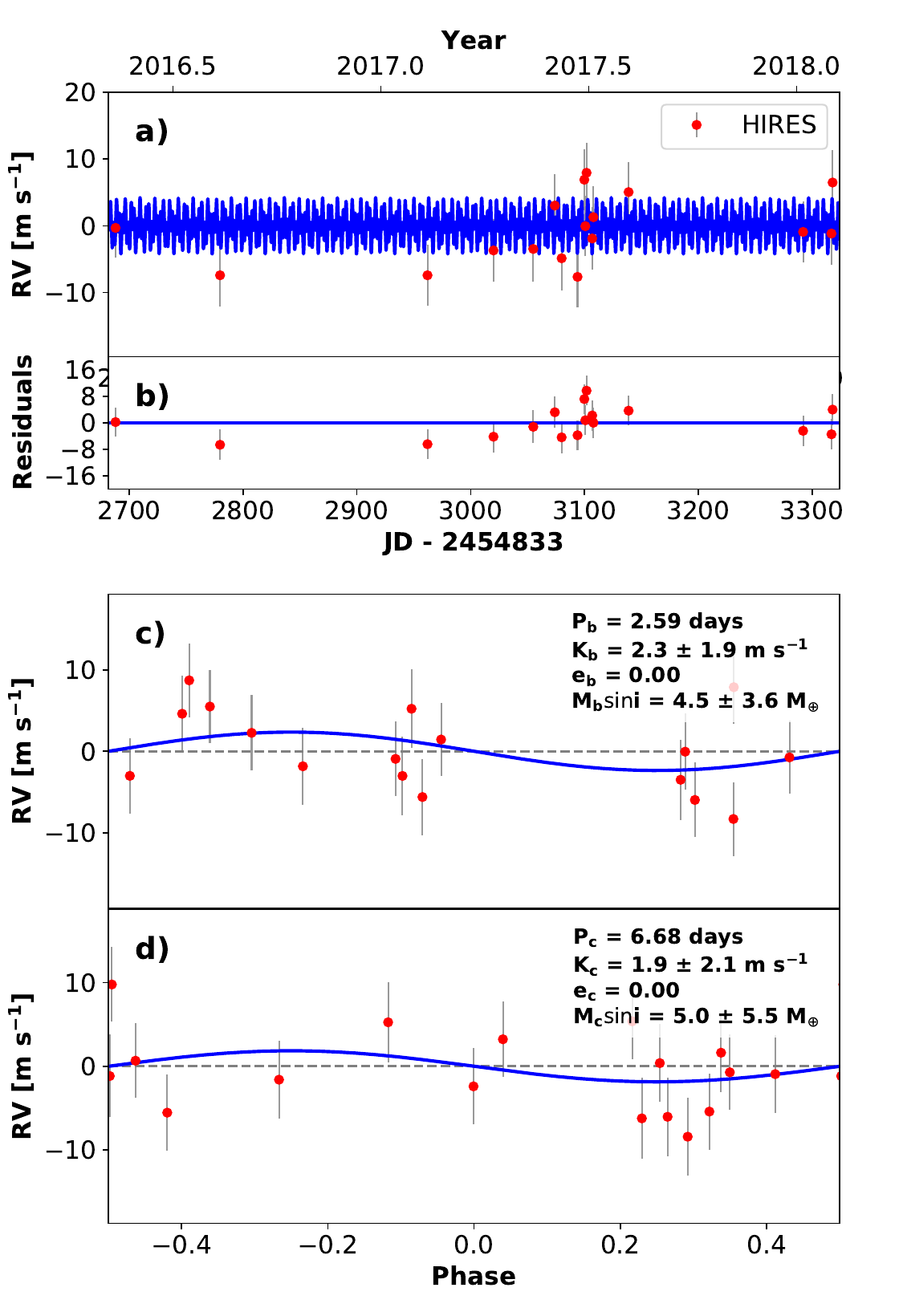}
\caption{RVs and Keplerian model for \egijSTNAME. Symbols, lines, and annotations are similar to those in Fig.\ \ref{fig:rvs_epic220709978}.}
\label{fig:rvs_k2-189}
\end{figure}

Our fit of the EVEREST light curve of the K2 photometry for \egijSTNAME is shown in Fig.\ \ref{fig:lc_epic212394689}.  We acquired \egijNOBSHIRES RVs of \egijSTNAME with HIRES, typically with an exposure meter setting of 60,000 counts.   We modeled the system as two planets in circular orbits with orbital periods and phases fixed to the transit ephemerides.  We rejected more complicated models with free eccentricity and/or a linear RV trend using the AICc statistic.  The results of our analysis are listed in Table \ref{tab:epic212394689} and the best-fit model is shown in Fig.\ \ref{fig:rvs_k2-189}.
Doppler signals from two planets are each detected with 1-$\sigma$ significance, which weakly favors a high density and a rocky composition for the super-Earth \egijPNAMEone and a low density and gas-dominated composition for \egijPNAMEone.


\begin{deluxetable}{lrrr}
\tablecolumns{4}
\tabletypesize{\scriptsize}
\tablecaption{ K2-189 System Parameters \label{tab:epic212394689}}
\tablehead{
  \colhead{Parameter} & 
  \colhead{Credible} & 
  \colhead{Maximum} & 
  \colhead{Units} \\
  \colhead{} & 
  \colhead{Interval} & 
  \colhead{Likelihood} & 
  \colhead{} }
\startdata
\sidehead{\bf{RV Analysis -- MCMC Step Parameters}} 
$P_{b}$ & $\equiv2.5881$ & $\equiv2.5881$ & days \\ 
$T_{\mathrm{conj},b}$ & $\equiv2457222.1466$ & $\equiv2457222.1466$ & BJD$_\mathrm{TBD}$ \\ 
$\sqrt{e}\cos{\omega}_{b}$ & $\equiv0.0$ & $\equiv0.0$ &  \\ 
$\sqrt{e}\sin{\omega}_{b}$ & $\equiv0.0$ & $\equiv0.0$ &  \\ 
$K_{b}$ & $2.3^{+1.8}_{-1.9}$ & $2.4$ & m s$^{-1}$ \\ 
$P_{c}$ & $\equiv6.6793$ & $\equiv6.6793$ & days \\ 
$T_{\mathrm{conj},c}$ & $\equiv2457223.4182$ & $\equiv2457223.4182$ & BJD$_\mathrm{TBD}$ \\ 
$\sqrt{e}\cos{\omega}_{c}$ & $\equiv0.0$ & $\equiv0.0$ &  \\
\enddata
\end{deluxetable}

\subsection{K2-10}  
\label{sec:k2_10}

\hadfSTNAME is a G dwarf in Field 1 with one transiting planet with a radius of 3.6 \rearth and an orbital period of 19 days.
See Tables \ref{tb:star_pars}  and \ref{tb:star_props} for stellar properties and Table \ref{tb:planet_props} for precise planet parameters.
The planet was discovered and validated by \cite{Montet2015}.  The planet also appears in the \cite{Crossfield2016}, \cite{Vanderburg2016-catalog}, and \cite{Schmitt2016} catalogs.
\cite{vanEylen2016a} measured a mass of $27^{+17}_{-16}$ \mearth and a density of $2.6^{+2.1}_{-1.6}$ \gmc based on 15 RVs from HARPS-N and 7 from FIES.

\begin{figure*}
\epsscale{1.0}
\plotone{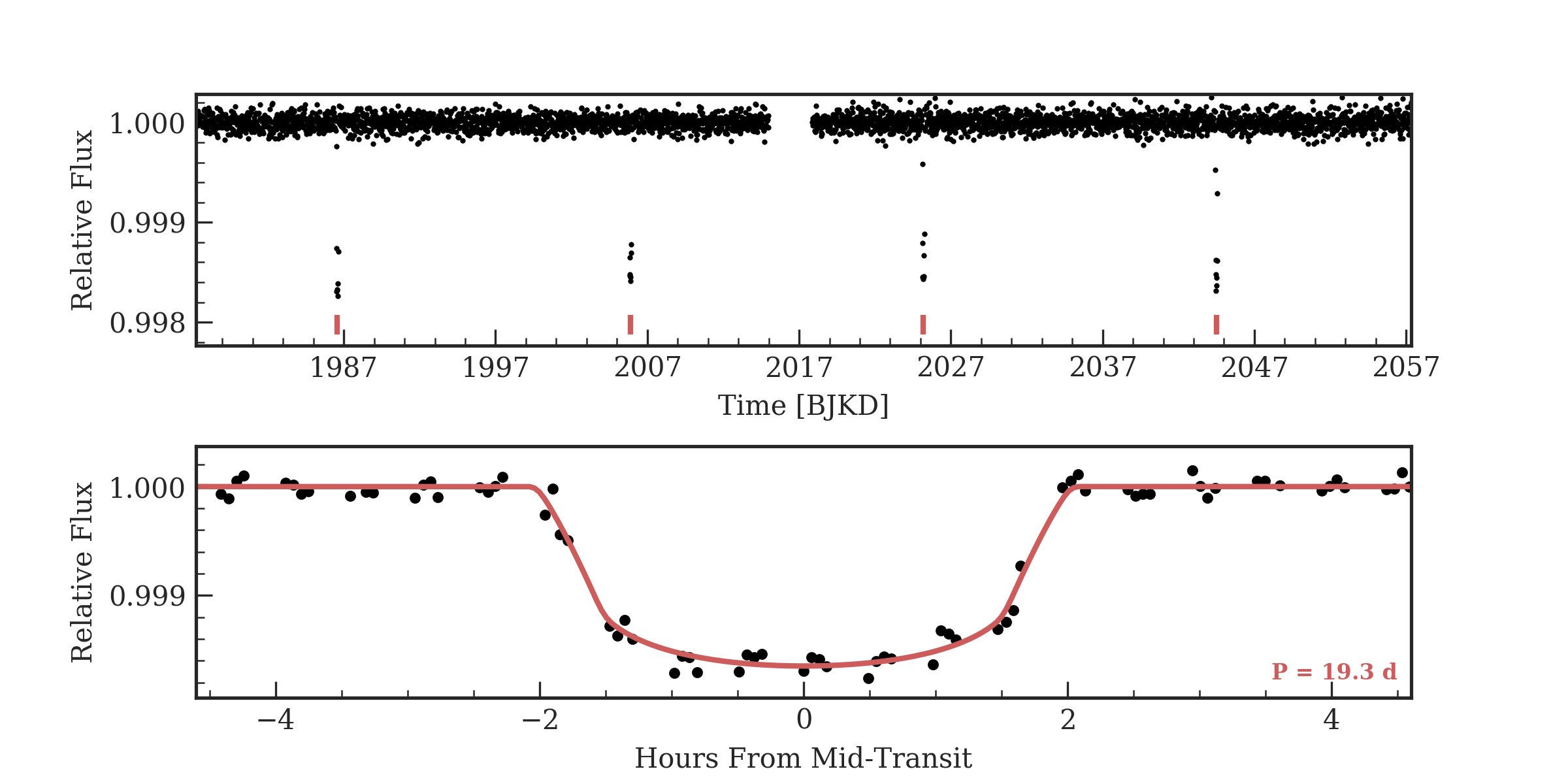}
\caption{Time series (top) and phase-folded (bottom) light curve for the planet orbiting \hadfSTNAME.  Plot formatting is the same as in Fig.\ \ref{fig:lc_epic220709978}.}
\label{fig:lc_epic201577035}
\end{figure*}

\begin{figure}
\epsscale{1.0}
\plotone{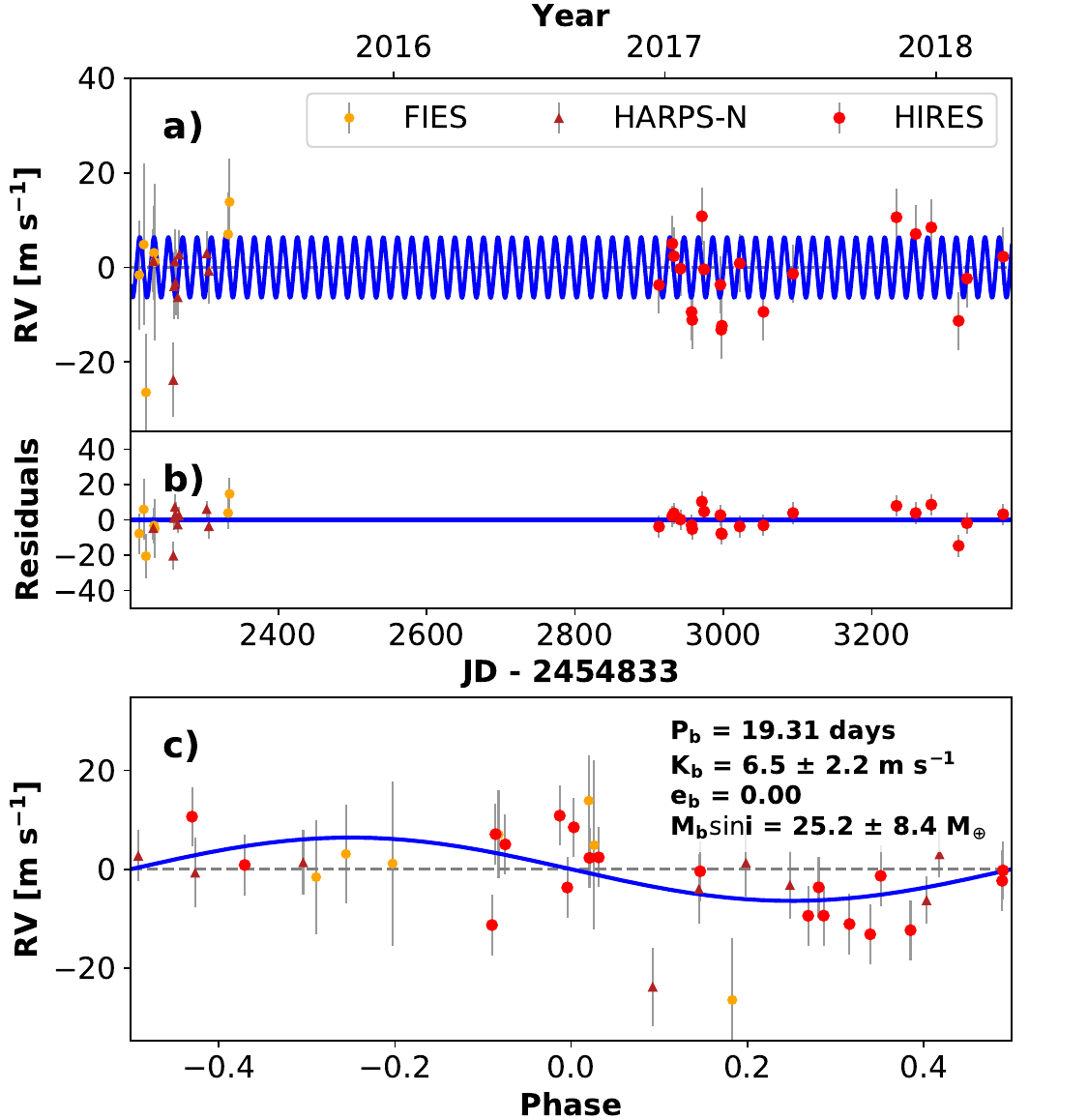}
\caption{RVs and Keplerian model for \hadfSTNAME. Symbols, lines, and annotations are similar to those in Fig.\ \ref{fig:rvs_epic220709978}.}
\label{fig:rvs_k2-10}
\end{figure}

Our fit of the EVEREST light curve of the K2 photometry for \hadfSTNAME is shown in Fig.\ \ref{fig:lc_epic201577035}.  We acquired \hadfNOBSHIRES RVs of \hadfSTNAME with HIRES, typically with an exposure meter setting of 50,000.  Using all of the available RVs, we modeled the system as one planet in a circular orbit with an orbital period and phase fixed to the transit ephemeris.  We rejected more complicated models with free eccentricity and a linear RV trend using the AICc statistic.  The results of our analysis are listed in Table \ref{tab:epic201577035} and the best-fit model is shown in Fig.\ \ref{fig:rvs_k2-10}.
The planet mass and density from our analysis are consistent with and more precise than those in \cite{vanEylen2016a}.  


\begin{deluxetable}{lrrr}
\tablecolumns{4}
\tabletypesize{\scriptsize}
\tablecaption{ K2-10 System Parameters \label{tab:epic201577035}}
\tablehead{
  \colhead{Parameter} & 
  \colhead{Credible} & 
  \colhead{Maximum} & 
  \colhead{Units} \\
  \colhead{} & 
  \colhead{Interval} & 
  \colhead{Likelihood} & 
  \colhead{} }
\startdata
\sidehead{\bf{RV Analysis -- MCMC Step Parameters}} 
$P_{b}$ & $\equiv19.3055$ & $\equiv19.3055$ & days \\ 
$T_{\mathrm{conj},b}$ & $\equiv2456819.5794$ & $\equiv2456819.5794$ & BJD$_\mathrm{TBD}$ \\ 
$\sqrt{e}\cos{\omega}_{b}$ & $\equiv0.0$ & $\equiv0.0$ &  \\ 
$\sqrt{e}\sin{\omega}_{b}$ & $\equiv0.0$ & $\equiv0.0$ &  \\ 
$K_{b}$ & $6.5\pm2.0$ & $6.0$ & m s$^{-1}$ \\ 
$\gamma_{\rm HIRES}$ & $1.1\pm1.6$ & $1.1$ & m s$-1$ \\ 
$\gamma_{\rm HARPS-N}$ & $8202.1^{+2.4}_{-3.1}$ & $8202.3$ & m s$-1$ \\ 
$\gamma_{\rm FIES}$ & $8061.9^{+5.1}_{-6.3}$ & $8062.4$ & m s$-1$ \\ 
$\dot{\gamma}$ & $\equiv0.0$ & $\equiv0.0$ & m s$^{-1}$ day$^{-1}$ \\ 
$\ddot{\gamma}$ & $\equiv0.0$ & $\equiv0.0$ & m s$^{-1}$ day$^{-2}$ \\ 
$\sigma_{\rm HIRES}$ & $6.2^{+1.9}_{-0.6}$ & $5.7$ & $\rm m\ s^{-1}$ \\ 
$\sigma_{\rm HARPS-N}$ & $1.9^{+4.0}_{-12.0}$ & $4.0$ & $\rm m\ s^{-1}$ \\ 
$\sigma_{\rm FIES}$ & $0\pm11$ & $0$ & $\rm m\ s^{-1}$ \\ 
\hline
\sidehead{\bf{Orbital \& Physical Parameters}}
$P_{b}$ & $\equiv19.3055$ & $\equiv19.3055$ & days \\
$T_{\mathrm{conj},b}$ & $\equiv2456819.5794$ & $\equiv2456819.5794$ & BJD$_\mathrm{TBD}$ \\
$e_{b}$ & $\equiv0.0$ & $\equiv0.0$ &  \\
$\omega_{b}$ & $\equiv0.0$ & $\equiv0.0$ & radians \\
$K_{b}$ & $6.5\pm2.0$ & $6.0$ & m s$^{-1}$ \\
$M_b$ & $25.2^{+8.9}_{-7.9}$ & $24.7$ & $M_\earth$ \\
$R_b/R_*$ & $0.03724^{+0.00120}_{-0.00010}$ & $0.03690$ &  \\
$\rho_b$ & $2.54\pm1.00$ & $2.00$ & g cm$^{-3}$ \\
$R_b$ & $3.773^{+0.020}_{-0.120}$ & $3.850$ & $R_\earth$ \\
\hline
\hline
\sidehead{\bf{Priors}}
Parameter & Prior \\
\hline
\\None \\
\enddata
\end{deluxetable}

\subsection{EPIC 201357835 (K2-245)} 

\begin{figure*}
\epsscale{1.0}
\plotone{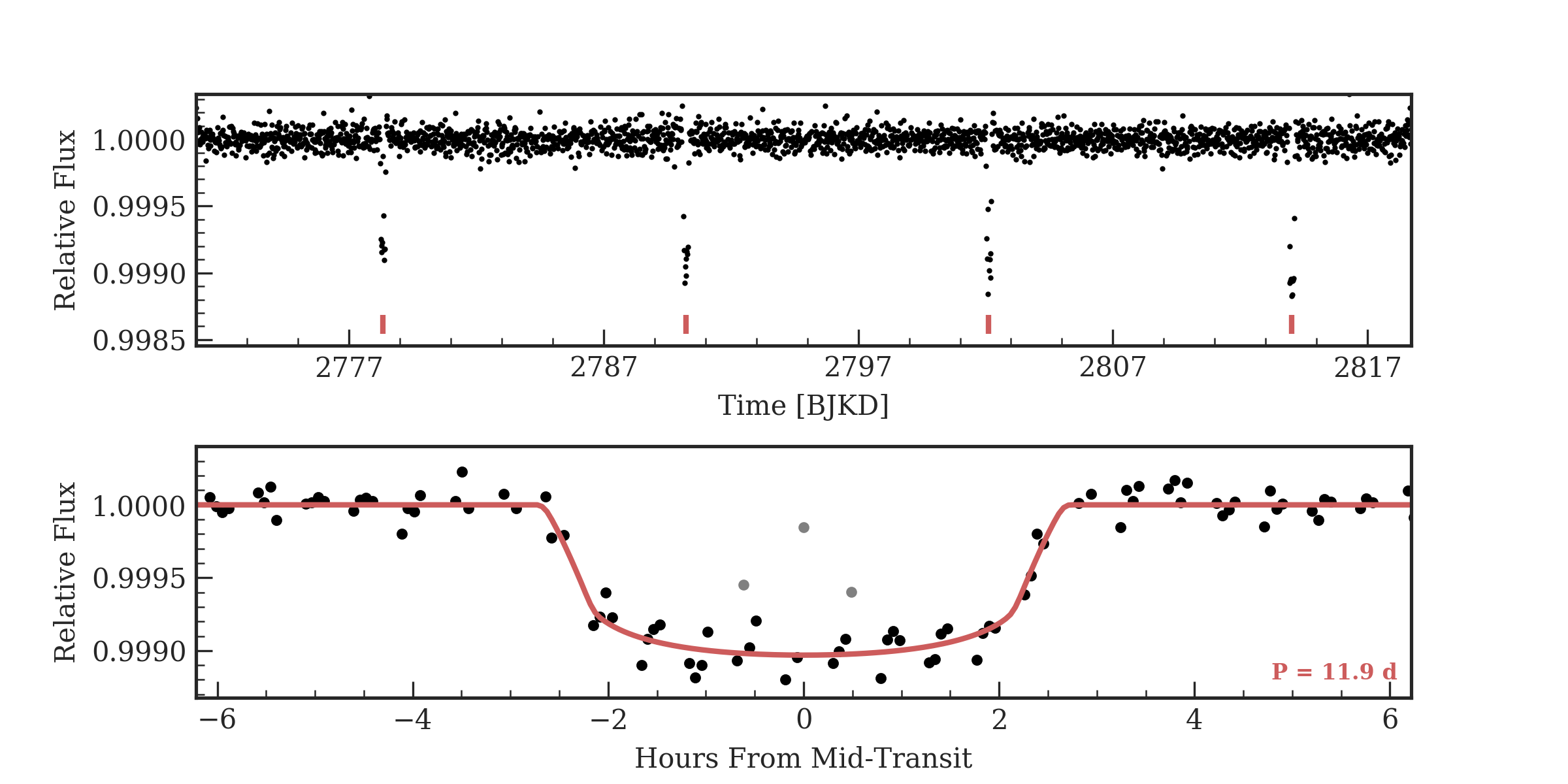}
\caption{Time series (top) and phase-folded (bottom) light curve for the planet orbiting \hidfSTNAME.  Plot formatting is the same as in Fig.\ \ref{fig:lc_epic220709978}.}
\label{fig:lc_epic201357835}
\end{figure*}

\hidfSTNAME is a low-metallicity star (\feh = \hidfFEH\ dex) with nearly solar temperature in Field 10.  It has one transiting planet with a radius of 3 \rearth and an orbital period of 12 days.
See Tables \ref{tb:star_pars}  and \ref{tb:star_props} for stellar properties and Table \ref{tb:planet_props} for precise planet parameters.  Our fit of the EVEREST light curve of the K2 photometry for \hidfSTNAME is shown in Fig.\ \ref{fig:lc_epic201357835}.  This system was first identified and validated as K2-245 by \cite{Livingston2018b}.  

We acquired \hidfNOBSHIRES RVs of \hidfSTNAME with HIRES, typically with an exposure meter setting of 60,000 counts. This star has low chromospheric activity with \lrphk\ = $-$5.19.
We modeled the system as a single planet in a circular orbit with orbital period and phase fixed to the transit ephemerides.  The results of this analysis are listed in Table \ref{tab:epic201357835}
and the best-fit model is shown in Fig.\ \ref{fig:rvs_epic201357835}.
Our limited RVs do not  {\referee measure a useful mass for this planet but we do successfully rule out massive eclipsing binary false-positive scenarios, increasing the likelihood that this signal is planetary in origin

\begin{deluxetable}{lrrr}
\tablecolumns{4}
\tabletypesize{\scriptsize}
\tablecaption{ EPIC 201357835 System Parameters \label{tab:epic201357835}}
\tablehead{
  \colhead{Parameter} & 
  \colhead{Credible} & 
  \colhead{Maximum} & 
  \colhead{Units} \\
  \colhead{} & 
  \colhead{Interval} & 
  \colhead{Likelihood} & 
  \colhead{} }
\startdata
\sidehead{\bf{RV Analysis -- MCMC Step Parameters}} 
$P_{b}$ & $\equiv11.8938$ & $\equiv11.8938$ & days \\ 
$T_{\mathrm{conj},b}$ & $\equiv2457611.3378$ & $\equiv2457611.3378$ & BJD$_\mathrm{TBD}$ \\ 
$\sqrt{e}\cos{\omega}_{b}$ & $\equiv0.0$ & $\equiv0.0$ &  \\ 
$\sqrt{e}\sin{\omega}_{b}$ & $\equiv0.0$ & $\equiv0.0$ &  \\ 
$K_{b}$ & $5^{+7.8}_{-7.9}$ & $4.5$ & m s$^{-1}$ \\ 
$\gamma_{\rm HIRES}$ & $-4.4^{+6.8}_{-6.7}$ & $-4.3$ & m s$-1$ \\ 
$\dot{\gamma}$ & $\equiv0.0$ & $\equiv0.0$ & m s$^{-1}$ day$^{-1}$ \\ 
$\ddot{\gamma}$ & $\equiv0.0$ & $\equiv0.0$ & m s$^{-1}$ day$^{-2}$ \\ 
$\sigma_{\rm HIRES}$ & $7^{+10}_{-27}$ & $9$ & $\rm m\ s^{-1}$ \\ 
\hline
\sidehead{\bf{Orbital \& Physical Parameters}}
$P_{b}$ & $\equiv11.8938$ & $\equiv11.8938$ & days \\
$T_{\mathrm{conj},b}$ & $\equiv2457611.3378$ & $\equiv2457611.3378$ & BJD$_\mathrm{TBD}$ \\
$e_{b}$ & $\equiv0.0$ & $\equiv0.0$ &  \\
$\omega_{b}$ & $\equiv0.0$ & $\equiv0.0$ & radians \\
$K_{b}$ & $5^{+7.8}_{-7.9}$ & $4.5$ & m s$^{-1}$ \\
$M_b$ & $15\pm26$ & $15$ & $M_\earth$ \\
$R_b/R_*$ & $0.0300^{+0.0017}_{-0.0001}$ & $0.0295$ &  \\
$\rho_b$ & $4.2^{+7.3}_{-7.2}$ & $4.1$ & g cm$^{-3}$ \\
$R_b$ & $2.67^{+0.07}_{-0.09}$ & $2.71$ & $R_\earth$ \\
\hline
\hline
\sidehead{\bf{Priors}}
Parameter & Prior \\
\hline
\\None \\
\enddata
\end{deluxetable}

\begin{figure}
\epsscale{1.0}
\plotone{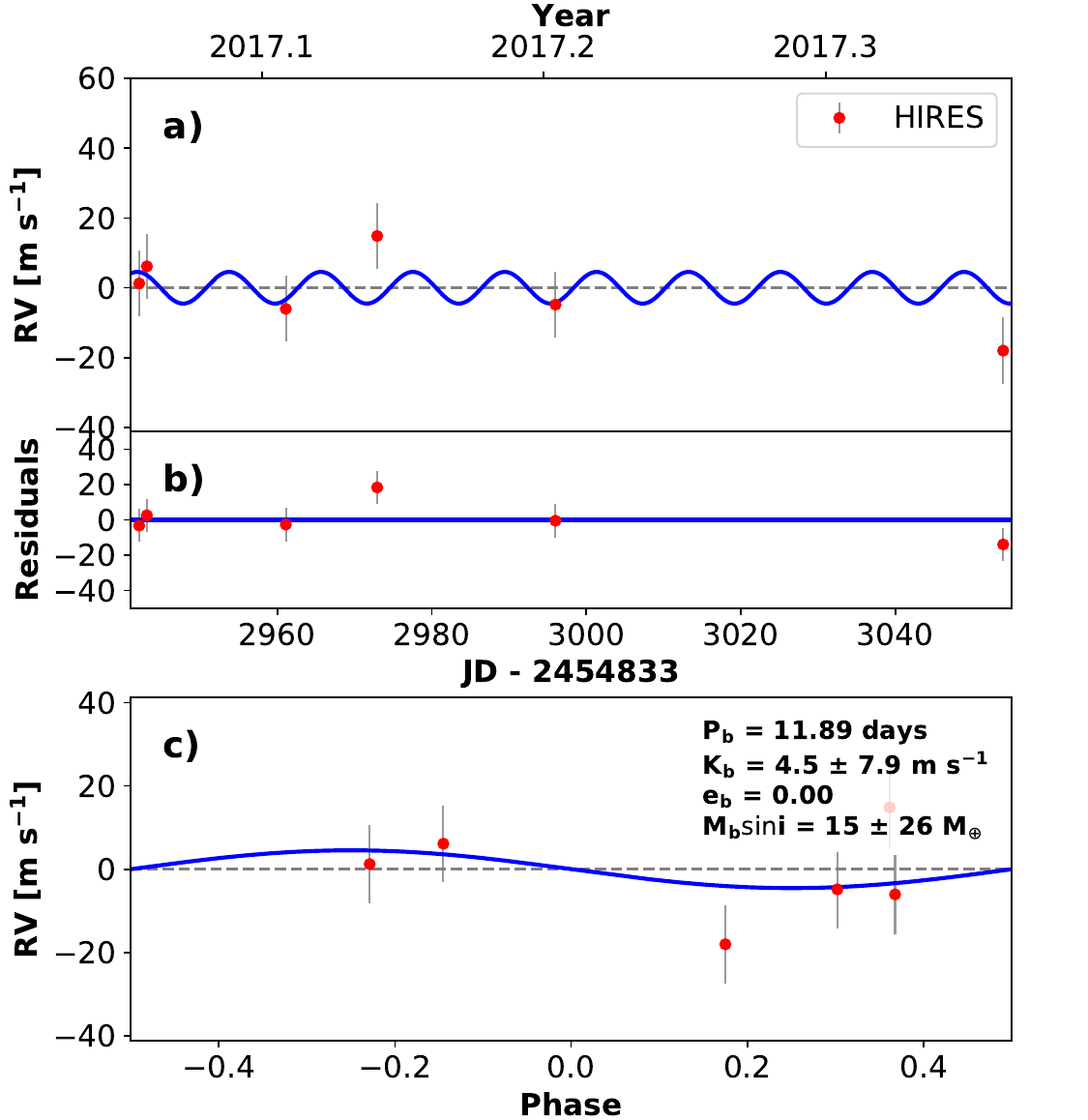}
\caption{RVs and Keplerian model for \hidfSTNAME. Symbols, lines, and annotations are similar to those in Fig.\ \ref{fig:rvs_epic220709978}.}
\label{fig:rvs_epic201357835}
\end{figure}

\subsection{K2-216} 


\bebbSTNAME is a chromospherically active (\lrphk = \bebbRPHK), late K dwarf in Field 8 with  one transiting planet with a radius of 1.7 \rearth and an orbital period of 2.2 days.  
See Tables \ref{tb:star_pars}  and \ref{tb:star_props} for stellar properties and Table \ref{tb:planet_props} for precise planet parameters.
The star was validated by \cite{Mayo2018} and appears in the \citep{Petigura2018} catalog.  \cite{Persson2018} {\referee reported that \bebbPNAMEone's mass is}  $8.0 \pm 1.6$ \mearth based on 8 FIES RVs, 9 HARPS RVs, and 13 HARPS-N RVs.  They accounted for stellar activity using a floating-offset model and also computed a model with a Gaussian process that gave similar results.

\begin{figure*}
\epsscale{1.0}
\plotone{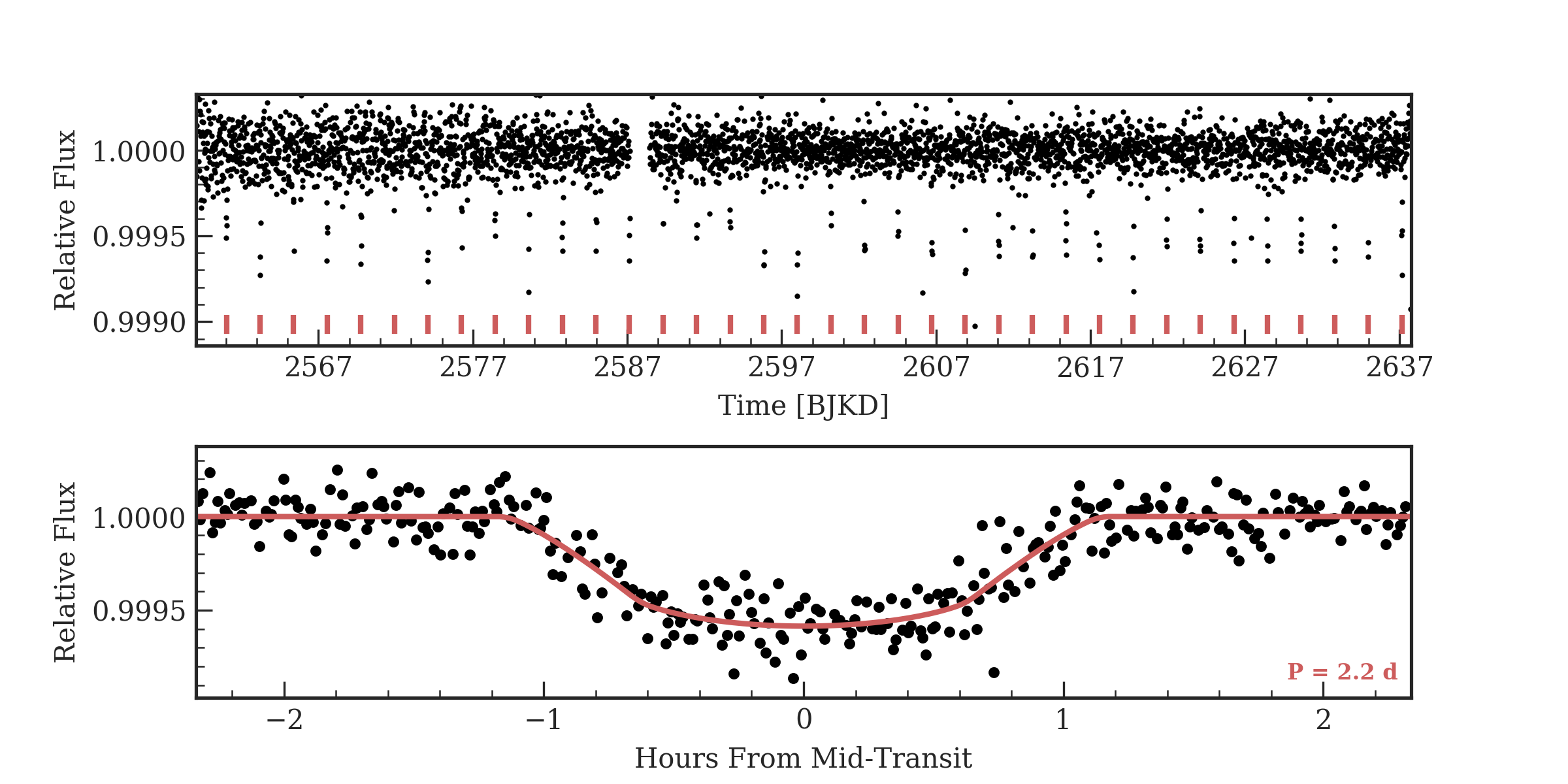}
\caption{Time series (top) and phase-folded (bottom) light curve for the planet orbiting \bebbSTNAME.  Plot formatting is the same as in Fig.\ \ref{fig:lc_epic220709978}.}
\label{fig:lc_epic220481411}
\end{figure*}

Our fit of the EVEREST light curve of the K2 photometry for \bebbSTNAME is shown in Fig.\ \ref{fig:lc_epic220481411}.  We acquired \bebbNOBSHIRES RVs of \bebbSTNAME with HIRES, typically with an exposure meter setting of 60,000 counts.  Since the star is active (\lrphk = \ddeiRPHK) and our $N_\mathrm{obs,HIRES}=$~\bebbNOBSHIRES measurements combined with RVs from the literature satisfies our prerequisites for a GP regression, we trained a GP on non-detrended Everest photometry (see Section~\ref{sec:gp_modeling}) before using it to compute RV orbit posteriors. We modeled the system as a single planet in a circular orbit with the orbital period and phase fixed to the transit ephemeris. We rejected more complicated models including eccentricity and linear trends based on the AICc statistic. For comparison purposes, we also perform a RV orbit fit using an untrained GP. The trained hyperparameters are clearly peaked and restricted to a portion of the parameter space allowed by the priors, while in the untrained GP the posteriors for $\eta_2$ and $\eta_3$ extend over the entire allowable parameter space. Both cases yield consistent planet parameters. 

Our GP analysis, combined with the higher precision HIRES RVs, yields a  smaller mass than \cite{Persson2018}. We find a $\sim 10\%$ smaller mass when considering models without a GP, which is consistent with the GP results given the uncertainties of the parameters. The results of our analysis are listed in Table \ref{tab:epic220481411}
and the best fit model is shown in Fig.\ \ref{fig:rvs_k2-216}.
\bebbPNAMEone is a short-period super-Earth with a density consistent with a rocky composition.


\begin{deluxetable}{lrrr}
\tablecolumns{4}
\tabletypesize{\scriptsize}
\tablecaption{ K2-216 System Parameters \label{tab:epic220481411}}
\tablehead{
  \colhead{Parameter} & 
  \colhead{Credible} & 
  \colhead{Maximum} & 
  \colhead{Units} \\
  \colhead{} & 
  \colhead{Interval} & 
  \colhead{Likelihood} & 
  \colhead{} }
\startdata
\sidehead{\bf{RV Analysis -- MCMC Step Parameters}} 
$P_{b}$ & $\equiv2.1748$ & $\equiv2.1748$ & days \\ 
$T_{\mathrm{conj},b}$ & $\equiv2457394.0417$ & $\equiv2457394.0417$ & BJD$_\mathrm{TBD}$ \\ 
$\sqrt{e}\cos{\omega}_{b}$ & $\equiv0.0$ & $\equiv0.0$ &  \\ 
$\sqrt{e}\sin{\omega}_{b}$ & $\equiv0.0$ & $\equiv0.0$ &  \\ 
$K_{b}$ & $3.77^{+0.80}_{-1.00}$ & $3.90$ & m s$^{-1}$ \\ 
$\gamma_{\rm HIRES}$ & $-2.2^{+1.3}_{-1.1}$ & $-2.3$ & m s$-1$ \\ 
$\gamma_{\rm HARPS-N}$ & $-25911.1^{+2.4}_{-3.5}$ & $-25910.8$ &  \\ 
$\gamma_{\rm HARPS}$ & $-25906^{+4.4}_{-3.7}$ & $-25906.2$ & m s$-1$ \\
\enddata
\end{deluxetable}

\begin{figure}
\epsscale{1.0}
\plotone{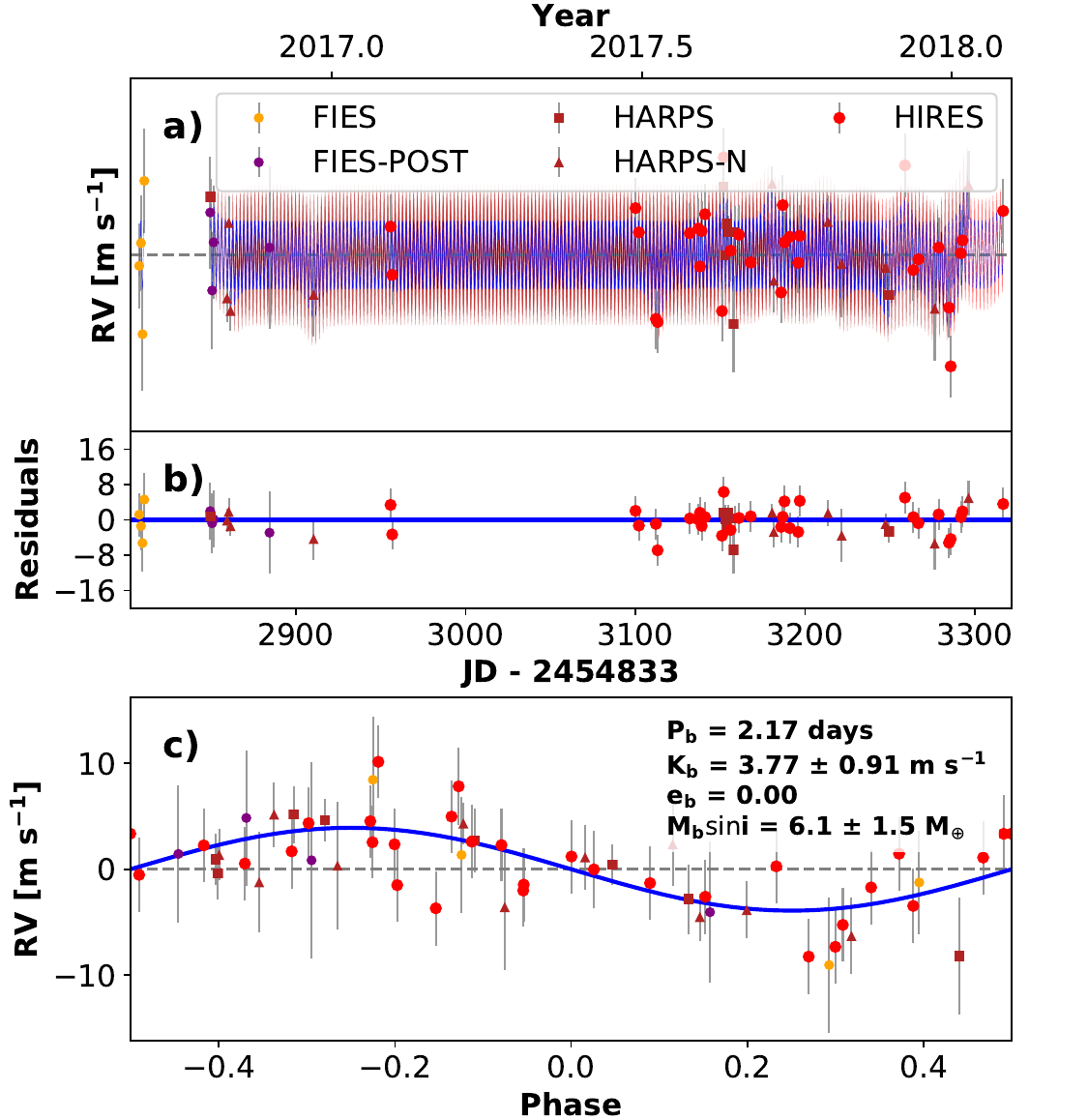}
\caption{RVs and Keplerian model for \bebbSTNAME. Symbols, lines, and annotations are similar to those in Fig.\ \ref{fig:rvs_epic220709978}.}
\label{fig:rvs_k2-216}
\end{figure}

\subsection{\ecdiSTNAME} 


\ecdiSTNAME is a metal-rich (\feh = \ecdiFEH dex), slightly evolved G star from Field 7 with one transiting planet with an orbital period of 20 days and a radius of 8.5 \rearth.
See Tables \ref{tb:star_pars}  and \ref{tb:star_props} for stellar properties and Table \ref{tb:planet_props} for precise planet parameters.  The object is listed as a planet candidate in the  \cite{Petigura2018} and \cite{Mayo2018} catalogs and \cite{Nowak2020} measure a planet mass of $37.1 \pm 5.6  M_\oplus$ and an eccentricity of 0.35.
As described below, modeling of our HIRES RVs also validates the planet.  Our fit of the EVEREST light curve of the K2 photometry for \ecdiSTNAME is shown in Fig.\ \ref{fig:lc_epic216494238}.  

\begin{figure*}
\epsscale{1.0}
\plotone{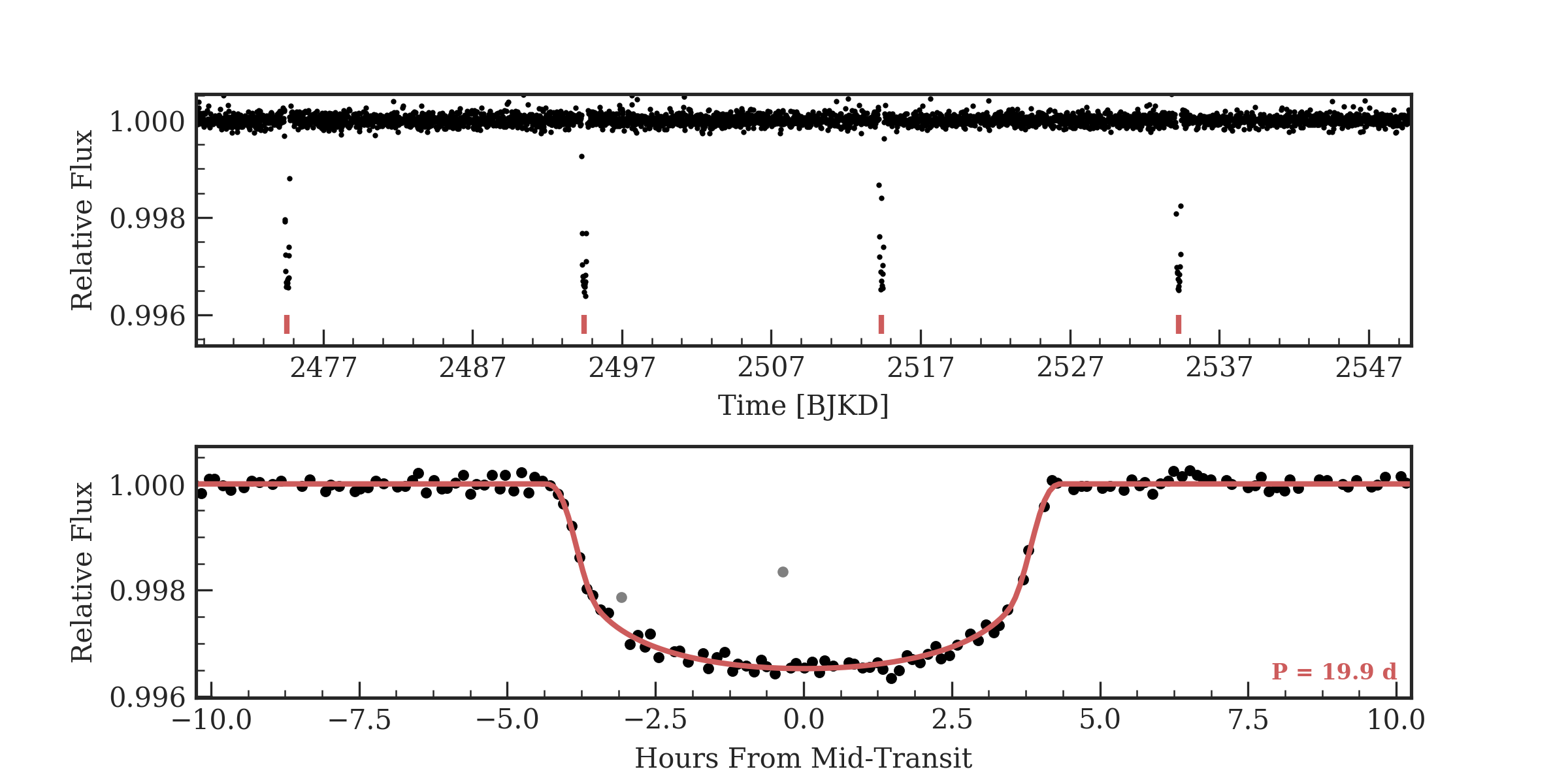}
\caption{Time series (top) and phase-folded (bottom) light curve for the planet orbiting \ecdiSTNAME.  Plot formatting is the same as in Fig.\ \ref{fig:lc_epic220709978}.}
\label{fig:lc_epic216494238}
\end{figure*}

We acquired \ecdiNOBSHIRES RVs of \ecdiSTNAME with HIRES, typically with an exposure meter setting of 80,000 counts.  
This star has low chromospheric activity with \lrphk\ = $-$5.24.
We modeled the system as a single planet in a Keplerian orbit with the period and phase fixed to the transit ephemerides.  We chose a model with a free eccentricity based on \dAICc = 18 compared to a model with a circular orbit. Using a similar comparison, we rejected a model with a linear RV trend.  The results of this analysis include a significantly eccentric orbit as listed in Table \ref{tab:epic216494238} and shown in Fig.\ \ref{fig:rvs_epic216494238}. 
We validated this planet with our 7$\sigma$ radial velocity detection.  (\cite{Livingston2018} had statiscially validated the planet.)
\ecdiPNAMEone is a giant planet with a short-period, eccentric orbit, and our values are broadly consistent with those of \cite{Nowak2020}.


\begin{deluxetable}{lrrr}
\tablecolumns{4}
\tabletypesize{\scriptsize}
\tablecaption{ K2-280 System Parameters \label{tab:epic216494238}}
\tablehead{
  \colhead{Parameter} & 
  \colhead{Credible} & 
  \colhead{Maximum} & 
  \colhead{Units} \\
  \colhead{} & 
  \colhead{Interval} & 
  \colhead{Likelihood} & 
  \colhead{} }
\startdata
\sidehead{\bf{RV Analysis -- MCMC Step Parameters}} 
$P_{b}$ & $\equiv19.895$ & $\equiv19.895$ & days \\ 
$T_{\mathrm{conj},b}$ & $\equiv2457327.4761$ & $\equiv2457327.4761$ & BJD$_\mathrm{TBD}$ \\ 
$\sqrt{e}\cos{\omega}_{b}$ & $-0.2\pm0.1$ & $-0.1$ &  \\ 
$\sqrt{e}\sin{\omega}_{b}$ & $-0.6\pm0.1$ & $-0.7$ &  \\ 
$K_{b}$ & $11.9^{+1.7}_{-1.5}$ & $12.8$ & m s$^{-1}$ \\ 
$\gamma_{\rm HIRES}$ & $-1.5\pm1.1$ & $-2.0$ & m s$-1$ \\ 
$\dot{\gamma}$ & $\equiv0.0$ & $\equiv0.0$ & m s$^{-1}$ day$^{-1}$ \\ 
$\ddot{\gamma}$ & $\equiv0.0$ & $\equiv0.0$ & m s$^{-1}$ day$^{-2}$ \\ 
$\sigma_{\rm HIRES}$ & $2.9^{+1.1}_{-0.9}$ & $2.0$ & $\rm m\ s^{-1}$ \\ 
\hline
\sidehead{\bf{Orbital \& Physical Parameters}}
$P_{b}$ & $\equiv19.895$ & $\equiv19.895$ & days \\
$T_{\mathrm{conj},b}$ & $\equiv2457327.4761$ & $\equiv2457327.4761$ & BJD$_\mathrm{TBD}$ \\
$e_{b}$ & $0.5\pm0.1$ & $0.5$ &  \\
$\omega_{b}$ & $-1.90^{+0.14}_{-0.21}$ & $-1.77$ & radians \\
$K_{b}$ & $11.9^{+1.7}_{-1.5}$ & $12.8$ & m s$^{-1}$ \\
$M_b$ & $49.0^{+6.5}_{-6.2}$ & $50.7$ & $M_\earth$ \\
$R_b/R_*$ & $0.05362^{+0.00078}_{-0.00038}$ & $0.05340$ &  \\
$\rho_b$ & $0.643^{+0.088}_{-0.083}$ & $0.670$ & g cm$^{-3}$ \\
$R_b$ & $7.46^{+0.11}_{-0.05}$ & $7.45$ & $R_\earth$ \\
\hline
\hline
\sidehead{\bf{Priors}}
Parameter & Prior \\
\hline
$e_{b}$ & $<0.99$ \\
$\sigma_{\rm HIRES}$ & $\mathcal{U}(0,1e+100)$ \\
\enddata
\end{deluxetable}

\begin{figure}
\epsscale{1.0}
\plotone{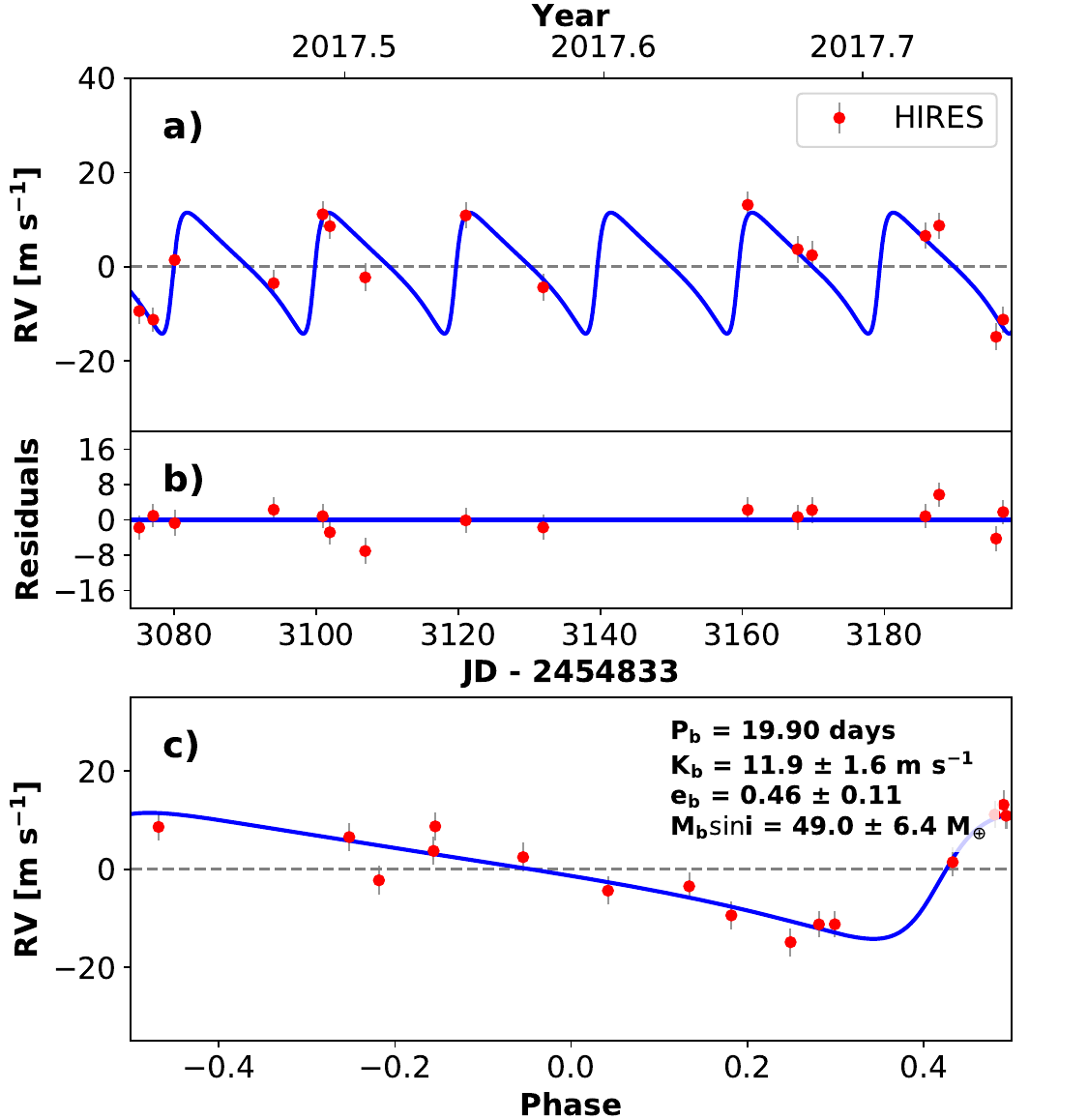}
\caption{RVs and Keplerian model for \ecdiSTNAME.  Symbols, lines, and annotations are similar to those in Fig.\ \ref{fig:rvs_epic220709978}.}
\label{fig:rvs_epic216494238}
\end{figure}

\subsection{K2-37} 


\gedgSTNAME is a late G dwarf with three transiting planets with sizes 1.6 \rearth, 2.5 \rearth, and 2.4 \rearth with orbital periods of 4.4 days, 6.4 days, and 14.1 days, respectively.
See Tables \ref{tb:star_pars}  and \ref{tb:star_props} for stellar properties and Table \ref{tb:planet_props} for precise planet parameters.
The planets are validated and in the \cite{Crossfield2016}, \cite{Vanderburg2016-catalog}, \cite{Sinukoff2016}, and \cite{Wittenmyer2018} catalogs.

\begin{figure*}
\epsscale{1.0}
\plotone{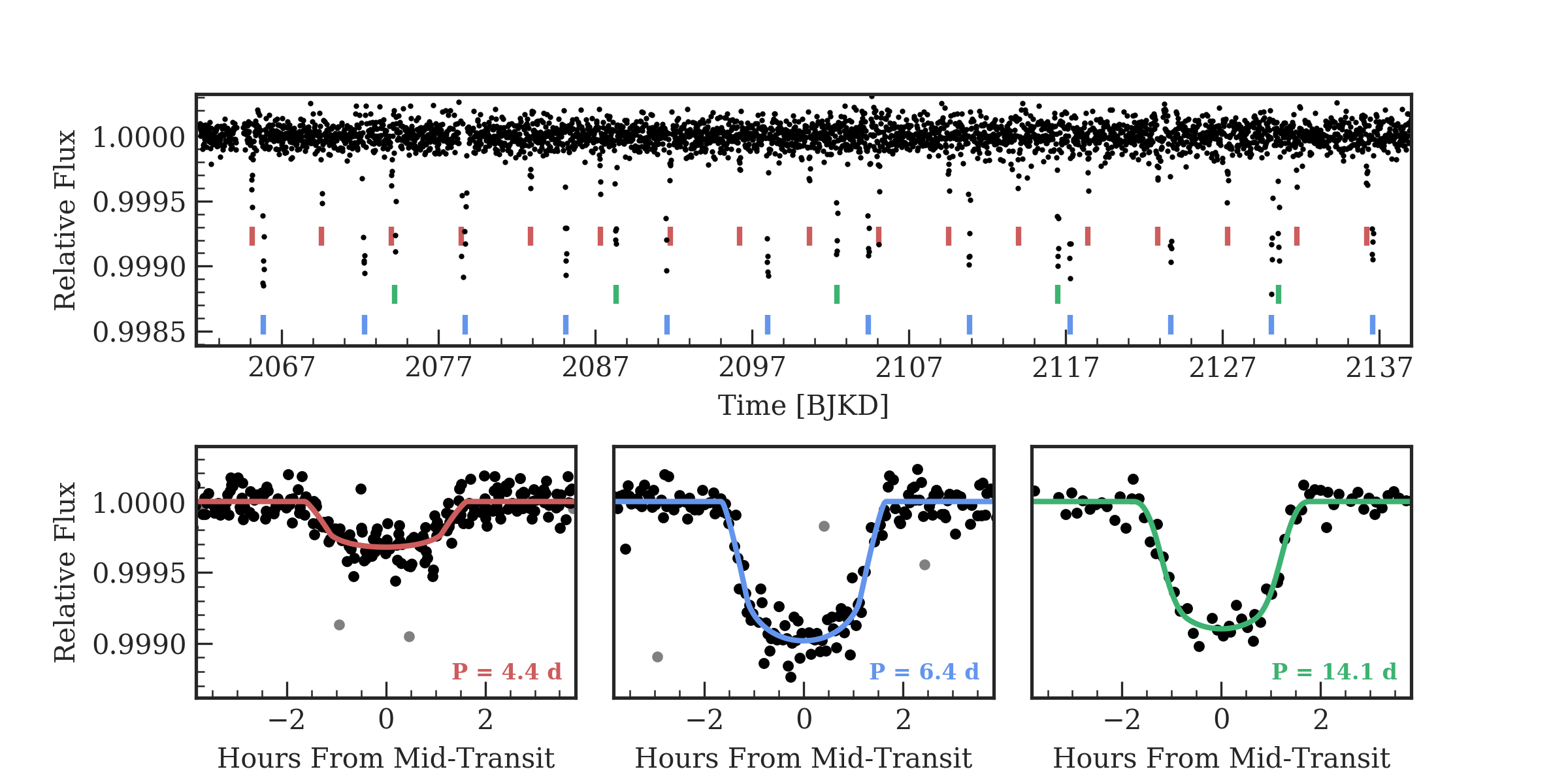}
\caption{Time series (top) and phase-folded (bottom) light curve for the planet orbiting \gedgSTNAME.  Plot formatting is the same as in Fig.\ \ref{fig:lc_epic220709978}.}
\label{fig:lc_epic203826436}
\end{figure*}

Our fit of the EVEREST light curve of the K2 photometry for \gedgSTNAME is shown in Fig.\ \ref{fig:lc_epic203826436}. We acquired \gedgNOBSHIRES RVs of \gedgSTNAME with HIRES, typically with an exposure meter setting of 80,000 counts.  
We modeled the system as three planets in circular orbits with orbital periods and phases fixed to the transit ephemerides.  We rejected more complicated models with noncircular orbits and/or a linear RV trend using the AICc statistic.  The results of our analysis are listed in Table \ref{tab:epic203826436} and the best-fit model is shown in Fig.\ \ref{fig:rvs_k2-37}.
The Doppler signals from planets b and c are not detected, while planet d is detected with $>$2-$\sigma$ significance. Continued monitoring of this multiplanet system is needed to constrain the properties of the two inner planets.


\begin{deluxetable}{lrrr}
\tablecolumns{4}
\tabletypesize{\scriptsize}
\tablecaption{ K2-37 System Parameters \label{tab:epic203826436}}
\tablehead{
  \colhead{Parameter} & 
  \colhead{Credible} & 
  \colhead{Maximum} & 
  \colhead{Units} \\
  \colhead{} & 
  \colhead{Interval} & 
  \colhead{Likelihood} & 
  \colhead{} }
\startdata
\sidehead{\bf{RV Analysis -- MCMC Step Parameters}} 
$P_{b}$ & $\equiv4.4434$ & $\equiv4.4434$ & days \\ 
$T_{\mathrm{conj},b}$ & $\equiv2456893.6753$ & $\equiv2456893.6753$ & BJD$_\mathrm{TBD}$ \\ 
$\sqrt{e}\cos{\omega}_{b}$ & $\equiv0.0$ & $\equiv0.0$ &  \\ 
$\sqrt{e}\sin{\omega}_{b}$ & $\equiv0.0$ & $\equiv0.0$ &  \\ 
$K_{b}$ & $-0.4\pm2.3$ & $-0.4$ & m s$^{-1}$ \\ 
$P_{c}$ & $\equiv6.4297$ & $\equiv6.4297$ & days \\ 
$T_{\mathrm{conj},c}$ & $\equiv2456898.8548$ & $\equiv2456898.8548$ & BJD$_\mathrm{TBD}$ \\ 
$\sqrt{e}\cos{\omega}_{c}$ & $\equiv0.0$ & $\equiv0.0$ &  \\
\enddata
\end{deluxetable}

\begin{figure}
\epsscale{1.0}
\plotone{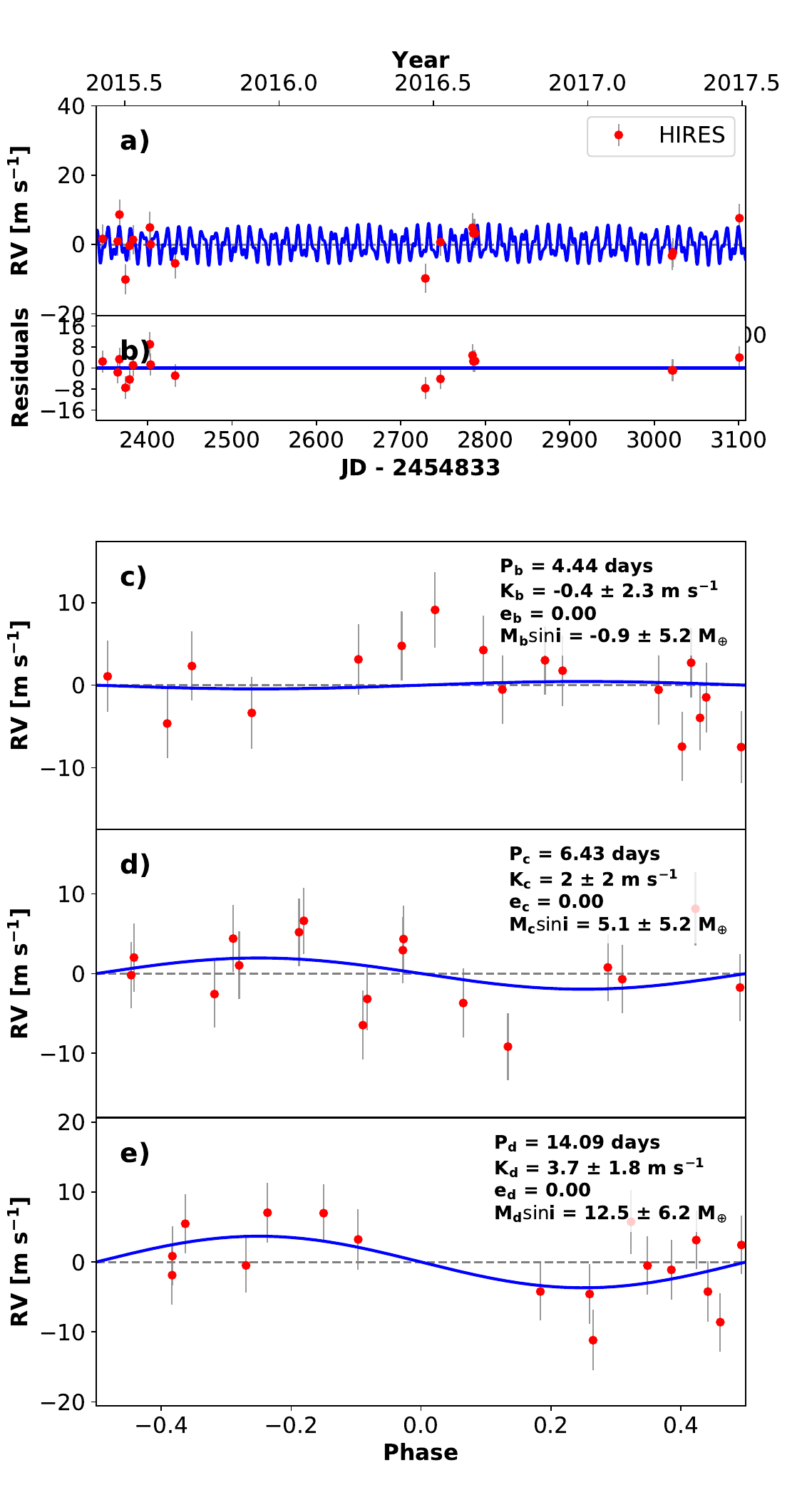}
\caption{RVs and Keplerian model for \gedgSTNAME.  Symbols, lines, and annotations are similar to those in Fig.\ \ref{fig:rvs_epic220709978}.}
\label{fig:rvs_k2-37}
\end{figure}
\subsection{K2-180} 



\jgbhSTNAME is a very metal poor (\feh = \jgbhFEH dex) K dwarf from Campaign 5.  It has one transiting planet with a radius of 2.4 \rearth and an orbital period of 8.9 days.
See Tables \ref{tb:star_pars}  and \ref{tb:star_props} for stellar properties and Table \ref{tb:planet_props} for precise planet parameters.
The planet was validated by \cite{Mayo2018} and appears in the \cite{Petigura2018} and \cite{Pope2016} catalogs. \cite{Korth2019} followed up this system with FIES and HARPS-N and measured a mass of $11.3\pm1.9$ \mearth suggesting a rocky composition with a density of 5.6$\pm$1.9 g cm$^{-3}$. 

\begin{figure*}
\epsscale{1.0}
\plotone{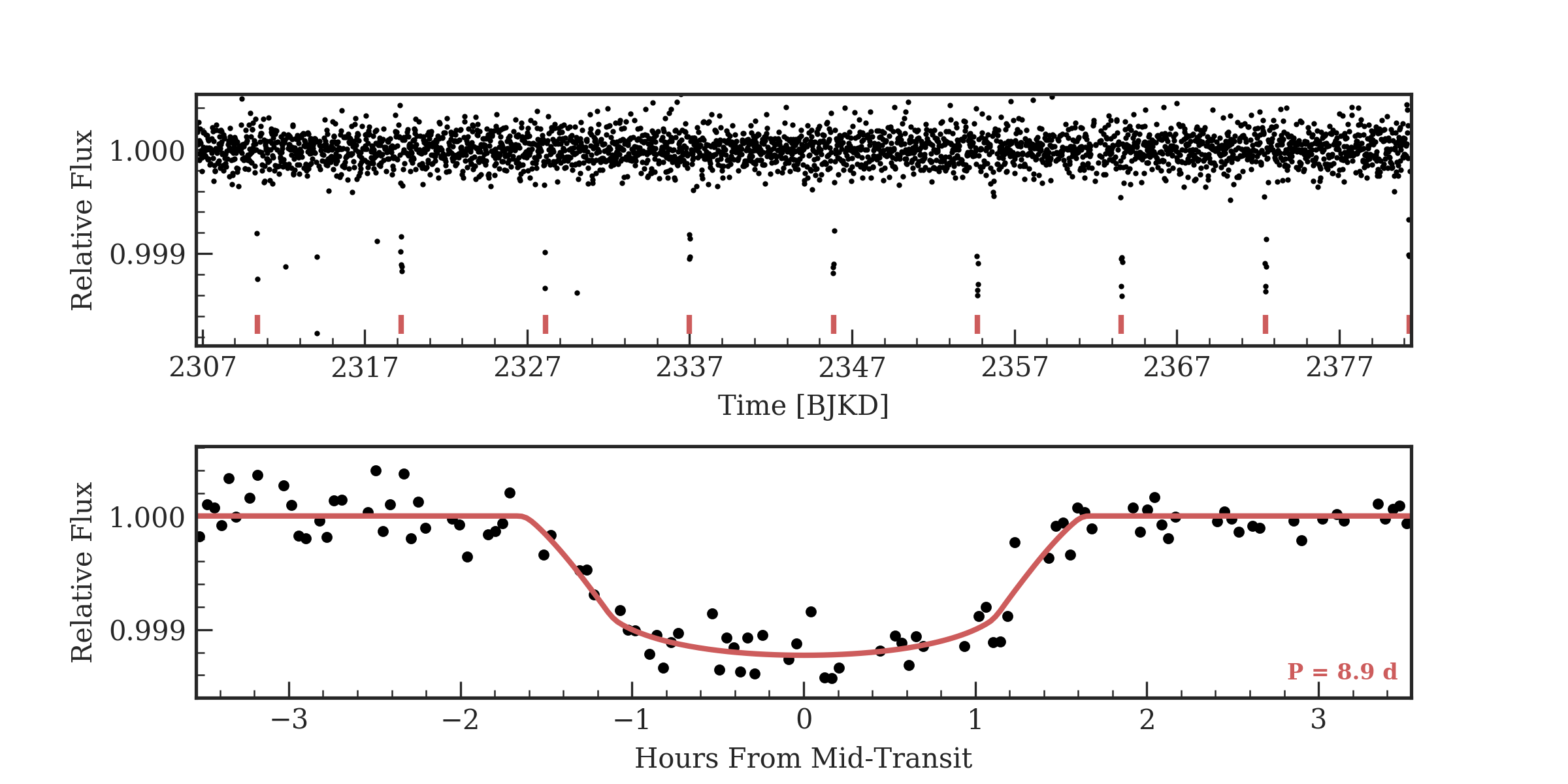}
\caption{Time series (top) and phase-folded (bottom) light curve for the planet orbiting \jgbhSTNAME.  Plot formatting is the same as in Fig.\ \ref{fig:lc_epic220709978}.}
\label{fig:lc_epic211319617}
\end{figure*}

Our fit of the EVEREST light curve of the K2 photometry for \jgbhSTNAME is shown in Fig.\ \ref{fig:lc_epic211319617}. We acquired \jgbhNOBSHIRES RVs of \jgbhSTNAME with HIRES, typically with an exposure meter setting of 50,000 counts.  
We modeled the system as a single planet in a circular orbit with an orbital period and phase fixed to the transit ephemeris.  We rejected more complicated models with a noncircular orbit and/or a linear RV trend based on the AICc statistic.  The results of our analysis are listed in Table \ref{tab:epic211319617} and the best-fit model is shown in Fig.\ \ref{fig:rvs_k2-180}.  \jgbhPNAMEone is a sub-Neptune with a low bulk density and orbits one of the most metal-poor planet hosts detected to date.

\begin{deluxetable}{lrrr}
\tablecolumns{4}
\tabletypesize{\scriptsize}
\tablecaption{ K2-180 System Parameters \label{tab:epic211319617}}
\tablehead{
  \colhead{Parameter} & 
  \colhead{Credible} & 
  \colhead{Maximum} & 
  \colhead{Units} \\
  \colhead{} & 
  \colhead{Interval} & 
  \colhead{Likelihood} & 
  \colhead{} }
\startdata
\sidehead{\bf{RV Analysis -- MCMC Step Parameters}} 
$P_{b}$ & $\equiv8.8656$ & $\equiv8.8656$ & days \\ 
$T_{\mathrm{conj},b}$ & $\equiv2457143.3945$ & $\equiv2457143.3945$ & BJD$_\mathrm{TBD}$ \\ 
$\sqrt{e}\cos{\omega}_{b}$ & $\equiv0.0$ & $\equiv0.0$ &  \\ 
$\sqrt{e}\sin{\omega}_{b}$ & $\equiv0.0$ & $\equiv0.0$ &  \\ 
$K_{b}$ & $3.8\pm0.9$ & $3.8$ & m s$^{-1}$ \\ 
$\gamma_{\rm HIRES}$ & $-1.63^{+0.92}_{-0.94}$ & $-1.60$ & m s$-1$ \\ 
$\gamma_{\rm HARPS-N}$ & $-76614.5\pm0.9$ & $-76614.4$ & m s$-1$ \\ 
$\gamma_{\rm FIES}$ & $-76854.4^{+5.4}_{-5.2}$ & $-76854.0$ & m s$-1$ \\
\enddata
\end{deluxetable}

\begin{figure}
\epsscale{1.0}
\plotone{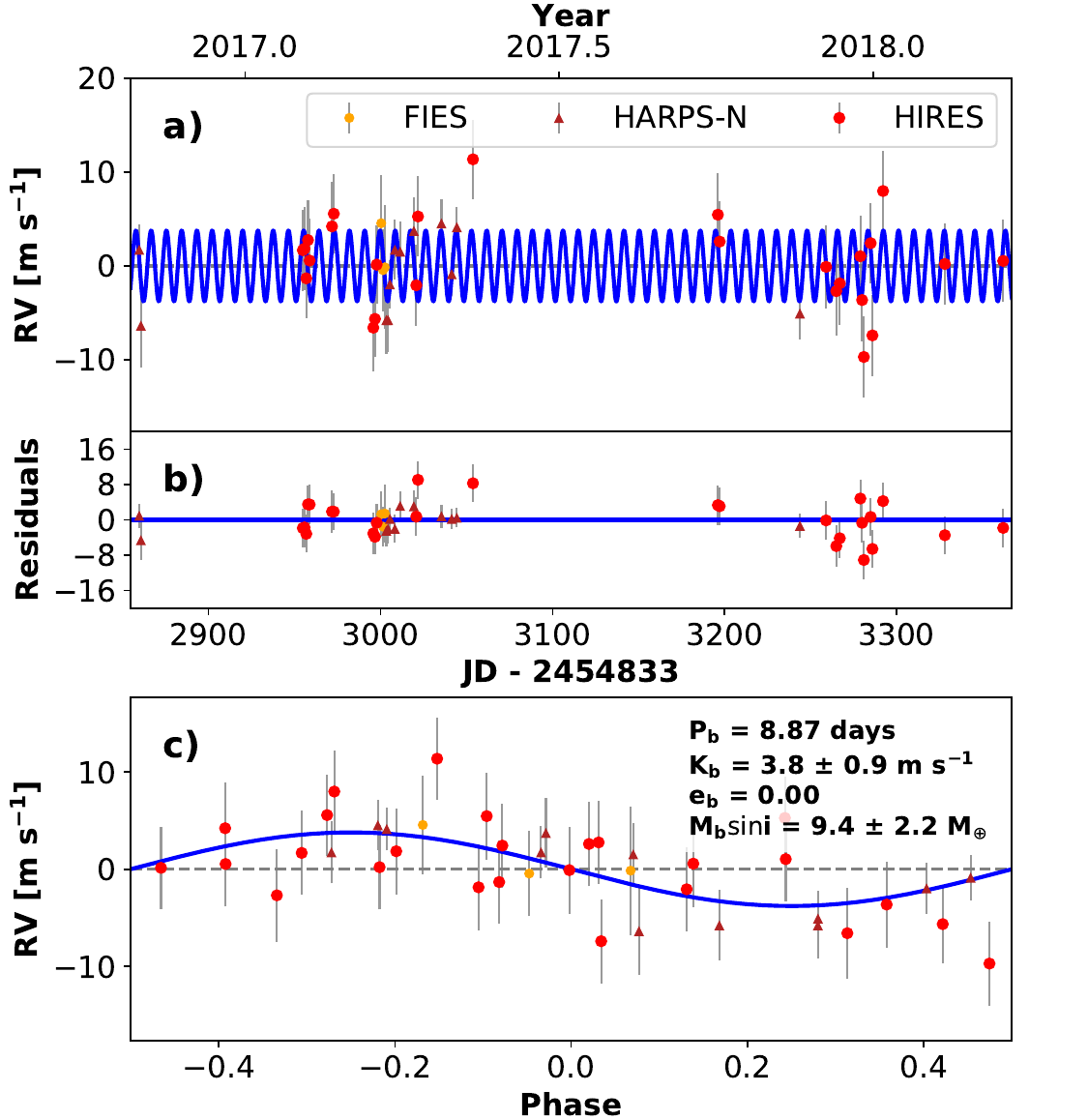}
\caption{RVs and Keplerian model for \jgbhSTNAME. Symbols, lines, and annotations are similar to those in Fig.\ \ref{fig:rvs_epic220709978}.}
\label{fig:rvs_k2-180}
\end{figure}

\subsection{K2-27} 


\ghijSTNAME is a late G dwarf with one transiting planet with a radius of 4.7 \rearth and an orbital period of 6.8 days. See Tables \ref{tb:star_pars}  and \ref{tb:star_props} for stellar properties and Table \ref{tb:planet_props} for precise planet parameters. The planet was noted in the catalogs of \cite{Crossfield2015}, \cite{Vanderburg2016-catalog}, \cite{Schmitt2016}, and \cite{Wittenmyer2018}. \cite{vanEylen2016a} measured a mass of $29.1^{+7.5}_{-7.4}$ \mearth based on 6 HARPS and 19 from HARPS-N spectra. \cite{Petigura2017} refined this mass measurement to $30.9 \pm 4.6$ \mearth using an additional 15 HIRES spectra.


\begin{figure*}
\epsscale{1.0}
\plotone{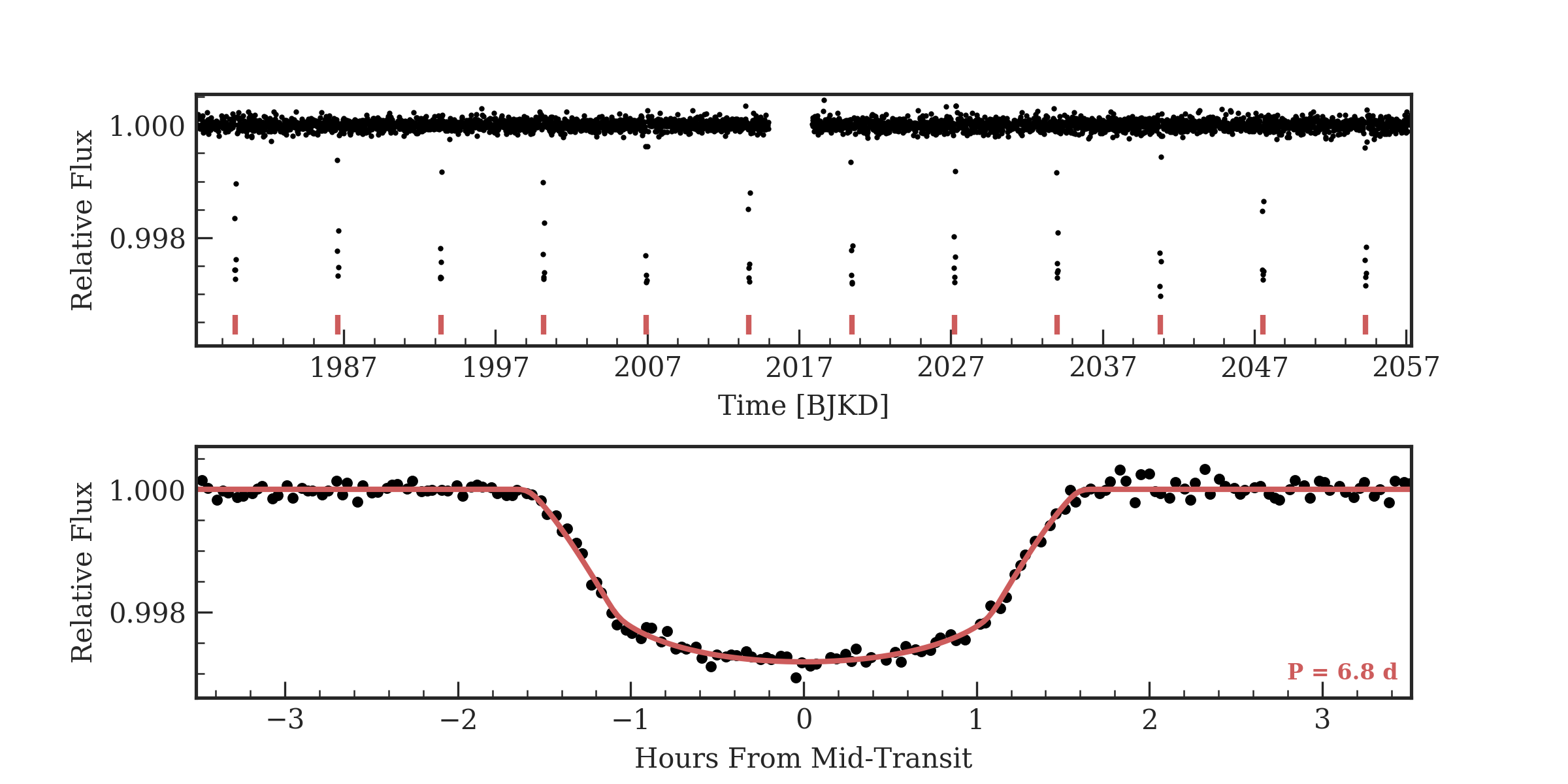}
\caption{Time series (top) and phase-folded (bottom) light curve for the planet orbiting \ghijSTNAME.  Plot formatting is the same as in Fig.\ \ref{fig:lc_epic220709978}.}
\label{fig:lc_epic201546283}
\end{figure*}

Our fit of the EVEREST light curve of the K2 photometry for \ghijSTNAME is shown in Fig.\ \ref{fig:lc_epic201546283}. Because we acquired only one additional RV of \ghijSTNAME with HIRES since \cite{Petigura2017}, our model only provides a slight update to their model. We modeled the system as a single planet with the period and phase fixed to the transit ephemeris. A model comparison based on the AICc statistic favors an eccentric orbit ($\Delta$AICc = 7.04) over a circular orbit. A linear trend is disfavored with a $\Delta$AICc of 4.20. The results of our analysis are listed in Table \ref{tab:epic201546283} and the best-fit model is shown in Fig.\ \ref{fig:rvs_k2-27}.
\ghijPNAMEone is an eccentric ($e = 0.24$) Neptune-sized planet with a density similar to Neptune.

\begin{deluxetable}{lrrr}
\tablecolumns{4}
\tabletypesize{\scriptsize}
\tablecaption{ K2-27 System Parameters \label{tab:epic201546283}}
\tablehead{
  \colhead{Parameter} & 
  \colhead{Credible} & 
  \colhead{Maximum} & 
  \colhead{Units} \\
  \colhead{} & 
  \colhead{Interval} & 
  \colhead{Likelihood} & 
  \colhead{} }
\startdata
\sidehead{\bf{RV Analysis -- MCMC Step Parameters}} 
$P_{b}$ & $\equiv6.7713$ & $\equiv6.7713$ & days \\ 
$T_{\mathrm{conj},b}$ & $\equiv2456812.8447$ & $\equiv2456812.8447$ & BJD$_\mathrm{TBD}$ \\ 
$\sqrt{e}\cos{\omega}_{b}$ & $0.37^{+0.09}_{-0.14}$ & $0.39$ &  \\ 
$\sqrt{e}\sin{\omega}_{b}$ & $0.31^{+0.17}_{-0.22}$ & $0.30$ &  \\ 
$K_{b}$ & $11.6\pm1.4$ & $11.6$ & m s$^{-1}$ \\ 
$\gamma_{\rm HIRES}$ & $-3.8^{+1.3}_{-1.2}$ & $-3.8$ & m s$-1$ \\ 
$\gamma_{\rm HARPS-N}$ & $-37772.9^{+2.3}_{-2.6}$ & $-37772.0$ &  \\ 
$\gamma_{\rm HARPS}$ & $-37775.4^{+1.6}_{-1.8}$ & $-37775.1$ & m s$-1$ \\
\enddata
\end{deluxetable}

\begin{figure}
\epsscale{1.0}
\plotone{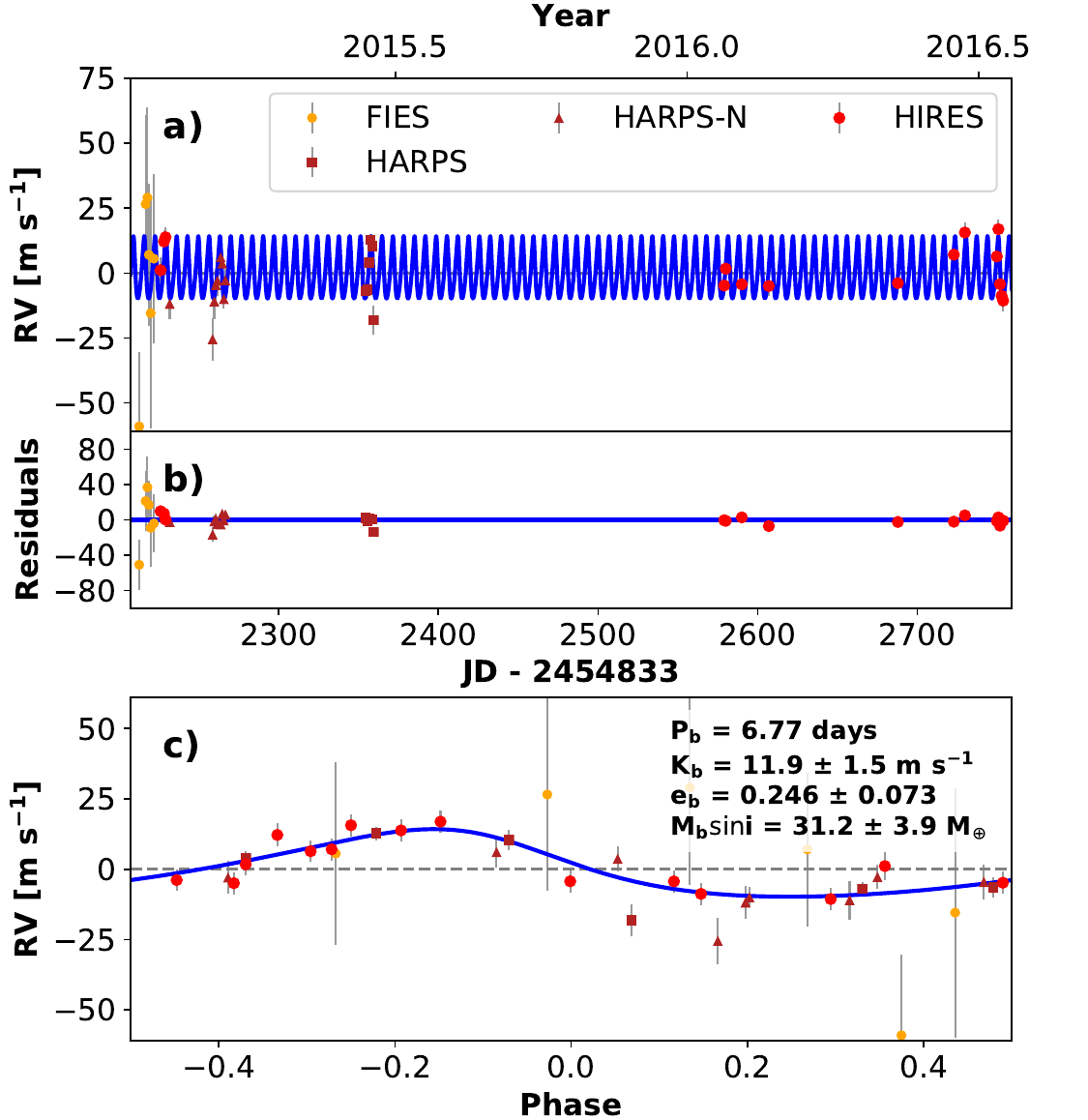}
\caption{RVs and Keplerian model for \ghijSTNAME. Symbols, lines, and annotations are similar to those in Fig.\ \ref{fig:rvs_epic220709978}.}
\label{fig:rvs_k2-27}
\end{figure}

\subsection{K2-181} 


\fdecSTNAME is a G dwarf in Field 5 with one transiting planet with radius of 2.3 \rearth and a 6.9-day orbital period.  
See Tables \ref{tb:star_pars} and \ref{tb:star_props} for stellar properties and Table \ref{tb:planet_props} for precise planet parameters. The planet is listed as a candidate in \cite{Barros2016} and \cite{Pope2016}, was listed as confirmed in \cite{Mayo2018}, but did not meet the validation criteria of \cite{Livingston2018} or \cite{Petigura2018}. Our fit of the EVEREST light curve of the K2 photometry for \fdecSTNAME is shown in Fig.\ \ref{fig:lc_epic211355342}.

\begin{figure*}
\epsscale{1.0}
\plotone{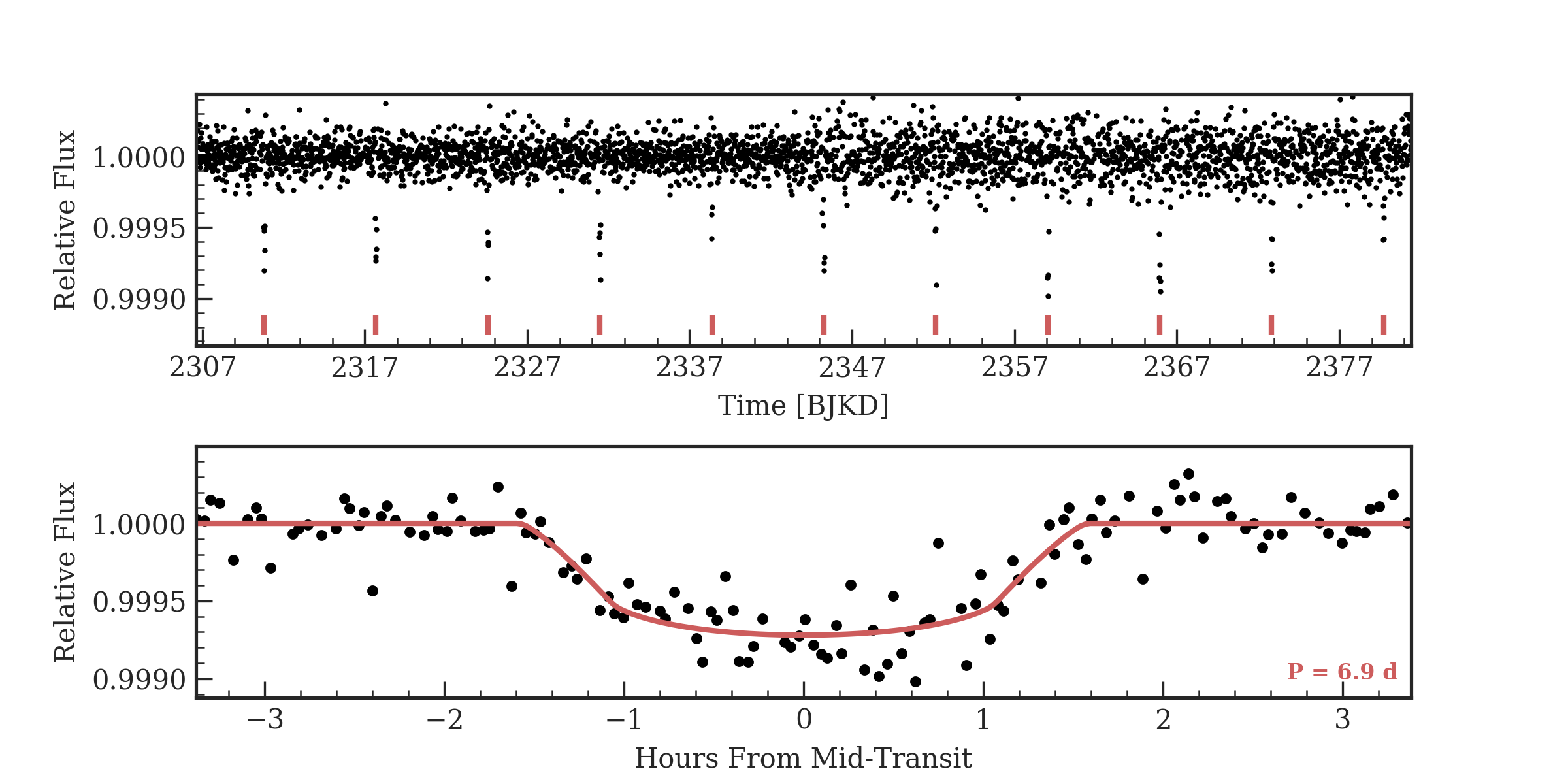}
\caption{Time series (top) and phase-folded (bottom) light curve for the planet orbiting \fdecSTNAME.  Plot formatting is the same as in Fig.\ \ref{fig:lc_epic220709978}.}
\label{fig:lc_epic211355342}
\end{figure*}

We acquired \fdecNOBSHIRES RVs of \fdecSTNAME with HIRES, typically with an exposure meter setting of 50,000 counts.  
We modeled the system as a single planet in a circular orbit with the orbital period and phase fixed to the transit ephemeris.  The results of this analysis are listed in Table \ref{tab:epic211355342} and the best-fit model is shown in Fig.\ \ref{fig:rvs_k2-181}.
We detected the Doppler signal at the 1-$\sigma$ level, which is insufficient to place meaningful constraints on the density or composition. This system would benefit from continued RV monitoring.

\begin{deluxetable}{lrrr}
\tablecolumns{4}
\tabletypesize{\scriptsize}
\tablecaption{ K2-181 System Parameters \label{tab:epic211355342}}
\tablehead{
  \colhead{Parameter} & 
  \colhead{Credible} & 
  \colhead{Maximum} & 
  \colhead{Units} \\
  \colhead{} & 
  \colhead{Interval} & 
  \colhead{Likelihood} & 
  \colhead{} }
\startdata
\sidehead{\bf{RV Analysis -- MCMC Step Parameters}} 
$P_{b}$ & $\equiv6.8941$ & $\equiv6.8941$ & days \\ 
$T_{\mathrm{conj},b}$ & $\equiv2457143.7948$ & $\equiv2457143.7948$ & BJD$_\mathrm{TBD}$ \\ 
$\sqrt{e}\cos{\omega}_{b}$ & $\equiv0.0$ & $\equiv0.0$ &  \\ 
$\sqrt{e}\sin{\omega}_{b}$ & $\equiv0.0$ & $\equiv0.0$ &  \\ 
$K_{b}$ & $4.1^{+6.3}_{-6.2}$ & $4.1$ & m s$^{-1}$ \\ 
$\gamma_{\rm HIRES}$ & $-0.7\pm4.0$ & $-1.0$ & m s$-1$ \\ 
$\dot{\gamma}$ & $\equiv0.0$ & $\equiv0.0$ & m s$^{-1}$ day$^{-1}$ \\ 
$\ddot{\gamma}$ & $\equiv0.0$ & $\equiv0.0$ & m s$^{-1}$ day$^{-2}$ \\ 
$\sigma_{\rm HIRES}$ & $9^{+5}_{-23}$ & $10$ & $\rm m\ s^{-1}$ \\ 
\hline
\sidehead{\bf{Orbital \& Physical Parameters}}
$P_{b}$ & $\equiv6.8941$ & $\equiv6.8941$ & days \\
$T_{\mathrm{conj},b}$ & $\equiv2457143.7948$ & $\equiv2457143.7948$ & BJD$_\mathrm{TBD}$ \\
$e_{b}$ & $\equiv0.0$ & $\equiv0.0$ &  \\
$\omega_{b}$ & $\equiv0.0$ & $\equiv0.0$ & radians \\
$K_{b}$ & $4.1^{+6.3}_{-6.2}$ & $4.1$ & m s$^{-1}$ \\
$M_b$ & $12\pm18$ & $12$ & $M_\earth$ \\
$R_b/R_*$ & $0.0248^{+0.0020}_{-0.0010}$ & $0.0250$ &  \\
$\rho_b$ & $3^{+5.3}_{-4.4}$ & $2.8$ & g cm$^{-3}$ \\
$R_b$ & $2.69^{+0.04}_{-0.25}$ & $2.86$ & $R_\earth$ \\
\hline
\hline
\sidehead{\bf{Priors}}
Parameter & Prior \\
\hline
\\None \\
\enddata
\end{deluxetable}

\begin{figure}
\epsscale{1.0}
\plotone{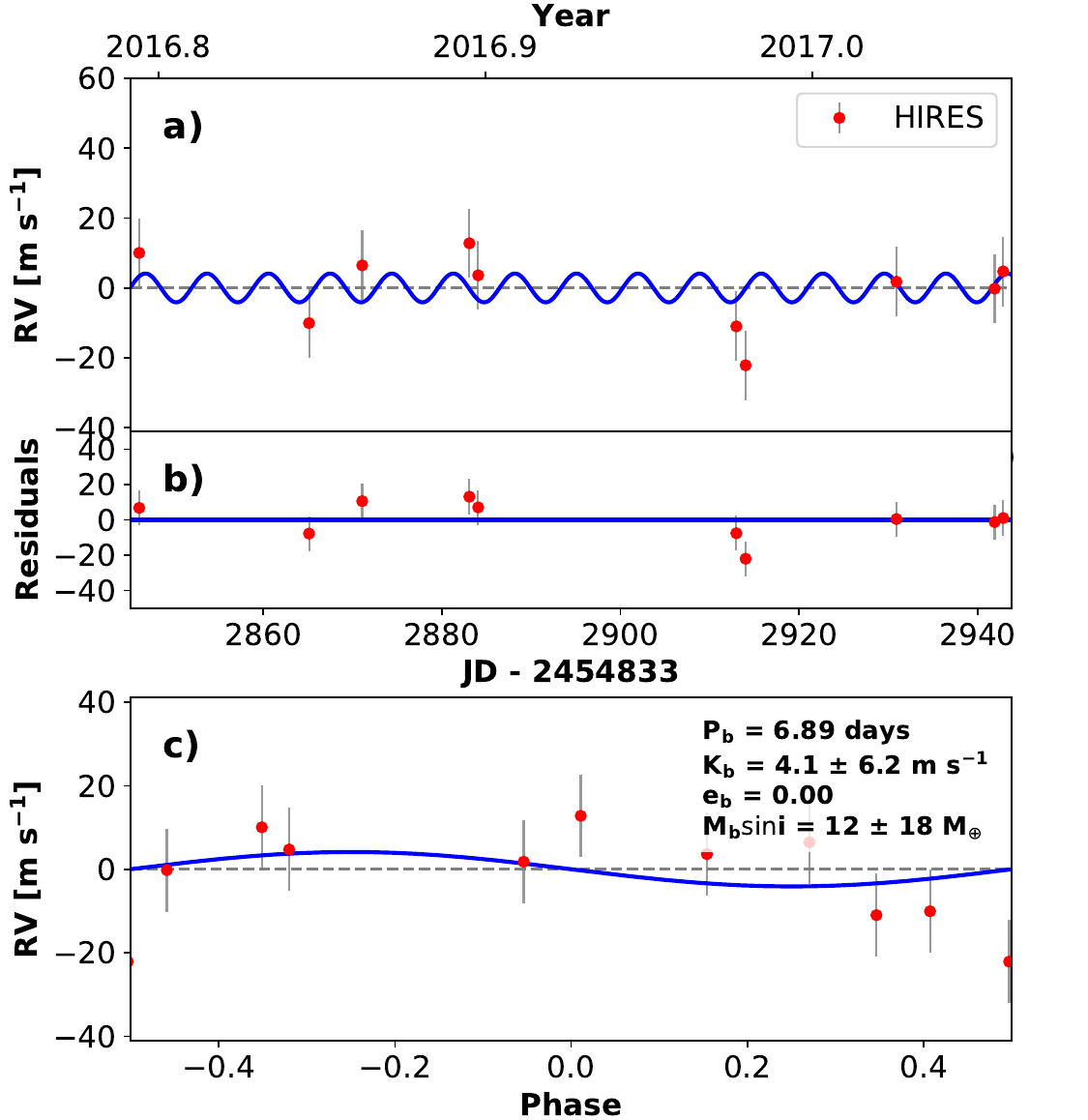}
\caption{RVs and Keplerian model for \fdecSTNAME. Symbols, lines, and annotations are similar to those in Fig.\ \ref{fig:rvs_epic220709978}.}
\label{fig:rvs_k2-181}
\end{figure}

\subsection{EPIC 245943455} 


\deffSTNAME is a G-dwarf from Field 12 with one transiting planet with a 4.1 \rearth radius and 6.3 day orbital period.
See Tables \ref{tb:star_pars}  and \ref{tb:star_props} for stellar properties and Table \ref{tb:planet_props} for precise planet parameters.
Our fit of the EVEREST light curve of the K2 photometry for \deffSTNAME is shown in Fig.\ \ref{fig:lc_epic245943455}. 
\cite{Dattilo2019} classify this object as a planet candidate.  Our observations described below are insufficient  {\referee to confirm the planet but we do successfully rule out massive eclipsing binary false-positive scenarios, increasing the likelihood that this signal is planetary in origin.}

\begin{figure*}
\epsscale{1.0}
\plotone{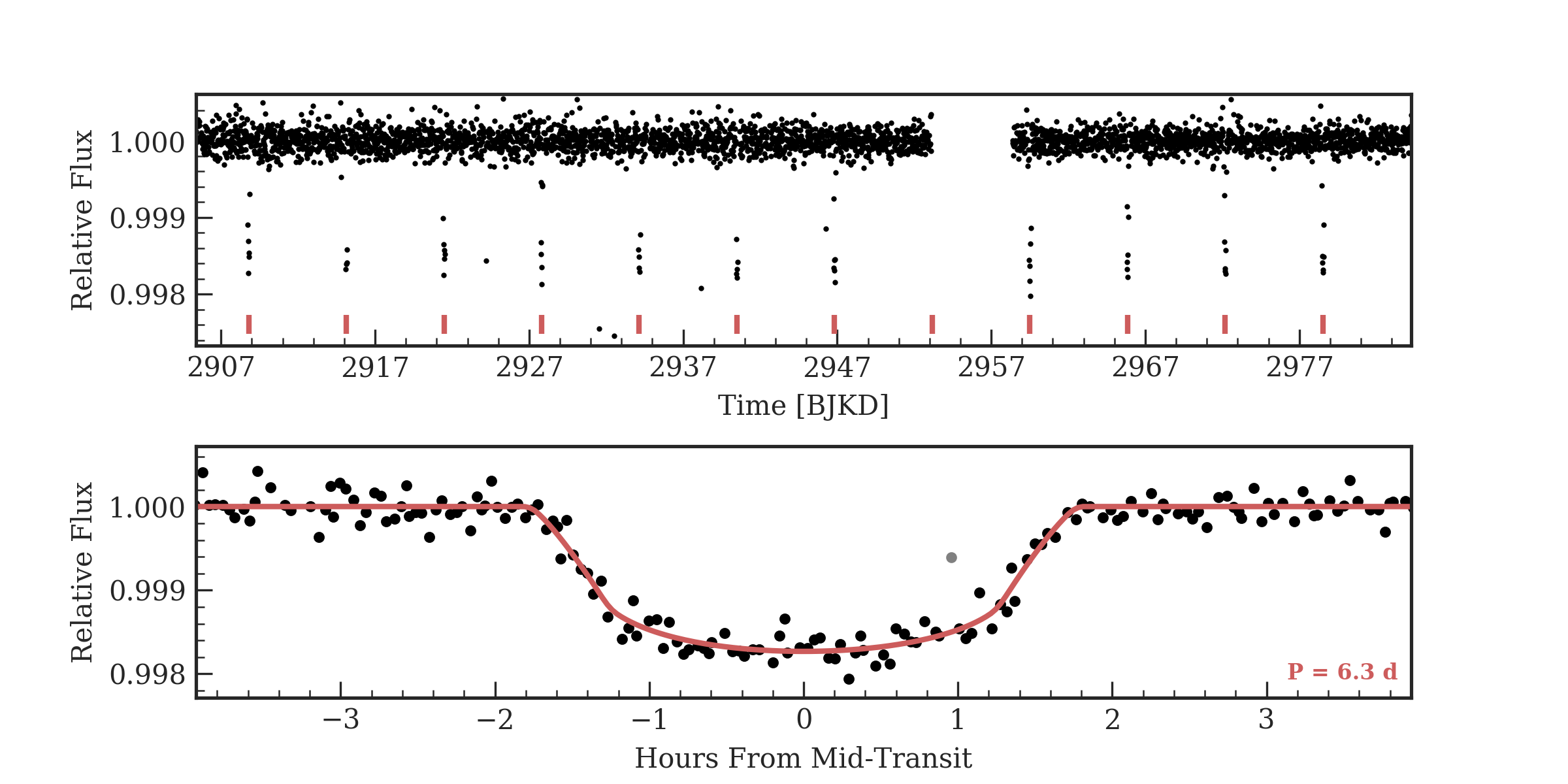}
\caption{Time series (top) and phase-folded (bottom) light curve for the planet orbiting \deffSTNAME.  Plot formatting is the same as in Fig.\ \ref{fig:lc_epic220709978}.}
\label{fig:lc_epic245943455}
\end{figure*}

We acquired \deffNOBSHIRES RVs of \deffSTNAME with HIRES, typically with an exposure meter setting of 50,000.  
We modeled the system as a single planet in a circular orbit with an orbital period and phase fixed to the transit ephemerides.  The results of this analysis are listed in Table \ref{tab:epic245943455}
and the best-fit model is shown in Fig.\ \ref{fig:rvs_epic245943455}.  
Including a linear trend, curvature, or non-zero planet eccentricity is not warranted due to the small number of measurements and the $\Delta$AICc between models. 
The RVs are few in number and cluster around the lower quadrature, making a definitive characterization of \deffPNAMEone\ difficult.  However, we can rule out high bulk densities.

\import{}{epic245943455_circ_priors+params.tex}

\begin{figure}
\epsscale{1.0}
\plotone{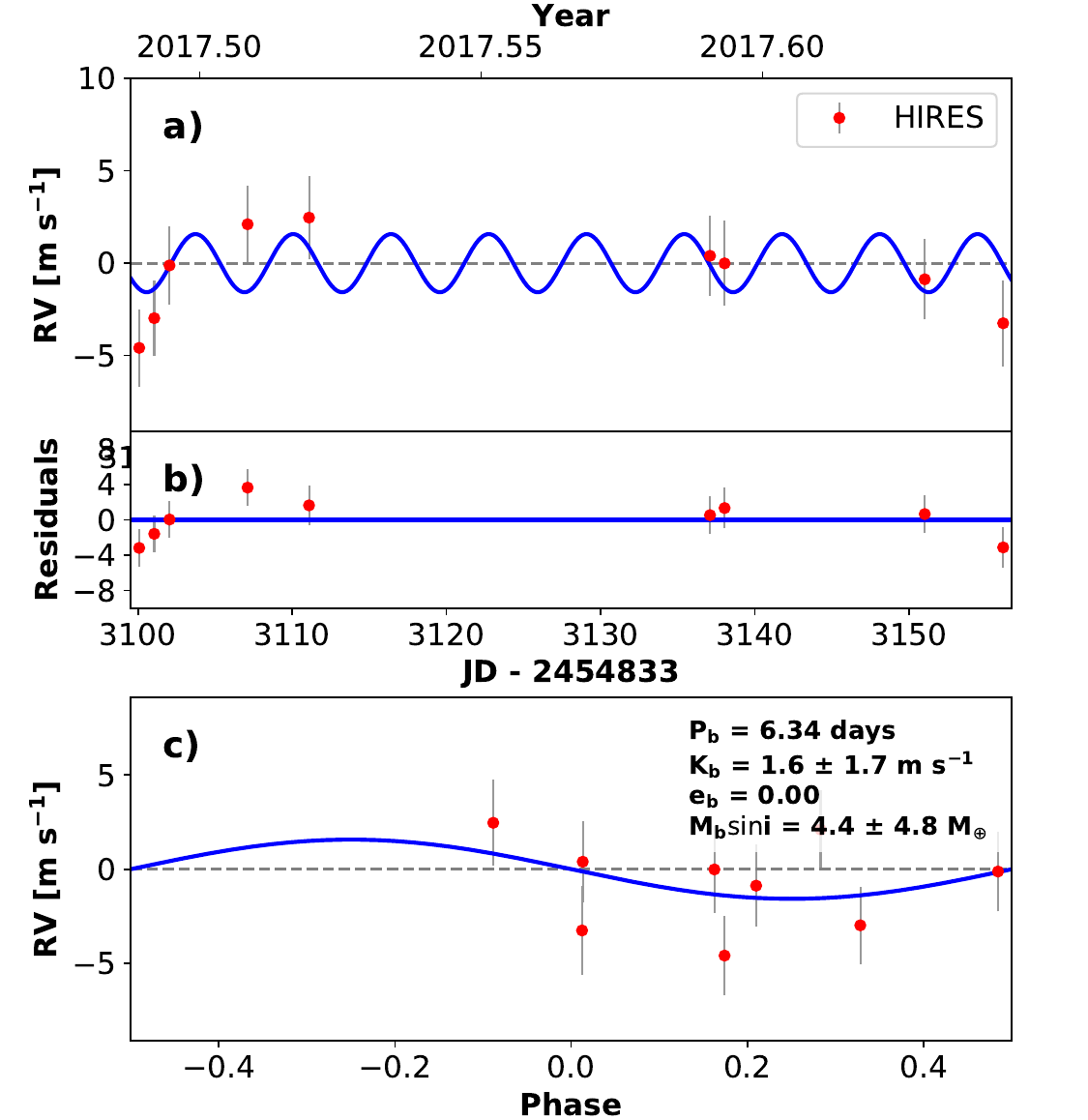}
\caption{RVs and Keplerian model for \deffSTNAME. Symbols, lines, and annotations are similar to those in Fig.\ \ref{fig:rvs_epic220709978}.}
\label{fig:rvs_epic245943455}
\end{figure}

\subsection{K2-61} 


\eiadSTNAME is a G star with properties similar to the Sun from Campaign 3.  The star has one transiting planet with a radius of 1.9 \rearth and an orbital period of 2.5 days.
See Tables \ref{tb:star_pars}  and \ref{tb:star_props} for stellar properties and Table \ref{tb:planet_props} for precise planet parameters.
The planet was discovered and validated in four catalog papers: \cite{Crossfield2016, Vanderburg2016-catalog, Barros2016, Mayo2018}.  Our fit of the EVEREST light curve of the K2 photometry for \eiadSTNAME 
is shown in Fig.\ \ref{fig:lc_epic206044803}.

\begin{figure*}
\epsscale{1.0}
\plotone{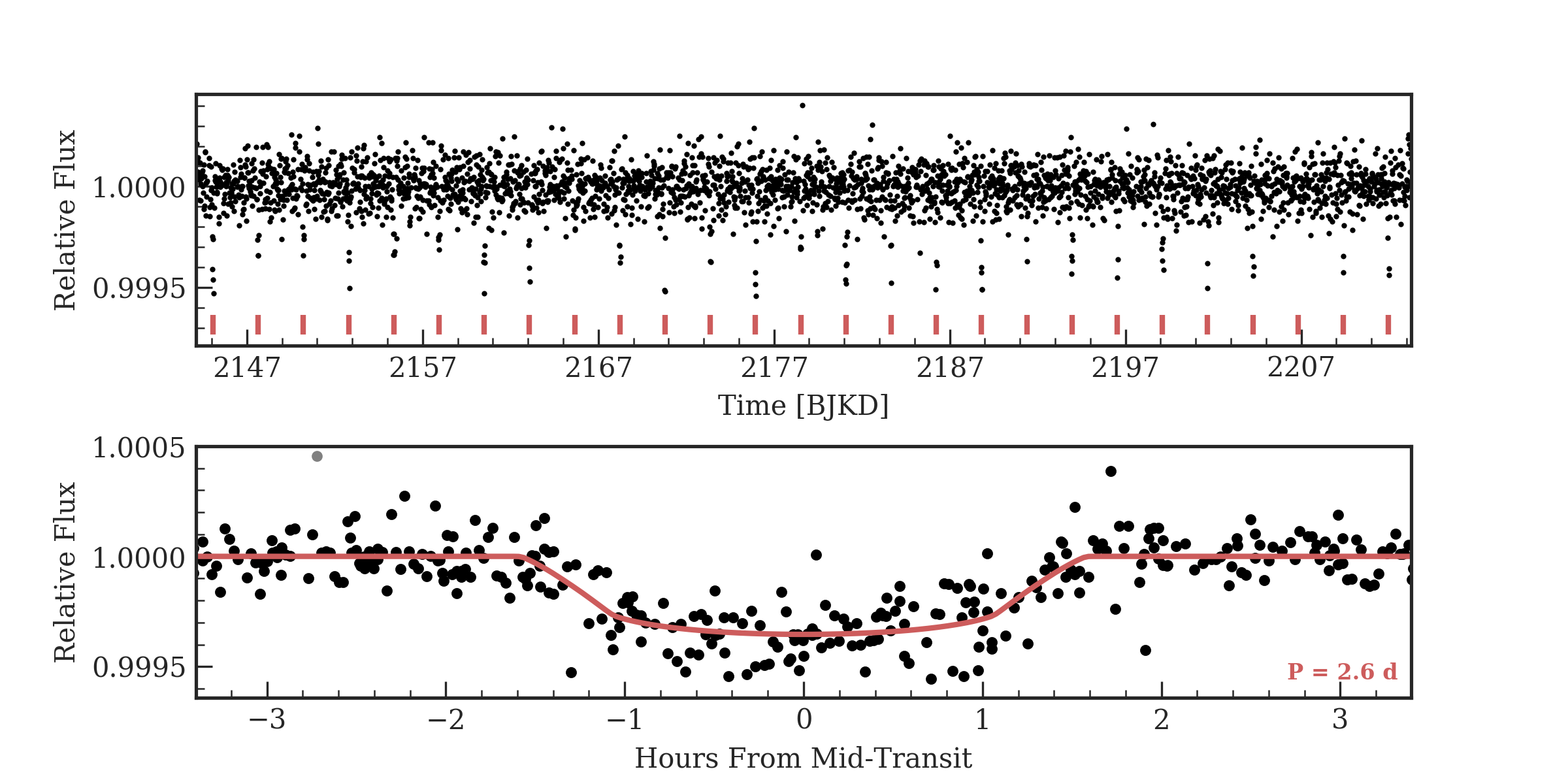}
\caption{Time series (top) and phase-folded (bottom) light curve for the planet orbiting \eiadSTNAME.  Plot formatting is the same as in Fig.\ \ref{fig:lc_epic220709978}.}
\label{fig:lc_epic206044803}
\end{figure*}

We acquired \eiadNOBSHIRES RVs of \eiadSTNAME with HIRES, typically with an exposure meter setting of 50,000 counts.  We modeled the system as a single planet in a circular orbit with the orbital period and phase fixed to the transit ephemeris.  The results of this analysis are listed in Table \ref{tab:epic206044803}
and the best-fit model is shown in Fig.\ \ref{fig:rvs_k2-61}.  This small number of RVs provides only weak constraints, but rules out very high planet masses.  


\import{}{epic206044803_circ_priors+params.tex}

\begin{figure}
\epsscale{1.0}
\plotone{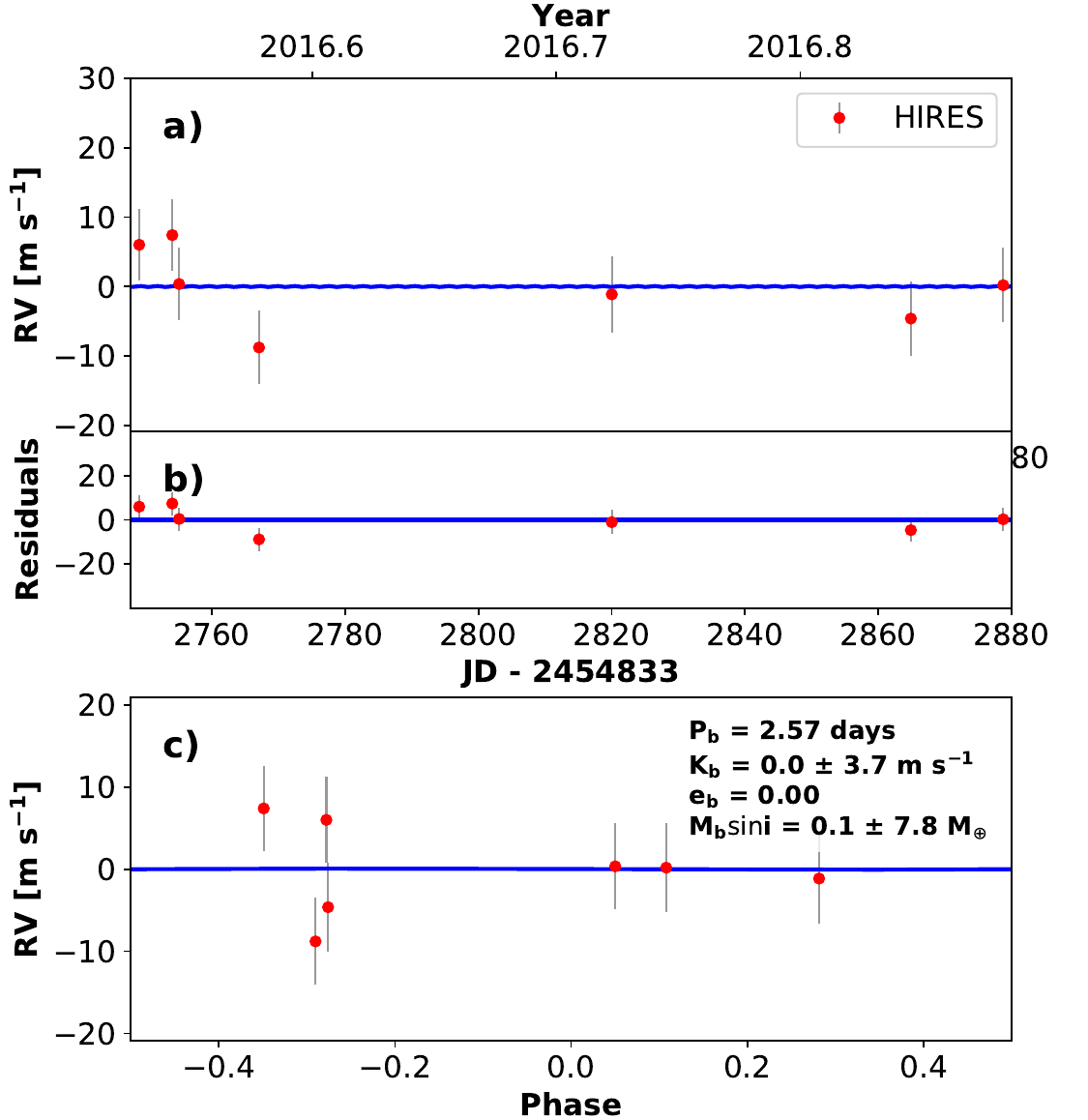}
\caption{RVs and Keplerian model for \eiadSTNAME. Symbols, lines, and annotations are similar to those in Fig.\ \ref{fig:rvs_epic220709978}.}
\label{fig:rvs_k2-61}
\end{figure}

\subsection{K2-121} 
\label{sec:k2_121}

\ifgjSTNAME is an active (\lrphk = \ifgjRPHK) K dwarf from Campaign 5 with one transiting planet with an orbital period of 5 days and a radius of 7.7 \rearth.
See Tables \ref{tb:star_pars}  and \ref{tb:star_props} for stellar properties and Table \ref{tb:planet_props} for precise planet parameters. The star was characterized by \cite{Dressing2017}.  The planet appeared as a candidate in the catalog of \cite{Barros2016} and was subsequently validated in the \cite{Petigura2018} and \cite{Mayo2018} catalogs.  Our fit of the EVEREST light curve of the K2 photometry for \ifgjSTNAME is shown in Fig.\ \ref{fig:lc_epic211818569}.

\begin{figure*}
\epsscale{1.0}
\plotone{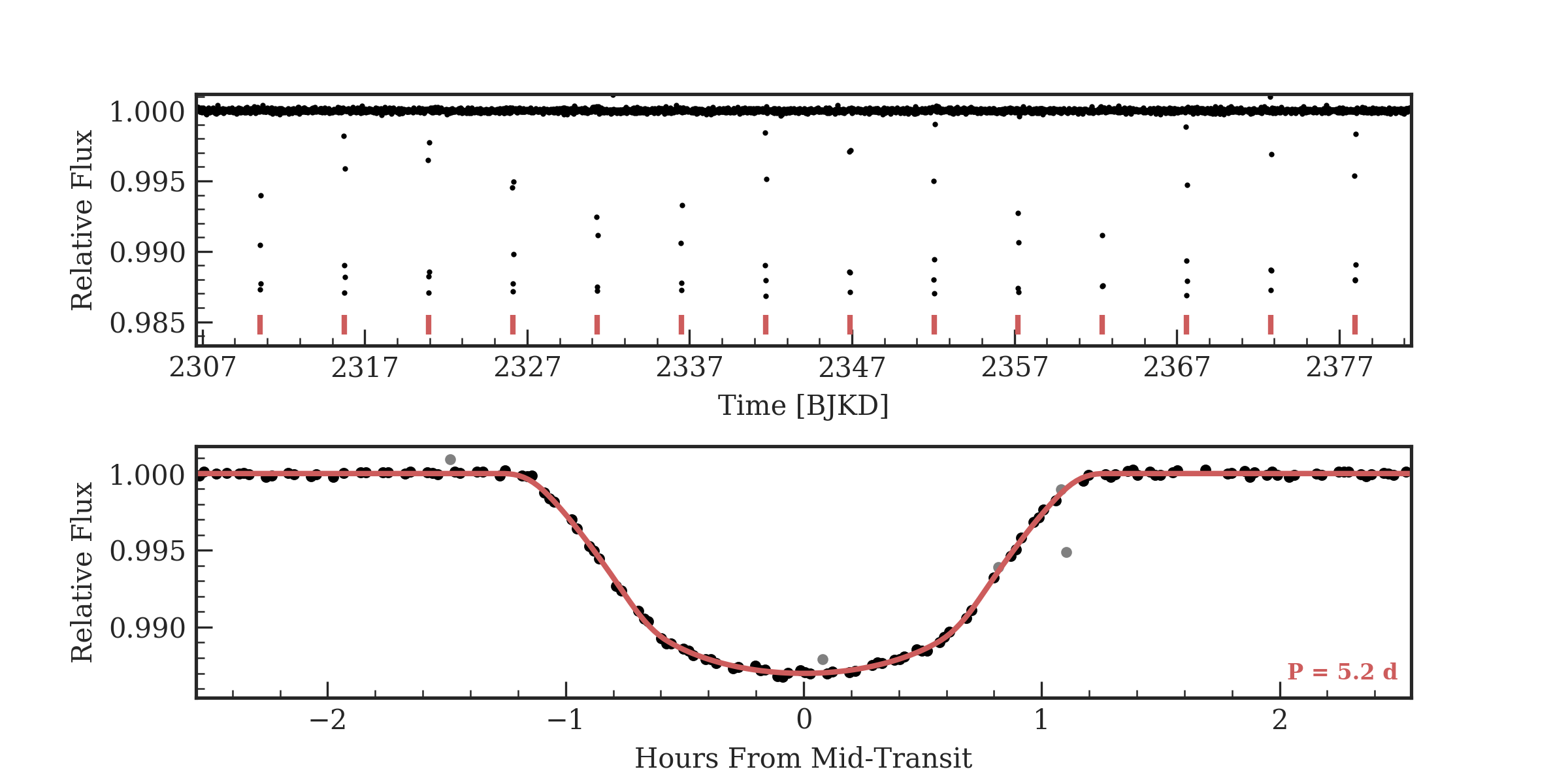}
\caption{Time series (top) and phase-folded (bottom) light curve for the planet orbiting \ifgjSTNAME.  Plot formatting is the same as in Fig.\ \ref{fig:lc_epic220709978}.}
\label{fig:lc_epic211818569}
\end{figure*}

We acquired \ifgjNOBSHIRES RVs of \ifgjSTNAME with HIRES, typically with an exposure meter setting of 40,000 counts.  
We modeled the system as a single planet in a circular orbit with the orbital period and phase fixed to the transit ephemeris.  The results of this analysis are listed in Table \ref{tab:epic211818569} and the best-fit model is shown in Fig.\ \ref{fig:rvs_k2-121}. A model with eccentricity is only slightly disfavored with $\Delta$AICc of just 1.46, but this may be from incomplete phase sampling.
We find a high stellar jitter for this star (Table \ref{tab:epic211818569}), likely due to the elevated chromospheric activity.  We did not apply a Gaussian process model because of the small number of RVs.  \ifgjPNAMEone is a giant planet with Saturn-like radius and density.

\import{}{epic211818569_circ_priors+params.tex}

\begin{figure}
\epsscale{1.0}
\plotone{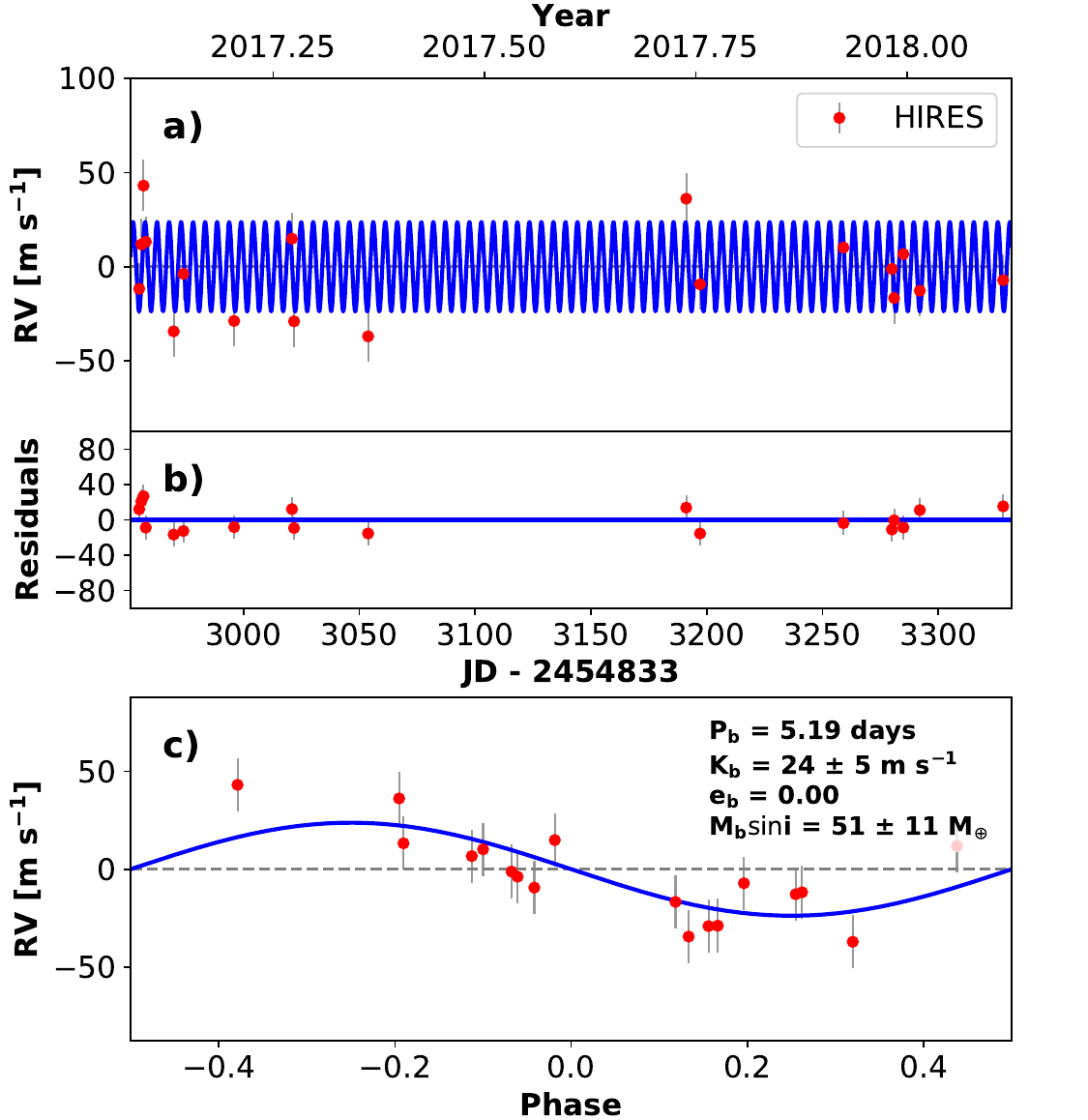}
\caption{RVs and Keplerian model for \ifgjSTNAME.  Symbols, lines, and annotations are similar to those in Fig.\ \ref{fig:rvs_epic220709978}.}
\label{fig:rvs_k2-121}
\end{figure}

\subsection{K2-18} 


\cffcSTNAME was an exciting early K2 system from Campaign 1 because the M dwarf hosts a small transiting planet in or near the habitable zone, opening the door to possible characterization by transmission spectroscopy.  The planet has a radius of 2.4 \rearth with an orbital period of 33 days.  See Tables \ref{tb:star_pars}  and \ref{tb:star_props} for stellar properties and Table \ref{tb:planet_props} for precise planet parameters.  
\cffcPNAMEone has been studied by \cite{Montet2015}, \cite{Foreman-Mackey2015}, \cite{Crossfield2016},  \cite{Vanderburg2016-catalog}, \cite{Schmitt2016}, \cite{Barros2016}, \cite{Wittenmyer2018}. 
Precise stellar properties of this star were determined by \cite{Martinez2017} and \cite{Dressing2019}. 

The original detection of \cffcPNAMEone was based on only two transits, leaving the transit ephemeris so uncertain that detailed characterization in the era of JWST would have been difficult because transit times with uncertainties of many hours to days.  \cite{Benneke2017} improved the ephemeris by an order of magnitude by observing a single transit with \textit{Spitzer} and removing an outlier from the analysis of the original K2 photometry.  

The first mass measurement was performed by \cite{Cloutier2017} using 75 HARPS RVs.  They found a mass for \cffcPNAMEone of $8.0 \pm 1.9$ \mearth and a density of $3.3 \pm 1.2$ \gmc, corresponding to a predominantly rocky planet with a significant gaseous envelope or an ocean planet with a water mass fraction $>$50\%. Their model also included a GP for stellar activity and favors a second, nontransiting planet with an 8.9 day orbital period and a mass of  $m_c \sin i = 7.5 \pm 1.3$ \mearth.

The possible second non-transiting planet was considered by \cite{Sarkis2018} based on CARMENES RVs.  They found that the signal varies in time and wavelength, and therefore interpreted it as being due to stellar activity. Their analysis found that the mass of planet b is $m_b$ = $8.4 \pm 1.4$ \mearth.

The system was revisited by \cite{Cloutier2019} with additional HARPS data. They investigated the effects of time-sampling and determined that the second signal is likely planetary. The revised radial velocity analysis gave planet masses of $m_b$ = $8.6 \pm 1.4$ \mearth and $m_c \sin i$ = $5.6 \pm 0.8$ \mearth.  Most recently, \cite{Radica2022} reported a mass and minimum mass for planets b and c of $9.5 \pm 1.7$  and $6.9 \pm 1.0 M_\oplus$, respectively.

Further observations characterized the atmosphere of \cffcPNAMEone with the Hubble Space Telescope. \cite{Tsiaras2019} and \cite{Benneke2019} detected spectroscopic modulation in the planet's transmission spectrum. Initially interpreted as H$_2$O, subsequent {\em JWST} transmission spectroscopy revealed a curious atmospheric composition dominated by CH$_4$ and CO$_2$ \citep{Madhusudhan2023}. \cite{Madhusudhan2020} used the observed bulk and atmospheric properties of \cffcPNAMEone to constrain the interior structure. They investigated the mass fraction of H/He given different core compositions and determined that there are a wide range of acceptable compositions, ranging from an iron core with 6\% H/He and 0.4\% water by mass to a majority water planet with a minimal H$_2$ rich atmosphere. Additionally, \cite{dosSantos2020} constrained the atmospheric escape of \cffcPNAMEone through Lyman-$\alpha$ measurements and determined that the inferred rate allows for a volatile-rich atmosphere throughout its lifetime. 


\begin{figure*}
\epsscale{1.0}
\plotone{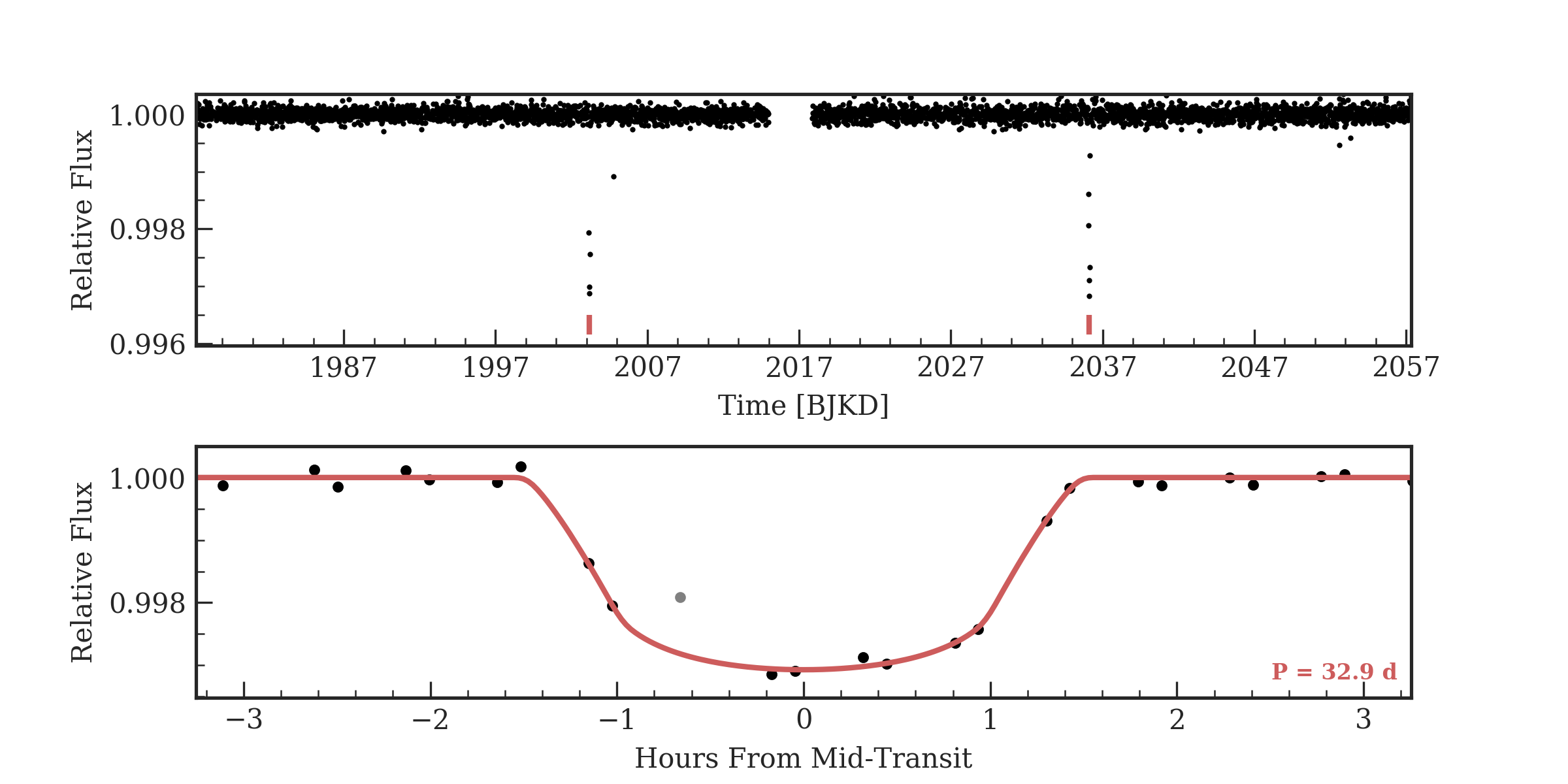}
\caption{Time series (top) and phase-folded (bottom) light curve for the planet orbiting \cffcSTNAME.  Plot formatting is the same as in Fig.\ \ref{fig:lc_epic220709978}.}
\label{fig:lc_epic201912552}
\end{figure*}

Our fit of the EVEREST light curve of the K2 photometry for \cffcSTNAME is shown in Fig.\ \ref{fig:lc_epic201912552}.
We acquired \cffcNOBSHIRES RVs of \cffcSTNAME with HIRES, typically with an exposure meter setting of 40,000 counts.  
We modeled the planet in a circular orbit with an orbital period and phase fixed to the transit ephemeris.  Our fit includes both the HARPS and HIRES RVs along with a GP for stellar activity and the non-transiting planet found by \cite{Cloutier2017}. The results are listed in Table \ref{tab:epic201912552}
and the best-fit model is shown in Fig.\ \ref{fig:rvs_k2-18}.  \cffcPNAMEone has a mass of 7.2 \mearth and density of 2.6 \gmc. Our results are consistent with those of previous studies. 


\begin{deluxetable}{lrrr}
\tablecolumns{4}
\tabletypesize{\scriptsize}
\tablecaption{ K2-18 System Parameters \label{tab:epic201912552}}
\tablehead{
  \colhead{Parameter} & 
  \colhead{Credible} & 
  \colhead{Maximum} & 
  \colhead{Units} \\
  \colhead{} & 
  \colhead{Interval} & 
  \colhead{Likelihood} & 
  \colhead{} }
\startdata
\sidehead{\bf{RV Analysis -- MCMC Step Parameters}} 
$P_{b}$ & $\equiv32.9411$ & $\equiv32.9411$ & days \\ 
$T_{\mathrm{conj},b}$ & $\equiv2456836.1719$ & $\equiv2456836.1719$ & BJD$_\mathrm{TBD}$ \\ 
$\sqrt{e}\cos{\omega}_{b}$ & $\equiv0.0$ & $\equiv0.0$ &  \\ 
$\sqrt{e}\sin{\omega}_{b}$ & $\equiv0.0$ & $\equiv0.0$ &  \\ 
$K_{b}$ & $3.13^{+0.44}_{-0.45}$ & $3.22$ & m s$^{-1}$ \\ 
$P_{c}$ & $9.194^{+0.008}_{-0.007}$ & $9.194$ & days \\ 
$T_{\mathrm{conj},c}$ & $2456836.20^{+0.72}_{-0.75}$ & $2456836.20$ & BJD$_\mathrm{TBD}$ \\ 
$\sqrt{e}\cos{\omega}_{c}$ & $\equiv0.0$ & $\equiv0.0$ &  \\
\enddata
\end{deluxetable}

\begin{figure}
\epsscale{1.0}
\plotone{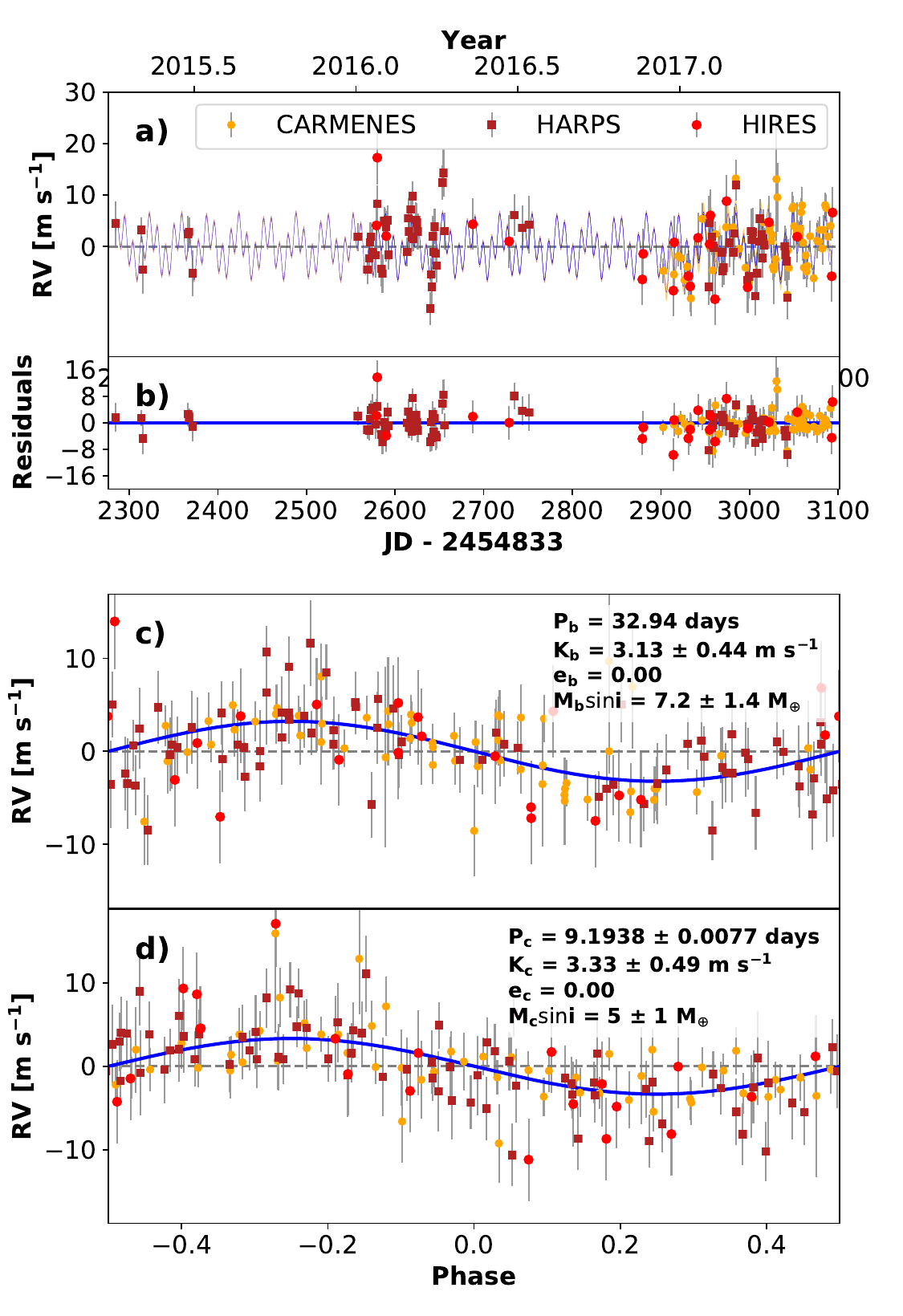}
\caption{RVs and Keplerian model for \cffcSTNAME. Symbols, lines, and annotations are similar to those in Fig.\ \ref{fig:rvs_epic220709978}.}
\label{fig:rvs_k2-18}
\end{figure}


\subsection{K2-55}
\label{sec:k2_55}

K2-55 is a K dwarf from Campaign 3 that hosts one Neptune-sized planet. This planet is cataloged in \cite{Crossfield2016,Vanderburg2016-catalog,Schmitt2016,Barros2016,Martinez2017,Dressing2017,Wittenmyer2018,Kostov2019,Wittenmyer2020}. Precise stellar parameters were determined in \citet{Martinez2017} and \cite{Wittenmyer2020}. 

\citet{Dressing2018} followed up this system with a Spitzer transit and 12 HIRES radial velocity measurements. We adopt their solution; see Tables \ref{tb:star_pars} and \ref{tb:star_props} for stellar properties and Table \ref{tb:planet_props} for the planet parameters. They found that a circular 1-planet model fits the data best, resulting in a planet mass of M$_b$ = 43.13$^{+5.98}_{-5.80}$ \mearth. This Neptune-sized planet is considerably denser (2.8$^{+0.8}_{-0.6}$ g cm$^{-3}$) than Uranus and Neptune in our own solar system, requiring a higher fraction of rocky material. \citet{Dressing2018} calculated that the planet is consistent with a 12\% H/He envelope surrounding a rocky core using grids from \citet{Lopez2014}. 


\subsection{K2-19} 


K2-19 (EPIC 201505350) is a late G dwarf with three transiting planets that have sizes of 7~\rearth, 4~\rearth, and 1.1~\rearth and orbital periods of 7.9 days, 11.9 days, and 2.5 days, respectively. See Tables \ref{tb:star_pars}  and \ref{tb:star_props} for stellar properties and Table \ref{tb:planet_props} for precise planet parameters.

Planets b and c were discovered by \cite{Foreman-Mackey2015}, who listed them as candidates, and were included in the \cite{Montet2015} catalog. \cite{Armstrong2015} independently discovered and validated planets b and c. K2-19 is also included in the catalogs of \cite{Vanderburg2016-catalog}, \cite{Crossfield2016}, \cite{Barros2016} and \cite{Schmitt2016} (Planet Hunters). \cite{Sinukoff2016} later discovered the third transiting planet at 2.5 days (K2-19 d).

K2-19 was the first K2 system to show transit timing variations (TTVs). Using three ground-based transits, \cite{Barros2015} obtained photodynamical masses of $44 \pm 12$ \mearth and $15.7 \pm 7.0$ \mearth.  Their analysis also included 10 RVs from Sophie. \cite{Narita2015} also characterized the system using high-dispersion spectroscopy, AO imaging, and TTVs. \cite{Dai2016} used 61 PFS spectra with 5~\ms uncertainties as well as eight HARPS RVs with 3.8~\ms uncertainties to measure planet masses of $28.5^{+5.4}_{-5.0}$ \mearth, $25.6 \pm 7.1$ \mearth, and $< 14.0$ \mearth (95\% confidence). A later analysis by \cite{Nespral2017} combining RVs from FIES, HARPS-N, and HARPS with TTVs found that including the TTVs resulted in a lower mass estimate than a fit to the RVs alone. Subsequent observations and joint TTV+RV modeling gave a detailed picture~\citep{Malavolta2017}, and as more RVs were acquired the mass discrepancy between the RV and TTV determined masses was relieved~\citep{Borsato2017}. 

We acquired 51 RVs of K2-19 with HIRES, typically with an exposure meter setting of  60,000 counts. We adopted the Keplerian model as described in \citet{Petigura2019}, which included these Keck-HIRES RVs and performed a photodynamical TTV analysis which included additional transit times from Spitzer and the Las Cumbres Observatory. The eccentricity for planets b and c was free, while that {\referee of planet} d was fixed to a circular orbit. From their analysis, the authors derived masses of $32.4 \pm 1.7$, $10.8 \pm 0.6$, and $<10$~\mearth for planets b, c, and d, respectively. The masses and eccentricities were most constrained by the photodynamical analysis, as the RV precision was limited by $\approx 7$~\ms stellar jitter. This system is an intriguing test case for planet formation theories, as planet b has $\sim 50\%$ of its mass in the form of a gaseous envelope, and interestingly while K2-19 b and c are in a 3:2 commensurability, the planets are just 0.1\% out of resonance~\citep{Petigura2019}. 



\subsection{HIP 41378} %
\label{sec:hip41378}


HIP 41378 (K2-93, EPIC 211311380) is a bright late G dwarf observed in Campaigns 5 and 18. HIP 41378 hosts five transiting planets that were discovered and validated by \cite{Vanderburg2016-hip41378} using observations from Campaign 5. The outermost planets had only a single transit measured and were then seen to transit again when revisited in Campaign 18, which reduced the period uncertainty to a set of discrete period solutions \citep{Becker2018, Berardo2019}. The transiting planets (b, c, d, e, f) have orbital periods of 16, 32, 278, 369, and 542 days,
and radii of 2.6, 2.7, 3.5, 4.9, and 9.2 \rearth respectively, adopting most likely solutions for the outer planets based on dynamical considerations.  The outermost planet f does not show clear spectroscopic characteristics in {\em HST}/WFC3 transmission spectroscopy \citep{Alam2022}.

Our adopted stellar parameters are shown in Tables \ref{tb:star_pars}, \ref{tb:star_props} and \ref{tb:planet_props}. We adopted the solution by \cite{Santerne2019}, who jointly modeled the radial velocities from PFS, HARPS, and HARSP-N, as well as 218 nightly HIRES observations from our team binned into 75 epochs, for a total of 464 RV epochs across the four instruments. 
They found masses of $6.89 \pm 0.88$, $4.4 \pm 1.1$, $< 4.6$, $12 \pm 5$, and $12 \pm 3$ \mearth, respectively. The ultra-low density inferred for HIP 41378 f, due to its Saturn-size yet sub-Neptune mass, may be indicative of a small core surrounded by an extended envelope, or even a planetary ring system that deepens the transit \citep{Akinsanmi2020}. 

\cite{Santerne2019} also discovered an additional non-transiting planet at 62 days with a mass of $7.0 \pm 1.5$ \mearth (assuming it is coplanar with the five transiting planets).

\subsection{HD 89345 (K2-234)} 







HD 89345 (K2-234, EPIC 248777106) is a bright ($V$ = 9.4, $K$ = 7.7) G star in Campaign 14. It is a slightly evolved star that exhibits solar-like oscillations.   
\citet{VanEylen2018} and \citet{Yu2018} both discovered one warm sub-Saturn-sized planet (R$_b$ = 6.7 \mearth) in an orbit of 11.8 days. 

\citet{VanEylen2018} measured the mass of planet b (M$_b$ = 35.7 $\pm$ 3.3 \mearth) with 46 RV measurements from a combination of FIES, HARPS, and HARPS-N. They find an eccentric orbit to best fit the data (e$_b$ = 0.2); however note that a circular orbit also reasonably fits the data given the small number of measurements. \citet{Yu2018} measured the mass of planet b (M$_b$ = 0.110 $\pm$ 0.018 M$_{Jup}$) with 12 HIRES and 9 APF radial velocity measurements. They also prefer an eccentric solution (e$_b$ = 0.22), in line with other sub-Saturns \citep[e.g.,][]{Petigura2017}.  

The \citet{Yu2018} paper is based on data from this project and contains all of our HIRES measurements on this system. We refer to the transit fit shown in \citet{Yu2018} and report the planet parameters in Table~\ref{tb:planet_props}. Because of the substantial data published in \citet{VanEylen2018}, we performed an updated radial velocity fit for HD 89345 including all of the data published so far. This fit includes 12 HIRES \citep{Yu2018}, 21 APF \citep[][and this work]{Yu2018}, 16 FIES \citep{VanEylen2018}, 18 HARPS \citep{VanEylen2018}, and 12 HARPS-N measurements \citep{VanEylen2018}.  The orbit of the planet is likely misaligned with its star's rotation axis \citep{Bourrier2023}, and the planet reveals no signs of mass loss via transit spectroscopy of the metastable 1083~nm Helium line \citep{Guilluy2023}.

We modeled the system as a single-planet fit. We test additional parameters including a trend, curvature, and planet eccentricity. We adopt the inclusion of planet eccentricity ($\Delta$AICc=20) and reject the other models based on model comparison using the AICc statistic.  The results of this analysis are listed in Table \ref{tab:epic248777106}
and the best-fit model is shown in Fig. \ref{fig:rvs_epic248777106}.

Our mass measurement is consistent with estimates from both \citet{Yu2018} and \citet{VanEylen2018}. With the inclusion of all RV measurements, we prefer an eccentric fit and find an eccentricity of 0.200$\pm$0.042. 

\begin{figure}
\epsscale{1.0}
\plotone{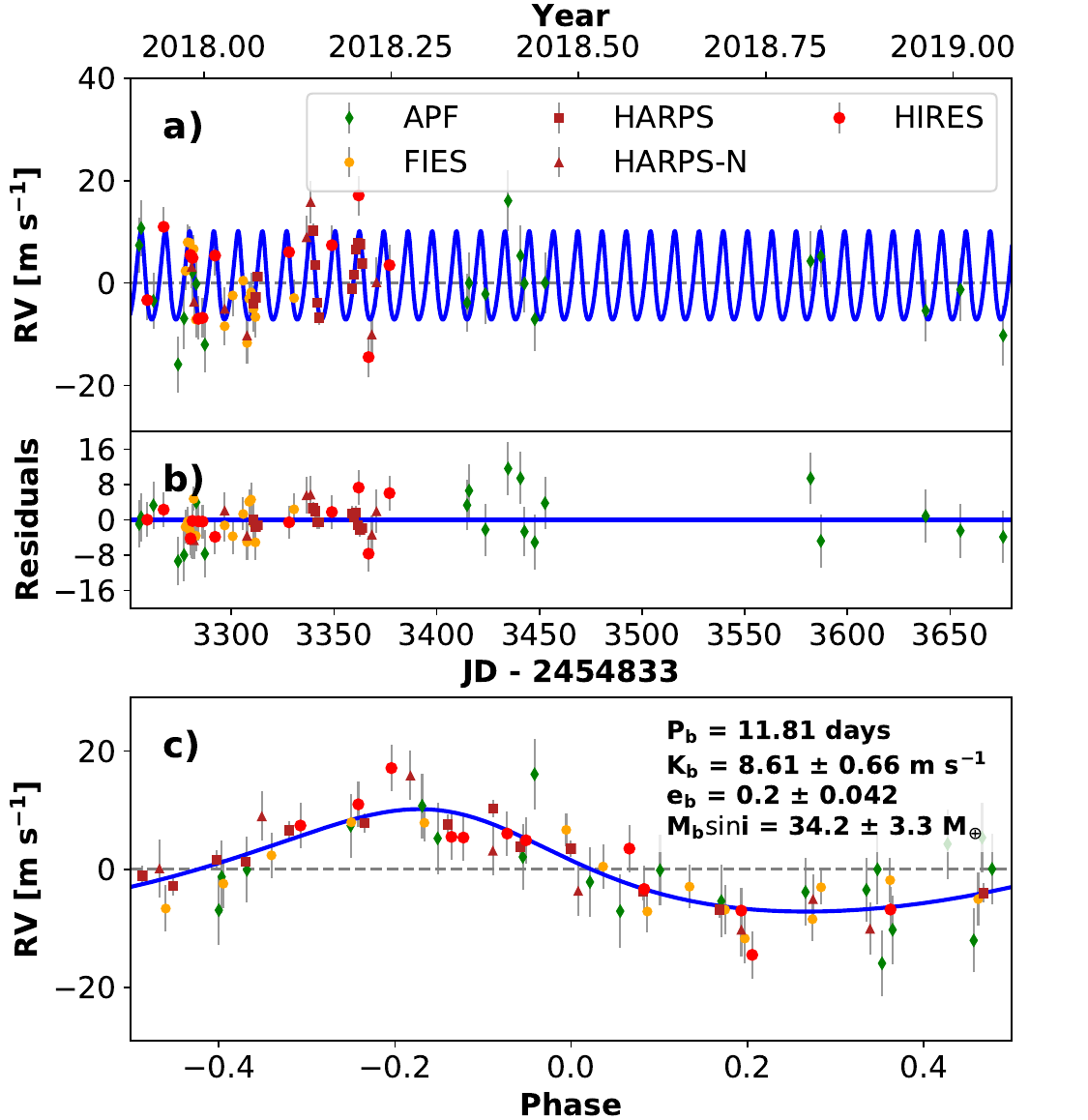}
\caption{RVs and Keplerian model for HD 89345.
Symbols, lines, and annotations are similar to those in Fig.\ \ref{fig:rvs_epic220709978}.}
\label{fig:rvs_epic248777106}
\end{figure}

\import{}{epic248777106_ecc_priors+params.tex}



\end{document}